# myPhysmatics
## Connected Mathematical Models in Physics

Sergey Pankratov

# Preface

"If you don't know where you are going, any road will get you there." Lewis Carroll

For many years, I have conducted a sort of scientific diary in which I put facts and results that I liked at the moment or found interesting and useful. My principal motivation was to learn how clever people construct beautiful mathematical models out of physical garbage - incongruent sets of data, apparently disjoined experimental facts and other multiple raw material. Having looked in hindsight through these notes, I could ascertain that nearly everything in them revolved around a bunch of more or less standard ideas, facts, techniques, and approaches which may be regarded as indispensable elements of the standard education of a physicist. Such elements can be networked to provide an apparently cohesive picture of a fair physical education and a mathematical culture, sufficient for a physicist. At least satisfactory to many of us.

The present book stems from that old diary of mine and to a certain extent retains its disparity. However, the surrounding world seems to be united, which manifests itself in the fact that new horizons are inevitably unfolded and unexpected links between apparently disjoined, isolated models, most of them being physics-based mathematical models, are continuously opened when one attempts to consider various aspects of stand-alone and ad hoc constructs. As a result, the book was not limited to physics alone, but also contains some rudimentary information on the situation in neighboring disciplines. The existence of hidden relationships between seemingly different domains has always been to many people a wonderful and astounding quality of physical and mathematical sciences. One can observe that the art of producing a good work in physics is essentially the art of revealing connections between seemingly disparate manifestations. The same applies to mathematics, sometimes in even larger extent. So, in my diary, I tried to pay attention to the links, analogies and similarities inside the mosaic of fragmentary mathematical models of physics. I rather register and describe those links than offer a rational explanation for the very phenomenon of linking. To illustrate such linking, I can bring a rather obvious example. Although general relativity is regarded by many people as an autonomous subscience, a discipline which is separate from the whole body of physics, I think that is erroneous. The study of general relativity helps to understand classical mechanics much better than while studying mechanics alone although general relativity and classical mechanics traditionally belong to different physical courses and are practically never taught together.

As one more example, one can recall Bohr's atom, which was basically an ad hoc model, but later this naive model has profusely stimulated the emergence of more sophisticated mathematical (quantum) approaches and theories. One can remember that during the conception of quantum mechanics, the model of Bohr's atom, quite unexpectedly, was connected with such distant subjects as the black-body radiation - also originally an ad hoc mathematical model, with the theory of adiabatic invariants, which is very close to the modern theory of dynamical systems, and with the spectral theory of linear operators. This and some other cross-disciplinary links will be described in the book where appropriate. For myself, long time ago, I called such unification of seemingly disjoint results a "physmatical effect" - pointing to the fact that multitudes of fine-grained physics-based mathematical models (PBMM) become inextricably linked and networked, and I called the corresponding network a physmatical one. Mathematics serves as a key code gluing together isolated physical models viewed as nodes of this network. Widely disparate and independently developed, subjects from physics and mathematics converge unexpectedly to become unified ingredients of a single theme. This is similar to a polyphonic construction of music, when a diversity of tunes combine to produce an interesting melody.

The corresponding discipline that is focused on ties and lateral associations between mathematical models in physics may, in this terminology, be called "physmatics" - in distinction to "mathphysics", or mathematical physics, which has traditionally been a well-structured discipline centered around partial differential equations. A considerable portion of mathematical physics is devoted to attempts of rigorous scrutiny of a selected bunch of mathematical models that involve mathematical aspects of the proof of these models' freedom from inconsistencies. In particular, proofs of existence and uniqueness of solutions, analysis of the properties of the chosen system of equations, construction of exact or self-similar solutions plays the major part in mathematical physics; nowadays a trend towards axiomatic constructs is more and more obvious. Although extremely important and comprising an indispensable element of any physicist's background, mathematical physics appears to be a realm of mathematicians, rather than physicists.

In contrast, "physmatics" treats mathematics as a service subject, a set of protocols carrying universal research techniques between physical models, the latter may be regarded as networked "nodes". Scientific journals (and, lately, servers) play the role of hubs, routers and, sometimes, switches in this network redirecting the research efforts. Here "switch" is a component of a cognitive network that connects two or more lines of thinking to complete a task or a model. Physmatics is a more or less a closed, "private" network, in the sense that it does not so far include social sciences or medicine - these possess their own cognitive networks. Focusing on common features provides exciting observations and deepens understanding of such models, like accidental recognition of common acquaintances stimulates mutual interest and fastens friendship ties. I think that people are better inclined to recognition than to hard-core *ab ovo* learning. The more unexpected the liaison, the deeper is the appreciation.

I concede of course that writing just about everything is a firm symptom of unprofessionalism: physicists are commonly encouraged to generate very focused articles. Today, however, in contrast with the 19th and the beginning of the 20th century, probably most fruitful period in physics, physicists too often explore minute problems. But, honestly speaking, papers such as - this is my fantasy, of course – "On the third correction to the second off-diagonal matrix element in the quasibound $X$-like to $\Gamma$-like states transition within the eight-band approximation in III-V heterojunctions" leave me unmoved, although such partial results form the very texture of physics and their authors should be utterly respected. Furthermore, the papers of the "Influence of Singing on Seeing"-type[1] serve as a publication multiplier, and the number of publications is, in practice, considered as the main indicator of success in science (although quite the opposite may be true). One can rapidly produce dissertations, which actually happens. In relation to this, I often recall J. W. Gibbs who may be considered an antipode to the current breed of prolific scientists. Since in 1970s I translated thermodynamical works by Gibbs into Russian, I had to study his scientific legacy and found out that he, being really a great figure, was quite reluctant to publishing his papers, at least in a rush. The American "publish or perish" approach seems to me a noisy impasse, more suitable to journalists than to scientists. It would be a mere truism to observe that rapid multiplication of scientific (and near-scientific) journals and other publications makes the entire physmatical network increasingly complex and noisy. This multiplication process is analogous to the unfavorable development of the Internet which can eventually result in its catastrophic transformation (e.g., split).

Furthermore, physmatics is constructed in such a way that one can, in principle, starting from each particular result (a physmatical node, in this terminology), reach any other point, even conceptually quite distant, say, one can travel from nanotechnology to topology. However, the time needed to drift from one physmatical concept (node) to another may well exceed the duration of human life. And what to do if so many things in physics and mathematics are equally interesting that it is one's desire to grasp them all? Then the networking agenda, with many routes between the concepts of interest would be the only solution. After all, intelligence is primarily the skill to interrelate seemingly different, heterogeneous things.

The great Russian theoretical physicist L. D. Landau considered theoretical physics a "small science", he used to say that a physicist can understand the whole of it. On the contrary, experimental physics, according to Landau, is a "big science", a single person is unable to know all its parts [51]. The famous "Course of Theoretical Physics" by L. D. Landau and E. M. Lifshitz was based on this idea - a physicist can understand the whole physics, no matter how far from each other its specific models might appear to be. But that was long ago. Today, unfortunately, physics and mathematics more and more remind us of the Tower of Babel, and the only efficient method to cope with such an overwhelming flood of information is to employ the networking approach (similar to hypertext or hyper-reference used in modern online encyclopedia). To a physicist trained in the late 20th century, theoretical physics constructed around the least action principle serves as a backbone for the entire physmatical network. To me personally, "physmatics" reminds me of a hybrid language[2] capable of activating the nodes of networked physical and mathematical knowledge. Network-oriented languages aimed at the analysis and depiction of data flows and network topologies have been known in networking technology for some time. Such languages enable us to see structures and links that may remain hidden at first glance. Creation of network-oriented languages is usually an interdisciplinary undertaking that merges network analysis, complexity theory, graph theory, communication theory and other disciplines elucidating the laws according to which new ideas, results, technologies, etc. are propagated. Networks are ubiquitous: just think of everyday networks like public transportation, communication, utilities, electrical engineering, etc. Physmatics represents a class of other, invisible networks that become observable when their objects are seen in relationship to each other. This all sounds unnecessarily abstract, but can be made transparent by simple examples. In the oscillator example, photons, phonons and many other quasiparticles are mostly described in the universal language of creation and annihilation operators, which is just an interpretation stemming from the fact that the energy spectrum of an oscillator is equidistant and can be obtained algebraically by representing the Hamiltonian through the product of ubiquitous raising and lowering operators. One might say that the oscillator model "radiates" its messages throughout the physmatical network and, in principle, receives feedback. Analogous exchange of messages can be observed, for instance, in nonlinear differential equations and dynamical systems, in geometrical properties of gauge models, in physics-based models beyond physics. Probably, in the near future, an algorithm may be built to produce a series of hierarchical - coarser - networks by searching highly dense subnets (or subgraphs) in each level of the entire network, and then a clustering multilevel algorithm can be applied. This networking approach would enable us to carry out a structured analysis of nowadays unweighted physical and mathematical networks. The networking paradigm fuels my conviction that physics and mathematics are a non-stop issue, with new nodes being constantly incorporated.

---

[1] In Russian, "Vliyanie peniya na zrenie".

Such an outlook is, of course, closer to expressing a dream than setting up a concrete problem[3]. My primary goal in this manuscript is much more modest - I was trying to make the book a good read as well as a good reference to some interesting facts thoroughly covered in the available literature. I also tried to combine attention to detail with personal recollections and the occasional anecdote (often in order to make my points), rather than the purely scholarly approach. The diary form implies by default a compendium of many topics, with not necessarily everything really derived. Thus, the book may also appeal to people who wish to "comprehend" numerous fields of physics without doing much work.

---

[2] Hybrid languages exist not only as artificial programming or document-oriented formal languages (with some distributed data model that allows addressing, searching and linking of content from documents), but also among human dialects. For instance, Yiddish, with its ex-territoriality and massive inclusion of Slavic elements, even sovietisms, is a typical hybrid language destined to consolidate a network of disparate enclaves.

[3] General features of representing science as a network or a map are discussed in the well-known book by John Ziman "Reliable Knowledge" [109], ch.4.

There is an obvious gap between the expert knowledge space and the educational one. The number of scientific papers grows so rapidly that the traditional methods of information dissemination (via journal papers, monographs, textbooks) become inefficient: even the specialists in narrow fields find it difficult to focus on the principal issues since there may be a lot of "noise" totally obscuring useful information. The Internet search, though very useful in skillful hands, often only aggravates the problem. Everyone is familiar with the difficulty of finding a precise piece of information, and the novices are completely lost in it. Strict monographs and detailed review papers provide little help: the monographs are usually too rigorous so that the reader must invest a lot of time to get to the essence, and rare overview articles are unbiased. The latter fact is natural: most of the authors extensively refer to their own work. Besides, people who are most active in the research seem to be reluctant to write long and detailed reviews about their field in general. Under these circumstances, it is likely that some free-style manuscripts might be required. Such free-style books would occupy an intermediate position between science[4] popularization and real monographs. The free-style genre provides an opportunity for the author to convey her/his point of view whereas for the reader it may serve as a means to fill the gap between the common university courses and the highbrow scientific monographs or research papers based on special mathematical techniques. The present text is an attempt to quilt such a free-style book. And I repeat, I am aware that loose reflections are perceived as a grave professional disadvantage as well as extensively using "I" instead of impersonal "we". This "I-ing" i.e., an obvious retreat from the academic style is neither puerilism nor exhibitionism - it is rather an expression of personal opinion. So, the book may appear heavily opinionated, which is quite natural for a diary.

Putting together seemingly diverse subjects takes time of course, so this book has been written in steps reflecting different levels of understanding. Some parts of the book are intended also for the people more attracted by arts, literature, and social studies than by science saturated with hard math. The obvious problem is that such people usually experience difficulties with mathematics. The book partially aims to reinforce basic mathematical knowledge, in particular, by favoring and interpreting mathematical models as well as by trying to criticize some of them. Mathematics viewed as an intellectual reality can be made instrumental not only in physics, but also in social life. As far as the scientific component of the manuscript goes (there are also unscientific ones), only the most basic models are discussed in this book. The style of the text - slow, ruminating, recurrent, sometimes light - reflects the tendency to study things while writing. Some readers might designate a number of fragments in the text as "philology" or as "pouring from void into empty". I, however, think this impression is deceptive. There exist at least three good reasons to ruminate about foundations of physics: firstly, by scrutinizing the fundamental ideas once again, one can learn a lot from ingenious individuals who originated them; secondly, one can reformat individual thought templates making them less susceptible to collective mantras; and thirdly, pondering over possible mathematical formulations of physical basics can catalyze the study of mathematics. We shall see later in this book that many valuable mathematical ideas have come through contemplating over various possibilities of exploring basic laws of physics. As an illustration one might recall that any successful physical theory has historically served as a great stimulus and rich source of ideas primarily for mathematicians. This was an exchange with instruments of understanding. Furthermore, we shall parse some customary notions; re-examining conceptual systems seems to be useful because otherwise attitudes pass ahead of knowledge.

---

[4] For definiteness, I shall only speak about physics.

Beside "philology", one may encounter some inconsistencies - often deliberate, as, e.g., in coordinate transformations - in the notation of different sections, largely because different parts of the book were written at different times. One may also notice that in special and general relativity different systems of notations such as $((x^1, x^2, x^3, x^4 = ict,)$ (pseudo-Euclidean) and Minkowski space with signatures $(+, -, -, -)$ or sometimes $(-, +, +, +)$, respectively, (see Chapters 3, 9) are in fact more convenient than a single one, although they are just parts of the same geometric theory (however, I use only one system of relativistic notations in this book with a small exception of writing coordinate indices below to simplify the notations in a couple of the most primitive cases). Usually, understanding the author's notations takes considerable time while reading physical and especially mathematical papers. To spare such superfluous efforts, I tried to keep notations as simple as possible. Yet, I would not consider this text as fully suitable for students since it would take a lot of precious time to read all my reminiscences which may be unnecessary for focused studies. I concede that this book is a kind of an occasional supplementary reading, a dubious introspective report rather than a textbook or a monograph, although I was much more interested in physical and corresponding mathematical problems than in discussions and gossips about these problems. Some parts of this book may be read by a general audience without difficulties. Because of numerous distractions this text is not exactly what one needs to get prepared for exams. Nevertheless, I disagree with those who regard all such distractions totally irrelevant - I hope they might produce helpful associations. There is, incidentally, a general principle known to many people studying martial arts, e.g., karate, but applicable for each kind of learning: absorb what is useful, reject what is useless, and add what is specifically your own.

As stated, the manuscript is intentionally defocused and lacks unity. In a number of fragments, I decided to sacrifice stringency to make the reading more comprehensible. In some respects, the book is closer to a collection of popular essays than to a scientific treatise. I can understand those readers who would be irritated by such style and organization of the book and would rate it as unsatisfactory. To such readers I might remark that I tried to assume the role of a generalist in order to see the forest behind the trees. It is my deep conviction that physics and mathematics are the playground for free-thinking, open-minded and versatile personalities.

As far as the interaction between physics and mathematics is concerned, one can, of course, better spend one's time reading refined and focused reflections about mathematics and physics by V. Arnold, G. H. Hardy, M. Klein, Yu. Manin, J. von Neumann, G. Polya, N. Wiener, a number of prominent physicists, even Bourbaki who seem to be a collective pseudonym for a group of mathematical extremists united in their disdain of physicists. The present book may only fuel such disdain since I never intended to meet the present-day mathematical standards. The book consists of twelve chapters which are in fact not independent:

1. Introduction
2. Principles of Mathematical Modeling
3. Mathematical Potpourri
4. Classical Deterministic Systems
5. Classical Fields and Waves
6. The Quantum World
7. Stochastic Reality
8. Radiation in Matter
9. What Remains to Be Solved
10. Climate as a Physical System
11. Made in Physics
12. Conclusion and Outlook.

I could not refrain from revealing cross-chapter associations and exploiting common models. There exist popular and pretty universal mathematical methods such as Fourier series, Green's functions, perturbation techniques, asymptotic expansions, etc. that can be applied in each field to handle models with totally different physical content. Thus, it seems indispensable to master these universal methods, if one is striving to work professionally with physical models.

Being essentially a diary and based on personal reflections, the book enjoys relaxed standards of modesty and customary humility. I did not try to make the text look maximally impersonal as well as to conceal subjective likings and dislikings. In this book I used to intersperse technical descriptions with some reminiscences of my personal interactions with physicists and mathematicians. The scientific content in some fragments and sections is nearly absent. Yet the book contains, apart from personal reminiscences and opinions, a pronounced objective i.e., a scientific drive whose aim is to demonstrate strength and versatility of modeling methods in physics. This scientific component may bring certain difficulties to an occasional general reader, so the book prerequisites are some standard university courses on algebra, analysis, and differential equations as well as familiarity with basic facts from physics. Honestly speaking, learning the models of physics presupposes a certain degree of maturity, since they usually involve tying together diverse concepts from many areas of physics and mathematics. One can observe that a good article on physics is in fact a collection of mathematical problems with solutions. Bearing in mind these mathematical necessities, the preliminary Chapter 3 ("Mathematical Potpourri") presents a recapitulation of known mathematical facts, with the notation adopted throughout the book being introduced. Tinkering with mathematical expressions seems to be a growing trend in modern physics, and the rows of such expressions are often overloaded with fancy notations and can be totally intransparent for a person who is not a narrow specialist in the field. Occasionally, when the calculations are too lengthy and tedious to be reproduced exhaustively, I have simply quoted the results with the respective references.

Many examples in the book can be traced back to the works of giants. Great physicists may be considered to come in two more or less pure varieties: deep thinkers (like e.g., N. Bohr, P. A. M. Dirac, F. Dyson, A. Einstein, J. W. Gibbs, or W. Heisenberg) and efficient problem solvers (like H. Bethe, R. Feynman, E. Fermi, L. D. Landau, A. Sommerfeld, or Ya. B. Zeldovich). There may be also intermediate types interpolating between the two pure varieties, e.g., S. Chandrasekhar, L. I. Mandelstam, W. Pauli. Besides, there exist great minds whose mathematical power equals physical considerations or even dominates over them, those are exemplified by V. I. Arnold, N. N. Bogoliubov, L. D. Faddeev, V. A. Fock, H. A. Kramers, J. von Neumann, E. Wigner, E. Witten. All these people have produced highly influential results to be consumed by a sea of less creative individuals. The present book reflects the position of such a consumer: my main motivation was rather to learn, understand and absorb than to create and excite. One should not, however, try to absorb new science or techniques in a gulp, it must be done gradually, bit by bit. Recall that a real connoisseur would never hastily gulp the precious old wine, he would rather enjoy each sip, feel the gradations of taste, nuances of scent, hues of color.

I realize that the title of this book may be misleading, since this is not a book on mathematical methods in physics. Many of such methods have been described in comprehensive textbooks and monographs. For myself, it was important to trace how the outstanding scientists worked, how they employed their intuition, set up concrete problems, guessed the answer and tried to corroborate it by developing the mathematical metaphors - pretty universal to be transferred to other models and to connect them. It was also utterly instructive to observe how the great inventive minds tried to adjust and sometimes distort mathematics in order to obtain the required, intuitively anticipated answer without, of course, crucially violating the strict rules of mathematical operations. Therefore, one may note that the book is mainly focused on examples of physics-based models and not on hard-core rigorous descriptions favored, in all their generality, mostly by professional mathematicians. It is in this sense that the present book may serve only as a supplementary material to existing textbooks, e.g., on theoretical and mathematical physics.

My interest lies not in mathematics as such but only in its use. I treat mathematics as a support stuff for physics or sometimes even as a part of physics. Therefore, despite some explicit calculations contained in this book, I wrote it mostly in a "do-it-yourself" manner so that in many places I give only drafts of well-known models or theories. This means that starting definitions and statements (theorems) are often only formulated, with basic ideas and formulas being provided. The importance of correct definitions of physical and especially mathematical concepts is irreducible. Nevertheless, physicists often believe that it is the image of the concept and not its definition which forms the most essential component of understanding, and mathematicians are inclined to unjustified exaggeration of partial results. To some extent, in this book I succumb to this popular physical stereotype, and some mathematically important details such as establishing consistence of definitions or proving existence/uniqueness theorems may be left to the careful reader.

To be really creative, one must solve problems and not just describe how other people did it - this is a second-hand science. Likewise, it is inefficient to study any physmatical subject of interest beforehand: much more useful would be to take some problem relevant to this subject. At first, I wanted to supply the book with a list of problems, some of them rather complicated (with solutions), but then I dropped this idea because many problems are inevitably scattered over the main text in the form of models and adding some artificial problems would make the manuscript look like a textbook, whereas it is basically a diary. In other words, most exercises that I conjured initially for this book would look very difficult not because they really are such, but because the theory included in the book is not systematized enough to solve them all. This is a book for reading, not for studying.

In many cases, I return to the problems and models I have already discussed at a different level of understanding and try to observe them from a different angle. Such a recursive method of viewing a problem helps to elucidate its sensitive points. I tried to write as freely as possible, often preferring to stop and once again scrutinize the well-known models, sometimes with slightly different notations, attempting to find new features or connections to other models in the formerly discussed problems.

Thus, despite a cacophony of subjects, I hope the book is not totally useless. In general, the book may be perceived as reflecting a kind of intuitive protest against the progressive compartmentalization of science, its artificial breakdown into narrow disciplines. There are no natural frontiers between disciplines - it is people who have established them, often for the purposes having nothing in common with science.

A few words about sources. Many results in physics are hard to ascribe to a single author, they may be perceived as an outcome of certain scientific evolution. This fact manifests itself in the cited literature. In view of the rather broad scope of the book, no attempt has been made to supply it with an exhaustive list of references. I tried to cite moderately but exactly. The general criterion was to point at sources that could supplement the information, contained in the present book, which I consider insufficient or "divergent" (too many chained references with branching). Although in general I tried to make all the calculations transparent and understandable, in many instances in this book explorations of ideas contained in many standard sources such as textbooks is not fully complete and thorough; to provide a detailed account in all cases would make the manuscript unobservable. Some references, especially related to the material to be readily found on the Internet, are given, for the reader's convenience, in the main text. I did not shy away from using "unscientific" sources and from citing popular science books and articles, even those written by journalists and appearing in newspapers and on the Internet i.e., the "gray literature" - the material not published in peer-reviewed scientific journals. I also repeatedly cite Wikipedia, although this source can hardly be called an authority (for instance, Wikipedia can be biased). The names of many great people are present in the text. I was lucky to listen and to talk to some of them: V. M. Galitskii, V. L. Ginzburg, L. V. Keldysh, D. A. Kirznitz, A. B. Migdal, Ya. B. Zeldovich, D. N. Zubarev. I would not dare to say that I know or knew them all: it is only narrow circles of close friends have the right to say so. I was also lucky to have very kind and understanding teachers: Professors N. P. Kalashnikov and M. I. Ryazanov.

Almost all the material in this book is not based on original results. Nevertheless, the presentation of the material is predominantly my own and consists of original or considerably modified explanations of known results, discussion of their relevance to the contemporary scientific or engineering practice. In this sense, although there are no fundamentally new results, the book is not a primitive compilation. I tried to acknowledge all borrowings of presentations, which I was aware of, in the notes scattered over the manuscript. Since I attempted to present the modeling techniques by developing them from the first principles and in a self-contained way, I did not provide a hundred percent comprehensive list of references. Yet, I have supplied the book with the minimal bibliography, which would help the reader to duly appreciate the work performed by many ingenious people.

After all that was said one can justifiably ask: why am I writing all this stuff? A variant of the answer lies in the following image I used to have in my mind: people who study physics and mathematics today often find themselves in a situation of a late passenger trying to catch the train that has already started. My vision is that it would be possible to spare the inconvenience of jumping into the last car of the train, and instead of hectic turns to be comfortably installed in a cozy lounge. To this end, one should not strive to be at the forefront of modern science, which is frequently dictated by fashion, but to understand thoroughly not so many crucial scientific patterns. And I would like to add that I wanted to write not to please the experts or the referees, but what people could remember and use.

# Acknowledgements

I would like to express my deep gratitude to all those colleagues and friends of mine who have helped me in discussions, in text and pictures preparation and in many other ways. I am grateful to Professor Christoph Zenger and Professor Hans-Joachim Bungartz of the Technische Universität München whose shrewd critical remarks made this text less vague. Special thanks to Dr. Dmytro Chibisov who was very patient to explain to me a lot of tricks in practical $\TeX$/$\LaTeX$ usage (I call such tricks $\TeX$-nicalities). Dr. Chibisov who is a well-known specialist in computer algebra techniques has also given me many good advice especially concerning computer usage in mathematics. I would also like to thank Professor Vladimir Preobrazhenski from l'Institut d'Electronique, de Microélectronique et de Nanotechnologie de Lille for many constructive comments and suggestions.

Finally, but not the least I would like to express my deep gratitude to Dr. Thomas S. Ligon, who overtook the burden of the scientific editing of this book and my wife, Tatjana Znamenski, for preparing this manuscript for publication.

# Contents

# 2      Contents

# 1 Introduction

By some experience I have noticed that many people do not really read the whole of a science book, actually they don't read anything beyond the foreword, introduction and conclusive chapter. Bearing this in mind, I decided to slightly exceed the commonly accepted scope of these ancillary parts, although the present manuscript is strictly speaking not a science book. It pursues two main goals: firstly, to refresh the standard repertoire of working physicists and, secondly, to emphasize the role of interdisciplinary links enjoying physical and mathematical inclination (I call the col- lection of all such links a physmatical network). The first goal implies only a superficial coverage of subjects, with the main intention here being to arouse interest to foundations, whereas the second goal was to demonstrate occasional fruitfulness of cross-disciplinary jumps and unexpected analogies, often regarded with disfavor as rather philosophical and speculative.

The very notion of "interdisciplinary approach" as well as the words "interdisciplinary", "cross-disciplinary", "transdisciplinary", or "multidisciplinary" seem to be strongly discredited although it is often asserted that highly institutionalized knowledge tends to limit understanding. Scientists habitually consider all such terms as a distinctive mark of unprofessionalism, when everything is swept into a single heap and it would be difficult to set up a clear-cut problem under blurred circumstances. As a consequence, no specialized knowledge appears to be required, and active charlatans with their childish babble or at least holistic half-professionals with their vague claims may profit. Although such a danger really exists, I still think that total refutation of the interdisciplinary approach (or even its occasional equalizing to pseudoscience) is erroneous. Interdisciplinarity is basically a good thing, allowing one to transcend artificial boundaries and to overcome excessively narrow specialization. More and more areas are becoming inherently interdisciplinary: biophysics, biomechanics, medical physics, other life sciences, robotics, nanoscience and nanotechnology, quantum computing, ecology, climate studies, complex systems and so on. Such renowned scientific journals as the Physical Review E and Physica D claim to be "interdisciplinary in scope" (PRE) or devoted to "nonlinear phenomena in general" (Physica D). Yet people are uncertain whether an interdisciplinary career is in principle feasible, with interdisciplinary studies only denoting a rubric.

People who are comfortable in multiple areas, e.g., in physics and social sciences, in mathematics and climate science, human perception and nuclear engineering, chemistry and geoscience, mathematics and genetics, or political analysis and scientific ecology are exceedingly rare, and I think one should create a favorite environment - institutional, intellectual and moral - that would encourage the emergence of a new breed of scientists feeling at home with unusual alliances of specialties. These are Renaissance persons[5] yet I don't think that today value of such scientists is properly appreciated: in spite of many words pronounced in favor of interdisciplinary research, the current reward system seems to impede flourishing of cross-field experts and projects. In this book, I group together things that appear to be utterly diverse partly in order to emphasize the value of interdisciplinarity.

Another curious observation of mine was that people - even very qualified persons - mostly do not think thoroughly about the basics. This is quite natural: people study the rudimentary things at a very young age when wonderful distracting factors surrounding them are abundant and afterwards, in adult life, there is a drastic shortage of time to return to foundations. One should not blame people for a few blank spots in their basic thesaurus. Many of us have probably met some angry, disturbed and humorless professors who were pinching students and younger colleagues, without any compelling reason, solely on the ground of having forgotten allegedly elementary facts. Bearing this situation in mind, I have devoted a considerable space to the subjects commonly considered too primitive to pay attention to. To me, they proved to be not at all primitive, and I often found, with utter amazement, that thinking about the basics rapidly develops into facing very intricate issues protruding into various fields of physics and mathematics. Although this remark about the importance of reviewing elementary concepts may sound trivial, I wanted to share my amazement with other people. This explains why I ruminate about pesky basics for so long in the manuscript.

This book is devoted to an informal discussion of patterns constructed for treating physical problems. Such patterns, when sufficiently formalized, are usually referred to as "models", and they tend to be applied today not only in physics, but conquer the fields traditionally occupied by other disciplines generally considered to be totally different from physics. Accordingly, in this book the word "physics" is understood in a broad sense as the general study of natural phenomena. A tiny part of the models related to natural phenomena may be set up as mathematical problems and solved using contemporary mathematical means, exactly or approximately (e.g., numerically), whereas a much larger part of the models can only be qualitatively described. These latter verbal patterns are typically regarded as imprecise low-level statements which, hopefully, will be formalized in future. A mathematical formalism, however, does not necessarily imply exact knowledge; rather it demarcates the frontiers of ignorance that are fuzzy in qualitative statements. Nevertheless, a large part of this book is devoted to qualitative statements and verbal patterns considered as a kind of raw material for building up satisfactory mathematical models.

Inevitably, the presentation of many topics may contain short excerpts from the courses of classical and quantum mechanics as well as from classical electrodynamics and quantum field theory. Much of this material may appear trivial and even unnecessary. I think this impression is false. The aim of including the textbook excerpts is to trigger the imagination. Well-known results start the intrigue, and afterwards the new and unusual things come along. The reader may also find it annoying when I use innumerable "possibly", "probably", "might be", "perhaps", "in principle" and so on. The reason for this approximativeness is not an obsessive carefulness but merely an inexact knowledge. Unfortunately, most of the knowledge in the world is inexact and cannot be reliably quantified, and one should not commit the adolescent sin of not admitting it.

---

[5] A renaissance person is a model of personality possessing a broad range of knowledge. One of such person was, for example, Alexander von Humboldt who had the ability to embrace nearly all scientific disciplines known at his time and traveled through all the continents (maybe except the Antarctic).

In the Middle Ages, there existed the so-called *"Exempla"*, i.e., collections of illustrative examples to be used by priests to save the parish people from falling asleep during the sermon. The wakening effect was reached by the presence of numerous vivid details in the "exempla", for instance, chilling minutes about hell, the devil, demons, etc. Imagining easy to understand, life-like pictures, the public heard prayers. In this book, I have tried to produce something of the kind of such *"Exempla"*, although I am not sure I can always keep the reader awake.

Being overloaded with personal details and opinions, this book nevertheless contains examples of the application of selected mathematical methods to mathematical models frequently encountered in various branches of science and engineering. Since physics seems to be the best-known collection of models, it is physics based mathematical models (PBMMs) that are predominantly discussed. Even when other disciplines come into play, most of the models under review have been imported from physics and adapted for scholarly or engineering problems arising in these disciplines. To promote cross-fertilization of ideas between a variety of disciplines, it is often advantageous to consider the so-called complex systems requiring a real exchange of concepts and techniques. In this transdisciplinary approach there is of course a danger of debating and arguing about too many things.

After a brief methodological prelude about general modeling principles (Chapter 2), some standard mathematical techniques are informally discussed in Chapter 3. This chapter is neither a micro-handbook nor a review of mathematical methods in physics. The main purpose of this "mathematical potpourri" is to introduce some mathematical concepts, mostly of geometrical nature, which have long become routine in mathematical texts but still are not quite easily acquired by physicists and engineers engaged in the business of mathematical modeling [6] . Chapter 4 is devoted to classical mechanics culminating in the theory of dynamical systems. This is perhaps the main part of the book; the emphasis in it is placed on dynamical systems - this is due to the fact that change is the most interesting aspect of models. Moreover, classical dynamics is probably the most developed part of science, it studies the evolution of systems of material points, bodies that are so small that their inner structure is disregarded and the only surviving characteristic is their position in space, $\mathbf{r} = \mathbf{r}_i(t)$. The dominant modeling concept exploited throughout Chapter 4 is the notion of local equilibrium. Mathematical modeling of complex systems far from equilibrium is mostly reduced to irreversible nonlinear equations. Being trained in such an approach allows one to model a great lot of situations in science and engineering. In Chapter 5, classical field theory is briefly outlined, with some accent being placed on wave motion and wavelike models. Since there is hardly any means to define what can be called a wave process in general, a very broad variety of problems is mentioned in this chapter. Physical systems can be roughly separated into two classes: particles and fields, correspondingly there are two basic classes of models. I used the word "roughly" because in fact there is an overlap, for example, particles serve as field sources. Moreover, the separation of matter into particles and fields is, strictly speaking, outdated and incorrect. It is used here only for convenience: the main difference between dynamical systems in classical mechanics, where particles are studied, and in field theory is in the number of degrees of freedom. Any classical mechanical system consisting of a finite number $N$ of particles has only a finite number of degrees of freedom, $n = 3N$, whereas fields possess an infinite number of them.

---

[6] I have written "business", but it is probably a wrong and compromised word; one should rather think of an art of mathematical modeling, even in its numerical stage.

Chapter 6 is devoted to quantum (and mostly relativistic) fields. No former knowledge of quantum field theory (QFT) by the reader is implied, although the corresponding mathematical problems are quite intricate. Quantum field theory has a reputation of being hard to study, but I think that such a reputation is mostly due to "user-unfriendly" expositions, heavily technical and overloaded with details. I hope that after scrolling this chapter one will be able to read the literature on QFT with the least amount of pain. Quantum field theory can be crudely interpreted as quantum mechanics of the systems with infinite number of degrees of freedom. The development of quantum field theory began with quantum electrodynamics whose main equations appeared in the 1920s, almost simultaneously with nonrelativistic quantum mechanics in the works of Dirac, Heisenberg and Pauli.

The primary bunch of quantum field models being just an extension of quantum electrodynamics may be called the "old" quantum field theory. As it is customary, we shall discuss quantum fields in the relativistic four-dimensional context, although the requirement of relativistic invariance is, strictly speaking, not necessary for quantum fields. After some recapitulation of the four-dimensional formalism, we shall subsequently observe quantum fields for spin 0, spin 1, and spin 1/2. Then we shall discuss a "new" physics of quantum fields, where geometric ideas are fully exploited. Thus, theories based on gauge invariance comprise a class of "new" quantum field theories. One can illustrate the bridge between "old" and "new" quantum field theories by an observation that all physical theories in four-dimensional spacetime are characterized by a number of common features. In particular, long-range forces should exist that require conservation of the corresponding charges. This fact provides a passage from "old" to "new" QFT based on gauge theories. It is widely known that methods of quantum field theory are applicable in other areas of physics; the most popular application area of such methods is statistical physics.

Mathematical models of quantum theory reviewed in Chapter 6 are partly connected with the wave problems discussed in the preceding chapter, so there are a number of natural cross-chapter references. This fact may be viewed as a manifestation of modularity and reusability of many mathematical models in physics. Symmetry issues that should necessarily be taken into account both in the orthodox quantum theory and in field theory are tackled in many sections of Chapters 5 and 6. Apart from a description of mathematical models of quantum mechanics, questions of its interpretation are discussed in this chapter. Although bearing a heavy philosophical load, these questions survived a renaissance in the 1990s and continue to be popular also among physicists dealing with quantum computing and quantum cryptography. For me, the so-called Copenhagen interpretation is good enough, although it leads to paradoxes and inconsistencies. It is presumed that N. Bohr understood causality as the Laplace determinism only, i.e., in a narrow sense, and juxtaposed to it "complementarity". Similarly to Laplace who extrapolated the successful solution by Newton of the Kepler problem on the entire universe, thus postulating a mechanistic model of the world, Bohr created an anti-Laplace model by a philosophical generalization of the Heisenberg "indeterminacy relations" which are in fact just trivial consequence of the Fourier integral. This juxtaposition led Bohr to the idea of incompatibility of quantum mechanics with determinism, at least of the Laplace type. Less conservative ideas lead to a negation of causality in some situations with participation of quantum particles, i.e., to acausal or even anti-causal quantum models. Causality as a principle is discussed more or less thoroughly in this chapter.

By the way, when studying quantum mechanics, it is only natural to acquire some interest in its historical developments. Physics professionals usually do not have a passion for the history of physics, mostly considering the human element behind the scene an unnecessary distraction. Yet people are different. As far as I am concerned, my interest in specific physical phenomena was quite often fueled by the desire to comprehend the corresponding historical occurrences. Besides, some evolutions of science contain a pronounced dramatic element and bringing it to light allows one to deepen the purely scientific understanding. For instance, I have found out that scrutinizing the historical facts which accompanied the development of superconductivity allowed me to understand its theoretical models easier and better.

Statistical models treated, maybe a little superficially, in Chapter 7 lead to problems that cannot be fully treated within the framework of classical theory only, despite a lot of classical concepts discussed previously in Chapter 4 being invoked. It would of course be difficult to treat the whole scope of statistical and stochastic issues permeating both physics and other disciplines using methods developed for physical systems (such as economics and sociology), thus many problems are only briefly mentioned in this chapter, with respective references being provided of course.

Chapter 8 is devoted to the loose description of theoretical approaches to the interaction of radiation, both electromagnetic and corpuscular, with matter. The chapter is divided into two parts, one related to the electromagnetic field (mainly radiation) in material media, the other to the passage of particles (mostly charged ones) through matter. The concept of material medium and its interaction with external agents, in its general setting, has a great many particular facets far beyond physics. For example, many social problems may be reduced to the depiction of an appropriate medium and its interaction with external influence such as immigration. One can also portray a human society in a steady communicational state affected by outside opinions. Mathematical models of immunology are conceptually close to these considerations. In physics, external agents provoking a response in material media are usually taken to be radiation fields or streams. The problem of material media response is very complex and is dissected - sometimes artificially - into many subproblems and models. Typical models are related to the interaction of pulsed electromagnetic radiation with material media. Such models are very relevant these days, specifically with regard to new laser technologies allowing one to produce extremely powerful ultrashort pulses.

The next Chapter 9 outlines "dark spots" - subjects that not only the author does not fully understand, but also the great minds of the current physmatical milieu seem to be unable to give exhaustive answers about. In this chapter and the next one, two large subsections can be found - on time reversal symmetry and about Earth's climate (the latter discussed in Chapter 10) - like "novels in a novel" or, say, dreams seen in a dream i.e., so to say, dreams of a second order. And like in a dream, the real problems are mixed with fantasies, speculations and pseudoproblems. Both subsections contain elements of a journalistic essay, and I do not shy away from discussing the fictive problems bordering on metaphysics or excessive philosophical generalities. In fact, a lot of current physics easily accommodates things that are little more than philosophy and cannot be falsified in any way. For instance, the problem of time-invariance violation discussed in this chapter is of semi-philosophical, worldview character. For centuries it was believed that all events are, in principle, predictable, time-revertive, and can be described by differential equations similar to the equations of motion for an isolated body. Seemingly unpredictable and irreversible phenomena, such as weather or human behavior, were believed unpredictable and irreversible only due to a very large number of variables. As to predictions, it is still hoped by some scholars that with the advent of the powerful computers all long-range predictions would be possible (which is probably wrong).

When thinking about mathematical models in physics, one cannot get rid of a thought that the whole army of researchers has won a great many battles but is losing the war. Indeed, there exist several diverse paradigms in physics which serve as a base for model construction, but taken together they produce an impression of great confusion. Which model of the particle is ultimately correct - pointlike, as prescribed by special relativity, or extended and spread over some finite domain, in accordance with quantum mechanics? Or a tiny string? Or containing an internal world? Is the reductionist approach, when one attempts to explain all phenomena as based on a fundamental set of elementary interactions, in general valid? One can notice that almost all cases of physical interest, except probably some exotic ultra-high-energy accelerator experiments, involve systems of a great number of particles. It may be impossible to directly calculate the behavior of complex systems from single-particle models. For example, reductionist arguments trying to derive properties of biological tissue starting with some fundamental model such as Newton's second law or the Schrödinger equation can hardly be called adequate. In Chapter 9 we shall discuss a typical reductionist problem: time reversal in complex physical systems. It is generally believed that each phenomenon in our reality should be, in the final analysis, invariant under time reversal on the ground that the most popular mathematical models of fundamental interactions are time-invariant. This, however, can be a delusion or at least a stereotype.

Chapter 10 "Climate as a Physical System" could be considered as a continuation of "dark spots". It has already been emphasized that the climate system is a very complex physical object and prognoses of its evolution must necessarily be an extremely refined issue. There are many discussions among professional climatologists, meteorologists, physicists and representatives of other sciences about how to approach - not even to solve - this problem. There seems to be more of humanities and politics than of natural sciences in the climate disputes. The matter is that humans - even the most enlightened climatologists do not know enough either about the Earth's climatic system or about the chaotic dynamic systems to produce accurate mathematical models containing thousands of entangled variables. Hence the prevailing role of the anthropogenic factor compared to the natural influencers on Climate such as solar activity, oceanic circulation or lunar motion is highly questionable.

Chapter 11 is devoted to mathematical models beyond physics. Physics is distinguished from other disciplines also employing mathematical modeling by the fact that models in physics are linked. This is an important property ensuring the success of the collection of mathematical models called a science, and that is a feature that makes physics "the splendid architecture of reason" [18]. This linked architecture appeared mostly due to firmly established laws of physics that can be expressed in the form of differential equations. As soon as a model is disconnected from the main architecture, its heuristic value becomes substantially reduced. One could see this diminishing value of standalone models - despite the generally flourishing physics - on the examples of numerous group (multiplet) models of elementary particles, being of fashion in the 1960s. Nowadays, after the Standard Model epoch, nobody seems to remember those quickly baked concepts which did not use the wealth of physical results. A similar situation is typical, for example, of economics where mainly ad hoc mathematical models are in use, characterized by a lot of arbitrary parameters and suggestions.

Now, let me make some additional comments about the subjects contained in the present book. This book is not about well-known mathematical methods in physics - there exist a great lot of sources on this subject, and I had no intention to compete. The reason for discussing foundational issues was to stir up the reader's interest, inducing one to address oneself to more precisely focused and professional sources. Nonetheless, in this manuscript I tried to evade repeating well-known results, numerously described in the vast literature, but whenever I still had to address such results - usually while starting to describe a particular model or a theoretical approach - I made an attempt to concentrate on those features which seemed to me fresh and unexpectedly connected to apparently far away subjects. Indeed, in physics and mathematics combined we primarily find out about the existence of wonderful connections between things which at first sight seem to be completely different. Some of these connections are really intriguing. Today, well-known links exist between the theory of dynamical systems and statistical mechanics, between the models of black holes and thermodynamics, although these subject pairs are apparently related to different fields of physics. There is now also a well-known interface between superconductivity, cosmology and high energy physics, based on the unifying idea of gauge invariance. Less known connections may appear quite striking. For example, can one apply the notion of wave function in classical statistical mechanics? What does classical signal analysis have to do with the foundations of quantum mechanics?  Another example of networked physmatical domains is the presence of thick links between heat and mass transfer (with countless engineering, geophysical, ecological, socio-demographic, etc. applications), quantum mechanics, and geometric topology, in particular the now fashionable[7] Ricci flows. The latter describe the spread of a tensor quantity, the Ricci curvature, which in distinction to scalar quantities governed by the heat, diffusion and Schrödinger equations manifests the behavior of a purely geometrical property.  But the idea behind all these models consisting in the dynamical smoothing out of high concentrations is basically the same. Or, I was astonished to find out that the Bogoliubov-Mitropolski expansion, well- known in the classical vibration theory [242], can be efficiently applied to nanostructure analysis. The burgeoning areas of biotechnology, nanotechnology, quantum engineering, and quantum information processing are today closely converging. This cross-border linkage of apparently diverse disciplines may serve as an example of a physmatical network in action. One might notice that physics is full of shared approaches and synthetic subdisciplines such as the interaction of radiation with matter, astrophysics, geophysics, cosmology, etc. Even the elementary study of physics and its relevant mathematics increasingly shows that apparently disparate topics are in fact closely related. For example, let us start from mechanical problems. Then we very soon come to a well-known fact that Hamiltonian systems determine a vector field and solving the Hamiltonian equations means finding the integral curves of this vector field. This is known as finding the dynamics of a physical system. Only a slight generalization of this simple physical situation leads to familiar linear algebra notions and geometrical concepts: a first-order differential equation on some manifold $M$ is a vector field on this manifold, and solving a differential equation for given initial conditions is translated into finding a curve or, rather, a family of curves. All this is of course trivial, yet there is an unexpected link here, namely that we naturally come to the notion of a flow for a vector field leading in its turn to quite sophisticated concepts of the theory of dynamical systems with a multitude of applications. In other parts of physics and associated to it mathematics, we have, for instance, the Hahn-Banach theorem and then the concept of separation of convex sets, which eventually leads to such ultra-practical things as the optimal control of missiles. Still in other nodes of this physical-mathematical network we locate the integral equations and Fredholm operators naturally connected with the scattering of waves and particles and, although it may seem strange, with the Navier-Stokes equations. Hence some people may find this text suitable for acquiring some knowledge on the interaction between different branches of mathematics and physics, starting from some seemingly familiar examples contained in the book. In physmatics, due to its networking properties, it is easier to think inductively or, roughly speaking, to generalize simple examples - in contrast to the famous Sherlock Holmes' method.

---

[7] Mostly due to the famous recent results by G. Perelman.

Although this book is about mathematical models and, predominantly, mathematical models in physics, it contains many notions that appear too inconcrete from a scientific point of view: nature, society, values, knowledge, science, politics, policymakers, bureaucracy, democracy, scientific tribes, history, intuition, conscience, God, faith, beauty, truth, freedom, etc. These descriptive terms usually serve to label some phenomena with the purpose to invoke associations and exploit human attitudes. Such objects form fuzzy sets or are high-level entities existing in dedicated spaces, multidimensional and of course not necessarily linear so that it would be difficult to define them exhaustively with words. These and a great deal of other badly defined concepts are appealing to unwritten laws. Moreover, such objects cannot be in general reduced to "black-white" dichotomies 0-1, but involve complex though essential details. More than that, the subsets of these fuzzy notions which describe certain groups of people are not only very unspecific but, on the one hand, impersonal while on the other hand too poorly defined to be studied by really scientific methods. By the way, fuzzy notions are often used not only in the humanities, but also in physics and even mathematics. For instance, such words as local, very small, negligible, immediate vicinity and their inverse i.e., global, very large, essential, far from, etc. are ambiguous words, but their use seems to be unavoidable in science. So it would be wrong to assume that "exact" sciences are based solely on numbers understood as ordered sets; by the way, one may recall that numbers themselves originally modeled physical (or rather biomechanical) objects such as human fingers.

Since this book is mainly devoted to building mathematical models in physics, allow me to expand a little bit on the subject of modeling in this introduction. Mathematical modeling in physics is a nontrivial interplay between mathematics, physics and engineering in the study of complex systems (see in the following chapter a slightly more extensive discussion of the properties of models in physics). Any model in general is a simplification of reality, with irrelevant details being ignored. It follows from here that when a model is used, it may lead to incorrect predictions when the neglected details cease to be irrelevant. Thus, it is important to estimate the limits of applicability of the model - the art largely mastered by theoretical physicists and often overlooked in more applied disciplines and, strange as it might seem, by computer scientists whose primary task is to appreciate errors. And appreciating errors is indispensable for validating models.

When we are approaching a problem, we tend to do it in a biased, limited way (recall the chrestomathic tale "The Blind Men and the Elephant"), with some feature seeming the most substantial according to previous experience, prejudices, stereotypes and other purely human factors. Understanding of where a particular model fits within the overall theoretical description enables one to estimate the model limitations. Moreover, it does not mean that we shall always succeed in giving a useful answer by employing mathematical tools. When one tries to solve a problem that one does not know thoroughly, there is always the risk of building castles in the air. For instance, it would be risky to model biomedical processes without elementary knowledge of the subject area. I think that if one is interested in applying mathematics, one should start first by studying carefully the problem one is interested in and then learning the necessary mathematical theory. The use of mathematics in physics is usually unavoidable, as well as in other areas which claim to be "exact" (chemistry, engineering, some parts of biology). Examples of disciplines that are definitely not "exact" are the major part of medicine, psychology and philosophy. The bad thing about such disciplines is that they are relying more on so-called "expert opinions" and too generally understood "human experience" than on a solid reproducible base so that it is difficult to controvert or falsify the vague statements of inexact disciplines. One often forgets the truism that even in "exact" sciences the theoretical solution to an actual problem must be confirmed by experiment. Experimental proof is the strongest of arguments whereas reference to an authoritative opinion is the weakest one.

# 2 Principles of Mathematical Modeling

The world available to us may be defined as the sum of manifestations of the systems having different levels of complexity. The systems whose complexity (defined in any manner) exceeds our abilities by orders of magnitude cannot be studied by exact methods, in particular, by mathematical means. In this case, two possibilities come out: either to dissect the observed system into subsystems of lower complexity, so that such subsystems could already be studied by available mathematical techniques, or resort to fuzzy cognitive representations. The first option is that of mathematical modeling, whereas the second is closer to philosophy. In this book, predominantly the first alternative is used.

Nonetheless, models of reality can be of vital importance also without an exact mathematical form. When the Black Death - the plague - swept relentlessly over Europe (it was several times, probably the most ferocious pestilence attack was in the 14th century), there were three main models for the cause of the plague: it was regarded as a medical event, as astrological misfortune, or as a God's punishment. Thus, there were roughly speaking three respective classes of models explaining the plague cause: terrestrial, celestial, and divine (religious). These three classes of models were not independent. Terrestrial models, for example, were based on the ancient Greek science represented, e.g., on the Aristotle's "Meteorology" stressing beside atmospheric phenomena the view that certain conjunctions of planets (e.g., Mars, Saturn, Jupiter) would bring disaster, and on the Hippocratic "Epidemics", where the importance of astrology for medical practice was discussed. We shall deal with astrological and religious models a little later; now it may be noticed that terrestrial models were closer to contemporary scientific concepts than the two other classes of models. Terrestrial models discussed atmospheric events, weather, climatic variations, comets, meteors, earthquakes as possible sources of poisonous gases spoiling the air and thus causing illness. Rain, thunder, storm, wet winds could disperse ominous vapors (for instance, produced by cadavers rotting in swamps). So the essence of the model for the cause of the plague consisted in the notion of the poisoned air entering the body, contaminating it and causing its organs to disintegrate - quite a viable model even from the contemporary viewpoint. It is clear that understanding the cause of the plague could substantially increase chances to prevent its spread. We shall discuss below some models of epidemics based on dynamical systems, here I wanted to emphasize the importance of non-mathematical models for the human race.

Many people think that real science and technology begins only where sophisticated mathematical methods enter the picture. I think it is just a stereotype. Refined mathematical models of the world are not necessarily destined to play an instrumental part in life practice. For example, there is no practical necessity to use geometric theorems in order to measure the plot area: the value of a particular estate is only very approximately determined by its area, "human factors" such as prestige or location are much more weighty. For practical purposes, the ancient Egyptian formula, $S = \frac{a+c}{2} + \frac{b+d}{2}$, crudely expressing the area of a rectangle as the product of half-sums of the opposite sides is quite sufficient. Moreover, using geometric theorems would be counterproductive for a land surveyor, since they require more precise measurement which does not produce a better estimate of the plot real value.

Actually, the dichotomy "mathematical models - cognitive models" is linked to the models of language and semantic scattering in it. Each word in any human language has a spectrum of different values (meanings). Such a spectrum can be represented in the dictionary[8], but if one reckons with a multitude of nuances and gradations (which may be represented as a fine structure of each semantic value) then one is inclined to think that discrete models of the language may be too deficient. Therefore, to build up a model of the language which would be adequate to its complexity one has to consider continuous spectra of values. Cognitive models of the world exploit such continuum spectra of verbal values, whereas mathematical models are centered around discrete notions which can be more precisely defined and thus expressed by equations.

We shall return below to cognitive models, some of them, as I have mentioned, may be quite valuable. Nevertheless, one can often encounter an utterly disdainful attitude to models which are not based on mathematical equations, especially from the young people studying mathematical, physical, or engineering disciplines. These mathematical extremists tend to decry anything not containing mathematical formulas as a "bla-bla-bla", not understanding that mathematical notation and models are only related to a comparatively low complexity level.

Therefore, the question arises, which seems to be common for all young people who study "exact" sciences but are in fact attracted by the humanities: is there in this latter part of the culture a concrete knowledge that must be considered necessary for understanding the world? This question may be rephrased in more operational terms: does the inductive method of scientific research bring the results whose value may be as high as obtained by formal deductive method favored by mathematicians? Traditional physics combines both methods, but which one is more important? The question may probably be formalized, but this is a prerogative of the philosophy or history of science. I intentionally put it in vague terms, because I do not know the answer. The only thing that seems to be certain: modeling requires a certain kind of people possessing crossover skills and an intensive dialog between specialists representing a variety of disciplines.

---

[8] In fact *only* a discrete semantic spectrum can be represented in the dictionary. One can imagine here a massive thesaurus such as Webster, Oxford, Roget's, Larousse, Duden, Meyers, etc.

Here, I would like to make a personal remark. For myself, ruminating about mathematical modeling in general is simply an occasion to babble on unspecified subjects, making idle recommendations. There are, however, people who possess a natural aversion to diffuse quasi-philosophical and other verbose texts. They can easily omit this chapter.

## 2.1  Basic Principles of Mathematical Modeling

This introductory section reiterates some traditional concepts which are basic to the overview of connected mathematical models of physics described in the subsequent chapters. The reader can find that I did not shy away from truisms and general statements. I hope I shall be forgiven for these elements of didactics. Moreover, this section contains standard methodological recommendations about rudimentary modeling principles. I tried to make all these observations a bit less boring by supplying them with some personal remarks. The reader not much interested in contemplative hints on unexpected interrelations between different domains of physics and mathematics may easily skip this section.

Since a large part of science is based on models, it would be useful to learn some general principles of their construction[9], no matter how declarative the enumeration of such principles may seem. To make their discussion less boring I rely not on terse quasi-scientific "proofs", but on simple examples - for myself, I call this anti-didactic procedure "descholastization".

In a scientific research or an engineering inquiry, in order to understand, describe or predict some complex phenomenon, we employ *mathematical modeling*, already reflected upon. In the mathematical modeling technique, we describe the state of a physical, biological, economical, social or any other system by time-depending functions of relevant variables. Then we attempt to formulate, in mathematical terms, a basic law governing the phenomenon. When we are interested in the system's behavior, the law to be provided is expressed by one or several differential equations relating the rate of temporal change of the system variables to the components of some vector field, analogous to a field of forces. Such a formulation of a modeling task leads to a *dynamical system*. Then the dynamical system models some piece of the world - of course, it may not model it very well, but the model's success or failure largely depends on the law that has been proposed to govern the phenomenon. Usually, mathematical modeling can be naturally connected with dynamical systems theory. It means that mostly the dynamical processes, that is those evolving in time, are considered. This restriction leaves out quasi-static models displaying the relationships between the system attributes close to equilibrium (in physics, for instance, the Gibbs thermodynamical equilibrium, in economics - the national economy models, etc.). In general, such equilibrium states when all distribution functions do not depend on time (see Chapter 4 of the book) are disregarded in dynamical modeling. In this book I am focused mainly on physical models and reduction of the latter only to dynamical systems is not necessarily relevant. One may note that, for example, the major part of quantum mechanics is devoted to describing the spectral states of a quantum system i.e. is static by nature.

---

[9] This was in fact the unique motivation to write the entire modeling chapter.

Present-time dynamic modeling of real systems such as weather forecast, climate viewed as a physical system, transportation networks, energy grids, genomics, etc. has become increasingly complex. To obtain trustworthy solutions to these problems, especially real time results, intensive use of computational resources (processors, memory, storage) is required. In general, modeling can be performed in a variety of ways, with different aspects of the situation being emphasized and different levels of difficulty standing out. This is a typical case in mathematical modeling, which implies that the process to be modeled is not necessarily described uniquely. It means that modeling is not reduced to a set of rigid rules but rather provides a vast field for transdisciplinary creativity. The reader I vaguely have in mind is a person who strives to transfer the methods developed within a certain scientific or engineering area to other branches of knowledge. Modern mathematical modeling is a synthetic discipline embracing mathematics, physics, computer science, engineering, biology, economics, sociology - you name it. Mathematical modeling enjoys enrichment due to interdisciplinarity. Although, as I have already mentioned, this book is mostly devoted to physical models, the term "physics" is to be understood throughout the book in the universal sense: most of the models in other fields of human knowledge are increasingly using methods developed in physics (see Chapter 11).

When computers were not yet abusing the most part of your time, the main tools of a mathematician were pencil, paper and a waste-paper basket. I don't remember who said that it was the waste-paper basket that distinguished a mathematician from a philosopher, but I still think this is very true even today, although mathematics has become much more scholastic than previously, when it was closely connected with physics. Unfortunately, as far as contemporary mathematical modeling goes, the waste basket should be used much more intensely as a working tool.

One might note that nearly all books on mathematical modeling are generally very eclectic and incorporate a variety of random problems as well as disjoint hat trick mathematical approaches. The very notion of mathematical modeling is hard to define - everything e.g., in physics is mathematical modeling. So, this term is infected with vagueness. Nevertheless, there exist certain common principles of modeling such as the just mentioned "divide and conquer" principle. To evade some methodological prolixity, which seems to be customary when speaking about modeling, I may refer the reader to many good books on principles of mathematical modeling ([1, 2, 13]) [10]. Here, I mention just the following counterpoints:

- Qualitative vs. quantitative

---

[10] The reader can also find some collection of modeling examples illustrating general principles in https://www5.in.tum.de/lehre/praktika/comp_mod/SS02/modeling.pdf, where modeling of dynamical processes is presented.

- Discreet vs. continuous

- Analytical vs. numerical

- Deterministic vs. random

- Microscopic vs. macroscopic

- First principles vs. phenomenology

In practice, these pure types are interpolated and, like in musical counterpoints, all parts of the model must be blended together as one, even if each is a feature of its own. The ultimate difficulty in mathematical modeling still persists: there cannot be a single recipe how to build a model. One always has to make a choice: this implies rather the art than the science of modeling. As in any art, some vague principles come into play, here e.g. the spirit of Occam's razor, which may be roughly formulated as "simple explanations are preferable to complex ones", should not be violated. This leads to a somewhat deplorable situation: many models may be rejected not because they are basically wrong but because they are too sophisticated. One can, in principle, construct any kind of mathematical model irrespective of their relationship to reality; such models may be non-contradictory and even beautiful (like in the modern string theory), but the requirement of their experimental validation may be substantially weakened. In this connection, I can recall an aphorism ascribed to J. W. Gibbs and quoted in "The Scientific Monthly", December, 1944: "A mathematician may say anything he pleases, but a physicists must be at least partially sane." Mathematical trickery alone rarely brings fruit in physics.

A mathematical model is not uniquely determined by the investigated object or considered situation. There are usually a plethora of conditions, and only one of them is selected. Moreover, selection of the model is dictated by accuracy requirements. For example, in motion and surface planning models that are very important in robotics a table may be represented as having a rectangular or oval shape; in road traffic models a car is typically represented either as a point or a rectangle; should one take into account atmospheric influence on a falling body or not? The answer depends on the required precision. When some difficulties arise due to an over-idealized mathematical model of a given physical situation, the model can be changed or improved. This is the approach most favored by physicists. The other approach, logically opposite and adopted by many mathematicians, would be to address the physical situation in the most general and possibly rigorous formulation right from the start imposing restrictions as one proceeds. Such methodics may be quite powerful (historical examples are the works by N. N. Bogoliubov, V. A. Fock, H. A. Kramers, L. A. Vainstein), but often makes the treatment unpalatable for physicists.

Core mathematics contained in the real physical models is not too abundant. By "real", I mean those models and theories that can be corroborated or disproved by some experiment or observation. In this sense, I do not think that a vast number of string theories are real, although some of them might be quite "beautiful". Incidentally, aesthetic criteria alone may well have nothing to do with reality: it may happen that physical phenomena are often described by simple and symmetrical equations which can be considered "beautiful", but the reverse is not necessarily true: not every "beautiful" (i.e. simple and symmetrical) expression pertains to physical phenomenon.

It is trivial but not always held that mathematical models should not contradict the fundamental laws of nature. Sometimes, however, this requirement is weakened. Thus, some numerical models, for example, based on the Runge-Kutta method may lead to energy non-conservation. The number of particles or mass should be conserved, which is usually formalized as the continuity equation; again, there exist numerical simulations, with the continuity equation being violated or, at least, not exactly fulfilled. It is not infrequent that the laws of physics are rejected in computer technology: one can recall "magical" fluid simulations and animations in computer graphics.

In general, however, mathematical models must be tested against basic laws of physics. Although an ever increasing number of people think that mathematics is totally different from physics and therefore must be free from constraints imposed by physical laws, I think nonetheless that mathematics is a service subject, like the art of hieroglyphic painting in the Oriental cultures, the content being still provided by natural sciences, in this case by physics. From here it follows that many principles successfully employed in physics, such as symmetry and scaling, should be taken into account in the domain of mathematical modeling and widely exploited in order to reduce the complexity of real-life models. In the spirit of our physmatics, one can also extensively use analogies, e.g., chemical reactions are analogous to competition models in biology and economics.

The latter fact points at the universality of certain techniques: different objects are described by the same mathematical model. The trivial example is given by the ubiquitous oscillator model which can be equally well applied in mechanical engineering, e.g., to explore the car body vibrations, in electrical engineering, e.g., to study the passage of signals through filters, and in physics - to build up models based on the concept of elementary excitations. On the other hand, the oscillator model participates in the intriguing duality between the Hooke's law and the Newton's or Coulomb $1/r$ forces. In general, the oscillator model seems to be the most favored by physicists, and I shall dwell on it many times in various contexts.

Universality of some models is a manifestation of numerous physmatical polymorphisms, somewhat mysterious analogies hinting at the fundamental unity of all parts of physmatics. This may imply that increasing specialization and fragmentation of physical and mathematical sciences into tiny domains can become the principal obstacle to their development, like bureaucratic subdivision of the world into nationalistic states has become the main impediment to global progress. Being universally interconnected, physmatical models almost always produce sub models, i.e., even apparently simple models are organized on hierarchical principle and can be step-like refined. It is here that the vague boundary between a model and a theory lies: a theory incorporates a large number of interconnected models, just as a manifold incorporates a number of coordinate patches being glued together. The presence of sub models naturally leads to modularity and reusability - concepts that are well-known for software developers.

When treated more or less seriously, mathematical modeling should be endowed with some unifying paradigm which would be operational throughout the whole domain (ideally) or, at least, constitute the backbone of it. For instance, the finite-dimensional linear algebra serves as a backbone for numerics and so-called scientific computing. In the domain of mathematical modeling, the dominant paradigm is a greatly simplified dynamical systems theory. We shall discuss dynamical systems in some detail in Chapter 4 largely devoted to classical deterministic mechanics; here I merely state some salient features of this, now more or less traditional, approach to mathematical modeling.

When speaking about mathematical modeling techniques, V. I. Arnold uses the terms "hard" and "soft" models giving an example of multiplication table as a hard model. One can find numerous examples of soft models in life, for instance, the statements of "the more the less" type as well as given by proverbs, aphorisms, and other common sayings. Thus the statement "A man with a watch knows what time it is, a man with two watches is never sure" may be considered an example of a soft mathematical model.[11] Nevertheless, one should remember the famous aphorism by Voltaire "A witty saying proves nothing", so the value of common sayings as soft models is very limited.

## 2.2  Mathematical Models in Physics

Almost a century ago, Lord Ernest Rutherford somewhat arrogantly subdivided science into physics and collecting stamps. This contemptuous remark can be interpreted in the following way: "physics" designates a group of experimentally-oriented sciences accompanied by a well-developed theoretical overlay. "Collecting stamps" denotes a group of disciplines that accentuate the descriptions and accumulation of data, for example, old-time biology, orthodox medicine, geography, and many others. Models accepted in such disciplines are mainly reduced to classifications, which seems to be the primary stage of a good theory, and one can observe that traditional disciplines that could be recently attributed to the "stamp collection" class continuously evolve towards "physics" (although with different speed). The theory wrapped over the "raw" reality is quantitative and endowed with predictive capabilities. Thus although the principal results are obtained through observations and dedicated reproducible experiments (as performed by Galileo, but in modern times relying on powerful scientific instrumentation), one should not think that theoretical models mostly fulfil a decorative function in physics as it was frequently claimed by administrators. In fact, experimental achievements and modeling results are difficult to separate, and it is this blend - not the experimental results alone - that can be converted into ubiquitous technologies. Unfortunately, the role of mathematical knowledge in technologies is often hidden.

---

[11] I like especially the saying "Power tends to corrupt, absolute power corrupts absolutely" which may be regarded as a soft model with a very large area of applicability.

Modeling is an art of simplifying a complex system, and mainly due to this, models play a very important part in physics and technology. A good model enables one to understand the most fundamental physical parameters of a complicated process. Mathematical modeling may be loosely defined as an idealized description of reality constrained by mathematical concepts. All approaches to study the reality, both in the form of natural or behavioral sciences, which are using mathematical tools may be declared as mathematical modeling. Many people even assert that there is no mathematical modeling, but just instances of applied physics and mathematics under a fancy - and sometimes fashionable - name. People claiming that they pursue some scientific purposes are just replacing a natural (or social) object by its model and studying the latter. There is an element of deceit in this process. No model fully describes the studied phenomenon, and models of physics are no exception. Therefore, by the way, the question of a model's applicability is usually highly nontrivial. The already mentioned successful solution by Newton of the mysterious Kepler problem inspired thinkers and philosophers, primarily Laplace, to make bold generalizations and to develop the mechanistic model of the universe. There was no room for randomness in this clockwork vision of the world. The biggest challenge of biology, medicine, society, and economics is that randomness leads to fine-tuned processes (in time) and structures (in space). It means that the notion of the world as a machine seems to be inadequate. In fact, fully mechanistic nature would be incompatible with life, where evolution gains order through fluctuations. Classical science mostly studied systems and their states close to equilibrium, and that allowed one to construct a beautiful collection of comparatively simple physical models for the world. Such models depicted the systems that reacted on perturbations more or less predictably: these systems tend to return to equilibrium (in the parlance of statistical physics, they evolve to a state that minimizes the free energy). However, systems close to equilibrium can describe only a small fraction of phenomena in the surrounding world; in fact, it is a linear model. In contrast, nonequilibrium systems are ubiquitous in nature. Any system subject to a flow of energy and matter can be driven in the nonlinear mode, far from equilibrium. For example, open systems such as the Earth, living cell, public economy or a social group exhibit highly complex behavior which is hard to model mathematically using the methods adapted mostly to mechanical patterns. Most of the processes in the open systems far from equilibrium are interrelated, nonlinear, and irreversible. Often a tiny influence can produce a sizable effect. One more typical feature of systems far from equilibrium is that they can lose their stability and evolve to one of many states. This behavior appeared so "unphysical" from the habitual viewpoint that many orthodox physicists were inclined to despise those colleagues who were trying to consider systems far from equilibrium, especially those beyond physics. Yet, to model the processes in the real world, one must learn how to describe the systems far from equilibrium. We shall deal with systems far from equilibrium many times in this book.

Physicists generally dislike the term "mathematical modeling" because of its overwhelmingly general applicability. Everything can be declared mathematical modeling. This is, of course, true. The most enthusiastic proponents of mathematical modeling were mathematical physicists, specifically those belonging to the well-known school of A. N. Tikhonov and A. A. Samarski. Unfortunately, I could not find any good cliometric work on mathematical modeling in physics where the popularity rises and falls of this concept would be monitored. Neither could I find a quasi-definition of what a mathematical model really is. Such a definition will be probably useless, yet I would prefer to have a unified consideration. What I understand by mathematical modeling in this book is building compartmental fragments of our representation of reality based on clear-cut assumptions and discussed in mathematical terms, predominantly with differential equations. One must remember that models are not reality and should not be perceived as such. In some extreme cases, models may have nothing to do with reality. There are two favorite mathematical models which play an outstanding role in physics: the harmonic oscillator in its linear part and the logistical model in its nonlinear part. These two privileged models are at the same time simple and rich, giving rise to many theories and connecting diverse domains. I shall discuss both of them thoroughly on a number of occasions. In the single-electron energy band theory of solid state, the favorite model is that of Kronig and Penney, which beautifully illustrates the spread of individual energy levels and zone formation. There also exist other favorite models which are applied in many areas of science and we shall play with them as well, however, not very thoroughly. Here I typically start from a qualitative discussion of a physical phenomenon to be modeled, then produce physical examples and arguments and only after that I attempt to provide the mathematical results.

By the way, one can easily find a situation when the qualitative discussion, though seemingly plausible, gives a wrong answer. For instance, there is a folklore question: you have a hot cup of coffee and want to cool it to the drinkable temperature, will the coffee be cooled faster if you leave it as it is and pour cold milk into it right before drinking than if you pour the same amount of cold milk right away? The usual qualitative answer is "yes", presumably because the difference of temperatures is higher if you leave hot coffee intact, but let us see what mathematics says. Let hot coffee have the initial temperature $\Theta_0$ which is higher than the environment temperature $T_0$ so that $\Theta_0 - T_0 > 0$. To produce a mathematical model corresponding to the coffee-with-milk situation we may assume that cooling subordinates to a linear law (usually ascribed to Newton):

$$-\frac{dT(t)}{dt} = k(T(t) - T_0),$$

where $T(t)$ is the coffee temperature at moment $t$. This is a simple first-order inhomogeneous differential equation with a parameter $k$. One can also consider for simplicity the milk to be in equilibrium with the environment and has therefore the ambient temperature $T_0$. This assumption is not essential and serves only to simplify the formulas. The general solution to the respective homogeneous equation is $T(k,t) = Ce^{-kt}$, and if the ambient temperature $T_0$ is constant, we have a physically natural particular solution to the inhomogeneous equation, $T(t) = T_0$. Then the general solution of the inhomogeneous equation corresponding to the initial temperature $T(0) \equiv \Theta_0$ is

$$T(t) = T_0 - (\Theta_0 - T_0)e^{-kt}$$

To further simplify our model we may assume that coffee and milk have the same density as well as specific heat values. Corrections in terms of the respective density and specific heat differences would be an unwarranted excess of accuracy. Thus, we have

$$TV_c + T_0 V_m - (V_c + V_m)\Theta - V\Theta,$$

where $V_c, V_m$ are the partial volumes of coffee and milk, respectively, $V = V_c + V_m$ is the total volume, and $\Theta$ is the temperature of coffee with milk. Let us now, to answer our first question, compare two cases.

1. We leave our coffee cooling by itself (according to the above equation for $T(t)$) and pour milk into it just before drinking. The coffee-with-milk has the temperature

$$\Theta_1(t) = \frac{V_c(T_0 + (\Theta_0 - T_0)e^{-kt}) + V_m T_0}{V_c + V_m}$$

2. We pour milk into the hot coffee right away, at $T = 0$, then the starting temperature becomes $\widetilde{\Theta}_0$ (a physicist might say is "renormalized") where

$$\widetilde{\Theta}_0 = \frac{V_c \Theta_0 + V_m T_0}{V_c + V_m},$$

and in the process of cooling the coffee-milk mixture reaches by moment $t$ temperature value $\Theta_2$

$$\Theta_2(t) = T_0 + (\widetilde{\Theta}_0 - T_0)e^{-kt}.$$

Now, the most astonishing thing happens. By simple algebraic manipulations one can show that $\Theta_1(t) = \Theta_2(t)$ for all moments $t$. Indeed,

$$\Theta_1(t) = T_0 + \frac{V_c}{V}(\Theta_0 - T_0)e^{-kt} = \Theta_2.$$

In other words, it does not matter whether you pour milk in your coffee first or last thing before drinking.

The model described above represents an example of semi-useless models. It is just a simple mathematical record of an everyday observation. There exist, however, totally useless models such as simulation of smoke rings puffed by a cigarette smoker. Constructing an exhaustive mathematical model for such a process would involve considerable difficulties, yet probably its analysis would produce a superfluous knowledge. There exist probably many instances of such extraneous knowledge, but the difficulty is to *a priori* diagnose it as such.

The reason why I am writing about mathematical modeling in physics at all is that, contrary to the general view that physics is centered around the laws of nature, I observe that it deals mostly with their mathematical models. Thus, Newton's law is a successful mathematical model that can be applied to a body moving under the influence of other bodies in the low-energy limit, rather than the universal law of nature such as, e.g., certain symmetry properties of the universe, more exactly, of its part available to observations. Likewise, the Schrödinger equation is a typical mathematical model, also quite successful when applied for description of atomic and subatomic scale particles in the low-energy limit. The wave function standing in the Schrödinger equation is not directly observable and is a typical attribute of a mathematical model. An extreme case of such a model - the wave function of the universe, for example, in the form suggested by J. Hartle and S. Hawking [67], see Chapter 9, is just an interesting speculation and hardly a verifiable model. In this sense, the model of Hartle and Hawking is more like a philosophical concept expressed in the mathematical language (untypical of philosophers) rather than a physical object, that is the one to be experimentally validated. This kind of model building reflects a current trend in physics, where the quest for experimental proof is largely subdued. A huge generation of armchair physicists, focusing almost exclusively on formalism and symbols, produce numerous pieces of writing without much attention to experimental techniques. For them there is no question of participating in and observing the mundane work of experimentalists, with all its tinkering and small cunnings. Armchair physicists are mainly motivated to express themselves so that the subjects are treated for their own sake and glory, with little attention paid to firm experimental knowledge or applications. The produced symbolic forms can also create powerful myths[12] and are closer to art than to physics. Moreover, it is still a philosophical extrapolation, a widely spread belief that the models developed by humans and rather arrogantly called the "Laws of Nature" should be necessarily applied to the entire universe. This statement may be considered a belief because it has never been proved, only examples are generally used to corroborate it, and isolated facts by themselves mean very little. Nevertheless, to say something against the belief in universality of our "Laws of Nature" is merely a *mauvais ton*, for example, in the community of astrophysicists, although nobody can deny that astrophysics in general achieved really great successes[13]. Yet, when one stops questioning the truth, it easily transforms into dogma.

---

[12] Examples of such powerful theoretical myths are, e.g., anthropic principle, broadly interpreted many-worlds interpretation of quantum mechanics, and possibly superstring/M-theories, to say nothing of such pseudoscience concepts as time travel, teleportation of macroscopic bodies, or military use of torsion fields.

Once we have mentioned the Schrödinger equation in the modeling context, let me make the following remark. The situation with the Schrödinger equation is very indicative: this equation works very well in a great number of situations so that the "shut up and calculate" approach to quantum mechanics has become quite popular among physicists. Yet we know that the Schrödinger equation is unsatisfactory in many respects. For example, it describes interaction between the particles purely phenomenologically, with a coefficient $U(\mathbf{r})$, it does not account for the quantum vacuum, it is not at all trivial to get the classical limit, there are still many questions in regard to the so-called Bohmian version of quantum mechanics, etc. We shall discuss the limitations of the Schrödinger equation in Chapter 6.

The greatest question of all is, probably: what are the ultimate laws of physics? And I don't think anybody has ever provided the final answer, although there have been many distinguished candidates. This question may be answered in the language of mathematics, like Newton's law was formulated as a mathematical model, but one cannot exclude that the answer will be provided in some other form. It is possible, as Professor John Wheeler, an outstanding physicist famous for his unexpected thoughts, once predicted (see section "Prognosis" below), that if and when people eventually happen to learn the ultimate laws of physics they will be astonished that these laws had not been known from the beginning - so obvious they will look.

I have already noted that nobody in fact can be certain that the ultimate underlying laws of physics from which everything we know can be derived do really exist. Even if such a set of final physical statements does exist, it is not obvious that it will be useful for concrete applications. The obvious example is thermodynamics as an engineering discipline. For engineering applications, thermodynamics can be based on the 17th century phlogiston model. Phlogiston, a hypothetical fluid conjured up only to explain the spread of heat, was a typical mathematical - not physical - model. Indeed, the heat propagates in matter in the same way as a fluid diffuses through other media: equations describing both processes are the same. Therefore, an *ad hoc* model of phlogiston could satisfactorily describe plenty of idealized thermodynamic processes. However, the fact that this model was only mathematical and not physical (or "physmatical" - in our fancy language) eventually backlashed. The phlogiston model encountered insurmountable difficulties. Nobody succeeded to observe the fluid called phlogiston or to measure its properties such as density, even in indirect experiments. A lot of heat-generating events, e.g., friction, dissipation in general, electric discharge, etc. could not be explained by the phlogiston model. So, the latter had to be abandoned in favor of the molecular-kinetic theory.

---

[13] Actually, it is in astrophysics that processes are abound when our conventional models for physical laws cease to hold. For instance, it is unlikely that one can apply Newton's law of motion to describe the phenomena accompanying the collision of black holes located in the centers of galaxies.

I think the phlogiston history is rather instructive, because a relatively successful mathematical model had to gracefully fade away, succumbing to a more physically compelling theory. Incidentally, the same process, but perhaps more painful, was observed in connection with the model of ether. The phlogiston model, a clever gadget which was nonetheless totally wrong from the physical point of view, would have been more adequate than molecular dynamics models of the future, when all molecular trajectories might have been computed. However, now we know that even if one can, in principle, predict the behavior of each molecule, the value of such a prediction for engineering applications will be very low - in engineering models based on thermodynamics people are interested in such quantities as temperature or enthalpy and not in molecular trajectories or quantum particles. For every phenomenon, one may have many possible levels of description, which makes the physmatic modeler's life more interesting but not easier. Physics looks as not simply a good reason - it is much less than that. It is reason constrained by experiment, so one must give pragmatic answers and withdraw from being carried away by mathematical possibilities. The goal of physically-based mathematical modeling (PBMM) may be formulated as to abstract from the tremendous wealth of variables leaving only those relevant to the research objectives. We can very well see the implementation of this principle in the works of masters of physics, a good example for me personally was the collection of papers by J. W. Gibbs ([71]), see more minutes in Chapter 7. This is in fact an intuitive construction of a multilevel hierarchical system of models. It is interesting to notice that modern - exponential - representation of numbers is successful in computational techniques just because it is hierarchical, in distinction to ancient Roman representation that makes arithmetical operations cumbersome. Nowadays the multilevel methods with different degree of detail and hierarchical ladder of abstraction are increasingly popular in computer simulations (integrated circuits, traffic, fluids, fusion plasma, weather, complex boundaries, etc.). One must, however, be very careful about which details are discarded. A single erroneous assumption about a detail that was arbitrarily assumed to be irrelevant can totally invalidate a model. One can see this effect in computer simulations where multitudes of beautiful images can be obtained - quite easily nowadays, but the physical or engineering value of such visualized artifacts might be quite doubtful, to the chagrin of many computer scientists. It is not a coincidence that computer science (informatics) departments at the universities are usually not in high demand among physicists, mathematicians or mechanical engineering people - the specialists in the respective fields prefer to develop their own computer simulations, often less artificial and based on solid knowledge of the subject matter.

One of the fundamental problems of classical physics was the stability of atoms. Why do the electrons not fall onto the nucleus? This notorious problem plagued physics in the beginning of the 20th century. We shall illustrate the instability of the "classical" atom as a good example of a local model, apparently free of contradictions but nonetheless producing wrong results. This wrong result was an indication that a more global model or the whole underlying theory should be modernized or abandoned. The point electron had already been discovered (by J. J. Thomson) and the model of point nucleus had been experimentally corroborated by E. Rutherford. To explain the stability of atoms, an entirely different approach should be summoned, and indeed, this fact was explained by quantum mechanics which is based on totally different mathematical models (see Chapter 6). Speaking very roughly, the stability of the atom is the consequence of noncommutativity of quantum mechanical operators, in this case of nonrelativistic kinetic energy of electrons and Coulomb potential energy between the charged particles. The stability of the atom may be formulated as the existence of a finite lower bound for the energy of electrons in an atom. Due to noncommutativity of kinetic and potential energy operators, any attempt to make the electrostatic energy very large negative would require that an electron be localized near the nucleus, but this would result in even larger increase of the positive kinetic energy. In short, one of the first successes of quantum mechanics was that it explained the stability of the atom. Now, let us see why classical models are inadequate even at the naive atomic level.

As mentioned before, physical systems can be roughly separated into two classes: particles and fields, these are two basic models. This separation is really very crude, primarily because there is an overlap, for example, when particles are treated as field sources or excitations of some gauge field. We shall discuss the difference between particles and fields more thoroughly when talking about fermions and bosons and their possible mixing within the supersymmetry framework. Although, to frequent complaints on the part of mathematicians, there seems to be no good definition of the notion "field" in physics, each physicist intuitively understands the main difference between particles and fields which lies in the number of degrees of freedom. Any physical system consisting of a finite number $N$ of particles has only finite number of degrees of freedom, $n \leq 3N$. Recall that the number of degrees of freedom is defined as dimensionality of the configuration space of a physical system, e.g., a system with 2 degrees of freedom is $\ddot{\mathbf{x}} = \mathbf{F}(\mathbf{x}, t)$ where $\mathbf{F}$ is a plane vector field, $\mathbf{x} \in \mathbb{R}^2$. The field, on the contrary, is characterized by an infinite number of degrees of freedom. We shall discuss the implications of this fact in Chapters 5 and 9.

Although in the foundation of mathematical model building lies the reduction of the complex initial real-life problem to some idealized scheme, typically with clearly defined input and output, which can be treated by comparatively simple mathematical means, many models in physics and technology are by necessity bundled - they incorporate concepts from different fields of physics. For instance, cosmological models as a rule cannot be compartmentalized to just general relativity; one has to include quantum mechanics, statistical physics, and sometimes high-energy physics considerations. In particular, different models of phase transitions involving cosmological phase transitions, which have given rise to inflationary models, may be reiterated. Superconductivity (which is also a model of phase transition) is also based on bundled models, mostly combining those of quantum mechanics, solid state, electrodynamics and statistical physics. These inter-field bundles make mathematical models in physics rich and flexible. We shall illustrate the richness of the model bundles, in particular, while discussing the interaction of radiation (both electromagnetic and corpuscular) with matter (Chapter 8).

As I have already mentioned, all physical laws are, in today's terminology, just mathematical models, although underlying ideas are not necessarily formulated in mathematical terms. The final questions are often put not by physicists, but by philosophers though maybe in a vague form. There is the reverse trend to canonize mathematical models, make a fetish of them. A model, even an accurate simulation, is only a rehearsal of some reality, there may be many conceivable realities and, in particular, those that people might not have devised without modeling. In practical terms, models aim to help technologists (in the broad sense) predicting the operations and evade possible pitfalls. Therefore, one should be reminded of some risks. Quite often, people construct would-be good mathematical models that have only a very distant relationship to reality. I - and probably many other persons - have seen it many times. In such cases, mathematical modeling gives a wrong idea of what it means to solve an actual problem. A mathematical model is not uniquely determined by the investigated object or situation, and selection of the model is dictated by accuracy requirements. Examples of this relativism of models are abundant. In our everyday life, if we look at a table, its representation as rectangular or not depends on our practical needs; in ballistics, we may take into consideration or totally disregard the influence of the atmosphere; in atomic physics, it is relatively seldom that we have to consider finite dimensions of the nucleus; in military planning, one may regard a variety of models with different degree of "hardness", for instance, harmonic oscillator without attenuation is "harder" than oscillator with damping and nonlinearity. More exotic examples may cite regular armies (linear models), and the presence of partisans or terrorists (nonlinear models). Some results can be easily obtained using one model, but very difficult within another approach. The fact that it would be possible to find a nice solution to a highly simplified model we have built does not at all mean that we were able to obtain a practical solution to the problem we actually face in the real world.

So, one must be careful in taking the current physical laws as something invariable and firmly established. Since the laws of physics are formulated as mathematical models for the current level of knowledge, they can be eventually corrected. Physics is an experimental science, and any of its theory, presumably rigorous and absolutely exact, must be from time to time experimentally validated. However, no experiment can have an absolute precision, an experiment's accuracy is inevitably limited (e.g., due to technological constraints, noise, etc.). One may always expect a violation of the established views, an indication at novel interactions [14], necessity of radically new mathematical (algebraic or geometric) representations. Even the important "correspondence principle" which is a statement that each new theory must incorporate the old one as the limiting case can be broken - nobody has ever proved its absolute inevitability, and nobody knows its area of application. This principle may well be just a belief "adopted by repetition" or an obscurantist stereotype, even though contemporary physics essentially uses it to test new theories.

## 2.3  Ten Worlds of Physics

In principle, the following topics comprise the main part of today's physics: quantum mechanics, with its applications in atomic, nuclear, particle, and condensed-matter physics; relativistic quantum theory including the concept of a photon, the Dirac equation, electron-photon interaction (QED), and Feynman diagrams; quantum fields; and general relativity. Elementary working knowledge of these topics would be sufficient for a solid foothold in the physical community. However, this standard repertoire is very limited, and I shall try to list the major concepts which, in my view, constitute the real bulk of physics. I call these major concepts "worlds of physics" to be organized as

1. The classical world

2. The thermal world

3. The nonequilibrium world

4. The continuum world

5. The electromagnetic world

6. The plasma world

7. The quantum world

8. The high energy world

9. The relativistic world

10. The cosmological world

These ten worlds constitute a backbone of physics. Some of them will be discussed in this book more or less thoroughly, others (e.g., the continuum world or the plasma world) will be touched upon only superficially. Of course, the above list is highly subjective and not exhaustive. Below I ventured to bring the main topics inside each item. All these internal sub lists are also open and far from being complete, anyone can complement them in accordance with their tree-like structure. In the following lists, I give only a telegraph-style account of the main items, in the respective chapters we shall discuss some of the concepts only mentioned here in more detail.

---

[14] I am not hinting at profound studies of notorious torsion fields of course, because they are small and their interaction with matter is negligible, but looking for new forces is a legitimate physical task. The flip side of the coin is always pseudoscientific fantasies.

### 2.3.1    The Classical World

The classical world of physics is based on the following key notions:

- The Galileo group (inertial systems)
- Newton's law of motion (classical limit of special relativity and quantum mechanics)
- Newtonian gravity (classical limit of general relativity)
- The Kepler problem (rotation of planets about the Sun)
- Potential fields, classical scattering
- The Euler-Lagrange equations
- Variational schemes
- Noether's theorems and conservation laws, conservative systems
- The Hamiltonian equations, Hamiltonian flows on symplectic manifolds
- The Hamilton-Jacobi equation
- Motion on manifolds, constraints
- The Liouville theorem
- Key figures: Galileo, J. Kepler, I. Newton, L. Euler, J. L. Lagrange, W. R. Hamilton.

### 2.3.2    The Thermal World

- Classical thermodynamics (equilibrium)
- The nature of heat, temperature, heat transfer
- Mechanical work and heat, interconversion, engines and cycles
- Heat capacity ($C = \frac{dQ}{dT}$)
- Laws of thermodynamics, thermodynamic potentials
- The concept of entropy, reversible and irreversible processes
- Entropy production
- Thermochemistry, chemical reactions
- Equations of state
- Phase transitions, Ginzburg-Landau model
- Low temperatures, superfluidity and superconductivity

- Heat as the particles motion, Maxwell distribution, statistical mechanics

Key figures: L. Boltzmann, S. Carnot, R. Clausius J.-B-J. Fourier, J. W. Gibbs, V. L. Ginzburg, J. P. Joule, L. D. Landau, A.-L. Lavoisier, J. C. Maxwell.

### 2.3.3 The Nonequilibrium World

- The Liouville equation, Gibbs distribution

- Open systems

- Kinetic equations, Boltzmann equation, Bogoliubov's hierarchy

- Diffusion, Langevin equation, Fokker-Planck equation

- Fluctuation-dissipation theorem (FDT)

- Linear response theory, Kubo formula, Onsager's reciprocal relations

- Multiple scattering theory

- Nonequilibrium phase transitions, time-dependent Ginzburg-Landau model

- Classical stochastic models, nonlinear regime, branching and bifurcations, stability of nonequilibrium stationary states, attractors

- The Poincaré map, logistic model

- Dynamical chaos, indeterminism (impossibility of predictions)

- Dissipative structures, order through fluctuations, Turing structures

- Chiral symmetry breaking and life

Key figures: N. N. Bogoliubov, L. Boltzmann, J. W. Gibbs, Yu. L. Klimontovich, N. S. Krylov, R. Kubo, L. Onsager, A. Poincaré, I. Prigozhin, D. Ruelle, Ya. G. Sinai, R. L. Stratonovich

### 2.3.4 The Continuum World

- The Euler equation

- The Navier-Stokes equation

- Hyperbolic flow equations, shock and rarefaction waves

- Compressible gas dynamics and supersonic flows

- Self-similar models and explosions

- Turbulent flows and the models of turbulence

- Elastic solid models

- Viscoelasticity, plasticity, composites

- Seismic ray propagation and seismic ray theory

- Acoustics, sound wave/pulse excitation, propagation and scattering

- Detonation and flames, propagation of fires
- Superfluidity

Key figures: Archimedes, D. Bernoulli, L. Euler, A. N. Kolmogorov, L. D. Landau, B. Pascal, Rayleigh (J. W. Strutt), G. Stokes, Ya. B. Zel'dovich.

### 2.3.5    The Electromagnetic World

- Maxwell's equations
- Laplace and Poisson equations
- Interaction of electromagnetic (EM) fields with matter
- Electromagnetic response of material media
- Linear and nonlinear susceptibilities
- Linear and nonlinear optics
- Atoms and molecules in the electromagnetic field
- Electromagnetic wave and pulse propagation
- Diffraction and scattering of electromagnetic waves
- Electromagnetic radiation
- Rays of light, asymptotic theories, coherence of light
- Photometry and colorimetry

Key figures: A.-M. Ampère, M. Faraday, C. F. Gauss, O. Heaviside, J. C. Maxwell, Rayleigh (J. W. Strutt).

### 2.3.6    The Plasma World

- The plasma dielectric function, linear waves in plasma
- Screening in plasma, correlations of charged particles
- Hydrodynamic models of plasma
- Distribution functions
- Kinetic models of plasma, collision integrals of Boltzmann, Landau, Klimontovich, Lenard-Balescu, etc.
- Collisionless plasma, a self-consistent field model (the Vlasov equation)
- Plasma in external fields, the magnetized plasma
- Landau damping
- Theory of plasma instabilities
- Quasilinear and nonlinear models of plasma

Key figures: P. Debye, Yu. L. Klimontovich, L. D. Landau, I. Langmuir, A. A. Vlasov.

### 2.3.7    *The quantum world*

- The particle nature of electromagnetic radiation, the Planck hypothesis
- The duality of light, photoelectric effect (A. Einstein, 1905)
- The Bohr atom
- The de Broglie hypothesis, hidden parameters discussion, quantum trajectories
- The Schrödinger equation, wave functions
- Observables, measurements, probability amplitudes
- Wave packets, Heisenberg uncertainty relations
- The Heisenberg-Weyl algebra, representations of compact Lie groups (H. Weyl)
- The theory of unbounded self-adjoint operators (J. von Neumann), Hilbert space
- Rotation group representations, spin
- Sturm-Liouville problem, discrete spectrum, eigenfunction expansions, Green's functions
- The density matrix, the Wigner function (E. Wigner, 1932), Husimi and tomographic representation
- Unitary evolution, semigroups
- Eigenvalue perturbation theory, iterative procedures
- Quantum-classical correspondence, decoherence, canonical quantization
- Asymptotic expansions, semiclassical limit
- Scattering theory, S-matrix, continuous spectrum
- Integral equations, inverse problems
- Decaying states, resonances
- Periodic potentials (F. Bloch, L. Brillouin, H. A. Kramers), Floquet and Hill equations
- Solid state physics, semiconductors, transistors, engineering applications of quantum mechanics
- Many-body problems, second quantization, elementary excitations, condensed matter physics, Coulomb energy, thermofield dynamics
- Ergodic potentials, Anderson localization
- New developments, EPR and hidden parameters debates, Bell's inequalities, Bohm version

- Quantum computing

Key figures: H. Bethe, N. Bohr, L. de Broglie, P. A. M. Dirac, A. Einstein, V. A. Fock, G. A. Gamov, W. Heisenberg, L. D. Landau, J. von Neumann, W. Pauli, M. Planck, E. Schrödinger, E. Wigner

### 2.3.8   *The high energy world*

- Strong (nuclear) interaction, $\pi$-mesons, exchange forces, Yukawa's model

- Resonances, super multiplets

- Baryons, mesons, hyperons

- *CPT*-theorem, group concepts in particle physics

- K-mesons, particle mixing, *C* and *P* non-conservation, *CP* and *T* violation

- Isospin, strange particles, "elementary particle zoo", SU(2), SU(3) - first attempts of Lie group classification

- Early quark models (1960s), color, flavor, charm, etc.

- Hypercharge, Gell-Mann - Nishijima relation

- Cross-sections, formfactors, *S*-matrix, current algebra

- $J/\psi$-meson, confirmation of charm, quark-antiquark system

- Quark-quark interaction through gluons, quark-gluon plasma

- Quantum chromodynamics (QCD) confinement, deconfinement, asymptotic freedom

- Electroweak interactions, the Standard Model, W- and Z- bosons, spontaneous symmetry breaking, Higgs particle

- Non-abelian gauge theories

- Grand Unification and new proposed theories and models: SO(10), left-right model, technicolor, SUSY, etc.

- String and M theories

Key figures: L. D. Faddeev, R. Feynman, M. Gell-Mann, S. Glashow, J. Goldstone, P. W. Higgs, G. t'Hooft, R. L. Mills, A. M. Polyakov, A. Salam, S. Weinberg, E. Witten, C. N. Yang

### 2.3.9   *The relativistic world*

- The Michelson-Morley experiment

- Lorentz transformations, relativistic kinematics

- Special relativity, Einstein's paper "Zur Elektrodynamik Bewegter Körper (On the Electrodynamics of Moving Bodies)", Annalen der Physik 17, pp. 891921

- Minkowski space, Poincaré group

- General relativity, Einstein's paper "Die Feldgleichungen der Gravitation (Field Equations of Gravitation)", Königlich Preussische Akademie der Wissenschaften, pp. 844847

- Redshift of spectral lines, deflection of light, time delay by gravitation

- Relativistic mechanics, $E = mc^2$, relativistic mass

- Accelerators, nuclear physics

- The Dirac equation, quantum vacuum, positron, antiparticles

- Relativistic quantized fields, particles as field excitations, bosons and fermions, spin-statistic theorem

- Quantum electrodynamics, particle-field interactions, Feynman diagrams, renormalization

- Path integrals, Feynman-Kac formula

- Gauge models, Yang-Mills theory

- Higgs particle, the Standard Model

- New quantum field theories, scalar, vector, tensor fields

- Gravitational radiation, gravitational wave detectors

- Strings and superstrings

- Controversies over relativity, tests of special and general relativity

Key figures: P. A. M. Dirac, F. Dyson, A. Einstein, R. Feynman, D. Hilbert, H. A. Lorentz, H. Minkowski, J. Schwinger, S. Weinberg

### 2.3.10   The Cosmological World

- Spacetime curvature, the spacetime of general relativity

- Solutions to Einstein's equations

- Early cosmological models, non-stationary metric, redshift, Hubble constant, FLRW (Friedman-Lemaître-Robertson-Walker) isotropic model

- The Big Bang, expanding universe, time asymmetry, the universe evolution

- Relic radiation, cosmic microwave background

- Black holes, escape velocity, Schwarzschild solution, Chandrasekhar limit, event horizon, spacetime singularities

- Astrophysics, radio, optical, infrared images, gravitational lenses

- Early universe symmetry breaking, cosmological phase transitions, topological defects, cosmic strings and structures

- Anthropic principle and other speculations (e.g., large numbers)

- Cosmological constant, vacuum energy, inflationary models

- Hartle-Hawking wave function

- Quantum gravity

- Strings and extra dimensions

- Universe: finite (Aristotle) or infinite (Giordano Bruno)

- Dark energy, dark matter (hidden mass), WMAP

Key figures: S. Chandrasekhar, A. A. Friedman, S. Hawking, E. Hubble, P. Laplace, G. Lemaître, R. Penrose, W. de Sitter.

Although these "worlds of physics" have links to each other, it is still difficult to combine current physical theories into a coherent picture. For instance, classical mechanics, quantum mechanics, classical field theory, quantum field theory, high energy physics (the Standard Model), general relativity, string/M theory are all essentially different. These are a basic set of theories that can be constructed independently of one another. Each of these theories may be regarded as a cluster of intrinsic models. One can, if desired, even find notorious contradictions between models belonging to different clusters (i.e., built up on the base of different theories). It has already been mentioned that particles in physics should be treated as point-like in classical relativistic models and as extended in quantum models. And in general, is the world finally classical or quantum? Should one take quantum mechanics seriously or is it a well guessed collection of calculational prescriptions? It is clear that, for example, the Schrödinger equation is a successful guesswork. Another contradiction is associated with fixed background metric of special relativity and "old" quantum field theory, which is badly compatible with the dynamic spacetime of general relativity. Some physicists, mostly trained in high-energy physics, are to these days reluctant to accept the geometric nature of general relativity [133] preferring to treat it as a usual quantum field theory (i.e., a Lagrangian theory on a fixed Minkowski background). This point of view obviously defies the Einstein guiding idea that spacetime has dynamical properties of its own and cannot be regarded as a passive background. We shall discuss this problem in Chapter 6 in some mathematical detail.

There is also a notorious irreversibility paradox, a contradiction between time-reversal invariant microscopic models of physics (the so-called laws of motion) and phenomenological time non-invariance observed in our everyday experience. This paradox stirs a lot of controversy up till now and does not seem to be ultimately resolved, although I think it belongs to an extensive class of pseudo problems. We shall devote some space to the discussion of the irreversibility paradox in Chapter 7, too.

The lack of a coherent picture can be tolerated in cookbook disciplines such as scientific computing, numerical mathematics, networking protocols or computer graphics, but is hard to be meekly accepted in physics which strives to provide a unified image of the world. Local experimental successes, albeit quite impressive as, e.g., in quantum field theory (QFT) do not soften the frustration, and typical compensatory reactions in the form of escape from empirical reality are more and more often observed. People easily accept fanciful anti-empirical speculations such as multiple universes, a great number of dimensions or going back in time. Nevertheless, I don't think it reflects a "trouble with physics" or some sort of a crisis. Simply a unique hierarchical principle or a dreamlike super-formula from which all physical theories could be derived as specific cases may not exist. The networking construction connecting a variety of mathematical models of physics seems to be more plausible, at least nowadays.

## 2.4  Physics-Based Mathematical Models (PBMM)

The world is, in principle, inseparable, and the ultimate purpose of physics is to find its meaning. It was presumably Einstein's dream - to grasp how God had designed the world. Contrary to that synthetic view, modern physicists (with very rare exceptions) never attempt to understand everything at once. One cannot be sure that there are in general universal, model-free structures in physics. Being by tradition strictly limited to the phenomena that occur in non-living nature, physics has always striven to produce models describing such phenomena in relatively simple terms. Indeed, while trying to model some entity, it is wise to think about the simplest at first. Nevertheless, one can apply the same approach to studying more complex phenomena than only those encountered in inorganic nature[15]. In this book, I treat physics not as a closed discipline, but rather as an intelligent approach to the entire human environment, that is to say this book may be considered as a form of a protest against the physical isolationism. And naturally, one can touch upon things that traditionally have nothing to do with physics in the narrow sense such as biology, economics, or even sociology. This extension often infuriates those physics professionals who are inclined to regard physics in the traditional "dead" sense as an ample arena for building even the wildest models. Representatives of other professions, primarily humanities and medicine, are also not always happy when physicists intrude into their "comfort zones". Of course, physics is so vast that one can easily spend one's professional life within any of its small and seemingly isolated subfields. However, our "physmatics" approach presupposes the study of links between different disciplines. Accordingly, in chapter 11, we shall discuss a few interesting mathematical models applied to the phenomena lying outside physics, if the word "physics" is still intuitively understood in the narrow sense, implying exclusively the subjects which I have classified for myself as "ten worlds of physics" (see above).

Worlds of physics are just clusters of suitable models. There are, as we have discussed, links between worlds invoking substructures with repeatable, reusable patterns. These patterns have lately been spread outside physics. The tree of mathematical modeling in physics, with branches, leaves and buds as individual models, provides the primary networking structure (called "physmatics" in our somewhat fancy terminology). This situation is reflected in new notions, for example "econophysics", "physical economics", "chaos in finance", "nonlinear dynamics in the financial markets", "fractals and scaling in finance", "percolation model of financial market", "dynamical model", "self-organized fragmentation and coagulation of financial markets", "stochastic PDE for option pricing", etc. One can observe the rising popularity of traffic flow models which are mainly constructed on the principles generally adopted in physics (e.g., conservation laws). All these model clusters are quite important and undoubtedly their potential future impact may be difficult to overstate, so they deserve a special treatment being out of scope of this book.

---

[15]This has nothing to do with the usual physicists' arrogance, when people, especially young and not quite experienced but trained in some physics, claim that they can excel in many other areas such as chemistry, Earth sciences, economics, history, politology, sociology, business, etc.

The very idea of easy linking different subjects seems to be close to the purely mathematical idea of compatibility lying, e.g., at the core of quantum mechanics. The concept of compatibility reflects the consistency of approaches or, as in the special case of quantum mechanics, of measurements (see Chapter 6). The idea of compatibility of different things is also manifest in mechanics where the notion of complete integrability of Hamiltonian flows has recently become of great value (partly due to some overcompensation for the oblivion of classical mechanics and nonlinear science during the most part of the 20th century). More specifically, the flow must be included into a class of compatible (commuting) Hamiltonian flows. Analogously, it is a characteristic feature of integrable PDEs (such as those encountered in the theory of solitons) that they should be organized in families of compatible i.e., commuting type, mostly having an hierarchical nature. So, the general idea of consistency of different subjects manifests itself in mathematics as integrability that is to say the possibility to find a closed solution to a complicate problem based, e.g., on nonlinear partial differential equations. Even in the realm of linear PDEs, one can trace many links at the level of mathematical models related to absolutely different fields of physics. It would be, for example, a truism to observe the connection between the one-dimensional Schrödinger equation

$$-\partial_x^2 \psi(x, E) + V(x)\psi(x, E) = E\psi(x, E)$$

and the one-dimensional scalar wave (acoustic) problem

$$-\partial_t^2 u(t, \lambda) = \frac{V(t)}{\lambda} u(t, \lambda).$$

It is a sad fact that a great deal of phenomena can be explained only at the level of hypotheses. For example, current science does not exactly know what is inside of the Earth. Geologists have dug the whole planet, but this is only a thin surface layer. One can only assume what happened on the Earth when life subsisted in the forms different from today's, e.g., in the form of viruses and bacteria. The human organism is an entire universe of its own, with the interplay of physical, mechanical, chemical, and biological processes in concerted action. Despite the successful pace of modern medicine (largely due to physical instrumentation), understanding of these processes is very approximate and primitive. A great number of serious illnesses remain unexplained, and biomedical models can only describe the symptoms intuitively correlating them to previous cases. This is an unsatisfactory procedure leading to frequently occurring medical errors dangerous for the patients. Moreover, inexact knowledge invokes pseudoscience such as healers and "biofield correctors" in paramedicine, parapsychologists and clairvoyants in other fields. In Chapter 10, we shall discuss climate viewed as a physical system. Unfortunately, the climatic system is so complex that no one seems to understand its functioning, although prognoses, pledges and political rigmarole are abound. In space science, despite a fast accumulation of observational data, almost everything is still at the level of hypothesis, rather than can be perceived as a reliably established result. To illustrate this pessimistic statement, it would be enough to recall the "dark side" of the universe problems (dark matter, dark energy, black holes). The cosmological constant problem (and the related problem of vacuum energy density), the idea of quintessence, inflationary models, the anthropic principle and so on are discussed, despite sophisticated mathematical tools, still on the hypothetical level. Everybody can prolong this list with topics of inexact knowledge. However, inexact knowledge is not necessarily bad: it has a provocative role, fascinates curious persons, and stimulates them to play with intelligent models.

## 2.5    Theory, Experiment, and Models

*"I take the positivist viewpoint that a physical theory is just a mathematical model and that it is meaningless to ask whether it corresponds to reality. All that one can ask is that its predictions should be in agreement with observation." Stephen Hawking*

    What is the relationship between these three components of human endeavor attaining physical knowledge? The interplay between theory and experiment induces the creation of models that are used to simulate the observations and predict new features that can be observed in new experiments. A theory in general may be defined as a cohesive system of concepts that was experimentally validated to the extent that there is no unclearness or contradictions within the limits of applicability for this system of concepts. A good theory contains a heavy load of knowledge. For example, the mechanical theory (discussed at some length in Chapter 4), together with the whole set of initial values, allows one to calculate the positions of the planets for many thousands of years into the future and into the past with sufficient accuracy (limited by the Newtonian approximation). The real mathematical modeling probably began with the nuclear bomb construction in 1940s in the USA and USSR (below, I shall reiterate some relevant facts that I was able to come across). Now, the trend of increasing complexity and resorting to expensive (sometimes prohibitively) experiments, initiated

during nuclear weapon projects, calls for simulation of both the theory and experiment. Modeling in this field usually gives the following results:

- The theory is insufficient and must be modified, revised, improved

- A new theory is needed

- The accuracy of experiments is insufficient

- New and better experiments are needed

One may note in passing that each nuclear test explosion is nothing else but a physical experiment, rather than a military event. A physical experiment is intended to measure certain physical quantities. In general, the main physical quantities to be measured in nuclear test experiments are the medium (gas) variable density and velocity - as in more conventional fluid dynamics measurements. For example, the so-called Richtmyer-Meshkov instability, which develops when the interface between two fluids having different densities is struck by the propagating shock wave, has been one of the prime objectives in nuclear explosion experiments. (This instability was predicted around 1960 by R. D. Richtmyer, a well-known mathematical physicist, and experimentally confirmed in the USSR by E. E. Meshkov.) Nowadays, due to the nuclear test ban, nuclear explosions are mostly simulated on computers, with validation provided by small-scale local laboratory experiments.

Now, how is a theory related to a model? This is again a philosophical question - one of those that can induce lengthy discussions with no outcome. Nonetheless, I shall try to answer it as I understand this relationship. Models are usually constructed within the framework of a certain theory, for instance, the Friedman cosmological model is built up within the general relativity theory or the BCS model is a limited fragment of the non-relativistic quantum theory.

The keyword for a model is the result, the keyword for a theory is a framework. The merit of a model is that it can be explored exhaustively. A theory cannot cover everything in the universe to the smallest detail and does not necessarily bring specific results, it only provides tools to obtain them. For example, the whole solid state theory may be considered as a particular case of quantum mechanics (also a theory but higher in the hierarchy), however, concrete results are produced when specific models within solid state theory are considered. There is a hierarchy both of theories and models, with inheritance of basic features down the ladder. Of course, the classification of models and theories is not absolute and may be subjective. Thus, the single-electron model is partly a model, partly a theory, whereas the Kronig-Penney model is just a model. One of the best examples of a model related to a theory is the E. Ising model of ferromagnetism [16] where the theory provides a simple framework and one can obtain ultimate results assuming a simple model of spin interactions. Later we shall see that the Ising model serves as a pattern for a number of derived models in the areas having nothing in common with a ferromagnetic.

---

[16] The Ising model is probably the simplest possible model of a ferromagnet in two directions.

In contrast to the broad mathematical theories, mathematical models are much more specific, illustrative and closed. The ultimate – and the most difficult – skill of modeling is to convert a seemingly complex problem into a much simpler one. Or to dissect a big problem into a multitude of small and sharply cut ones. This modelization of reality does not go without penalty: some reality must be sacrificed in order to gain more understanding of the problem being modeled. Even worse: the scope of reality left outside of a mathematical model is in many cases unknown, so it is rather nontrivial to quantitatively establish the model's applicability unless the model is formulated in terms of analytical functions or asymptotic expansions when one can evaluate and quantify the corrections. More than that: a model may be understood as a set of constraints which is bound to be replaced by a new set in the further study, and one can often foresee how the old model would be discarded.

When a mathematical model is set up and processed, it results in the outcome of mathematical statements. Here, one must not forget that the validity of such statements is very limited - largely conditioned by the model. Nowadays, one can often observe the tendency to treat the model's outcome as absolute, which leads to a confusion and even grave mistakes. As an example, one might recall the deeply rooted stereotype that all physics should be time-reversal invariant, which is not true. It is just an arbitrary extrapolation of time-invariant models onto all observed phenomena. However, this simple belief is sometimes used as an instrument, for instance, to discard solutions containing terms with odd powers of frequency.

Mathematical theories, to the extent they are not part of physics, are considered true when they are accepted by a certain social group - a community of mathematicians, with only a negligible part of this community really understanding a theory in question. In distinction to such public acceptance, physical theories model reality and are on the one hand a product of observations and on the other hand a source of predictions in new series of observations. Ideally, a theory is based on measured results, which are not precisely known. Then the predictions made from such a theory are also not exact.

Scientific measurements - not necessarily in physics - have one profoundly lying common problem: one is limited in them by the type of experiments one is capable of performing. Even in the case of comparatively pure physical systems, the applied experimental techniques severely constrain our ability to investigate things. Take as an example one of the most precise experimental techniques, a scanning probe microscopy/spectroscopy (SPM/SPS) applied to explore solid surfaces. Even in such highly sensitive experiments, one has to admit the drastic information loss: to explore quantum localized states on the surface one has to apply voltage and produce the tunneling current which would then strongly perturb and modify such states. Thus, a genuine "non-demolition" experiment intended to study quantum (in this example) states in their pure, native form becomes highly problematic. This situation is typical not only of the quantum world. In the case of mental processes, for example, to produce a non-demolition measurement may be more difficult than in physics: there are a lot of components and interactions between them.

One can see that such disciplines as physics and mathematics, though incorporating considerable applied parts, are centered around basic research. So, the question naturally arises: is basic research in physics and mathematics a luxury for certain countries or a necessity for all the countries?

## 2.6    On the Relationship Between Physics and Mathematics

"I am so clever that sometimes I don't understand a single word of what I am saying." (Oscar Wilde)//

The relationship between physics and mathematics is, of course, a subject for philosophical discussions, with scientific content in this section being close to zero. I would only state the following observation: complicated models are seldom useful - maybe solely to produce the texts exhibiting scholarly merits of an applicant, e.g., in theses. Strangely enough, mathematics continues to stir powerful emotions not only in mathematical circles, but also among non-mathematicians including physicists. For the old guard, using more modern mathematical techniques such as differential forms or geometric methods in general seems to be similar to undergoing an unpleasant medical procedure. An appropriate use of mathematics appears to be unclear even to theoretical physicists to whom mastering math is an absolute must. This element of obscurity and ill-digested trials is more and more evident in contemporary papers on theoretical physics, being strongly aggravated by an obsessive drive towards abstraction and maximal generality fashionable among "pure" mathematicians (and being servilely acquiesced by some physicists). Mathematics, among the so-called exact science, is the only discipline that enjoys being abstract. Philosophy is also abstract, but it is in no way an exact science. The feature of abstraction results in the unique power of mathematics - generality that works. In distinction to physics and other exact sciences, the underlying content is immaterial in mathematics. For example, if one considers a rotation group and properties of its operations, it is irrelevant to ask whether one implies an electron, an atom, a nucleus or a molecule of some chemical substance, although all of them are totally diverse physical objects. This working generality of mathematics permits it to be applied throughout all other sciences. It permits thus to perform cross-disciplinary mathematical modeling.

The property of working generality leads to a special role of mathematical models in other sciences: these models combine the insight accumulated in specialized disciplines, primarily in physics, with the powerful abstraction of available mathematical structures - at least available to the developers of mathematical models. So far, such a combination worked amazingly well, especially in physics. Analogies stemming from generalization attempts are very important, they sharpen mathematical instincts and encourage one to look for counterexamples. Many intricate things in physics appear simpler due to analogies. For example, non-commutative algebra based on space-time variables $(x, p) \leftrightarrow (t, \omega)$ have the same mathematical framework and may be treated similarly, although the physical content is totally different.

The main ideas of using mathematics for model construction in physics are better grasped by discussing analytical models. Once we understand them, we can proceed to computational models and to specific cases, collectively named physical engineering. One can say that all this is the so-called applied mathematics. Of course, it does not matter how you denote your field of activities, but personally I don't think there is such a field as applied mathematics, there are applied mathematicians. For instance, quantum mechanics may be considered as applied functional analysis, cryptography as applied number theory, signal theory as applied theory of functions, and so on. Even a very abstract mathematics may turn its application side to physicists. Does a physicist really need topology? Traditionally, the work of a physicist consisted in describing models in terms of differential, rarely integral or integro-differential equations and then trying to solve them. However, to really produce the solutions to physical-based mathematical equations, applied to specific cases, is typically quite tiresome and may even prove impossible, even despite the fact that the starting-point equations of physics are as a rule well known and all their nice mathematical properties - smoothness of solutions, their localization, compactness, etc. - have been explored. It does not help much. Therefore, a number of ingenious methods have been developed, with attempt to analyze the solutions of complicated equations without solving them. There has appeared lately a number of human activities aimed at doing something without really doing it. Management, for example - it's the art of doing something with others' hands. Or software engineering - a popular movement under the slogan "how to produce a software system without knowing how to write a code". A very cunning strategy. I would call such an approach "creative alienation". One of the best examples of productively using this approach in science is group theory, which consequently exploits symmetry. Topology in physics serves more or less the same purpose of creative alienation from the honest process of directly solving physical equations (see below).

It seems only natural that a progressively widening chasm between mathematically minded physicists and physically minded physicists is to be observed. Mathematically minded people are the real insiders in math, they must know what is going on in mathematics and are attentive to all fine mathematical stitches, all those smooth and nice properties that fasten mathematical constructions that initially appear totally abstract but later foster the progress of physics. We shall see examples of such fostering in this book. In contrast to this refined approach, physically minded physicists prefer broad strokes, largely intuitive, not meticulously founded or even completely unfounded (we shall see some salient examples of mathematical inaccuracies in physics). This approach is typical of engineers for whom it is important to obtain a tangible result than to prove that it can (or cannot) be in principle obtained. Thus, the physically minded physicists, even theoretical physicists, are just mathematical engineers. And these two breeds of physicists are submerged in different environments: while mathematically minded people deal mostly with books and papers, those physically minded still have to understand experiments - physics never ceases to be an experimental science, despite all theoretical extremism. For a physically minded person, it is harder to deal only with books than for a mathematically minded one. There are almost no drawings in most mathematical books, which is uncommon for an engineer, even mathematical engineer. And the basic questions for these two breeds of people are different: "how would I prove it?" - for a mathematician and "how would I do it?" - for a physicist.

There has been written a lot about the interaction between mathematics and physics. One of the best statements about this relationship I (as well as many friends and colleagues of mine) have read was the inscription left by some anonymous thinker on one of the student tables at the Central Physical Audience in the Physical Department of the Moscow State University: "Physics without mathematics is the same as a naked person in the metro: possible but indecent". Here, one can sense a hint at a somewhat troubled relationship. Personally, I still think that mathematics stems from physics, at least even today physics serves as the greatest source of mathematical ideas. Professional mathematicians are usually infuriated by this statement claiming that mathematics is totally different from physics. Nowadays there are new sources of inspiration for the mathematicians originating from other disciplines such as economics, telecommunications, networking, computer science, social studies, defense and military planning. We shall review some mathematical models related to these disciplines and we shall see that all such models to a large extent bear the imprint of approaches developed in physics. It is surprising to me that vocal objections to contributions from theoretical physics as the main supplier of mathematical problems seems to be a *bon ton* among mathematicians. "Theoretical physics is a science locally isomorphic to mathematics" is the most favorable saying I have heard recently from mathematicians. Mathematicians and physicists are not always friendly tribes. Is a new kind of religious cold war pending?

This is my hyperbolization, of course. Sometimes, however, human incompatibility of physicists and mathematicians reaches a rather high degree. To illustrate this strange feud one can recall the following episode from the history of the Soviet nuclear weapons project. Needless to say, what importance was assigned to this project by the Soviet authorities. In the 1950s, the theoretical laboratory headed by I. M. Khalatnikov, a well-known physicist and one of the closest Landau disciples [17] was performing calculations of the design and physics of the Soviet hydrogen bomb. The lab employees worked at the Institute of Physical Problems but, due to political intrigues, the whole laboratory was transferred to the Institute of Applied Mathematics formed in 1953 specifically for nuclear weapons and missile computations (now the Keldysh Institute of Applied Mathematics). Rather acute conflicts between newcomer physicists and aboriginal mathematicians followed, and the physical laboratory could survive in the milieu of mathematicians for only half a year. Owing to the personal order of I. V. Kurchatov, the head of the Soviet nuclear weapon project, the entire lab of Khalatnikov was transferred again - without mathematicians, this time to the institute headed by I. V. Kurchatov (now the Russian Scientific Centre "Kurchatov Institute").

---

[17] I. M. Khalatnikov is mostly known for his classical works on superfluidity, superconductivity and other issues of low-temperature physics. He has been a long-time director of the famous L. D. Landau Institute for Theoretical Physics.

Let us conjecture a little bit about what physicists (and other natural scientists) might think regarding mathematical methods in their respective sciences. Mathematics is a wonderful language, but does it really produce exact results? Mathematics can correctly - without singularities, infinities, and unsurmountable difficulties - describe only very simple and artificial physical situations. Our everyday phenomena like the transition to turbulence in the flow of water streaming down from the tap or the friction force are beyond the reach of mathematical description. One can produce more examples where mathematics fails than where it admits a successful theoretical description. Just try to construct a mathematical theory of a flag fluttering in the wind. Or of a living cell irradiated by a laser pulse. Or even of an isolated living cell. Complexity of biomedical, social, economical or cultural systems is merely of a different level than that attained in contemporary mathematics, and it is not incidentally that the so-called serious scientists usually express astonishment mixed with contempt when hearing that some of their colleagues had ventured to attack a "weird" social problem by mathematical means[18]. In the kingdom of Soviet physics, such people were outlaws.

It is because of this complexity level unreachable for contemporary mathematics that we have to focus primarily on particular mathematical models of reality than on new theories[19]. Yet, frankly speaking I think we already have enough theories. Mathematics in its actual form appears to be perfectly adapted to describing general principles such as motion equations and conservation laws. The study of symmetries on a certain complexity level can also be achieved by mathematical means. Nevertheless, I doubt that such general objects as spacetime, universe, quantum background, mass or - even worse - complex sociocultural systems can in principle be investigated by theoretical mathematics. It may simply not work, that is to say, standard mathematical methods may be compromised by infinities and singularities. Then probably only ad hoc "cookbook" methods and local mathematical models (e.g., not based on any regular theory) would be possible. This would be a very unfortunate development, of course.

"A mathematician may say anything he pleases, but a physicist must be at least partially sane", this statement is ascribed to J. W. Gibbs. In that sense, string theory, for example, is more mathematics than physics. Probably today it would be better to say "string/M theory", which sounds even more mysterious, is even farther alienated from experimental roots of traditional conservative physics and based almost uniquely on purely mathematical considerations. Where is the demarcation line between physics and mathematics in the new theoretical physics, especially after the so-called second string revolution, I don't know. Although I devoted a considerable amount of time trying to understand new concepts in physics, of course, I am familiar with the string/M theory only extremely superficially. Like many other amateurs, I failed to find any experimental confirmation or astronomical observations in favor of string theory (except some attempts to predict the value of the cosmological term, see Chapter 9). Besides, does anyone know exactly what the M-theory is? The rapidly increasing number of string theorists is, to my mind, a very weak indicator of physical soundness of the string theory (see a well-known book by L. Smolin [74] on this subject).

---

[18] The disdainful attitude towards these persons exceeded that semi-openly manifested in regard to those who became interested in the history or (even worse) philosophy of science. Such persons were regarded as just impotents and weaklings, whereas the misguided thinkers over fanciful social mathematics were simply considered inadequate.

[19] It does not necessarily mean that the models should be phenomenological - they may be based on first-principle equations.

I am not going to even try to formulate the axiomatics of the current string/M theory. As near as I understand, this theory may be regarded as a speculative area of theoretical physics beyond the traditional theoretical or even mathematical physics developed in the course of the 20th century and closely related to experimental real-life problems. In general, theoretical physicists try to describe the world inventing models of reality which are used to explain, rationalize and, in the best case, predict physical phenomena within a certain "physical theory". As for physical theories, one usually distinguishes between three sorts of them: basic or mainstream theories - such as classical or quantum mechanics; proposed but not validated theories - such as loop quantum gravity; and marginal or fringe theories - such as torsion fields. On the one hand, some of the physical theories simply cannot be confirmed by an experiment or even by observation (although it was always required of classical theories and generally would be highly desirable) and, on the other hand, they cannot be deduced from a non-contradictory system of axioms like a mathematical theorem. This specific position of modern physical theories leads to regarding theoretical physics as a chimeric discipline between physics and mathematics, a cross of physics without experiment and mathematics without rigor[20].

Both physics and mathematics can be vital, not optional for survival. Indeed, it could take, for example, just one comet to demolish the Earth. To ensure the survival of the biosphere, humans need to learn how to avoid this danger which is basically a physical and mathematical task.

## 2.7   Mathematical Physics and Physmatics

What is the traditional mathematical physics? Some prominent physicists and mathematicians used to say that it is also, like theoretical physics, neither mathematics nor physics. When I was studying physics at the university, our course of mathematical physics was largely reduced to a linear theory of partial differential equations, covering the respective theorems of existence and uniqueness, some concepts of functional analysis including the spectral theory of linear operators in Hilbert space, and the usual tricks of solving boundary value problems (variable separation, Green's functions, eigenfunction expansions, etc.) Such university courses were typically named "The Equations of Mathematical Physics" or "Methods of Mathematical Physics", with the traditional textbooks such as those by A. N. Tikhonov and A. A. Samarski ([3]), R. Courant and D. Hilbert ([7]) and sometimes P. Morse and H. Feshbach ([4]) being recommended as standard. (All these books are even now a right place to look, anyway.)

---

[20] « La physique théorique est l'alliance de la physique sans l'expérience, et des mathématiques sans la rigueur ». Jean-Marie Souriau, a French mathematician.

In traditional mathematical physics, rather simplistic models are generally constructed, with actual physical details being reduced to an absolute minimum. In this way, the conventional PDEs of mathematical physics have been obtained and investigated. This is a typical model-building approach, and although it enabled one to study the standard set of equations and the properties of their solutions - so-called special functions - physicists usually consider the canonical mathematical physics inappropriate or at least insufficient for their opulent circumstances.

Physmatics stands to mathematical physics approximately in the same relationship as, e.g., physical biology to biophysics. During its evolutionary development and due to the accumulation of vast experimental material, biophysics has become a more or less crystallized discipline, and people usually understand - although intuitively - what one is talking about when the term "biophysics" is used. Contrariwise, physical biology[21] is a synthetic discipline devoted to the application of general physical principles and models to biological systems, and understood as such it may even include biophysics as a particular subdiscipline.

## 2.8    The Role of Human Communication

"Unthinking respect for authority is the greatest enemy of truth." Albert Einstein

Science as the refined extension of the intrinsic human curiosity instinct necessary for survival evolved as the increasingly specialized attempts to answer the "old human questions" (OHQ) such as why is the sky blue, why is the heart on the left side, why do birds fly, why is the water wet and liquid, etc. Over the last several hundred years, more and more specialized science has succeeded to answer many old human questions in terms of light and atoms. As a by-product, new weaponry has been invented and deployed, which justified the separation of scientists into an isolated (in extreme cases, physically) urban group, not all of professional scientists and not continuously privileged but always under direct control of the top authorities. During and after the World War II, a distinct and influential scientific component of the human society emerged, instead of separate dispersed individuals driven by curiosity and exchanging letters. Inside this coherent scientific milieu, light and atoms have been described in progressively microscopic terms of quarks, leptons, gauge bosons, etc. That was a chain of downsized multilevel models about the human environment that appeared synchronously with the stabilization of the social institute of science, with all attributes of human groups: dominant bosses who are forced to surround themselves by intellectual specialists, self-assertive leaders who are afraid most of all to lose face, small number of compulsive fanatics fostering conflicting evidence, and the large mass of the weaker members of the community supporting off-the-shelf enthusiasms. All like in other hierarchical groups.

---

[21] The originator of physical biology was probably the mathematician Alfred Lotka whose book "Elements of Physical Biology" (1925) seems to be unfairly forgotten.

One of the salient phenomena of the 20th century science was the establishment of semi-closed scientific communities, so-called scientific schools. This is neither bad nor good, it has just happened. Scientific schools provide people with great strengths as well as grave weaknesses. It is a common observation that science nowadays is done by big collaborations which may be called scientific tribes (scitribes). There are countries where more accent is put on such tribes, as, e.g., the former Soviet Union[22], as well as the countries where individualistic traditions were more pronounced. There are also paradoxical countries with a combination of collectivist trends and individual science. In Germany, for example, the culture of scientific schools is not highly developed, although communities - also scientific ones - are highly praised.

In this section, I shall offer some personal observations about scientific schools [167], predominantly in the former USSR where they were very powerful. By their very nature, these remarks of mine are highly subjective and can be met with skepticism. They reflect that tiny chunk of experience that I had while observing scientific as well as bureaucratic mechanisms from inside. However, I cannot be counted as a genuine insider. Honestly speaking, I know very little about the whole complicated texture of the relations between scientific schools and between people inside each of them which, in the former Soviet Union, was developed against a background of communist party and KGB control, state supported antisemitism and ideological doublethink. I know the situation with the two major Soviet schools of physics (these of Landau and Bogoliubov, see below) only by attending their respective seminars and from my sporadic informal contacts with some of the established participants of these two schools. My observations are related to 1970s when I was able to attend seminars especially often, first as a student of the theoretical department of the Moscow Engineering Physics Institute and then as a young research scientist. Later, in 1985-1989, while working in the popular science magazine "Science and Life", where I was responsible for physics and mathematics, I have had a second round of observations, but by that time the "Perestroika" began already distorting the pure Soviet patterns. So, my knowledge of the situation with these schools is very superficial or, rather, tangential. Nevertheless, while filtering my memories, I tried to refrain from inadvertent confabulations or vague, verbose fantasies that could have only a hazy relationship to reality. And I don't want to name concrete persons. Indeed, can one remember all the details after 30-40 years have elapsed?

---

[22] There were social reasons for it since the ruling authorities preferred to communicate with an appointed leader.

One might then justifiably ask: if you don't know the situation exactly from inside, why the hell are you writing about it? My answer is: that was my subjective vision of the situation with the so-called scientific schools, and there is nothing wrong in presenting it. The only grave sin would be a deliberate lie, which is not the case here. Furthermore, this "sociological" context is inseparable from a "scientific" one, at least for me - and I have strong suspicions that is the case for any person immersed in the scientific milieu.

The prevalent effect of scientific schools is sociological, not scientific, even despite wonderful achievements obtained by the school members. It means that social forces can act against the purely scientific process. In such a case science can be easily betrayed and even sacrificed. The cold war of thinking or not thinking was dominating my beginner's studies of theoretical physics when I was under twenty. This "war" was waged between the adepts of the well-known "Landau school" and those who considered themselves to belong to the "Bogoliubov school"[23]. Both school leaders were great scientists and there were probably equally outstanding scientists among the both school members, but it was a rare occasion that they openly interacted, and I don't remember any joint cross-school publications related to that period (end 1960s - beginning 1970s). When recollecting my experience as a physics student, I remember an odd sense of irrationality when I heard at the famous theoretical seminar in the Institute for Physical Problems that one should not read papers on low-temperature physics unless they stem from this institute, or when some prominent people in Dubna asserted that all what is and will be (!) said about elementary particles and high energy physics in "Physproblemy"[24] is a priori crap and should not be even listened to. We, physics students, were of course too inexperienced to make our own judgment, and the situation of uncritical choice between two mainstreams (either theoretical physics in the Landau school or mathematical physics formalism in the Bogoliubov school) produced sort of a schizophrenic effect on a number of young persons. I suspect that some of them were unable to recover before today. As far as I am concerned, I experience the same strange sense of irrationality when I hear, for instance, contemptuous remarks about the scientists belonging to the Princeton Institute for Advance Studies. I simply cannot rationalize this attitude nowadays, in the same way as I could not at my student time comprehend derogatory remarks exchanged by the representatives of the two great Soviet schools of physics. This is a pure sociology whose role in physics and mathematics is, regrettably, enormously important but is mostly left unnoticed by the historians of science. At least, I failed to find any good accounts of the impact of group prejudices and near-scientific stereotypes on functioning of physics or mathematics communities.

---

[23] Every school of thought is like a man who has talked to himself for a hundred years and is delighted with his own mind, however stupid it may be. (J.W. Goethe, 1817, Principles of Natural Science)

[24] The colloquial name of the Institute for Physical Problems

Unfortunately, I was too young to see and hear L. D. Landau himself, as well as N. N. Bogoliubov, so my impression of their schools is by necessity secondary from sporadic contacts with their participants, therefore my remarks about the Bogoliubov and Landau schools should be taken only as personal impressions. There is a lot of literature depicting these two great persons, but I don't think the information in this literature is sufficiently accurate. As far as L. D. Landau is concerned, some of the publications bear a scandalous character. The figure of N. N. Bogoliubov is not so controversial. I hope that maybe Professor Igor Landau[25], son of L. D. Landau, will write his own authentic recollections, which could probably remove a lot of controversies. I grew up in the USSR, the country characterized by what may be called a monocentric evolution: all important scientific and bureaucratic institutions were concentrated in Moscow. As a reaction to this top-heaviness, people in many regional scientific centers disliked the Moscow scientists intensely. (Dwellers of Moscow are in general nearly hated in the rest of Russia.) The scientists in provincial cities formed their own "schools" where Muscovites were rarely welcome unless, of course, they were influential scientific or bureaucratic bosses. My personal observations were that scientific workers from the city of Gor'ki (now Nizhni Novgorod) were marked by specific arrogance towards rank-and-file Moscow colleagues. Even the Leningrad (now St. Petersburg) and Novosibirsk scientists, which had created really world-class scientific schools in many directions, were more tolerable to the Muscovites. This parochial miscommunication rooted in regional prejudices considerably influenced the scientific scene. At least, group attitudes in scientific communities have always interfered with answering the simple question of research: "Is it true or false?"

Nevertheless, even in countries characterized by polycentric evolution such as Germany, group stereotypes in local scientific communities often prevail over rational responses. There are attempts to solve this problem by rotating research staff between various universities, but such rotation may be painful for individuals and destroys scientific schools that need a prolonged period of stability to be firmly established.

Presumably, the "Landau school" was originally based on the so-called theoretical minimum or, shortly, "theorminimum" that was in effect a test to be passed by any physicist who wanted to work with L. D. Landau. On the other hand, the program of "theorminimum" later provided the contents for the famous "Course of Theoretical Physics" by L. D. Landau and E. M. Lifshitz. Already from here one can see that a sheer preparation to the obligatory "theorminimum" test ensured that an applicant would acquire rather extensive qualifications in physics and accompanying it mathematics. The Landau school produced physical universalists.

---

[25] By a sheer coincidence, I studied with Igor Landau in the same class in the Moscow high school No. 5

Now the times of universal physicists like L. D. Landau are gone forever[26]. Today, universality became feasible only within large collectives or institutions.

But in such cases the problem of the common language appears. Such a common language is hard to develop in a loose assembly of physicists unless they are carefully picked up and uniformly trained.

It is my firm belief that extremely sharp specialization, a typical present-day phenomenon, has a pronounced detrimental effect on the development of physics. Strictly speaking, it is not obvious that the laws of nature do exist. These laws are invented by humans - in modern times in the form of mathematical models - to approximately describe the processes in nature. In fact, these laws and their numerous consequences are present only on paper and mostly serve as tools to increase the status of the "professional scientists". It is sometimes astounding that these mathematical laws and especially their corollaries really do work not on the paper of status seekers, but to manufacture real products. There is a kind of complicity among the professional scientists bordering on tribalism, with socially significant slangs, jargons and dialects purposed to insulate professional scientists from common folk, usually under disguise of scientific precision. Members of scientific tribes calling themselves academics are relying on internal communication more than on anything else, with somewhat ambivalent attitude towards colleagues: rivalry mixed with support against outsiders. It is curious that if the specialized jargon becomes widely spread and partly adopted by non-scientific groups as, for example, in cosmology, computer science and chaos theory, then the jargon's isolating function is lost, and new slang expressions are created restoring the anti-communication barriers.

These barriers are erected between scientists and "ordinary folk" as well as between scientific groups. Life, however, by putting forward real problems, in most cases forces us to transcend the frontiers between different sciences, which have been established - purely artificially - by people. Nature does not care at all what mental schemes people are trying to impose on its manifestations. Indeed, why should we look for an artificial way to structure data and procedures, when the real world has already organized them for us? Nevertheless, "professionals" tend to safeguard artificial frontiers invented by them with jealousy bordering on aggression[27]. People tend to form closed groups declaring them "elite". I have just given above examples of zealotic behavior in Bogoliubov and Landau schools and I shall try to describe some salient features of these elitist establishments a bit later. The Landau school was especially indicative: theoretical physicists belonging to the "Landau school" created in the 1960s rather high barriers, mostly of psychological character, around themselves. Young proselytes were specifically aggressive: some of them seemed to be always ready to apply clichés to the work of everyone not too eager to join their elitist circle, denouncing physical papers containing much math as "a disordered set of formulas" (that label was often applied to the works of competitors from "Bogoliubov school") whereas papers based mainly on physical reasoning were labeled as "philology" or "philosophy". The latter label was considered especially humiliating: it was considered a shame not to be able to saturate paper with equations (mostly differential), inequalities delineating limiting cases, integrals and, in the final part, formalized statement of results. Frankly speaking, papers based on "physical considerations" usually produce weaker results and contain vague fragments. Yet, the folklore saying that each discipline contains the amount of science exactly equal to its mathematical content is, to my mind, a mathematical extremism.

This is all history now, and perhaps of no interest to the new generations of scientists but, to my understanding, a very instructive history. There have been no similar mechanisms of scientific development in the West after the World War as scientific schools in the USSR. Great scientists in Western countries did not form schools around themselves. It has been noticed that neither Einstein nor Feynman, nor Wigner - in one word nobody has created scientific schools that could rival in impact with those existing in the Soviet science. University professors may have a couple of students, graduates, or postgraduates, or in comparatively rare cases a chair. That is all.

What personally I did not like about elitist scientific establishments - not only "Physproblemy", there were some other institutions and not only in Moscow, usually calling themselves "firma" (a firm) where the research employees saw themselves as local prim donnas. There was even an element of ritualized aggression mixed with inferiority in the behavior of these researchers - against intruders and under the slogan "I am great". By the way, really bright people such as V. L. Ginzburg, L. V. Keldysh, D. A. Kirzhnitz, A. B. Migdal or A. D. Sakharov never emphasized their top position. Both the Landau and the Bogoliubov schools were really very advanced and powerful, yet there were certain things which I found appalling. What one could observe, sometimes painfully, about both schools was their "national orientation"[28]. As for the Landau school, this effect was of course a natural defensive reaction to the disgusting antisemitic politics of the Communist Party of the USSR in approximately 1965-1985, but sometimes one could not get rid of an impression that the "national orientation" extended far beyond the compensatory focus. This may probably be called an overreaction. However, such an overreaction had some inherent dangers distorting the delightful construction of scientific schools - an informal community of highly qualified people. In particular, the "national orientation" tended to attract stray persons (I met some of those). The second thing that I could not unconditionally accept was the principle of problem selection according to the leadership taste. Take the Landau school, for example. L. D. Landau was indisputably a genius, and he raised a bunch of super professionals who inherited, to some extent, his arbitrariness and prima donna habits. One of the principles of problem selection was according to their potential solvability. That amounted to the situation when there existed a class of physical problems which deserved some consideration and an antagonist class of problems not worthy to spend time on. Technically speaking, this meant that the problems where there was no distinct small parameter should be neglected. Unfortunately, some physically interesting problems landed in the "unworthy" class and had to be rejected. For example, the theory of liquids, as near as I remember, was totally uninteresting - at least for young adepts of the Landau school of that time (beginning 1970s). Incidentally, a somewhat haughty attitude towards the theory of liquids was reflected in the textbook "Statistical Physics" by L. D. Landau and E. M. Lifshitz [24], beginning of §76. I remember that I tried to discuss the situation with liquids with Professor D. N. Zubarev whom I respected very much and who was a very kind and knowledgeable person. D. N. Zubarev was generous and absolutely neutral as far as group prejudices were concerned, and there were, as I remember, no limitations in topics discussed at his seminar in the Steklov Mathematical Institute devoted mostly to statistical mechanics. I also remember approximately what he said to me about the attitude of the "Physproblemy" people to liquids: this problem is outside the customary mathematical repertoire of young Landau scholars, therefore liquids are not interesting to them; it is rather our problem.

---

[28] This expression belongs to A. A. Rukhadze, a well-known plasma physicist[148]. Quite naturally, I risk invoking indignation and derogatory arguments from the remnants of the both schools, if they condescend of course.

I had an impression that collective norms, likings and dislikings, were obvious and nobody tried to conceal them. The sense of solidarity was more important to school members than personal aspirations and even individual achievements. This could be easily explained: belonging to a school meant security, a conviction that the personal and at least professional career is mapped out in advance. The pressure for working commitments and rapid fulfillment, if one had already established oneself within the school, was lessened.

Despite the pressures exerted on ordinary members by the elite scientific community, the benefits are great. The community and its influential leaders protect the community members from the critic of rival schools, even in the case of scientific mistakes. Thus, the basic problems of survival in a scientific milieu, very acute for an individual researcher, become inessential for a member of an established community. Formally, this quasi-paternal protection is due to severe filtering of scientific mistakes inside the community, but in fact it is the very stamp of the renowned scientific school that matters, especially in highly complex subjects.

There is nothing unnatural or unexpected about such protective mechanisms and cooperative behavior in science, all this is a basic part of human social and biological inheritance. Bygone are the unhurried days of individual, alienated scientists. Today, collective norms dominate. Even

Einstein was known to be largely neglected and almost ridiculed by the herd of physicists engaged in so-called elementary particle physics which in 1950s - 1960s was in fact hardly physics, rather multiple attempts to cope with the bewildering number of particles by classification. Now that period is often referred to as the "zoological" one, but at that time it was of fashion, with crowds of people being attracted. In today's parlance, that was the period of an extensive build-up of mathematical models based on the Lie group theory - a decent pastime in its own right, but having little in common with physics. One can see the remnants of this approach in http://pdg.lbl.gov/2007/download/rpp-2006-book.pdf.

Scientific schools, despite all their drawbacks and emphasis on human communication, have become, as stated, a powerful development mechanism of basic science. They ensured a very encouraging, even protective atmosphere which is vital for the young members. Such a coaching and mutual assistance, a fundamental urge to cooperate within the "scientific tribe" - this was basically provided by the school - existed only in the Soviet Union of those years. Now this scientifically productive period has gone forever.

To make things clear, I don't have any nostalgia for the Soviet times, as many people do. Nostalgia for one's childhood does not necessarily mean that the childhood was a happy one. The communist regime commited a lot of atrocities, and any society under communist rule is completely deformed. There is no point to be attracted to this regime. But science was excellent, maybe due to some compensatory merit selection - the best people went to science and not in business. Besides, radically different criteria were applied in Soviet scientific schools as compared with the current situation. For example, nobody paid much attention to the number of publications. The quality of the work, the degree of understanding of scientific ideas by a given person (for concreteness, I speak of physics here) were most valued - even more than his or her nearness to the leader. Nothing of the kind exists today. The "publish or perish" principle of career development resulted in the multiplication of scientific and quasi-scientific journals, their number nowadays exceeding the human capacity to observe. Worse than that, scientific literature has become cluttered up by a heap of immature, wrong or "not even wrong" papers. Unfortunately, these statements are not empty lamentations.

I am not saying that a disintegration of science into "schools" or scientific tribes as major units - a 20th century phenomenon - is necessarily a drawback. This is probably an answer on the increased complexity of problems, especially in nuclear weapon creation. What I am trying to say is the increasing price people have to pay for the breakdown of science into groups, usually in the form of community norms, attitudes, strict subordination, prejudices and stereotypes. I have a few personal reminiscences how powerful these community norms can be. In 1970s, together with one good experimental physicist, we produced a theory and made a device aimed at exploring radiation properties of modulated electron and electron-ion beams. This work was supported by some leading Soviet physicists including Prof. E. P. Velikhov and Prof. M. A. Leontovich. Now I think that it was not a bad work elucidating some interesting microscopic properties of coherent mechanisms of electromagnetic radiation. We were recommended to submit the work to the competition for the so-called Komsomol Prize. In the final part of the competition, however, we were rejected with the explanation that the Prize that year should go to the city of Gorki (now Nizhni Novgorod) because the powerful leaders of the local radio physical school complained that Moscow had taken all the prizes in the previous years. My colleague-experimentalist was so disappointed that in a short time ceased to work on his experimental installation. Later, he completely changed his field of activities and became the managing director of one of the most famous restaurants in Moscow. I also remember an extreme case of a brutal verbal attack on a person from the Moscow Engineering Physics Institute (MEPhI) during his presentation of the doctor of sciences' thesis. The attack carried out by a member of a rival scientific organization was scientific only in the form. The genuine reason was, as I understood, to demonstrate a supremacy of the scientific institution the attacker represented. The unhappy applicant died of a heart attack the same evening. I have also had an experience of being attacked by some narcissistic member of a well-known scientific school during a presentation. His argumentation was ridiculously incompetent and not in the form of questions, but even if he was understanding this, it did not matter at all. It was the social act of community defense.

How important it was to belong to a certain "school" was manifested in the group attitudes of even the most prominent physicists. I was constantly reminded of this. When in autumn 1985 I wanted to publish a very interesting and fresh article by D. N. Klyshko on quantum optics in the journal "Science and Life" (David Klyshko was a modest university professor and little known to physics grands at that time; besides he was a very modest person and an individualist). The editor-in-chief and his servile deputy were harshly against this publication on the ground that Klyshko did not belong to any known scientific community. It was only much later that D. N. Klyshko's works became almost classic (unfortunately, he died prematurely in 2000). In the same 1985, I was at a very interesting MePHI summer school on condensed media physics (later I wrote a report on this school in "Science in Life", No.1, 1986). One of the leading presenters there was Professor D. A. Kirzhnitz, a super qualified physicist and a very good person, whom many of us, school participants, liked very much. He seemed to possess a special charm of firmness - a feature of independent and unfairly traumatized person. I think Professor Kirzhnitz was not fully estimated as he deserved it. Once we - Professor Kirzhnitz and me - accidentally met near the news stand, and his first words to me were: "And you, Sergey, to which school do you adhere?" I did not know what to reply. What astonished me then was that even such a prominent physicist as D. A. Kirzhnitz tended to automatically classify people according to their adherence to a certain school. At least I interpreted his question in such a way[29].

So there are many tribal signs in scientific communities. No one apart from your scitribe dares using the community resources such as library, computers, Internet access, discussion notes, etc. From time to time some members of the community are set off assailing other communities. Younger members produce as much noise as possible around the community, imitating the aggressive manners of their senior leaders. In case the community is successful, it attracts new members and grows in size. Then splinter groups typically appear establishing themselves inside new institutions [30]. This process reminds me of colonization of new territories, in this case in socio-bureaucratic space.

## 2.9    Antimodels

The problem with science is that it is an extreme manifestation of the human curiosity and exploratory activity, thus it tends to be alienated from everyday needs on which "common wisdom" is based. Therefore, scientific truth must be safeguarded, developed, improved and continuously explained to people who, mostly due to personal circumstances, do not necessarily have an appropriate background - the frame of reference required to appreciate scientific statements. Science is developed to be approved by colleagues, non-science is mainly approved by consumers - bureaucrats, businessmen, journalists, and other lay public. Both science and non-science can be good or bad, interesting or dull, useful or useless, as well as approvers can be fair or foul, competent or incompetent, disinterested or biased. Many prominent philosophers were engaged in analyzing the phenomenon of pseudoscience (see, e.g., [141]). In my terminology, non-science incorporates scholarship, fiction and pseudoscience. Scholarship is a collection of disciplines based on study of documents and in this respect it is close to science. Fair and unbiased investigative journalism, in my understanding, is akin to scholarship. Fiction does not pretend to be based on accurate documentary investigation, it constructs models of life by depicting a series of specific situations, in which presumably free-will personages develop some action to embody the author's presumed ideas. Fiction borders on philosophy - in fact philosophy is a kind of fiction where the presumed ideas become so valued to the author that he/she does not resort to specific situations to convey these ideas, using abstractions and generalizations instead[31]. Philosophers tend to argue and declare instead of calculate, in this sense philosophical works are closer to theoretical systems than to specific models. Pseudoscience is based neither on documents nor even on philosophical generalizations of observed factoids; it uses beliefs and vague resemblances - as in, e.g., telepathy - instead of analyzing facts. In this sense, pseudoscience is a falsified science which substitutes the knowledge about the real world by speculative prescriptions and arbitrary rules.

---

[29] Perhaps incorrectly, but now it is impossible to find out the truth: unfortunately, D. A. Kirzhnitz died in 1998.

[30] Remarkably, not necessarily scientific, which means that status drive is stronger than curiosity that drives science.

[31] One can see an interesting example of the transition from fiction to philosophy in the novel "War and Peace" by L. N. Tolstoy.

Scientific models, in particular, mathematical models, in distinction to others, operate with quantities that may be measured with quantified precision. To measure a quantity means to compare it with a homological quantity taken as a unit and express the obtained result of this comparison by a real number. One must remember that there exist many "quantities" that can be ordered, i.e., to them the notions "more" (>) or "less" (<) may be applied, but still not measurable, for example, beauty and ugliness, courage and cowardice, cleverness and stupidity, wittiness and dullness. I am not sure that there are universal units to measure, e.g., creativity and orthodox ineptitude or wit and dumbness, etc. Despite all the efforts of psychologists and other social scholars, such quantities cannot be so far expressed by numbers. Therefore, it is difficult to build mathematical or really scientific models for the major dichotomies expressed by vague antonyms in our everyday language.

One should not, however, conclude that modeling should by necessity be scientific in character. In our intuitive modeling of reality, the scientific component is customarily negligible. There exist also para-scientific and even anti-scientific models, and they may be quite popular, e.g., astrology, esoterics, medieval magic, or creationism (now being taught in schools). In such models of reality, the scientific component, even if it is present, is sinking in the spicy bouillon brewed of disinformation, mysticism and deliberate charlatanry. If scientific facts contradict such esoteric beliefs, too bad for the facts. A good example of such an approach is delivered by the "castrated" history model advertised by a group around the well-known Russian mathematician A. Fomenko. Even the Chinese medicine with its representation of the human body as a bundle of channels, although frequently quite successful, is based on a bunch of para-scientific models not properly corroborated by experiments and easily adopted by charlatans. An interesting example of para-mathematical modeling are torsion fields as an explanation of telepathy/telekinesis (see below). Thus, some sectarian views might even be put in a mathematical form. Although many such models, frequently appealing to awe and prejudice, may be intriguing and it is sometimes exciting to disclose their inconsistencies - I shall not discuss them. Life is very short, and normal, non-esoteric mathematical models provide ample material to deal with, see however "The PEAR Proposition", [278].

There is usually a great demand for pseudoscience in society, and one of the reasons for explosive growth of pseudoscientific beliefs lies in indifferent and condescending attitude, although perhaps mixed with contempt, of absolute majority of scientists to all that "esoterics". It is curious that esoteric beliefs often go side by side with science making them hardly distinguishable from the latter in the layperson's eyes. One can easily notice that a vast number of newspapers and magazines regularly publish astrological assertions and report on various "miracles" alongside observing scientific facts and discoveries. It is clear that astrology is pragmatically used as a tool to increase circulation, which is a sad evidence of mercantile populism. Despite the fact that many millions of people believe in astrology (there exists even a TV channel "Astro TV"), I have no respect for it. Devoted people usually say that astrology is not a discipline being capable of fulfilling the demands for a scientific testing at the present time, just like, e.g., string theory. Yet the obvious difference between astrology and string theory is the fact that the latter is mathematically sound, at least some versions of it. Astrology does not seem to be mathematically sound. I cannot understand the main statement of astrology that natal horoscopes ensure definite prediction about what a person's life will hold for each individual. This implies that there must be a bijective mapping between a set of time points, say $h$ hours $m$ minutes $s$ seconds on dd mm yy and a set of individuals, each of them born at this time point, with their totally different life trajectories described to the minutest details. Of course, there is no such mapping. In other words, how many people are born each second? Do you really think they will have the same fate? Each day about 300000 people are born to the world, which gives approximately four humans pro second. Do all of them have the same fate? Or the temporal resolution of 1 second is insufficient for astrological predictions? Then one can remark that in the absolute majority of horoscopes accuracy is limited by a day. Even if the time of birth is known with hourly precision, it means that approximately 12500 people are born simultaneously within this accuracy. Dividing by two (accounting for gender) does not improve the situation. This is an obvious objection to astrology, and I could not find any response to it in astrological texts picked up at random in media and Internet. There were numerous statistical tests of the validity of astrological prognoses ([11]). Researchers failed to find in the available literature any statistical match in astrological interpretations. It is remarkable that astrologers dislike such tests. They typically assert that scientific methods are too crude and intrusive whereas the alleged astrological effects are so subtle and hard to detect that astrologers can only arrive at the correct prediction (interpretation of an astrological chart) by using paranormal forces or by tuning to the "cosmic order". This tuning can solely occur during authentic (one-to-one) consultations. As soon as inquisitive scientists interfere by posing their primitive questions, the occult tuning disappears. Presumably, scientists are not a part of the cosmic order.

An astonishingly high potential for pseudoscience is not an uncomplicated socio-psychological phenomenon. General public strives to miracles, magic manifestations, and all kind of mystical rubbish whereas scientific standards require considerable intellectual efforts. Scientific truth is not for couch-potatoes. It has been noticed that striving to mysticism is amplified during periods of social instability and confusion - it's a kind of escapism. One can also notice another socio-psychological phenomenon related to "esoteric" models of reality consists in a sharp division on "pro" and "contra". The proponents of "all that" (i.e., pseudoscientific beliefs) make absurd statements and their opponents - mostly scientifically literate persons - tend to produce emotional response: see, what crap! Such polemics is based on stereotypes - unfortunately for both sides. Only rarely the "scientific" side suggests constructive approaches to objectively estimate the claims promoted by the "pseudoscience" side. The Godik-Gulyaev group attempting to study a full set of physical fields that could in principle account for paranormal phenomena (this research was conducted in the Soviet Union in 1980s [41]) was a notable exception.

Instead of precise scientific evaluation of facts, propaganda and counterpropaganda are flourishing. In my opinion, it's a shame primarily for real science, not pseudoscience, since by now a rather extensive observational material related to rare natural events (such as light emitting objects, UFOs, rare biomedical phenomena, etc.) has been accumulated. Of course, not all of these observations are trustworthy but authenticity of some of them may be confirmed. The question is how can one correctly interpret such phenomena[32].

It is exactly here that orthodox science is very passive. Scientists consider it "indecent" to scrutinize incomprehensible observations, construct intelligent models and propose validating experiments. Naturally, the niche is filled up with charlatans disproving current science, offering "new physical principles" and "scientific pluralism", while ignoring the entire scientific experience and the current state of knowledge. The increasing activity of charlatans is promptly catalyzed by the falling level of education in natural sciences[33]. One can see a distorted logic in the "scientific pluralism", which can be demonstrated each time on examples but it is a tedious procedure to point out mistakes of inventors of just another perpetuum mobile. Let us illustrate this distorted logic by a physically neutral example of creationists. The typical statement of liberal anti-Darwinists (there exist also militant ones) is as follows: "Evolution is only a model. It is not a fact. Creationism is also a model." Therefore, we can present these two models in schools together as just two different models.

However, there is an implicit logical mistake in this statement: a replacement of the ∩ quantor by the ∪ one. evolution and creationism are incompatible so that they cannot be presented side by side - it's like 0 and 1 in logical schemes, true and false. Speaking truth to the public is not always appreciated, but if one claims to be a scientist one should not allow ignorance to prevail, even if comes under the guise of common sense or is supported by the majority. One should leave succumbing to the majority's urging to politicians.

---

[32] I like scientific skepticism, but I don't understand stubborn rejection of facts, even in case they lack proper bureaucratic registration. I don't think all the people (including a lot of military personnel and even ex-President Carter) who claim to have seen UFOs are lying or confabulating. On the contrary, it is very interesting why the other party, namely high military commanders, special service officials and governmental investigators in many countries are covering up the facts.)

[33] Nowadays, we are observing the Great Bureaucratic Revolution in the whole world, which requires more soft-skill persons than scientists and engineers. So, there is a redistribution of people over specialties, favoring administrators, managers, lawyers, marketing and PR agents, etc.

Forewarned is forearmed. "New physical principles" are attracted by not very qualified but enthusiastic people to explain the phenomena of telepathy, clairvoyance and other "biofield"-driven type. Some philosophers stipulate, for example, that mysterious particles with imaginary mass are responsible for information transfer between humans, and the respective fields may be called the "conscience fields". But what is the imaginary mass? To answer this question we must recall, firstly, what is mass and, secondly, what is "imaginary". We shall talk a lot about mass in the subsequent chapters, in particular in Chapter 8, let us now recollect basic facts about imaginary numbers. First of all, there are no imaginary numbers, only complex numbers that play a great role in physics. There exists only one imaginary number: $i$ defined as $i^2 = -1$. Of course, there are decent complex mass models in serious physics, for example to describe unstable (decaying) particles and resonances (see, e.g., http://en.wikipedia.org/wiki/Particle_decay). The complex mass concept, by the way, may work well for Higgs bosons (and in general for all gauge bosons, see below) in the upcoming investigations on LHC (Large Hadron Collider).[34] The LHC investigations, specifically tests of the Standard Model, see Chapter 9, are based on scattering and decay processes with the participation of many particles, and one of the suitable mathematical tools to describe such processes is exactly the complex mass scheme. But this real physics has nothing to do with philosophical speculations about "imaginary mass" as the foundation for telepathy.

Incidentally, the imaginary mass can be ascribed to tachyons, hypothetical particles whose energy grows with decreasing velocity - quite an unconventional behavior. Tachyons most probably do not exist, since otherwise a lot of physical principles that have been experimentally confirmed would be violated, for example, causality in which people strongly believe (see, e.g., http://en.wikipedia.org/wiki/Grandfather_paradox; see also Chapter 9). In this sense, tachyons are "parasitic" quantum field excitations rather indicating at some instabilities in the theory than providing real means for telepathic information transfer.

Some other people believe that telepathic signals are carried out by gravitational waves that presumably affect us as the people denoted as "lunatics". So, there may be links between various fanciful clusters (nodes) of models such as ESP (extra-sensory perception) and, e.g., astrology. There is probably a subnetwork of speculative pseudoscience with a gateway to real scientific research. Astrology for example, as well as some other occult models, is based on exaggerated ancient fascination with the relationship between the motion of heavenly bodies and earthly phenomena such as regular variations of illumination and temperature, as well as of plant growth and animal behavior. Extrapolating this on human life, one gets a would-be system of regularity patterns. According to astrology, distant heavenly bodies exert a profound influence on you at and after the moment of birth. This is an example of focusing on resemblances rather than cause-effect relations. From the physical viewpoint, it has been clear for over two centuries that astrology has nothing to do with reality. Astrology (together with other occult disciplines) always reminded me of a smoke-and-mirror trick to distract people's attention. Physics has left no place for astrology[35]. Indeed, physical understanding allows one to separate the true influence of heavenly bodies on the objects on the Earth from fanciful speculations. Of the four known forces in nature - electromagnetism, strong interaction, weak interaction, and gravity - only gravity and electromagnetism could theoretically have such influence over long distances. Could the force of gravity carry astrological influences? Obviously not, and already the secondary school course in physics would be sufficient to understand it. There were many objects close by when you were born that exerted much larger gravitational force on your infant body (it is remarkable that astrology is silent about them). For example, one can easily calculate that one's mother would produce the gravitational (tidal force) effect approximately $10^6$ stronger than the Moon:

$$\mathbf{F} \approx -G\frac{mM}{r^3}\mathbf{r} + G\frac{2mM}{r^3}\left(\frac{a}{r}\right)\mathbf{r} \qquad (2.1)$$

where $G$ is the gravitational constant, $m$ is the mass of the newborn, $a$ is its dimensions; for $M$ one should take concurrently the mass of the Moon ($7.36\times10^{22}$ kg) and the typical mass of a human body (70 kg), so the comparison gives

$$\frac{F_{Moon}}{F_{mother}} = \frac{M_{Moon}}{M_{mother}}\left(\frac{a}{R}\right)^3 \sim 10^{21}\left(\frac{0.4m}{4\times10^8 m}\right)^3 \sim 10^{-6} \qquad (2.2)$$

---

[34] This text was written in 2007, before LHC was put in operation.

Likewise, the other planets of the Solar System cannot exert any considerable influence on a newborn child against the background of neighboring gravitating objects. The effect on the biosphere of cosmic bodies located outside the Solar System is so negligible that no microscopic scales typical of living species are adequate to describe it.[36] It is curious that such simple estimates are not (or seldom) produced at schools. So, astrology is a pseudoscience because there is no known physical mechanism accounting for the influences it postulates. Astrology claims that the very weak forces exerted on us by the planets can strongly affect our lives and characters, but science allows for no such physical possibility.

Specifically, it can be neither the gravitational nor the electromagnetic force that carries astrological influences, which means that none of the known forces of nature is responsible. Of course, there may be new long-range forces in Nature, and there are some experiments aimed at finding new interactions, but physics imposes very stringent limitations on their magnitude (coupling constant) and range. For instance, considering some new hypothetical interaction, it is indispensable to verify its compatibility with conservation laws, with celestial mechanics (which is very thoroughly studied), astronomical and astrophysical data, with the data from high-energy physics about the lifetime of elementary particles, with other empirical information such as chemical element distributions, with the equivalence principle that has been tested with record accuracy ([12]), etc. All these constraints accumulated by modern physics make the room for such fundamental discoveries as new forces extremely narrow and the requirements to the skills of the researchers who venture to look for fundamental discoveries inordinately high. I strongly doubt that such requirements could be fulfilled by anyone in the astrological milieu.

---

[35] Nevertheless, I personally know some physicists, even of rather high rank, who are fond of astrology and even hire "professional" astrologers.

[36] By the way, tidal forces are the manifestations of spacetime curvature, see Chapter 8.

I would dare to give the people who try to discover new and unorthodox phenomena a good advice: first to learn and then to make discoveries. It applies also to a large number of people who childishly believe in various fanciful or occult models of reality: microlepton, torsion and unmeasurable bio-energetic or psi fields, telekinesis, cosmic influence, astrology, chaos magic, biorhythms, and "all this". These are in fact not models nor theories, but a number of verbal patterns. Occult (lat. secret, hidden, clandestine) or mystical models are unscientific or, rather, anti-scientific. Indeed, one can define mystics as something that violates the laws of physics. The rational component in these models is that they serve as a consolation for unsuccessful scientists or social losers. Honestly speaking, I do not think that it makes much sense to intellectually fight or scientifically disprove astrology. This strange sin is as inevitable as drunkenness, depravity, wickedness and other human vices. Astrology is supported by fatalism and belief in superior leadership, both had been used for long centuries by feudals and authorities to exercise control. I do not discuss here the so-called business astrologers - the majority of them are probably mere swindlers, so these activities have nothing to do with any modeling. Today, astrological business is flourishing owing to the ubiquity of personal computers allowing one to run simple astrological software. The common feature of "not quite scientific models" is their vagueness: such models make rather unspecific predictions, just as astrology makes weak, general and highly inconcrete predictions about an individual's character and life. A standard set of vague statements can hardly be called a model of reality. However, a number of scientists somewhat uneasily confess that they are studying astrology as a possible base to explore the "cosmic influence on the biosphere". This position of scientists testifies to the fact that there is no strong immunity to pseudoscience in different social groups, even among professional scientists and natural science students. Ironically, one of the first commercial astrologers openly practicing for money was a great scientist Johann Kepler. Nevertheless, Kepler apparently treated astrology with disdain as an ordinary superstition ([11]).

Usually, the astrologers respond to the scientific criticism that "it is obvious that astrology works in practice, but you do not know why. Likewise, the gravity existed and worked long before people formulated the laws of gravity. You cannot find now scientific evidence for astrology; many scientific claims are made in the absence of any known underlying physical mechanism. For example, few doubt that there is a strong causal connection between smoking and lung cancer, but no one knows the precise causal mechanism involved - because no one knows the details of carcinogenesis. The fundamental error in this criticism of astrology is to look at it only in terms of a part of science, the part that explains by means of laws and theories. But there's another part: discovery of new phenomena that have yet to be explained".

That is a deliberate confusion. Statistical tests show that there is no empirical evidence that astrology works in practice, in contrast to gravity. For a rational person, only statistical data are enough to disprove astrology. The kernel of astrology is characterological prognosis and forecasts of an individual's fate. It is exactly when medieval astrology was engaged in these activities that it diverged from astronomy regarded as a science and became a kind of a social drug. Many people promised to donate considerable sums of money, in one extreme case the entire fortune and property to a person who in a correct, scientific test could prove that the astrology functioned. Large prizes were destined also for adepts of other occult or pseudosciences. To my knowledge, nobody has lost a cent.

## 2.10  Topological Models

We characterize topology as a general means to analyze differential equations without solving them. This is only a local task for topology. In general, this discipline deals with those properties of manifolds (in particular surfaces) that are preserved under continuous transformations. For example, a triangle and a disk are equivalent topological objects since any two points located within these figures may be connected by a continuous curve. However, a ring (circle) and a disk are not equivalent in the topological sense: in contrast to the disk, there are closed paths (loops) on a ring that cannot be contracted into a point by any continuous transformation. This fact is a manifestation of topological non-equivalence of these two figures: the disk is simply connected whereas the ring (or $S^1$-circle) is not - it may be separated into two totally disjoint open sets.

One can readily recall some advantages arising from the application of topological techniques in physics. The first thing that comes to mind is the analysis of critical points of a function (see Chapter 4). Recall that the critical point $x_c$ is the one in which $\partial_i f \equiv \partial f \partial x^i = 0, i = 1, \ldots n$, $n$ is the number of variables (dimensionality of the domain space or manifold). If $det(\partial_i \partial_j) \neq 0$, then we can represent the final difference $\Delta f$ near the critical point as the competing sums of positive and negative squares:

$$\Delta f = -\sum_{i=1}^{m} (\Delta y_i)^2 + \sum_{j=m+1}^{n} (\Delta y_i)^2,$$

where[37]

---

[37] Here we retain the summation symbol as it is customary in topological texts

$$y_i = \sum_{k=1}^{n} b_{ik} x^k,$$

$b_{ik}$ is the matrix which diagonalizes $\partial_i \partial_j f$. The number $m$ of negative squares determines the type of a critical point $x_c$: if $m = 0$ then $f(x), x = (x^1, \ldots, x^n)$ has a minimum in $x_c, f(x_c) = min$ ; if $m = n, f(x_c) = max$ , and in intermediary cases $x_c$ is a saddle point. If we define as $p_m$ the number of critical points of the $m$-type, then we can observe that the minimal possible values of $p_m$ depend on the topological properties of the manifold on which $f$ is defined. Thus, the qualitative properties of solutions to the physical equations, which can be nonlinear and quite complicated, may be studied by topological methods. Such methods are especially valuable when it is necessary to establish the relationship between the structure of critical points of the considered differential equation and of its solutions (see Chapter 4 for more details).

In Chapter 9, we shall briefly discuss superconductivity - an astonishing phenomenon that has not been, in my opinion, fully understood yet in spite of many loud claims and a lot of hype. If one produces a ring-like superconductor (experimentally, a frequent case), then the magnetic flux trough the ring associated with the superconducting current becomes quantized. It is a nontrivial fact that the flux quantization in a superconductor is determined by the latter's topology: for a simply connected superconductor the magnetic flux is zero whereas for a multiply connected one the flux $\Phi = nh/2e$, with integer $n$.

## 2.11  Engineering Models

Engineering is not an exact science. During studies of engineering, one grows little by little accustomed to finding empirical patches adjusted by the hand to make things work, and indeed, they work. For a theoretical physicist or, even worse, for a mathematician, it may be rather unpleasant to make engineering calculations based on coefficients which may seem arbitrary, or on relations presumably stemming from experiments. "Pure" science seems to be underemployed - maybe down to several per cent in some profane engineering applications. It is a truism to say that life is very complicated, so even best science cannot always provide us with powerful engineering tools and off-the-shelf solutions. That is why scientific models always have a comparatively narrow applicability area and always require experimentally-based correction.

It is natural that both in a scientific research and in an engineering inquiry, in order to understand, describe or predict some complex phenomenon, we employ mathematical modeling, already reflected upon. Mathematical models range from a simple formula (such as, e.g., the Aristotle model of motion) to an extremely sophisticated system of mathematical concepts such as superstring/M-theory or climate research. For an engineer, designing a concrete technological system in contrast to a general researcher, it is desirable that a model should be as simple as possible still revealing all essential properties of an engineering system. But what does the word "simple" mean in the operational engineering context? In my view, engineering simplicity is not an antonym of mathematical complexity which is, by the way, still a poorly defined concept (see, e.g., [279]) closely tied with the interconnection of elements within the system and not necessarily with the system's functioning. The latter is the primary focus of an engineering endeavor. One can formally construct, e.g., on paper, a mathematical model based on highly nonlinear partial differential equations for which conventional engineering control structures, such as inputs and outputs, will be hard to define and still more difficult to design.  In this way, we would have a kind of a mathematical model giving very little to an engineer from the practical viewpoint (such models are in reality produced on paper in large numbers). It is thus a matter of considerable practical interest to know how to build simple models - from the engineering viewpoint. Here, by "simple" I understand models involving the least possible number of variables and phenomenological adjustment parameters that ought to be considered. Quite naturally, this minimal number is a function both of the system type to be modeled and of modeling purposes[38].

Although personally I admire good - and sometimes beautiful - engineering solutions, I still think that engineering education lacks a certain mathematical hygiene. For specific problems, engineers have to rely on their experience in order to choose the most essential parameters destined to serve as the dynamic state variables. Then one may pose the important question: would it be possible to develop a systematic procedure for the model reduction, which would be intuitively prompted and, at the same time, would preserve the most relevant features of the system to be designed? One must remember that an engineering system in question may be mathematically represented by complex non-linear dynamical equations. The answer is probably "yes, it would be possible", although such a statement, of course, cannot have the rank of a mathematical theorem - at least so far. Some systematic procedures to obtain viable engineering schemes from scientifically-oriented mathematical models do exist, but they have a local character. One can recall the examples of the Kotelnikov-Nyquist-Shannon-Whittaker theorem or of the Karhunen-Loève expansion.  However, the lack of universal systematic procedures for transforming highly sophisticated mathematical models into engineering schemes is the main reason for the chasm between the world of science and that of technology.

Speaking honestly, engineering models are not quite mathematical; they are rather ad hoc solutions cooked for a particular real-life problem but cooked also by great people. We are using a lot of wonderful things which are engineered and manufactured but are not based on solid mathematical models. The only purpose of engineering models is to help technologists with arranging the operations, predicting their outcome, and evading inherent pitfalls. This purpose does not need mathematical stylishness.

---

[38] In this spot, I could have produced a number of simple examples of engineering systems differing by the set of variables and parameters to be taken into account depending on the required accuracy and design goals, but I rather refrain from stating these trivial observations.  Everyone can readily do it.

## 2.12  Mathematical Models in  Biology

By biology in this section, I understand the study of living organisms i.e., of structures possessing the property of life, namely capable of growth and reproduction. For instance, modern humans arrogantly calling themselves Homo sapiens are a particular species in the branched hierarchy of living organisms. Biology of humans is bordering with medicine, and one often unifies these two disciplines under a compound biomedical area. Medicine, however, is so far a descriptive discipline not largely based on reliable data. Therefore, accompanying myth creation is still spread among medical doctors, often in the form of allegedly vital advice. (One should of course be skeptical in applying medical advice, especially if they contradict one another.) It is interesting that people usually attribute medicine to sciences although it is more concentrated on practical services and less on research in background fields such as biology, chemistry, and physics. Medical clinics bear more similarity to a repair shop staffed with engineers than to research institutions with scientific personnel.

In contrast with inorganic matter, dynamics of biological systems is determined not only by the cause, but also by the goal. In general, goal setting and selecting among future alternatives, completely nonexistent in the inorganic part of the universe, serve as efficient survival tools for biological systems. One can say that not only retarded (causal) solutions but also advanced (non-causal) ones can be realized in the biological world. Modeling equations can have not only time-delayed coefficients, but also forward-shifted. In other words, biological dynamics, from cells to complex organisms, requires some goal-oriented mathematics.

In biophysics, one has long attempted to study cells as systems interacting with the environment. The living cell contains a series of naturally evolved complex nano systems which can be studied by quantum-mechanical methods and probably reproduced with the help of quantum engineering. However, given the extreme complexity of even the simplest cellular functions, comprehensive mathematical models of living cells can likely be developed primarily for experimentally well-studied biological systems such as Drosophila, yeast, E. coli, etc. Such models can then be used to investigate and analyze the genotype-phenotype relationship in general.  This study of phenotypes, provided the genotypes are already known, will thus have a pronounced theoretical component because of the mathematical modeling and computer simulation techniques. The main functions of the living cell seem to be replicating and processing information on the quantum level, therefore special techniques of quantum biology and quantum information processing probably should be developed. Besides, the available (at least to me) techniques of mathematical results are not yet sufficiently matured to describe the living objects.

The biggest challenge of biology is that the great diversity of data and their randomness can nonetheless lead to fine-tuned processes (in time) and structures (in space). It means that the ideal physical notion of the world as a machine seems to be inadequate so that a paradigm shift is required. In fact, fully mechanistic nature would be incompatible with life, where evolution gains order through fluctuations. Biological evolution is typically understood as a descent accompanied by slight modifications. Diversity of biological components increases viability and resilience of a biological system. From the point of view of biological esthetics, diversity is the foundation of beauty since it produces outstanding samples against a uniform background. Variability and the ensuing diversity arise as replication errors: in the now fashionable language of code, such errors are insertion, deletion and replacement of code fragments. Anyway, biological evolution basically occurs due to errors, and this picture is poorly consistent with deterministic classical mechanics.

Living systems are, in general, complex open systems that constantly develop and strongly interact with the environment. Each biological entity, however small, is a complex system in itself. The interplay of its various parts e.g., cells, proteins reveal new emerging properties which initially may not be a part of single subsystem. They are typically found in non-equilibrium states compatible with the environment so that living systems must retain their inner stability (often known as homeostasis). Preservation of inner stability and compatibility with the environment is the principal strategy for staying alive. Nonlinear equations modeling biological systems reflect their self-organizing properties that are in most cases manifested without any external organizing agent or aligning force.

The human organism can be regarded as a complex open system in a state of delicate equilibrium exchanging chemical substances, energy, and entropy (information) with the outer world. Since the whole organism consists of a number of strongly coupled and cohesively working subsystems, the failure of any one of them tends to provoke an avalanche of other failures. Therefore, treatment of a single organ or a separate disease may very well lead to complications involving other organs – an effect well-known to any practicing doctor. Mathematically speaking, the health state is a manifold in the space of all possible physiological parameters. From the perspective of new technologies, in particular, mobile technologies such as wearable body networks, the human organism can be conveniently treated as a nonlinear unstable dynamical system incorporating a number of coupled subsystems.

Nonlinearities play a crucial role in biology and, in particular, in the physiology of the human organism. In the organism consisting of interacting subsystems (organs), changes tend to affect the entire system, therefore in a functioning organism considered as a system of highly integrated parts, changes are in general deleterious. Yet if the homeostasis is stable and the stability is robust, then alterations do not jeopardize the whole organism, with the eventual outcome that changes will be lost. If the stability limits of the organism have been transgressed, unstable states emerge, interpreted as diseases. For instance, the transition to unstable states (diseases) from the health state can be based on the hierarchy of instabilities developed in a nonlinear system, which is a far analogy with the transition from regular fluid motion to turbulence. Stability can be monitored, in particular, by the mHealth technologies [182].

A rather primitive illustration of health (stability) margins is the ubiquitous reaction of the human organism to alcohol (ethanol). Small doses of this substance i.e. blood alcohol concentration (BAC) less than 30 milligrams of ethanol in 100 milliliters (0.1 l) of blood or 0.3 per mil (0.3 g/l) do not appear to seriously impair cognitive abilities or locomotion of an "average" person. This value might be roughly taken as the boundary of the stability (health) domain. Returning to this stability domain occurs with the estimated rate of 0.15 per mil/hour, although of course this relaxation process may vary from person to person. Both the health state and the restoring rate depend on other variables such as age, gender, BMI, etc., not only on BAC, which illustrates the fact that the health (stability) domain and the restoring path cannot be uniquely defined for all persons and conditions. In general, most chemical substances in the body have their specific threshold levels with respect to their concentration in blood.

Classical science mostly studied systems and their states close to equilibrium, and that allowed one to construct a beautiful collection of comparatively simple physical models for the world. Such models depicted the systems that reacted on perturbations more or less predictably: these systems tend to return to equilibrium (in the parlance of statistical physics, they evolve to a state that minimizes the free energy). Remarkably, however, systems close to equilibrium can describe only a small fraction of phenomena in the surrounding world – it is in fact a linear model. Any realistic system subject to a flow of energy and matter will be driven to the nonlinear mode i.e., far from equilibrium. For example, open systems such as Earth, climate, living cell, public economy or a social group exhibit highly complex behavior that is, firstly, hard to be replicated in the laboratory and, secondly, almost impossible to model mathematically using the methods adapted mainly to mechanical patterns. In contrast with the closed mechanical models which are a drastic simplification of reality, open and nonequilibrium systems are ubiquitous in nature. Most of the processes in the open systems far from equilibrium are interrelated, nonlinear, and irreversible. Often a tiny influence can produce a sizable effect, which is a universal property of nonlinear regimes, and in the real world almost any system is nonlinear.

We see that from perspective of physics living organisms demonstrate classical properties of open, complex and highly dynamic systems that would primarily be described by nonlinear equations that capture the dynamics such as differential or time dependent ordinary of partial differential equations. However, while designing mathematical modelling of the living structures one should always consider its biological nature. One can also observe less time constrained events or quasistatic process that allow the system to slowly preserve its inner equilibrium such as e.g., metabolic procedures. Mathematical Modeling of Complex Biological Systems [181].

If living creatures are a certain form of physical objects, they can be treated by physical means and the latter by the adequate mathematical instruments.

Take for example the effect of scaling. Scaling properties of Newton's law show that if we increase the mass of the body four times, it will pass the same trajectory twice as slow. Here we implicitly assumed that the potential $U$ does not depend on the inertial mass $m$. This fact contradicts our intuition, e.g., if we recall falling bodies. However, in the gravitational field $U$ is proportional to gravitational mass (in Newtonian approximation), so the period does not depend on $m$ . The scaling function $f(\theta)$ can be determined either numerically, or by ODE integration, or from experiment (measuring the period or frequency, all other parameters being fixed).

Let us apply scaling to biological modeling. Many people still remember the "Jurassic Park" movie – this efficient manifestation of mathematical zoology. The group of scientists had to escape from the hungry Tyrannosaurus Rex. Would it be easier to run across a flat plane or up the hill? The speed of a living creature while crossing hills is inversely proportional to its size, since in such conditions it is not the air resistance that the animal should overcome but the force of gravity, so $P \sim Mv \sim L^3 v \sim L^2$, which means that $v \sim L^{-1}$. In order to not be compacted by a promenading dinosaur, it is better to escape uphill: big size makes its move difficult for a creature to catch you.

The year 2020 showed us that the world is becoming increasingly vulnerable to outbreaks of sudden epidemics. Infectious diseases are caused by a variety of pathogens (viruses, bacteria, fungi, protozoa) that can pass from host to host. Mathematical and computer modeling of this process is usually based on subdivision of the total population into three classes (stocks).

1. Individuals susceptible to the disease but not yet infected (S)
2. Persons who are infected and capable of transmitting the disease (I)
3. Removed persons: those who are no longer capable of transmitting, due to recovery, quarantine or death (R)

This 1,2,3 model is often called the SIR model. The disease is transmitted horizontally by interaction between the S and I classes, the transfer of I to R is assumed to occur at constant rate.

To model the spread of infectious disease one can apply a spatially homogeneous (ODE) model:

$$\frac{dS}{dt} = -\alpha I(t)S(t)$$
$$\frac{dI}{dt} = \alpha I(t)S(t) - \beta I(t)$$
$$\frac{dR}{dt} = \beta I(t)$$

The spatial spread of infectious disease can be modeled by reaction-diffusion systems. These are PDE equations that include the reactions (the time derivatives as above) and also space derivatives for diffusion.

A simplified model of the infection spread can be produced by using the logistic equation. Let $p$ be the infected fraction of the population, then $(1 - p)$ is the susceptible fraction, and the time variation of $p$ can be modeled as

$$\frac{dp}{dt} = \alpha p(1 - p) - \gamma p$$

where $\alpha$ is the transmission rate of the disease and $\gamma$ reflects the recovery (removal) of the infected population. These parameters are specific for the disease and, in general, for different populations.

Investigation of this logistic equation may be performed by separation of variables, integration and inversion. It will be discussed in more detail later. Rates $\alpha$ and $\gamma$ are control parameters.

Discrete analogs and critical parameters can be obtained, describing the transition to catastrophic amplification of a disease, in the same way as for the general logistic model. The transmission rate $\alpha$ can be reduced by quarantining; $\gamma$ can be raised by immunization.

The logistic model, despite its simplicity, was able to adequately represent population dynamics in various countries, e.g., in England, Scotland and USA. In biology and ecology this model is used to describe various evolutionary scenarios when the future population of species depends linearly on the present population:

$$x_i(t + 1) = \sum_j M_{ij} x_j(t), x_i(t) \geq 0$$

The diagonal terms of the matrix $M$ represent individual growth rates, while the off-diagonal terms represent the interaction between species. Fixed amount of available resources corresponds to conservation laws. Evolution gets really competitive when the total population reaches its maximum value,

$$\sum_j x_j(t) = X$$

Let us not forget that any mathematical model is just an abstraction and not an exact replica of its living prototype. It does not capture all the diversities and dynamic changes of nature. Often the initial assumption of a researcher may erroneously ignore some of the essential parameters critical to reconstruct the behavior of biological species in question. The good thing is that unlike laboratory experiments mathematics modeling can almost endlessly verify initial assumptions, finetuning the model and narrowing its approximation to reality.

Recently, the basic research of living organisms has noticeably expanded. The front of such a research is advancing along many directions that seem to be converging. Indeed, physics and chemistry are studying the functions of organisms on the molecular level; mathematicians, physiologists, electronics engineers and medical doctors are working together trying to exhaust various options. Although there is a rising perception that contemporary medicine is approaching a state of crisis characterized by bureaucratic inertia, rapidly growing number of errors, averaging out creativity and stifling originality[39], innovative life science segments spread far from biology as the basic life science discipline spanning a wide range of subjects: from purely biomedical topics to the theory of dynamical systems, evolutionary models, physics, mathematics, engineering, even history and philosophy. [181]

## 2.13  Cognitive Models

It may be considered an unfortunate fact that our world mostly consists of non-quantifiable things. One says: "he is a good man" or "this book is interesting", how can you naturally make a numerical - ordered - set of such statements? Human communication in general consists of only a tiny fraction of mathematically sound messages, the absolute majority of them being purely intuitive (you can prove this thesis by listening attentively to your own everyday communication). The human language is largely adapted to such intuitive messaging and has very little in common with formal mathematical systems - it is here that the main difficulties of man-machine communication using natural languages seem to be hidden.  Take, for example, rhetoric - an important art of influencing other people by messages: how can you judge who is the better orator? Are there formal, quantifiable means to measure eloquence? Hardly. So, in most cases, a person must develop highly subjective algorithms to make a judgment and intuitive patterns to appreciate the truth in human statements. It means that each of us constantly makes up cognitive models making it possible for a person to separate true from false, thus recognizing reality. In many cases, cognitive models are purely based on intuition, insight, and former experience and can be "silent" i.e., not explicitly formulated in words; in other situations, such models are verbalized as opinions. I do not know the statistical distribution between verbalized and purely intuitive - pre-verbal - judgments in typical human behavior (what is typical?), and I doubt that any one does know.  Here, I shall try to focus only on cognitive models based on verbal messaging since silent insight is too complicated for me to discuss.

In such a subclass of cognitive models, people use verbal constructions rather than formal mathematical statements.  Verbal constructions are probably more appropriate for the human brain than thinking in numbers or even in formulas. Our conscience does not in general function as a digital computer, although there are such hypotheses as well. It would be interesting to understand, why?

The incalculable complexity of human society results in the circumstances when it is practically impossible to make sharp, clear-cut and rational decisions required by mathematical thinking. The data continuously obtained by a person are typically so diverse and quite often conflicting that if one tries to apply mathematical reasoning to the received evidence one would be confused and compelled to long hesitation. This would be an obvious disadvantage for survival in the condition of delicate balance between harsh competition and unstable cooperation in the human tribe. So, it is more beneficial to base the decisions on incomplete information, assumptions, and not necessarily correct prognoses - mathematical pondering would be a great luxury or even dangerous not only for an ordinary person, but even for a leader. No wonder, many people have developed an aversion to mathematical formulas.

---

[39] See, e.g., Bartens, W. Das Ärztehasserbuch [243].

One of the reasons for replacing mathematical statements by verbal constructions, apart from a lack of qualifications, is possibly the fact that, as Stephen Hawking put it in his famous "Brief history of time" [78], each equation contained in the book would halve the sales. I could never understand it. Take one of the best popular science magazines "Scientific American". Mathematical formulas have always been forbidden in this journal. As a result, some articles in it are more difficult to read than in "Physical Review". I am not speaking about "Scientific American" articles on biology, which typically require some professional competence in this field and only rarely win lay readership. Why should one forbid a limited use of formulas in popular articles? It is true that many people are afraid of mathematical expressions, mostly because of lousy teaching methods in high school, but such people would hardly buy "Scientific American" anyway. Banning mathematical expressions out of everyday life is counterproductive and, to my mind, leads to preservation of illiteracy, not only mathematical. School-time ambivalence and fear of math suppress mathematical instincts badly needed for survival[40], and instead of encouraging their development, powerful social forces still further inhibit these instincts.

Models play an important part not only in physics, but generally in life. Those who find mathematical tools in modeling the human environment unnecessary tend to take refuge in the alleged human experience or "expert knowledge". The authors of cognitive models concentrate on verbally extracting the ideas for modeling out of the history of thinking and culture. The statistical and mathematical problems are out of the scope in this approach. It is quite natural since, even if there are some laws governing human behavior and social processes, the mathematical structure of these laws is unknown and the information contained in them cannot as yet be condensed in formulas. In cognitive models, there is no means to controvert a vague statement, therefore verbose discussions are habitual. Nevertheless, cognitive models, sometimes taking the form of common maxims or aphorisms, may be perceived as rather cute and precise statements, a kind of "soft" model. Or take poetry as an example. Poetry may be hard to read, and some people feel alien to it, which fact likens it to math. Poetry is similar to mathematics in various aspects: for instance, in careful selection of teasing words to compose the text, in anti-verbosity, in the tendency to reach the distinct result by some sort of textual proof.

---

[40] For instance, frequently putting the question "What would be if...?"

My interest in mathematical structures lies not in mathematics *per se*, but rather in its possible use. In this sense, it is a pragmatic interest. I think that mathematics, due to its abstract nature and the resulting generality, allows one to avoid boring lists, enumerations, tedious classifications and verbose descriptions which is, to my mind, an obvious advantage of the so-called exact sciences. Cognitive models studied, e.g., in philosophy, psychology, theoretical sociology, etc. are not even expected to be "exact", in the sense that the issue of the accuracy is never being discussed in these models, nor in the respective disciplines. Contrariwise, mathematics and physics as well as a number of areas in engineering, chemistry and biology are supposed to be exact since they are preoccupied with quantitative precision. It is the question of accuracy that mainly distinguishes exact sciences from the disciplines based on cognitive models, where proper approximations and reasonable answers are not quantified.

One of such disciplines is literature, although it is conventionally not related to anything scientific. Nonetheless, literature is also exploring reality by model-building. One of the best cognitive models, in my opinion, was produced by George Orwell (Eric Blair) in "Animal Farm" [11] which may be regarded as an essential model of socialism. Another famous book by George Orwell, the novel "1984", represented a rather exact model of a totalitarian future: the world is divided between three despotic states with absolutely identical behavior - they lie, kill, and try to dominate. There is no place for democracy in this dystopian extrapolation of current social patterns. Luckily, Orwell's prognosis was erroneous, mostly due to the necessity of sustainable economic development. Economies are striving to be open systems; economies subjected to the overwhelming bureaucratic control are foredoomed to failure. The model of evolutionary history was given in the famous book by William Golding "The Lord of the Flies", where a bunch of modern English schoolboys, having suffered a shipwreck, reverted to savagery and primitive dominance, reinventing the entire package of basic ancient forms: hunting-gathering, ritual dances and sacrifices. The bestseller novelist Arthur Hailey has conducted a number of case studies modeling certain complex systems (Airport, Hotel, The Evening News, The Final Diagnosis, The Moneychangers, Strong Medicine, Wheels, etc.). Chekhov has produced a gallery of almost mathematically exact patterns, especially manifested by the personages of his plays. Maybe because of this nearly mathematical modeling, Chekhov's plays are often considered by many people boring and schematized. For instance, the main female characters in Chekhov's plays are in fact the same personage standing out under different names.

Literary writers design situations i.e., they are trying to lie as truthfully as possible about what might have happened but in reality, never occurred. Many successful literary works are in fact constructed as the models of physics. For instance, a satirical dyad, "The Twelve Chairs" and "The Golden Calf", immensely popular in the former Soviet Union, was designed by its authors who wrote under the pseudonyms E. Ilf and E. Petrov[41] around a hero without internal states, a dimensionless figure of a swindler Ostap Bender which serves as a probe submerged into the Soviet reality. The function of this hero-probe was only to illuminate the stupidities of the Soviet bureaucratic society. It moves through the medium as a structureless elementary particle, with the medium being excited and its hidden absurdities becoming manifest. The same or close models have been exploited in other well-known works of the world literature. To begin with, Gogol's classics, the play "The Inspector" and the novel "Dead Souls"[42] were also built around featureless swindlers functioning as a probe. To a great extent, the same framework of a dimensionless hero (in this case not a con man) applies to Nabokov's famous "Lolita", Solzhenitsyn's "The Cancer Ward" and "The First Circle"[43], etc. One can provide one's own multiple examples of the same cognitive model: a structureless hero-probe producing excitation of the medium in order to exhibit its properties, as well as other archetypal literary schemes. The model of biological survival by Daniel Defoe ("Robinson Crusoe") has a clear message of creative individualism against the forces of nature. Witty fantasies by Jonathan Swift and Lewis Carrol were pointed narrative models mirroring the contemporary society and patterns of human behavior. More obvious as magnifying tools of some pieces of social and physical reality are the science-fiction works: they are just physical (e.g., in stories of Jorge Luis Borges) and in some cases even mathematical models (such as the famous Bradbury butterfly, see Chapter 4) expressed through specially constructed human situations. Physical models of A. Clarke or R. Heinlein (in particular, "Skyhook") have been sources for serious scientific investigations.

One of the obvious deficiencies of cognitive models is the tendency to operate with poorly defined notions. Take, for instance, models in psychology (see, e.g., http://www.psychology.org/), even those where mathematical symbolics and techniques are beginning to enter the scene. Such terms as conscience, subconsciousness, emotion, attitude, perception, self-identity, personality, mental health, and mental disturbance are constantly used, but hardly any psychologist - applied or even academic - can provide a fair definition for all these terms with which she/he so readily operates. And even when such a definition is offered, there is no guarantee that it would be accepted by fellow psychologists. Take any of the key notions, for instance, conscience. Using this term, we do not understand what it is and are referred to vague intuitive images. Such branch of psychology as psychoanalysis operating with specially constructed models of "personality" seems to be especially unsatisfactory from the scientific point of view, primarily because of relying on essentially unobservable notions (such as superego, "id", etc.).

---

[41] Even a small planet (registered under the number 3668) that was discovered by a Soviet astronomer L. S. Karachkina was named "Ilfpetrov".

[42] Gogol himself named this novel "a poem".

[43] The 1970 Nobel Prize winner A. I. Solzhenitsyn, as a former mathematician, seemingly used quasi-scientific methods to create his works. He used to dissect the source material into tiny pieces, reassembling them afterwards under the guidance of a leading idea.

Current interpersonal psychoanalysis is thus inconsistent with physical or mathematical techniques. Nowadays, there exists "global psychology", attempting to explore large-scale phenomena in psychological terms, with the entire human population being taken as the object of investigation. Such psychological texts may be amusing to read, but their positive value is highly questionable.   Can one predict or calculate anything with a prescribed accuracy using psychological techniques?

There is nothing wrong in using poorly defined notions in everyday talk, but this talk does not claim to be a science. It is just a means to state a person's position or to pass simple messages. Ambiguity and lack of precision are partly compensated for by the redundancy of human speech. It is these three qualities: ambiguity, lack of precision, and redundancy that make human languages unsuitable for computer programming, and one has to invent artificial machine languages dedicated for coding.

With further development of neurological science and machine learning, one can notice an attempt to apply nonlinear computational methods to cognitive processing in order to understand psychological functions of different areas of the brain. In this sense, cognitive models are frequently distinguished from conceptual frameworks; the latter largely adhere to natural language (verbal) descriptions of theoretical assumptions. [173]

Neuroscience was always struggling to understand the unique ability of the brain to perform instructions controlling human physical and mental behavior. However, experiments on humans are rather costly and difficult to replicate. Artificial neural networks (ANN) aimed to resemble the structure of their biological prototype. Just like biological neural networks, ANN consist of several layers of neurons with processing nodes, the so called neuronodes, arranged in layers.  After receiving the external data, nodes on the input layer pass signals to the output layer responsible for problem solving. The ANN training algorithm analyzes collected data sets asserting weight to neurons depending on the error rate between target and actual output, thus generating communication patterns. [174]

Benefits of cognitive process computation are manifold, especially in those areas involving large amounts of data, such as natural science e.g., biology or medicine. Health data are exploding, expected to grow by 48 percent annually, already reaching over 150 Exabytes in 2013. If the trend continues, (and there is no evidence that it will stop), no human brain alone will be able to cope with such an avalanche. "The ability of Artificial Neural Networks for "self -training" repeatedly comparing input and output data using "deep learning" algorithms proved to be effective in various branches of medicine e.g., cardiology, neurology, and oncology, where physicians have to deal with complex images containing multiple physiological data.  In order to achieve a degree of complexity necessary for these applications, deep learning systems always need to process a large number of nodes and nonlinear components in certain cases surpassing the analytical capabilities of a human brain. However, the technology is mostly applied while dealing with imaging and cognitive models clearly defined (e.g., stimulus-reaction). For example, in image analysis, purely linear models are only capable of recognizing straight lines and planes, but not curves, as explained in a recent review of deep learning. [175].

Unfortunately for scientists, but perhaps still fortunate for humanity, not all cognitive models could be computationally simulated having distinct measurable variables. There will always remain "soft models" e.g., perceptions, biases and concepts based not only on theories and verifiable assumptions, but on emotions, beliefs, religious dogmas and ideologies (the latter are almost synonyms).

Therefore, cognitive models alone do not allow one to answer the important questions which continuously appear like the heads of the Lernaean Hydra. Is global warming real or is it a political hoax? Should one build nuclear power plants? Does psychoanalysis deserve to be considered a helpful medicine? Should one invest heavily in the research of the physical basis of cognitive (mental) processes?

### 2.13.1   Religious Models

Since the emergence of the notion "exact sciences", the role of religion in science and technology has drastically decreased. Scientists, the best of them being obsessed with scientific truth, have been inclined to assert that it is very unlikely that God does exist - at least as a biblical character, and religious claims that the universe contains God are not indispensable to explain natural disasters such as earthquakes, hurricanes, tsunamis and the like. Introduction of extensive mathematical modeling in every branch of science, hopefully capable of predicting and analyzing such cataclysms, is going to make one more serious blow to religious beliefs. The head of the Russian Orthodox Church likes to claim that the earthquakes (in particular, the 2010 Haiti earthquake) are the punishment for human sins. All traditional religions typically associate such disasters with human misdemeanors. One might recall in this context the death of Pompeii and Herculaneum in 79 A. D. destroyed by volcano eruption allegedly as a punishment for massive sins.

Scientists tend to be less religious than the public in general. This is natural, since scientists could observe (starting probably from Laplace) that the God hypothesis does not help much to calculate, find experimental values with an appreciated precision or produce engineering applications of science. This hypothesis has nothing to do with mathematical or physical models, and its "scientific" meaning is reduced to an attempt to verbally - i.e., without including mathematical formalism - explain certain foundational issues pertaining to the world around us. This apparent uselessness notwithstanding, there are still a considerable number of scientists who believe in God, and some are probably devout believers. Why are they not atheists, if scientific knowledge seems to exclude religion - at least to the extent that science declares the existence of God very unlikely? Is not it a paradox?

One can explain this "schizophrenic" attitude of believing scientists by the human ability to place contradictory concepts into non-intersecting domains of conscience in order to evade a cognitive dissonance. So, a person may be a deeply religious thief (examples abound) or a highly superstitious intellectual. In the same way, a scientist may believe in God implying that it does not interfere with her/his scientific activities. I, however, still think that such split of conscience is a painful syndrome leading to substantial mental disorders. But one can try to reconcile believers and atheists within the scientific milieu. If the term "God" designates the presence of poorly understood - "mysterious" and low-probability - connections between events, then its usage is justified. Yet it is hard for an honest person to imagine God as a biblical character, more or less anthropomorphic looking down at each person and busy recording all minor sins. The idea that God takes a special interest in our petty affairs seems to be the manifestation of an extreme egocentricity. Such a model of God is most probably a product of the human brain, the latter being a very complicated object constantly performing simulations of the surrounding world. The models of God and other religious characters form a class of simulations which may be studied, in particular by scientific methods. In the process of such studies, one can find some results, expected or not, for example, that religious rituals affect human health indicators (see, e.g., https://www.oecd.org/els/health-systems/49105858.pdf) for a given region or the distribution of wealth in the population. Then the whole ladder of further questions arises such as what exactly is the effect of religious rituals on health, do they influence the human immune system, improve the transport of oxygen, or else? In case the wealth distribution varies as a function of, say, frequency and intensity of prayers, one can establish the optimal timing and tone. All this can be precisely measured, and quantified, medical or economic prescriptions would be elaborated without any mysticism. And this is not a blasphemy but a typical science with practical output (see more on that in [178], chapter 2; see also [179]).

The rationalization of religion is in the well-known thesis: absence of evidence is no evidence of absence. It is in general always very difficult to prove an absence of some entity. That is to say, absence of evidence of presence of God is no evidence of God's absence. There is, however, an obvious logical error here: if you assert that some entity does exist, it is your business to provide the proofs of its existence. Otherwise, one can postulate anything, for instance, that some invisible blue pigs are floating in the stratosphere and produce global warming. So logically there is nothing wrong in being an atheist, although in some cultures you have to be very careful with it. Crowd effects often bring about forces directed against scientific truth, or simply against truth. One can observe how the wake of religious fundamentalism, of an uncritical faith, is instigating human aggression, setting people against each other. Of course, this crowd effect [44] is used by politicians for their purposes having nothing in common with scientific truth. But if scientists who are often passionate about scientific truth still yield to the religious pressures of the crowd, then it tends to result in a serious damage for a personality and a society.

---

[44] To have a feeling of the local crowd effects, one might ask an arbitrary number of individuals, e.g., in Russia: "Are you a racist?" or in Germany: "Are you an antisemite?" and the answer would be in most cases an indignant "No", despite the multitude of racist manifestations in Russia and antisemitic ones in Germany.

Thus, religion is not a rational concept, or a point of view based on empirical information meticulously collected and calibrated in an objective manner. It is not even an idea. Although being a product of brain simulation, religion is primarily a passion, often a deep passion. One of the popular justifications of religion is that it makes people happy, contrariwise to science. "Will a person come to an academician when she/he is suffering? No, he will come to a priest." I have heard this sentence several times. This argument makes religions similar to drugs: using them may also offer a temporary consolation, which is of course deceptive. Likewise, in religious faith, one is comfortable not with truth, but with dreams - the fact exploited by the terrorists setting up suicide bombers. In this respect, "war on terror" is effectively the war waged against the abuse of an uncritical religious faith, although the politicians are reluctant to admit it.

Religious models are difficult to make compatible with science already because they operate with undefined notions. Can one define, for instance, the term "soul"? The concept "God" itself is hard to define, which has long been a subject of theological discussions. The concept of God can symbolize human weakness - this statement might sound banal, but it does not cease to be true in spite of banality[45]. Specifically, in ancient times, the notion of God had manifested a collective intellectual deficiency and attempts to resort to the ultimate superstition instead of analysis. There exist in human history several written formalizations of such superstitions, mostly taking the form of loose collections of legends. One of the most influential such collections - the Bible - is a set of beautiful tales, most of them directly contradicting experimental facts of science. Even the Christians (who are today not the most intolerant believers) still do not always accept the theory of evolution, the origins of life, cosmological models and even some astrophysical facts. It is well known that contradictions with observational data produced acute conflicts in the Middle Ages and many outstanding people suffered, were condemned or murdered by the political bureaucracy called the church. This is a typical exploitation of cognitive - and occasionally even of mathematical - models by ideological institutions relying on the conformist mainstream of mass conscience. Nevertheless, the set of fancy narratives, partly childish, contained in the Bible and other codified belief formats had been compiled by extremely creative persons. One may notice that religion is not necessarily focused on the belief in God, especially in its quasi-anthropomorphic form; it may be an abstract set of hypotheses intended to describe the world and to express one's feelings about it. The real meaning of the Bible, for example, lies in social models (which are abundant in this book) such as the collision of a small, enlightened elite and a vast uncultured mass of oppressed population - a conflict typical of the societies based on ideological coercion and "oriental despotism". There may be tolerant religions, but they are rather an exception, and the representation of reality fixed in a dominating religion in the given society tends to be obligatory for its members. Even relatively modern societies such as Russia can be clericalized to the extent of religious intolerance "us" against "them".

---

[45] If something is banal, it does not necessarily imply that it is wrong. It may be just simple. For instance, it may seem banal that one should exit through the door and not through the window. Nonetheless, it is true and, in most cases, more convenient.

As already mentioned, the world we live in is, almost as in ancient times, largely incomprehensible to the most of us, and the absence of any coherent picture produces a profound frustrating effect requiring some psychological defense. In such a seemingly inconsistent world where no easily understandable order can be observed by an ordinary person, untrained in physically sophisticated modeling, nothing is incredible. So, one could without much cognitive difficulties accept the statement that, e.g., exactly 2 $10^4$ angels were sitting on the tip of a pin. Why not? According to the religious concepts, the universe is populated by angels, after all. More and more people, disoriented, backward, and self-opinionated are taught by the popes to vaunt their ignorance as a virtue, like Roman matrons.

However, it is not at all obvious that religion helped humans to survive as a biological species, spreading their DNA (if we assume that the propagation of DNA is the ultimate purpose of the species' existence). At an individual's level, religion may help dying but it interferes with living. "I do not want to hear what is true, but only good news", this is the motto of religious doctrines. From the cultural side, if science can be figuratively called a kitchen of the future, religion may be likened to a cellar of the past. As almost all ideological doctrines, religion tends to preserve the yesteryear. This is natural because religion and its self-appointed carrier - the church - claims to possess the ultimate knowledge so it must use a selective filtering of scientific data to ensure the invariability of this old knowledge. If a person believes in six days of creation or that the Earth is flat and rests on something (three whales, elephants, or tortoises), one cannot resist these beliefs, in spite of the fact that they clearly contradict the evidence. There have recently been numerous debates about teaching religious models at schools, and this would amount to overviewing ancient beliefs about flat Earth, tortoises and so on. It would be a very entertaining read, of course.

The fact that religious models of reality are close to superstitions may be seen from the fact that almost all religious beliefs are labeled "superstitious" by representatives of the rival religion. Omens, supernatural events, miracles, the concept of afterlife, etc. - in other words, miracles that contradict the laws of physics are present in all religions but addressing them in various religions is different and cross-religious accusations of superstitions are commonplace. If we take the Orthodox Church as an example, we shall see that it is tacitly (and even openly) declared that all other Christian denominations, except Orthodox, are not genuine churches. Even rival Orthodox flavors can be anathemized. This is a manifestation of the fact that churches like other ideological bureaucracies cannot get along with one another. Intolerance of extraneous and new ideas is a key signature of the church who always claims to possess the ultimate truth. Whereas bundled antimodels such as astrology, mysticism or postulating immaterial parts of a person such as soul, spirit, ghosts, etc., which are distinguished from physical objects and biological life, are propagated by hostile ignoramuses, religious models are mostly imposed by obscurants.

## 2.14  Science and Arts

Most people can observe the crucial difference between the practitioners of the arts and the sciences, pointed out by a prominent British writer C. P. Snow in his famous book "The Two Cultures" [171]. Honestly speaking, I don't quite understand why this book has stirred such a controversy: it contains quite obvious observations and some mild criticism of the education system. The main idea is that cultural differences between "artists" and "scientists" obstruct the solution of the world problems. "Artists" take it for granted that there is no absolute truth, for example in literature, painting, music, etc. The public and other artists, the peers or the reference group, make value judgments which can be modified with time, so the truth in "arts" is only relative.

"Scientists", in contrast, are inclined to loathe the concept of relative truth and it is considered suitable in the scientific milieu to hope that there is much more in science than public opinion or even peer approval, no matter how influential, predominant or well-intentioned this opinion may be.

It is not so easy to apply the category of correctness to the conflict of these two groups. Both may be consistent within their own cognitive frame of reference. There are of course extremists on both sides, similar to religious extremists, who can be infuriated by the opponents' views. In this extreme case there is no communication between the two camps. Each of us can attribute herself/himself to each of these large groups: either to the camp of arts and humanities or to the one of science and engineering, irrespective of profession or place in the society. However, the thesaurus accumulated during the lifetime is important, and it is this load that determines as a rule the respective frame of reference. When I describe a particular mathematical model or an experimental fact, I presume by a prospective reader an a priori understanding of terminology, basic ideas, and assumptions. It saves time because I don't have to provide verbose preliminary explanations. But the flip side of the coin is that the same thesaurus hinders the understanding of how the world works. One cannot readily apply this knowledge to other important subjects, and the ability to accumulate new knowledge is limited, e.g., by the memory capacity and available time. So, the knowledge gaps are natural.

It would be still completely respectable to say, even among the people who are considered educated: "I know nothing about physics, and I never liked it" or "I can't do math", but it is only seldom that one would dare saying: "I have never read any novel" - that would single out such a person as dumb. In the popular science journals, mathematical formulas are typically forbidden in order not to frighten lay public, which would reduce the circulation[46].

As I have mentioned, it would be difficult to prove in general that nature is rigidly governed by some impersonal laws, and at least these laws, if those exist, are not necessarily condensed in a simple mathematical model like, for example, Newton's law or the Schrödinger equation. The "objective reality" is, unfortunately, a philosophical belief and as such it can hardly be taken as a guiding principle for science. At least this belief should not be a sacred cow of science. From here, it would be just one step to "hidden parameters". One could see in many experiments, and not only in physics, that the observer and the observed can be interdependent. This very fact leads to the idea that knowledge is a construction created by the observer, e.g., in the form of a mathematical model, rather than an unveiling of a pre-existing hidden reality. It is, by the way a perennial controversy among mathematicians: is mathematics a purely mind construction or does it discover some hidden facts. I don't think all this is important. Good science, to my understanding, should not be opinionated and a priori dismissing, but always prepared to amend the views of the world by continuously creating new models induced by new observations. Imposing the hidden reality principle as opposite to the "artistic" view of cultural relativism seems to be a poor defense for "scientific truth". It is totally unnecessary because perpetually accommodated facts and constructed models do not depend on philosophical beliefs - is there an objective absolute truth or not. It is known that P. A. M. Dirac criticized his colleague J. Robert Oppenheimer for his keen interest in poetry: "The aim of science is to make difficult things understandable in a simpler way; the aim of poetry is to state simple things in an incomprehensible way. The two are incompatible." [14]

---

[46] Biological schemes, in my view, are much more complicated, but they are allowed.

Although physics and mathematics are very important sciences, there exist also other disciplines, in particular those that are located in the area between science and arts. These disciplines, such as medicine, architecture, sociology, computer science[47], to some extent biology, have also accumulated rules, principles and models which are often called scientific laws within respective communities. These laws, however, distinctly reflect cultural influences and there is no certainty whatsoever that these disciplines will eventually become free from the cultural load. It does not mean of course that one should incorporate anti-science such as voodoo, astrology or creationism, support "proletarian" rejection of genetics and quantum mechanics or study Lysenko's biology because of the pronounced element of cultural relativism. Science may be neutral with respect to philosophical beliefs, but it can hardly be "naked" with respect to social influences.

## 2.15  Physics and Philosophy

The most appealing cognitive models take the form of verbose philosophical discussions. Honestly speaking, I dislike philosophy and fully agree with the aphorism often ascribed to L. D. Landau that physics ends exactly where philosophy begins. Nevertheless, I must admit that some philosophical systems or generalizations can be quite clever. For example, I enjoyed reading Karl Popper [141], partly H. Reichenbach [58], and when philosophical concepts are essentially based on hard facts and records rather than being merely speculative spiritual products, they may become a source of inspiration for more precise research. An example is the philosophy of history by A. J. Toynbee [142], sometimes harshly criticized by his fellow historians, but still a forerunner of a modeling approach to history, sort of a theoretical history. Apart from some brilliant think-stars - mostly due to their eloquent style - the rank-and-file philosophers do not produce anything interesting. At least, however hard I tried, I could find nothing attractive, on the Internet or otherwise. Having grown up with obligatory Marxist-Leninist philosophy, which is just a dubious form of ideological aggression, I thought at first that free-thinking Western philosophers, unavailable or even forbidden in the Soviet Union, would be less prejudiced and offensive. I was bitterly disappointed, especially in the philosophical discussions of scientific concepts. I found that the typical style of such discussions was a mixture of arbitrary assertions and pompous terminology. I also discovered pretty soon that it is probably the custom to make philosophical polemics as intellectually aggressive as possible, which reminded me of good student times in the USSR. I am not going to give references, although it is not difficult despite the fact that I am not a professional philosopher, but each time I encounter philosophical discourse of scientific issues I have a feeling of being submerged in some viscous medium. Scrolling philosophical papers, one can see that philosophers typically dislike mathematical models, favoring texts "not contaminated" with mathematical expressions. This makes polemics confused and ambiguous. To me, a right mathematical model, simple and clear, speaks for itself, without the need of verbose philosophical defenses. "Some people speak from experience, others, from experience, don't speak".

---

[47] Software engineering, for example, in contrast to "hard" sciences, incorporates and accentuates a considerable human component.

A witty person - not a very frequent quality amidst professional philosophers, Bertrand Russel once said that philosophy lies between science and religion, being attacked from both sides. One of the typical reprimands is that philosophy and cognitive models in general may be only good for writing papers, they do not result in any products - it contrasts to, for example, numerical simulations. This is simply not true. Recall, for instance, such a bestselling product as "Civilization", a very successful computer game entirely based on the historical philosophy model by Arnold J. Toynbee. It would be difficult to assert that strictly technical products such as those stemming from discretization of equations more stimulate the advancement of science and technology than, e.g., cognitive representations of reality[48].

## 2.16  Prognosis

Many people are seduced by an apparent simplicity of making forecasts. Every wild and speculative projection is a model of the future. Unfortunately, most forecasts of the future usually take the form of unconditional qualitative statements without indicating their accuracy. In such a form, they are just irresponsible cognitive models. Even outstanding scientists have produced odd and erroneous forecasts. Thus, Lord Kelvin (Sir William Thomson) was known to predict in relation to the Marconi experiments in 1897 that "radio has no future" or, while being the president of the Royal Society, that "heavier-than-air flying machines are impossible" or that "X-rays will prove to be a hoax" (see, e.g., http://zapatopi.net/kelvin/quotes/). These curious facts indicate that making prognoses is a difficult business. A witty sentence attributed to Mark Twain emphasizes the hardships of dealing with the future: "The art of prophecy is very difficult, especially with respect to the future" (see, however, http://bancroft.berkeley.edu/MTP/.)

---

[48] I dare to recommend in this respect my own article in the Russian journal "Priroda" (Nature), No. 6, 1997, p. 43-48 (in Russian).

Prognoses border on science or, rather, antiscience fiction. They are almost always extremely inexact and almost never try to establish their error margins, which is much worse than inexactitude. Unless one deals with more or less well-understood phenomena such as, for example, that heat will flow from a hotter to a colder body until the two bodies get the same temperature there seems to be no general scientific principle for prognostic activity. One might justifiably ask about prognoses: on what principle can they be found? I would answer that prognoses are always based on the same thing: clear understanding of the process which evolution one is trying to forecast. In other words, you must have a firm model of at least a non-contradictory hypothesis about the causal links in this process i.e., what consequences may be expected if certain strings are pulled. Besides, one must accumulate hard facts, not just opinions nor even expert considerations, and be able to quantitatively assess those facts, in particular using mathematical techniques on statistical data. The latter are, in most cases, available, e.g., in economics or social studies, but one usually lacks a good model.

It is in this lack of reliable models that the problem with trustful prognoses may be located. Regretfully, since time machines and crystal balls are not produced in ample quantities so far, there is currently no other way to gain insight into what is lying ahead of us than mathematical modeling, albeit with highly imperfect models.

One might notice two major things about prognoses: firstly, sophisticated statistical methods and software codes seem to give no advantages over intuition; secondly, there is no responsibility for making false forecasts. This is especially obvious in ecological catastrophism. I recall that around thirty years ago doomsday forecasts were abound predicting acid rains over the entire Europe, perishing forests, and steppe-forming everywhere. Who cares now about acid rains? Carbon dioxide projections are now on the agenda. The dire prognoses centered around the anthropogenic global warming (AGW) concept predict that many countries will suffer from either terrible heat waves or terrible cold, that Europe is likely to experience very low winter temperatures, and humans will be on the verge of total extermination. There are many attempts to substantiate such prognoses with computer models, but unfortunately the value of these models is indeterminate since there is no clear understanding of the physical processes controlling the evolution of the climate system.

Demographic prognoses: can you say with - not certainty but at least some convincing probability - what can we anticipate, demographic collapse in many countries or inadmissible population explosion? As far as sophisticated scientific techniques vs. human intuition in forecasts go, it seems to be an interesting feature of the European culture that only numbers - no matter how inaccurately evaluated - are accepted, which automatically discards intuition as a "serious" forecasting method.

Overcoming the shortcomings of commonplace models is a long process.

Typically, during half a life we are building stereotypes, clichés, and dogmas - almost automatically, without paying much attention to it - then we find out that the accumulated stereotypes begin constricting us, and finally, when they collapse, we feel relief and are happy to be free. Stereotypes prevent us from building bold models on which prognoses can be founded. Besides, one tends to confuse adherence to stereotypes with common sense. Moreover, science has become so dispersed that it is difficult to imagine successful forecasting results in the real - transdisciplinary - world. Prognostic activities seem to be horizontal (a philosopher might say "Locke-like") rather than vertical ("Hobbes-like"), within a specific discipline. Making intelligent prognoses requires a lot of knowledge and may appear unacceptably high-brow for a great number of people who believe in astrology and divine, e.g., Bible-based, creationism. Politicians, in their continual pursuit of gaining popularity, use to attract these people's voices to promote the solutions having unfavorable impact on future development. Should one construct nuclear power plants? Can astrology be taught at schools? Should one use genetically engineered food? How can governments regulate nativity (birth rate), if at all? There are many important issues having a direct impact on the future, about which people cannot form intelligent judgments if these people are scientifically illiterate. It means that prognostic activities, unless they are restricted to a narrow circle of bodies which may be biased in favor of the technologies they have been supporting or developing, would need to increase the general level of scientific education.

The most interesting thing for the people who long for prognoses is their questionable usefulness. There exist customers for whom long-term prognoses, extending over one thousand years, are of vital practical importance. For instance, there is an urgent problem of radioactive waste management and disposal, especially of high-level waste such as spent nuclear fuel. Some of the radioactive elements contained in spent fuel have long half-lives, for example isotope Pu-240 has the half-life of 6800 years and the half- life of Pu-239 is 24000 years. Besides, plutonium is highly toxic chemically. Therefore, radioactive waste containing spent nuclear fuel must be isolated from the environment and controlled for many thousands of years. Ideally, an impenetrable barrier of radiation protection must be placed between high-level radioactive waste and the biosphere, which would be intact for such a long period. Nuclear waste is supposed to be buried in special mines (disposal in crystallite rock) or salt mines as in Gorleben, Germany, and one must be sure that nothing will happen to these sites in several thousand years, e.g., their walls will not be destroyed due to intensive heat and radiation released by decaying high-level waste. If, in other example, the nuclear waste repositories are constructed in permafrost zones, one must be sure that permafrost does not melt in $10^3$ - $10^4$ years. In general, determining whether the site would be suitable for permanent disposal of the high-level waste is a very challenging scientific and engineering problem.

Many now fashionable dogmas may become obsolete in the near future. The same will probably apply to today's specialized training such as in computer science and information technologies (IT). The currently precious degrees and certificates which required years of hard work may be declared worthless and give no advantages for available jobs[49].

Furthermore, attitude to prognoses drastically changes with time. In the 13th-14th centuries, people who tried to make prognoses were declared witches or warlocks and burned alive. Now such people exhort from nearly all TV channels. People who make plausible forecasts, e.g., of oil or stock option prices, earn a lot of money irrespective of the accuracy of their prognoses. Prognoses over 1-10 years are extremely important for economies and highly politically charged so that the scientific component in such prognoses may easily succumb to political pressure. Moreover, politicians typically use prognoses for their purposes: an obvious example is climate forecasts. Climate is a complex physical system, and its space-time variations must be studied by scientific means. However, in this issue political interests take over. For example, the well-publicized Kyoto Protocol is a political action based on presumption that global warming of the near-surface atmosphere is, firstly, a proven fact and, secondly, is only due to human industrial activities. The fact that the climate has undergone many noticeable variations, comparable or exceeding current changes, during the existence of the Earth is ignored or rebutted as willful accusations - largely on political reasons.

In reality, there exist powerful natural factors affecting the climate, and human-induced changes in the gaseous composition of the atmosphere are not necessarily the dominating ones. Unless one makes serious modeling of physical processes, people will never get ahead of climate changes - one can only be reactive. However, current projections of the future climate are largely a matter of faith. One must admit that today's science knows too little about the Earth's climatic system to make the models which are built today a reliable foundation for meaningful predictions. However, making catastrophic prognoses is a sheer pleasure, a sort of a sweet candy. Doom, nemesis, downfall have their trade cycle, and fears can be better sold (recall Y2K or the recent - unsuccessful - panic about the black hole formation during the LHC experiments). Yet the catastrophes, to our great luck, seldom materialize. As a rule, non-cataclysmic, routine development is statistically more probable. So, prognoses are closely connected with risk assessment, and the word "assessment" implies a quantitative approach. Here, I would only like to remark that there are no zero risk endeavors. One can only compare the risks.

In the world of physics and mathematics, prognoses are obtained as a statement having the form `IF ¡conditions¿ THEN ¡prediction¿` i.e.,

---

[49] Who remembers nowadays the Novell or Digital certificates highly praised at the beginning of 1990s?

as an answer to a problem corresponding to some mathematical model. Therefore, prognoses can be only as accurate as the model employed and input data used, and one should always interpret the predicted values in the context of conditions. The situation is quite opposite in the social disciplines and political space. Here, the hidden dangers of exceeding the validity of a model or of unduly extrapolated results are typically ignored, and prognoses usually have an unconditional character. For instance, the language of environmental catastrophism is used to proclaim a universal prophecy, physical conditions notwithstanding.

The ability to predict future behavior and even events depends on our understanding of the laws determining such behavior. If we restrict ourselves to physical phenomena and natural events, then one can speak of physical laws. Nevertheless, even for some physical systems which are subordinated to well-understood physical laws the long-term prediction can be impossible, as in the case of chaotic systems. Here, one usually brings the example of weather forecasts illustrating the impossibility of a reliable long-term prediction because of intrinsic instability, but there are other important examples whose value is often underestimated. As mentioned above, it has been lately of fashion to make alarmistic predictions of the imminent climate catastrophe stating that the Earth surface temperature will rise by 2-10 degrees centigrade in 50 years exclusively due to human activity[50]. Based on such predictions essential political decisions affecting the lives of many million people have been made. Many European politicians are probably sure that industrial development directly leads to occasional summer heat waves and because of the excessive number of cars in private hands there is not enough snow at ski resorts. The truth, however, is that the climatic system, being influenced by a lot of factors outside human control, is at least as unstable as the weather. It means that the quality of long-term climate forecasts is by necessity very low and should be carefully scrutinized by non-engaged experts before being put as a starting point for political moves.

Certain forecasts[51] are bordering on dreams, for instance, intergalactic space travel or time loops. Such concepts belong to the realm of science fiction rather than science - at least at today's level of knowledge. The prognoses in the real world are limited by what one can expect for the future, based on scientific extrapolations. Nevertheless, most of the prognoses take the form of an informal study - not in the form of mathematical models - of a likely future development. Interestingly enough, this lack of mathematical description relates both to "soft" issues (e.g., social unfolding) and technological development.

There may also be "chaotic" trajectories or at least parts of trajectories, when the future or some of its elements such as technological development cannot be envisaged in principle. Such parts of world trajectory start after passing a singularity in the space of parameters determining the world as a dynamical system. However, I am not sure that such a space of parameters and, in particular, a phase space can be correctly stated for the biological subsystem of the world i.e., the biosphere. Of course, one can formally define the phase space of all the positions and momenta for all N particles constituting the biosphere, but this physical procedure would not have much sense. Firstly, it would be an uncontrollable idealization since in such a construction a closed classical system would be assumed (considering an open quantum system operating with myriads of particles involved in biological processes would probably be a hopeless task). Secondly, one can hardly proceed further from formally defining a phase space since there seem to be no criteria to select the subsets of biologically relevant collective variables (similar to elementary excitations in inorganic physics) which mainly determine the evolution. So far there is no algorithm for biological dynamical systems in terms of microscopic variables. Already this limitation of knowledge, which may have nothing to do with instabilities or chaos, leads to essential unpredictability in foreseeing the path of the biosphere (a subset of the total world trajectory). Thus, unknown species may emerge. A very similar situation can be observed in the technological evolution: the standard example is the Internet. Could professional futurologists predict its explosive development fifty years ago?

---

[50] I failed to find explicit mathematical relationships giving these figures as an outcome, so I do not know the accuracy of such prognoses. Is the uncertainty 10 percent, 100 percent or not quantified at all?

[51] The term "prognosis" sounds more scientific than "forecast", the latter invoking associations of clairvoyants and other charlatans; however, I do not see a substantial difference between the two terms and use them interchangeably.

Nonetheless, making forecasts can be a lucrative business if these forecasts do not contradict current political attitudes. Of course, the question whether the politically motivated prognoses have anything to do with reality is totally irrelevant. For example, in already mentioned climate forecasts there are billions of dollars endorsed by decision-makers and invested in "independent" research as a reward for the scientific backup of political moves. It may be very tempting to make forecasts, but when one dares to ask about their accuracy the predictor is usually abashed and does not know how to reply.

One may count five mainstream directions of science and technology development in the 21st century: (1) information and communication technology (ICT); (2) biotechnology (BT); (3) nanotechnology (NT); (4) space research (SR); (5) nuclear power and fusion (NPF). Since all of these sectors are technologically oriented, one must confess that there is not much room in the 21st century for great individual scientists, their time seems to be gone. Besides, the general public displays less and less esteem to individual minds, this phenomenon, together with rising infantilism and what might be called Africanization[52], can be observed everywhere in the world. Technological

---

[52] By Africanization I understand not only the intensive immigration from Africa to Europe, but the entire emerging system of values, to some extent opposite to the so-called protestant ethics.  Once in hot summer, I saw a barefooted university teacher in the canteen area, and nobody seemed to be astonished. The trend in dress is exactly opposite to that which dominated, say, in 1960s or 1970s when a person felt uncomfortable without coat and tie at a conference (see old conference photographs). Today, it is perceived almost abnormal when a person attends a conference wearing

evolution fosters not Einsteins, but efficiently managed scientific-technological collaborations, and indeed such groups are taking over. Macrosocially, creeping socialism in the 21st century, infantile striving of the population for governmental control, and increasing bureaucratization, alongside with higher taxation and lessened personal freedom are favorable to the development of group science versus individual achievements.

## 2.17  Some Tricks of the Trade

What mathematical techniques should we use for modeling? The common modeling triad is: constructing, solving, and interpreting the results. For constructing a model, usually elements of the systems science are employed, e.g., basic concepts of dynamical systems, control theory, linear systems and the like. For solving, simple mathematical methods such as calculus, linear algebra, and theory of differential equations are predominantly used. More sophisticated mathematics, honestly speaking, is rarely needed; more useful techniques than, say, working on complex manifolds seem to be model reduction, e.g., by dimensional analysis (which is, however, a peripheral part of the group theory). With the advent of more computer power, the principal art of model building has become to determine what is needed to synthesize an insightful computer simulation. This involves a careful selection of relevant numerical methods, designing or utilizing an algorithm, doing the programming and using Web resources.

One may notice that all the so-called laws of nature have the form of equations. It is, in principle, not at all obvious that equations must play such a principal part - laws of nature might be expressed as inequalities, long codes or concise verbal statements. But they are, to our current understanding, expressed by equations which may be considered a hint at a primary pattern for efficient mathematical modeling. So, assume that we must select some equations to describe processes or phenomena. As a rule, real - not too much idealized - systems and processes are described by nonlinear equations, e.g., nonlinear partial differential equations (PDE) with respect to time and spatial coordinates. Such systems are distributed in space and correspond to an infinite number of degrees of freedom. Nonlinear PDE are usually very complicated and only very specific types of such equations admit observable analytical solutions. However, models - by their very definition - can be drastically simplified, for instance, by considering first spatially-homogeneous situations. In this case, the equations modeling the system do

---

jacket and tie. Although the tie is in general a redundant attribute, I prefer the jacket-and-tie style to the T-shirted one in a conference room. Besides, this old style appears to be more compatible with scientific discipline, e.g., in performing calculations, than the "creative" T-shirted eccentricity. Africanization promptly intrudes into people's lifestyle; it aggressively manifests itself in numerous love and gay parades, overwhelming popularity of rhythmical recitatives, expressive collective dances, and whining tremolos. I have noticed that young people with tattoos and piercing are met more often than without such symbols of infantile exhibitionism. Key notions of this Africanized colonization of Europe are "fun" and "sexy", both symbolizing hedonistic protest against the parochial protestant ethic, boring but  productive.

not contain spatial derivatives and become ordinary differential equations (ODE-based models). Such a system is called point-like or having null-dimension (0d). In other words, it is the transition to homogeneous (point) models or, more exactly, ignoring spatial distribution of quantities, $\partial/\partial \mathbf{r} = 0$, that leads to ODE instead of PDE. A good example of such reduction to ODE-based (0d) modeling is the treatment of atmospheric processes in terms of radiation balance and global near-surface temperature dynamics (see Chapter 10). It is always a favorable situation when one can reduce the equations to the most primitive form, retaining the essence of the model.

Many equations of motion for classical mechanical systems have the form of ODE.  Such systems are point-like or lumped. In modeling lumped mechanical system with ODEs, each degree of freedom is described by a second-order ODE. Differential equations of the first order correspond to 1/2 degrees of freedom. Here one must be careful, since in the theory of dynamical systems the notion of a degree of freedom may be defined differently: each degree of freedom corresponds to a phase space coordinate (see Chapter 4 for more details). For example, instead of denoting a practically important case in chaos theory as having 1.5 degrees of freedom one can say "a system having 3 degrees of freedom".

The equation $dx/dt = f(x,t)$ gives an example of a dynamical system, i.e., the one whose behavior is uniquely determined by its initial state (deterministic behavior). Strictly speaking, one may also consider such dynamical systems whose future states are not uniquely determined by the evolution of initial ones, but we shall not discuss them in this book.

What are the typical ways to simplify mathematical models provided they have been written in the equation form? The usual tricks of the trade are:

- disregarding small terms (in fact expansion in power series)

- using small or large parameters (as a rule, it is an asymptotic expansion)

- replacing geometrical forms by more symmetrical ones

- substituting constants for functions (in fact applying the average value theorem)

- linearization

- scaling

- transition to dimensionless units (a special case of scaling)

- discretization and introduction of lattices (lattice models)

In this book, we shall employ all of these tricks to simplify the problems. Now, for a brief demonstration, let us see how, e.g., scaling works. If the modeling situation is rather complicated and one has no a priori idea what system of equations can describe it, one may choose from intuitive  physical considerations some parameters that could in principle determine the quantity we are primarily interested in.  One can assume that such an interesting quantity has some physical dimensionality (this is an overloaded term, please don't confuse it with the dimensionality of mathematical space) and can be represented as the product of undefined so far powers of parameters determining the interesting quantity. By equaling the exponents in the dimensions of the quantity we are interested in and in its expression as a product of determining parameters, we obtain some simple system of algebraic equations specifying the dimensions. This description of a simple trick sounds a little abstract, but later I shall show on examples how it is done. The most salient example is probably the so-called point explosion problem solved by J. von Neumann [299] in the USA and L. A. Sedov [300] in Russia[53] using just this method. The principal idealization in the point explosion models is the assumption that the energy release is instantaneous and in dimensionless point (explosive material has zero mass and occupies zero volume). One can find the solution details to the point explosion problem in the textbook by L. D. Landau and E. M. Lifshitz [85], §106.

The dimensionless form of mathematical models plays a somewhat special role among other tricks: the matter is that numerical values are independent of measurement units. Thus, one can consider specific cases of the model just by choosing numerical limits and not by comparing physical quantities. This fact is especially important in numerical modeling: in numerical techniques being applied to dimensionless models, one can readily neglect terms that are small in numbers as compared to errors in other terms (but one in general cannot disregard terms that are small compared to other terms). And we know that estimating errors is essential for validating the model.

Historically, scaling and dimensionless combinations appeared first in the problems of heat and mass transfer, these problems represent complex interactions of thermodynamical, fluid dynamical and electrodynamical processes. In the so-called multiphysics processes in general, such as the study of turbulence, multiphase system dynamics, scaling and dimension analysis is a powerful heuristic tool. For example, in the problem of high practical importance such as Chernobyl heavy accident modeling (motion of the fluid with internal heat sources) scaling considerations and dimension analysis alone enabled one to produce tangible results (see [54]). The well-known computer code "Rasplav" designed to model heavy nuclear accidents was based to a large extent on scaling and dimensional analysis.

This chapter could have carried the title "Models of the World and the World of Models" because the set of mathematical or other cognitive representations reflecting the "reality" are forming the parallel world which sometimes appears totally different from the world we live in. Had we limited ourselves to just the evident features as ancient people had, we would be satisfied with the picture of the flat Earth (everyone knows how difficult it is to walk over a ball's surface!), small Sun rotating around the Earth (it is but totally obvious!), little shiny stars, and the God who made all the things that we see around including ourselves.

---

[53] As well as by J. Taylor in USA and K. P. Staniukovich in Russia.

Luckily for the species who call themselves "human sapiens" due to developed capabilities for abstractions, the perception of the Universe is not so direct: it is mediated by mental images being eventually transformed into models of the world, the most advanced of which are mathematical ones. Models reveal hidden mechanisms of the world allowing one to understand and exploit them. Some of them are so deeply hidden that they initially induce massive protests (quantum theory, relativity, and genetics are standard examples). Nevertheless, the achievements of technology are mostly due to such "crazy" models which were perceived at first as fantasies (Maxwell's electrodynamics, quantum mechanics, genetic engineering, etc.).

Mathematical modeling teaches us how we can rely more on mathematical techniques rather than on gut instincts. At the same time some mathematical modeling is intended to quantify things which typically are not quantified.

Science in general works like brain - by model-fitting. The brain constantly fits models which are, as we have discussed, crude and partial representations of reality to the data obtained from the outside world.

Paradoxes and discrepancies[54] provoke people to invent new models, many of them are still awaiting their mathematical polishing (e.g., climate, social and economic systems).

---

[54] One can recall such well-known problems as those of the cosmological constant (see Chapter 9) or biological evolution which stimulate scientists to constantly generate new theories.

# 3 Mathematical Potpourri

In this book, only that portion of mathematics which is directly - in some rudimentary sense - connected to physics is discussed. The preceding sentence is an example of a statement deprived of any exact meaning. I try not to be afraid of making such statements - intuitive and logically irresponsible. Sometimes vague formulations invoke vivid associations and provide more cognitive freedom than rigid pedantry. It is a wonderful delusion that mathematics is based on strict axiomatic foundation. Try, for instance, to construct a fair definition to the notion "trapezium", and you will find out that there is no consensus among professional mathematicians even about such a simple object. It is well known that the fundamental notion of set is not defined properly. The same applies to the not less fundamental notion of field (in the physical sense), of space and time, spacetime, and a great lot of other, intuitively embraced notions. There are a lot of verbal constructions in modern physics which are not defined at all and can be understood only at the hand-waving level. In classical physics, the term "pseudovector" - a very useful notion - hardly has a precise definition. The list of poorly defined but widely used concepts is rather long, and we shall have to discuss a number of them, so I am prepared to take the criticism of "verbosity" and "philosophy". And although, frankly speaking, I like hard technical definitions, I try to evade algebraic gymnastics in order to put an accent on physical systems and not on mathematical structures. Thus, only the facts and statements relevant to physics will be presented in this introductory mathematical chapter.

One can easily find an extensive treatment of the subjects touched upon in this introductory chapter in a large number of sources including those Internet-based. As a matter of fact, only basic definitions and notations used throughout the book are presented here. Some well-known mathematical facts to be used further in the text are only briefly mentioned, with the references for a more detailed acquaintance being provided. In principle, this chapter is not indispensable for further reading.

Another goal of this chapter is to overcome rather acute cultural differences between the mathematical and physical communities and to share an informal understanding of the mathematical subjects I happened to deal with. Although the subjects I propose to discuss in this book may be only tangential to the main matters dealt with by professional mathematicians, these subjects would enable us to handle physical (and related to physics) models contained in subsequent chapters in a free and flexible manner.

Classical mathematics has always been a language and an intrinsic part of physics, and many eminent scientists, even mathematicians, for example V. I. Arnold, think that it is a pity that mathematics revolted and diverged from physics in the second part of the XX century. Nowadays more and more mathematicians consider mathematics to be totally different from physics. In this chapter, I shall attempt to revert this trend and discuss mainly those mathematical techniques that are relevant for physics. I tried to evade formally correct but long, boring and not fully physically comprehended mathematical derivations, with formulas used mainly as illustrations serving to clarify the whole picture of the modeled phenomena. Such an approach reflects the general philosophy of physmatics where math is considered as a service subject. Bearing some risk of being reprimanded for triviality, I tried to explain mathematical formulations in detail, unless they were really very simple.

Since I wrote this book mostly for mathematically oriented physicists and engineers, I was motivated by the applications - to find an appropriate mathematics for them. I selected physically based mathematical models not only by their value for specific engineering or scientific problems but also by their intrinsic methodical interest. Nevertheless, revering some scientific or engineering usefulness, I largely omitted works of pure mathematicians - although I admit their extreme importance. References to these authoritative works are mostly relegated to the notes and are also contained in the literature I cite. There are strong elements of translation work in this chapter. Mathematical insiders may know a lot, but few from the physical or engineering community would understand them. At least the time required to dig into notations and terminology may be too excessive. I am fully aware that I would probably be executed by mathematicians for some vulgarization of mathematical discussions. My physmatical level implies a certain primitivizing.

Mathematics is a process rather than a fixed scope of knowledge, and the process of math is a very living thing: people visit each other, attend the conferences, talk and communicate. However, I shall try to give only a short account of standard mathematical subjects, and this account is far from being rigorous or complete. Indeed, each matter I touch upon here can be adequately treated in a fair lecture course given by an expert mathematician. However, honestly speaking I would not recommend supplementing books on physics that appear not sufficiently rigorous by some authoritative Bourbaki-style treatises. The point is that the latter are, as a rule, too abstract to be useful. The Bourbaki group endeavored the mission of unification and standardization of mathematics, at least of its major part, and many people think (even those who did not read the volumes) that the Bourbaki authors have succeeded. Nevertheless, even though this unification and standardization has been performed, the work of the Bourbaki group now turns out to be largely useless, at least for the majority of physicists. The matter is that when we handle a problem, especially in physics or applied mathematics, we rarely need that high level of abstraction which was used by the Bourbaki authors. Moreover, what we really need - at least what I know from my own experience - are specific techniques or results to deal with the details of our models and solutions. Many mathematicians copying the Bourbaki patterns pay very little attention to practical needs; they prefer keeping to the most general possible setting. Such "ivory tower" (and partly schizoid, i.e., extreme alienated) treatment devoids abstract mathematics of any perspective of use by a physicist or an engineer because she/he misses completely the purpose of abstraction. The power of abstraction should manifest itself when dealing with concrete situations. Are the Bourbaki volumes widely read by physicists or engineers?

In modern physics, various mathematical concepts and methods are used, e.g., differential equations, phase spaces and flows, manifolds, maps, tensor analysis and differential geometry, variational techniques, groups in general and, in particular, Lie groups, ergodic theory, stochastics, etc. In fact, all of them have grown from the practical needs - mostly those of mechanical engineering - and only recently acquired the high-brow and somewhat pathological axiomatic form that makes their study and direct application so difficult. Of course, the level of mathematical sophistication should always be adequate, otherwise one may get absorbed in mathematical prerequisites instead of solving physical or engineering problems. In some respects, modern mathematics diverges from physics and tends to philosophy, since, like philosophy, mathematics increasingly juggles with concepts invented just for this purpose. Mathematics cares very little about the relations of these concepts to physical observations. For example, although geometry lies at the core of mathematics, it has many practical applications, but mathematicians focus only on symbolic forms and typically could not care less about any practical use. Some of them are flaunting their superiority stating that "mathematicians do not like applied stuff" - a sentence I heard many times since my school years. Physics also deals with mathematical concepts, and ideally the latter are meaningful only upon demonstrating their relations to observations. At least, the physical traditionalists would cheer these relations. Nowadays, when string theories and some highly mathematized models are in fashion, despite the fact that they are unsupported by experiment, the role of mathematical sophistication increases. Pragmatically speaking, a person possessing extra mathematical skills can more easily generate new versions of currently popular theories in order to gain reputation and grants. Rather few physicists think about foundational problems; most scientists make their careers by mathematically treating specific problems. For example, quantum theory, which is maybe the most fundamental science, allows one to produce both a vast number of applied mathematical models and speculative interpretations. Mathematical performance can be sufficient for career achievements in modern physics, irrespective of the agreement with reality. Partly because of that I decided to devote a considerable place to mathematical preliminaries and the discussion of mathematical techniques.

I hope that owing to that preliminary chapter all parts of this book will be understood by physics or engineering students, the precise direction of studies or working experience being immaterial. The book does not contain really advanced mathematics, so nobody will sweat blood over fanciful notations and lengthy definitions. Proofs are more intuitive rather than technical, to a possible disgust of mathematicians. Nevertheless, I don't think that the readership would consist of mathematicians alone; for many uninitiated, e.g., for engineers and physicists, numerous technical proof details are hardly essential. I would be happy if I could share with mathematically not necessarily well-informed readers some working knowledge of certain subjects of mathematics that can be applied to construct physics-based models. I am inclined to think that an obsessive drive to extremely great generality and precision accompanied by a scholastic pedantry, which are becoming customary in contemporary mathematics, build up a considerable barrier between mathematics and modern science. Making their language available to other scientists is not an easy task, many professional mathematicians usually neglect it. My observation is that to preserve a certain separation between themselves and the rest of the scientific/engineering community is considered a certain chic among the mathematical community, especially among young mathematicians. This group norm does not actually serve as an eye-opener for physicists and engineers, and I call it for myself the tedium barricade, and to avoid it I tried to mix up mathematical rigor in homeopathic doses.

An example of the detrimental effect of fashion mixed with the trend to obsessive rigor is an attempt to formulate everything in the language of differential forms - a wonderful instrument *per se*. They have been introduced nearly a century ago and I think they are indispensable for many applications in physics (and maybe in some derived disciplines). This is an example when "classic" - classic vector analysis - does not necessarily mean the best. Now, just try to ask physicists who of them is using differential forms in her/his regular work. I did it, although my sampling was neither very extensive, nor representative. Nonetheless, my primitive sociological experiment showed that only a very tiny percentage of working physicists are familiar with this tool, which may be considered a convenient alternative to vector calculus, the latter being ubiquitous in physics and engineering. Some respondents have even become aggressive: "Differential forms? What are they good for? To formulate the Stokes' theorem? All this new stuff - differential forms, bundles, connections, multilinear algebra - is just ostentatious propaganda tricks devised by some clique of mathematicians striving to become known." Needless to say, I surveyed only theoreticians; it would be impertinent to ask experimentalists or engineers about forms - I would have been heartily laughed at.

This is not accidental. I could not find in the abundant literature a good exposition of differential forms for physicists. Most of the sources I came across were written by mathematicians with utter neglect to the physicist's needs. Maybe I was unlucky[55]. But the fact remains: mostly because of modern mathematics, "physics is becoming so unbelievably complex that it is taking longer and longer to train a physicist. It is taking so long, in fact, to train a physicist to the place where he understands the nature of physical problems that he is already too old to solve them". These words belonging to E. P. Wigner, a prominent mathematician and at the same time a great physicist - a rare combination, reflect an unfortunate tendency that, according to D. Hilbert, "physics is becoming too difficult for physicists". The sheer variety of physical models combined with newly obtained experimental facts, multiplied by the abundance of mathematical methods and channeled to clumsy computers to obtain numerical results make the whole phsymatical construction unobservable and lacking any consistent interpretation. Any person who studied physics must somehow cope with his/her ignorance, and persons claiming to have mastered the most part of physics, including the relevant mathematics, are just arrogant hypocrites.

---

[55] In this book, I am using mostly the more ubiquitous formalism of traditional tensor analysis (one might call it the Riemann-Ricci calculus) [154], rather than the less widespread techniques of differential forms (Cartan calculus) [103, 155, 104]. This does not reflect my personal preferences, but I do not think they are relevant. Sometimes I give both versions, which amounts to some kind of translation.

On the other, mathematical, side, it may be a common opinion among the majority of modern mathematicians that the mathematical methods used by physicists would not stand up to the current level of rigor. I have heard several times that the methods the physicists use appear mysterious, ad hoc or merely incorrect to many mathematicians. The latter claim to have a natural aversion to inexact statements. Honestly speaking, I don't care. Moreover, I think that the majority of these mathematicians would also derogate, for example, many of Euler's methods on the same ground. But somehow there appeared no new Eulers amidst such rigorists. Certainly, mathematical rigor is extremely important, but it cannot replace the intuitive and imaginative line of thought that Euler shared e.g., with Einstein, Dirac, Feynman. There were not many people who could combine extended mathematical skills and deep physical insight. Interestingly enough, the latter may not necessarily help in practical work, e.g., when treating quantum models. To keep a balance between physical model-building and mathematical abstraction is probably the most intricate art in the work of a physicist. So instead of threading one abstraction after another, I was trying to comment on some contemporary developments that might be useful for physicists.

I have already mentioned that the questions of rigor are only of secondary importance for an average working physicist. Even nowadays, when physical problems are frequently formulated in the language more palatable for mathematicians than for physicists of older generations. One might recall that during the most fruitful period of physics in Russia (1950s-1970s) dominated by the "Landau school" (see Chapter 2) it would be highly improbable to encounter the word "theorem" in the leading Russian physical journal JETP.[56] I suspect that this word was informally prohibited in JETP. Published papers practically never began with formal definitions, and terminology was quite different from that of the mathematical literature. This difference of cultures was emphasized by physicists and mathematicians alike. I remember very well that when I was studying analytical mechanics our professor strongly advised us against using the textbook "Mechanics" by L. D. Landau and E. M. Lifshitz [23] as a source, allegedly because of the lack of rigor in this book.

One more cultural difference between the physical and mathematical communities is related to notations. In distinction to mathematical texts where notations are often specially designed, notations in physics are not used to stress the logical structure of the notions involved, but rather to make the heuristic physical calculations as transparent as possible.

---

[56] One could meet the word "theorem" in the text of an article as a reference, e.g., "according to the Wigner's theorem", but the formal structure usual for mathematics "definitions theorem - proof" was probably inadmissible in JETP.

## 3.1    Sets

The notion of set is known to survive without a formal definition, at least I have never encountered any. A set can be loosely understood as a collection of objects usually called elements. Other terms for a set may be a variety, a class, a family, an assemblage or an assortment. Synonyms for the things called elements may be, e.g., members, constituents or even points. Here, I shall denote a set of elements $x_1, x_2, x_3, \ldots$ by curly brackets $\{x_1, x_2, x_3, \ldots\}$. A set containing one element is called a singleton. In case each element $x_i$ belonging to a set $A$, $x_1 \in A$, is also an element of a set $B$, then $A \subseteq B$, i.e., $A$ is a subset of $B$. Obviously, when both relationships $A \subseteq B$ and $B \subseteq A$ are valid, then $A = B$. Some sets have standard notations, e.g., $\emptyset$ is an empty set (containing no elements), the set of all real numbers is denoted $\mathbb{R}$, the set of all complex numbers is $\mathbb{C}$, the set of all integers is $\mathbb{Z}$, the set of all natural numbers is $\mathbb{N}$.

By a domain, I shall mean a connected open subset in $\mathbb{R}^n$ or $\mathbb{C}^n$ for some positive integer $n$. Throughout this book, I shall use $x = (x^1, x^2, \ldots, x^n)$ or $z = (z^1, z^2, \ldots, z^n)$ for the coordinates of a point in $\mathbb{R}^n$ ($\mathbb{C}^n$). Sometimes it is notationally more convenient to consider domains in $\mathbb{R}^{n+1}$ ($\mathbb{C}^{n+1}$) rather than in $\mathbb{R}^n$ ($\mathbb{C}^n$) starting coordinate sets from $x^0$ ($z^0$).

There are simple rules to operate with sets. For example, one can easily prove the following identities for sets $A, B, C$ (see any textbook on the set theory):

$$(A \setminus B) | C = A \setminus (B \cup C), \tag{3.1}$$

$$A \setminus (B \setminus C) = (A \setminus B) \cup (A \cap C), \tag{3.2}$$

and

$$(A \cup B) \setminus (C \setminus B) = (A \setminus C) \cup B. \tag{3.3}$$

This theory seems very abstract, yet it may help solving practical problems, specifically in computer science (informatics). One can easily construct an example of a real-life situation when the set theory reasoning can help. Suppose you have to analyze the test results for a group of students being examined on three subjects: mathematics, physics, and French. There were $N$ persons to undergo the tests, with $n_m$ successful in math, $n_p$ in physics and $n_F$ in French. We know that $a_{mp}$ failed both in mathematics and physics, $a_{mF}$ in mathematics and French, $a_{pF}$ could pass neither physics nor French; $a_{mpF}$ failed in all three subjects. The question is, how many students, $n_{mpF}$, successfully passed all three tests? In this example, just to draw the usual pictures (set diagrams) would help to solve this problem.

Now, to make the discussion of sets more concrete, let us review basic algebraic structures. A vague notion of a "mathematical object" does not necessarily denote a set with a certain algebraic structure, but if it does then a mathematical object can be understood as some algebraic structure, for example, group (a set with a single operation), ring (a set with two operations), or a vector space. A little below we shall briefly discuss mappings (or maps) between sets; now I just mention that in algebra the sets endowed with operations are typically studied which makes the following question relevant: what happens with the algebraic structure i.e., with the set elements and operations on them when one set is mapped to another?

## 3.2  Maps and Operators

In the study of sets, the most important concept seems to be mapping. Assume there are two sets, $X$ and $Y$, then a rule $F$ assigning $y \in Y$ to each $x \in X$, $F : X \to Y$ i.e., $x \mapsto y$ is called a mapping or a map. In the general algebraic context one can define such mappings as isomorphisms which are typically understood as bijective homomorphisms. For the above examples of the most popular algebraic structures such as groups, rings, or vector spaces, homomorphisms are defined as some specific, context-dependent mappings, e.g., a linear operator for the vector space, group and ring homomorphisms respectively for groups and rings. In any case, a homomorphism is one of the basic morphisms, it is a mapping that preserves the algebraic structure of two sets or algebraic objects to be mapped from one to another, in particular, from the domain into the target image of a function.

The term "eigenfunction" refers to a function that preserves its original form (perhaps up to a multiplicative complex constant) after having passed through some system, e.g., an engineering or measuring device.

## 3.3  Groups

One can observe that the concept of a group is the simplest and the most basic among all algebraic structures used in physics. A group is an algebraic structure consisting of a set of objects together with a binary operation on this set. That is an operation which produces a third object of the same kind from any two. Historically, the notion of a group has originated from studying the symmetry transformations in geometry on a plane (what we used to call planimetry in high school). Algebraic generalizations of such transformations have led to classifying symmetry properties through the operations regarded as elements of certain sets. After the invention of matrices[57] and especially following their systematic study by A. Cayley in the 1850s, matrices began playing the key role in linear algebra. Accordingly, matrix representations of groups emerged, a tool especially widely used in quantum mechanics. The discovery of Lie groups, named after Sophus Lie, an outstanding 19th century Norwegian mathematician, has led him to the idea of applying continuous groups to the solution of differential equations [190]. Recall (and we shall see it below) that a Lie group is simultaneously a differential manifold obeying the group properties so that one can study Lie groups by using ordinary calculus of infinitesimals. Incidentally, S. Lie himself was known to call such groups infinitesimal ones.

Mathematical models of physics are always constructed in some spaces which are sets of elements, in general of any kind, endowed with an appropriate mathematical structure. Standard mathematical structures typically have an algebraic or a topological nature. An example of an algebraic structure is the group; this structure may be considered one of the simplest. (Some elementary topological structures will be briefly discussed below.) A set $G$ of elements of arbitrary origin is known as a group if a group operation is defined and the following rules are fulfilled.

---

[57] It is difficult to trace who really has introduced matrices in mathematical calculations, see [191].

1. For any two elements $a, b \in G$, there exists an element $c = a \otimes b, c \in G$.

2. This operation is associative: for any three elements $a, b, c \in G$,

   $(a \otimes b) \otimes c = a \otimes (b \otimes c)$.

3. There exists a neutral element $e$ such that for any $a \in G$, $a \otimes e = e \otimes a = a$.

4. For each element $a \in G$ there exists an inverse (symmetric) element $a^{-1}$ such that $a \otimes a^{-1} = a^{-1} \otimes a = e$.

Note: if the group operation $\otimes$ is called multiplication (usually denoted as $\times$ or $\cdot$), the element $c$ is called a product of $a$ and $b$, the neutral element is called unity, and the group itself is known as multiplicative. If the group operation is understood as addition, the element $c$ is usually written as $c = a + b$, the neutral element is called zero, and the inverse element to $a$ is also called the opposite one, being written as $-a$. The group itself is known as additive. If for any two elements $a, b \in G$, $a \otimes b = b \otimes a$, the group is called Abelian or commutative.

The simplest transformation group is probably the one of rotations of a circle, we shall look at it below in some detail. Although one can probably find even simpler examples, rotation of a circle is interesting to us since it is a trivial analog of the $SO(n)$ group which corresponds to rotations of $n$-dimensional Euclidean space $E^n$.

### 3.3.1. Semigroups

One can weaken some of the imposed rules for the group, and then other algebraic structures arise. For example, a set of elements without a compulsory neutral (identity) element $e$ and without an inverse element $a^{-1}$ is known as a semigroup. Semigroups became important, in particular, in the study of dynamical systems described by ordinary and partial differential equations. The matter is that any evolutionary problem modeled by a dynamical system may be formalized as an initial value problem for some ordinary differential equation (ODE), in general for a vector function $f$ defined on some space $V$:

$$\frac{df(t)}{dt} = Af(t), f(t_0) = f_0,$$

where $A$ may be interpreted as an evolution generator. The formal solution to this problem is $f(t) = \exp\left\{\int_{t_0}^{t} A\,dt\right\} f_0$ or, in the simpler case of an autonomous problem, $f(t) = \exp(t - t_0) A f_0$. This expression can be heuristically interpreted as an action of a semigroup of operators (neither the neutral nor the inverse elements are in general required), $\exp(t - t_0) A$, taking the initial state $f_0 = f(t_0)$ to $f(t)$. However, one should give a precise meaning to the exponential $\exp(t - t_0) A$. We shall deal with this issue a number of times in different contexts.

## 3.4    The Rotation Group

Studying the rotation group is necessary for many quantum-mechanical applications such as the theory of angular momentum, atomic spectra and nuclear physics. There is one stronger motivation to become familiar with the rotation group: it is a wonderful example of general Lie groups, and the latter serve as a natural tool for treating symmetry in physics (for instance, the Lie group $SU(3) \times SU(2) \times U(1)$ plays the most essential part in the Standard Model of elementary particle physics).

Let us start, as promised above, from a very simple example of the group of rotations of a circle. Let $g_\alpha$ denote the rotation by angle $\alpha$ so that the representation $R(g_\alpha)$ of this operation i.e., operator $R(g_\alpha)$ acting on a space of functions $F(\varphi)$ of the form $(\varphi) := f(\cos\varphi, \sin\varphi)$ is defined by the expression

$$R(g_\alpha)F(\varphi) = F(\varphi + \alpha).$$

This operation symbolizes a rotation by angle $\alpha$. One can slightly generalize this expression as $R(g)f(x) = f(g^{-1}x)$.

To be more specific, imagine for a moment that there exists a single point which may be regarded as a center of universe (it is not necessary that God is sitting exactly at this point). Then we may consider this point as a universal coordinate origin so that the entire space can be rotated about it. Let us denote such rotations by $g_i$. It is clear that the $g_i$ form a group, i.e., all the axioms stated above are fulfilled, with the group unity corresponding to the rotation by zero angle. Let $\mathbf{r}$ be the vector connecting our origin with a point $(x^1, x^2, x^3)$.

The rotation group and its affine extensions such as the Galileo group with elements $g(t, \mathbf{r}) = t + b, R\mathbf{r} + \mathbf{v}t + \mathbf{a}$, together with Lorentz and Poincaré groups, play a fundamental role in physics determining the invariant structures in geometric transformations.

## 3.5    Lorentz and Poincaré Groups

One can find a number of common features in the Galileo and Lorentz groups. Primarily, they both are intended to define admissible reference frames for an inertial observer with respect to any event in spacetime. As to the Poincaré group $P(\mathbb{R}, 3)$, it is just an affine extension of the Lorentz group, a semidirect product of the Lorentz group $\Lambda(\mathbb{R}, 3)$ and the abelian translation group $T(4)$ acting on a Minkowski spacetime, $\Lambda(\mathbb{R}, 3) \rtimes T(4)$. Here, I shall remind us that a semidirect product $C = A \rtimes B$ of two groups defines a way to form a new group $C$ from two groups $A$ and $B$ which then become subgroups of the new group. In particular, an affine group of all non-degenerate motions (isometries)[58] on a plane $\mathbb{R}^2$ contains transformations defined by the pair $(A, \xi)$ where $A \in O(2)$ i.e., the matrix $A$ belongs to the group of orthogonal $2 \times 2$ matrices describing rotations and reflections while keeping the origin fixed, and $\xi \in \mathbb{R}^2$ is a plane vector i.e., belonging to the abelian translation group $\mathbb{R}^2$ so that $\mathbf{x} \mapsto A\mathbf{x} + \xi : \mathbf{x}' = A\mathbf{x} + \xi$ with $\det A \neq 0$. The composition rule in this simple affine group has the form:

$$(A, \xi) * (B, \eta) = (AB, \xi + A\eta) \coloneqq A \rtimes B.$$

We shall observe some features of the Poincaré group as the affine extension of the Lorentz group i.e., the semidirect product of the translations and the Lorentz transformations a little later. Let us now discuss some formalism pertaining to the most basic relativistic groups, primarily the Galileo (mainly homogeneous) and Lorentz group. These groups give rise to the two types of theories that are known as Newtonian and Einsteinian physics so that one might justifiably ask: do all the differences between these two types of physics, including interacting particles and not only inertial motion, stem from the differences in the structures of the Galileo and Lorentz (Poincaré) groups? I think this is a rather complicated question, and we shall discuss it - rather superficially - also a little later.

Recall the Galilean transformations (see more in Chapter 3 under Geometry of Classical Mechanics): $t \to t', \mathbf{r} \to \mathbf{r}'$ which we can symbolically write as

$$\begin{pmatrix} t' \\ \mathbf{r}' \end{pmatrix} = \begin{pmatrix} 1 & 0 \\ \mathbf{v} & \mathbf{I}_3 \end{pmatrix} \begin{pmatrix} t \\ \mathbf{r} \end{pmatrix} \tag{3.4}$$

Here $\mathbf{I}_3$ is an orthogonal $3 \times 3$ matrix which intuitively symbolizes rotations in the Euclidean 3d space i.e., $\mathbf{I}_3 \in SO(3)$. Let us choose, in particular, Galileo transformations along the $x$-axis, $\mathbf{v} = v_x \mathbf{e}_x = v\mathbf{e}_1$, then $\mathbf{I}_3 \to 1$ and we have

$$\begin{pmatrix} t' \\ x' \end{pmatrix} = \begin{pmatrix} 1 & 0 \\ v & 1 \end{pmatrix} \begin{pmatrix} t \\ x \end{pmatrix}$$

i.e. the simple Newtonian kinematics with an absolute time, $t = t'$. This is, of course, a very particular case even in the class of linear spacetime transformations, and we can find some generalizations of the Galileo transformations. For example, special relativity gives one possible generalization of Newtonian kinematics, when time and space become interconnected so that $t = t(x)$ and $t \neq t'$. Restricting ourselves to the linear relationships, we get in the general one-dimensional case

$$\begin{pmatrix} t' \\ x' \end{pmatrix} = \begin{pmatrix} a & b \\ c & d \end{pmatrix} \begin{pmatrix} t \\ x \end{pmatrix}$$

where matrix elements $a, b, c, d$ of the transformation (boost) matrix are so far unknown functions of the relative velocity $\mathbf{v} - v\mathbf{e}_x - v\mathbf{e}_1$. The standard way to obtain the Lorentz transformations i.e., to specify the above matrix elements $a, b, c, d$ is to consider the simple kinematic situation when two inertial systems move with respect to each other with constant velocity $\mathbf{v} = v\mathbf{e}_x$. This is a textbook derivation, and I shall briefly reproduce it so that we can later discuss some less trivial facts and ideas. So, the hyperbolic (light) form $s^2 = x_0{}^2 - x_1{}^2 = (x_0 + x_1)(x_0 - x_1)$ must be invariant[59] which naturally gives

$$x_0' + x_1' = f(v)(x_0 + x_1), \qquad x_0' - x_1' = \frac{1}{f(v)}(x_0 - x_1), \qquad f(v) \neq 0.$$

---

[58] Recall that an isometry is usually defined as a map that preserves the distance; for example, the plane isometry $f : \mathbb{R}^2 \to \mathbb{R}^2$ ensures that $d(x, y) = d\big(f(x), f(y)\big)$ where $d(x, y) = ((x^1 - y^1)^2 + (x^2 - y^2)^2)^{1/2}$ is the Euclidean distance corresponding to the norm $\|x\| = (x, x)^{1/2}$ where $(x, y)$ is an inner product in the Euclidean space $E^n$ (in this example $n = 2$).

One can assume the unknown scalar function of relative velocity $f(v)$ to be strictly positive so that, e.g., the difference $(x_0 - x_1)$ should have the same sign in each coordinate system. Obviously, $f(v) \to 1$ with $v \to 0$. Solving this elementary linear system of equations, we get

$$x_0' = \frac{1}{2}\left(f + \frac{1}{f}\right)x_0 + \frac{1}{2}\left(f - \frac{1}{f}\right)x_1, \qquad x_1' = \frac{1}{2}\left(f - \frac{1}{f}\right)x_0 + \frac{1}{2}\left(f + \frac{1}{f}\right)x_1.$$

If we now regard, for instance, the coordinate origin of the "moving" system i.e., the point $x_1' = 0$ with respect to the "laboratory" system, we get for $(x_0', 0)$, using the relationship $\frac{x}{1} = \frac{dx}{dt}$ in the "laboratory" system

$$x_1 = -x_0 \frac{f - \frac{1}{f}}{f + \frac{1}{f}} = \frac{v}{c}ct.$$

From this equation we can find the unknown function $f(v)$:

$$\frac{f^2 - 1}{f^2 + 1} = -\frac{v}{c} \equiv -\beta, \qquad f^2 = \frac{1 - \beta}{1 + \beta}$$

so that

---

[59] Here, to make the formulas ultimately simple I temporarily write coordinate indices below ($x_i$) and not above as it is conventional in geometry. I hope it will not result in a confusion.

$$x_0' = \frac{1}{2}\frac{1+f^2}{f}x_0 - \frac{1}{2}\frac{1-f^2}{f}x_1 = \gamma x_0 - \beta x_1,$$

$$x_1' = \frac{1}{2}\frac{1-f^2}{f}x_0 - \frac{1}{2}\frac{1+f^2}{f}x_1 = -\beta x_0 + \gamma x_1$$

since

$$\frac{1+f^2}{f} = 2\frac{1}{(1-\beta^2)^{1/2}} \equiv 2\gamma, \qquad \frac{1-f^2}{f} = 2\frac{\beta}{(1-\beta^2)^{1/2}} \equiv 2\beta\gamma.$$

Notice that $0 \le \beta \equiv v/c \le 1$ and the Lorentz factor $\gamma \equiv (1-\beta^2)^{-1/2} \ge 1$.

So, the elementary Lorentz transformation ("Lorentz rotation") in the $(x_0, x_1)$ plane takes the form

$$\begin{pmatrix} x_0' \\ x_1' \end{pmatrix} = \frac{1}{2}\begin{pmatrix} f+1/f & f-1/f \\ f-1/f & f+1/f \end{pmatrix}\begin{pmatrix} x_0 \\ x_1 \end{pmatrix} = \begin{pmatrix} \gamma & -\beta\gamma \\ -\beta\gamma & \gamma \end{pmatrix}\begin{pmatrix} x_0 \\ x_1 \end{pmatrix}$$

or in $(t, x)$-variables

$$\begin{pmatrix} t' \\ x' \end{pmatrix} = \begin{pmatrix} \gamma & -\beta\gamma \\ -\beta\gamma & \gamma \end{pmatrix}\begin{pmatrix} t \\ x \end{pmatrix}.$$

The word "elementary" implies in this context the simplest possible case of relativistic kinematics. If we now assume that Lorentz rotation corresponding to inertial motion occurs not in $(x_0, x_1)$ plane, but in the pseudoeuclidean (Minkowski) $\mathbb{R}^{3,1}$ space, the velocity $\mathbf{v}$ still being directed along $x^1$-axis, $\mathbf{v} = v\mathbf{e}_1$, then we get the following matrix representation of the Lorentz transformations

$$\Lambda^1(\mathbf{v}) = \begin{pmatrix} \gamma & -\beta\gamma & 0 & 0 \\ -\beta\gamma & \gamma & 0 & 0 \\ 0 & 0 & 1 & 0 \\ 0 & 0 & 0 & 1 \end{pmatrix}$$

Notice that matrix $\Lambda^1$ is symmetric, and the upper index 1 signifies that the motion occurs along $x^1$-axis. Matrices $\Lambda^2$ and $\Lambda^3$ i.e., corresponding to the case when spatial coordinates normal to the direction of motion are left intact, have obviously a similar form.  One can easily verify that $\det\Lambda^i(\mathbf{v}) = 1, i = 1,2,3$.  If we do not assume the velocity to be directed along any of axes $x, y, z$, we shall get a rather clumsy but still symmetric matrix (see, e.g., [192], B.1, Ch. 4) representing the arbitrary Lorentz transformation of special relativity:

$\Lambda(\mathbf{v})$

$$= \begin{pmatrix} \gamma & -\beta^x\gamma & -\beta^y\gamma & -\beta^z\gamma \\ -\beta^x\gamma & 1+(\gamma-1)\left(\frac{\beta^x}{\beta}\right)^2 & (\gamma-1)\frac{\beta^x\beta^y}{\beta^2} & (\gamma-1)\frac{\beta^x\beta^z}{\beta^2} \\ -\beta^y\gamma & (\gamma-1)\frac{\beta^y\beta^x}{\beta^2} & 1+(\gamma-1)\left(\frac{\beta^y}{\beta}\right)^2 & (\gamma-1)\frac{\beta^y\beta^z}{\beta^2} \\ -\beta^z\gamma & (\gamma-1)\frac{\beta^z\beta^x}{\beta^2} & (\gamma-1)\frac{\beta^z\beta^y}{\beta^2} & 1+(\gamma-1)\left(\frac{\beta^z}{\beta}\right)^2 \end{pmatrix}.$$

This matrix still reflects only the Lorentz boost i.e., a transition between two reference frames $K$ and $K'$ whose axes $x, y, z$, and $x', y', z'$, respectively, remain parallel in the transformation. It is only in this case that the Lorentz matrices are symmetric. In the most general case of Lorentz transformations, the $x, y, z$ axes are also rotated (by an arbitrary real angle), and then the corresponding matrices are not restricted to the class of symmetric ones. The matrix of the general Lorentz transformation may be obtained as the product $L(1,3) = O^{-1}(1,3)\Lambda(\mathbf{v})O(1,3)$ where $O(1,3)$ is the matrix of Galilean rotation

$$O(1,3) := \begin{pmatrix} x_0' \\ \mathbf{r}' \end{pmatrix} = \begin{pmatrix} 1 & 0 \\ \mathbf{v} & \mathbf{I}_3 \end{pmatrix}$$

leaving time $t$ invariant (see equation (3.4)).

The above elementary derivation of the Lorentz transformations, directly exploiting the invariance of hyperbolic bilinear form $s^2 = x_i x^i = (x^0)^2 - (x^1)^2$ (recall the appearance of the scaling function $f(v)$ ), is contained in numerous textbooks on special relativity (see, e.g., [192], B.1, Ch. 4)) so that I do not need to discuss it any further. One can, however, mention an elegant way to obtain the Lorentz transformation formulas in the textbook by Landau and Lifshitz ([193], 4), which is in fact a special case of the above classical derivation corresponding to the exponential mapping $f(v) = exp(-\theta(v))$ . Parameter $\theta$ corresponds in this representation to hyperbolic rotation in pseudoeuclidean space (Minkowski spacetime), in the simplest case in a complex $(x^0, x^1)$ plane[60]. Here $\tanh\theta = \beta$. Thus, using such $\theta$-parametrization, one can emphasize the notion of Lorentz rotation straight away.

Let us now discuss some features of the Lorentz transformations, a few of them not being very conspicuous. Apparently the Lorentz group, as presented above, is of a purely kinematic (i.e. geometric) nature - indeed it is the group of invertible linear transformations of $\mathbb{R}^4$ that preserves the quadratic form $(ct)^2 - \mathbf{r}^2 = c^2 t^2 - x^2 - y^2 - z^2$ where $c$ is a constant. The form $(x, y) = x^0 y^0 - x^i y^i, i = 1,2,3$ usually known as the Minkowski metric so that the Lorentz group is the one of linear isometries of this metric. Nevertheless, to be absolutely honest one has to admit that the Lorentz transformations for a single particle in a given spacetime point - and it is usually only a single particle that is considered in kinematic theories - may also depend on the interaction between particles, which makes Lorentz boosts sensitive to the presence of other particles or fields. One might pick up the hint about the necessity of possible dynamic corrections to purely kinematic Lorentz transformations already in the fact that time evolution of a system interacting with the environment is different from the time evolution of an isolated system (see more on time evolution in Chapters 4 and 7).

---

[60] We may call hyperbolic rotations about the coordinate origin in pseudoeuclidean space all linear transformations that do not change the distance to the coordinate origin and taking the cone domains, e.g., $s^2 = x_i x^i > 0$, into themselves. One can easily see that hyperbolic rotations form a group just like the usual rotations in $\mathbb{R}^n$. Indeed, a composition of two hyperbolic rotations is again a hyperbolic rotation as well as an inverse hyperbolic rotation, see more on it in [188].

However, the phenomena described by special relativity are not reduced to just kinematics. Recall that Lorentz transformations were used by Einstein in his famous first paper of 1905 on special relativity to express the transition between two frames of reference: one is considered to be fixed i.e., at rest with respect to an observer ("laboratory system") whereas the other is instantly co-moving with the observed electron, and the latter can be in general i.e., accelerating motion. It means that Lorentz transformations may contain the *instantaneous* velocity $\mathbf{v}(t)$ as a parameter, which fact manifests itself in relativistic mechanics considering not only kinematic but also dynamical situation. For example, the particle 4-momentum $p_i = \frac{\partial L}{\partial u^i}, u^i := \frac{dx^i}{d\tau}$, is usually expressed through the derivatives of coordinates over the proper time, $p_i = \gamma_{ij} p^j, p^j = mcu^j$ (in this particular case we restrict the dynamics to the Galilean plane metric), and the dynamical 3-momentum is differentiated with respect to laboratory time, as in nonrelativistic Newton's law,

$$\frac{d\mathbf{p}}{dt} = \mathbf{F} + \frac{\mathbf{v}}{c} \times \left( \frac{\gamma}{\gamma + 1} \left( \frac{\mathbf{v} \times \mathbf{F}}{c} \right) \right),$$

where $\mathbf{F}$ is the force field in the classical (Newtonian) sense, for instance, the Lorentz force. However, the question of applying special relativity to dynamical situations that imply an accelerated motion is not at all trivial (see more on that below in the "Relativistic Mechanics" section).

In special relativity, spacetime is considered plane (flat) so that one can describe it globally by the pseudo-Euclidean coordinates with the Galilean diagonal metric $\mathrm{diag}(1, -1, -1, -1)$ i.e., in the general metric tensor $g_{ik}$ only the terms with $i = k$ are left and the length element (usually known in relativistic physics as interval) is simply

$$ds^2 = (dx^0)^2 - (dx^1)^2 - (dx^2)^2 - (dx^2)^2.$$

All inertial (Lorentz) frames leave this metric invariant. This is of course a very special choice of the metric which, by the way, justifies the term "special relativity". One cannot introduce such a plane metric globally on a general manifold $M^4$ because the latter may be curved. In such cases the metric tensor $g_{ik}(x)$ has in general all off-diagonal components, $i \neq k$ and, besides, depends on spacetime point $x$ on the manifold $M^4$. Yet one can always[61], by an appropriate coordinate transformation, bring tensor $g_{ik}$ locally i.e., in the vicinity of $x$ to a diagonal form, in other words, to introduce the locally Galilean metric diag$(1, -1, -1, -1)$. Physically it means that the spacetime can be regarded as locally Euclidean. By the way, if such a coordinate transformation brings the metric tensor to a diagonal (Galilean) form in each point of $M^4$ then $g := \det g_{ik} < 0$ everywhere in $M^4$ - the fact that has a significant importance in general relativity.

Let us briefly discuss the meaning of the relativistic (Minkowski) interval $ds = (dx_i dx^i)^{1/2} = cd\tau$, where $\tau$ is the "proper time" (see [193], §§2,3). From the geometric perspective, this interval is just the line element whereas physically it manifests the constancy of the light speed and its independence on the frame of reference - an experimental fact verified many times with very high accuracy (see, e.g., [198], see also a comparatively recent account of Michelson-Morley type experiments in [199]. One can interpret the Minkowski interval for a physical body, say a particle, as the distance counted in units of proper time. For example, if a particle had a clock the latter would register the passage of time experienced by the particle. If the time is measured in seconds, this interval will be measured in light-seconds; if the unit of time is years, interval will be counted in light-years.

We can see that the Lorentz group is six-dimensional i.e., the Poincaré group $P(\mathbb{R}, 3)$ is the semidirect product of the Lorentz rotations $O(1,3) = \Lambda(\mathbb{R}, 3)$ and pseudoeuclidean spacetime translations $T(4) = \mathbb{R}^{1,3}$.

One can formulate a more general question: do *all* the differences between classical and relativistic physics, not restricted to inertial motion and special relativity, follow from the differences in the structures of the Galileo and Lorentz (or Poincaré) groups?

## 3.6   Rings and Fields

Other important algebraic structures apart from groups are rings and fields, although they are rarely mentioned in physical literature. Both structures are the sets of elements of any kind on which not one, as for the groups, but two operations - e.g., addition and multiplication - are defined. One understands a ring as a set which is a commutative group with respect to addition and a semigroup with respect to multiplication, with the distribution property, $a \cdot (b + c) = a \cdot b + a \cdot c, (a + b) \cdot c = a \cdot c + b \cdot c$. This means that the presence of unity and an inverse element with respect to multiplication is not required in rings. A field is a ring in which the set of all elements without 0 forms an Abelian group with respect to multiplication. For example, the real numbers $\mathbb{R}$ represent such algebraic structures as a field and a group whereas the set $\mathbb{Z}$ of integers forms a ring. If, however, the requirement of commutativity of multiplication is omitted i.e. the set of all elements without 0 forms a non-commutative group with respect to multiplication, such a ring is called a skew field; it is sometimes also called a division ring (an often cited example is the ring of quaternions). The ring of integers $\mathbb{Z}$ is a ring with unity 1, the ring of even integers $2\mathbb{Z}$ is a ring without 1. One might notice that generally in mathematical axiomatics it is important to pay main attention not to what is present, but to what is absent. Thus, in the ring axioms, commutativity of multiplication is not required as well as the presence of an inverse element, although it is not explicitly stated in the sequence of axioms. Examples of fields: the rational numbers, the integers modulo a prime, e.g., integers $3\mathbb{Z}$ of the form $3n$. For physics, the most important structures are the ring $\mathbb{Z}$ of all integers, $\mathbb{R}$ of real numbers, the field $\mathbb{Q}$ of rational numbers, and the field $\mathbb{C}$ of complex numbers. One can find plenty of other examples of rings and fields in any course of modern algebra.

---

[61] Provided some natural conditions are fulfilled, see [188] §87.

One can naturally define "subthings": subgroups, subrings and subfields as subsets of, respectively, groups, rings and fields retaining all their properties.

## 3.7    Morphisms

In abstract mathematics, morphism is a notion generalizing structure-preserving maps between two mathematical structures. This is, of course, not a definition but a free interpretation. To correctly define morphism, one has to make an excursus into category theory, which would take us too far away from physically motivated mathematics. My task in discussing morphisms is rather modest: primarily, to refresh the concept of mapping and, secondly, to eliminate the confusion that is frequently encountered as to what is an isomorphism, a homomorphism, an automorphism, and an endomorphism. Moreover, it is worthwhile to elucidate the relationship of all these morphisms to such fundamental notions as bijection, injection and surjection. Assume, for example, that there exists a bijection between two sets endowed with some algebraic structures, and if this bijection preserves all the operations defined for such algebraic structures, then it is known as an isomorphism whereas the sets between which an isomorphism is established are called isomorphic. If we now do not require a bijection between two sets endowed with algebraic structures, but an injection or a surjection, the operations on such algebraic structures still being preserved, we shall have a homomorphism, and the two sets between which there exists a homomorphism are called homomorphic. For instance, if two groups, $F$ and $G$ are homomorphic, the elements of $F$ corresponding to the identity (neutral) element $e_G$ of $G$ form a subgroup $H$ of $F$. Such a subgroup is known as a normal (or invariant) subgroup, whereas the group isomorphic to $G$ is denoted as $F/H$ and is called the quotient (or factor) group. It is also known as $F \, mod \, H$, $mod$ being a shorthand for modulo. To gain an understanding of a similar factoring and the respective terminology for rings and fields, consider two homomorphic rings, $\mathcal{P}$ and $\mathcal{R}$. The elements of $\mathcal{P}$ corresponding to zero in $\mathcal{R}$ form a subring $\mathcal{I}$ of $\mathcal{P}$ called an ideal. Then the ring isomorphic to $\mathcal{R}$ is called the quotient ring (rarely the factor ring) $\mathcal{P}/\mathcal{I}$.

## 3.8   Algebras

One of the algebras important for physics, although studied mostly not by the physicists but by pure mathematicians, is the so-called Jordan algebra invented by the Göttingen physicists Pascual Jordan in 1933 [208]. P. Jordan[62], together with his teacher M. Born [194] ("Dreimännerarbeit") as well as with other creators of quantum mechanics, W. Heisenberg, J. von Neumann, W. Pauli, and E. Wigner [195], was trying to develop algebraic tools making the manipulations with self-adjoint operators more convenient. The core idea of P. Jordan was quite simple: if we have two self-adjoint operators $A$ and $B$ on a Hilbert space $\mathbb{H}$, representing, for example, quantum "observables", then the product $AB$ is not necessarily self-adjoint[63], but the symmetrized (later called Jordanian) product $A * B := (AB + BA)/2$ is already self-adjoint.

## 3.9   Lie Groups and Lie Algebras

The rotation group is a typical example of a Lie Group, where the set is not a small, discrete set, but a continuum. The mathematical description of Lie Groups is often done by the Lie algebra related to the group, so it is worth taking a look at that.

In 1960s, Lie group representations became a dominant paradigm for the construction of the theory of elementary particles. In fact, it was not the theory, but attempts to classify the particles, a conventional framework rather than a physical theory. Some physicists did not address this classification other as "zoology".

Lie algebras are rather important for physics, although they are rarely covered in textbooks, even on theoretical physics (an exception is a useful course by F. Scheck [280]). The motivation to study Lie algebras in connection with physics stems from the necessity to systematically study binary, more specifically bilinear skew-symmetric operations which are abundant in physics. For instance, the Poisson brackets in classical mechanics, commutators in quantum mechanics, and the usual cross (vector) product in linear algebra - physical examples are torque and angular momentum - are specific cases of such binary operations. Recall that a binary operation defined on a set $G$ is understood as a map $f$ taking the elements of the Cartesian product $G \times G$ to $G$ i.e., $f: G \times G \to G$. A Lie algebra is a vector space $V$ (see the next section) over some field $K$ together with a skew symmetric bilinear operation $V \times V \to V$, often denoted by square brackets $[,]$ or simply $[\,]$, provided the Jacobi identity

$$\big[a[bc]\big] + \big[b[ca]\big] + \big[c[ab]\big] = 0$$

---

[63] Indeed, $(AB)^+ = B^+ A^+ \neq AB$.

holds. Thus, a vector product which is bilinear, skew-symmetric, and satisfying the Jacobi identity[64] turns our Euclidean vector space into a Lie algebra. Another example: it is not difficult to see that a set (space) $M(n, \mathbb{R})$ of all $n \times n$ matrices with real entries becomes a Lie algebra if the bilinear skew-symmetric operation is defined as commutator $[A, B] = AB - BA$. Bilinearity in the definition of the Lie algebra is understood as the property

$$[\alpha a + \beta b, c] = \alpha[a, c] + \beta[b, c], \qquad [c, \alpha a + \beta b] = \alpha[c, a] + \beta[c, b],$$
$$\alpha, \beta \in K, a, b \in V,$$

whereas skew-symmetry is merely understood as anticommutativity $[a, b] = , [b, a],\ a, b \in V$.

## 3.10  Vector Spaces

This is the basic stuff for great many mathematical models, and although it seems to be quite simple, its importance does not diminish because of simplicity. It is a great delusion that right things in mathematics or physics must be complicated - quite the opposite. For instance, superstring/M theory may be far too complicated, despite the generally inspired awe, to be the ultimate truth (see Chapter 9). So, in spite of being comparatively simple, the notion of vector spaces plays a fundamental role in physics - and not only in physics of course - and this may be a justification for a somewhat extensive, though not very rigorous, treatment of this subject. In fact, in the previous chapter, we have already dealt with vector spaces without giving an explicit definition.

There are numerous motivations to consider vector spaces in mathematics, classical and quantum mechanics, computer physics, computer graphics - you name it, and also in "soft sciences" such as economics, ecology or sociology. One of the obvious motivations is related to systems of linear equations

$$a_{ik}x^k = b_i, \qquad i = 1, \ldots, m, \qquad k = 1, \ldots, n.$$

---

[64] V. I. Arnold [196] has recently proved that the Jacobi identity has in this case the following physical meaning: it ensures that the three altitudes of a triangle have a common point (called the ortho-center). For a plane triangle, this fact was known already to Euclid and is, accordingly, discussed in geometry lessons in high schools (provided such lessons still exist), of course without mentioning the Jacobi identity. But in the case of non-Euclidean geometry, e.g., for hyperbolic triangles, the situation is less trivial. In his paper, V. I. Arnold used the Jacobi identity for the Poisson brackets of quadratic forms over $\mathbb{R}^2$ endowed with a canonical symplectic structure (see [197], see also section "Geometry of Classical Mechanics" in this book.)

Such systems naturally arise in a great number of applications, e.g., in numerical modeling, electrical engineering, industrial planning, even in agriculture.[65]

In general, a system of linear algebraic equations may, for instance, be represented as a linear combination of matrix columns

$$x^1 \begin{pmatrix} a_{11} \\ \vdots \\ a_{m1} \end{pmatrix} + x^2 \begin{pmatrix} a_{12} \\ \vdots \\ a_{m2} \end{pmatrix} + \cdots + x^n \begin{pmatrix} a_{1n} \\ \vdots \\ a_{mn} \end{pmatrix} = \begin{pmatrix} b_1 \\ \vdots \\ b_m \end{pmatrix},$$

where the parameters $a_{ij}$ (coefficients of the system) form the rectangular matrix

$$A := \begin{pmatrix} a_{11} & \cdots & a_{1n} \\ \vdots & \cdots & \vdots \\ a_{m1} & \cdots & a_{mn} \end{pmatrix}$$

called matrix of the system. If we denote the columns of matrix $A$ as vectors $\mathbf{a}_1, \ldots, \mathbf{a}_n$ where the lower index $i = 1, \ldots, n$ corresponds to variables $x^i$, then the question whether the system of linear equations can be solved would be equivalent to the question whether vector $\mathbf{b} = (\mathbf{b}_1, \ldots, \mathbf{b}_m)^T$ $\mathbf{b} = (b_1, \ldots, b_m)^T$ $\in \mathbb{R}^m$ can be represented as a linear combination of vectors $\mathbf{a}_i$ or, in more current terminology, $\mathbf{b} \in \langle \mathbf{a}_1, \ldots, \mathbf{a}_n \rangle$, where angular brackets denote the so-called span of $\mathbb{R}^n$ (see below).

If we consider first a homogeneous system of equation (vector $\mathbf{b} = \{b_i\}$ = 0), then we may notice that any two solution vectors, $\mathbf{x} = \{x^i\}$ and $\mathbf{y} = \{y^i\}$ produce also solutions $\mathbf{z} = \mathbf{x} + \mathbf{y}$ with $z^i = x^i + y^i$ and $\mathbf{w} = \lambda\mathbf{x}$, where each $w^i = \lambda x^i$ for any real $\lambda$. Thus, the set of solutions to a homogeneous system of linear equations has a vector space structure - vectors can be added (e.g., using the rule of parallelogram) and multiplied by a number. One can see that

---

[65] A typical problem known from ancient times: one has two fields with the total area 1 (in arbitrary units). One field supports crop output $a$ per unit area, the other $b$. The crop collected from the first field exceeds that produced by the second field by $m$. What are the areas of field 1 and 2? This problem is obviously formalized as a linear system

$$\begin{aligned} &x + y = 1 \\ &ax - by = m \\ &a > 0, b > 0, x > 0 \; y > 0 \end{aligned},$$

with the solution

$$x = \frac{b+m}{a+b}, y = \frac{a-m}{a+b}.$$

It would not be easy to obtain the field partial areas in a purely thoughtful logical way, without solving the system of equations.

a real vector space $V$ is just an additive group [66] whose elements can be multiplied by real numbers so that

$$\lambda(x + y) = \lambda x + \lambda y, \qquad \lambda \in \mathbb{R}, \qquad x, y \in V,$$

$$(\lambda + \mu)x = \lambda x + \mu x, \qquad \lambda, \mu \in \mathbb{R},$$

$$\lambda(\mu x) = (\lambda \mu)x,$$

$$\text{and } 1x = x.$$

So, vectors are elements of a linear (vector) space. If a linear space $V$ is exemplified by a coordinate space $\mathbb{R}^n$, then vectors are typically identified with $n$-tuples of numbers and written as $a := (a^1, \ldots, a^n)^T = a^1 \mathbf{e}_1 + \cdots + a^n \mathbf{e}_n \equiv a^i \mathbf{e}_i$. The vectors $\mathbf{e}_i$ form a coordinate basis in $\mathbb{R}^n$, the most convenient choice of the basis vectors is when they are orthonormalized i.e., their scalar (inner) product gives a Kronecker symbol, $\mathbf{e}_i \mathbf{e}_j = \delta_{ij}$. We shall see soon that the generalization of this scalar product naturally leads to the concept of metrics.

In mathematics, one usually defines vector spaces over some "field". This amounts to choosing $\lambda$ not necessarily from $\mathbb{R}$ but from another set with similar properties, e.g., $\mathbb{C}$ of complex numbers, or in general any set in which the operations of addition, subtraction, multiplication, and division (except division by zero) are allowed and the same rules that are familiar from the school-time arithmetic of ordinary numbers can be applied. In such more general cases the considered additive group ceases to be a real vector space.

One may notice by association with numbers that natural numbers were obviously the first used by the humans, these numbers could be easily added and multiplied, but the inverse operations - subtraction and division - required some head-scratching. Such inverse operations became possible only with the invention of negative and fractional numbers.

Quite often the definition of a vector space consists of a long list of axiomatic statements which may produce the impression that the vector space is a complicated object. In fact, just the opposite is true: objects that comply only with some parts of these rules are more complicated. For instance, human beings that obey none of such rules are not described by any satisfactory theory. But an important thing to be stressed is that vectors can be multiplied by numbers from some field called scalars and added according to the "rule of parallelogram" and these operations retain vectors within a clearly defined set which is the vector space. The possibility to use such operations means that vector spaces are supplied with a linear structure. The preservation of a linear structure is not a trivial requirement: for instance the maps $f : \mathbb{R}^m \to \mathbb{R}^n$ implemented by differential functions do not in general preserve the linear structure, although such maps are ubiquitous in geometry and in the theory of dynamical systems.[67] In vector spaces with a linear structure, linear maps are naturally the most important ones, just because they preserve the linear structure. A map $F: U \to V$ (map, mapping, operator, transformation are all synonyms. In quantum mechanics, the term "operator" is mostly used in connection with maps $F: U \to U$.) A map is linear if for all $x, y \in U$ and $a \in \mathbb{R}$ (here for simplicity we restrict ourselves to scalars from $\mathbb{R}$ ), $F(x + y) = F(x) + F(y), F(ax) = aF(x)$ , i.e., $F(ax + by) = aF(x) + bF(y)$ for all $x, y \in U$, and $a, b \in \mathbb{R}$. One may note that the set of all linear maps $F: U \to V$ forms itself a vector space with respect to addition and multiplication by a scalar. Symbolically, one may write $F: U \to V + G: U \to V = (F + G): U \to V$ or $(F + G)(x + y) = (F + G)x + (F + G)y$ and the composition of linear maps $F$ and $G$ is also linear, $GF(ax) = GaF(x) = aGF(x)$. Such a transitivity of linear maps is well reflected in the matrix theory: once a basis has been fixed, every linear map $F$ is represented by a matrix $F_{mn}$ and every matrix corresponds to a map (isomorphism). This fact is extensively used in quantum mechanics (see Chapter 6); actually, it is the mathematical reason for the equivalence of the Schrödinger and Heisenberg formulations, although both great physicists presumably did not wish to accept this equivalence in the initial period of quantum mechanics (1924 - 1928).

---

[66] A commutative (Abel) group with respect to addition.

There is a long list of familiar linear maps in mathematics and physics. For example, integration may be interpreted as a linear map or a linear functional. Let us take as $U$ the vector space of all real-valued continuous functions (with compact support) defined at $[0,1]$ and $V = \mathbb{R}$. Then

$$F(f) = \int_0^1 f(x) dx$$

A linear functional may be given for the same choice of $U, V$ also by, e.g., $F(f) = f(0)$. And of course, differentiation also produces a linear map: let $U = V$ be the vector space of all differentiable functions on $\mathbb{R}$, then $F(f) = f'$ is a linear map. In simple geometry in $\mathbb{R}^2$, rotation or shift

$$\begin{pmatrix} x' \\ y' \end{pmatrix} = f \begin{pmatrix} x \\ y \end{pmatrix}$$

with

$$x' = x\cos\varphi - y\sin\varphi$$
$$y' = x\sin\varphi + y\cos\varphi$$

or

---

$$x' = x + ay$$
$$y' = y$$

Here, it would be worthwhile to recall what the isomorphism is (in simple terms). A linear map $F: U \to V$ is an isomorphism if there exists a map $G: V \to U$ such that both $FG = E$ and $GF = E$ where $E$ is the identity map, $Ex = x$ for all $x \in U$. Strictly speaking, we must write $E_U$ and $E_V$ for spaces $U$ and $V$, respectively, but we shall neglect this difference. One usually says: $F: U \sim V$ is an isomorphism from $U$ to $V$ (or between $U$ and $V$). Since $G$ is the inverse of $F$, $F$ is invertible and $G = F^{-1}$. An invertible map is nonsingular, i.e., from $F(x) = 0$ follows $x = 0$ and if $F(x) = 0$ for $x \neq 0$ then $x$ is a singular element (a singular vector, if one considers a vector space) of map $F$. A set of such singular elements $\{x \in U: F(x) = 0\}$ is called the kernel of map $F$, denoted $KerF$. One can easily prove that $KerF$ forms a linear subspace in $U$ and if $U$ is finite-dimensional then $KerF$ is also finite-dimensional. It is exactly a subspace and not just a subset. In other words, the kernel, $KerF$ of $F: U \to V$ denotes the subspace of all singular vectors of $F$. It is easy to prove (see below) that $F$ is injective if and only if $KerF = 0$, i.e., map $F$ is non-singular.

The finite-dimensional ( $dimV = n$ ) vector space allows a simple geometrical or rather graphical (for $n \leq 3$) interpretation, with elements of the space being directed arrows connecting the origin and the point $x^1, \ldots, x^n$. In some different cases, e.g., for dual spaces (see below), other pictorial representations are more adequate. Usually, only the real (as in classical mechanics) and the complex (as in quantum mechanics) vector spaces, $\mathbb{R}^n$ and $\mathbb{C}^n$ respectively, are considered but nothing prevents us from using other number fields as the source of scalars over which the vector space is defined. Different fields are important in various scientific contexts, for example, finite fields consisting of 2,4,8,16 or in general $2^n$ elements are important in digital technology and theory of communication.

Allow me some remarks on the terminology and sources. Algebra seems to be a discipline which provides a certain freedom for each author to demonstrate his/her "mathematical culture". Therefore, there exists a variety of expositions of essentially the same basic facts. When I was first studying linear algebra, the contemporary exposition of maps in terms of injection, surjection and bijection was not yet accustomed. Honestly speaking, I had some difficulties in translating old notions related to mappings into this new (and in many situations very convenient) terminology. For instance, one can see that a map $F: U \to V$ is an isomorphism when and only when it is a linear bijection so that in the linear case the "old" notion of isomorphism and a comparatively "new" notion of bijection are equivalent. Indeed, a linear map $F: U \to V$ is by definition an isomorphism from $U$ to $V$ if there is an inverse map $G: V \to U$ such that $FG = E$ and $GF = E$. Thus, an isomorphism is a linear map by definition and a bijection since its inverse does exist. Conversely, if $F$ is a bijection, then there exists $G: V \to U$ (which need not be necessarily linear) so that $FG = E$ and $GF = E$. If, additionally, map $F$ is linear, then it is easy to prove that its inverse $G$ is also linear. Indeed, let us take two elements $v_1, v_2 \in V$. Then since $F$ is a bijection and, consequently, a surjection, there exist such $u_1$ and $u_2$ that $v_1 = F(u_1)$ and $v_2 = F(u_2)$. Since $F$ is linear, we have $v_1 + v_2 = F(u_1 + u_2)$ and $G(v_1 + v_2) = GF(u_1 + u_2) = u_1 + u_2 = GF(u_1) + GF(u_2) = G(v_1) + G(v_2)$. Therefore, an isomorphism implies that the linear map $F$ is surjective and nonsingular. In fact, non-singularity means that $F$ is an injection, i.e., its inverse contains not more than one element. Here, I am using the standard convention: the simple right arrow $U \rightarrow V$ denotes mapping from set $U$ to set $V$ whereas mapping of point $u$ into point $v$ is usually denoted by the "$\mapsto$" arrow.

Why is all this stuff important for physics? Primarily because we can in practical calculations - physical modeling - identify any finite-dimensional space $U$ with $\mathbb{R}^n$. More precisely, if $U$ is an $n$-dimensional vector space then there exists an isomorphism $F : U \rightarrow \mathbb{R}^n$. This fact enables us to easily change spaces by specifying a map $F$ from one $n$-dimensional space $U$ to another $n$-dimensional space $V$. Technically in practice this is done by fixing a basis and writing vectors as $n$-tuples of numbers. A map to another space is then defined by its action on the basis vectors. Let $U$ and $V$ be two finite-dimensional vector spaces ($dim U = dim V = n$) and let $\alpha = (\alpha_1, \ldots, \alpha_n)$ be a set of basis vectors that we have fixed in vector space $U$ and $\beta = (\beta_1, \ldots, \beta_n)$ the basis in $V$. Then we have $\mathbf{u} = u^i \alpha_i$ and $\mathbf{v} = v^i \beta_i$, with

$$v^j = [F(u)]^j = [F(u^i \alpha_i)]^j = u^i [F(\alpha_i)]^j$$

or $v^j = \alpha_i^j u^i$ where $\alpha_i^j = [F(\alpha_i)]^j$. It is important and highly convenient that linear maps can be represented by matrices but for this, as we saw, one needs to fix a basis (in the above example $\alpha = (\alpha_1, \ldots, \alpha_n)$ in $U = U_n$). According to the definition of a linear map $F$, the latter is determined by images $F(\alpha) = (F(\alpha_1), \ldots, F(\alpha_n))$. An inverse map is represented by the inverse matrix. See in this connection, e.g., the classical book by P. K. Rashevski [154], where one can find a lot of clearly explained details.[68]

We have just asked ourselves, what do we need all this algebraic stuff for? Apart from playing with bases, which is one of the favorite games for physicists and not particularly appreciated by mathematicians, elementary algebraic notions are really needed to explore the quantum world (Chapter 6). The concept of a linear operator that plays a key role in quantum mechanics, is conventionally defined as a linear map from a vector space $V$ to itself. In case this map is an isomorphism, it is called (as well as an operator on vector space $V$) an automorphism. A frequently used concept of the general linear group of $V$, $\mathrm{GL}(V)$ which will be from time to time encountered in this text[69], may be interpreted as merely the set of all automorphisms on $V$.

---

[68] In spite of being very old and out of fashion, this is one of my favorite books. When reading it, you get an impression of being in a good company, talking with a very clever interlocutor - a feeling which never emerges when dealing with a computer program, although one can also learn a lot in this latter case.

[69] We shall deal with the general linear group especially in the context of Lie groups which are of utmost importance to physics.

One can find an exhaustive treatment of the formal linear algebra in many modern courses and books. Personally, I studied this subject using the book by G. E. Shilov [153] which I found at that time excellent. There are of course more modern textbooks, from which I especially like the one by A. I. Kostrikin and Yu. I. Manin [155].

For completeness, let us now recall the notion of an additive group using the form commonly given in relation to vector spaces. In principle, groups are easier to understand not algebraically but geometrically, e.g., through physical transformations such as motion. We shall exploit this geometrical representation later in this book; now we shall stick to the more traditional algebraic definitions. A group may be simply defined as a pair of a non-empty set $X$ and a map $*$ (composition law, $X * X \to X$) such that this map is associative, invertible, and possesses an identity element. An additive group $G$ is a set together with a map such that to each pair of elements $x, y \in G$ corresponds a new element from $G$, denoted as $x + y$, with the following rules being valid:

1.   $(x + y) + z = x + (y + z)$  (which means that operation "+" is associative)

2.   $x + y = y + x$ (commutativity of the group, which is a very strong requirement)

3.   There exists an element $0 \in G$ such that $x + 0 = x$ for all $x \in G$

4.   For any $x$ there is $y$ such that $x + y = 0$ (existence of the inverse element)

This short list of axioms is usually supplemented by some simple consequences (see any textbook on algebra). These consequences look very much like simple exercises, but algebra teachers are fond of them considering them good logical exercises (at least such was my university experience). The primary consequence is that there may be only one 0 element. Indeed, assume that there are two zeros, $0_1$ and $0_2$, then we would have $x + 0_1 = x$ and $x + 0_2 = x$ for all $x \in G$. By putting, e.g., $x = 0_1$, we get $0_1 = 0_1 + 0_2 = 0_2 + 0_1 = 0_2$. The second consequence of axioms is usually formulated as "uniqueness": for any pair $x, y \in G$ there exists only one $z \in G$ such that $x + z = y$. This property also can be easily proved: according to axiom 4, to each $x$ we can find $x'$ so that $x + x' = 0$. Then for $z := x' + y$ we have

$$x + z = x + (x' + y) = (x + x') + y = 0 + y = y + 0 = y$$

A variation of this "uniqueness" statement: let $x + z = y$ and $x + w = y$. Then $z = w$. Indeed,

$$z = (x + x') + z = x' + (x + z) = x' + y = x' + (x + w) = (x' + x) + w$$
$$= 0 + w = w$$

The inverse element $y$ in Axiom 4 is determined by $x$, $y(x)$. It is denoted as $-x$, which is a kind of a mathematical standard. Like any good standard (and standards are not necessarily good, an example of a bad standard is the SI system of units), this one makes life easier. So instead of $x + (-y)$ we can simply write $x - y$, which is interpreted as the difference between elements $x$ and $y$.

A remark on zero. Strictly speaking, there are two zeros in vector spaces - one (denoted as $0$) is inherited from the number system (field), the other (denoted as $\mathbf{0}$) from the vector space $V$ itself. However, it is easy to see that $0x = \mathbf{0}$ for all elements $x \in V$. Indeed, $0x + x = 0x + 1x = (0 + 1)x = x$, which exactly means that $0x = \mathbf{0}$. As far as the inverse element in $V$ goes, one can see that $(-1)x = -x$ for any element $x \in V$. Indeed, $-x$ is the only element from $V$ with the property $x + (-x) = 0$. But $x + (-)x = (1 - 1)x = 0x = \mathbf{0}$, thus $(-)x = -x$. Thus, we can disregard the logical disparity of zeros from different algebraic structures and always write only $0$.

Let us now give some illustrations to this very simple though slightly abstract algebraic theory. Sets (fields) $\mathbb{R}$ and $\mathbb{C}$ are examples of an additive group $G$ with respect to ordinary addition. Yet the group operation "+" is, of course, not always the ordinary addition. Let us now look at a simple but less common example, the even-odd algebra $A$ with a binary operation $\{,\}$, $A = \{e, o\}$: $e + e = e$, $e + o = o$, $o + o = e (e - \text{even}, o - \text{odd})$. Here the role of zero is played by $e$ and $-x = x$ for all elements $x$. We can generalize this algebra by introducing a cycle. Take any integer $n > 1$ and the set $A = \{0, 1, 2, \ldots, n - 1\}$ with the binary operation

$$x \oplus y = \begin{cases} x + y, & \text{if } x + y < m, \\ x + y - n, & \text{if } x + y \geq m. \end{cases}$$

where "+" denotes the usual addition of numbers. Here zero is the usual $0$ and the inverse element $\ominus x$ equals $n - x$ if $0 < x < n$ and $0$ if $x = 0$. This construct leads to the concept of point groups which are very useful in the classification of molecular structures (see [84], §93).

There are several other standard examples of real vector spaces (see, e.g., [153]) such as the vector space of all $n$-tuples $(x^1, \ldots, x^n)$ of real numbers with component-like addition. If one writes such $n$-tuples as vector columns, then the vector space property becomes quite obvious since

$$\begin{pmatrix} x^1 \\ \vdots \\ x^n \end{pmatrix} + \begin{pmatrix} y^1 \\ \vdots \\ y^n \end{pmatrix} = \begin{pmatrix} x^1 + y^1 \\ \vdots \\ x^n + y^n \end{pmatrix}$$

and

$$\lambda \begin{pmatrix} x^1 \\ \vdots \\ x^n \end{pmatrix} = \begin{pmatrix} \lambda x^1 \\ \vdots \\ \lambda x^n \end{pmatrix}$$

It is this vector space that is denoted as $\mathbb{R}^n$. By the way, the rule of parallelogram and multiplication (scaling) by a scalar familiar to us from the schooltime physics are just a geometric equivalent of these two rules. In principle, any field can be used for coordinates in geometry, with properties of the field being reflected in the geometry features [70]. Other standard examples are the vector spaces of all single-variable polynomials and of all real-valued continuous functions with compact support, e.g., defined on [0,1].

One might notice that it is possible to consider other structures in $V = \mathbb{R}^n$, not only the vector space. For example, in classical mechanics such geometric structures as the Euclidean space $\mathbb{E}^n$ and affine space $\mathbb{A}^n$ and of course symplectic manifolds are of essential importance. The Euclidean 3d space of classical physics $\mathbb{E}^3$ admits a global Cartesian system i.e., defined on the entire space. Writing physical quantities in the Cartesian coordinates is very convenient unless the studied phenomenon has some complex symmetry: in such cases the equations written in Cartesian coordinates may become rather involved and the transition to curvilinear coordinates consistent with the symmetry type of the problem is advantageous (the hydrogen atom is an archetypal example, see [84], 36,37).

The simplest example of a vector space is probably the straight line, which represents a one-dimensional vector space uniquely determined by any non-zero vector. Another simple example of a vector space is the familiar set of complex numbers. Here the basis consists of two elements $(1, i)$, with the generating operations being addition and multiplication by real numbers. Thus, we can easily produce the representation of complex numbers by $2 \times 2$ matrices with real entries, i.e. of the form

$$Z = \begin{pmatrix} a & -b \\ b & a \end{pmatrix}$$

One can show that matrices of such type form a field, which is isomorphous to the complex field. In other words, the set of these $Z$-matrices is closed with respect to addition, subtraction, multiplication and division (except by zero). Moreover, all the rules familiar from the school-years arithmetic such as associativity, commutativity and distributivity are valid. Indeed, we can represent every $Z$-matrix as the sum

$$Z = a \begin{pmatrix} 1 & 0 \\ 0 & 1 \end{pmatrix} + b \begin{pmatrix} 0 & -1 \\ 1 & 0 \end{pmatrix} = aE + bI$$

where $E$ is the unit (identity) matrix and the symplectic matrix $I$ can be identified with the imaginary number $i$, $i^2 = -1$ (one can easily verify that $I^2 = -1$). One can see that $2 \times 2$ $Z$-matrices are fully defined by ordered pairs $(a, b)$ of real numbers, with the pair $(a, 0)$ being itself a real number $a$. The matrix $I$ transforms any plane vector $(x, y)^T$ into $(-y, x)^T$, i.e., rotates the vector $(x, y)^T$ counterclockwise just like the complex operator $i = \exp(i\pi/2)$. If one inverts a $Z$-matrix, one gets again a matrix of the same class. Indeed, for any invertible $2 \times 2$ matrix

$$A = \begin{pmatrix} \alpha & \beta \\ \gamma & \delta \end{pmatrix}, \det A \neq 0$$

the inverse is

$$A^{-1} = \frac{1}{\det A} \begin{pmatrix} \delta & -\beta \\ -\gamma & \alpha \end{pmatrix}$$

---

[70] This material is usually discussed in modern geometry courses, see e.g., the Pappus's theorem and the Desargues's theorem.

Our matrix $Z$ is obtained from the generic $A$-matrix by putting $\alpha = \delta = a, -\beta = \gamma = b$. Then

$$A^{-1} = \frac{1}{a^2 + b^2} \begin{pmatrix} a & b \\ -b & a \end{pmatrix}$$

is of the same type as $A$, i.e.

$$A^{-1} = \frac{a}{a^2 + b^2} E + \frac{b}{a^2 + b^2} (-I)$$

and corresponds to $Z^{-1} = \frac{1}{|z|^2}(a - ib)$ if $Z = a + ib$. Here we denote $\lfloor z \rfloor^2 = a^2 + b^2$. The operator $Z^{-1}$ implements the clockwise rotation of a plane vector $(x, y)^T$.

The algebra of $Z$-matrices is closed under ordinary operations characterizing the mathematical field, addition and multiplication. It is easy to verify that sums and products of $Z$-matrices are also of the same type. Complex conjugation corresponds to transposition of $Z$-matrices, which is not a linear operation. It is actually the antilinear operation, and in quantum mechanics, for example, it manifests the time inversion (see Chapters 5, 9). The conjugate $Z^* = (a, b)^*$ of $Z$ is given by

$$Z^* = \begin{pmatrix} a & -b \\ b & a \end{pmatrix} = (a, -b)$$

with $Z^* Z = \lfloor z \rfloor^2 (E + 0I) = (a^2 + b^2, 0)$.

One can define a norm with the help of the operation of conjugation, since the quantity $a^2 + b^2$ is always a non-negative real number being zero if and only if $a = b = 0$. Therefore, matrices representing complex numbers form a two-dimensional real normed vector space, which is reflected in the above matrix operations. Due to isomorphism, the field of complex numbers $\mathbb{C}$ may be also regarded as a vector space. It is, however, easier to operate with complex numbers than with their matrix representation. Nevertheless, the latter may be interesting from the theoretical viewpoint. We have seen that matrices representing complex numbers are determined by two independent real parameters $a, b$. Thus, one can generalize the matrix construction for complex numbers by taking matrix elements not necessarily from the real field. For example, in case these matrix elements are complex numbers, one arrives at quaternion algebra. Or the $Z$-matrix entries may be themselves matrices, then we obtain a sequence of algebras.

One might ask, why are the mathematicians so fond of abstract vector spaces, would not it be enough to work only with simple $\mathbb{R}^n$? The confusing point here is that one can find in many algebra textbooks every $n$-dimensional vector space $V_n$ is isomorphic to $\mathbb{R}^n$. However, this isomorphism is, in the mathematical terminology, not canonical - simply speaking it depends on the choice of the basis. Therefore, modern mathematicians and mathematical physicists prefer using abstract vector spaces instead of $\mathbb{R}^n$ because they consider it desirable to operate quantities in a coordinate-free manner, without choosing some arbitrary basis. As a matter of fact, this is a well-grounded reasoning since many vector spaces naturally arising in contemporary physics do not have a hint on a preferred choice of a basis. Nevertheless, the bases in vector spaces are of utmost importance not only in physics but also to building mathematical models in different applied areas, in computer science, etc., therefore we shall devote some attention to their selection.

## 3.11  Basis Vectors

In the preceding section, we had an example of the homogeneous system of linear equations. The set of solutions of such a system (expressed as $n$-tuple, $(x^1, \ldots, x^n)^T$) gives an example of a subspace $U$ of $\mathbb{R}^n$. In general, a subspace $U \subset V$, where $V$ is the vector space, is called the vector subspace, if for $x, y \in U$ also their sum $x + y \in U$ and also for $x \in U$ and any $\lambda \in \mathbb{R}, \lambda x \in U$. For instance, all linear combinations, viz. vectors of the form $\mathbf{u}_k = \alpha_k^j \mathbf{v}_j, \alpha_k^j \in \mathbb{R}, \mathbf{v}_j \in V, j = 1, \ldots, n; k = 1, \ldots, m$, comprise a subspace of $V$, sometimes denoted as $\langle \mathbf{v}_1, \ldots, \mathbf{v}_n \rangle$ and called span (denoted also by $\mathrm{Sp}(\mathbf{v}_1, \ldots, \mathbf{v}_n)$). Span is in some sense the smallest subspace of $V$ containing $\mathbf{v}_1, \ldots, \mathbf{v}_n$. To see whether a set of vectors spans a vector space, one must check that there are at least as many linearly independent vectors as the dimension of the space. It is clear that if the $\mathbf{u}_k$ are linearly independent, then $m \leq n$. For example, in the $n$-dimensional space $\mathbb{R}^n$, $n + 1$ vectors are never linearly independent, and $n - 1$ vectors never span. It follows from here that each vector space $V$ contains a finite subspace $V_n$ of linearly independent vectors, namely of dimension $n$. In other words, there exists a number $n \in \mathbb{N}$ so that any $n + 1$ vectors are linearly dependent. If we neglect some mathematical pedantry, then any linearly independent system of $n$ vectors in $V_n$ may be used to generate all other elements of $V_n$ and then it is called "basis". Thus, a basis of a vector space is a linearly independent spanning set. In particular for $V_n = \mathbb{R}^n$, any linearly independent system of $n$ vectors in $V_n$ may be used to generate all other elements of $V_n$[71]. The term "generate" means in this context

---

[71] In general, in abstract algebra where the concept of span is defined as a module $M$ over a ring $R$, a linearly independent subset which generates the whole module may not necessarily

that each vector $\mathbf{x} \in V_n$ may be represented as a linear combination $\mathbf{x} = a_i \mathbf{u}^i, i = 1, \ldots, n$. It is clear that all bases in $V_n$ have the same number of vectors, namely $n$[72]. Indeed, if we have two bases, $\mathbf{e}_1, \ldots, \mathbf{e}_m$ and $\mathbf{f}_1, \ldots, \mathbf{f}_n$, then we may take $\langle \mathbf{f}_1, \ldots, \mathbf{f}_n \rangle$ to be a span and $\mathbf{e}_1, \ldots, \mathbf{e}_m$ a system of linearly independent vectors. Then it follows from the above considerations that $m \leq n$. But of course, we may exchange these two systems of vectors with the consequence that $n \leq m$ - the argumentation is symmetric with respect to $m$ and $n$. Putting it all together, we get $m = n$. It is the common length $n$ of all bases in $V$ that is called the dimension of $V$, $n = \dim V_n$.

One can define a basis $B$ in the vector space $V$ defined as a subset $B \subset V$ of linearly independent vectors[73]. Then the set of all linear combinations of $B$ (the linear hull) coincides with $V$. We have just seen that if $B$ consists of a finite number $n$ of vectors, then $V$ is finite-dimensional ( $\dim V = n$ ). The standard basis for $V = \mathbb{R}^n$ is conventionally the set of $n$ vectors $\mathbf{e}_1, \ldots, \mathbf{e}_n, \mathbf{e}_i = \{0, \ldots i = 1, \ldots, n\}$ (i.e., 1 stands in the $i$-th place). It seems to be clear that if $B$ is a basis in $V$, then no subset of $B$ (with the trivial exception of $B$ itself) may also be a basis in $V$.

It has already been mentioned that there exist vector spaces that do not possess a finite system of vectors with the help of which one can generate all other vectors belonging to the same vector space. In other words, one cannot find a natural number $n \in \mathbb{N}$ so that any $n + 1$ vectors from $V$ are linearly dependent. In this more complicated case, vector space $V$ is called infinite dimensional. In other words, a vector space is infinite-dimensional if it does not have a finite basis [74]. For instance, the vector space of all polynomials is infinite-dimensional since vectors (functions) $1, t, t^2, \ldots, t^n$ are linearly independent for each $n$. The same applies to the vector space of all continuous real-valued functions on $[0,1]$ - this vector space is infinite-dimensional.

On the other hand, $\mathbb{R}^n$ is finite-dimensional, $\dim \mathbb{R}^n$ since the unit vectors

$$\mathbf{e}_1 = \begin{pmatrix} 1 \\ 0 \\ \vdots \\ 0 \end{pmatrix}, \mathbf{e}_2 = \begin{pmatrix} 0 \\ 2 \\ \vdots \\ 0 \end{pmatrix}, \ldots \mathbf{e}_n = \begin{pmatrix} 0 \\ 0 \\ \vdots \\ 1 \end{pmatrix}$$

are linearly independent and form the basis in $\mathbb{R}^n$. As a simple but important example, one may consider the vector space $V$ of all differentiable functions $f = f(x)$ on $\mathbb{R}$ with the property $f' = f$. This property may serve as one of the definitions for the exponent, $f(x) = \exp(x)$, and being its own derivative

---

exist. In other words, even if module $M$ is generated by some $n$ elements, it is not necessarily true that any other set of $n$ linearly independent elements of $M$ spans the entire module $M$. In effect, this means that a basis may not always exist. The example usually produced in higher algebra in relation to this statement is that $\mathbb{z}$ is generated by 1 as a $\mathbb{z}$-module but not by 2.

[72] One must note that it is only in finite-dimensional vector spaces, where any vector may be represented as the finite linear combination, that the number of basis elements does not depend on basis. If the vector space is infinite-dimensional, $\dim V = \infty$, then, although it possesses a basis, the latter is always infinite and, consequently, one cannot compare the number of basis elements.

[73] One colloquially says that $B$ is linearly independent.

[74] And is not zero-dimensional i.e., does not consist only of a zero, which is a trivial zero-space.

selects the exponent as the only function having this property. Another way to express it is to state that we have here a fixed point of the differential operator (regarded as a functional). The exponential function is an element of $V$ and is non-zero, which means that $\dim V \leq 1$. Let us now consider some other element of $V$, $u \in V$. From the differentiation formula and the property $(.)' = (.)$, we get

$$\left(\frac{u}{f}\right)' = \frac{fu' - uf'}{f^2},$$

therefore $u = cf$, where $c = const$ and $\dim V = 1$.

Physicists and mathematicians are forced to systematically visualize objects in four, five or even infinite-dimensional spaces. Thus, thinking over other dimensions led B. Riemann in 1854 to getting out of the Euclidean geometry and showing that space can be curved - nowadays a trivial notion, but not at all in Riemann's time[75]. General relativity operates on curved spaces and perhaps our universe is a kind of a $3D$ spherical surface - a closed three-dimensional manifold carrying a metric of positive Ricci curvature. These concepts are closely connected with modern mathematical topics such as Einstein manifolds, Ricci flow, and G. Perelman's famous proof of the so-called Poincaré conjecture [159]. The new physics implies that our 3+1 dimensional space is just what we see in our primitive everyday dullness, boring low-energy province; in fact our world has 10, 11, 26 or some other number of dimensions (so far undetermined). Anyway, one has to learn to deal with any number of dimensions, even infinite as in quantum mechanics, and the first step in this direction is working with vector spaces.

Allow me to make some general observations here that might look as distractions. One usually extends and applies well-known ideas from finite dimensional linear algebra to infinite dimensions. This motivates us to carry out the difficult calculations whose purpose is to connect various models and cross-link different levels of abstraction in the studied problem. Even though the basic ideas and mathematical structure have been developed in linear algebra, the motivation for using them stems from physics and engineering. In particular, one can recall multidimensional mathematical tools for electromagnetics and vibration theory for engineers and quantum mechanics for physicists. This may be nontrivial mathematics from a cross-disciplinary perspective. Furthermore, the main ideas of linear mathematics in infinite dimensions are already present in waves, signals, and fields, even in those whose domains are one-dimensional. The transition to higher-dimensional domains comes rather smoothly once the one-dimensional models have been explored. Interestingly enough, the transition to higher dimensions leads to surprising new features such as the appearance of special functions - envoys of the world of symmetries. The very existence and properties of special functions are a consequence of the group properties [70]. For instance, the group of motions of the Euclidean plane leads to the families of Bessel functions, the rotation group of the Euclidean three-dimensional space produces the Legendre and Jacobi polynomials. In physical terms, the symmetry properties giving birth to these classes of special functions are mostly interpreted as the required invariance under translations and rotations of measured distances and of the shapes of propagating waves. More complicated groups such as those of motion of an $n$-dimensional Euclidean space, $n$-dimensional sphere or $n$-dimensional Lobachevsky space lead to some new classes of polynomials, not necessarily reduced to special functions generated by the motions in lower-dimensional spaces. This reflects the general observation: increased dimensionality has its own specifics.

---

[75] Bernhardt Riemann lived a rather short life, 1826 - 1866, but his impact is enormous. Some people consider that Riemann geometry is hopelessly outdated, I think they are hopelessly wrong. I do not know which name is more suitable for the famous equations in the theory of analytical functions - the Cauchy-Riemann equations or D'Alembert-Euler equations, but it does not diminish Riemann's contribution to this beautiful theory, extremely powerful in physical applications. And without Riemann surfaces, one would be totally confused when dealing with multivalued functions.

We shall return to a discussion of special functions several times in this book; I think this is a fascinating subject although many people consider it mundane and boring. In fact, special functions may always be considered as realizations of some basis in an appropriate functional space - exactly in the sense of linear algebra - and thus they connect three disciplines: algebra, geometry and physics.

As for bases in general, there exists a mathematical recommendation (largely ignored by physicists): use coordinate systems only when needed. Of course, one has to introduce a specific coordinate system to perform computations to the end, but what is meant by mathematicians in the above maxim is that one should resist the temptation to fix an arbitrary basis until it is absolutely necessary. What is a basis? The word "basis" is often used both in physics and mathematics, but I don't think there exists a fully satisfactory general definition (we have seen that such a situation is not infrequent in mathematics). We can use the following description as a substitution for the lacking definition. One usually understands a basis for a set $X$ as some subset $A \subset X$ that can generate $X$. What does "generate" mean? Here this term means that it is possible to get any element $x \in X$ by applying operations from some class $U$ to elements $a \in A$, i.e., $x = Ua$ for any $x \in X$. The basis is typically thought of as a minimal subset $A$, with which it is possible to generate the whole set $X$. The term "minimal" in this context means that no subset of $A$ can generate $X$, i.e., $\exists \tilde{A} \subset A$ such that $x = U\tilde{a}, \tilde{a}, \in \tilde{A}, x \in X$. One also implies that all elements $a$ from the subset $A$ of $X$ are independent, i.e., none of them can be obtained by means of operations $U$ applied to other elements from this subset. Contrariwise, other elements from $X/Y$ become dependent on elements $a \in A$ via operations $U$.

It is curious that there seems to be sort of a fight (reminding us of Jonathan Swift) among mathematicians, at least at the folklore (not textbook) level. Some of them argue that metric spaces should be given a priority, whereas others state that the concept of metric spaces is an old-fashioned framework, and what people need is normed spaces or even more - the normed algebras. Besides, one may notice that both mathematicians and physicists are sometimes quite loose with terminology - we shall find some examples of this looseness below.

## 3.12  Dual Spaces

Dual spaces is an extremely important concept for physics, although many physicists do not use it in their everyday language. I shall try to demonstrate the importance of dual spaces on some examples. Please be lenient if some constructions will lack rigor and completeness.

Usually, dual vector spaces are defined through linear functions[76] on a vector space $V$. Recall that a map $L$, for instance, between two vector spaces $V_1$ and $V_2$ is called linear if it satisfies the relationship $L(a_1 x_1 + a_2 x_2) = a_1 L(x_1) + a_2 L(x_2)$ for any $a_1, a_2 \in K$ where $K$ is the field, usually understood as the field of scalars (please do not confuse with the scalar field in physics which we shall study in Chapters 4 and 9). Any linear map $L: V_1 \to V_2$ (we shall deal exclusively with the fields $K$ of real or complex numbers that is $K = \mathbb{R}, \mathbb{C}$) may be written as a linear combination $\mathbf{L} = l^i \mathbf{e}_i$ where coefficients $l^i, i = 1, 2 \ldots n$ are real or complex numbers. It is clear that a linear combination of linear functions makes a linear function, therefore the set of linear functions also forms a vector space and this linear vector space is called the dual space $V^*$ to $V$. In modern mathematics one usually calls the duality operation $(*)$ a "functor", more exactly it is an example of a functor.

To write out the dot product of two vectors $\mathbf{a}, \mathbf{b}$ in terms of abstract linear functions or linear functionals is probably not the most convenient way. To make the expressions less abstract one usually introduces two sets of basis vectors, sometimes called the "direct" and the "dual" basis vectors that satisfy the relations $\mathbf{e}_i \mathbf{e}^{*j} = \delta_i^j$. Since the "dual basis" vectors also form a basis, one can write any vector as a linear combination of them. For example, we can rewrite the vector $\mathbf{b} \in V^*$ as $\mathbf{b} = b_i \mathbf{e}^i$ and vector $\mathbf{a} \in V$ as $\mathbf{a} = a^i \mathbf{e}_j$. This allows us to write the dot product of $\boldsymbol{a}$ and $\boldsymbol{b}$ in a much simpler way:

$$(\mathbf{a}, \mathbf{b}) = a^i b_j (\mathbf{e}_i, \mathbf{e}^j) = a^i b_i = b_i a^i = (b_1 \quad \cdots \quad b_n) \begin{pmatrix} a^1 \\ \vdots \\ a^n \end{pmatrix} = (\mathbf{b}, \mathbf{a}) = \mathbf{b}\mathbf{a} = \mathbf{a}\mathbf{b}$$

$$= a_i b^i = (a_1 \quad \cdots \quad a_n) \begin{pmatrix} b^1 \\ \vdots \\ b^n \end{pmatrix}$$

One can notice, by the way, that the scalar product of basis vectors $(\mathbf{e}_i, \mathbf{e}^j)$ serves as a tool for raising and lowering indices. We shall see later that the scalar product of basis vectors can be generalized to the so-called metric tensor. In the case of dual spaces, the formula for a scalar product is fully analogous to that for the Euclidean space, the difference is that such a formula connects vectors living in different spaces.

One should remember that it is only in the simplest case that a dual vector space is defined by the available basis. If we managed to fix a basis in a vector space $V$, with a set of basis vectors $\mathbf{e}_i$, we may define the dual space $V^*$ as the vector space spanned by the basis vectors $\mathbf{e}^{*j}$ with the requirement $\mathbf{e}_i \mathbf{e}^{*j} = \delta_i^j$ where the right-hand side is the Kronecker symbol, i.e., the dot product of the basis vectors is zero for $i$ not equal to $j$[77].

---

[76] These constructions are also called linear maps or linear functionals.

This construction means that if $V$ is finite-dimensional, then $V^*$ has the same dimension as $V$. The mathematicians often say that each $\mathbf{e}^{*j}$ represents a scalar linear map on the set of basis vectors $\mathbf{e}_i$. In the language of tensor analysis and tensor geometry, elements of vector space $V$ are called contravariant vectors and elements of the dual space $V^*$ are called covariant vectors or covectors. In the coordinate-free language of differential forms, these dual space elements are called one-forms (see below a discussion of differential forms). One-form (or 1-form) is just another name for a covector. The set of $\mathbf{e}^{*j}$ provides a basis for linear functions (maps) $L$, and we can represent any linear map on $V$ in the form $\mathbf{L} = l_j \mathbf{e}^{*j}$ $\mathbf{L}$ where $l_j$ are real (or complex) coefficients [78]. A linear map in this context may be called a dual vector.

So, we see that dual vectors are important spin-offs of conventional vectors i.e., dual spaces are spin-offs of vector spaces. As dual vectors when multiplied by their corresponding vector produce a number, one may notice a peculiar feature of duals: if the vector space - the space a vector lives in - is diminished, a conventional (contravariant) vector shrinks whereas a dual (covariant) vector, roughly speaking, becomes larger.

Let us, as an example of usage of dual spaces, consider the special case interesting for classical mechanics and dynamic systems theory, $V = T_q M, V^* = T_q^* M$ . These reciprocally dual spaces are called tangent and cotangent spaces adjoining to each point $q$ of the configuration manifold $M$. We shall more or less thoroughly discuss these constructions in the next chapter, in connection with Lagrangian mechanics. It will be clear that $T_q M$ is the space of directional derivatives, which are actually operators acting on functions defined on the manifold $M$. We may choose a basis for the tangent space $T_q M$, for instance, as the set of partial derivatives, $e_i = \frac{\partial}{\partial x^i}$. Note that here basis vectors are differential operators. Let us now introduce the dual basis using the above definition, $\mathbf{e}_i \mathbf{e}^{*j} = \delta_i^j, i, j = 1, 2 \dots n$ (it is not necessary to mark up vectors with boldface; later I shall omit the asterisk denoting the dual basis vectors, leaving only contravariant index does not result in any ambiguity). One can easily guess that the dual basis is $e^j = dx^j$ so that $\partial_i dx^j = \delta_i^j, \left( \partial \equiv \frac{\partial}{\partial x^i} \right)$. This choice of basis allows us to write any vector

[77] It is customary today to denote the dual basis vectors with an asterisk. Previously, the dual basis was designated only by upper indices. I shall use both notations depending on the context, unless it produces a hopeless clarity violation.

[78] In physics, one is mainly interested in $K = \mathbb{R}, K = \mathbb{C}$. There exist, of course, both algebraic and geometric extensions of the real and complex number systems (e.g., with many roots of $-1$). Yet, despite the existence of quaternionic quantum mechanics [86] and some geometric algebra applications [115], I am unaware of unbeatably successful attempts in physics to identify the scalars with anything different from the real or complex numbers.

$u(q) \in T_q M$ as $u = u^i \partial_i$ and any dual vector $v(q)$ as $v = v_j dx^j$. Then we get a general expression for the linear map (functional) $L$ acting, e.g., on $u$:

$$Lu \equiv L[u] = l_j dx^j [u^i \partial_i] = l_j u^i \delta_i^j = l_j u^j \qquad (3.5)$$

It is this linear functional representing a dual vector in $T_q^* M$ that is called a 1-form (see above). We see that there is a well-defined field of 1-forms on every cotangent space $T_q^* M$ (and on every cotangent bundle $T^* M$). It is important to note that this linear functional acting on each tangent vector takes it to the $2n$-dimensional manifold $T^* M$ and not to $M$. We shall discuss this point in some detail in connection with Lagrangian mechanics.

If the basis has been fixed, we can associate mutually dual vectors by their components, $u^i \leftrightarrow v_j$. This is an important trick in solid state physics (see below). However, this association is not, as mathematicians use to say, a natural mapping between two sets of components and, therefore, between two spaces. This mild mathematical objection is based on the fact that components are mutable, they change when we transform the coordinate system - and we always have the right to do it for our convenience. This volatility of components of a vector was one of the main motivations for mathematicians to use the differential form calculus instead of "old-fashioned" coordinate formulation usually preferred by physicists. Today, differential forms are gradually penetrating into many areas of physics, at least one can observe that they have become the dominating tool in Hamiltonian mechanics, in the physical part of differential geometry and even in field theory.

The concept of dual spaces seems to be a fair platform to start studying differential forms, since, roughly speaking, differential forms are dual to vectors. However, before we proceed a bit later to discussing differential forms, their formal definitions and how they are used in physics, I shall review some preparatory points addressing once again the ubiquitous examples of tangent and cotangent spaces of Lagrangian mechanics. So, we have selected the basis $\partial_i$ in the tangent space $T_q M$ and the basis $dx^j$ in its dual, cotangent space $T_q^* M$. Actually, one can define the cotangent space through its basis $dx^j$. One might be astonished by this latter notation, since the quantity $dx^j$ is extensively used in classical calculus and integration. Is such a coincidence a symptom of negligence or is it due to a lack of symbols, frequently encountered in physical and mathematical texts? Neither the first nor the second. There exists a deep connection between classical calculus and more modern[79] techniques of differential forms. To feel this connection, we may do the following. Let us take a differentiable function $f : M \to \mathbb{R}$[80] defined on a manifold $M$. Now let us define some special element $v(f) \in T_q^* M$: $v(f) = \partial_j f dx^j$ or, in terms of ordinary vector analysis, $v(f) = \nabla f d\mathbf{r}$. Now if we apply $v(f)$, regarded as an operator, to $u \in TM$

---

[79] By "modern" one can understand methods developed over last hundred years.
[80] An exact meaning of the word "differentiable" is not essential here.

$$v(f)[u^i\partial_i] = \partial_j f dx^j[u^i\partial_i] = u^i\delta_i^j\partial_j f = u^j\partial_j f \qquad (3.6)$$

That is by acting with $v(f)$ from the cotangent space on element $u$ from the tangent space we get, due to

$$L(f) \equiv v(f)[u^i\partial_i] = u^j\partial_j f = u(f)$$

the directional derivative of $f$ in the direction of $u$. If we denote, in the spirit of differential forms, $v(f) := df$, we shall have[81] $df = \partial_j f dx^j$ and $df[u] = u(f)$. In general, any $d$-operator in $T_q^*M$ can be written as $d = \partial_j dx^j$, and we find a general formula for $\partial$ acting on $u$:

$$\partial[u] = \partial_j dx^j[u^i\partial_i] = \partial_j u^i\delta_i^j = \partial_i u^i$$

These exercises with linear maps in dual spaces may appear trivial and not deserving the space devoted to them, yet these forms have some hidden profound meaning, which will be gradually unfolded in our studies of mechanics and geometrical aspects of the field theory.

So far, to exemplify the relation between dual spaces $T_qM$ and $T_q^*M$ we have used some unspecified differentiable function $f$. Let us now take the simple specific case of $f = x^j$. Then

$$df = d(x^j) = \partial_i x^j dx^i = v(x^j)dx^i = \delta_i^j dx^i = dx^j$$

So $d(x^j) = dx^j$, which expression connects "new" notations with "old". The meaning of "new" notation $d(x)$ invokes the concept of exterior derivative, whereas the "old" notation $dx$ traditionally understood as a small increment of the quantity $x$ is in fact an element of the dual space $T_q^*M$ - a linear map from vectors to numbers (linear functional). The old derivative studied in classical calculus and vector analysis as well as vector operations such as gradient, rotor (curl) and divergence are specific cases of the exterior divergence. The latter is an operator that maps an n-form $\theta$ defined on a manifold $M$ to an ($n$+1)-form $d\theta$, which is also defined on the same manifold. The notion of exterior derivative may be regarded as a generalization of the traditional derivative introduced first by Newton and Leibniz and then substantiated in classical calculus. The ordinary classical derivative maps a scalar function (scalar field), which is in this parlance a 0-form, to 1-form $df$. See more on derivatives in section "Notes on Derivatives" below.

Let us now return to our discussion of dual spaces. We could have felt that it might be very convenient to know whether a quantity belongs to the main or base space such as, for example, a velocity-like and displacement-like vector in mechanics or to its dual such as momentum or wave vector, the first being vectors whereas the second are covectors.

---

[81] in the mathematical texts on differential forms such linear forms are customarily denoted by $\omega$ or $\theta$.

Thus, if we have fixed a set of basis vectors, its corresponding dual set can be determined by a simple procedure. Just write a matrix whose columns are the Cartesian coordinates of the chosen basis vectors. Now, the Cartesian coordinates of the dual set of vectors are represented by the rows of the inverse matrix. It follows from here that if the chosen basis vectors happen to be orthonormal, as in the case of the Cartesian coordinate systems, then the given set of basis vectors coincides with the set of their dual basis vectors. In other words, if we simply interpret the space of all sequences of real numbers $\mathbb{R}^n$ as the space of columns of $n$ real numbers, its dual space may be written as the space of rows of $n$ real numbers. Again, in the terminology of mathematicians, one may say that such a row represents a linear functional on $\mathbb{R}$ with respect to ordinary matrix multiplication. Here we may see a hint on some deficiency of this conventional approach: it restricts the transition to dual space to finite-dimensional spaces since it is only then that duality is well defined. Another mathematical saying in relation to dual spaces is that there is no canonical (i.e., independent of coordinate choice) isomorphism between a vector space and its dual: isomorphism here depends on a particular choice of a basis. Nevertheless, there exists a canonical isomorphism between a finite-dimensional vector space and its double dual.

One may note in conclusion of this section that duality plays a nice part in a number of geometrical applications. Take projective geometry (see below) as an example. In this discipline, projective planes are self-dual: if one switches the words "point" and "line", one doubles results, so to say, for free.

## 3.13  Some Remarks on Indices

The upper and lower indices are used to distinguish two sets of components: those of a vector in a given basis (more or less arbitrarily fixed by us) and in the corresponding dual basis. In many textbooks on physics where exclusively the orthonormal bases are used, the authors typically make no distinction between upper and lower indices. The reason for such a simplification is that in orthonormal bases the sets of components of a vector in a given coordinate system and in that with the respective dual basis are the same. In more general situations, raising and lowering indices expresses a natural duality between tangent vectors and covectors or 1-forms. Specifically for vector bundles of mechanics (see Chapter 1), lowering indices corresponds to mapping of the tangent to cotangent bundle $TM \to T^*M$, whereas raising indices reflects a reciprocal map $T^*M \to TM$. Both operations are examples of a duality isomorphism.

## 3.14  Operators in Quantum Mechanics

While studying quantum mechanics, one might notice that the mathematicians seem to talk mostly about self-adjoint operators whereas the physicists prefer to talk about Hermitian operators, the use of operators for practical quantum-mechanical calculations being basically the same. Is there any real difference between self-adjoint operators and Hermitian operators? It is interesting that in some modern linear algebra courses, where I wanted to check how exactly the term "Hermitian" is defined, I could not find the word even in the index. (I just checked two Linear Algebra texts to see exactly how they defined it and the word is not even in the index!)

Of course, one can a priori assume that mathematicians would tend to talk more abstractly than physicists, but let us try to explore the substance of these two classes of operators leaving aside "sociological" considerations. Let $U$ and $V$ be any inner product spaces and $A$ a linear map from $U$ to $V$, then the adjoint of $A$ denoted as $A^*$, is a linear map from $V$ to $U$ such that, for any $u$ in $U$ and $v$ in $V$, $(Au, v) = (u, A^*v)$. The brackets here denote the inner product[82], and the two inner products are taken in $V$ and $U$, respectively. In the particular case when $U = V$ and $A^* = A$, i.e., if $(Au, v) = (u, Av)$, then operator (map) $A$ is self-adjoint. In other words, an operator is Hermitian if it possesses the above symmetry property, $(Au, v) = (u, Av)$ for all $u, v$ in the domain of $A$, and an operator is self-adjoint if $A^* = A$ everywhere. The subtle difference here is that the domains of $A$ and $A^*$ in general may not coincide. When dealing with arbitrary normed spaces, the adjoint map, by the construction of scalar product, acts between the duals, $A^*: V^* \to U^*$ (see above). Therefore, there may be difficulties in identifying self-adjoint operators with Hermitian ones at least for unbounded operators when it is not easy to define a dual space.

## 3.15  Dualities in Physics

This section is a digression from a traditional mathematical exposition of an introductory material. I am making this running-ahead digression intentionally in order to demonstrate how rather abstract mathematical concepts can work in physics. Although the term 'duality' might appear a bit overloaded, contemplating about various dualities leads us to beautiful extensions. Duality is not only a high-brow theoretical concept, it also enables one to solve concrete problems. Duality enters physics under various guises, for example, electromagnetic duality may be regarded as a four-dimensional representation of duality.

Take, for instance, the Maxwell equations (Chapter 5). In vacuum and without external sources they have the form

$$\nabla \times \mathbf{H} = \frac{1}{c}\frac{\partial \mathbf{E}}{\partial t}, \nabla \mathbf{E} = 0, \qquad (3.7)$$

$$\nabla \times \mathbf{E} = -\frac{1}{c}\frac{\partial \mathbf{H}}{\partial t}, \nabla \mathbf{H} = 0 \qquad (3.8)$$

where $\mathbf{E} = \mathbf{E}(\mathbf{r}, t), \mathbf{H} = \mathbf{H}(\mathbf{r}, t)$ denote electrical and magnetic components of the electromagnetic field. Obviously, these equations possess a symmetry $\mathbf{E} \to \mathbf{H}, \mathbf{H} \to -\mathbf{E}$ which is now termed as duality.

---

[82] For simplicity, I denote the inner product by round brackets and not by angular brackets as customary in mathematical literature.

This symmetry exchanging electric and magnetic fields has been probably known since the 19th century[83], but only recently has it been exploited as a source of inspiration for constructing new theories and models. The first question to be discussed while trying to generalize the duality of the Maxwell equations is: does the symmetry observed while interchanging electric and magnetic fields still hold for inhomogeneous Maxwell equations, i.e., in the presence of external charges and currents? This is not a simple question, since the symmetry of the inhomogeneous Maxwell equation seems to be broken by the observational fact that while the electric charge does exist, nobody has yet succeeded to detect the magnetic charge (commonly called magnetic monopole). The $\mathbf{E} - \mathbf{H}$ symmetry of microscopical electrodynamics appears also to be broken in material media whose response to an incident field is actually due to induced charges and currents (Chapter 5). Even in vacuum, when one treats the electromagnetic response naively, without a consistent relativistic consideration, electrical charges respond to $\mathbf{E}$ and $\mathbf{H}$ differently: electric fields provide a force accelerating the charge parallel to the field lines, whereas magnetic fields result in the Lorentz force ensuring only a normal acceleration. Furthermore, magnetic materials exhibit their highly useful properties due to macroscopic ordering of magnetic dipoles, but isolated magnetic charges, even if they could be observed, seem to play no part in magnetic interactions.

From the mathematical viewpoint we know that one can represent the magnetic field, which is a solenoidal vector field, with the help of the vector potential, $\mathbf{A}(\mathbf{r}, t)$. This is a great convenience. On the other hand, one represents the electric field, which is a potential field in a static frame of reference, as the gradient of a scalar potential, $\varphi(\mathbf{r}, t)$. This difference of representations of the two would-be symmetric components of the unified electromagnetic field also seems to break the duality.

In the four-dimensional formalism (see Chapter 5), the dual Maxwell equations

$$\partial_\mu F^{\mu\nu} = 0, \qquad \partial_\mu F^{*\mu\nu} = 0 \tag{3.9}$$

where the dual field is defined as

$$F^{*\mu\nu} = \frac{1}{2} \varepsilon^{\mu\nu\varrho\sigma} F_{\varrho\sigma}$$

Recall that the electromagnetic field tensor $F_{\mu\nu}$ is expressed through the potentials $A_\mu$ as $F_{\mu\nu} = \partial_\mu A_\nu - \partial_\nu A_\mu$. We postpone the discussion of the electromagnetic field properties till Chapter 5 (see also Chapter 9) and make now some more remarks about dualities.

A simple way to preserve the symmetry between $\mathbf{E}$ and $\mathbf{H}$ in the presence of external electrical charges and currents would be to add magnetic charges and currents to the Maxwell equations. This would also entail a symmetry between electrical and magnetic charges and currents in the quantum field theory (QFT). Although it appears difficult to ensure such a duality in the traditional version of QFT, there exist a number of quite consistent field theories containing both types of charges (Polyakov-t'Hooft monopoles) [97], [208]. Magnetic monopoles in this class of theories appear as collective excitations of elementary particles. Recall in this connection that collective excitations in condensed media physics are characterized by the wave vector **k** that is dual to the position vector **r** or velocity **v**. In other words, electrical and magnetic charges emerge in the respective quantum field theory in a totally disparate manner, at least in the so-called weak coupling mode. The term "weak coupling" denotes in this context the domain where the coupling constant is small. Applied to electrodynamics, this simply means that the fine structure constant $e^2/\hbar c \approx 1/137$ is small.

---

[83] The first who explicitly used this symmetry was probably O. Heaviside [37].

A simple form of duality may be modeled by the conformal transformation of a complex variable, e.g., $z \rightarrow -1/z$. Roughly speaking, by a duality we may understand some transformation of variables or parameters in a given domain of a theory or model into a physically related (not necessarily completely equivalent) theory or model with a different set of variables or parameters that define it. For instance, a duality may exchange a string theory with a field theory. Or a duality may exchange a strong coupling mode of a given theory with the perturbative regime of the dual theory - the most advantageous case from the physical viewpoint.

Duality in physics is usually defined in rather simple terms: if in some theory $A$ function $f_A$ depends on variable $x$, i.e. $f_A = f(x)$, and in another theory $B$ there appears a function $f_B := f_B(1/x)$, then in case . $f_A(x) = f_B(1/x)$ such theories are known as dual. For $f_A(\xi) = f_B(\xi) := f(\xi)$, a theory is called self-dual. As I have just mentioned, an often given example of duality is the monopole theory, in particular the Dirac monopole introduced by P. A. M. Dirac in 1931 [183], with the monopole "charge" $m$ having the property $m \cdot e = const \cdot n$, $n = 1,2,\dots$ i.e. the product of electric and magnetic charges is quantized. In other words, the conjectural magnetic monopoles, should they exist, are quantized in units inversely proportional to the elementary electric charge. From here, $m \sim 1/e$ so that the monopole coupling constant $\beta \leftrightarrow 1/\alpha$, where $\alpha$ is the electromagnetic coupling constant (the fine structure constant). This means that if $\alpha = e^2/\hbar c \approx 1/137$ is small, then $\beta \sim 1/\alpha$ is large and no perturbation theory of the quantum electrodynamics type seems to be possible. Therefore, one can deduce that point-like monopoles probably do not exist, in contrast with electrons, and one has to consider extended objects from the very beginning. One of the popular versions for such extended monopoles is a soliton.

Typical dualities in modern physics are related just to the coupling constants. More specifically, suppose there are two theories[84], each being characterized by a coupling constant $g_i, i = 1,2$. A coupling constant is typically an arbitrary parameter. Assume that we change coupling $g_1$ of the first theory so that eventually it becomes inverse to coupling $g_2$ of the second theory. If in this case both theories provide similar regimes, e.g., nearly identical relationships of state probabilities, then such theories are called dual. In string theory this situation is known as S-duality, S standing for "strong-weak" ($g$ and $1/g$). More accurately one can say that the S-duality maps states and vacua arising in one theory (with coupling constant $g_1$) to those in the dual theory (with coupling $g_2$) when $g_1 = 1/g_2$. Duality here is a cunning trick allowing one to apply the perturbation theory - the main tool of quantum mechanics which normally can be used only for weak coupling, i.e., for $g < 1$ - to strongly coupled theories ($g > 1$), by mapping strongly coupled to weakly coupled states. In the context of usual four-dimensional quantum field theory (QFT), which is much simpler than string theory, S-duality exchanges the electrical and magnetic field components and, respectively, charged particles with already mentioned magnetic monopoles (see, e.g., [134], [135]).

---

[84] It is customary nowadays to call pieces of theoretical endeavors "theories", I would prefer to name them "models" because of their limited scope.

Discussion of magnetic monopoles, as well as of T-dualities (topological, see, e.g., [https://en.wikipedia.org/wiki/Superstring_theory](https://en.wikipedia.org/wiki/Superstring_theory)), would lead us far astray, e.g., to scrutinizing the solutions of nonlinear PDEs and topological properties of space. Topology is a highly advanced discipline requiring a profound treatment, even in the physical context. More or less detailed theory of nonlinear PDEs is a special and serious topic being outside the scope of this book - otherwise the text would be swollen beyond the perceptive ability of a normal person. Nevertheless, we shall discuss later some symmetry properties of the main equations of physics and see how the duality appears. Now I shall only note that a symmetry of the Maxwell equations permitting us to exchange electrical and magnetic fields appears as a duality between elementary charges and collective excitations, since in a weak coupling case electrical charges look like quanta whereas magnetic charges emerge as collective excitations. Moreover, it seems that in general, quantum dualities exchange particles with collective excitations, in QFT with quasiparticles represented as solitonic solutions.

Furthermore, one can often hear about the "wave-particle duality" (see Chapter 6). However, the concept of wave-particle duality as it is widely taught to students is intuitive, unspecific, somewhat ambiguous and imprecise. This is essentially the same kind of duality as between dual spaces in algebra, because to each displacement or velocity-like vector in classical mechanics we have a momentum or wavenumber-like vector in quantum mechanics. Due to this duality, quantum mechanics becomes very similar to classical field theory by semiclassical correspondence between particles (mechanics) and waves (fields) - the same wave equations are used since the mechanical Hamiltonian becomes the kinetic energy operator of the corresponding wave theory. The fact that differentials and derivatives are dual to each other can be a good introduction not only into differential geometry but into quantum mechanics as well. So, examples of dualities are abundant both in physics and in mathematics. One of the popular recent examples is the Veneziano dual model [184], see also [185]. The Veneziano model was later reformulated and included into a set of modern string theories. For the students of numerical mathematics and so-called scientific computing, nowadays of fashion, it would be interesting to know that, for instance, matrix columns, which may be treated in Euclidean space as forming a vector subspace, possess the dual space which is the subspace of rows. In solid state physics, one uses dual (reciprocal) basis to describe the crystallographic lattice and specify the electron states. If we take the Dirac notation in quantum mechanics (Chapter 6), the ket vectors form a subspace of a Hilbert space and have the dual space of bra vectors. For any vector field we have a dual field, however sometimes not so easily represented in the form of simple geometrical images. Likewise, the dual space for the tangent vector space, $T_x(M)$, is the cotangent space $T_x^*(M)$ which, unfortunately, also does not have a simple geometrical representation like the tangent space - we can mathematically define but cannot easily visualize the cotangent space even for the low-dimensional cases $d = 1,2,3 \dots$ So it seems that in this case we must leave the visual patterns of physics aside and stick to mathematical definitions.

We have already seen that for each vector space $V$ there exists a dual space $V^*$ - here I may refer, e.g., to the section on coordinate transformations where the difference between contravariant and covariant vectors and coordinates was discussed. For convenience, I shall recall some simple geometrical interpretations. For example, the velocity vector for each smooth curve passing through point $x \in M$ is a contravariant tangent vector lying in tangent space $T_x(M)$ at a point $x$ of a manifold $M$. It is an element of the vector space tangent to the manifold in this point and one can view it as being located in a hyperplane analogous to the tangent plane for dimension $d = 2$. Another example is a displacement vector **r**. We have seen that a covariant vector in a point $x$ of a manifold is an element of another - dual - vector space called cotangent space in the point $x$ of the manifold. A covariant vector is a geometrical object that transforms like the basis vectors, whereas a contravariant vector is an object that transforms "against" the basis vectors. By the way, one should not confuse covariant vectors and Lorentz covariance, this is just an abuse of the language. For instance, a contravariant Lorentz vector is different from a covariant Lorentz vector, although the former is a vector that is Lorentz covariant (see Chapter 5). Now, however, we confine ourselves to simple examples needed only to illustrate the concept of duality.

Duality, in human terms, is just two different ways of looking at the same phenomenon.[85] In physics, duality in fact appeared long before many other concepts, when in the 17th century Huygens and Newton proposed competing wave and corpuscular theories of light's behavior. For over a century Newton's corpuscular theory was dominant, mainly, it seems, due the author's authority - one more example of sociological effects in science. It is only in the early 19th century when diffraction has been reliably observed, e.g., in Thomas Young's double slit experiments, that people apprehended grave complications for the Newton's corpuscular theory of light. Many observations clearly demonstrated an obvious wave behavior, the light waves showed interference patterns - here the principle of superposition for linear waves is manifested (see many interesting details in [106]). All this seemed to prove that light traveled in waves so that the Newton's theory had to be replaced by Huygens' theory. But if light existed as waves, this fact would imply, according to the standard wave theory accepted at that time, a medium of some kind through which the wave must propagate. This invisible medium was suggested by Huygens and called by him "luminoforous aether", in a more customary terminology, ether. Nobody could observe this medium despite a number of experimental attempts throughout the 19th century. Finally, the search for ether culminated in the famous and direct Michelson-Morley experiment, which led to creation of the relativity theory. Then the photoeffect was studied (in particular, in the Einstein's work) and again a particle theory of light took dominance, eventually leading to mathematical models of quantum mechanics (see Chapter 6). Thus, the old concept of duality was plainly related to today's physics, the latter possessing more modern dualities. Now, it would be a truism to say that light manifests itself as both a particle and a wave, depending on what observations are made and how the experiment is set up. The same duality applies to matter, e.g., to elementary particles. To manifest dual properties of matter, different conditions are necessary. Incompatible external conditions reveal either wavelike or corpuscular properties of microparticles, e.g., electrons or neutrons. In other words, depending on imposed external conditions (such as observation means - experimental setup) quantum particles display wave or particle features, just like light. The meaning of duality is that there is a potential possibility for every object to display totally diverse - even opposite and complementary - features depending on premeditated settings. It would be probably wrong to understand duality too literally, for instance, in the quantum-mechanical case, to model the particle as a singular point of a certain field or represent the particle motion as being carried by the wave. Such mathematical models, of course, can be constructed but their heuristic value seems to be too low (see more details in Chapter 6).

---

[85] This principle has been long exploited by authors in classical literature. For example, in in polyphonic novels by F. M. Dostoyevsky such as "The Karamazov Brothers" there exists no 'objective' description of the world, but only different manifestations of reality depending on the perception of acting personages. A brilliant illustration of duality has been produced in the film "Rashomon" by the famous Japanese film director, Akira Kurosava. This film was based on the short story "In a Grove" by Akutagawa Ryunoske (although it was a far adaptation); both the story by Akutagawa and the movie depict several characters offering different "true" accounts of rape and murder. Thus, it is an artistic exploration of the nature of "truth".

Here, I would like to make a few comments. One must notice that such circumstances may exist when dual features can be displayed simultaneously. For example, the bound state of an electron in atoms is described by a standing wave, usually with an amplitude that is rapidly diminishing with the distance to the center (nucleus). This fact means that the electron is approximately localized (a corpuscular feature), sometimes with a good accuracy, but it is nonetheless a wave. A typical feature of such combined manifestations of the dual (complementary) properties is that they are less distinctly exhibited.

competing wave and corpuscular theories of light's behavior. For over a century Newton's corpuscular theory was dominant, mainly, it seems, due the author's authority - one more example of sociological effects in science. It is only in the early 19th century when diffraction has been reliably observed, e.g., in Thomas Young's double slit experiments, that people apprehended grave complications for the Newton's corpuscular theory of light. Many observations clearly demonstrated an obvious wave behavior, the light waves showed interference patterns - here the principle of superposition for linear waves is manifested (see many interesting details in [106]). All this seemed to prove that light traveled in waves so that the Newton's theory had to be replaced by Huygens' theory. But if light existed as waves, this fact would imply, according to the standard wave theory accepted at that time, a medium of some kind through which the wave must propagate. This invisible medium was suggested by Huygens and called by him "luminoforous aether", in a more customary terminology, ether. Nobody could observe this medium despite a number of experimental attempts throughout the 19th century. Finally, the search for ether culminated in the famous and direct Michelson-Morley experiment, which led to creation of the relativity theory. Then the photoeffect was studied (in particular, in the Einstein's work) and again a particle theory of light took dominance, eventually leading to mathematical models of quantum mechanics (see Chapter 6). Thus, the old concept of duality was plainly related to today's physics, the latter possessing more modern dualities. Now, it would be a truism to say that light manifests itself as both a particle and a wave, depending on what observations are made and how the experiment is set up. The same duality applies to matter, e.g., to elementary particles. To manifest dual properties of matter, different conditions are necessary. Incompatible external conditions reveal either wavelike or corpuscular properties of microparticles, e.g., electrons or neutrons. In other words, depending on imposed external conditions (such as observation means - experimental setup) quantum particles display wave or particle features, just like light. The meaning of duality is that there is a potential possibility for every object to display totally diverse - even opposite and complementary - features depending on premeditated settings. It would be probably wrong to understand duality too literally, for instance, in the quantum-mechanical case, to model the particle as a singular point of a certain field or represent the particle motion as being carried by the wave. Such mathematical models, of course, can be constructed but their heuristic value seems to be too low (see more details in Chapter 6).

Here, I would like to make a few comments. One must notice that such circumstances may exist when dual features can be displayed simultaneously. For example, the bound state of an electron in atoms is described by a standing wave, usually with an amplitude that is rapidly diminishing with the distance to the center (nucleus). This fact means that the electron is approximately localized (a corpuscular feature), sometimes with a good accuracy, but it is nonetheless a wave. A typical feature of such combined manifestations of the dual (complementary) properties is that they are less distinctly exhibited.

Another remark is rather trivial: diverse experimental settings often may properties. The resulting wave function (amplitude) is constructed as a linear superposition of all interfering probability amplitudes and, consequently, reveals the same wave-like features. One can interpret this construction as follows: the probability of finding a particle in some location is determined by a wave, but the actual physical observation detects a corpuscle.

The so-called physical meaning of the wave-particle duality has long been a focus point of heated debates in the early years of quantum physics. Nowadays, it seems this issue remains a hot topic only for amateurs. Nevertheless, attempts to explain what the wave-particle duality "really means" has resulted in a number of interesting interpretations of quantum mechanics. To my knowledge, all these interpretations produced no new results and no fresh experimental predictions as compared with the set of wave equations. The mathematical models associated with the latter may appear to some people rather complicated, but they mostly provide accurate experimental predictions.

One can also note that duality as a principle exists not only in mathematics and physics. In the preceding chapter, I have briefly mentioned cognitive models, in particular those developed in so-called cultural anthropology. One such model, comprehensively discussed by the famous philosopher Karl Popper in his highly interesting two-volume manuscript "The Open Society and Its Enemies" [186], was designed to explain the stability of human societies[86]. The main statement here is that any stable society must have a dual structure, in particular support two parties, e.g., "left" and "right". This bipartite system was found to be typical even of the ancient and primeval societies, and up till now the majority of societies stick to the primeval pattern. The society was divided into two parts, and each member of society was exactly aware to what part she/he belonged. Even today, one can tell whether a person sympathizes with the liberal attitudes or identifies oneself with a strong state domineering over its citizens. Ancient people even formulated social duality in the form of a slogan: "Always take your wife from the other half of the tribe." This principle made intra-tribal aggression highly ritualized and, consequently, non-destructive. The dual notions are especially evident in oriental - mainly Chinese - traditions (Yin-Yang etc., see, e.g., http://en.wikipedia.org/wiki/Yin_and_yang). Social duality was also popular and extensively depicted in literature (R. L. Stevenson, J. Swift).

## 3.16  Manifolds

One may ask: why did we devote so much attention to a seemingly trivial subject of vector spaces? The main motivation was the following: $n$-dimensional linear (vector) spaces, e.g., $\mathbb{R}^n$, and their open subsets $U$ are the simplest sets on which one can define functions as well as vector and tensor fields. Recall that $\mathbb{R}^n$ may be thought as consisting of vectors $(x^1, \dots, x^n)$, with dimensionality $n$ being arbitrary large, but still finite.

---

[86] Periods of instability, e.g., revolutions, are assumed significantly less durable than periods of stability.

The most useful feature of such simple sets is the possibility to establish on them a unique coordinate system. For instance, one can represent a map $f(x)$ on $x \in U \subseteq \mathbb{R}^n$ as a function of $n$ coordinates $x^i, i = 1, \ldots, n$.

However, already very simple examples demonstrate that such a compelling-looking scheme is a somewhat provisional way of looking at physical spaces, and it may easily become inadequate. If we take, for instance, the unit sphere, $S_{n-1} \subset \mathbb{R}^n$, which may be described by the equation $x_i x^i = 1, i = 1, \ldots, n$, then we quickly get into trouble. To be more specific let us take $n = 3$ i.e., $S_2 \subset \mathbb{R}^3$. Then we have the familiar equation from the school-time geometry:

$$(x^1)^2 + (x^2)^2 + (x^3)^2 = 1$$

(hopefully, there will be no confusion between coordinate and power indices). Writing this equation in spherical coordinates

$$x^1 = \cos\varphi\sin\theta, \qquad x^2 = \sin\varphi\sin\theta, \qquad x^3 = \cos\theta,$$
$$0 < \varphi \leq 2\pi, \ \ 0 \leq \theta < \pi.$$

we may immediately observe that there is no coordinate system that can describe the whole $S_2$ surface.

Cartesian (rectangular) or affine coordinates naturally reflect geometric properties of Euclidean and affine spaces. However, in many situations it is much more convenient to work in curvilinear coordinates than in Cartesian ones, e.g., by dealing with curvilinear objects. It is enormously easier, for example, to integrate differential equations of mathematical physics for a spherically symmetric problem in spherical coordinates than in Cartesian those. (It is remarkable that this adaptation of coordinate frame to symmetry and respective geometrical techniques are often disregarded in numerical treatment of respective problems.) A systematic study of the curvilinear coordinate systems naturally leads to the merge of geometry and physics, which became obvious after the construction of general relativity after Einstein and Hilbert (actually D. Hilbert submitted an article containing the correct field equations for general relativity several days before Einstein, yet Hilbert never claimed priority for this theory.

One may notice that the idea of a manifold is so simple that it is astonishing that the physicists rarely used this concept when physics was especially fruitful, say, in the period of 1900-1970. Working with manifolds simply means that one should be able to view the surrounding space as locally Euclidean (see above the definition and discussion of a Euclidean space). In other words, a manifold looks like $\mathbb{E}^n$ (or $\mathbb{R}^n, \mathbb{C}^n$) locally, but not necessarily globally. This is basically a geographic idea: the Earth looks like plane near your home, and when mathematically generalizing this illusion, we can approximate a rather small domain around a point on practically any surface by the tangent plane at this point - an idea close to linearization.

I apologize for using such vague notions as locally, near, small domain, practically and the like. Later, little-by-little a concrete, but context-dependent meaning will be assigned to these words. One might note in passing that in contrast with geometry which mostly deals with the local properties of manifolds (for instance, differential geometry), there exists a discipline known as topology where the manifolds are studied as whole entities i.e., at the global level.

What is the fundamental difficulty in doing classical differential calculus on manifolds? One may remember that differential calculus is the basic tool of physics, starting from Newton and Leibniz. The main difficulty is to extend the results obtained in one coordinate system to other coordinate systems on a manifold making sure that these results remain valid in all of them. The traditional way, extensively used in physics, to circumvent this problem is to apply the language of tensor analysis. In this manner, the laws of physics to be used in physical models on manifolds are formulated through tensor derivatives. Then the requirement of coordinate invariance in handling physical models on manifolds naturally leads to similarity or isomorphism between the vector spaces attached to neighboring but distinct points, in the limiting case between the points which are infinitely close to each other (in the spirit of classical differential calculus). Such isomorphisms form a mathematical structure commonly known as the Levi-Civita connection.

But before we study the standard concepts of differential geometry, we have to discuss the possibility of confidently using the tools of classical calculus on manifolds. I have already mentioned that the concept of a manifold is a simple extension of a surface in $\mathbb{R}^3$. Then the notion of a differentiable manifold is a generalization of a regular surface in $\mathbb{R}^3$. The surface $S$ is considered regular if for each point $x \in S$ there exists a neighborhood $U_x$ of $x$ in $\mathbb{R}^3$ and a map of an open set $V$ of $\mathbb{R}^2$ onto $U$ i.e. $f : V \in \mathbb{R}^2 \to U \cap S$, this map being a differentiable homeomorhism (simply a diffeomorphism, although I know that there exist pedants who distinguish these two notions). Intuitively, a regular surface is perceived as a union of two-dimensional ($\mathbb{R}^2$) open sets making up an "atlas" of a number of "charts" so that such open sets overlap, the transition from one to another can be made in a differentiable (smooth) fashion. This perception signifies that one can painlessly go from one parametrization of a surface to another, and this change of parametrizations or, in simple words, coordinate transition is a diffeomorphism. Use of differentiable maps for the change of coordinates on the one hand requires and on the other hand ensures applying the powerful techniques of classical calculus.

One can, however, immediately notice that the presence of $\mathbb{R}^3$ in the concept of a regular surface - in fact of any surface - is totally irrelevant. This ambient space in no way affects the surface; its role is only reduced to a physical container or a vessel.

Viewing from the physical positions, differentiable manifolds have a number of extremely useful applications. First of all, a differentiable manifold may be considered a rudimentary model of our spacetime. Why rudimentary and why only a model? Primarily because a differentiable manifold in general does not yet possess the customary and convenient, sometimes even vital for us geometric features such as, e.g., distance, length, angle, area[87]. Thus, a differentiable manifold may in fact serve only as a prototype for our spacetime. To make this prototype closer to the version of spacetime we think we know, one must introduce some additional properties such as metric and affine connection on the tangent space at each point.

A vector bundle is a manifold created by assigning a vector space to each point of the manifold. In particular, the tangent bundle consists of a manifold plus the tangent space at each point. The cotangent bundle is formed analogously. We will see that tangent bundles are used for the Lagrangian formulation of classical mechanics, and cotangent bundles are used for the Hamiltonian formulation.

## 3.17　Notes on Derivatives

Let us recall some basic facts from classical analysis. The main object of classical calculus is the function $f$ taking $x \in X$ to $f(x) \in Y$ where $X$ and $Y$ are some vector (affine) spaces. Now, what is a derivative of a function? One can say that it is a very simple object, yet thinking about it in association with other concepts gives rise to interesting observations. Let us take the simplest example of a single-variable function $f \colon \mathbb{R} \to \mathbb{R}$. Differentiating it produces another function, $d_x f \equiv df/dx \colon \mathbb{R} \to \mathbb{R}$, so the differentiation is a map on a set of functions defined on $\mathbb{R}$ with the range also in $\mathbb{R}$. Actually, it is one of the simplest examples of a linear operator that acts on a set of functions forming a vector space with respect to ordinary addition and multiplication. If the function $f$ has a derivative at $x$, it is uniquely defined and referred to as *the* derivative of $f$. In nearly all physical models, differentiability is assumed beforehand, and non-smooth functions rarely exist.[88]

Imagine now a function of more than one variable. Geometrically, we have in this case more than a single number to characterize the derivative, i.e., the tangent to the graph of a map $(x, f(x)) \colon X \to Y$. For instance, in the $2D$ case, $\mathbb{R}^2 \to \mathbb{R}$ which still can be easily visualized, two numbers are required to specify the tangent plane to a surface in each point ( $x^1, x^2 \in \mathbb{R}$ ). Differentiation does not produce the same kind of map, $\mathbb{R}^2 \to \mathbb{R}$, and if we fix a basis and write the map of differentiation in coordinates, we get partial derivatives. Now, remembering that basis and coordinate representation is a convenient but not a universal one - there is often no natural way to fix a basis - we may say that taking the derivative gives a map from the tangent space at $x \in X$ to the tangent space in $f(x) \in Y$. There are a number of such maps and we shall discuss them later one by one.

In a slightly different view, the derivative of $f$ at $x$ is a linear functional taking functions to $\mathbb{R}$. In case $f$ is defined on some vector (affine) space $X, f: X \to \mathbb{R}$, the derivative is a linear functional defined on the tangent space, $T_x X \to \mathbb{R}$, to the vector (affine) space $X$. Here we may recall that $\mathbb{R}$, while being algebraically a field, is understood as a source of real scalars [89]. Generalizing this construction, we obtain the derivative of $f$ at $x \in X$ as a linear map $d_x f: T_x X \to T_{f(x)} Y$ where $f: X \to Y$ is a map between vector (affine) spaces $X$ and $Y$. From here on, we can easily understand the concepts of partial derivatives and of a directional derivative, the latter being very important in Hamiltonian mechanics, the theory of dynamical systems and relativity. It is remarkable that tensor quantities naturally emerge from differentiation, even when one deals with scalar functions. This fact is, in particular, manifested by the matrix representation $d_{x^j} f^i$ for the linear map $d_x f$, with partial derivatives $\partial f^i / \partial x^j \equiv \partial_j f^i(x)$. Here $x^j, j = 1, \dots, n$ are coordinates on $X, f^i, i = 1, \dots, m$ are components of the vector function $f \in Y$ in basis $\beta$ on $Y$. This manner of writing the derivatives as a map of a vector-function is common in dynamical systems theory and we shall employ it later.

---

[87] These are the examples of local structures: a distance between points specifies a metric structure, an angle between curves manifests a conformal structure (angles are left unchanged by a conformal mapping), area on a surface corresponds to a symplectic structure.

[88] Recall the first-year calculus terminology: if $f$ has a derivative at $x$, $f$ is called differentiable at $x$, and if $f$ is differentiable wherever it is defined, it is simply called differentiable.

Thus, the apparently familiar concept of a derivative leads us to rather delicate subjects - covariant derivatives, metric tensors, geodesics, etc. For instance, the simple notion of a directional derivative leads to rather nontrivial entities in variational calculus. A "directional derivative" in $\mathbb{R}^3$ i.e., the derivative of function $f(\mathbf{r}), \mathbf{r} = (x^1, x^2, x^3)$ along the direction determined by vector $\mathbf{l} = (l^1, l^2, l^3)$ is defined in calculus as

$$\partial_1 f = \lim_{\epsilon \to 0} \frac{1}{\epsilon} [f(x^1 + \epsilon l^1, x^2 + \epsilon l^2, x^3 + \epsilon l^3) - f(x^1, x^2, x^3)]$$

(see any course of elementary calculus). It can be easily verified that $\partial_1 f = (1, \nabla f)$. Let us now consider a functional $S[f]$, in particular defined as an integral of Lagrangian $L = L(x, f, f_x)$, where $x = (x^1, x^2, x^3) \in D$, $f \in V, V = V_m$ is a linear space of smooth functions $f(x) \coloneqq f^i(x^\mu), i = 1, \dots, m$. The Lagrangian $L$ is also considered a smooth function of $x^\mu, f^i$ and $f^i_{x^\mu}$. One might note that the requirement of smoothness is sufficient for nearly all variational problems having practical importance in physics, but can become too restrictive in certain singular cases. We, however, shall not consider such cases. So let us write the functional $S[f]$ in the usual form

$$S[f] = \int_D d^n x L(x^\mu, f^i, f^i_{x^\mu}), i = 1, \dots, m.$$

Thus, the functional $S[f]$ is defined on $V$. The domain $D$ of variables $x^\mu, \mu = 1, \dots, n$ which may be regarded as parameters of a variational problem is, for simplicity, supposed to be finite in $\mathbb{R}^n$ i.e., delimited by a border $\partial D$, assumed smooth or at least piecewise smooth. Notice that the shorthand

---

[89] Scalars used in physics are not necessarily real, other number systems can also underlie a vector space. For example, in quantum mechanics vector spaces are built over complex scalars (which, by the way, may be treated as vectors in the context of complex analysis)

$d^n x$ actually denotes $n$-dimensional volume in $\mathbb{R}^n$, $d^n x \equiv d\Omega_n := dx^1 \wedge dx^2 \ldots \wedge dx^n$.

Let us now perturb $S[f]$ i.e., displace, as we have already done, $f^i(x^1, x^2, x^3) \in V$ to $f^i(x^1, x^2, x^3) + \epsilon h^i(x^1, x^2, x^3)$, where $h := h^i(x^\mu), \mu = 1, \ldots, n$ also belongs to $V$ and is chosen in such a way as $h|_{\partial D} = 0$; $\epsilon$ is a small parameter. Now we may construct an expression very similar to a directional derivative well-known from classical calculus:

Let us briefly review what we have done. First of all, we introduced the basis, i.e., the possibility to label each point in the vector (affine) space by coordinates $x^j \in \mathbb{R}^n$, and then identified the point $x \in X$ with this set of coordinates. One must, however, remember that the set of $x^j, j = 1, \ldots, n$ are just labels for a point in the given coordinate system. If you choose another basis, these labels are changed, they do not have absolute meaning. Besides, there is often no natural way to choose the basis, the sphere being the favorite example. Nevertheless, once the concrete identification of the points has been accepted, we may write the usual limit that is considered, in the traditional courses of calculus, as the definition of partial derivative:

$$\frac{\partial f^i}{\partial x^j} \equiv \partial_j f^i(x) = \lim_{\delta x^j \to 0} \frac{f^i(x^1, \ldots, x^j + \delta x^j, \ldots, x^n) - f^i(x^1, \ldots, x^n)}{\delta x^j}$$

An important thing about this limit is that it may exist, although the map $d_x f$ does not exist.

The matrix $\partial_j f^i(x)$ representing the map $d_x f$ in some specified basis is the Jacobian matrix of the map $f(x): X \to Y$. We shall encounter this matrix many times in various contexts.

There are some particular cases studied in analysis courses. If $X = \mathbb{R}$, we go back to the simplest case when the variable $x$ has only one direction (and its reverse), so the derivatives become 'ordinary', $df^i/dx = df^i/dx(x), i = 1, \ldots, m$ and the Jacobian matrix is reduced to a column vector. If, on the other hand, $Y = \mathbb{R}$, then vector $f^i, i = 1, \ldots, m$ has a single component that may be denoted $f(x)$ and the Jacobian matrix is degraded to the row vector of partial derivatives, $\partial_j f$.

## 3.18  Notes on Calculus

While discussing derivatives, one can naturally recollect the basic concepts of calculus. I totally disagree with the point of view that calculus, or classical analysis, is an old-fashioned and "completed" discipline, so there is little hope to invent something new in it. This snobbish attitude towards classical disciplines is a symptom of exaggerated attention to fashionable parts of science. It is a sociological effect encountered not only in mathematics. In physics, some vociferous proponents of current outfit treat classical electrodynamics and even classical mechanics with a poorly concealed disdain. As I have mentioned in the preceding chapter, when I was very young, I often came to the so-called "Landau seminar" at the Institute of Physical Problems in Moscow. There, it was probably considered trendy to derogate all the directions that were outside the scope of local theoreticians (the "Landau School", L. D. Landau himself had unfortunately died by this time). I remember how it stunned me when I occasionally heard from the persons for whom I held a great respect such fanciful statements as "classical electrodynamics is a pastime for assiduous idiots", "what they do in mechanics has nothing to do with physics", "irreversible thermodynamics is irreversible stupidity" (the last statement was ascribed to Landau). Probably, such fads and fancies exist also in the biological community - at least I have heard some rumors from biologists about the subjects that do not even deserve of being studied.

I have already touched upon sociological effects in physics (see Chapter 2), and in Chapter 9 ("What Remains to Be Solved") we shall have to discuss them once more; now, reverting to calculus I would like to note that there is a lot of room for improvement in both understanding its conceptual roots and in effective exposition of the subject. When working with students, I could observe that, in spite of its beauty, official calculus is too hard for the majority of them, especially for those who are more inclined to use computers than to learn numerous theorems and proofs.

## 3.19  Basic Geometry for Physics

When considering this subject, one can get a feeling that there are different cultures in physics: intuitive, algebraic, geometric, computational, visionary, and so on. In particular, one can understand the term "geometry" as the study of sets $X$ consisting of elements $x$ called points (irrespective of their physical nature, e.g., they may actually be straight lines) together with some fundamental relations between them. Points $x$ are all peers i.e., they are usually considered homogeneous with respect to such fundamental relations, and a group of automorphisms that is of one-to-one mappings of set $X$ onto itself, $x \mapsto x'$ acts on $X$ leaving all fundamental relations intact. In many instances, we shall have to use geometrical language to understand physical models. I have read long ago that one of the great mathematicians, David Hilbert, considered physics and geometry to be actually the same subject, since both deal with the objects of the real world. At least, modern physics extensively uses geometrical concepts. For me personally, it has always been really fascinating how deep is the role of differential geometry in physics. Certain things, even very simple facts from classical mechanics, cannot be properly understood without elementary notions from differential geometry. For instance, the whole of kinematics is indistinguishable from elementary differential geometry. Yet, some physicists consider reducing physics to geometry a grave crime, since spatial and temporal scales are essentially physical quantities.

Fundamentals of differential geometry include such important and in general nontrivial concepts as flows, manifolds, bundles, connections, differential forms, Lie groups, jets and so on. One can see that discussing them even in moderate scope would require a lot of time and pages, so nolens volens I have to be superficial while addressing the subject of differential geometry. Trying to cover this subject in a very lapidary mode, I will persistently refer to a couple of my favorite books [187] and [188].

Physical geometry starts from discussing spaces in which physical processes can evolve and the coordinate systems that can be introduced in such spaces. Nowadays, coordinate-free methods (e.g., those based on differential forms) are becoming increasingly popular. The coordinate-based approach is, of course, less general but provides a clear and intuitive introduction to many delicate topics such as covariant derivatives, metric tensors, geodesics, connections, etc. Therefore, it is still very valuable even now, when the differential form approach seems to have won the battle. I shall try to use both approaches, in many cases giving translations from one language into the other.

There exist many kinds of geometries. The most well-known and natural from the viewpoint of everyday life is the Euclidean geometry on which classical physics was based. One usually models point spaces considered in Euclidean geometry as vector spaces (see the respective section above), but the entire Euclidean ideology is not always an adequate choice. A slight generalization of school-time Euclidean geometry produces affine geometry, also very important for physics, studying geometric properties that are invariant under the action of the group of affine transformations. When the Renaissance painters started using pictures as a synthetic language to express or model reality, projective geometry became essential, although it was not mathematically formulated in the Renaissance epoch. Projective geometry studies projective transformations distorting angles and distances but preserving lines. From the painter's viewpoint, projective geometry is that of a perspective. Further development of human civilization quickly brought new mathematical ideas serving to depict reality, however not in the conventional naive form. These ideas were quite often formulated as new geometries such as Riemann geometry (now serving as the mathematical base for general relativity), Bolyai-Lobachevsky hyperbolic geometry (needed for spacetime concepts based on special relativity), old (with coordinates and indices) and modern (coordinateless and based on forms) differential geometry widely used in many fields of physics, Hermitian geometry (serving quantum mechanics with its Hilbert space), symplectic geometry (needed for modern formulations of classical mechanics), convex geometry (primarily studying the Radon transformations that are very important for modern technology), computational geometry (the main objects of study being convex sets, Delaunay triangulations and Voronoi diagrams), non-commutative geometry (modern microscopic physics), and some other geometry kinds. In the traditional geometry that makes up unnecessary long courses in high schools, such structure quantities as distance, angle, segment, triangle, parallel lines, etc. are the principal notions. Actually, it is these notions that form the whole school-time geometry. In contrast to this traditional approach, geometry may be defined by transformation properties, i.e., the transformation (symmetry) groups can serve as fundamental principles determining the geometry. This idea was put into the foundation of the so-called "Erlangen Program" formulated by Felix Klein in 1872. According to this concept, ordinary geometrical structures, like those studied at schools, are only of secondary nature and can be derived from the action of the respective transformation group. In this manner, one can specify a number of standard geometries by their fundamental groups. For example, the principal geometries of physics based on transformation groups are as follows: Euclidean geometry, affine geometry, Riemann geometry, differential geometry, symplectic geometry, noncommutative geometry. This is just a list - a short one, and there is no hierarchy between these geometries. They are, however, connected with physically interesting links - thus momentum signifies transition from ordinary Euclidean geometry to Lee groups and symplectic geometry. We are comfortable with the phase space of classical mechanics as well as with the four-dimensional pseudo-Euclidean geometry describing the flat spacetime in special relativity and, perhaps to a somewhat lesser extent, with the conformal geometry of quantum field theory. Later we shall observe the possible geometrical interpretation of quantum mechanics, which provides a number of curious connections between different variants of geometry.

Why is affine geometry important for physics? This issue was quite clearly explained in [14]. Simply speaking, our world is a $4D$ affine space $A^4$ whose elements are events (world points). Any affine space "unites" points and vectors, in this case $A^4$ includes the four-dimensional vector space $\mathbb{R}^4$ which works as a group of parallel translations: for any two points $a, b \in A^n$ one can find $\mathbf{v} \in \mathbb{R}^n$ so that $a = \mathbf{v} + b$. In other words, the action of the vector space $V^n$ in $A^n$ is a map $V \times A \to A$. One can find the full definition of affine space in any textbook on geometry, that is why I don't bring it here.[90] For classical mechanics we can take a particular model of an affine space $\mathbb{A}^4$ with $\mathbb{R}^4$ as a point space and $\mathbb{R}^4$ as a vector space of all translation vectors (in this case, of course, $n = 4$). The time is interpreted as a linear map $T: V \to \mathbb{R}$ of the vector space of parallel translations to the real axis. This map implements a projection onto one of the coordinates, a temporal one, $T(a, t) = t$ for $a, t \in V$, i.e., a split of $V$ into $\mathbb{R}^3 \times \mathbb{R}$ - a specific feature of classical mechanics distinguishing it from relativistic theories. The subspace of simultaneous events is obviously $A^3$, and the kernel of $T: V_0 = \text{Ker} T = \{v \in V : T(v) = 0\}$ includes all parallel translations linking all simultaneous events. In classical mechanics, $V_0 = \text{Ker} T$ is simply the spatial part $\mathbb{R}^3$. Simply speaking, a set of all simultaneous events constitutes a $3D$ affine space. The time difference for events $a, b \in A^4$ is $T(ab)$ and is zero for vectors $\mathbf{v} = \{a - b\} \in V_0$. The mathematician would say that fibers $V_t = T^{-1}(t)$ describe in $V$, and consequently in $A^4$ the layers of simultaneous events.[91]

An important feature of the affine space distinguishing it from the ordinary vector (linear) space is that no coordinate origin is specified in the affine space (and no preferred directions). For creationists' models of the world not the affine, but the linear structure would be more appropriate.

---

[90] See, for example affine coordinates in the Encyclopedia of Mathematics (https://encyclopediaofmath.org/wiki/Affine_coordinate_system).

[91] One may recall here that in mathematics notations and terms are often specially designed; in traditional (pre-string) physics very seldom.

An affine map of a plane onto a plane is of the form

$$x' = a_{11}x + a_{12}y + c_1, \qquad y' = a_{21}x + a_{22}y + c_2$$

Let us denote by $\mathbb{A}^n$ an affine space of finite dimension $n$ based on the vector space $V^n$.

A very important example of a set is $\mathbb{R}$ which is a set of all reals. Then $\mathbb{R}^n$ which was previously known as an arithmetic space and is a set of all real $n$-tuples i.e., ordered sets of $n$ real numbers (arithmetic points). So $\mathbb{R}^n$ is an $n$-dimensional space of points $\mathbb{R}^n = (x^1, \ldots, x^n)$, where each $x^i, i = 1, \ldots, n$ is a real number. It is an $n$-dimensional real vector (linear) space so that its elements are called vectors. It means that two operations, addition and multiplication by a scalar from some field are defined in $\mathbb{R}^n$, with all standard axioms delineating a vector space being fulfilled (see the section "Vector Spaces" below). In fact, however, points and vectors are different objects so that one needs to introduce another set called an affine space, $\mathbb{A}^n$, in addition to $\mathbb{R}^n$. From the physical perspective, the affine space $\mathbb{A}^n$ differs from vector space $\mathbb{R}^n$ by an absence of a fixed coordinate origin (see also the respective section on the affine space below). If, additionally, the vector space $\mathbb{R}^n$ is endowed with a Euclidean structure i.e., a positive-definite bilinear symmetric form $(x, y)$ called an inner product (also known as scalar product) is defined on $\mathbb{R}^n$ (or on $\mathbb{A}^n$), then such a space is called a Euclidean space $\mathbb{E}^n$. Thus, a Euclidean space is an inner product space. Moreover, it is a normed space since inner product $(x, x) \geq 0$ for any $x \in \mathbb{E}^n$ so that this product allows us to define the length of a vector $x$ as a real nonnegative function of $x$, $\|x\| = (x, x)^{1/2} = (x_i x^i)^{1/2}$ satisfying all the properties required of a norm. These properties are (see any textbook on linear algebra or functional analysis): (1) $\|x\| \geq 0$, with $\|x\| = 0$ if and only if $x = 0$ which then means that the norm is the positive-definite function of its argument $x$; (2) $\|\alpha x\| = \alpha \|x\|$, $\alpha \in \mathbb{R}$, $x \in \mathbb{E}^n$; (3) $\|x + y\| \leq \|x\| + \|y\|$, $x, y \in \mathbb{E}^n$    ;    (4)    $\|x - y\| \leq \|x - z\| + \|z - y\|$, $x, y, z \in \mathbb{E}^n$ (properties (3) and (4) are usually known as the triangle inequalities).

One can symbolically write an $n$-dimensional Euclidean space as $\mathbb{E}^n = \{(x^1, \ldots, x^n) | x^i \in \mathbb{R}, i = 1, \ldots, n\}$. Thus, $\mathbb{E}^1$ represents the real line, $\mathbb{E}^2$ is the ordinary plane that we study at school (recall the Pythagorean theorem), and $\mathbb{E}^3$ is our three-dimensional space where all classical nonrelativistic events take place. When studying classical mechanics, we shall mainly deal with the Euclidean space. Note that the inner product allows one to define a distance between two points $x$ and $y$ of $\mathbb{R}^n$ (or $\mathbb{A}^n$), $s(x, y) = \|x - y\| = (x - y, x - y)^{1/2} = (x_i - y_i, x^i - y^i)$ which is nothing more than the Pythagorean theorem. This distance is a particular case of a metric (usually called the Euclidean metric), $s(x, y)$, so that Euclidean space is also a metric space.

It is curious that there seems to be sort of a fight (reminding us of Jonathan Swift) among mathematicians, at least at the folklore level i.e., not necessarily in the textbook expositions. Some of the mathematicians argue that metric spaces should be given a priority, whereas others state that the concept of metric spaces is an old-fashioned framework, and what people need is normed spaces or even more - normed algebras. For instance, the Euclidean space $\mathbb{E}^n$ is more conveniently treated as a normed vector space such that its norm, namely a positive definite function $\|x\| \to \mathbb{R}$ is the square root of inner product $(x, x) \to \mathbb{R}$. Accordingly, Euclidean space is known as an example of an inner product space. Besides, one may notice that both mathematicians and especially physicists are sometimes quite loose with terminology - we shall find some examples of this looseness below.

By a domain, I shall mean a connected open subset in $\mathbb{R}^n$ or $\mathbb{C}^n$ for some positive integer $n$. Throughout this book, I shall use $x = (x^1, x^2, \dots, x^n)$ or $z = (z^1, z^2, \dots, z^n)$ for the coordinates of a point in $\mathbb{R}^n$ or $\mathbb{C}^n$, respectively. Sometimes it is notationally more convenient (for example in relativistic theories) to consider domains in $\mathbb{R}^{n-1}$ ($\mathbb{C}^{n-1}$) rather than in $\mathbb{R}^n$ ($\mathbb{C}^n$) starting coordinate sets from $x^0$ ($z^0$).

For discussing tensors, let us start from recapitulating very simple facts some of which have already been partly touched upon. Consider the transition between two coordinate systems, $X$ and $Y$: $x^1, \dots, x^n \mapsto y^1, \dots, y^n$. Here, assuming bijective mapping and in order to avoid some distracting complications, I implicitly assumed that both coordinate systems have the same dimensionality $n$. It would of course be more accurate to talk about coordinate patches comprising an $n$-manifold, where each patch having $n$-coordinates is an open domain in $\mathbb{R}^n$.

If we consider now a simple generalization of the gradient i.e., a similar operation that will be performed not over a scalar function $f(x): \mathbb{R} \to \mathbb{R}$, but over a tensor field $T^{i_1, \dots, i_r}_{j_1, \dots, j_s}(x^1, \dots, x^n)$ we shall at first assume for simplicity the point $x \in \mathbb{R}^n$ to be labeled by Cartesian coordinates $x = (x^1, \dots, x^n)$. The derivative

$$T^{i_1, \dots, i_r}_{j_1, \dots, j_s}(x^1, \dots, x^n; k) := \frac{\partial T^{i_1, \dots, i_r}_{j_1, \dots, j_s}(x^1, \dots, x^n)}{\partial x^k} \equiv \partial_k T^{i_1, \dots, i_r}_{j_1, \dots, j_s}(x^1, \dots, x^n)$$

is in general not a tensor with respect to arbitrary coordinate transformations $y^i = f^i(x^j)$ unless this is a linear (affine) transformation $\mathbf{y} = \mathbf{Ax} + \mathbf{D}$ or, in components, $y^i = a^i_j x^j + d^i$ where $\det A \neq 0$ that is the transformation is considered reversible. To make the formulas less clumsy I shall restrict myself to linear transformations $y = Ax, x = By$ i.e., $B = A^{-1}$  or  $y^i = a^i_j x^j, x^j = b^j_k y^k, a^i_j b^j_k = \delta^i_k$ , where all matrix elements $a^i_j$ and $b^i_j$ are constants. Indeed, for linear transformations, second derivatives $\frac{\partial^2 y^i}{\partial x^j \partial x^k} \equiv \partial_j \partial_k y^i = 0$ and tensor $T^{i_1, \dots, i_r}_{j_1, \dots, j_s}$ transformed to new coordinates $y^1, \dots, y^n$ (we may denote this transformed tensor as $\tilde{T}$)

$$\tilde{T}^{k_1,\ldots,k_r}_{l_1,\ldots,l_s}\left(y^1(x^1,\ldots,x^n),\ldots,y^n(x^1,\ldots,x^n)\right)$$
$$= \frac{\partial y^{k_1}}{\partial x^{i_1}}\cdots\frac{\partial y^{k_r}}{\partial x^{i_r}}\frac{\partial x^{j_1}}{\partial y^{l_1}}\cdots\frac{\partial x^{j_s}}{\partial y^{l_s}}T^{i_1,\ldots,i_r}_{j_1,\ldots,j_s}(x^1,\ldots,x^n)$$
$$= a^{k_1,\ldots,k_r}_{i_1,\ldots,i_r}b^{j_1,\ldots,j_s}_{l_1,\ldots,l_s}$$

for $y^i = a^i_j x^j, x^j = b^j_k y^k$. Note that it is a trivial generalization of the notion of contra- and covariant transformation of a vector, only the notations are very clumsy. Since all the coefficients $a^{k_1,\ldots,k_r}_{i_1,\ldots,i_r}$ and $b^{j_1,\ldots,j_s}_{l_1,\ldots,l_s}$ are constant, we have after differentiation

$$\tilde{T}^{k_1,\ldots,k_r}_{l_1,\ldots,l_s};p \coloneqq \partial_p \tilde{T}^{k_1,\ldots,k_r}_{l_1,\ldots,l_s}$$

## 3.20  Vector Fields

Speaking very crudely, in the case of vector fields, the coefficients $a'_i, a_j$ are no longer numbers but functions of $x$. The idea of a field in $\mathbb{R}^3$ being applied to vectors amounts to attaching a vector to each point of $\mathbb{R}^3$, this vector changing smoothly from point to point. The same may be said about the vector $\mathbb{R}^4$ and affine $\mathbb{A}^4$ spaces of classical mechanics. If one considers some general manifold instead of $\mathbb{R}^3$ or $\mathbb{R}^4$, the concept of vectors attached to each point may be based on tangent vectors. Now the question arises: how can we ensure smoothness while constructing a vector field on a manifold? Will it be the smoothness of a map $x \to V(x)$ where $V(x)$ represents some tangent vector at point $x$? We know that smoothness always implies some differential structure, therefore we shall try to introduce and analyze such a structure. But before we proceed with it, let us recall some elementary geometric notions.

The school definition of a vector introduces the quantity having both absolute value and direction. In a differential context, to some extent discussed above, this traditional description, where the term "quantity" is evasively indistinct, one usually defines a vector in $\mathbb{R}^n$ as a set of $n$ numbers $a^i, i = 1, \ldots, n$ that transform according to the rule

$$a'^i = \frac{\partial x'^i}{\partial x^j}a^j \tag{3.10}$$

(see section "Vector Spaces" for details). Two sets of $n$ numbers, $a'^i$ and $a^j$, actually represent the same vector, provided they are related by this formula. In more formal terms, a vector field $F(x)$ on a manifold $M$ is a smooth section of the tangent bundle $TM$, i.e., for each point $p \in M$ a choice of a tangent vector $F(x) \in T_p M$ is possible so that a map between manifolds $F \colon M \to TM$ is smooth.

## 3.21  Geometry and Physics

In this section, one can find a non-technical overview of main ideas and approaches when physicists are trying to adapt contemporary geometry for their purposes.

I have already remarked that there exist many interrelationships, sometimes striking, between many domains of physics and mathematics that are traditionally regarded as completely autonomous. Probably the most obvious example is given by countless links between geometry and physics: the two disciplines are in many fields so interpenetrating that it would be hard to tell the difference between geometry and physics, for instance, in classical mechanics, theory of dynamical systems, relativity theory, gauge models of fundamental interactions and nearly all other concepts or modern quantum field theory. In this section, one can find a brief non-technical overview of main ideas and approaches used by the physicists who are trying to adapt contemporary geometry for their purposes. The standard evidence of intimate ties between physics and geometry is provided by relativity theory. Special relativity, for example, is a purely geometric theory in which three habitual dimensions of space $r = (x, y, z) \in \mathbb{R}^3$ are combined with a single-dimensional time $t \in \mathbb{R}$ forming a four-dimensional manifold usually called a spacetime. Adding one more dimension to Euclidean space changes its basic symmetry: the symmetry group of our $3d$ Euclidean space is the Euclidean group $E(3)$ whereas the symmetry group of the Minkowski spacetime is the Poincaré group.

We shall define metric in the Minkowski space by the diagonal tensor $g_{ik} \equiv \gamma_{ik}$ with the following signature: $g_{00} = -g_{11} = -g_{22} = -g_{33} = 1$, i.e., the scalar product is $(a, b) := ab = a_0 b^0 - \mathbf{ab}, a_\alpha = -a^\alpha, \alpha = 1,2,3$.

Based on this metric, we can define the D'Alembert operator $\Box := \Delta - \partial_0 \partial^0 = -\partial_i \partial^i$, which is wave operator used in special relativity. It is used for formulating wave equations of electromagnetic fields.

## 3.22  Geometry of Classical Mechanics

A correct formulation of classical mechanics is based on the notions of geometry and symmetry, being in fact inseparable from them - we shall see it many times shortly. In particular, classical mechanics and differential geometry are close relatives, being affectively devoted to each other and even inseparable. The reader could probably get a feeling from the previous material that even nonrelativistic classical mechanics may be viewed as a specific field theory over a one-dimensional base (time). We shall return to the discussion of the differential-geometric and field aspects of classical mechanics in a number of places of this book. I have already mentioned that mathematics has traditionally served as a glue binding together a variety of physical subjects; in the specific case of classical mechanics, it is Lie groups and algebras that play a fundamental role of the main gluing substance.

Typically, geometric formulation of mechanical problems faces some conceptual difficulties, at least during the first study. These difficulties are mostly due to the counter-intuitive transformation of velocities and accelerations between generic curvilinear and non-inertial systems. More specifically, the source of embarrassment is hidden in the fact that in the Euclidean (or Cartesian, or Galilean) coordinates differentials of some vector $a^i$ also form a vector, together with the time derivatives $da^i/dt$, and the partial derivatives over coordinates $\partial_k a^i \equiv \partial a^i/\partial x^k$ form a tensor. However, one cannot uncritically spread this intuitive perception of differentiating a vector over the arbitrary curvilinear or non-inertial coordinate systems, when the metric tensor $g_{ik}$ is some (in general non-linear) function of coordinates: in this case $da^i$ is not a vector and $\partial_k a^i$ is not a tensor, because in different spatial points vectors transform differently. It means that velocity, for example, is characterized by a distinct vector in each coordinate system; moreover, all such velocity vectors will have diverse sets of components in every system.

Recall in this connection that a vector, from a geometric viewpoint, is a linear operator acting on functions (see the above discussion of vector, affine, and dual spaces). The output of the operator is the derivative of the input function in the direction defined by the vector. For instance, the tangent vector $a_q := (\mathbf{a}, \mathbf{q})$, where $\mathbf{q} \in \mathbf{U}$ is a point in an open set $U \in \mathbb{R}^n$, operates on any differentiable function $f(\mathbf{q})$ according to the formula $a_q(f) = a^i \partial_i f(\mathbf{q})$ i.e., the tangent vector $a_q$ is a differential operator. In elementary textbooks, vectors are conventionally identified with their components, which may bring some confusion. In classical mechanics, vectors such as position, velocity, or acceleration live in the tangent bundles whereas forces belong to the cotangent ones. So, there must be a natural duality between vectors and forces, and in the traditional mechanics (e.g., of constrained systems) such a duality is ensured by the principle of virtual work.

The fundamental concepts of space and time are essentially defined through affine spaces and parallel transport. The Galilean spacetime structure defining the class of inertial systems serves as a fundamental symmetry (the Galileo group) for the classical world[92]. One may, however, notice that the affine structure alone is not sufficient to fully describe the physical space. One must also provide means to measure lengths and angles. This necessity leads to introducing the Euclidean structure, i.e., the Euclidean scalar product on the vector space $V$, which is a symmetrical bilinear form, $(, ): V \times V \to \mathbb{R}$ with $(v, v) > 0$ for all $v \neq 0, v \in V$. This scalar product results in the distance between two simultaneous points $a, b$ (on fibers $V_t$)

$$\rho(a, b) = (a - b, a - b)^{\frac{1}{2}} = \|a - b\| \qquad (3.11)$$

and makes each fiber $V_t \subset \mathbb{R}^4$ a $3D$ Euclidean space (fig.3.1) or, more generally, a manifold $M$ which is not necessarily Euclidean (Galilean). Roughly speaking, the $n$-dimensional Euclidean space $\mathbb{E}^n$ is the vector space $\mathbb{R}^n$ endowed with the Euclidean metric.

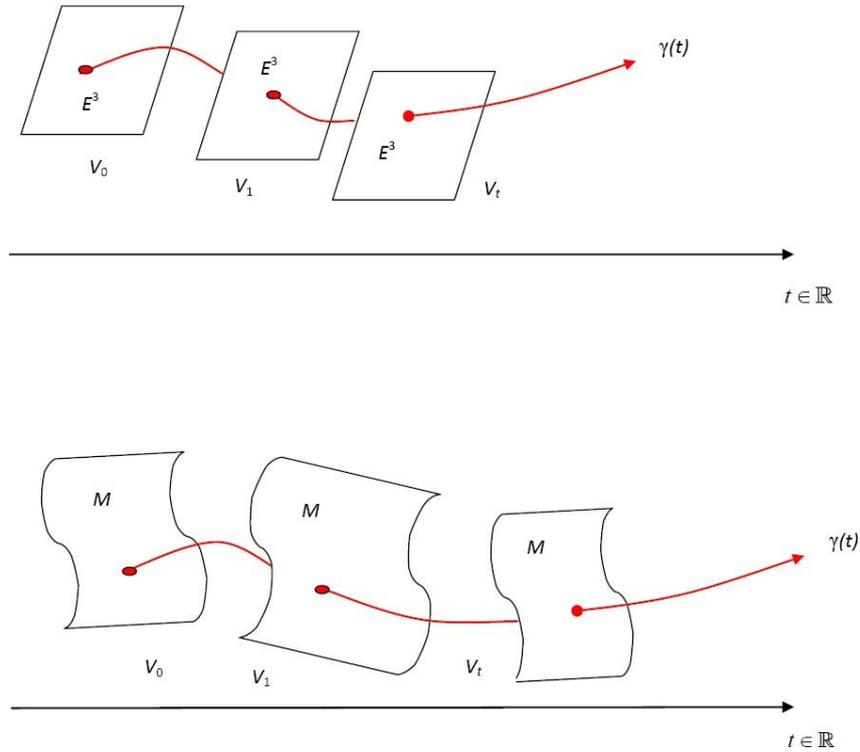

Figure 3.1.: Galilean and curved spacetime fibrations. Here $\gamma(t)$ is a path of a material point (world line) represented as a curve in spacetime $\mathbb{R}^3 \times \mathbb{R}$.

Slightly generalizing this picture, we might say that a global time mapping should exist on the spacetime, i.e., a function $t\colon M \to \mathbb{R}$ whose gradient is everywhere timelike. The existence of this mapping signifies that the spacetime can be foliated into time-ordered spacelike hypersurfaces, each corresponding to $t_i = const, i = 1, 2, \dots$. The mapping $t\colon M \to \mathbb{R}$ guarantees simultaneity of all events contained in $t$-hypersurfaces (see, e.g., [189]). Each spacelike hypersurface $V_t$ corresponding to some value $t$ of the global time splits the entire spacetime into two halves, being in the Galilean-Newtonian picture the mirror copies of one another. Initial conditions to evolutionary equations are set up on any hypersurface $V_t$. In a curved spacetime this time-reflection symmetry can be broken: there may be no spacelike hypersurfaces $V_t$ from which the spacetime observed in two opposite directions of time looks identically. This temporal asymmetry is expressed by the spacetime metric $g_{ik}$.

The Galileo group expressing the invariance properties of classical mechanics, together with the Euclidean, Lorentz and Poincaré groups, may be all considered the relativity groups. Mathematically, all these groups are particular cases of semidirect group construction, which is often used in group theory and its applications. Take as a basic example the Euclidean group $E(3)$ or $SO(3)$ acting on $\mathbb{R}^3$ - it is a semidirect product of rotations and translations. Generalizing this construction a little bit, we get some group $G$ acting on a vector space $V$ (and on its dual space $V^*$)[93]. In a nonrelativistic framework (Chapter 4), the Galileo group $G$ is the natural kinematic model. Adding independent space and time dilations, we get the affine Galileo group. It is interesting that by imposing certain constraints on these dilations we may get other invariance groups important for physics, for example, the Schrödinger group (the invariance group of the Schrödinger and heat equation) and the Poincaré group essential for relativity theory. To better understand all this terminology let us write down the generic element of the affine relativity group, $A_R$. The latter is the Galileo group $G$[94] combined with independent spatial and temporal dilations. Recall that dilations (also called dilatations) are, in the physical language, just scaling transformations whereas from the mathematical standpoint these transformations represent a specific case of the conformal group. Dilatations are usually understood as a collection of maps from a metric space into itself that scale the distances between each two points. Thus, dilatations give the result of uniform stretching or shrinking, perhaps accompanied by rotations. In school geometry we study similarity transformations, which are dilatations in Euclidean space. Dilatations comprise a subgroup of the affine group (see above), with matrix $A$ of a linear transformation is an orthogonal matrix multiplied by a scalar (dilatation ratio). So roughly speaking, a dilatation is a transformation that expands or contracts all points with respect to a given central point by some ratio, which may be greater or smaller than unity. Therefore, in order to specify a dilatation, one has to fix the central point and the ratio. A well-known physical example of dilatations is the $\gamma$-factor, which appears in relativity. It is always greater than unity but very close to it for nonrelativistic velocities. If the velocity (of a particle, for example) equals the velocity of light $c$, $\gamma$-factor is infinite and for velocities greater than $c$ it is, strictly speaking, undefined.

Let us take a look for a moment at the affine relativity group $A_R$. A generic element of $A_R$ may be denoted as $g = (\varphi, t_0, x_0, v, R, \alpha, \beta)$ where $t_0 \in \mathbb{R}, x_0 \in \mathbb{R}^n$ are time and space translations, $v \in \mathbb{R}^n$ in the boost parameter, $R \in SO(n)$ is a rotation, $\alpha$ and $\beta$ are time and space dilatations, respectively.

---

[93] $G$ may be, e.g., a Lie group or a group of automorphisms.

[94] Some authors consider the group of Galilean transformations not being the affine group since in general one cannot write $x \mapsto x' = \mathbf{A}x + b$ with matrix $\mathbf{A} \in O(n)$, but it depends on the definition of the affine group. Here, we simply consider any Euclidean group $E(n)$ to be a subgroup of an affine group $A(n)$ in $n$ dimensions.

It is interesting, by the way, that dilatations provide a link to quantum theory where there the quantity $\mathbf{r}\nabla V = x^i \partial_i V$ ($V$ is the potential) known as virial is considered. In fact, operator $x^i \partial_i$ is the generator of the dilatation group. The virial theorem in quantum mechanics states that when wave function $\psi$ is the eigenfunction of the Schrödinger operator, then the expectation value of the virial is twice the kinetic energy i.e., $(\psi, \mathbf{r}\nabla V\psi) = 2(\psi, -\Delta\psi)$.

Now, let us return to affine geometry of classical mechanics. The fundamental group for this geometry is the group of all affine transformations, i.e., the composition $u = P(\mathbf{a}) \circ v$ where $P(\mathbf{a}) := x + \mathbf{a}$ is the translation by vector $\mathbf{a}$, $V$ is the vector space over $\mathbb{R}$, $v \in GL(V)$ is the linear and bijective mapping, $GL(V)$ is, as usual, the general linear group. Simply speaking, the group of affine transformations or affine group is just a group of linear maps supplemented by translations. The transition from the Euclidean group $E(n)$ to the affine group $A(n)$ exemplifies a mathematical structure which is called a semidirect product (see some details in [187], 4). So, the Euclidean group $E(n)$ is a subgroup of affine transformations.

The affine group $A(n)$ in $n$-dimensional space works over the general linear group and is isomorphic to the group of matrices of order $n + 1$ of the type

$$L = \begin{pmatrix} \mathbf{A} & \mathbf{a} \\ 0 & 1 \end{pmatrix}$$

- a fact extensively exploited also in computer graphics (where $n = 3$ of course) [95]. To put it another way, we may specify each element of the group of affine transformations in an $n$-dimensional space in two ways: either by a pair $(\mathbf{A}, \mathbf{a})$, with $\mathbf{A}$ being an $n \times n$ orthogonal matrix and $\mathbf{a}$ being a real $n$-dimensional vector, or by a single $(n + 1) \times (n + 1)$ square matrix.

One may notice that affine groups are simple specific cases of Lie groups. An affine group on $\mathbb{R}^n$ is a combination of orthogonal (in general pseudoorthogonal) transformations and translations in the vector space $\mathbb{R}^n$.

Now we might try to discern at which point geometry enters classical mechanics. To begin with, we are often reminded of *geometric* optics in textbooks on mechanics. This allusion actually means that the ray of light selects the path between each two points, $x_1$ and $x_2$, requiring the shortest time to travel. In our everyday life we usually say that light travels in straight lines. In simple mathematical terms, this ray theory of geometric optics as well as of classical mechanics is formulated by the expression

$$\delta \int_{x_1}^{x_2} n(x)ds = 0, \tag{3.12}$$

where function $n(x)$ is in optics the refraction index, $n(x) = c/v(x)$, $ds = vdt$, $dt = dsn(x)/c$, $c$ is the velocity of light in vacuum, $v(x)$ is its velocity in the medium. An analogous expression holds also for mechanical particles, and now we shall find this analog. The above expression is an elementary path integral construction, where time does not appear explicitly - usually a very convenient presentation, a predecessor of the Feynman path integrals [44], see below "Path Integrals in Physics". It is interesting that the ideas of replacing time when considering the mechanical motion by other variables circulated already in the 18th century (Euler, Fermat, Jacobi). Excluding the time as a parameter leads directly to geometrization of mechanics. Indeed, let us consider, for simplicity, a single free particle whose kinetic energy is

$$T = \frac{1}{2} m_{ik} \frac{dx^i}{dt} \frac{dx^k}{dt}.$$

Here, the quantities $(x)$ are the components of the mass tensor. (A little later we shall see that the notion of mass is highly nontrivial and stirs a lot of controversy up to now.) In the simplest (e.g., isotropic) case $m_{ik} = m\delta_{ik}$. Introducing the length element by the usual expression $ds^2 = m_{ik}(q^j)dq^i dq^j$, where $q^i$ are the generalized coordinates in the three-dimensional coordinate space[95], we get for the kinetic energy

$$T = \frac{1}{2} \frac{ds^2}{dt^2}.$$

We may observe that the mass tensor $m_{ik}$ begins to play the role of metric tensor $g_{ik}$. This simple analogy provides a direct link from mechanics to geometry. We see that the attained geometry is non-Euclidean since the components of mass (metric) tensor are position dependent. For instance, in cylindrical coordinates we have

$$m_{ik} = diag(m, I, m) = \begin{pmatrix} m & 0 & 0 \\ 0 & I & 0 \\ 0 & 0 & m \end{pmatrix},$$

where $I := mr^2$ is the moment of inertia of a point mass, and

$$T = \frac{1}{2} m \frac{dr^2 + r^2 d\varphi^2 + dz^2}{dt^2}.$$

Mass tensor models are frequent attributes of condensed matter physics where the anisotropic response of quasiparticles (electrons, excitons, polarons, etc.) to an applied force should be described. They are an important concept in condensed matter, particularly in energy band theory. The mass tensor is also encountered in other fields of physics such as relativistic theories, nuclear physics, studies of particle motion in a magnetic field (e.g., in accelerators), theory of elasticity, soft tissue and polymer physics, mechanics of the rigid body, robotics, study of mobility and frictional effects, etc. In general, in the non-relativistic picture mass is a covariant tensor to be contracted with two contravariant velocities producing a scalar kinetic energy, $T = (1/2)m_{ik}\dot{q}^i\dot{q}^k$. The corresponding inverse mass tensor is given by $(m^{-1})^{ik}$. In general, two tensors $a_{ik}$ and $b_{ik}$ are called inverse[96] if $a_{ij}b^{jk} = \delta_i^k$. We can define the contravariant mass tensor $m^{ik}$ to be inverse to $m_{ik}$ which means that $(m^{-1})^{ik} = m^{ik}$, analogously to the metric tensor $g_{ik}$. One might observe that the mass tensor, at least in many examples, is an involution, i.e., its own inverse.

---

[95] In this simple example, the coordinate manifold is three-dimensional, in a more general case of $N$ particles it obviously has $n = 3N$ dimensions.

## 3.23  Transformation of Affine Coordinates

Let us start from the most primitive notions. Affine coordinates are defined as rectilinear coordinates in an affine space. If we have an $n$-dimensional affine space, in which we have fixed a basis, with each point $x$ in it has coordinate labels $x$ which are transformed as

$$x'^j = a_i^j x^i + b^j \tag{3.13}$$

when we move from one affine coordinate system $(K)$ to another $(K')$. The coefficients $a_i^j$ forming elements of a matrix $A$ are, in principle, arbitrary; the only condition on this matrix, $\det a_i^j = \det A \neq 0$, ensures that one can express the old coordinates $x^i$ through the new ones $x'^j$. This reversibility of the coordinate transformation $(K \leftrightarrow K')$ in fact means that the systems $K$ and $K'$ (which are arbitrary affine coordinate systems) are equivalent.

Affine coordinate unit vectors (formerly often called repère) in $K'$ can be expanded over unit vectors in $K$

$$\mathbf{e}'_j = \alpha_j^i \mathbf{e}_k \tag{3.14}$$

---

[96] Inverting a tensor or a matrix requires in general some tedious calculations usually performed on a computer. A more or less simple formula is practical only for small dimensionality such as, for example $3 \times 3$. Thus, for matrix (tensor)

$$A = \begin{pmatrix} a & b & c \\ d & e & f \\ g & h & i \end{pmatrix}$$

$$A^{-1} = \frac{1}{\det A} \begin{pmatrix} -fh + ei & ch - bi & -ce + bf \\ fg - di & -cg + ai & -db + ae \\ -eg + dh & bg - ah & -bd + ae \end{pmatrix}$$

where coefficients $\alpha_j^i$ represent a matrix that must be related to the previously encountered matrix $a_i^j$. This relationship is well-known and can be found in any textbook on linear algebra. For the sake of completeness, I shall bring here some main formulas. Vectors $\mathbf{e'}_j$ and $\mathbf{e}_k$ may be chosen arbitrary, the sole requirement being that they are linearly independent, which results in $\det \alpha_j^i \neq 0$. Due to this condition, the inverse transform

$$\mathbf{e}_i = \beta_i^k \mathbf{e'}_k \tag{3.15}$$

does exist, with matrix $\beta_i^k$ being inverse to $\alpha_j^k$, i.e., $\alpha_i^k \beta_j^i = \delta_j^k$ and $\alpha_i^k \beta_k^j = \delta_i^j$. These formulas express a simple identical transformation $\mathbf{e}_i = \beta_i^k \alpha_k^l \mathbf{e}_j$ (one may recall that the expansion on repère vectors is unique). Now let us write the components of the same vector $\mathbf{x}$ in two different coordinate systems, $K$ and $K'$:

$$\mathbf{x} = x^i \mathbf{e}_i = x'^j \mathbf{e'}_j = x^i \beta_i^k \mathbf{e'}_k = x^i \beta_i^k \delta_k^j \mathbf{e'}_j \tag{3.16}$$

Comparing these two expressions, we have

$$\begin{cases} x'^j = \beta_i^k x^i \delta_k^j = \beta_i^j x^i \\ \qquad \mathbf{e'}_j = \alpha_j^i \mathbf{e}_i \end{cases} \tag{3.17}$$

One may see that when one moves from $K$ to $K'$, the vector coordinates are transformed with the matrix $\beta_i^j$ which is the inverse transposed of $\alpha_i^j$.

This is the traditional linear algebra techniques. Later we shall see that the basic facts from linear algebra are related to the spectral properties of Hermitian matrices in finite-dimensional spaces. Here the connectivity with other physmatical domains, e.g., quantum theory, is reflected in the faith that unbounded self-adjoined operators in Hilbert spaces are characterized by similar properties.

## 3.24  General Coordinate Transformations

The above example shows some simple properties of coordinate transformations, namely the connection between matrices relating the change of basis vectors for two affine coordinate systems with the matrices governing the transformation of vector components in the transition between these different coordinate systems. The problem of a change of coordinate systems is ubiquitous in physmatics and may be considered one of the most important. We shall encounter it practically in all fields of physics: relativity theory is essentially the discipline studying the invariance properties of physical quantities under coordinate transformations; quantum mechanics is based, to a large extent, on a mathematical model of a unitary state space understood as an infinite-dimensional Hilbert space, with state vectors represented as linear combinations of basis states being transformed from one representation to another (see Chapter 6); in classical mechanics, canonical transformations (the group) preserving the measure in the phase space are a powerful tool to simplify problems by making a canonical change of variables (see Chapter 4); the Fourier transform, which is one of the main mathematical instruments of physics (see Chapters 5, 6), is nothing more than a transformation between different bases (transition to a dual basis); change of coordinates (substitution) is a favorite trick in integration, and so on. Therefore, it would be worthwhile to study some general principles of general coordinate transformations.

It is curious that this subject, even in its not mathematically refined and abstract version, seems to be traditionally difficult for the students. So, I shall try to leave no ambiguities in telling the story of coordinate transformations and especially as far as their relationship with various physical models is concerned.

## 3.25  Variational Methods

The most demonstrative example of the usefulness of variational calculus is traditionally given by the Lagrangian formulation of classical mechanics (see the next chapter). In the variational formulation of classical mechanics, the system (e.g., particle) trajectories $q(t)$ are extremizers (minimizers), or at least critical points, of the action integral with fixed endpoints

$$S = \int_{t_1}^{t_1} L(t, q, \dot{q}) dt \qquad (3.18)$$

where $L(t, q, \dot{q}) : \mathbb{R}^2 n + 1 \to \mathbb{R}$ is the Lagrangian - the difference between kinetic and potential energy. One usually assumes the Lagrangian to be smooth and strictly convex in $v \equiv \dot{q}$ , i.e., $\partial_{vv}^2 L > 0$ (physically this last condition may be wrong for the systems with negative mass). The minimizing trajectories are then the solutions to the Euler-Lagrange equations

$$\frac{\delta S}{\delta q(t)} = \frac{d}{dt} \frac{\partial L}{\partial \dot{q}_i} - \frac{\partial L}{\partial q_i} = \qquad (3.19)$$

Here we have used the symbol of functional derivative (see below).

In the classical monograph by R. Courant and D. Hilbert [7], chapter 7, one can find a very interesting and, in my opinion, quite actual discussion of the relation between the calculus of variations and the boundary value (eigenvalue) problems. This relationship has been thoroughly worked out ever since and served as a foundation for the so-called direct methods of variational calculus.

In physics, especially in quantum field theory (QFT), it is customary to introduce the notion of a so-called functional derivative. To elucidate the idea of such a derivative, let us consider the system with a single degree of freedom. By definition, the functional derivative of a quantity $S$ with respect to $q(t)$ is written as

$$\frac{\delta S[q,\dot{q}]}{\delta q(s)} = \lim_{\epsilon \to 0} \frac{S\left[q(t) + \epsilon\delta(t-s), \dot{q}(t) + \epsilon\frac{d}{dt}\delta(t-s)\right] - S[q(t), \dot{q}(t)]}{\epsilon} \quad (3.20)$$

Let us expand the quantity containing delta-functions:

$$S\left[q(t) + \epsilon\delta(t-s), \dot{q}(t) + \epsilon\frac{d}{dt}\delta(t-s)\right] =$$

$$\int_{t_1}^{t_1} dt L\left(q(t) + \epsilon\delta(t-s), \dot{q}(t) + \epsilon\frac{d}{dt}\delta(t-s)\right) =$$

$$\int_{t_1}^{t_1} dt L(q,\dot{q}) + \epsilon \int_{t_1}^{t_1} dt\left(\frac{\partial L}{\partial q}\delta(t-s) + \frac{\partial L}{\partial \dot{q}}\delta(t-s)\right) + o(\epsilon) =$$

$$S[q,\dot{q}] + \epsilon\left(\frac{\partial L}{\partial q} + \frac{d}{dt}\frac{\partial L}{\partial \dot{q}}\right)_{t=s} + o(\epsilon) \quad (3.21)$$

So the Euler-Lagrange equations may be written as

$$\frac{\delta S}{\delta q(t)} = 0 \quad (3.22)$$

(where we have put $t = s$).

## 3.26 Differential Equations

In fact, we have already encountered differential equations many times, but this subject is so important for physics that I shall try to briefly describe a few methods of obtaining solutions to differential equations, both in the two-variable domain (ordinary differential equations, ODE) and many-variable domain (partial differential equation, PDE). Differential equations may be without exaggeration called the principal tool of physics.

The term *æquatio differentiale* was probably first coined by Leibnitz to designate the relationship between variables $x, y$ and their infinitesimal increments (differentials) $dx, dy$.

In the linear case, $\dot{\mathbf{x}} = A(t)\mathbf{x}, \mathbf{x} \in \mathbb{R}^n$, matrix function $A(t): \mathbb{R}^n \to \mathbb{R}^n$ may be considered smooth, but in a number of physically interesting applications one can only require this matrix to be continuous in $t \in I \subseteq \mathbb{R}$. Physicists usually don't pay attention to such sophistry, but would it be mathematically correct? Curiously enough, yes, because the proof of the famous Cauchy-Peano (or Picard-Lindelöf) theorem stating existence and uniqueness of a solution as well as its continuity with respect to initial data of a differential equation, $\mathbf{x}(t_0) = \mathbf{x}_0$ uses only the Lipschitz condition or, at maximum, differentiability over $x$ for a fixed $t$ (see above more on this theorem and also below the section on dynamical systems) so that continuity in $t$ would be sufficient.

A second-order linear differential equation

$$y'' + py' + q = 0,  \tag{3.23}$$

where $p(x)$ and $q(x)$ is extensively used in the wave-mechanical (Schrödinger) version of quantum mechanics.

In this chapter, well-known mathematical facts and results are rendered. I have provided only a cursory coverage of geometric methods in physics since, firstly, I am in no way a specialist and, secondly, there are excellent texts on modern geometry with physical applications such as [187]. The main purpose of mathematical potpourri is to ensure a more or less firm ground to painlessly read the major part of the literature on current physics, in particular where geometric methods are used. I was hoping that this section will at least help to unite physical and mathematical communities overcoming corporate arrogances and isolationism.

# 4 Classical Deterministic Systems

Deterministic systems are those for which the full dynamic description can be achieved without introducing the concept of probability. Ideally, in deterministic systems the state of the system at an initial moment $t_0$ uniquely determines the future behavior of the system, i.e., for any $t > 0$. The study of continuous deterministic systems may be reduced to mathematical models based on differential equations for the functions giving the state of a system. The first - and the most famous - model of this type was the one of Newton and is usually called Newtonian mechanics. It studies the motion of a system of material points in $\mathbb{R}^3$ and may be regarded as a special case of classical dynamics.

Classical dynamics is probably the most developed part of science, it studies the evolution of systems made of material points - bodies that are so small that their inner structure is disregarded and the only characteristic is their position in space, $\mathbf{r}_i = \mathbf{r}_i(t), i = 1,2, \dots, N$ where $N$ is the number of points incorporated into the system. Typically, in classical mechanics the so-called generalized coordinates $q^i(t)$ are used, which correspond to the degrees of freedom of a mechanical system. Generalized coordinates have been introduced by Lagrange and are especially useful in Lagrangian mechanics (see below). In modern terminology using generalized coordinates means that one should consider possible configurations of a mechanical system as points on a differentiable manifold. Naturally, one can also use any system of local coordinates which are convenient from a practical point of view. We have already observed in Chapter 3 the general change of coordinates; below we shall see in some detail how to pass from one system of curvilinear coordinates to another. In modern terminology this means that one should consider possible configurations of a mechanical system as points in a differentiable manifold, and, of course, then use any system of local coordinates which are continuously differentiable.

Here, one may notice that the number of degrees of freedom is defined in classical mechanics as dimensionality of the configuration space of a physical system. For example, a system with two degrees of freedom is $\ddot{\mathbf{r}} = \mathbf{F}(\mathbf{r}, t)$ where $\mathbf{F}$ is a plane vector field, $r \in \mathbb{R}^2$. If we convert this equation into the dynamical system form, we shall get four first-order equations and, respectively, four degrees of freedom as the dimensionality of the vector system of differential equations (defining a vector field) corresponding to a dynamical system. In other words, there may be a certain confusion in counting the number of degrees of freedom when one passes from conventional classical mechanics to the dynamical systems theory. The set of possible paths $q^i(t)$ in the configuration space of generalized coordinates completely describes the system. Of course, these paths should be single-valued functions, usually with a compact support, i.e., defined on a compact time interval $[t_1, t_2]$. Time $t$ is clearly a very important variable; in the static models the role of dynamics - not only evolution – it has been excluded. One can of course take the dynamics into account employing a parameter which changes with time (such as the growing size of some object). This approach is used, for example, in self-similar models.

## 4.1   Main models of classical mechanics

The primary feature of models based on classical deterministic (dynamical) systems is causality, i.e., in such models the effect cannot precede the cause and the response cannot appear before the input signal is applied. One may note that causality does not follow from any deeply underlying equation or a theory, it is simply a postulate, a result of human experience. Non-causal systems would allow us to get the signals from the future or to influence the past.

Causality is closely connected with time-reversal non-invariance (the arrow of time). The time-invariance requires that direct and time-reversed processes should be identical and have equal probabilities. Most mathematical models corresponding to real-life processes are time non-invariant (in distinction to mechanical models). There is a wide-spread belief that all real processes in nature, in the final analysis, should not violate time-reversal invariance, but this presumption seems to be wrong (see Chapter 9).

Now, let me say a few words about bibliography. There exist tons of books on classical mechanics, but most of them must be updated by adding a number of modern subjects such as nonlinear phenomena, dynamical systems and differential geometric methods. The first book on classical dynamics that I studied was a comparatively thin book by F. R. Gantmacher [21], the Russian (Soviet) mathematician well-known for his monumental volume "The Matrix Theory" [22]. Gantmacher's book on analytical dynamics was simple and rigorous, so one could easily trace all the conventional transformations and derivations comprising the major part of traditional analytical dynamics (e.g., transition to generalized coordinates, canonical transformations, etc.). Difficult topics were treated by the author in a comprehensive manner. After the nice Gantmacher's "Lectures", the obligatory course of L. D. Landau and E. M. Lifshitz [23] at first seemed too lapidary and full of declarative prescriptions. Only much later did I realize that the whole classical multi-volume course by Landau and Lifshitz was intended not for students, but rather for professionals working specifically in the field of theoretical (not mathematical!) physics. Classic books are invariably valuable, but they are not always the best ones for a concrete person. Despite some strange but luckily rare mistakes, it is an undeniable fact that every physicist (at least in Russia) has learned a lot from these books and specifically from "Mechanics". Now I think that as far as classical dynamics goes, the book by V. I. Arnold [14] is fully sufficient to provide the basic knowledge. Despite being over thirty years old, this is the kind of book belonging to the narrow class I appreciate most of all: one that contains the information you need. The meticulous text of Arnold covers almost every subject of classical mechanics (and some extra), starting from elementary Newtonian mechanics to semi-classics (short-wave asymptotics) and perturbation theory. In my opinion, this book, like some volumes of the course by Landau and Lifshitz, may be considered an example of connected sciences approach.

Physicists often treat classical mechanics with an element of disdain or at least as something alien to "real" physics, say condensed matter physics. Personally, I think that classical mechanics is extremely useful for any part of physics, for many great ideas are rooted in classical mechanics. Examples of Gibbs and Dirac who persistently looked for opportunities to transfer methods of classical mechanics to other fields are very persuasive. One may notice that the ideas from dynamical systems theory have been present in mechanics from the time of Lyapunov and Poincaré, but could not overcome the barrier between classical mechanics and physics until the 1970s when nonlinear dynamics all of a sudden became of fashion.

Why is classical mechanics so important? The matter is that classical mechanics has been the primary collection of mathematical models about the world, mostly about the motion of its objects. The principal aim of classical mechanics is to describe and explain this motion, especially under the influence of external forces. Ancient thinkers were enchanted by the celestial clockwork and devised a number of interesting models, the most well-known among them was that of Aristoteles (Aristotle), which can be formulated as the statement that the motion of bodies is possible only in the presence of external forces produced by other bodies. This verbal construct of Aristoteles can be translated into the mathematical language as the first-order differential equation, with the state of a moving body (or particle) being described by three coordinates $(x, y, z)$ that change under the influence of an external force

$$\frac{d\mathbf{r}}{dt} = \mathbf{f}(\mathbf{r}), \mathbf{r} = (x, y, z), \mathbf{f} = \left(f_x, f_y, f_z\right) \qquad (4.1)$$

This is really the simplest model of the motion. One can immediately see that the crucial difference of Aristotle's model based on the first-order vector equation with the second-order system corresponding to Newton's model is, primarily, in the irreversible character of the motion.

Aristotle's considerations were rooted in everyday experience: the trolley should be towed to be in motion; if the muscle force stops acting, the trolley comes to rest. In such a model the state of a system would be given by positions alone - velocities could not be assigned freely. One might observe in this connection that even the most fundamental models reflecting everyday reality are not unique and often controversial.

The contrast of Aristotle's model to that of Newton is readily seen when one starts thinking about acceleration (as probably Einstein did). In Newton's model of single-particle dynamics, the general problem is to solve the equation $\mathbf{F} = m\mathbf{a}$, where

$$\mathbf{a} := \frac{d^2\mathbf{r}}{dt^2}, \mathbf{r} := (x, y, z), \mathbf{r} \in \mathbb{R}^3$$

when the force $\mathbf{F}$ is given.

One might, however, ask: was Aristotle always wrong? The trolley stops due to friction: $\mathbf{F}_R = -\alpha\mathbf{v}$. The Newton's equations for this case may be written in the form

$$\frac{d\mathbf{r}}{dt} = \mathbf{v}, m\frac{d\mathbf{v}}{dt} = \mathbf{F} - \alpha\mathbf{v}, \tag{4.2}$$

where $\mathbf{F}$ is the towing force and $m$ is the mass of the trolley. When $\mathbf{F}$ is nearly constant, i.e. the towing force, as it is frequently the case, slowly varies with time, the mechanical system is close to equilibrium, so the inertial term $m\frac{d\mathbf{v}}{dt}$ is small compared to other terms. In this case, we get the equilibrium (stationary) solution $\mathbf{v} = \frac{d\mathbf{r}}{dt} = \frac{\mathbf{F}}{\alpha}$, which has Aristotle's form. This solution form is valid only when the motion is almost uniform, i.e., the acceleration is negligeable, and the friction is sufficiently large, $\left|\frac{d\mathbf{v}}{dt}\right| \ll |\alpha\mathbf{v}|$. One can, in principle, imagine the world constructed on Aristotle's principles: the Aristotle model would correspond to the Universe immersed in an infinite fluid with a low Reynolds number.

It is curious that Aristotle's model is in fact extensively used in contemporary physics, engineering, and even in everyday life. An example is Ohm's law, $\mathbf{j} = \sigma\mathbf{E}, \mathbf{v} = \mathbf{E}/ne\rho$, where $e$ is the electron charge, $\mathbf{E}$ is the electric field (acting force), $n$ is the charge density, and $\mathbf{v}$ is the average velocity of the charge carriers. Ohm's law is a typical macroscopic stationary model, when the driving force is compensated by resistance. Stationary models are typical of classical physics: in fact, classical physics dealt only with slow and smooth motions, e.g., planetary movement. Models describing rapid and irreversible changes, resulting in multiple new states, appeared only in the 20th century. Stationary and quasi-stationary models serve as a foundation of thermodynamics (Chapter 7) whose principal notion - temperature - may be correctly defined only for equilibrium. Another example of such models is steady-state traffic flow simulation.

## 4.2    Newtonian Mechanics

Let us now get back to the Newton's model. Although the stuff below looks trivial, there are several reasons to discuss it once more in order to better understand why, despite the fact that Newton's law of motion may be considered a complete statement of classical mechanics, one would desire other mathematical formulations for this discipline. As is well known, the general problem in Newtonian mechanics is to solve the system of equations

$$m_a\frac{d^2\mathbf{r}_a}{dt^2} = \sum_i \mathbf{F}_i(\mathbf{r}_a, \dot{\mathbf{r}}_a, t),$$

where index $a = 1, \ldots, N$ enumerates the particles and the sum goes over all the forces acting on the $a$-th particle. Forces $\mathbf{F}_i$ are regarded as given functions of $\mathbf{r}, \dot{\mathbf{r}}, t$ in every point $\mathbf{r}_a, \dot{\mathbf{r}}_a, t$. A mathematical solution of this system of second-order differential equations is a $3N$ vector-function $\mathbf{r}_a(t)$, $a = 1, \ldots, N$. If one considers the simplest case of the motion of a point mass (material particle) in $3D$ space, the position vector $\mathbf{r}$ may be expressed as $\mathbf{r} = x\mathbf{i} + y\mathbf{j} + z\mathbf{k}$, where $\{x, y, z\}$ are projections that are varying when the particle moves, three quantities $\{\mathbf{i}, \mathbf{j}, \mathbf{k}\}$ (also denoted $\{\mathbf{e}_1, \mathbf{e}_2, \mathbf{e}_3\}$) are unit vectors which are assumed constant in the laboratory system.

Consider a particle in a potential force field, namely the one where force $\mathbf{F}$ is the (negative) gradient of some potential energy function $V$:

$$m\ddot{\mathbf{r}} = -\nabla V(\mathbf{r}) \tag{4.3}$$

where $V: \mathbb{R}^n \to \mathbb{R}$ is some function on $\mathbb{R}^n$. The total - kinetic plus potential - energy is given by the expression $E = \frac{m}{2}|\dot{\mathbf{r}}|^2 + V(\mathbf{r})$ and is easily verified to be constant on any solution. To show it, one can just differentiate $E(t)$ and use Newton's equation:

$$\frac{dE}{dt} = m\frac{d\mathbf{r}}{dt}\frac{d^2\mathbf{r}}{dt^2} + \frac{d\mathbf{r}}{dt}\nabla V(\mathbf{r}) = 0 \tag{4.4}$$

In principle, dimensionality $n$, $\mathbf{r} = (x^1, \ldots, x^n)$ may be arbitrary. For the sake of simplicity and to make the treatment more intuitive, let us confine ourselves to $n = 3$. Assume that the potential $V(\mathbf{r})$ is spherically symmetric, i.e., that $V$ depends only on the distance from the origin - this is usually called the central field case. We may put e.g., $V(r) = (1/2)v(|r|^2)$, then the equation of motion becomes

$$m\ddot{x}^i = -v'(|r|^2)x^i \tag{4.5}$$

The energy $E$ is still a conserved quantity. The term "conserved" here means that the function $E(x^i, \dot{x}^i, t), t \in \mathbb{R}$ remains constant on any solution of the equation of motion.

## 4.3    Lagrangian Mechanics

Although Newton's formulation of classical mechanics has been very successful, it still has some drawbacks. First of all, it requires that all forces acting on a mechanical system should be explicitly known. In practice, however, forces are often given implicitly, for example, by constraining the trajectory to belong to some manifold such as to lie on a surface. Newton's law is a vector equation, and it may be cumbersome to transform its components into curvilinear coordinate systems. Furthermore, the analysis of symmetries is not easy to perform as long as we stay within the framework of Newtonian mechanics. Constraints are also difficult to take into account. The Newtonian model is essentially a local one: it was designed to describe the motion in terms of the local velocity change caused by the local force value. Thus, it is difficult to gain an insight into the global features of motion, its mathematical structure. Mathematically speaking, it is reflected in the fact that Newton's equation is a second-order one, so the global properties of the system cannot be figured out as easily as for the first-order equations, for which an extensive dynamical systems theory has been developed. Furthermore, Newton's model is difficult to apply to fields - it is basically suitable for classical systems which consist of particles. Because of this, Newtonian mechanics cannot serve as a starting point for quantization of distributed systems. Thus, Newton's equations, despite their ubiquity, are seldom used in modern condensed matter physics. And in general, the quantization of a classical system can hardly be based on Newton's formulation of mechanics.

In contrast to Newtonian mechanics, the Lagrangian mathematical model provides a global and, in principle, coordinate-free formulation of classical (i.e., trajectory-based) motion. Let us first give some elementary considerations. Imagine a system of many particles having coordinates $q_i(t)$ and velocities $\dot{q}_i(t)$ (here, for simplicity, we make no distinction between co- and contravariant coordinates), although the usual summation convention omitting the sum symbol will be used. The quantities $q_i(t), \dot{q}_i(t)$ labeling the particles' positions and velocities are not necessarily Cartesian (rectangular) coordinates. For $N$ particles, $i$ runs from 1 to $3N$, so the motion of these particles in $\mathbb{R}^3$ may be represented by a trajectory $\gamma: [t_1, t_2]$ in $\mathbb{R}^{3N}$ usually called the configuration space. This trajectory starts from the point in the configuration space corresponding to the initial time $t_1$ and terminates at the point corresponding to the final time $t_2$. The main idea of Lagrangian mechanics is to characterize the actual motion of a mechanical system by a single trajectory, against all others joining the initial and final points in the configuration space, which extremizes some functional $S$ called action. The latter is usually written as

$$S = \int\limits_{t_1}^{t_2} L(q, \dot{q}, t) dt \qquad (4.6)$$

We already know that the function $L$ is called the Lagrangian. Mathematically speaking, a Lagrangian is defined as a smooth function $L: \mathbb{R} \times TM \to \mathbb{R}$ on a manifold $M$. Nowadays people say that $L$ is a function on the tangent bundle $TM$ of $M$. The action $S = S_L$ associated with $L$ is defined for any smooth curve $\gamma: [t_1, t_2] \to M$. The set of all such smooth curves may be thought of as an infinite-dimensional manifold, with the action to be a function on it. Simply speaking, the classical action is the integral of the Lagrangian along the trajectory. We allow the Lagrangian $L$ to depend on time $t$, i.e., $L$ is a function on $L: \mathbb{R} \times TM \to \mathbb{R}$, although often $L$ is only required to be defined on some open set in $TM$. Also, for some models it might be important to weaken the smoothness requirement for $\gamma$ and $L$.

In other words, one is usually interested in determining the curves $\gamma: [t_1, t_2] \to M$ with given endpoint conditions, for which the action functional reaches a minimum among nearly lying curves. Such a set of curves are typically called variations. Given a curve $\gamma$ with fixed endpoints $t_1, t_2$ and some $\varepsilon > 0$, one can define a smooth variation of it as a smooth map

$$\gamma_\varepsilon(t, \sigma) := \gamma_\varepsilon[t_1, t_2] \times (-\varepsilon, \varepsilon) \to M, \tag{4.7}$$

$\sigma \in (-\varepsilon, \varepsilon)$ with the properties $\gamma_\varepsilon(t, \sigma) = \gamma(t)$ for all $t \in [t_1, t_2]$ and $\gamma_\varepsilon(t_1, \sigma) = \gamma(t_1), \gamma_\varepsilon(t_2, \sigma) = \gamma(t_2)$ for all $\sigma \in (-\varepsilon, \varepsilon)$. In standard texts on Lagrangian mechanics, this variation is typically denoted as $\delta q(t)$.

Now, it is easy to produce once again the Euler-Lagrange equations, but before that I would like to note that the Lagrangian $L$ depends on two independent sets of $3N$ variables denoted $q_i(t)$ and $\dot{q}_i(t)$, but the latter are not necessarily the $t$-derivatives of the former. One should not be confused by the fact that in these coordinates an arbitrary curve in the tangent bundle has to be denoted as $\big(q_i(t), \dot{q}_i(t)\big)$. Geometrically, the second set of variables coincides with the time derivative of the first only when the curve in $TU$ is the so-called tangential lift of a curve in $U$ where $U \subset M$ is e.g., an open set, on which we may establish a coordinate chart $x: U \to \mathbb{R}^n$ (here $n = 3N$).

From the physicist's perspective, only the elementary methods of the calculus of variations are needed to derive the Euler-Lagrange equations. The latter are produced by the elementary integration by parts. Assume $q_i(t)$ to be a path for which the action $S$ is stationary, i.e., under the path variation $q_i(t) \to q_i(t) + \delta q_i(t)$ where $\delta q_i(t)$ is an arbitrary smooth function of $t \in [t_1, t_2]$ vanishing at the endpoints of the curve (for simplicity, we are using here the customary physical notations), the variation $\delta S$ of $S$ must be zero in the first order in $\delta q_i(t)$. This means

$$S = \int\limits_{t_1}^{t_2} dt \left( \frac{\partial L}{\partial q_i} \delta q_i + \frac{\partial L}{\partial \dot{q}_i} \delta \dot{q}_i \right) = \int\limits_{t_1}^{t_2} dt \left( \frac{\partial L}{\partial q_i} - \frac{d}{dt} \frac{\partial L}{\partial \dot{q}_i} \right) \delta q_i + \left[ \frac{\partial L}{\partial \dot{q}_i} \delta q_i \right]_{t_1}^{t_2}$$
$$= 0 \tag{4.8}$$

where integration by parts has been carried out. At the endpoints of the path $\delta q_i = 0$, and we get the Euler-Lagrange equation in its elementary form for the classical extremal trajectory

$$\frac{d}{dt} \frac{\partial L}{\partial \dot{q}_i} - \frac{\partial L}{\partial q_i} = 0 \tag{4.9}$$

It is in general the system of second-order ODEs and according to the Cauchy-Peano theorem from the theory of differential equations, this system should have a unique solution, provided the initial values of $q_i, \dot{q}_i$ are given.

How can we establish connection with the Newtonian model? Let us take, for simplicity, a single particle:

$$L(q_i, \dot{q}_i, t) \equiv L(x^i, \dot{x}^i) = \frac{m}{2} \dot{x}^i \dot{x}^i - V(x^i) \tag{4.10}$$

Then the Euler-Lagrange equation takes the form of Newton's law

$$\frac{d}{dt}(m\dot{x}^i) = -\frac{\partial V}{\partial x^i} \tag{4.11}$$

This simple example in fact demonstrates the generality of the Lagrangian approach. Taking other Lagrangians, one can produce a set of mathematical case studies. For instance, one may consider a Riemannian metric $L = g_{ij}(x^k)\dot{x}^i \dot{x}^j$ or $L$ may be a linear form on each tangent space, $L = a_i(x^k)\dot{x}^i$. The system under consideration is nearly always specified in physics by choosing an appropriate model Lagrangian. As we shall see in the chapter devoted to quantum field theory, the same methods can be applied to physical systems with an infinite number of degrees of freedom which are not constituted of moving particles, such as fields, by simply stipulating a suitable Lagrangian function.

We have mentioned that it is not so easy to take into account symmetries in the context of the Newtonian mechanics. Let us try to consider symmetries in the Lagrangian formalism (one can consult the book "Mechanics" by L. D. Landau and E. M. Lifshitz for an elegant treatment of this subject, [23]). Assume that the Lagrangian $L$ does not depend on a certain coordinate $q_j$ (such a coordinate is historically called cyclic). Naturally, $L$ may depend on $\dot{q}_j$, otherwise the variable $q_j$ is of no interest at all. Then one can immediately see from the Euler-Lagrange equations that the generalized momentum $p_j$ conjugate to a cyclic coordinate, $\frac{\partial L}{\partial q_j}$, is conserved:

$$\frac{dp_j}{dt} = \frac{d}{dt}\frac{\partial L}{\partial \dot{q}_j} = \frac{\partial L}{\partial q_j} = 0 \tag{4.12}$$

It is from here, by some simple generalization in the spirit of Lie group theory, that one of the two famous Noether's theorems can be obtained [14]. Assume that the Lagrangian has a symmetry that may be continuously parameterized, i.e., the action $S$ is invariant under an infinitesimal symmetry operation applied to the path $q_j(t), q_j(t) \rightarrow q_j(t) + \delta q_j(t)$. It is important that the symmetry is continuous: in this case it is always possible to define an infinitesimal symmetry operation (infinitesimal displacement leaving the action invariant). Another way to put it is to notice that discrete groups do not in general result in conservation laws. The reason to this fact is that discrete symmetries cannot in general imply infinitesimal variations. Since the action $S$ is invariant under the displacement $\delta q_j(t)$, we have for an arbitrary number of degrees of freedom

$$S = \int\limits_{t_1}^{t_2} \sum_j \delta q_j \left( \frac{\partial L}{\partial q_j} - \frac{d}{dt} \frac{\partial L}{\partial \dot{q}_j} \right) + \sum_j \left[ \delta q_j \frac{\partial L}{\partial \dot{q}_j} \right]_{t_1}^{t_2} = 0 \qquad (4.13)$$

The first term here vanishes since all $q_j(t)$ are the solutions of the Euler-Lagrange equations. Thus, we get

$$\sum_j \delta q_j(t_1) p_j(t_1) = \sum_j \delta q_j(t_2) p_j(t_2) \qquad (4.14)$$

where $t_1$ and $t_2$ are arbitrary initial and final time-points on the trajectory $q_j(t), t \in [t_1, t_2]$. It is clear that $\delta q_j(t_1)$ and $\delta q_j(t_2)$ do not in general vanish. Since $t_1$ and $t_2$ are arbitrary, the above equation may be interpreted as the conservation of the quantity

$$\delta S = \sum_j \delta q_j(t_1) p_j(t_1), \qquad (4.15)$$

i.e., independence of this quantity on $t$ (see the textbook "Mechanics" by L. D. Landau and E. M. Lifshitz, [23], §43 "Action as a function of coordinates", for a discussion and examples). A connection of this invariant with the customary conservation laws may be illustrated by a classical particle moving in a central potential $V(r)$. The Lagrangian written in spherical coordinates $(r, \theta, \varphi)$ is

$$L = \frac{m}{2} \left[ \dot{r}^2 + r^2 \left( \dot{\theta}^2 + \dot{\varphi}^2 \sin^2\theta \right) \right] - V(r) \qquad (4.16)$$

Since $\varphi$ is here cyclic, the invariant $(\partial L / \partial \dot{\varphi}) \delta \varphi = const$ results in the conservation law:

$$m(r\sin\theta)^2 \dot{\varphi} = const, \qquad (4.17)$$

which is the conserved component of the angular momentum about $z$-axis.

Let me now, as usual, make a few comments on the Lagrangian framework and its general validity in physics. The famous "Course of Theoretical Physics" by L. D. Landau and E. M. Lifshitz is based on the assumption that the whole physics may be derived from the least action principle[97]. Action and action density have long been the well-established concepts in classical mechanics and field theories. However, in fluid dynamics the corresponding quantities have been introduced only comparatively recently. One of the reasons for this delay is probably owing to the fact that fluid dynamics had traditionally been developed in the Eulerian coordinates whereas the action principle is more conveniently written and treated in the Lagrangian frame.

---

[97] An exception is "Fluid Mechanics" volume [85], see below.

The Lagrangian framework has many other advantages and spreads well beyond classical mechanics. The generalized coordinates $q^i(t)$ do not necessarily describe the motion of a material point or a set of material points. For instance, in scalar field theory (we shall study some models based on it later) each $q^i(t)$ may correspond to the field value regarded as a function of time at various field points. Certain versions of quantum mechanics (e.g., a novel sum-over-histories formulations) also use the Lagrangian framework and generalized coordinates. Moreover, such a description is not restricted to nonrelativistic physics only. Thus, physical systems in quantum field theory are customarily specified by their Lagrangians. The Lagrangian approach provides a global description of the trajectory in a configuration space as well as a fair understanding of the conservation laws. One trivial example of a conservation law is obvious already from the form of the Euler-Lagrange equations. If the function $L$ does not depend on a particular variable $q_i$, then the quantity $\frac{\partial L}{\partial \dot{q}_i} \equiv p_i$ is a constant of the motion - a conserved quantity. In this case, it is the conserved momentum conjugate to the cyclic coordinate $q_i$. In general, there exists a conserved quantity corresponding to each generator of the Lie algebra of a Lie group of continuous symmetries. Moreover, there exists a close relationship between Lagrangian mechanics, which is typically based on positive-definite symmetrical bilinear forms, and Riemannian geometry. One can even say that Lagrangian mechanics is a subdomain of Riemannian geometry.

## 4.4   Hamiltonian Mechanics

There are some practical inconveniences in Lagrangian mechanics. First of all, the Euler-Lagrange equations, like those of Newton, are second-order ODEs. This fact makes it more difficult to provide a simple geometric interpretation for their solutions than for a system of the first-order differential equations, which readily admit geometric interpretation as flows along vector fields. Possibly the main motivation underlying the Hamiltonian formulation of mechanics was the intention to interpret the time evolution of a physical system as such flows. This nice interpretation is especially convenient when studying the properties of dynamical systems (see below). Moreover, in the Hamiltonian formulation there is practically no difference between coordinates and momenta - they are just variables in a system of an even number $(2n)$ of equations, with the possibility to treat all of them equally, make changes and transformations between them and construct some kind of geometry in this $2n$-dimensional phase space. Such a geometry is usually formulated in terms of symplectic manifolds, with a group of diffeomorphisms acting on them. However, before we proceed to using this geometrical language, I shall make some general observations and try to illustrate the main ideas on simple examples.

One more important motivation for some other description, different from the Lagrangian formulation, consists in the desire to get explicit expressions for the right-hand side of the dynamical system vector equation, $dx/dt = F(x,t)$ where in the Lagrangian formulation variables are $x =$ $(q^i, \dot{q}^i)$, so the first half of equations in the dynamical system is simply $dq^i/dt = \dot{q}^i$, but there is no regular method to extract the right-hand side for $d\dot{q}^i/dt$ from the Euler-Lagrange equations. This is related to the geometric properties of the tangent bundle $TM$, which is too simple to support dynamics, but luckily not the only one. The tangent bundle $TM$ is formed by tangent spaces (fibers) attached to each point of the configuration manifold $M$, which contain the velocity vector fields of the same contravariant type as vectors $q^i \in M$. One may note in passing that the Hamiltonian equations are covariant whereas the Lagrangian ones are contravariant. The tangent bundle $TM$ is insufficient to form the phase space of a physical system, one needs also a dual space (see below). We can interpret the Hamiltonian equations as a dynamical system on a cotangent space to configuration manifold $M$.

There exist of course many important books on classical mechanics where the geometry of phase space is thoroughly discussed. I have already remarked that for me, personally, the best one is still the book by Arnold [14], in spite of the fact that it has been written over thirty years ago. Yet, I have heard several times from the students an astounding declaration that the book by V. I. Arnold was highly unorthodox and was viewed a bit skeptically by their professors. (A similar statement was related also to another book by V. I. Arnold [15] which personally I found quite fresh when I first read it in the beginning of the 1970s.)

Before we proceed, one may notice that the Hamiltonian systems are only a very restricted class of dynamical systems, since they are specified by a single scalar function - the Hamiltonian. The description of an autonomous (time-independent) Hamiltonian system possessing $N$ degrees of freedom, i.e., evolving in a $2N$-dimensional phase space[98], is reduced, due to the energy integral of motion, to ($2N$-1) variables - one can say that the Hamiltonian dynamical system is restricted to ($2N$-1)-dimensional energy shell (manifold).

Physicists traditionally prefer the coordinate-dependent notation, whereas in contemporary mathematics the coordinate-free geometric language is nowadays of fashion. For me, it was useful to compile sort of a dictionary translating the notions of Hamiltonian (and, to some extent, Lagrangian) mechanics into the language of modern geometry. In fact, this language is not that modern - it has been used in mathematics for about a century. As usual, I avoid hard technical definitions.

The particle coordinates $x^i, i = 1, \ldots, n$ on the configuration space $\mathbb{R}^{3n}$, $n$ is the number of particles in the system, are usually replaced by "generalized" coordinates $q^i$ which are understood as global variables, but more generally they may be interpreted as local coordinates on a configuration space $M$, which can be any manifold whose points coincide with all possible configurations of the system. The variables $\dot{q}^i$ represent all tangent vectors to all possible curves (paths) lying in $M$. These variables, viewed as "generalized" velocities, are usually called by mathematicians "fiber coordinates on the tangent bundle $T(M)$", the fiber itself is sometimes denoted as $T_x(M)$. Tangent bundles and fibers are often denoted with no brackets, i.e., as $TM$ and $T_xM$, respectively (see more details in Chapter 3 on mathematics).

---

[98] Recall that variables in Hamiltonian systems appear as canonically conjugated pairs, which results in an even dimension of the phase space. This is not true for the case of generic dynamical systems.

The paradigmatic examples of a vector bundle are the tangent and cotangent bundles $TM$ and $T^*M$ of a smooth manifold $M$. These are the bundles whose fibers at $x$ in $M$ are the tangent space $T_xM$ and cotangent space $T_x^*M$ respectively. In this case the bundle charts and transition maps are induced by the derivatives of the coordinate charts on $M$ (or their adjoints, for the cotangent bundle).

The momenta $p_i$ in this language are called fiber coordinates for the cotangent bundle $T^*(M)$, which is in fact the phase space. In Lagrangian mechanics, the momenta are defined as $p_i = \partial L / \partial \dot{q}^i$ which implies that $p_i$ and $q^i$ would be transformed with mutually inverse matrices under a coordinate change, similarly to co- and contravariant vectors in linear algebra. These inverse matrices correspond to dual bundles. The concept of the cotangent bundle, or simpler the cotangent space, is constructed by using the base notion of a manifold and its associated tangent space (see Chapter 3 on vector spaces and other geometrical definitions). Let us now recall some basic facts.

A manifold $M$ may be regarded as a direct generalization of the concept of a surface. To put it simply, a manifold can be treated as an $n$-dimensional "curved" space where one desires to define vectors as in usual Euclidean space. In principle, there are several ways to introduce vectors, one of them being with the help of a tangent space. At each point $x$ of a manifold $M$ (we assume it to be finite-dimensional, in the infinite-dimensional case one may encounter some complications, see below) we can define the tangent space of vectors, $T_x(M)$, for example, as the set of all vectors tangent at point $x$ to smooth curves passing through $x$. Such a construction produces the space of all vectors defined at $x$ with the same dimension $n$ as $M$, which is also a manifold. By taking the union over all points $x \in M$, we may define a tangent bundle of $M$

$$T(M) = \bigcup_{x \in M} \{x\} \times T_x(M)$$

Later we shall often use these (and a little extended) geometric concepts; now, before we proceed with simple analysis and examples more commonly used in traditional physics textbooks, e.g., in [23], allow me to make a small digression.

In general, given any $n$-dimensional vector space $V$, we might look at the space of all real-valued linear maps on $V$. This space of linear maps forms itself a vector space, which we can call, under certain conditions of orthogonality, the dual space $V^*$. This may appear, at the first sight, an extremely abstract space, but in fact it is not drastically different from the initial space $V$. If, for example, the space $V$ is formed by the contravariant vectors like radius-vector or velocity, then $V^*$ contains covariant vectors such as gradient, wave vector or quantum-mechanical momentum. One can, by the way, rather abstractly define a covariant vector as a linear functional or, rather, a polylinear operator mapping contravariant vectors into scalars, yet I don't think this level of mathematical abstraction would bring new understanding.

One may ask: what was the main idea behind the introduction of tangent and cotangent spaces? One idea probably was to bypass the traditional coordinate representation of classical differential geometry. In fact, choosing a basis is an arbitrary operation: once you choose a basis in a vector space, it becomes isomorphic to $\mathbb{R}^n$. Then, unless you take some precautions, you risk establishing an isomorphism between tangent and cotangent spaces, because $\mathbb{R}^n$ is isomorphic to its dual. Moreover, you may be tempted to introduce a usual inner (scalar) or canonical dot product via coordinate entries as in $\mathbb{R}^n$, but this object does not exist in general in arbitrary vector spaces.

When we discussed differential manifolds in Chapter 3, we observed that for them a similar problem exists: although one may introduce local coordinates, there is no regular way to select them. On a sphere, for example, which is one of the simplest realizations of a manifold, one cannot find a natural place for the coordinate origin. Thus, intrinsic properties of manifolds must be studied without any special choice of local coordinates. The latter cannot be found by physics - they do not occur in nature and we introduce them only for convenience. One of the reasons some people identify tangent and cotangent (dual) spaces e.g., in the form of contravariant and covariant objects may be that they implicitly assume the existence of local coordinates and bases isomorphic to those in $\mathbb{R}^n$.

Let us repeat the main points once again. The concept of a cotangent space is built upon the already introduced structures of a manifold and an associated tangent space. In a manifold $M$, which actually is an $n$-dimensional space, we would like to define "vectors" as we usually do in the ordinary Euclidean space. There are different manners to do it, and one of the simplest ways is to introduce the tangent space of vectors at a point $x$ in $M$, denoted $T_x(M)$, which is essentially the space of all vectors defined at $x$. By the way, this construction is reasonable, because the tangent space is a manifold in itself.

Now, the linear algebra - the most powerful discipline in handling linear vector spaces - teaches us that for each vector space $V$ there is a naturally associated dual space $V^*$. Thus, we may assume that in this special case there exists a dual space to the tangent vector space $T_x(M)$, which is what we may call the cotangent space $T_x^*(M)$. Dual vectors in $T_x^*(M)$ are often referred to as one-forms. We may note that, even though $T_x(M)$ and $T_x^*(M)$ may have the same size and more or less the same structure, there is in general no unique map between these two spaces. In other words, given an arbitrary vector $x$, there is no natural way to associate it with a unique one-form $\omega^1$. One can try to put these vectors in mutual correspondence using components in some fixed basis (see Chapter 3) but, as I have several times mentioned, the components of a vector are not fundamental quantities: they change under coordinate transformations. Nevertheless, we shall see later that it is possible to choose some particular structure helping to identify vectors and one-forms, and this convenient additional structure is referred to as metric.

Given the local cotangent space $T_x^*(M)$, one can also define the cotangent bundle in the same way as with the tangent bundle:

$$T^*(M) = \bigcup_{x \in M} \{x\} \times T_x^*(M).$$

One may observe (and the reciprocal lattice in solid state physics is a good physical illustration to this observation) that if one has an $n$-dimensional vector space, one can build up a space of all linear maps on it (e.g., comprised of nontrivial linear combinations with real coefficients). This space of linear maps is also a linear space, which can be defined through basis vectors $\mathbf{e}_j^*$, each of the latter being a linear combination of the full set of basis vectors $\mathbf{e}_i$ for the vector space $V$. If we require, as when we considered above the transformations between affine coordinates, that $\mathbf{e}_i^* \mathbf{e}_j = \delta_{ij}$, then we get a basis for a dual space formed by linear maps in the initial vector space $V$. Any linear map in $V \in \mathbb{R}^n$ may be written as a linear combination of the basis vectors $\mathbf{e}_j^*$ with real coefficients (see Chapter 3). In other words, given a basis $\mathbf{e}_i$ of a vector space $V$, we may define the dual basis of $V^*$ using the rule $\mathbf{e}_i^* \mathbf{e}_j = \delta_{ij}$, i.e., any basis in $V$ may produce a dual basis in $V^*$. The Kronecker symbol (as well as delta-function in the case of functional spaces) ensures a one-to-one correspondence between the two sets of basis vectors, but does not guarantee orthonormality, if one deals with vector spaces and not necessarily with inner product spaces (see about the hierarchy of spaces in Chapter 3).

A brief note about infinite-dimensional spaces: this case may be quite nontrivial, since vectors or functions dual to those forming a basis in $V$ do not in general provide a basis in $V^*$. In particular, the usual consequence of isomorphism between vector spaces $V$ and $V^*$ and the Euclidean space $\mathbb{R}^n$, $\dim V = \dim V^*$ becomes meaningless in the infinite-dimensional case. Moreover, if you try to find the double dual, i.e., dual of dual vector space $V^{**}$, it will not necessarily be isomorphic to the starting vector space $V$. See e.g., http://planetmath.org/encyclopedia/DualSpace.html.

After all these digressions let us get back to Hamiltonian mechanics. We know that it is essentially based on the notion of a Hamiltonian function or Hamiltonian. One usually defines the Hamiltonian as a function $H(p_i, q^i, t)$ of $6N+1$ variables $p_i, q^i, i, j = 1, \ldots, 3N$ plus parameter $t$ which is interpreted as time. The role of time is not always simple, and we shall specially discuss it several times in various contexts (see e.g., Chapter 4). The Hamiltonian is related to the Lagrangian as

$$H(p_i, q^i, t) = p_i q^j \delta_j^i - L(q^j, \dot{q}^j, t). \tag{4.18}$$

In curved spaces one should in general write, instead of $\delta_j^i$, some non-diagonal tensor $g_j^i$, e.g., the metric tensor. One usually assumes that the defining expressions $p_i = \partial L(q^i, \dot{q}^i, t)/\partial \dot{q}^i$ are invertible and can be solved as equations with respect to $\dot{q}^i$ giving the generalized velocities in terms of $p_i, q^i$. A sufficient condition for such invertibility is the nonsingularity of the Hessian matrix

$$D_{ij} = \frac{\partial^2 L}{\partial \dot{q}^i \partial \dot{q}^j} = \frac{\partial p_i}{\partial \dot{q}^j}, \det D_{ij} \neq 0$$

This condition is almost always fulfilled in simple mechanical problems but is not necessarily valid in more complicated physical problems. This fact may be a source of difficulties in quantum field theory (QFT), and we shall discuss this issue later (Chapter 6).

The procedure of inverting the momentum equations, $p_i = \partial L/\partial \dot{q}^i$, and arriving at the Hamiltonian from the Lagrangian is usually called in mechanics the Legendre transformation, although it is only a particular case of the Legendre transformation in general. Now we may notice that the Hamiltonian $H$ is a function of its variables on the cotangent bundle $T^*M$ of a $d = 3N$-dimensional manifold with the local coordinates $q^i, i = 1, \ldots, 3N$, whereas the Lagrangian of the system is a function on the tangent bundle $TM$ of $M$. Considering, for simplicity the case when the Hamiltonian does not explicitly depend on time (this case corresponds to a closed system, which is not driven by any external agent), we have

$$dH = p_i d\dot{q}^i + \dot{q}^i dp_i - \frac{\partial L}{\partial q^i} dq^i - \frac{\partial L}{\partial \dot{q}^i} d\dot{q}^i = \dot{q}^i dp_i - \frac{\partial L}{\partial q^i} dq^i, \qquad (4.19)$$

which means that the differential $dH$ is conveniently expressed only through the differentials of its own (cotangent bundle) variables with

$$\frac{\partial H}{\partial p_i} = \dot{q}^i, \qquad \frac{\partial H}{\partial q^i} = -\frac{\partial L}{\partial q^i} \qquad (4.20)$$

We know that trajectories of mechanical particles are described by the curves $q^i(t)$ on $M$, with the dynamics being specified by the Lagrangian $L(\dot{q}^i(t), q^i(t), t)$. These trajectories obey the Euler-Lagrange equations resulting from extremizing the classical action defined as an integral of the Lagrangian along the trajectory (see above)

$$\frac{\partial L}{\partial q^i} = \frac{d}{dt} \frac{\partial L}{\partial \dot{q}^i_{\ i}} = \dot{p}_i$$

Thus, using these simple transformations, we arrive at an alternative description of the extremal path - this time in terms of other equations of motion:

$$\dot{q}^i = \frac{\partial H}{\partial p_i}, \qquad \dot{p}_i = -\frac{\partial H}{\partial q^i}. \tag{4.21}$$

This system of the first-order ordinary differential equations (ODEs) is called the Hamiltonian's system of equations. We may write the least action principle in the Hamiltonian formulation as

$$\delta S[p_i, q^i] = \delta \int_{t_1}^{t_2} dt \left( p_i \dot{q}^i - H(p_i, q^i) \right) = 0, \tag{4.22}$$

which gives the above written Hamiltonian equations of motion. Denoting $\eta := p_i dq^i$ (often called the Poincaré form) so that the first term in the action may be written as an integral over path $\gamma$

$$\int_{t_1}^{t_2} p_i \dot{q}^i dt = \int_{\gamma} \eta$$

One can associate with this integral a so-called symplectic form, $\omega = dp_i \wedge dq^i$ (see e.g., [14], Ch.8). It is sometimes said that the action written in the Hamiltonian form characterizes the symplectic structure, with the cotangent bundle $T^*M$ on which the Hamiltonian $H$ is defined being the symplectic manifold associated with the considered mechanical system, i.e., the particle motion on manifold $M$. Since this geometrical language seems to be universally accepted nowadays, I shall use it systematically even in simple examples in order to get accustomed to it. Imagine, for instance, a particle moving over the real line $\mathbb{R}$ so that $\eta = pdq$ [99]. The associated symplectic manifold $T^*\mathbb{R}$ has local coordinates $(p, q)$. Then the corresponding symplectic form is simply $\omega = dp \wedge dq$. Let us now try to generalize this construction a little bit. The local coordinates $(p, q)$ may be denoted, for instance, as $(z^1, z^2)$, and if a symplectic manifold has an arbitrary finite dimensionality, then local coordinates can be written as $z := (z^1, \dots, z^n)$ so that the Poincaré form is $\eta = B_\alpha(z)dz^\alpha$ and the symplectic form may be written as

$$\omega = \frac{1}{2}\omega_{\alpha,\beta}(z)dz^\alpha \wedge dz^\beta$$

where $\omega_{\alpha,\beta}(z) := \partial_\alpha B_\beta(z) - \partial_\beta B_\alpha(z)$.

This last equation provides the coordinate expression of an exterior derivative of the form $\omega$. One can define here the general Poisson brackets (some authors prefer a singular - the Poisson bracket)

---

[99] It is common now to omit the symbol $d$ in differential forms, in particular in the form $d\eta$; frankly speaking, I do not like this "modern" style (although I use it), since I belong to the old school in which the number of integrals was always equal to the number of differentials. Besides, there is a lot of confusion about this modern notation.

$$\{F, G\} = \omega^{i,j}(z)\frac{\partial F}{\partial z^i}\frac{\partial G}{\partial z^j}$$

where $\omega^{i,j}(z)$ is the inverse matrix to $\omega_{i,j}(z)$, $\omega^{i,j}(z)\omega_{j,k}(z) = \delta_k^i$; here to simplify notation, I changed indices from Greek to Latin - I hope it will not produce any confusion. One can encounter the Poisson brackets in various contexts, and we shall discuss their properties ad hoc. Now I would like only to emphasize their connection with quantum mechanics. Indeed, quantization of classical mechanics on a manifold $M$ was originally performed as replacing the Poisson bracket algebra by the commutator algebra, i.e., by the transition from the classical phase space (cotangent bundle $T^*M$) to the quantum mechanical Hilbert space $\mathbb{H}$. In the conventional textbook example when the manifold $M$ is simply $\mathbb{R}^3$, the Hamiltonian coordinates $q^k$ and $p_i$ are treated as operators $\hat{q}^k$ and $\hat{p}_i$ on $\mathbb{H} = L^2(\mathbb{R}^3)$, with $[\hat{p}_i, \hat{q}^k] = i\hbar\delta_i^k$ whereas the classical Poisson brackets are $\{p_i, q^k\} = \delta_i^k$. Operator $\hat{q}^k$ is simply the multiplication by $q^k$, whereas $p_k$ is a covector differential operation, $p_k :=$ $-i\,\partial/\partial q^k$. Then one needs to replace "classical observables" represented by real functions $F(p, q)$ defined on $T^*M$ [100] by quantum self-adjoint operators $\hat{F}$ on $\mathbb{H}$. Then, by definition, the expectation value $\bar{F}$ of a "quantum observable" $\hat{F}$ at some state $\Psi \in \mathbb{H}$ is the scalar product $\bar{F} = (\Psi, \hat{F}\Psi)$ (we assume the unit norm here). See Chapter 6 for details of quantization, in particular canonical quantization. Now we may just state that orthodox quantum mechanics based on the Schrödinger equation finds its direct analog in the Hamiltonian picture of classical mechanics. In Chapter 6 we shall see that the Schrödinger equation itself is the quantum analog of the Hamiltonian equations. Thus, one may boldly say that Hamiltonian systems play a central part in mathematics and physics. They have brought forth symplectic geometry methods, canonical quantization and the concept of integrable systems. They also allowed us to better understand time evolution and conserved quantities as Noether symmetries. We have also seen that if the Lagrangian of a physical system is known, then one can obtain the Hamiltonian by the Legendre transformation (see about the general Legendre transformation in Chapter 4). However, for an arbitrary system of equations, it may be not an easy procedure.

From the viewpoint of a theory of differential equations, the Hamiltonian form of a system of differential equations has a specific structure, namely the symplectic structure.

To use all the advantages of the Hamiltonian structure it would be desirable to know whether an arbitrary system of differential equations has a Hamiltonian structure, and if it has to be capable of writing the system of equations in the Hamiltonian form, i.e., finding such a structure explicitly.

Thus, when we use the Hamiltonian formulation in classical mechanics, we have to study symplectic geometry in the phase space and, correspondingly, the algebra of differential forms. When turning to quantum mechanics, the classical Hamiltonian formulation naturally mutates into the Hilbert space paradigm using the language of operators and the techniques of boundary value problems for the Schrödinger equation. In distinction with the Hamiltonian formulation, the Lagrangian mechanics requires, as we have seen, studying variational calculus, and when we pass from the Lagrangian description to quantum theory, we have to learn the language of path integrals (developed by P. A. M. Dirac and R. Feynman).

---

[100] We have already noticed that in classical mechanics functions are assumed to be smooth, e.g., $F \in C^\infty(T^*M)$.

Let us now try to illustrate how Hamiltonian mechanics works on several simple examples. We have already briefly described the most trivial example possible - that of a particle moving along a line. If we try to formulate this one-dimensional problem in the jargon of symplectic geometry, we must first write the necessary forms. The corresponding symplectic manifold (the cotangent bundle $T^*\mathbb{R}$) has local real coordinates $(z^1, z^2) = (q, p)$, with the following forms:

Poincaré: $\eta = p\,dq$ symplectic: $\omega = d\eta = dp \wedge dq$ exterior derivatives: $\omega_{1,2} = \omega^{2,1} = -1; \omega^{1,2} = \omega_{2,1} = 1$ Poisson brackets: $\{F, G\} = \omega^{1,2} \frac{\partial F}{\partial q} \frac{\partial G}{\partial p} + \omega^{2,1} \frac{\partial F}{\partial p} \frac{\partial G}{\partial q} = \frac{\partial F}{\partial q} \frac{\partial G}{\partial p} - \frac{\partial F}{\partial p} \frac{\partial G}{\partial q} = 1; \{q, p\} = 1$. Here, as we have noted a couple of times, $F$ and $G$ are typically considered smooth.

One may also notice that the operation of taking the Poisson bracket is skew-symmetric and satisfies the Jacobi identity. This means that the Poisson brackets are simultaneously Lie brackets (see Chapter 3) on $C^\infty(M)$. We can write the Hamiltonian equations of motion for any "classical observable" $F(q(t), p(t), t)$, which is again considered smooth, $F(q, p) \in C^\infty(T^*M)$ (for simplicity, I have omitted the explicit dependence on parameter $t$):

$$\frac{dF\big(q(t), p(t)\big)}{dt} = \big\{F\big(q(t), p(t)\big), H\big(q(t), p(t)\big)\big\}$$

This equation is quite important, since it is equivalent to the Hamilton equations and states that the evolution of an "observable" $F$ - i.e., the rate of change of the observed value of the function $F(q, p)$ - is equal to the observed value of the Poisson brackets $\{F, H\}$.

For many applications of classical mechanics, it is sufficient to assume the Hamiltonian function $H: \mathbb{R}^2 \to \mathbb{R}$ to be a $C^2$ function of $2n$ variables $q^i, p_j, i, j = 1, \dots, n$. In the $1D$ case ($n = 1$) the integral $H\big(q(t), p(t)\big) = const$ fully describes the trajectories in the phase space (phase plane). One can easily illustrate the meaning of the "energy surface" $H(q^i, p_j) = E$ on this simple example. Since we have excluded an explicit dependence on time (such a case is usually called autonomous), we obviously have a time-translation invariance. This property can be easily verified by replacing $t$ with $\tau = t - t_0$ where $t_0 = const$ is a translation along the time axis. Consider, e.g., the Hamiltonian equations or, in general, a vector equation of the form $dz/dt = F(z)$ where, as before, $z = (z^1, \dots, z^n)^T$ are local coordinates. The motion equations written in this form are usually called a dynamical system (see below). Consider the initial value problem (IVP) for the above vector equation, $dz/dt = F(z), z(0) = z_0$, and assume that it has the solution $z = \varphi(t)$. Then making a substitution $\tau = t - t_0$ we see that the IVP $dz/dt = F(z), z(0) = z_0$ has the solution $\psi(t) = \varphi(t - t_0)$. This second solution is obtained by merely translating the first one along the time axis. It is important to understand that these two solutions to our dynamical (IVP) problem, $\varphi(t)$ and $\psi(t)$, are different, but belong to the same energy. In our $1D$ case, these solutions lie on the same phase curve (orbit).

This situation is commonly illustrated by the oscillator equation, $\ddot{z} + \omega_0^2 = 0$, which has solutions $z_1 = \sin \omega_0 t$ and $z_2 = \cos \omega_0 t$, the latter being obtained from the first by the transformation $t \to t - \pi/2$. The oscillator equation can be cast in the form of the equivalent vector equation with $x = x^1, \dot{x} = x^2$.

So, in a fairly unorthodox manner, the notation has led us to a map from functions into forms; $d$ is known as the exterior derivative. You might ask, "What about our notion of $dx$ as a small change in $x$?" Well, we have to throw that picture out of the window, because the notation "$dx$" does not actually mean a small change in $x$. "$dx$" isn't even really a number. It's a linear map from vectors to numbers. It can act on small vectors to produce small numbers, but it isn't a small number in itself; it's not even an element of $\mathbb{R}$. It's an element of $T_p^* M$. So what does it really mean when we see "$dx$" in an integral?

To finalize this section, we can emphasize the link between the general theory of dynamical systems and classical mechanics describing the motion of material bodies. Namely, one can show (see, e.g., [14]) that any dynamical system $dx^j/dt = v^i(x^j)$ admits, at least not in the vicinity of its critical points, a symplectic structure (i.e., Hamiltonian vector field) $\omega_{ij}(x), x = (x^1, \dots, x^n)$ so that one can write the dynamical system as $v^i \omega_{ij}(x) = \partial_j H(x)$, where $H(x)$ is the respective Hamiltonian function.

## 4.5 Oscillations

In many cases, the forces appearing in a mechanical system, when it is driven from the equilibrium, strive to return the system to the equilibrium position, when there are no forces producing motions in the system. Return to equilibrium is the general tendency of almost all natural systems (see Chapter 7). For small deviations from equilibrium, the returning forces are proportional to such deviations (Hooke's law), which is simply a manifestation of the fact that each smooth (or analytical) function is locally linear. This linearity leads to the equation of small oscillations in a $1d$ mechanical system, $m\ddot{x} + kx = 0, k > 0$, which is the simplest possible model of a finite analytical motion. This motion is obviously periodic.

One may notice that although the motion equations are purposed to describe evolution, systems in purely periodic motion are difficult to call evolving since they repeat the same states a countless number of times. In an evolving system, each state is, in general, unlike any other. A little later we shall specially discuss time evolution in classical mechanics in connection with dynamical systems theory.

The case of small oscillations with a single degree of freedom is such a common mathematical model that it is rather boring to discuss it once again. onetheless, I shall do it in order to have a possibility of returning to this ubiquitous model in future chapters and, perhaps, to find some not so common features in it. One can of course find careful exposition of a $1d$ oscillation theory in any textbook on mechanics so that I won't care much about details. Let $x = x_0$ be a stable equilibrium i.e., $\partial U(x_0)/\partial x = 0, \partial^2 U(x_0)/\partial x^2 > 0$.[101] Then the solution to the mechanical system with kinetic energy $T(p, x) = T(\dot{x}, x) = (1/2)x(a)\dot{a}^2, U = U(x)$ is periodic for $(p, x)$ near $(p = 0, x_0)$. The first question that arises here is: what is the period of the motion? The answer is: the period $T_0$ near equilibrium position with the decrease of the amplitude $x(t)$ tends to the limit $T_0 = 2\pi/\omega_0$ where

$$\omega_0^2 = \frac{b}{a}, \qquad b = \frac{1}{2}\frac{\partial^2 U(x_0)}{\partial x^2}, \qquad a = a(x_0),$$

since for a linearized system $U(x) = (1/2)b(x - x_0)^2$ and the solution is $x(t, \omega_0) = A\cos\omega_0 t + B\sin\omega_0 t$ . The qualitative picture of $1d$ small oscillations (e.g., in the $(\dot{x}, x)$-space) can be plotted even without solving differential equations of motion: the relationship for the "invariant manifold" (see below the section on dynamical systems)

$$\frac{m\dot{x}^2}{2} + \frac{bx^2}{2} = const$$

represents the ellipse.

The most common example of a periodic, i.e., oscillating, $1d$ mechanical system is a pendulum. The simple mathematical pendulum consisting of a point mass $m$ and of a massless thread (constraint) of length $l$ is described by the Hamiltonian $H = p_\varphi^2/2ml^2 - mgl\cos\varphi$ where $p_\varphi = ml^2\dot{\varphi}$ is the angular momentum and $\varphi$ is the declination angle. In this section, we shall mostly deal with small oscillations. For small oscillations $\varphi \gg 1$ around the equilibrium position $\varphi = 0$ , the pendulum motion is described by the linearized (harmonic) Hamiltonian

$$H = \frac{p_\varphi^2}{2ml^2} + \frac{mgl}{2}\varphi^2 = \frac{ml^2\dot{\varphi}^2}{2m} + \frac{mgl}{2}\varphi^2 = \frac{m\dot{x}^2}{2} + \frac{m\omega^2 x^2}{2},$$

where $\omega = (g/l)^{1/2}$.

One can of course describe oscillations in all possible versions of mechanics: Newtonian, Lagrangian, Hamiltonian, with the help of Hamilton-Jacobi equations and all other possible methods. I prefer to attach oscillations to the most intuitive version of mechanics, the Newtonian one, because geometric language, more adequate in other formulations than in Newtonian, may obscure the basic notions of the oscillation theory. Oscillations are usually understood as finite motions that occur in the vicinity of equilibrium points. Recall an equilibrium point.[102] If, for instance, point $a$ is a local minimum of the potential $U(x, \lambda)$, where $\lambda$ is some parameter, then $x = a(\lambda)$ brings the Lyapunov stability (see the section on dynamical systems), i.e., for initial conditions $\{p(0), x(0)\}$ sufficiently close to $\{0, a\}$ the whole phase trajectory $\{p(t), x(t)\}$ is close to $\{0, a\}$. Mathematical models of almost all multidimensional vibrating systems are often just generalizations of the $1d$ case: $x_{n+1} = Q(x_1, \dots, x_n)$, and if $Q$ is positive definite, then any small motion can be represented as a superposition of oscillations along the main axes.

---

[101] I write these expressions for brevity not in a fully correct form; one should write, e.g., $\partial U(x)/\partial x \mid_{x=x_0}$ and analogously for second derivatives.

## 4.6  Harmonic Oscillator

The model of harmonic oscillator is not only the favorite toy model for physicists, both in classical and quantum domain, but it possesses certain unique features that manifest themselves especially conspicuously in quantum mechanics and quantum field theory (QFT). We shall see in Chapters 6 that the harmonic oscillator in quantum mechanics has very distinguished spectral properties. The energy spectrum of a harmonic oscillator is noteworthy due to at least three reasons. Firstly, its lowest energy is non-zero which means that the quantum oscillator performs ground-state motions. The average kinetic energy of the quantum oscillator in a ground state is positive. This fact leads in quantum field theory to formally infinite vacuum energy. In QFT, non-zero energy in the lowest state implies that the electromagnetic field exists even when there are no photons. All these ramifications of harmonic oscillator treatment irritated theoretical physicists to such an extent that eventually a bunch of new theories emerged out of this seemingly primitive model, and we shall later discuss some of them. Secondly, the energy levels of the harmonic oscillator are quantized i.e., they may only take the discrete values of elementary energy $\hbar\omega$ times $(2n + 1)/2$, where $\omega$ is the classical oscillation frequency, $\hbar$ is the Planck constant, $n = 0, 1, 2, \dots$. Taking only discrete values is a typical feature of many (but not all!) quantum systems and, in general, of boundary value problems, but the energy levels of a harmonic oscillator are equally spaced unlike for other quantized systems. The equidistant character of the harmonic oscillator spectrum is the third and maybe the most important feature. Besides, scrutinizing the solutions to the quantum oscillator problem one can observe the wavelike tunneling effect non-existent in classical mechanics.

   Within the classical Newtonian framework, it will be for the beginning sufficient to know that oscillations are just finite motions occurring in the vicinity of equilibrium points. In principle, Newton's equations may be highly nonlinear which makes their solution a complicated problem, but harmonic oscillations are the result of linearization of any smoothly changing force near the state of equilibrium. This fact makes the model of t h e harmonic oscillator very general. In the Newtonian version of classical mechanics, one usually considers coordinates $x^i$, corresponding velocities $v^i = dx^i/dt \equiv \dot{x}^i$ and forces $F^j(x^i, v^i, t)$, not necessarily potential ones. The kinetic energy is defined simply as a quadratic form $T_{kin} = \sum_i m_i v_i^2$. We have already seen in connection with a brief discussion of geometric properties of classical mechanics that the kinetic energy is in general a bilinear form defining a metric on the tangent space $TQ$ where $Q$ is the configuration manifold, $T_{kin} = \frac{1}{2} g_{ik} \dot{q}^i \dot{q}^k$ , but these geometric concepts are not indispensable for discussing simple models of Newtonian mechanics. By the way, since geometric transformations are usually of minor importance in the Newtonian formulation of classical mechanics, we shall temporarily not distinguish between contra- and covariant coordinates, writing vector indices below. Let us start, for simplicity, from a one-dimensional motion of a point mass: some practical experience shows that in Newtonian mechanics mathematical models for a great many systems are just multidimensional generalizations of $1d$ situations. In a single dimension we have the second-order motion equation $m\ddot{x} = F(x, \dot{x}, t)$ or, writing it in the dynamical system form,

$$\dot{x} = \frac{1}{m} p, \qquad \dot{p} = F(x, p, t),$$

we may define vectors $x := \{x_1 = x, x_2 = p\}$ and $f := \{f_1 = p/m, f_2 = F\}$ and then write Newton's equation in the vector form $\dot{x} = f(x, t)$. Here $x$ represents points in the phase space (phase manifold) $M$ of dimension two ( $\dim M = 2$ ), with both variables $x, p$ being regarded as independent coordinates. Solutions to the motion equation or, equivalently, to the corresponding dynamical system define a family of phase curves $\gamma(t) = (\gamma_1, \gamma_2)$ or, in "physical" variables, $x_1(t) = \gamma(t), x_2(t) = m\dot{\gamma}(t)$. If the function $F$ does not depend on time $t$ (one usually assumes this function to be also independent of $p$ and continuous in $x$), then one may introduce the potential [103] $U(x) = -\int_{x_0}^{x} F(x') dx'$ so that $F(x) = -\frac{d}{dx} U(x)$. Then the total energy $E = T + U$ is conserved i.e., $\frac{dE}{dt} = \frac{d}{dt}(T + U) = 0$ in the process of motion, in other words, energy $E(p, x) = E(\dot{\gamma}(t), \gamma(t)) = const$ along the phase curves $\gamma$. Note that along such solution curves $p$ is a function of $x$ (and vice versa). The solution "flows" through the phase space, being restricted to constant energy curves (or surfaces if $\dim M > 2$) if energy is conserved.

---

[102] One more time trying to infuriate the mathematicians, I shall make in the present context no distinction between equilibrium, fixed, stationary and critical points. In general, however, these notions may be different.

---

[103] The independence of $F$ on time and velocity is a condition at any rate sufficient, but in certain cases may be too restrictive. One can sometimes introduce a potential for time and velocity dependent forces.

All these trivial reminders serve as a preparatory material for the models of concrete physical systems such as oscillator. We have seen that the harmonic oscillator is characterized by the simplest possible force law, $F(x, p, t) \coloneqq F(x) = -kx, k > 0$, which means that the force pulls the point mass back to the equilibrium position (origin of coordinates). Here the quantity $k$ is the model parameter which physically describes the oscillator rigidity: the more the numerical value of $k$, the more is the force driving the point mass to the origin of coordinates. One, however, typically introduces another parameter: the frequency $\omega = (k/m)^{1/2}$ and writes all the formulas pertaining to the oscillator in terms of this parameter. For example, using the definition of potential, we get $U(x) = \frac{1}{2} m \omega^2 (x^2 - x_0^2)$, where $x_0$ stems from the integration and we can put $x_0 = 0$ without any loss of generality[104]. Then using the above notations, we have $\dot{x} = f(x)$ with $x_1 = x, x_1 = p$ and $f_1 = p/m = x_2/m, f_2 = F(x) = -m\omega^2 x_1$ so that the motion equations in the dynamical system form look as

$$\dot{x}_1 = \frac{1}{m} x_2, \qquad \dot{x}_2 = -m\omega^2 x_1.$$

## 4.7   Symmetries and Conservation Laws

Symmetries play a fundamental role in understanding the physical process considered. Usually, the importance of symmetries for physics is related to invariance properties of differential equations modeling the physical situation. Symmetries associated with conservation laws are specific cases of such invariance properties leading to integral combinations for the respective differential equations. One is tempted to think in this connection that each symmetry of differential equations leads to a conservation law and, conversely, each conservation law is the consequence of some symmetry. Both statements are wrong: there exist symmetries of differential equations of physics that do not provide conservation laws[105] as well as there are conservation laws which are not related to any symmetries of such equations (e.g., conservation of mass).

A natural question would be: how many integrals of motion are there in a mechanical system? A classical system is described by $2n$ first-order differential equations[106] whose solution is determined by $2n$ independent constants. It may be natural to choose the initial coordinates and momenta of the classical system, i.e., coordinates of the initial point of the system's path in its phase space, as such independent constants. One can also arrange these initial coordinates and momenta in various combinations, not all of them independent. Corresponding to the number of initial coordinates and momenta as independently chosen constants, a classical system admits exactly $2n$ independent integrals of motion. By the way, one often forgets that initial coordinates and momenta are also integrals of motion.

---

[104] One performs this transformation $(x - x_0 \to x)$ every time in physics often forgetting that we are justified to do it owing to affine properties of our space.

[105] Point or discrete symmetries.

[106] This is the typical dynamical systems form, see the following section.

Here, it would be probably worthwhile to make the following remark. There are two interpretations of symmetry transformations: active and passive. Under the active interpretation one understands an actual change of a physical state such as rotation or reflection whereas the passive interpretation consists in changing the viewpoint of an observer such as transition to a different coordinate system. In other words, the passive interpretation does not imply any real physical transformation, and maybe because of that it is mostly favored by mathematicians. In contrast, physicists generally prefer the active interpretation, and this diversity of preferences is of course curious. As a standard example, one usually takes the rotation of a vector: under the active interpretation vector is actually rotated and a new vector is obtained, while under the passive interpretation vector remains intact and only the basis vectors are rotated.

## 4.8    Relativistic Mechanics

In almost all the textbooks, it is customary to treat relativistic mechanics together with classical field theory, probably because of the unifying concept of Lorentz invariance. This is a matter of taste of course, but personally I think that mechanics is just mechanics and can be treated as a quasi-isolated cluster of models, without the concept of fields. Relativistic mechanics of a material point, on a primitive level of a $3D$ geometry may be viewed as a classical mechanics with a velocity-dependent mass (see [39], §9). As R. Feynman put it in his "Feynman's Lectures on Physics", "The theory of relativity just changes Newtons laws by introducing a correction factor to the mass." [138], p.15-1. Simply speaking, one can substitute $\gamma m$ instead of $m$, where $\gamma = (1 - \beta^2)^{-1/2}, \beta = v/c$ is the relativistic factor, $v$ is velocity with respect to a laboratory frame (an observer being at rest). For many practical calculations this is sufficient. So, one sees that the "relativistic mass" increases with velocity. In popular science interpretations of special relativity, the statement that the mass of a body rises with its velocity is ubiquitous. This statement is supported, although more accurately, in many textbooks on relativity (see [158]). On a more profound, geometric level the simple heuristic concept of "relativistic mass" may lead to some difficulties. Let us take a closer look at relativistic mechanics, using, in the spirit of connected models in physics, this opportunity to discuss the properties of mass in general.

It is interesting that the concept of variable mass, e.g., depending on the body velocity, emerged before the creation of special relativity, in particular, in the paper by O. Heaviside [38]. Moreover, there were experiments set up in the beginning of the 19th century which demonstrated the dependence of the quantity that the experimenters identified with the mass of a moving body on its velocity.

## 4.9    Dynamical Systems

Historically, the term "dynamical system" was applied to a mechanical system with a finite number of degrees of freedom. The state of such a system was usually characterized by its position, for example, by the location of a center mass point or by the configuration of a number of points, whereas the rate of change of this position (more generally, of the system's state) was given by some law of motion. That was all in the original meaning of the term. In general, the state of a dynamical system may be characterized by some quantities, not necessarily of a mechanical origin, which may assume arbitrary real values, for instance in chemistry, biology or ecology. Mathematically, of course, complex values are also admitted. If these quantities are treated as coordinates $x^i$ of a point in an $n$-dimensional space, $i = 1,2,\dots,n$, then such a space is usually called the phase space of the considered dynamical system and the point $x^i$ representing the state of a system is usually called its "phase". This terminology is probably due to J. W. Gibbs who called the state of the system its phase. The phase space $U \subset \mathbb{R}^n$ is usually considered an open domain; we shall limit ourselves to Euclidean domains, although it is of course possible - and in many cases necessary - to consider more general differential manifolds as the phase space, e.g., circle, cylinder, or torus. We have already seen that the phase space of a particle is a six-dimensional Euclidean space, the six components of the phase velocity vector being the components of the ordinary velocity and of the force, whereas the projection of the phase trajectory on the space $T_p X$ (parallel to the momentum space) is the trajectory of the particle in the ordinary sense of the word. The evolution of the system with time is represented as a motion of the phase point in the phase space over some curve - the phase trajectory. As we have seen on the example of vector spaces, to each point $x^i$ a vector with components $x^i$ may be associated.

The usual type of a dynamical system is given by a map $F: U \to U$. As particular cases of maps $F$, we have homeomorphisms, which are continuous maps admitting a continuous inverse, and diffeomorphisms, which are continuously differentiable maps admitting a continuously differentiable inverse. In other words, by a dynamical system one can mean a diffeomorphism of a compact differentiable manifold without boundary or a one parameter group such that $\varphi_{t+s} = \varphi_t \circ \varphi_s$. This is a one-parameter transformation of a phase space - a phase flow, which may be both causal, $dx^i/dt = f^i(x^1, \dots, x^n)$ and non-causal, $dx^i/dt = f^i(x^1, \dots, x^k(t+\tau), \dots, x^n)$.

## 4.10   Dynamical Systems and the Cauchy Problem

Let us consider the elementary dynamical systems theory in simple terms of the theory of ordinary differential equations. One often writes the system of differential equations corresponding to a dynamical system in the abridged (vector) form, $dx/dt = f(t,x)$, where $f$ is a vector function $f: S \to \mathbb{R}^n$, with $S$ being an open subset of $\mathbb{R}^{n+1}$. The parameter $t \in \mathbb{R}$ is as a rule identified with time, $x \in \mathbb{R}^n$. More specifically, the vector function $x: T \to \mathbb{R}^n$ is a solution to the equation $dx/dt = f(t, x)$ on an interval $T \to \mathbb{R}^n$, and we shall assume it to be at least continuously differentiable at this interval. The vector function $f(t, x)$ then is supposed to be at least continuous in both $t$ and $x$. In elementary dynamical systems theory, all the variables are considered real.

It is clear that the general scalar equation of the $n$-th order

$$\frac{d^n x}{dt^n} = F\left(t, x, \frac{dx}{dt}, \dots, \frac{d^{n-1}x}{dt^{n-1}}\right),$$

$F: \mathbb{R}^{n+1} \to \mathbb{R}$, can be represented in the vector form.

One usually requires the vector function $f(t, x)$ defined in $x \in D \subset \mathbb{R}^n, t \in T \subset \mathbb{R}$, to satisfy the Lipschitz condition

$$\|f(t, x_1) - f(t, x_2)\| \le L\|x_1 - x_2\|$$

with respect to $x$ for all $(t, x) \in T \times D$. Here $\|.\|$ denotes the norm, $\|f\| = \sum_{i=1}^{n}|f_i|$. One often calls such functions to be "Lipschitz-continuous in variable $x$". It is easily seen that Lipschitz-continuity in $x$ leads to ordinary continuity but the reverse is not true. However, continuous differentiability - absence of breakpoints - is sufficient for Lipschitz-continuity.

The notion of Lipschitz-continuity is an important condition for establishing uniqueness of the solution to the initial value problem (the Cauchy problem or IVP):

$$\frac{dx}{dt} = f(t, x), x \in D \subset \mathbb{R}^n, t \in T \subset \mathbb{R}, \tag{4.23}$$

$$x(t_0) = x_0. \tag{4.24}$$

One can see the proof and commentaries to this essential theorem in any good textbook on ordinary differential equations (ODE), e.g. [15], [19]. It may be important to note that in the context of this uniqueness theorem the domain $S = T \times D$ is usually specified as $T := |t - t_0| \le a, D := \{x: \|x - x_0\| \le d\}$, where $a > 0$ and $d > 0$ are constants. The statement of the uniqueness theorem is typically formulated as follows: if $f(t, x)$ is continuous in $S = T \times D$ and is also Lipschitz-continuous in $x$, then the Cauchy problem has one and only one solution in $|t - t_0| \le inf(a, d/G)$ where $G := \sup_S \|f\|$. It is interesting that the existence and uniqueness of a solution $x(t) := x(t, t_0, x_0)$ is ensured not very far from initial "time" $t_0$, the respective domain effectively decreasing with the supnorm $G = \sup_{(t,x)\in S}\|f\|$. However, the time domain of existence and uniqueness is determined by a sufficient condition that may become too restrictive if the actual solution can exist beyond the guaranteed domain. Therefore, it may be more practical in a specific case to analyze a concrete problem finding the time domain of existence ad hoc from the available $a, d, G$, rather than to apply straightforward the general theorem.

## 4.11 Autonomous Dynamical Systems

Let us now consider a particular but very important case when the variable $t$ is not explicitly present in the vector equation of a dynamical system. Such systems are called autonomous[107]. We have already seen, in connection with energy conservation issue, that autonomous systems are invariant with respect to time translations, viz. if $x(t)$ is a solution in $D \subset \mathbb{R}^n$ then $x(t - t_0)$ is also a solution, in general a different one (as, e.g., $\sin t$ and $\cos t$). Here $t_0$ is assumed to be a constant time shift.

In the theory of dynamical systems, the domain $D \subset \mathbb{R}^n$ is regarded as the phase space for the vector equation $\dot{x} = f(x), x \in D$. This concept may be considered as a certain generalization of the traditional physical terminology where phase space is understood as a direct product of coordinate and momentum spaces. In modern classical mechanics, phase space is typically defined as a cotangent bundle $T^*M$ where $M$ is a configuration manifold (see section on Hamiltonian mechanics for some comments). However, when dealing with dynamical systems there are some other features and accents which are important as compared to the traditional exposition of classical mechanics. For instance, in mechanics it is often convenient to consider the phase space as a Euclidean space whereas in the theory of dynamical systems the phase space is, in general, not a Euclidean space but a differential manifold, on which a vector field corresponding to the vector equation $\dot{x} = f(x)$ is defined. This equation means that, in the process of evolution described by the dynamical system, to each point $x$ a vector $f(x)$ is ascribed determining the velocity of the phase point (the vector $f(x)$ belongs to the tangent space of the manifold at point $x$). This is a kinematic, in fact a geometric interpretation of the above vector equation.

Nevertheless, all this are just nuances: one usually knows exactly in what space one finds oneself and operates. In all cases, phase space is a geometric concept embracing the total number of all states of a dynamical system and convenient to describe the evolution of state points $x = (x^1, \ldots, x^n)^T$ parameterized by variable $t$ (usually time). The state points $x(t) = \left(x^1(t), \ldots, x^n(t)\right)^T$ for a fixed $t$ are, as we know, called phase points[108], and they move through the phase space with changing $t$, each of them traversing the phase manifold along its own phase trajectory. The term "phase" was probably coined by J. W. Gibbs who referred to the state of a system as its phase. In physical literature, one often talks about ensembles of dynamical systems - the set of non-interacting systems of the same type differing from one another only by their state at any given moment, that is by initial conditions. There is an illustrative analogy that is customarily exploited in physics namely the picture of a stationary flow of some fluid, in which every fluid particle moves from point in phase space to another during time $t - t_0$ according to equation $\dot{x} = f(x), x(t_0) = x_0$. Later we shall see that this equation may be interpreted as a mapping of the phase space into itself so the "flow" of the "phase fluid" implements a transformation (in fact a family of transformations) of the phase manifold into itself, in the considered case of a smooth vector field described by differential equations - a diffeomorphism i.e., continuously differentiable invertible map. One may notice that the analogy between the "phase fluid" and some physically observable continuous media - liquid or gas - is not complete: there is no interaction between particles of the phase fluid.

---

[107] Scalar equations of the $n$-th order corresponding to the autonomous case take the form $\frac{d^n x}{dd^n} = F\left(x, \frac{dx}{dt}, \ldots, \frac{d^{n-1}x}{dt^{n-1}}\right)$

[108] Rarely, representing points.

Excluding $t$ from parametric equations $x^i(t), i = 1, \dots, n$ (if we can do it), we get a projection onto phase space $D$. The domain $S = T \times D$ is conventionally called (mostly in old sources) an extended phase space. Although it is not always easy or at all feasible to get rid of parameter $t$[109], it may be possible to obtain differential equations directly describing the trajectories in the phase space. Indeed, from the equation $\dot{x}^i = f^i(x), i = 1, \dots, n$ we get, for example, the following system of $n - 1$ equations:

$$\frac{dx^2}{dx^1} = \frac{f^2(x)}{f^1(x)}, \dots, \frac{dx^n}{dx^1} = \frac{f^n(x)}{f^1(x)}$$

whose solution gives the trajectories parameterized by $x^1$. Conversely, we can turn any non-autonomous system into an autonomous one by introducing a new variable $x^{n+1}$, thus producing a system of $n + 1$ equations instead of $n$ which corresponds to increasing the dimensionality of the phase space. In this sense, autonomous systems may be considered general enough to focus mainly on their study.

Applying the uniqueness theorem to the above system of $n - 1$ equations, we may conclude that the phase trajectories do not intersect. Of course, here it is assumed that $f^1(x) \neq 0$. If $f^1(x_a) = 0$ in some points $x_a, a = 1, 2 \dots$, we can take any other $f^j(x), j = 1, \dots, n$ instead of $f^1(x)$ provided $f^j(x) \neq 0$, which means that we are taking $x^j$ as a parameter. There may be, however, difficulties in the so-called critical points - zero points $\bar{x} = (\bar{x}^1, \dots, \bar{x}^n)$ where $f(\bar{x}) = 0$ i.e., $f^i(\bar{x}) = 0$ for all $i = 1, \dots, n$. We shall discuss this problem as soon as we learn a little more about the dynamical systems in general.

The above-mentioned uniqueness theorem[110] states that under rather weak assumptions about the properties of vector field $f(x)$ (usually it is assumed differentiable or just Lipschitz-continuous) there exists for each point $x \in D$ exactly one solution $x(t)$ of the law of motion $\dot{x} = f(x)$ with initial value $x(t_0) = x_0$. In other words, the evolution of a dynamical system - its future states at $t > t_0$ - is completely determined by its initial state.[111] It is a fundamental model of determinism: what we can say about tomorrow (more correct - about the properties the system in question will have tomorrow) is uniquely determined by what we can say about the system today. This is not true for quantum or statistical mechanics, although some philosophers contend that quantum mechanics is a fully deterministic theory (because of time evolution features). In my opinion, this is just a cunning word usage typical of philosophers since it is hard to imagine a deterministic scheme where observations affect the system.

---

[109] One might recall in this connection that the parametric form is the most general one in representing curves and surfaces.

[110] This theorem is usually known as the Cauchy-Peano theorem, although probably it is more correct to name it the Picard-Lindelöf theorem. The matter is that the first theorem states only existence whereas the second one requires less and guarantees uniqueness.

[111] It may also be the final state that can be taken to determine the backward evolution - a retrodiction setting.

When speaking about a dynamical system, one can totally forget about its mechanical origin. It is completely irrelevant whether the vector equation for a dynamical system describes a mechanical or any other evolution. Mechanical systems are commonly attracted in textbooks as convenient examples of dynamical systems. More important, however, is the fact that some mechanical systems possess specific properties narrowing the entire class of dynamical systems to a clearly defined distinguished subclass, e.g., that of Hamiltonian systems. Nevertheless, it would be a mistake to think that only the Hamiltonian systems are considered in classical mechanics. For instance, non-holonomic systems of mechanics are also dynamical systems of course. The notion of a dynamical system is a generalization of classical mechanics.

Thus, the term "dynamical system" can be applied to any vector field described by a first-order vector differential equation of the form $\dot{x} = f(x), x(t_0) = x_0$ (or any equivalent form, see below), irrespective of its natural or behavioral content. This abstraction serves as a background for mathematical modeling based on dynamical systems.

Similarly, for all $r > 1$. See also Topological dynamical system; Bendixson criterion (absence of closed trajectories); Poincaré-Bendixson theory that has the unfortunate tendency to explode. A rough system is sometimes called a structurally stable system or a robust system. A well-documented survey on (mainly differentiable) dynamical systems is in [146]. Many recent developments are discussed in the various volumes of [147].

In general, an attractor of a dynamical system is a non-empty subset of the phase space such that all trajectories from a neighborhood of it tend to this subset when time increases. An attractor is also called a domain of attraction or basin of attraction. A repelling set, or repellor in a dynamical system is a subset of the phase space of the system that is an attractor for the reverse system. If an attractor, respectively repellor, consists of one point, then one speaks of an attracting, respectively repelling, point. For details (e.g., on stability of attractors) see [146]. It should be noted that in other literature the definition of an attractor is what is called a stable attractor in [146]. For discussions on the "correct" definition of an attractor see [205], Sect. 5.4, and [147].

## 4.12  Non-autonomous Systems

We have already partly discussed the non-autonomous case in the theory of differential equations, now we shall mainly focus on some aspects of non-autonomous dynamical systems i.e., with more focus on system evolution properties. One must confess that even today, in the beginning of the 21st century, one does not have fully rigorous foundations of non-autonomous (as well as relativistic) mechanics. Indeed, even such basic notions as force, energy, power, frame of reference, etc. often need to be reassessed for non-autonomous systems. We have probably felt from the brief overview of differential equations that it seems to be appropriate to treat autonomous and non-autonomous cases differently - so different they are despite the fact that each non-autonomous dynamical system can be reduced to an autonomous one merely by introducing a new variable thus increasing the dimensionality of the vector field constituting the dynamic system. This additional dimension of the phase space is the price paid for the transition from a non-autonomous to an autonomous system.

The situation with non-autonomous systems is much more complicated from both the physical and mathematical perspective as with autonomous systems. Physically, a non-autonomous dynamical system corresponds to an open system placed in a time-dependent external field. This fact is reflected by the explicit dependence of vectors $f^i, i = 1, \dots, n$ in the corresponding vector differential equation $\dot{x}^i = f^i$ on the independent variable $t$ (usually interpreted as time), $\dot{x} = f(t, x)$ which makes the solution time-noninvariant in distinction to those for autonomous systems (see above). Of course, in the non-autonomous case the energy integral in general does not exist (see [23], §6) which makes the system of equations significantly more difficult to integrate than in the autonomous case. In classical mechanics, explicit dependence of the coefficients in motion equations on time are usually interpreted as the presence of an external field ([23], §5). On an intuitive physical level one can illustrate additional difficulties by allowing the coefficients of a differential equation which could be explicitly solved to depend on time. Even in the most primitive linear models, this dependence will immediately produce new types of solutions (e.g., parametric oscillations in elementary vibration theory, see [23], §27).

One can sometimes encounter the assertion that the equation $\dot{x} = f(t, x)$ can be "solved" implying that the solution for $x(t)$ can be represented as an explicit expression consisting of, e.g., power functions, exponentials, trigonometric functions, etc. This is not true: even in the scalar case, $n = 1$ i.e., $x \in \mathbb{R}$ and $f \in \mathbb{R}$, equations of the type $\dot{x} = f(t, x)$ can be explicitly (analytically) integrated only in a relatively small number of very exceptional cases. To illustrate this point for themselves, just try to integrate a rather innocent-looking first-order equation $\dot{x} = \exp tx$. One should not, however, think that substantial difficulties and - quite often - a sheer impossibility to integrate non-autonomous equations in terms of elementary or even special functions mean that such equations do not have solutions.

Let us try to "feel the difference" between autonomous and non-autonomous dynamical systems. One may notice that explicit dependence on time in a dynamical system (non- autonomous case), although not essential from purely theoretical viewpoint [112], can lead to substantial technical difficulties. Indeed, the usual concept of attractor (see below) may become inadequate since the natural notion of time asymptotics is hard to apply: any object in the phase space is "drifting" with time. To understand the situation better let us recall some basic facts from the theory of ordinary differential equations (ODE). The vector ODE $\dot{x} = f(t, x), x \in V_n$, $V_n$ is an $n$-dimensional vector space (manifold), defines a smooth field of directions in the domain $T \times V_n, T \subset \mathbb{R}$ of the extended phase space of arbitrary finite dimension $1 + n$.      Any equation of motion has this form, with $f = \{f^1, \ldots, f^n\}$ being composed of the vector field components for a dynamical system. If one deals with distributed systems described by partial differential equations (PDE), as in the case of fluid dynamics or climate modeling, then the vector space $V$ can be infinite-dimensional. Physically speaking, the system becomes non-autonomous when a time-dependent driving force is acting on it, or when time-dependent constraints are applied, or if the system parameters are varying with time. Equilibrium (stationary) states of non-autonomous systems are, in general, no longer fixed points in the phase space, but rather extended orbits. One might recall from the courses of classical mechanics that the concept of a phase portrait is not as helpful for non-autonomous, e.g., driven, systems as for autonomous ones. One can also recall that attractors play a vital role in assessing the long-term behavior of a dynamical system. [113]    In the non-autonomous case attractors are sets extending in the phase space so that attractors become global objects which are typically difficult to study with standard mathematical tools (e.g., by classical analysis). Strictly speaking, one cannot associate a non-autonomous ODE with a dynamical system interpreted as a vector field acting on $V_n$. Nevertheless, if it is known that the initial value (Cauchy) problem has a unique solution (see above), one can introduce, in a standard fashion, a two-parameter family of evolution operators, $T(t, s), t \geq s$, acting on $V_n$ in such a way that $T(t, s)x(s) = x(t)$, where $x(t)$ is the solution of the Cauchy problem with initial conditions $x(s)$. This family of operators satisfies obvious relationships, $T(s, s) = I(V_n)$ and $T(t, \tau)T(\tau, s) = T(t, s)$ for all $\tau \in [s, t]$, where $I(V_n)$ is the unity operator on the vector space $V_n$.

---

[112] As I have mentioned above, one can easily make an autonomous system from non-autonomous by increasing the dimensionality of the corresponding vector field by introducing the variable $x^{n+1} = t$ and adding one more equation, $dx^{n+1}/dt = 1$. For example, in the case of the undriven $1d$ pendulum, the phase space has two dimensions whereas in the case of a driven pendulum, $\ddot{x} + \omega^2 x = F(x, t)$, one may consider that it has three i.e., $\dot{x}^1 = x^2, \dot{x}^2 = -\omega^2 x^1 - F(x^1, x^3), \dot{x}^3 = 1$.

[113] Everyday examples of attractors are the planetary climate or the national character.

## 4.13  Dynamical Systems in Mathematical Modeling

Today, one is inclined to think that modeling can be only computer-based i.e., in the form of computer simulations. Unfortunately, computer simulations almost never provide the modeler with an insight why the physical mechanism or, say, a biological system operates in a particular way. Computer mathematicians usually retort that the great Carl Friedrich Gauss and other prominent mathematicians were busy calculating the motion of celestial bodies in the 18th-19th century during many years whereas nowadays it takes seconds. Indeed, mathematicians of that time solved algebraic and differential equations, calculated integrals without reaching the efficiency of modern computers. But computers have not created much really new in physically motivated mathematics, rivaling e.g., calculus, complex numbers and functions, Riemann geometry and Lie groups.

So, computer simulations although often accompanied by a great hype and even serving as a base for political decisions (as, e.g., in climate modeling or universal flight bans following volcano eruptions) only have a limited validity. It would be interesting, firstly, to understand this validity and, secondly, to gain an insight when without the computer one would definitely fall short of modeling targets. One of the main questions in modeling an evolving physical situation is: "What happens as time flows from now to infinity?" And in reality, all situations are evolving, steady-state ones are rare and very approximate.

In fact, under modern conditions war is the most economically inefficient, politically destabilizing, and militarily counterproductive mechanism to gain control over any kind of resources. As an example, one can mention the US war in Iraq. If the objective was to gain control over oil resources, then it is at least a poor calculation in military planning since at least USD 100 million/day was spent by the USA to wage this war. One can easily calculate the amount of oil that could have been bought for this money. Given a total unproductiveness of military solutions, one can assume that starting a war is typically the outcome of a crisis in the country's governing model.

The solutions in such models show the temporal evolution of the military conflict and may, in principle, predict which party can be defeated. In the simplest model of military planning, the Lancaster model, the state of a dynamical system describing two interacting (fighting) armies is given by a 2D-point $(x, y)$ located in the upper right (positive) quadrant on the plane, $(x > 0, y > 0)$. Models in economic and military planning are usually unified by this common feature: an essentially non-negative character of variables, which may be regarded as a supplementary constraint. Now imagine two adversary armies, $X$ and $Y$, counting respectively $x$ and $y$ soldiers. Soldiers are by default professional killers, so we can assume in the simplest model that each soldier of the $X$-army kills per unit time $a$ soldiers of the $Y$-army and, conversely, each soldier of the $Y$-army destroys per unit time $b$ soldiers of the $X$-army. Then we can obtain a linear system of equations (parameters $a$ and $b$ are assumed constant so far)

$$\frac{dx}{dt} = -by$$

$$\frac{dy}{dt} = -ax$$

One can interpret parameters $a > 0$ and $b > 0$ as characterizing the weapon power of opposing armies $X$ and $Y$. We can of course explore this linear system using standard linear algebra techniques (see Chapter 3 about linear differential equations), but in our simple analysis it is convenient to integrate this system directly. Dividing the second equation by the first, we get

$$\frac{dy}{dx} = \frac{ax}{by}$$

or $ax^2 - by^2 = c$ where $c = \text{const}$. Thus, integration gives a family of hyperbolas depending on the integration constant $c$ that should be determined by the initial state of military confrontation. One can see that the phase point describing the temporal evolution of the conflict is moving along the hyperbolic phase trajectories separated by a straight line $\sqrt{a}x = \sqrt{b}y$ (one can use, e.g., Maple to draw a simple picture of phase trajectories). Phase trajectories cannot cross this straight line (a symptom of a "hard" model) distinctly dividing two regimes: army $X$ is defeated or army $Y$ is defeated. If the initial state lies above the neutral line separating two outcomes, army $X$ is doomed: quantity $x$ (the number of soldiers in $X$) is reduced to zero in a finite time, whereas quantity $y$ remains positive - a complete victory of $Y$. Of course, the $x$ and $y$ curves have certain symmetry properties whose meaning is that army $X$ can be interchanged with army $Y$ with simultaneous replacement $a^{1/2} \leftrightarrow b^{1/2}$. For $a = b$ hyperbolas lie symmetrically with respect to the line $y = x$. The meaning of the straight line dividing the phase plane into two distinct areas may be clarified in the following way: this is a neutral line on which both armies will equally run out of manpower and eventually be destroyed (asymptotically $x, y \to 0$ for $t \to \infty$). The military conflict persists even when both armies are almost totally demolished. The equation of the separating straight line $\sqrt{a}x = \sqrt{b}y$ implies that to counterbalance the $n$-fold manpower advantage of one adversary the other has to possess the $n^2$-fold more powerful weaponry. Thus, to resist the Mongol invasion in the 13th century, the Mongols' combined forces presumably surpassing those of Kievan Rus at least three times, the Russians should have been about an order of magnitude more efficient warriors.

However, the above mathematical model is too simplistic, and its applicability is utterly questioned. One can treat it only as a toy model which, nevertheless, may serve as a backbone for more realistic approaches. So, in accordance with general methodological recommendations exposed in Chapter 2, we can try to improve this model.

For economic planning, analogous models allow one to divide the produced output into consumed and accumulated parts one of the crucial problems of economic growth. Similar models appear in economics and in military planning, specifically in the case when non-regular armies (e.g., partisans) are participating in military activities.

## 4.14  Nonlinear Science

We have seen that the theory of dynamical systems is just another name for nonlinear dynamics, at least these two fields are inseparable. Nonlinear dynamics is, in fact, as old as classical mechanics. In general, the dynamics of planetary motion turns out to be nonlinear, but, fortunately, two-body systems are integrable and can be treated exactly. Three-body problems are of course also nonlinear, but in general non-integrable and very complicated. For example, the question originally posed in the XIX century: "is the Solar System stable?" naturally leads to a nonlinear description of the paths of three bodies in mutual gravitational attraction. It is this ancient problem which was simple to formulate but extremely difficult to solve that resulted in many modern ideas of nonlinear science, e.g., chaos. It is curious that with the advent of quantum mechanics and displacement of accent on atomic and nuclear physics nonlinear dynamics almost disappeared from physical books up to the 1970s. There was very little or no discussion of nonlinear science in the popular courses on methods of mathematical physics during the prewar and postwar (WW-2) period, when quantum mechanics and linear electrodynamics dominated science and engineering. It is also curious that in the 1980s nonlinear science gained a compensatory extreme popularity, to the extent that many people in the physics community began considering linear models as too primitive and unrealistic.

The great difference that exists between linear and non-linear problems is one of the most important and, perhaps, subtle features of mathematics. One always tries to linearize whenever possible, because linear problems are enormously easier to solve. Unfortunately, the world is not linear, at least to a large extent, so we have to learn how to deal with non-linear problems.

Nonlinearity in general can be well understood as the opposite of linearity - actually its negation. The main features of linearity are additivity and homogeneity, which result in linear superpositions. In linear theories such as quantum mechanics or classical (pre-laser) electrodynamics superposition is the key feature: an infinity of solutions may be constructed provided a finite set of solutions is known. It is true that pure linearity rarely occurs in the mathematical description of real-life phenomena, including physical systems. The example of quantum mechanics, which is an essential linear theory, although extremely important, may be considered an exception - yet to a certain extent, since the superposition principle for the wave function implies the infinite dimensionality of the respective vector space (see Chapter 6). We shall see that many finite-dimensional nonlinear systems can be mapped, with the help of some transformations, to linear systems with infinite dimensionality. In other words, nonlinear systems, i.e., those which should be modeled by nonlinear differential (integro-differential) equations or nonlinear discrete maps are ubiquitous, while linear ones really seem to be exceptions or approximations.

One obvious and strong motivation to study nonlinear dynamical systems is the rich variety of applications of such studies covering such apparently different areas in mathematics, physics, chemistry, biology, medical sciences, engineering, economics, political and military planning, financial management, etc. Yet, the ubiquity of nonlinear systems is counterbalanced by their complexity, so another motivation to study nonlinear dynamics is their intricate behavior. Examples are bifurcations, solitons, strange attractors, and fractal structures. There are no counterparts of these manifestations in the linear world; one can say that the extreme complexity of nonlinear structures marks an essential difference between linear and nonlinear phenomena.

The main source of difficulties in the nonlinear world is the fact that it is in general not feasible to describe a nonlinear system by dividing it into parts which are treated independently or blockwise - a favorite trick in linear system theory. As near as I know, no general techniques have been invented so far to foresee even the qualitative properties of a nonlinear system. For instance, it is rarely possible to predict a priori whether a dynamical system would exhibit regular or chaotic behavior. We shall illustrate this difficulty even on the simplest example of the logistic equation.

There exist, of course, a multitude of textbooks, journal articles and other sources where nonlinear dynamics in general and chaotic behavior in particular are beautifully described. We shall try to discuss only some cases of nonlinear dynamics which I consider quite important. There are, of course, other cases, e.g., of nonlinear PDEs (elliptic, parabolic and hyperbolic), which are very important in modern mathematical physics and, besides, present an intrinsic theoretical interest. However, the theory of these equations as well as related functional spaces and operator theory for nonlinear analysis are not included in this book. We shall discuss Euler and Navier-Stokes equations for incompressible fluids, but this topic seems to be inexhaustible, so the discussion may be considered superficial and much of it is relegated to Chapter 7. I shall also consider in Chapter 9, specifically in association with some unsolved problems, Einstein's equations, some aspects of which tend more to nonlinear science than to customary field theory.

It is widely believed that the 20th century was the century of physics and the 21st is the century of biology. The latter deals mostly with nonlinear phenomena, and the respective models should by necessity be nonlinear. On a macroscopic scale, it has been illustrated above by Lotka-Volterra model of the struggle for existence between two competing species. This dynamic situation (a predator-prey model) is described by nonlinear differential equations giving the time rate of evolution. This is a very simple model, of course, as compared with the level of complexity typically encountered in biology, but it provides a good foundation for more sophisticated mathematical modeling in this field.

We are immersed in the natural world of nonlinear events. For instance, our emotional reactions and our likes and dislikes are probably highly nonlinear. Our behavioral reactions to heat and cold, to colors and sounds, to local pressure and other stimuli are mostly subordinated to the Weber-Fechner law: the magnitude $R$ of psychological response is proportional to the logarithm of magnitude $J$ of physical stimulus, $R = A\ln(J/J_0), J = J_0, R = 0$ for $J < J_0$ (threshold).

At the end of this section, I would like to give a few references to the books I found interesting and useful in the field of nonlinear science. Some of these sources deal predominantly with mathematical concepts of nonlinear dynamics, such as [16, 73], whereas others accentuate the physical aspects [17, 56, 57, 55]. Considering the abundant literature on dynamical systems and nonlinear dynamics, I would recommend the curious reader to start from simple, physically motivated examples which can be found in any textbook or in the Internet (see, e.g., http://www.faqs.org/faqs/sci/nonlinear-faq/). My own discourse is by necessity compilative and remains far from covering even the small portion of this vast field.

## 4.15  The logistic model: the bugs are coming

Imagine a bunch of insects reproducing generation after generation so that initially there were $B$ bugs, and after $i$ generations there will be $N_i$ of them (to bug us: the total mass of insects on the Earth grows faster than that of humans). When the population in a given region is sufficiently large, it can be represented by a real continuous variable $N(t) > 0$. If we assume that there is a maximum sustainable population in the region (an equilibrium state) and that that the population dynamics can be described by a single autonomous equation, $\dot{N}(t) = \varphi(N)$, then we can produce a simple model assuming that the iteration function has a quadratic polynomial form, $\varphi(N, a) = aN(1 - kN)$. This is a real-valued function of two variables, $a > 0$ and $0 \leq N \leq 1/k$. Notice that the logistic model $\varphi(N, a)$ is not invertible with respect to $N = N(\varphi)$ since all the states in $(0, N)$ have two sources in $\varphi$-domain (preimages) so that each value of $\varphi$ corresponds to a pair of different values of population. Therefore, information is lost in the course of inversion. In the Bourbaki nomenclature, the logistic model is not injective; topologically-oriented people prefer calling such maps non-continuous because they do not take an open set into an open set.

Equation $\dot{N}(t) = aN(1 - kN)$ is called the logistic model. Here parameter $a > 0$ is called a control or growth parameter (in theoretical ecology, this parameter, interpreted as the growth rate per individual, is usually denoted as $r$) and parameter $k > 0$, determining the competition for some critical resources, is often called in ecology and population biology an inverse "carrying capacity" of the environment: it defines the equilibrium population when the competition decreases the growth rate to such an extent that the population ceases to rise. In electrical engineering, the same equation describes the energy of $E$ of a nonlinear oscillator in the self-excitation regime (in the first Bogoliubov-Mitropolsky approximation). A discrete alternative to the continuous (differential) logistic model is the so-called logistic map: $N_{i+1} = aN_i(1 - kN_i)$, $0 \leq N_i \leq 1/k, i = 0,1,2,\dots$ , where the discrete variable $i$ plays the role of time in the continuous model. This variable can be interpreted as years or other characteristic temporal cycles, e.g., naturally related to the reproductive behavior of respective species. Recall that the term a "map" or "mapping" $\varphi$ usually refers to the deterministic evolution rule with discrete time and continuous state space $X, \varphi: X \to X$. Then evolution is synonymous with iteration: $x_{n+1} = \varphi(x_n)$. The logistic map, in contrast with the logistic equation, is a model with discrete time i.e. corresponds to snapshots taken at time points $t = n\tau, n = 0,1, \dots$ (the elementary time step $\tau$ may be put to unity). It is interesting that discrete-time dynamical systems can be produced from flows described by continuous-time differential equations, an example is a stroboscopic model provided by the Poincaré sections.

A discrete version of the logistic model can also have a direct physical or biological meaning, for example, in the cases when generations are separated (non-overlapped) in time. Thus, some insects just lay their eggs and die; they do not interact with the next generations. One more famous (appeared around 1200) discrete-time dynamical system, which was initially also a mathematical model of biological reproductive behavior, is the Fibonacci sequence, $a_{k+1} = a_{k-1} + a_k, k = 1,2, \dots, a_0 = 0, a_1 = 1$. To describe this process in terms of evolution, we can introduce matrices

$$x_k = \begin{pmatrix} a_{k-1} \\ a_k \end{pmatrix}, A = \begin{pmatrix} 0 & 1 \\ 1 & 1 \end{pmatrix}$$

so that the Fibonacci sequence will be represented by a discrete-time cascade $x_{k+1} = Ax_k = A^k x_1$. Here the phase space is $X = \mathbb{R}^2$. The Fibonacci map, in distinction to the logistic map, is a linear transformation, and we can directly apply the standard prescriptions of linear algebra. The characteristic equation of the Fibonacci map is $\det(A - \lambda I) = \det \begin{pmatrix} -\lambda & 1 \\ 1 & 1-\lambda \end{pmatrix} = \lambda^2 - \lambda - 1$ so that eigenvalues $\lambda_{1,2} = \frac{1 \pm \sqrt{5}}{2}$ give eigenvectors $v_{1,2} = (1, \lambda_{1,2})^T$. Then $x_1 = C_1 v_1 + C_2 v_2$ and $x_{k+1} = \lambda_1^k C_1 v_1 + \lambda_2^k C_2 v_2$. Using initial conditions $x_1 = \begin{pmatrix} 0 \\ 1 \end{pmatrix} = C_1 \begin{pmatrix} 1 \\ \lambda_1 \end{pmatrix} + C_2 \begin{pmatrix} 1 \\ \lambda_2 \end{pmatrix}$ we get $C_1 = -C_2, C_1\lambda_1 + C_2\lambda_2 = 1$ so that $C_1 = \frac{1}{\lambda_1 - \lambda_2} = \frac{1}{\sqrt{5}} = -C_2$ and $x_{k+1} = \begin{pmatrix} a_k \\ a_{k+1} \end{pmatrix} = \frac{1}{\sqrt{5}} \left[ \begin{pmatrix} 1 \\ \lambda_1 \end{pmatrix} \lambda_1^k - \begin{pmatrix} 1 \\ \lambda_2 \end{pmatrix} \lambda_2^k \right]$ which gives for Fibonacci numbers $a_k = \frac{1}{\sqrt{5}}(\lambda_1^k - \lambda_2^k)$.

One can notice that the logistic map is a simple case of polynomial maps. Of course, the logistic map can be represented in dimensionless form $x_{i+1} = ax_i(1 - x_i)$, where all $x_i$ are numbers interpreted as the population density (or expectation value), and it is required that $0 \le x_i \le 1$, although certain values of intrinsic growth parameter $a$ and initial data $x_0$ may lead to negative population densities, which fact indicates that the logistic map should not be taken literally as a demographic model. In other words, the logistic map takes the interval [0,1] into itself. Despite its apparent simplicity, the logistic map is very general since any function having a nondegenerate extremum behaves in the latter's neighborhood like this map near $x = 1/2$. An obvious extension of the logistic map depending on parameter $a$ is the relationship $x_{i+1} = f(x_i, a)$ usually called the Poincaré map; here generations corresponding to $i = 0,1,2, \ldots$ can be interpreted as periods of motion. Notice that in the case of maps with discrete time, phase trajectories are given by a discontinuous sequence of points. The fixed point of the map ($x = f(x)$) does not change with generations, $x_{i+1} = x_i$ so that it would be natural to define $\mu_i := dx_{i+1}/dx_i$ as a multiplier. It is clear that the maximum value in the sequence $x_i$ is reached when $\mu_i = a(1 - 2x_i) = 0$ and is $x_i = 1/2$. Therefore, the maximum iterated value $x_{i+1}(a) = a/4$, which means that to hold the logistic map in the required domain $0 \le x_i \le 1$ one must restrict the growth parameter to segment $0 \le a \le 4$.

In general, for one-dimensional Poincaré recurrences, the iteration function $f$ and control parameter $a$ can be selected and normalized in such a way as when the initial data $x_0$ (known as the seed) are taken from a finite interval $P = (\alpha, \beta)$, the iterates $x_1, \ldots, x_n$ also belong to $P$ (in the logistic map $P = (0,1)$ for small values of the control parameter, see below). It means that function $f$ maps interval $P$ into itself i.e. this map is an endomorphism (the term "into" means that the iterations may not fill the whole interval). Single-dimensional endomorphisms are often not invertible, which physically means that the past cannot be uniquely reconstructed from the current data, so that one might say that in such cases we are dealing with the systems having an unpredictable past (isn't it typical of human history?). When, however, an endomorphism has a smooth inverse, then the map is a diffeomorphism.

The discrete logistic model may exhibit a somewhat unusual behavior of the population. Thus for small values of the growth (multiplication) parameter $a$, the initial value of the population produces a rather little effect on the population dynamics, the latter being mostly controlled by parameter $a$. Nevertheless, when this parameter is increased, the population (e.g., of bugs, mosquitoes or other insects and, apart from the latter, of rats, mice, reptiles, leeches, etc.) start to change chaotically, and in this chaotic regime one can observe an extreme sensitivity to the concrete number of initially present members $N(t_0) = B$ (or to the "seed" $x_0$). For large enough values of the growth parameter, e.g., for $a > 3$ in the simple logistic map $x_{i+1} = ax_i(1 - x_i)$, the population bifurcates into two, so that the colony of insects acts as if it were "attracted" by two different stable populations (such asymptotic solutions are called attractors). This is a primary example of the so-called period doubling. As this process has become very popular in nonlinear dynamics (starting from approximately 1976) and has generated a great lot of papers most of which are easily available (see the list of literature) and because of the lack of space, we shall not reproduce here the respective computations, restricting the current exposition to a catalogue of the basic facts. We only mention here that period doubling occurs in many models and in a variety of disciplines: in the Navier-Stokes equation (turbulence), in meteorology (the Lorenz system), in chemical reactions, in nerve pulse propagation, etc. In all such cases, we see that for different values of the control parameter $a$ the system's behavior alters drastically: from settling down to a point (the population dies out or tends to a non-zero fixed value), through quasi-regular oscillations, then irregular oscillations, and finally to chaos i.e. totally decorrelated process. Recall that the dynamical systems, described by ODEs, in general may have four types of solutions: equilibrium states, regular (periodic) motion, quasi-periodic motion and chaos. These solution types are associated with four attractor varieties: stable equilibrium, limit cycle, d-dimensional torus and chaotic attractor.

One usually studies the properties of the logistic map with the help of a computer, giving the seed $x_0$ and the growth parameter $a$ as inputs. The task is to find the asymptotic value of $x_i$ for $i \to +\infty$; one should of course bear in mind that computer modeling gives no genuine asymptotics, but only the values corresponding to large finite $i$-numbers. Yet one can obtain the graph $(x, a)$ which is known as a bifurcation diagram for the logistic map. This diagram plots a sequence of generations (population densities $x_i$) vs. growth parameter $a$, actually representing the limit solutions i.e. the same and cyclically repeated (after some number $m$ of steps) values of $x$. Such cycles appear following the attempts to find the already mentioned fixed (also known as stationary) points of the map $x_{i+1} = \varphi(x_i, a)$ i.e. limit solutions of equation $x = \varphi(x, a)$ mapping point $x$ onto itself. Fixed points can be both stable and unstable, in particular, depending on the values of the control parameter $a$. Specifically, for the logistic map, when $0 < a < 1$ the only limit value is $x = x^{(1)} = 0$, and at $a = 1$ the first bifurcation occurs: now there are two solutions i.e. besides $x^{(1)} = 0$ a new solution $x^{(2)} = 1 - 1/a$ appears. Indeed, the search for fixed points gives $x = ax(1 - x)$ which produces $x^{(2)}$ for $x \neq 0$. In the range $1 < a < 3$, the computed limit value corresponds to this solution i.e. fixed points coincide with $x^{(2)} = 1 - 1/a$ since it is stable whereas the first solution $x^{(1)} = 0$ is unstable. We can test the stability of both solutions using the standard linearization procedure. Putting $x = x^* + \delta x$, where $x^*$ is either $x^{(1)}$ or $x^{(2)}$, we have after linearization $\delta x_{i+1} = a(1 - 2x^*)\delta x_i$, and we see that when $x^* = 0$, this solution is stable for $a < 1$ and unstable for $a > 1$. If $x = x^{(2)}$, then $\delta x_{i+1}/\delta x_i = 2 - a$ i.e. $\delta x_i = (2 - a)^i \delta x_0$ and $\delta x_i$ converges to zero for $|2 - a| < 1$ and diverges for $|2 - a| > 1$. Thus, solution $x^{(1)}$ becomes unstable for $a > 1$ and solution $x^{(2)}$ for $a > 3$; within interval $1 < a < 3$, $x^{(2)}$ remains stable. At point $a = 3$, the second bifurcation occurs: apart from solutions $x^{(1)} = 0$ and $x^{(2)} = 1 - 1/a$, both of which become unstable, a stable two-cycle ($m = 2$) fixed point emerges. One can predict the population for this two-cycle attractor e.g. by requiring that generation $(i + 2)$ has the same number (density) of individuals as generation $i$: $x = x_i = x_{i+2} = ax_{i+1}(1 - x_i)$. Combining this relationship with the logistic map, we get $x = a^2x(1 - x)(1 - ax + ax^2)$. One may notice that solutions $x^{(1)} = 0$ and $x^{(2)} = 1 - 1/a$ satisfy this algebraic equation. To find two other solutions, one can divide the equation by $x - x^{(1)} = x \neq 0$ and $x - x^{(2)} = x - (1 - 1/a)$ to obtain the quadratic equation $x^2 - (1 + 1/a)x + (1/a)(1 + 1/a) = 0$ whose solutions are

$$x = \frac{1 + a \pm \sqrt{a^2 - 2a - 3}}{2a}.$$

Putting here the value $a = 3$ that delimits the new bifurcation interval in the diagram, we get the exact solution $x = 2/3$ (which can also be verified directly from the logistic map). It is remarkable that a simple computer iteration procedure demonstrates rather slow convergence to this exact solution. It is also interesting that one can already observe the onset of oscillations: when $2 < a < 3$, the population density $x_i$ begins to fluctuate near the solution $x^{(2)} = 1 - 1/a$. With the growth parameter exceeding $a = 3$, oscillations become more and more pronounced, at first between two values (depending on $a$) and then, with a further increase of the growth parameter, between 4,8,16,..., $2^m$ values. The size ratio of subsequent bifurcation intervals on the diagram converges to the so-called Feigenbaum constant $\delta = 4.6692 \ldots$ that has been computed to more then $10^3$ decimal places. So with the growth parameter being increased, period doubling bifurcations occur more and more often, and after a certain value $a = a^*$ (sometimes called an accumulation point) has been reached, period doublings transit to chaos. One can also notice that different parts of the bifurcation diagram contain similar plots i.e. they are self-similar.

The physically limiting case $a = 4$, i.e. $x_{i+1} = 4x_i(1 - x_i)$ is especially often encountered in the literature on logistic maps because one can find an exact solution in this case. One usually makes the substitution $x_i = (1/2)(1 - \cos 2\pi\theta_i) \equiv \sin^2 \pi\theta_i$ to obtain the following form of the logistic map: $\cos 2\pi\theta_{i+1} = \cos 4\pi\theta_i$. One can then define the principal value of $\theta_i$ as belonging to interval $[0,1/2]$ so that $\theta_{i+1} = 2\theta_i$ and $\theta_i = 2^i\theta_0$. Thus depending on the initial value $\theta_0$, the map admits a countable set of cyclic (periodic) solutions and a continuum of aperiodic solutions. Periodic solutions can be found by putting successively $\theta_0 = 1$ ($x = 0$); $1/2$ ($x = 0$); $1/3$ ($\theta_i = 1/3,2/3,4/3,8/3, \ldots$ i.e. stationary solution $x = 3/4$ ); $1/5$ (double cycle $\theta_i = 1/5,2/5,4/5,8/5 \ldots \to 1/5,2/5$ i.e. $x_1 \approx 0.345, x_2 \approx 0.904$); $1/7$ (triple cycle $1/7,2/7,4/7$ i.e. $x_1 \approx 0.188, x_2 \approx 0.611, x_3 \approx 0.950$), etc. Using the standard trick of linearization, one can see that all such solutions are exponentially unstable, $\delta\theta_i = 2^i\delta\theta_0$ so that the actual solutions fluctuate between such unstable ones: the situation known as chaos (or at least quasi-chaos, when some solutions still remain stable).

This example is quite instructive since one can observe on it the emergence of simple statistical concepts. Indeed, although chaos is a dynamical notion arising in deterministic systems (there are no random variables in the starting equations), the continuum chaotic states can only be adequately described by introducing a probability measure i.e. probability to find the population between $x$ and $x + dx$. It is this indispensable emergence of probabilistic description that is usually known as stochasticity. To find the respective distribution function, let us, in accordance with the archetypal model of statistical mechanics, consider the simplified logistic map $\theta_i = 2^i\theta_0$ as an ensemble of solutions labeled by different initial conditions. A dynamic map $x = g_t x_0$, where $g_t$ is a one-parameter group (or semigroup) of translations along vector field trajectories, takes these initial conditions into the actual state of the system. For example, in the case of continuous time and smooth vector field $v(x)$ on an $n$-dimensional manifold $M$, $g_t$ is a measure-preserving diffeomorphism (see the above section on dynamical systems) i.e. if $\mu$ is a measure that in any local coordinate system is represented through the probability density, $d\mu = \rho(x)dx, x = \{x^1, \ldots, x^n\}$, then measure $\mu$ is invariant under the action of $g_t$ if density $\rho(x)$ satisfies the Liouville equation, $\partial_i(\rho v^i) \equiv \text{div}(\rho\boldsymbol{v}) = 0$ (the Liouville theorem, see sections 4.1, 4.5.3.). Such a measure is an integral invariant of flow $g_t$ (i.e. of dynamical system). A similar probability distribution for the logistic map with growth parameter $a = 4$ for fixed $i$ is $\rho(\theta_i) = d\theta_i/d\theta_0 = 2^i = \text{const}$. But from substitution $x_i = (1/2)(1 - \cos 2\pi\theta_i)$ we have

$$\frac{d\theta_i}{dx_i} = \frac{1}{\pi \sin 2\pi\theta_i} = \frac{1}{\pi\sqrt{1 - \cos^2 2\pi\theta_i}} = \frac{1}{\pi\sqrt{1 - (1 - 2x_i)^2}} = \frac{1}{2\pi\sqrt{x_i(1 - x_i)}}$$

and, in particular, $\frac{d\theta_0}{dx} = \frac{1}{2\pi\sqrt{x(1-x)}}$. Defining $d\mu = \rho(x)dx := \rho(\theta)d\theta$, we get $\rho(x) = \frac{\text{const}}{\sqrt{x(1-x)}}$, where the constant can be determined from the normalization condition, $\int_0^1 \rho(x)dx = 1$ which gives const $= 1/\pi$. Finally, we have the distribution function for the probability to find the logistic system (e.g., the population) between $x$ and $x + dx$

$$\rho(x) = \frac{1}{\pi\sqrt{x(1 - x)}} \tag{4.16.1.}$$

Notice that this distribution function does not depend on the starting value $0 \leq x_0 \leq 1$ i.e. is universal. This is a specific feature of the physically limiting case of the logistic map with $a = 4$.

One can show that the logistic map for $a = 4$, which is in this case chaotic for almost all initial conditions, may be related to the Lorenz attractor appearing in the three-dimensional meteorological model constructed in 1963 by E. Lorenz. This mathematical model is represented by a dynamical system with 3d phase space and is fully deterministic since it is represented by three ODEs $\mathbf{x} = \mathbf{f}(\mathbf{x}, a), \mathbf{x} = (x^1, x^2, x^3)$ (with quasilinear vector field $\mathbf{f}$). Nonetheless, the model demonstrates chaotic behavior i.e, abrupt and apparently random changes of state for some set of control parameters $a$.

We have already noted that there are many interesting things about chaos and one of them is that it had not been discovered much earlier. The logistic map, for instance, could well have been explored by the brilliant Enlightenment mathematicians, but probably, after the creation of Galileo-Newton's mechanics in the late 17th century, scientists were preoccupied with continuous-time mathematical analysis and differential equations describing the objects that change smoothly with time. Even today, in the epoch of digital technologies, many "old-guard" scientists are much better familiar with differential equations than with their discrete counterparts.

Another chrestomathic example of a discrete-time system $x_{n+1} = f(x_n), n \in \mathbb{Z}$ besides the logistic map is given by piecewise linear (!) maps known as the Bernoulli shift $B(x)$

$$f \colon [0,1), f(x) \equiv B(x) = \begin{cases} 2x, 0 \leq x < 1/2 \\ 2x - 1, 1/2 \leq x < 1 \end{cases} = 2x \bmod 1.$$

This simple map produces a rather complex dynamics which is characterized by an extreme sensitivity to initial conditions which can be demonstrated, as usual, by computing the Lyapunov exponents. Indeed, for two paths beginning in two nearby points $x_0$ and $x_0' = x_0 + \varepsilon_0$ with displacement $\varepsilon_0 \ll 1$ we shall have $\Delta x_n \coloneqq |x_n' - x_n| = 2\varepsilon_{n-1} = 2^2 \varepsilon_{n-2} = \cdots = 2^n \varepsilon_0 \equiv e^{n \ln 2} \varepsilon_0$. We see that two closely located points diverge with the rate $\lambda = \ln 2 > 0$, which is the Lyapunov exponent for map $B(x)$. Since $\lambda > 0$ the map exhibits an exponential instability and can be viewed as chaotic.

Discrete time maps $x_{n+1} = f(x_n), n \in \mathbb{Z}$ are also known as fixed point iterations. In general, single-dimensional discrete time maps $f(x) \colon x \in X \subseteq \mathbb{R}, X \to X$ that may be represented as $x_{n+1} = f(x_n)$ (an obvious generalization of the logistic map), even very primitive ones, can also exhibit an exponential dynamical instability and thus may be called chaotic (in the Lyapunov sense). Expression $x_{n+1} = f(x_n)$ supplied with initial condition $x(0) = x_0$ (as the initial population in the logistic map) can be viewed as an equation of motion for our 1d deterministic dynamical system with discrete time.

Let us now return to the continuous-time case (such systems are the most interesting for physics and technology). In accordance with the idea of linearity, the model of exponential growth, discussed in the preceding section, can be applied when the number $N$ of individuals is relatively small so that the correlation effects between them can be disregarded. Correlations between individuals can be observed on both the pairwise and collective level[114]. Pair correlations give rise, e.g., to the hyperbolic or explosion model leading to the sharpening regime with vertical asymptotes meaning that the population will tend to infinity in finite time (this model was briefly discussed in 4.15. and 4.16.), whereas collective correlations manifest themselves, for example, in the competition for resources (such as food, water, living space, energy, information, position in the hierarchy, political power, etc.) within the population. With the rising number of individuals, competing for resources tends to impede the growth rate i.e. growth factor $b = b(N), db/dN < 0$. The simplest model of an impeded growth would be putting rate $b$ to fall linearly with growing population, $b = a - kN, a > 0, k > 0$. This mathematical model is quite natural since one can approximate any smooth function by a linear one for sufficiently small values of its argument (this is, by the way, the main idea of calculus and differentiation), in the present case, for small enough $N$. Then we have the equation expressing, in the continuous case, the logistic model:

---

[114] The patterns of pairwise vs. collective correlations is quite often encountered in many-particle physical models, e.g., in statistical and plasma physics.

$$\frac{dN}{dt} = (a - kN)N = aN\left(1 - \frac{N}{N_0}\right), \qquad (4.16.2.)$$

where $N_0 \equiv a/k$ is the equilibrium population (in ecological literature this parameter is usually denoted as "carrying capacity" $K$). One-dimensional equation (4.16.2.) is known as the logistic equation; this is an apparently primitive, but very rich and important mathematical model which may be interpreted as a combination of the exponential and hyperbolic models. The meaning of the logistic model is that the reproductive rate is assumed to be proportional to the number of individuals, whereas the mortality rate is proportional to the frequency (probability) of pair encounters (collisions). Of course, by properly scaling time $t$ and number of individuals $N$, e.g., $t \to \tau := at, N \to kN/a = N/N_0 := x$, we can reduce (4.16.2.) to a dimensionless form $\dot{x} = x(1-x)$ which is convenient to study the main dynamical properties of the model in the $(x, \tau)$ plane, but we shall rather stick to the form (4.16.3.) in order to better understand the role of parameters $a$ and $N_0$.

The logistic equation, though being a nonlinear ODE, is simple in the sense that it can be easily integrated (since it is autonomous and variables in it can be separated). There are many ways to integrate the logistic equation; for demonstration purposes, we can choose probably the simplest one, representing the logistic equation as

$$\frac{dN}{dt} = (a - kN)N = aN - kN^2, N(t_0) = B, a > 0, k > 0, \qquad (4.16.3.)$$

$B, a, k$ are constants. For $N \neq a/k$, $t - t_0 = \int_B^N dN(aN - kN^2)^{-1}$. This integral exists only when both $N$ and $B = N(t_0)$ i.e the current and the initial values of the population lie in intervals $(0, a/k)$ or $(a/k, +\infty)$. In other words, there exist no solutions that cross the straight line $a - kN = 0$. Integration gives

$$t - t_0 = \frac{1}{a}\int_B^N dN\left(\frac{1}{N} + \frac{k}{a - kN}\right) = \frac{1}{a}\log\frac{N(a - kB)}{B(a - kN)}$$

Solving this equation with respect to $N$, we have

$$N(t) = \frac{aB\exp[a(t - t_0)]}{a - kB + kB\exp[a(t - t_0)]} = \frac{aB}{kB + (a - kB)\exp[-a(t - t_0)]}$$

One can see that for an initial population lower than the equilibrium value, $N(t_0) < N_0$ i.e. $0 < B < a/k, N(t)$ is defined for all $t, 0 < t < +\infty$, while for $N(t_0) > N_0$ i.e. $B > a/k, N(t)$ is only defined for

$$t > t_0 - \frac{1}{a}\log\frac{kB}{kB - a}$$

In the case $N(t_0) \equiv B = a/k$, the solution is a constant, $N(t) = a/k$, since in this case $dN/dt = 0$.

One can also see that the solution converges to the constant value $a/k$. For $B < a/k, N(t) < N_0 \equiv a/k$ for all $t$ so that $a - kN(t) > 0$ and $dN/dt > 0$ which means that if the initial population is lower than the equilibrium one, the number of individuals monotonicly increases. In the opposite case, when the starting population exceeds the equilibrium one, $B > a/k$, we have $N(t) > N_0$ for all $t$ and $dN/dt < 0$ which means that the population is monotonicly shrinking.

The linear regime of the logistic model corresponds to the case when one can neglect the second term (proportional to $N^2$) in the logistic equation. This is correct for small probabilities of pair correlations and for short observation times. More accurately, $(kB/a)(\exp[a(t - t_0)] - 1) \ll 1$ which gives for the short period of observation, $a(t - t_0) \ll 1$, the following restriction on the death rate in the population, $kB(t - t_0) \ll 1$. An obvious drawback of the logistic model is that it does not contain spatial variables, which means that it cannot be applied to spatially inhomogeneous situations.

The most frequently used form (4.16.2.) of the logistic equation introduces explicitly the asymptotic value $N_0 = k/a$. Then the solution with the initial condition $N(t = t_0) \equiv B$ can be written as

$$N(t) = \frac{BN_0 e^{a(t-t_0)}}{N_0 + B(e^{a(t-t_0)} - 1)} = N_0 \frac{N(t_0)}{N(t_0) + (N_0 - N(t_0)e^{-a(t-t_0)})}  \quad (4.16.4.)$$

So, by directly integrating the logistic equation (which is a rare occasion in the world of nonlinear equations), we get the explicit expression for integral curves that are usually called the logistic curves whose family depends (in the deterministic case) on three parameters, $a, N_0$ and $B = N(t_0)$. The graph of this solution produces an S-shaped curve which can be represented in dimensionless units by the function $N(t) = 1/(1 + \exp(-at))$, in which one can easily recognize the Fermi distribution of fermions over single-particle energy states in quantum statistical physics. The process described by the logistic model (the logistic process) has two equilibrium points, $N = 0$ and $N = N_0$, with the first being unstable, since small population $\delta N$ grows near $N = 0$ ($\dot{N} > 0$), and the second stable ($\dot{N} < 0$ for $N > N_0$ and $\dot{N} > 0$ for $N < N_0$ near $N = N_0$ (which can be seen already from equation (4.16.2.) i.e. without even integrating it). In other words, the population dynamics evolves to the equilibrium value $N = N_0$. More exactly, the process tends asymptotically for $t \to +\infty$ to the stable equilibrium $N = N_0$ at any starting value $N(0) = B > 0$. For $t \to -\infty$, the process asymptotically converges to the state $N = 0$. Thus the logistic model describes the transition from the unstable state $N = 0$ to the stable state $N = N_0$, occurring in infinite time. We shall see the examples of similar transitions when addressing quantum-mechanical models below. The integral curves have vertical asymptotes $t = $ const for any $t > 0$. The logistic process is very close to the exponential (Malthusian) growth for small $N, (N \ll N_0)$, but begins to fall behind approximately at $N \approx N_0/2$. This saturation effect is a manifestation of correlations within the population.

One usually considers a single control parameter in the logistic model, the growth parameter $a$. In fact, however, there are at least two control parameters which can be both important, especially if the logistic model is not necessarily applied to describe the population growth, when all the parameters and variables are positive by default. Renaming the constants, e.g., $a = kN_0^2, b = a/kN_0$ (see (4.16.3.)-(4.16.4.)), we arrive at equation $\frac{dx}{dt} = f(x, a, b) = ax(b - x)$ whose solutions depend on two parameters $a$ and $b$, not necessarily strictly positive. In other words, the dynamical evolution occurs in 3d space $(x, a, b)$ without restricting the motion to domain $x > 0, a > 0, b > 0$ as in population models. For $b > 0$, point $x = 0$ is unstable whereas for $b < 0$ it is stable; at $b = 0$ stability is changed to instability both at the stable point $x = b$ and the unstable point $x = 0$ of the traditional logistic model – a simple example of the bifurcation.

One can emphasize that there is a signifcant contrast between the solutions to differential equations and the ones to the respective difference equations (the logistic map). If, e.g., we take differential equation $\dot{x} = ax(b - x)$, its solution (an S-curve) can be easily obtained:

$$x(t) = \frac{x_0 e^{at}}{1 - x_0(1 - e^{at})}, x_0 = x(0)$$

(see above, here $t_0 = 0$), whereas the solution to the logistic map $x_{i+1} = ax_i(1 - x_i)$ obtained by iterating this difference equation has a totally different character reflecting much more complicated behavior. This behavior has been studied by many researchers and is still hardly totally understood.

From a more general viewpoint, the logistic model is a very particular case of the evolution of an autonomous system described by vector equation $\dot{x} = \mathbf{f}(\mathbf{x}, a)$, where vector field $\mathbf{f}$ depends on parameter $a$ (it can also be a vector). As we have seen, in certain cases, variations of the control parameter $a$ can radically change the system's motion, for instance, result in chaotic behavior. Note that the logistic model is not necessarily identical with the population model. When the logistic equation is not interpreted as describing the population growth, one can explore the behavior of solutions as parameter $k$ (see (4.16.2.)-(4.16.3.)) is varied.

### 4.15.1. Extensions of the logistic model

Although the logistic model was initially devised to describe mathematically the population dynamics, in particular, the survival conditions, this model has a more general meaning. We have already mentioned the similarity between the continuous-time logistic model and the self-sustained oscillations in laser generation. Indeed, one often writes the logistic equation in the extended form $dN/dt = aN(1 - kN) = (\alpha - \gamma - \beta N)N$, where parameters $\alpha$ and $\gamma$ define the birth and mortality rates, respectively, whereas the nonlinear term $\beta N$ corresponds to population shrinking due to intraspecific competition for resources. This form of the logistic equation is analogous to one of the forms of the Van der Pol equation written for the oscillation energy $E = \left(\frac{m}{2}\right)(v^2 + \omega_0^2 x^2)$ (in fact the Lyapunov function)

$$\frac{dE}{dt} = (\alpha - \beta E)E, \qquad \alpha = \alpha_0 - \gamma,$$

where $\alpha_0$ is the feedback parameter, $\gamma$ and $\beta$ are friction coefficients (linear and nonlinear). Thus, the birth rate in the logistic equation characterizes the positive feedback, while the mortality rate accounts for friction.

This analogy enables one to link the logistic model with the nonlinear theory of Brownian motion. This latter theory is quite universal and has a number of important physical, chemical and biological applications, in particular, in electrical and chemical engineering (e.g. catalysis), radiophysics, laser technology, biological self-organization, etc. Mathematically, the analogy between the logistic model and the nonlinear Brownian motion may be expressed on the level of the Fokker-Planck equation with the nonlinear diffusion coefficient, $D(N) = \gamma + \beta N$:

$$\frac{\partial f(N,t)}{\partial t} = \frac{\partial}{\partial N}\left(D(N)N\frac{\partial f(N,t)}{\partial N}\right) + \frac{\partial}{\partial N}[(-\alpha + \gamma + \beta N)Nf(N,t)],$$

where $f(N,t)$ is the distribution function characterizing the probability of the population to be in a state with $N$ individuals at time $t$. Notice that the description of population dynamics both in terms of the continuous-time logistic equation and the Fokker-Planck equation is valid for a large number of individuals, $N \gg 1$. One can read about the relationship of the logistic equation to stochastic differential equations and closely connected theory of nonequilibrium phase transitions in the often-cited book [301].

Both the logistic map and the logistic model can be extended and improved. Thus, the growth (control) parameter $a$ in the logistic map may depend on discrete time i.e. iteration step $i$ so that we shall have $x_{i+1} = a_i x_i (1 - x_i)$. This is a discrete-time version of the optimal control problem, and the corresponding extremal properties similar to Pontryagin's maximum principle in continuous-time dynamical systems can be established. In a more general situation than the one-dimensional logistic map, the discrete-time iterated process may have a vector character, $\mathbf{x}_{i+1} = \boldsymbol{\varphi}(\mathbf{x}_i, \mathbf{a}_i)$, where $\mathbf{x}_i = \{x_i^1, \ldots, x_i^p\}$, $\mathbf{a}_i = \{a_i^1, \ldots, a_i^q\}$ ($\mathbf{a}$ is known as a decision vector), and index $i$ enumerates iteration steps. The simplest logistic model with scalar control is of course linear, e.g. $a_i = a_0(1 + \sigma i), i = 1,2,\ldots$ One can rewrite the logistic map for the variable growth parameter as $x_{i+1} - x_i = a_i x_i (X_i - x_i)$, where $X_i \equiv 1 - 1/a_i$ (recall that this quantity coincides, for $a_i = a = $ const, with one of the main fixed points of the logistic map, $X = x^{(2)}$). If we assume that $\Delta a_i = a_{i+1} - a_i \ll a_i$, then we can approximate the logistic map by the continuous-time equation, $x_{i+1} - x_i \equiv \frac{x_{i+1} - x_i}{1} \approx \frac{dx}{dt}$ and $a_i \approx a(t)$ so that $\frac{dx}{dt} =$ $a(t)x(X(t) - x)$ or, in the form (4.16.2.) $\frac{dx}{dt} = a^*(t)x\left(1 - \frac{x}{X(t)}\right)$, where $a^*(t) \coloneqq a(t)X(t)$. We see that the logistic map under the assumption of small variations of the growth parameter between two successive generations is reduced to the continuous-time logistic model with variable growth parameter and drifting stable equilibrium point. The drift on a slow time scale of equilibrium population $X$ is quite natural since the ecosystem carrying capacity may change with time due to human technological impact, biological evolution or geophysical variations (such as climate change).

One can consider two subcases here: one corresponding to adiabatic variations of parameters $a^*(t)$ and $X(t)$: i.e. they both change little during the characteristic time $t \sim 1/a^*$ of the process i.e. $|da^*(t)/dt| \ll \left(a^*(t)\right)^2$, $|dX(t)/dt| \ll a^*(t)X(t)$ on some set of temporal values $t$; the other reflects the opposite situation of abrupt changes. In the linear model of growth parameter variation, $a^*(t) = a_0^*(1 + \sigma t)$ (one can omit the asterisks for simplicity), the adiabatic regime can be described by a series over small parameter $\sigma$ (the characteristic time of slow control parameter variation is $1/\sigma$). Writing the logistic equation in the form $x = X - \frac{\dot{x}}{ax}$ and noticing that $\dot{x} \sim (a_0\sigma t)x$, we may obtain the expansion $x = X + \delta x^{(1)} + \delta x^{(2)} + \cdots$, where $\delta x^{(1)} = -\frac{\dot{X}}{a_0 X} \lesssim \sigma t$, $\delta x^{(2)} = -\frac{1}{a_0 X}\left(\dot{X}\frac{\delta x^{(1)}}{X} + \frac{d\delta x^{(1)}}{dt}\right) = \frac{\ddot{X}}{(a_0 X)^2}$, etc. The meaning of the adiabatic regime is that the stable fixed point $X$ (e.g., interpreted as the eventually reached population) is slowly drifting and the solution $x(t)$ (current population) adjusts to this smooth evolution of parameters. In this adaptation process a band of values is formed instead of pointlike solution $x$. The positions of bifurcation points also change a little so that the initial bifurcation diagram is slightly distorted.

In the other limiting subcase of abrupt changes i.e. occurring at times $\tau \ll 1/a$, one can consider all the model parameters $(a, X, \dots)$ to remain constant during the change so that the solution will jump between two levels (or two bands). This situation is close to the ones treated within the framework of perturbation theory in quantum mechanics (recall that the quantum-mechanical perturbation theory had its origins in the methods of finding solutions to ordinary differential equations).

One can of course consider the system of two or more interacting populations, each being described by the logistic model. For two populations we might have the following system of coupled logistic equations

$$\frac{dx_1}{dt} = a_1 x_1\left(1 - \frac{x_1 + x_2}{N}\right), \qquad \frac{dx_2}{dt} = a_2 x_{21}\left(1 - \frac{x_1 + x_2}{N}\right). \qquad (4.16.1.1.)$$

We can explore the stability of solutions to this model (with a two-dimensional phase space in the same way as for a single population) and, e.g., draw the phase portrait. The model of two coupled populations belongs to the class of competition models which we shall discuss shortly.

We have already noted that the logistic model does not account for spatially inhomogeneous processes. In general, mathematical and computer models that are constructed by averaging out the spatial[115] information such as territorial population distribution and migration and keeping only the time variable as the vector field parameter hide many essential characteristics. To overcome this obvious drawback of the logistic model, one can use a simple trick of introducing one more parameter. Namely, we can write the initial condition $N(t_0) \equiv B$ (see equation (4.16.3.)) in the form $N(x, t_0) \equiv B(x) = 1/(1 + \exp(-px))$, where $x$ is interpreted as a spatial variable and treated as a supplementary parameter. One can regard variable $x$ as completely hidden due to some averaging, with only the time variable remaining directly accessible. The spatially dependent initial condition is assumed to restore the hidden variable and to evolve with time into a space-dependent solution

$$N(x, t) = \frac{1}{1 + e^{-p\xi}}, \qquad \xi = x + wt$$

(here, we put for simplicity $t_0 = 0$). In this extension of the logistic model, one can interpret $N(x, t_0) = N(x, 0)$ as a spatially inhomogeneous distribution that spreads with time as a kinematic wave: its propagation is due to nonlinear terms and is not caused by dynamical interactions. If we put $\xi = 0$, we get $x = wt$ i.e. a steady profile. In other words, condition $\xi = 0$ corresponds to choosing the coordinate frame moving with transfer velocity $w = x/t$.

The mentioned shortcoming of the model - absence of spatial information – can also be surmounted by using the logistic model combined with partial differential equations, $Lu = \varphi(u)$, where $L$ is some operator (not necessarily linear) acting on the space on which function $u$ is defined, $\varphi(u) = au(1 - u)$. Here for brevity the dimensionless form is used. Operator $L$ is assumed to contain a spatial part, for instance, $L = \partial_t - \partial_{xx}$ (diffusion) or $L = \partial_{tt} - \partial_{xx}$ (linear waves), $L = \partial_t - (\eta(u))_{xx}$ (nonlinear diffusion), etc. In such models, each zero of $\varphi(u)$ corresponds to a stationary solution of the associated PDE, $Lu = 0$. When $L$ is the diffusion operator, one can interpret the problem simply as supplementing the logistic model with a spatially dependent spreading part (which reminds one of the Navier-Stokes equation)

$$u_t = a\Lambda^2 u_{xx} + \varphi(u) \qquad\qquad (4.16.1.2.)$$

so that solutions now depend on both space and time variables, $u = u(x, t)$. Here quantity $\Lambda$ is the so-called diffusion length, by an appropriate choice of scaling one can set $\Lambda^2$ to unity. Models similar to (4.16.1.2.) are rather popular in biological sciences and biomathematics, for example, Fisher's model of biological adaptation[116]. One can interpret this diffusion model as a strategy to compute varying distributions of gene frequencies within the population – an approach known today as population genetics. Equation (4.16.1.2.) is also said to describe nonlinear diffusion and is known as the reaction-diffusion equation. One often ascribes the probabilistic meaning to the right-hand side $\varphi(u)$ (it is interpreted as being proportional to the product of probabilities $p$ that an event occurred and $q = 1 - p$ that it did not), but mathematically this is still the function forming the logistic equation. The corresponding PDE has two stationary solutions $u = 0$ and $u = 1$ and a traveling wave solution, $u(x,t) = F(x - wt)$. Note that equation (4.16.1.2.) is invariant under reflection $x \to -x$ so that velocity $w$ can be both positive and negative. It is also clear that (4.16.1.2.) is invariant under affine translations of $x$ and $t$, therefore one can add any arbitrary constant to $z := x - wt$. Inserting the ansatz $u = F(z)$ into PDE (4.16.1.2.), we get the second-order ODE

$$F'' + \frac{w}{a\Lambda^2} F' + \frac{1}{a\Lambda^2} \varphi(F) = 0, \qquad (4.16.1.3.)$$

where the prime denotes differentiation over $z$. This ODE is of the nonlinear damped oscillator type (sometimes equations of this type are called the Liénard equations), and after solving it, we can find, if possible, asymptotic solutions for $x \to \pm\infty$ and the transitions between two stationary points $u_1 = u(-\infty) = F(-\infty)$ and $u_2 = u(+\infty) = F(+\infty)$ (in the case of the underlying logistic model, $u_1 = 0, u_2 = 1$). Such asymptotic requirements play the role of boundary conditions for (4.16.1.3.); now the problem is to determine the values of $w$ (one may consider them eigenvalues) for which solution $F \geq 0$ satisfying these asymptotic conditions exists, and if it does, then for what initial states $0 \leq u(x,0) \leq 1$. Will the solution $F$ take the form of a traveling wave for any initial distribution $u(x,0)$? This question proved to be rather nontrivial, and stimulated an extensive research. One of the natural ways to treat this problem is by representing (4.16.1.3.) in the form of a dynamical system

$$F' = y, y' = -\frac{w}{a\Lambda^2} y - \frac{1}{a\Lambda^2} \varphi(F),$$

then we can investigate the solution in the phase plane $(y, y')$ or, more conveniently, $(u, y) \equiv (y_1, y_2)$. The phase trajectories are obtained, as usual, by excluding the vector field parameter (here $z$), and we have[117]

$$\frac{dy}{du} = -\frac{1}{a\Lambda^2 y}\big(wy - \varphi(u)\big) \equiv -\frac{1}{a\Lambda^2 y_2}\big(wy_2 - \varphi(y_1)\big). \qquad (4.16.1.4.)$$

---

[115] Also, information provided from other underlying spaces, not necessarily of geometrical nature, such as biological, ecological, epidemiological, immunological, physiological, social, etc. character.

---

[116] R. Fisher was a prominent British biologist and statistician. Fisher's model was explored in detail by A. N. Kolmogorov, I. G. Petrovsky and N. S. Piskunov and is sometimes called the Fisher-KPP model.

[117] Here, to avoid confusion we are placing vector indices below. The difference between vectors and covectors is usually immaterial in the context of two-dimensional dynamical systems.

This equation has two equilibrium (also known as fixed, singular or critical) points in the $(u, y) \equiv (y_1, y_2)$-plane: (0,0) and (1,0). In the vicinity of point (0,0), one can linearize equation (4.16.1.4.), which corresponds to the transition from logistic to exponential growth model, obtaining

$$\frac{dy}{du} \approx -\frac{1}{a\Lambda^2 y}(wy - au) = \frac{u}{\Lambda^2 y} - \frac{w}{a\Lambda^2} \qquad (4.16.1.5.)$$

or, identically,

$$\frac{dy_2}{dy_1} = \frac{y_1}{\Lambda^2 y_2} - \frac{w}{a\Lambda^2}$$

When exploring the behavior of solutions, one can observe here the competition between two terms in the right-hand side, therefore the phase portrait changes its character for velocity $w$ greater or smaller some critical value ( $w = w_c$ ). By solving (4.16.1.5.), one can show that $w_c \sim a\Lambda$ . To determine the type of fixed points and thus to characterize the phase flow, we may proceed according to the standard prescriptions of two-dimensional dynamical systems. Writing $y_1' = y_2, y_2' = -\frac{1}{\Lambda^2}y_1 - \frac{w}{a\Lambda^2}y_2$ and looking for a solution of the form $\exp(\mu z)$, we obtain the system matrix

$$A = \begin{pmatrix} 0 & 1 \\ -\dfrac{1}{\Lambda^2} & -\dfrac{w}{a\Lambda^2} \end{pmatrix}$$

(notice that quantity $w/a\Lambda^2$ is formally analogous to the dissipation coefficient). The trace of matrix $A$ is $\text{Tr } A = -w/a\Lambda^2$ and $\det A = 1/\Lambda^2$ so that the characteristic equation is $\mu^2 - \mu \text{Tr } A + \det A = \mu^2 + \mu w/a\Lambda^2 + 1/\Lambda^2 = 0$, and the roots are real and different when discriminant $D = (w^2 - 4a^2\Lambda^2)/4a^2\Lambda^4 > 0$ i.e. $|w| > 2a\Lambda$. This last value can be identified with $w_c$. Assume at first that the roots of the characteristic equation have the same sign, then the real solutions are $y_1 = y_{10}e^{\mu_1 z}, y_2 = y_{20}e^{\mu_2 z}$ , where $y_{10}, y_{20}$ are arbitrary constants. One can, as before, get rid of parameter $z$ and obtain either the family of parabola-like[118] orbits $|y_1| = c|y_2|^{\mu_1/\mu_2}$, where $c$ is some constant (independent of $z$), or $y_1 = 0$. Recall that critical point of this kind is called a node; if $\mu_1, \mu_2 < 0$ i.e. $w > 2a\Lambda$, then point (0,0) is a positive attractor i.e. there exists a neighborhood of (0,0) such that all the paths starting at this neighborhood at some $z = z_0$ finish at (0,0) with $z \to +\infty$ (this property is almost obvious due to the exponential character of solutions). Likewise for $\mu_1/\mu_2 > 0$, the critical point (0,0) is a negative attractor. From the physical viewpoint, condition $w > 2a\Lambda$ signifies that the disturbance for $t > 0$ moves in the direction $x > 0$ (recall that $u = F(x - wt)$ ). If $0 < w < 2a\Lambda$ , the singular point (0,0) becomes an unstable focus, and the degenerate case $w = 0$ (i.e. a stationary disturbance; from the positions of dynamical systems, this fixed point is a center) leads to "unphysical" solutions $u < 0$ (recall that we need to have solutions in the band $0 \leq u \leq 1$). One may note that in general equilibrium near the coordinate origin, when one can disregard nonlinearity, is an unstable focus.

---

[118] The ratio $\mu_1/\mu_2$ not necessarily equals 2 or ½.

In the same way one can explore the other singular point (1,0). In the vicinity of this point $u \approx 1$ so that the linearized equation, instead of (4.16.1.5.), is

$$\frac{dy}{du} \approx -\frac{1}{a\Lambda^2 y}\big(wy - a(u-1)\big) = \frac{u-1}{\Lambda^2 y} - \frac{w}{a\Lambda^2}.$$

Proceeding just as before, we see that point (1,0) is a saddle for all $w \geq 0$ as long as we consider growth factor $a > 0$ and constant and diffusion length $\Lambda$ real.

A more complicated mathematical setting of the Fisher-KPP model is the Cauchy problem for a nonlinear parabolic equation (parabolic equations express the most popular models arising from biological studies)

$$u_t(x,t) = (\eta(u))_{xx} + \varphi(u)$$

in domain $D := \{x \in \mathbb{R}, 0 \leq t < +\infty\}$ with initial condition $\lim_{t \to +0} u(x,t) = u_0(x)$ and, in some versions, asymptotic boundary values $\lim_{x \to -\infty} u(x,t) = 0$, $\lim_{x \to +\infty} u(x,t) = 1$, with a non-negative and continuously differentiable on [0,1] source function $\varphi(u)$ such as $\varphi(0) = \varphi(1) = 0, \varphi'(0) > 0$. The initial function $0 \leq u_0(x) \leq 1$ can be piecewise continuous in $\mathbb{R}$. When $\eta(u) = u$, we have a linear diffusive process with a nonlinear source.

### 4.15.2. Applications of the logistic model

Both the logistic model and the logistic map have many applications in science, society and engineering. The general idea leading to the logistic model – simple growth limited by self-interaction – may be applied to many real-life processes. For instance, epidemics and spread of rumors can be modeled by the logistic equation. We can take a typical microeconomic situation as another example: what will be the output of certain goods produced by a company (e.g. car manufacturer) over several years? The simplest model describing the annual growth of the output, with the imposed constraints of limited resources and market saturation would be a logistic model. What would be the total revenue (and respectively the profit) of a company over several years, if the annual revenue growth is $a$? The revenue for the $(i+1)$-th year will be $x_{i+1} = ax_i$. However, for rather high revenues the latter are restrained, e.g., by the market share, and we arrive again to the logistic model. These examples suggest a natural generalization: any human activity subordinated to the imposed constraints of external factors and/or limited resources can be described by a logistic model. The nonlinear term that limits growth is sometimes metaphorically interpreted as "influence of the future".

If we make an affine transformation $x_i = pz_i + q$, where $p, q$ are as yet undetermined parameters, we get from the logistic map a quadratic recurrence equation

$$z_{i+1} = -a\left[pz_i^2 - (1-2q)y_i - \frac{q}{p}(1-q)\right],$$

and putting $p = -1/a, q = 1/2$, we obtain the quadratic map $z_{i+1} = z_i^2 + c, c \equiv a/2 - a^2/4$ which produces Julia sets. Quadratic Julia sets are probably the best known examples of fractals that are generated by this quadratic map for almost any value of $c$ (although $c = 0$ and $c = -2$ are exceptional: the produced sets are not fractals). It is interesting that the above quadratic map was commercially used in the graphical industry to obtain rather beautiful ornaments which are fractals: this is an example of direct market value of mathematics.

It is remarkable that the logistic map can serve as a rough model of the transition to turbulence, when the regular (laminar) or almost periodic character of fluid motion is destroyed after some critical value of the flow parameter (the Reynolds number $Re$) has been reached. Like the eventual population in most logistic models, turbulence practically does not depend on the initial state of the fluid. The Reynolds number is a control parameter in the models of turbulent flow and plays the role of intrinsic growth parameter $a$. Multiplier $\mu$ passes in the transition to turbulence the value +1.

The logistic map manifests such common features of discrete-time algorithms as stability and chaotic behavior. From this viewpoint, it is interesting for numerical techniques and generally for computational science and engineering. As far as engineering applications of the continuous-time logistic model go, we have already mentioned that the equation describing the energy evolution of a nonlinear oscillator in the self-excitation mode has the form $\dot{E} = aE(1 - E)$ (in dimensionless units). Here the growth parameter $a$ is close to 1.

One can consider an example of estimating the population growth with the help of the logistic model. We may assume the total current (2010) human population to be $6.9*10^9$, the growth factor to be $a = 0.029$, the annual population growth to be 0.011 $year^{-1}$ (more or less standard demographic data). Then we have

$$\frac{dN(2010)/dt}{N(2010)} = \frac{d}{dt}\log(N(2010)/N(t_0 = 2010)) = 0.011 = a - kN(2010)$$
$$= 0.029 - k \cdot 6.9 \cdot 10^9$$

which can be considered an equation to find the attenuation factor $k \approx 2.32 \cdot 10^{-12}$. Then the projection for the equilibrium population will be $N_0 = a/k \approx 11.1 \cdot 10^9$ i.e., the world population tends to converge to approximately 11 billion people. This result is not very sensitive to the slight changes of the constant $a$ and the annual population growth rate.

We can comment here on the general concept of the control parameter whose variation results in altering the evolution regime of a dynamical system. Choosing the control parameter may present a special problem when modeling a complex (multiparametric) process, in particular, when the transition to chaos in an open system should be explored. For instance, in medico-biological studies the concentration of the prescribed medicine can play the role of control parameter. In surgery, the control parameter may be identified with the pre-planned invasion path. In fluid motion problems, one can define the control parameter as the pressure gradient between the flow boundaries, e.g., pipe ends. In laser technology, the control parameter can correspond to the pumping level i.e., energy input producing the population inversion. Using the engineering language, one can define an analogous control parameter as the feedback level in classical generators.

One often uses the logistic model in practical ecology to control the population. For example, a slight modification of this model (harvesting) is applied in fisheries. Harvesting means mathematically a transition from the population freely evolving in accordance with internal processes (balance of the birth, death, and migration) to introducing an external pressure $q(N, t)$ i.e. the model equation will be $\dot{N}(t) = \varphi(N) - q(N, t)$, where term $q(N, t)$ signifies the removal of $q$ individuals per unit time (say, each year). In fisheries, for example, this external influence corresponds to fishing quotas that can be constant, $q(N, t) \equiv q$ or differential, $q(N, t) = q(N)$. The latter case may be interpreted as the simplest manifestation of a feedback, which is a milder model than the one corresponding to $q = \text{const}$. Indeed, $q = q(N)$ depends on the actual state of the system. It is usually important to select quotas $q$ in such a way as to ensure sustainable fishing. Sometimes, as e.g. during genocides, external influence $q(N, t)$ cannot even be exactly known and should be treated as a perturbation, not necessarily small. One can in such cases figuratively call function $q$ the murder rate.

More complicated population models arise when pair bonding is considered. The effectiveness of survival mechanisms tends to decrease as the population density falls since it becomes increasingly difficult for sexually reproducing species to find appropriate mates. The respective mathematical models are in general spatially dependent, at least at the level of any given individual, since extended mobility and higher mate detection capability (greater identification distance) can to a certain degree compensate low population density. However, there are stringent physiological limitations for excessive mobility since it requires higher metabolism rates that are only possible under the conditions of ample resource availability (just as increased consumption in the rich population groups of human society enhances mobility and selectiveness). Considering pair stability issues further complicates the model.

To focus on the role of harvesting we can use, for simplicity, the dimensionless form of the logistic model that can be, in particular, achieved by scaling time $t$ as $\tau = aN_0 t$ (see equation (4.16.3.) and below). Formally, we can put the growth parameter $a$ and the "equilibrium population" $N_0$ equal to unity (it is trivial to notice that for $a \neq 1, q \to q/a$). Then we have $\dot{x} = x(1-x) - q \equiv f(x, q)$, and when quote $q$ is constant, its critical value is $q = 1/4$. In fisheries, this critical point is usually called maximum sustainable yield (MSY), and its evaluation is vital for sustainable fishing. For $0 < q < 1/4$, there are two equilibrium points ($\dot{x} = 0$) corresponding to two roots, $x_1, x_2$ of the resulting quadratic equation. The lower equilibrium point $x_1$ is unstable which means that if, for some reason, the population falls under $x_1$, then the entire population dies out in finite time. When $q > 1/4$, both equilibriums disappear, but what is more important, $f(x, q) < 0$ for all values of population $x$ i.e. *the population necessarily becomes extinct*. On the contrary, for $q < 1/4$ the population never dies out, and at the bifurcation point $q = 1/4$, the vector field $f(x, q) = 1/2$ i.e. equilibriums $x_1$ and $x_2$ merge, which, for sufficiently big initial population, can ensure that it will asymptotically end up near the $x = 1/2$ value. However, just a small drop of the population below this value would result in its extinction in finite time. In this sense, bifurcation point $q = 1/4$ is unstable. When the growth parameter $a \neq 1$, bifurcation occurs at critical value $q = a/4$.

The population evolution accompanied by forceful elimination of some part of the individuals represents simple feedback mechanisms. For example, one can exercise the ecological control over the quantities of certain species (e.g. mosquitoes), in such cases $q = q(x) > 0$ and $dq/dx > 0$ for all $x$. One can assume $q(x)$ to be a polynomial with positive coefficients. Let us consider the simplest case of a feedback scenario, $q(x) = bx$. In this case there are two stationary values, $x_1 = 0$ (unstable) and $x_2 = 1 - b$, $0 < b < 1$ (stable). We see that equilibrium point $x_2$ corresponds to elimination, e.g. harvesting or hunting, quota $bx_2 = b(1-b)$ with sustainable maximum (optimum) $b = x_2 = 1/2$. When $b \to 1, x_2 \to 0$ i.e. both equilibrium points merge, and the population becomes extinct in finite time ($\dot{x} = -ax^2 \to x(t) = 1/a(t - t_0)$). Physically, this signifies excessive harvesting, hunting or killing and geometrically to the disappearing crossing point of parabola $y = x(1 - x)$ and straight line $y = bx$. For $b > 1$ equilibrium point $x_2$ becomes negative. It is easy to see that when the growth parameter $a \neq 1$, we must scale $b \to b/a \equiv \tilde{b}$ so that a stationary state corresponding to an optimum is reached with $b = a/2$.

The case of quadratic elimination quota is analyzed similarly to the case of the linear one i.e. $\dot{x} = ax(1 - x) - bx - cx^2$ so that the stationary points are $x_1 = 0$ and $x_2 = \left(1 - \frac{b}{a}\right) / \left(1 - \frac{c}{a}\right)$, $0 < b < a$, $0 < c < a$. The optimum point is $b = a/2$.

The same model describes bankruptcy of companies and, with slight modifications, the downfall of political entities such as groups, parties, unions, states, etc. In certain models, e.g., taxation models in some economies, where $x$ denotes a tax and $\varphi(x)$ the taxation base, quotas $q(x)$ can be negative and piecewise continuous. The meaning of the model is that one should not harvest more than a certain threshold, otherwise the system will devour itself in finite time, regardless what it is: fisheries, small businesses, emigration of scientists and specialists. One can also interpret this model as the manufactirer-consumer equilibrium, where the logistic part $ax(1-x)$ corresponds to production and the harvesting part $-q(x) = bx + cx^2 + \cdots$ to the consumption of goods.

The lesson learned from studying the logistic model and logistic map is that apparently simple and completely deterministic dynamical systems can exhibit very complex motion that can be perceived as chaotic and stochastic. It may be instructive to make the computer implementation of the logistic model, in particular, trying to produce some computer codes corresponding to it. In Supplement 2, the Java code is given, where for simplicity the form (4.16.3.) is used i.e., with $N_0 = a/k$. One can also readily apply Euler's method to the logistic equation (see section on scientific computing below).

## 4.16 Instabilities and chaos

The most outstanding fact in the theory of dynamical systems is that fully deterministic systems depending on only a few variables can exhibit chaotic behavior which is similar to that formerly encountered only in many-body systems. Many nonlinear deterministic systems although looking quite simple are observed to behave in an unpredictable, seemingly chaotic way. The term "chaotic" is commonly attributed to evolutions exhibiting an extreme sensitivity to initial data, but this is not the whole story. Chaos is also understood as an aperiodic – close to random – behavior emerging from a totally deterministic environment i.e., described by a dynamical system involving no random parameters or noise. Such a behavior appears only in nonlinear systems and manifests an extreme sensitivity to initial conditions (in general, also to external parameters). Therefore, nonlinear systems can be viewed as one of the most interesting subjects in mathematics since even in the fully deterministic case they envisage a certain unpredictability.

Thus, classical chaotic systems, although they may have only a few degrees of freedom and be mathematically represented through deterministic equations, have to be described by probabilistic methods. Nonetheless, a distinction from traditional probability theory is one of the main issues of chaos theory. Chaos is also sometimes called "deterministic noise".

The notion "aperiodic" mathematically means that the flow paths do not converge for $t \to +\infty$ to fixed points or to periodic (quasiperiodic) orbits such as limit cycles. We have just mentioned that extremely sensitive dependence of the system trajectories on initial conditions is the distinguishing mark of deterministic chaos. But this is just a verbal expression that has to be quantified. One of the possible ways to estimate this sensitivity is to find a variational derivative $\delta x(t, x_0)/\delta x_0$, where $x(t) = x(t, x_0) = g_{t,t_0} x_0, x_0 \equiv x(t = t_0)$ is the motion generated by flow (the dynamical system) $g_t$. In simple cases, we can replace the variational derivative by the usual one to have

$$\frac{dx(t, x_0)}{dx_0} = A\exp\left(\frac{t}{\tau_0}\right),$$

where $\tau_0 \approx \Lambda_0^{-1}$ is the predictability horizon, $\Lambda_0 > 0$ is the greatest positive Lyapunov exponent which expresses instability in the form of "stretching" (and, perhaps, also "folding" leading to chaos). Thus, the necessary (not sufficient!) condition for deterministic chaos is that dynamics should be unstable. In particular, instability means that small perturbations in the initial conditions bring large uncertainties in dynamics after the predictability horizon.

By remarking that extreme sensitivity to initial data is not the whole story, we wanted to hint that such sensitive dependence can be found in very simple, e.g., linear systems. Take, for example, the map $x_{n+1} = 2x_n, x_n \in \mathbb{R}, n \in \mathbb{Z}$, which has an unfortunate property to explode (see also below, "The logistic model: the bugs are coming"). However, the explosive divergence of nearby trajectories in this case alongside the blow-up of an initial data discrepancy to infinity has nothing to do with deterministic chaos. The latter, combining the sensitivity to initial data with unpredictable behavior, appears only if the trajectories are bounded which is possible only in nonlinear dynamical systems: in linear ones there can be either bounded trajectories or sensitive dependence on initial conditions but not both, so that nonlinearities are necessary to have both effects.

The word "chaos" is intuitively associated with a disorganized state, completely without order. This is deceptive: chaos in dynamical systems is not a total disorder but corresponds to irregular variations of the system's variables controlled by rather simple rules. Probably, no mathematical definition of chaos has been universally accepted so far, but the following descriptive explanation what chaos is can be used to work with this phenomenon. Chaos is a durable irregular (i.e., aperiodic) behavior emerging in deterministic systems which become extremely sensitive to slight variations of parameters (in particular, initial conditions). Notice that this verbal description of chaos contains three components:

1. Durable irregular (aperiodic) behavior implies that the phase paths do not asymptotically ($t \rightarrow \infty$) stabilize either to a point or to periodic orbits.

2. The term "deterministic" implies that the system is not described by stochastic differential (or other) equations i.e., there are no random parameters or noise present.

3. Sensitivity to slight variations of parameters (in particular, initial conditions) implies that the integral trajectories, at first very close to one another, diverge exponentially fast with time, the divergence rate being governed by the Lyapunov exponents (with at least one of them positive).

Thus, if one abstracts oneself from certain fine points such as irregular trajectories, uncorrelated behavior in close time points and so on, chaos can be basically viewed as an absence of Lyapunov stability. Although "chaos" has become the code word for nonlinear science, there is nothing particularly exotic or intricate about chaos. The fact that chaotic phenomena practically had not been studied until 1970s, when chaos suddenly came into fashion together with its interdisciplinary applications, can only be attributed to an absolute dominance, both in science and technology, of successful linear theories such as classical electrodynamics, quantum mechanics, linear oscillations, plasma instabilities, etc. Ubiquitous linear input-output models in electrical engineering that have resulted in many practically important technologies also left little room for studying somewhat exotic nonlinear models. The main feature of chaos in finite-dimensional dynamical systems (known as deterministic chaos to distinguish it from molecular chaos in many-particle systems) is the exponential divergence of trajectories. The quantitative measure of this divergence is the so-called K-entropy (the Kolmogorov-Krylov-Sinai entropy [281]). The value of K-entropy is positive for chaotic states, which corresponds to mixing and exponential decay of correlations.

Many people are still inclined to think that one should not treat chaos in deterministic systems as a special subject, and the word itself is just a poetic metaphor for the long familiar instability. This is wrong: chaos is not completely synonymous with instability; it incorporates also other concepts (such as irregularity of behavior and decorrelation in time series). In distinction to instabilities, chaos points at unpredictability of behavior in the systems described by completely deterministic and even primitive looking equations. Instability in dynamical systems can be viewed as a symptom of the possible transition to chaotic motion. One can observe on some examples, in particular on deterministic models of growth that may exhibit instability (i.e., have diverging integral trajectories), but cannot be viewed as chaotic; the simplest example of such a system is the model of exponential growth $\dot{x} = ax$. Conversely, one should not think that the chaotic state is always visibly unstable: for example, turbulence is a stable chaotic regime. Thus, the often repeated view that chaos is just an extreme case of instability i.e. in chaotic regimes paths in the phase space that start arbitrarily close to each other diverge exponentially in time whereas in regular (nonchaotic) regimes two nearby trajectories diverge not faster than polynomial, typically linear in time, is not quite accurate. If we compute the distance $d(t)$ between two phase paths whose initial separation is $d_0 \equiv d(0)$, then exponential divergence of trajectories, $d(t) = d_0 e^{\Lambda t}$, where $\Lambda$ is the Lyapunov characteristic exponent, is a necessary but not sufficient condition for chaos.

Notice that the concept of dynamical systems as evolving quasi-closed parts of the world does not exclude their chaotic behavior, when a system's evolution (i.e., dependence of its observable quantities $x^i$ in time) looks like a random process; at least it is totally unpredictable beyond a certain "horizon of predictability" (recall weather forecasts). One should not, however, think that unpredictability in chaotic systems is identical to the true randomness, but the difference is rather academic since chaos looks quite like a random process and can also be described by a probability measure (distribution functions). One can crudely say that there are at least two types of randomness: one is due to an unobservable quantity of interacting subsystems (as, e.g., in gas), the other - usually known as deterministic chaos – is due to our limited ability to formulate the rules of behavior under the conditions of drastic instability. Such poorly known rules must govern the processes that basically arise from irreversibility in dynamical systems.

The main feature of chaos in dynamical systems is an extreme sensitivity to small perturbations, e.g., to tiny inaccuracies in input data such as initial conditions. It is this property that makes it impossible to forecast the state of a dynamical system for the time exceeding some characteristic predictability horizon amenable to the state-of-the-art numerical computation. The time scale for exponential divergence of nearby trajectories, $\Lambda^{-1}$, may serve as an estimate for this predictability horizon. It is important that this time scale usually does not depend on the exact value of initial conditions. Anyway, large Lyapunov exponents are symptomatic of the onset of chaos. Recall that the Lyapunov characteristic exponent manifests the expansion rate of linearized dynamical system along its trajectory.

Nevertheless, one can distinguish between dynamical chaos in deterministic systems and physical chaos in many-particle models. For example, evolution to thermodynamic (statistical) equilibrium is directed to the most chaotic and disordered state characterized by maximal entropy. The difference between "dynamical" and "physical" chaos has been reflected in long-standing debates about the origin of stochasticity in physical systems. There were historically two distinct trends of thought: (1) stochasticity arising due to dynamic instability of motion in nonlinear deterministic systems and (2) necessity of statistical description due to the enormous number of degrees of freedom i.e., huge dimensionality of the phase space in realistic (many-particle) physical systems, resulting in practical irreproducibility of solutions (integral trajectories). These two trends were poorly compatible because they required different approaches to physical statistics. In the section devoted to statistical physics and thermodynamics, we shall comment on the possibility to reconcile the two manners of description.

One can produce many examples illustrating the importance of chaotic behavior in real life. Thus, when an asteroid or a comet approaches a planet (e.g., Earth), the planet's gravity perturbs the comet's trajectory so that small changes of the latter can be amplified into large and poorly predictable deflections. Since a comet or an asteroid trajectory is affected by numerous close encounters with planets, small variations in the initial parameters of the trajectory may result in practically complete unpredictability of the body's eventual motion (for large enough time – outside the predictability horizon). One can call this few-body process a trivial chaotization of the comet or asteroid path. Because of such sensitivity to small variations of parameters, trajectories of small celestial bodies can diverge and cross the planetary orbits, possibly with devastating consequences. The information about the fate of a comet or an asteroid is practically lost after several Lyapunov time scales, and the resulting unpredictability may be the source of meteorite hazard for the Earth.

One can show that the logistic map for $a = 4$ which is in this case chaotic for almost all initial conditions may be related to the Lorenz attractor appearing in the three-dimensional meteorological model constructed in 1963 by E. Lorenz. This mathematical model is represented by a dynamical system with 3d phase space and is fully deterministic since it is represented by three ODEs $\dot{\mathbf{x}} = \mathbf{f}(\mathbf{x}, a), \mathbf{x} = (x^1, x^2, x^3)$ (with quasilinear vector field $\mathbf{f}$). Nonetheless, the model demonstrates chaotic behavior i.e., abrupt and apparently random changes of state for some set of control parameters $a$.

From a more general viewpoint, the logistic model is a very particular case of the evolution of an autonomous system described by vector equation $\dot{\mathbf{x}} = \mathbf{f}(\mathbf{x}, a)$, where vector field $\mathbf{f}$ depends on parameter $a$ (it can also be a vector). As we have seen, in certain cases variations of the control parameter $a$ can radically change the system's motion, for instance, result in chaotic behavior. Note that the logistic model is not necessarily identical with the population model. When the logistic equation is not interpreted as describing the population growth, one can explore the behavior of solutions as parameter $k$ is varied.

The logistic map manifests such common features of discrete-time algorithms as stability and chaotic behavior. From this viewpoint, it is interesting for numerical techniques and generally for computational science and engineering. As far as engineering applications of continuous-time logistic model go, we have already mentioned that the equation describing the energy evolution of a nonlinear oscillator in the self-excitation mode has the form $\dot{E} = aE(1 - E)$ (in dimensionless units). Here the growth parameter $a$ is close to 1.

One can bring an example of estimating the population growth with the help of the logistic model. We may assume the total current (2010) human population to be $6.9*10^9$, the growth factor to be $a = 0.029$, the annual population growth to be 0.011 year$^{-1}$ (more or less standard demographic data). Then we have

$$\frac{dN(2010)/dt}{N(2010)} = \frac{d}{dt}\log(N(2010)/N(t_0 = 2010)) = 0.011 = a - kN(2010)$$
$$= 0.029 - k \cdot 6.9 \cdot 10^9$$

which can be considered an equation to find the attenuation factor $k \approx 2.32 \cdot 10^{-12}$. Then the projection for the equilibrium population will be $N_0 = a/k \approx 11.1 \cdot 10^9$ i.e., the world population tends to converge to approximately 11 billion people. This result is not very sensitive to the slight changes of the constant $a$ and the annual population growth rate.

### 4.16.1    Chaos in dissipative systems

Chaotic behavior is quantified by the presence and the value of Lyapunov exponents that must be positive in order for the chaotic regime – and the complexity associated with it - to emerge. Recall that the notion "chaotic" is related to the bounded motions displaying an extreme sensitivity to initial data. If there is no dissipation i.e., the dynamical system is conservative, the sum of all Lyapunov exponents $\Lambda_i$ must be zero – to ensure that a volume element of the phase space remains intact along the phase trajectory (see 8.3.1. "Phase space and phase volume"). If the system is dissipative, the sum of all Lyapunov exponents should be negative, and if $\sum_i \Lambda_i < 0$, at least one Lyapunov exponent must exist in a dissipative system[119].

We observed from expression (1) that a free pendulum eventually stops: it is damped down to a stable state corresponding to the minimum of potential energy in the gravitation field. Such a stable state is referred to today as an attractor. Although there seems to be no universally accepted definition of attractor, we may intuitively understand it as a limit set for which any nearby orbit regardless of its initial conditions ends up (has its limit points) in the set. The simple example of a pendulum has, however, a deep meaning. Firstly, it demonstrates that a dynamical system can radically change its behavior when energy is withdrawn from the system. Secondly, for large times ($t \rightarrow \infty$), trajectories of a dissipative system tend to a low-dimensional subset of the phase space which is an attractor. Thirdly, the evolution of a dissipative system, after a certain time, is restricted to this attractor forever.

There may be more than one attractor in a dynamical system, and the latter can wander between its attractors as the value of some control parameter varies (we shall observe such transitions shortly on a simple example of the logistic map). Many physical systems exhibit such a behavior. For example, addressing again the meteorite hazard, we may note that small celestial bodies such as asteroids or comets (sometimes called "planetesimals") move around the Sun on almost Keplerian orbits experiencing a slight drag due to the presence of interplanetary gas and solar radiation. This small resistance forces the small body trajectories to be damped down to the Sun, and the ensuing radial component of the motion results in crossing the planetary orbits. Then small bodies can be trapped by the gravity field of a major planet such as Earth or Jupiter, usually in some resonance motion – a far analogy to Bohr's model of the atom. The drag force plays the role of control parameter: with the increased drag, equilibrium paths of asteroids or comets can bifurcate (as in the logistic map, see below). For large enough drag, the system may become chaotic so that an asteroid or a comet no longer stays on a fixed trajectory relative to the planet, but wanders between seemingly random distances, down to a number of close approaches to the planet within a short time interval on an astronomical scale. So, the dissipative chaotic motion may bring celestial bodies dangerously close to the Earth's surface. However, exact mathematical modeling of the dissipative motion of small celestial bodies is rather involved and requires numerical integration on high-performance computers rather than analytical calculations: the cascade of chaotic returns to close encounters produces a complex pattern that can hardly be explored analytically. The uncertainty cloud of the asteroid or comet initial conditions, taken from observational data, must be propagated throughout decades to detect possible intersections with the Earth's orbit (see, e.g., [302]).

---

[119] Contrariwise, a positive Lyapunov exponent reflects stretching along the corresponding axis.

In conclusion to this section, one can make the following nearly obvious remark about instabilities and chaos. We have seen that the theory of continuous-time differentiable dynamical systems is largely based on the linearization of the respective differential equations near equilibrium points. Therefore, this theory may fail while attempting to explain the global behavior of highly nonlinear systems, in particular, the complexity of chaotic motions. Notice that, in the phase portrait, chaotic motion is typically imaged as "clouds" of the phase points.

# 5 Classical Fields and Waves

One may easily notice that in the Newtonian picture of classical mechanics particles interact through instantaneous forces, $\mathbf{F} = m\ddot{\mathbf{r}}$. Forces $\mathbf{F}$ are always produced by other bodies or particles. But writing down the differential equation with forces is clearly not the whole story. So already in the first half of the 19th century, mostly due to experiments of M. Faraday, it became obvious that there was something else in the physical world besides bodies and particles. It is presumed that Faraday was the first to call this something a "field", although I failed to find this term in sources available to me on Faraday's works, see however http://en.wikipedia.org/wiki/Field (physics). Then J. C. Maxwell and O. Heaviside constructed the classical theory of the electromagnetic field.

   The classical theory of fields is a remarkable subject because it unifies extremely abstract and mathematically advanced areas of modern theoretical physics and down-to-earth engineering. This intermediary, bridging position of classical electrodynamics is even more pronounced than in the case of classical mechanics. Although there exist a number of very good books on electromagnetic theory [5,6], I still prefer the textbook by Landau and Lifshitz [39] and will often cite it. Almost all of the material contained in this chapter can be found in [39] except for some comments of mine, still I write out the main expressions - for the reader's convenience and because some of them may be needed further for more complex subjects. One may think that the classical field theory (CFT) in empty space is not a difficult subject - indeed, a great many of its results and occasional derivations appear boring and trivial, but this impression is deceptive. There exist a lot of refined implications of classical field theory and classical electrodynamics is full of understatements. But the most important thing is that CFT provides a solid foundation, a perfect model for all other field theories, and while touching upon them we shall recall CFT with gratitude.

## 5.1    The Maxwell Equations

The classical theory of fields [120] is based on Maxwell's equations which we have already discussed in other contexts. Since these equations are very fundamental, I write them once more. The inhomogeneous pair:

$$\nabla \mathbf{E} = 4\pi\rho, \qquad \nabla \times \mathbf{H} - \frac{1}{c}\frac{\partial \mathbf{E}}{\partial t} = \frac{4\pi}{c}\mathbf{j};$$

and the homogeneous pair:

$$\nabla \mathbf{H} = 0, \qquad \nabla \times \mathbf{E} + \frac{1}{c}\mathbf{H} = 0$$

---

[120] Here, we are limiting our discussion to an electromagnetic field.

When discussing dualities in Chapter 3, we have mentioned the magnetic monopole. Modern physics does not exclude the possibility that the magnetic monopole might exist, and eventually it may be discovered. If the magnetic monopole really exists, then the right-hand side of the first equation from the homogeneous pair namely that expressing the solenoidal character of the magnetic field, $\nabla \mathbf{H} = 0$, should be not zero, but proportional to the density of magnetic monopoles. At present, however, this possibility is highly hypothetical, and we shall not take monopoles into account.

The electric and magnetic components of the electromagnetic field can be more conveniently written using the potentials[121]. By introducing the vector and scalar potentials, $A$ and $\varphi$, with

$$\mathbf{E} = -\nabla \varphi - \frac{1}{c}\frac{\partial \mathbf{A}}{\partial t}, \qquad \mathbf{H} = \nabla \times \mathbf{A} \tag{5.1}$$

we get the wave-type equations for the potentials

$$\Delta \mathbf{A} - \nabla\left(\nabla \mathbf{A} + \frac{1}{c}\frac{\partial \varphi}{\partial t}\right) - \frac{1}{c^2}\frac{\partial^2 \mathbf{A}}{\partial t^2} = -\frac{4\pi}{c}\mathbf{j}$$

$$\nabla \varphi + \frac{1}{c}\nabla \frac{\partial \mathbf{A}}{\partial t} = -4\pi\rho \tag{5.2}$$

Below, we shall often use the symbol $\Box := \Delta - \frac{1}{c^2}\frac{\partial^2}{\partial t^2}$ (the D'Alembertian).

These are four coupled linear partial differential equations. Nevertheless, this system of equations with respect to four scalar functions is obviously simpler than the initial system of Maxwell equations for six field components. In fact, the matter is even more complicated, since one has to account for the motion of charged particles in the electromagnetic field. This is a self-consistency problem because these particles, on the one hand, are experiencing forces from the field and, on the other hand, themselves contribute to the field. The charge and current densities on the right-hand side are defined as

$$\rho(\mathbf{r}, t) = \sum_a e_a \delta\big(\mathbf{r} - \mathbf{r}_a(t)\big)$$

and

---

[121] It is worth noting that the four expressions that we know as the Maxwell equations are probably the creation of O. Heaviside who reduced the original system of J. C. Maxwell consisting of twenty equations to four vector equations, see, e.g., http://en.wikipedia.org/wiki/Maxwell's equations. I would also highly recommend the beautiful book about O. Heaviside by a prominent Russian physicist and a very gifted writer, Prof. B. M. Bolotovskii, [282]. Unfortunately, this book is in Russian and I don't know whether its translation into other European languages exists.

$$j(\mathbf{r}, t) = \sum_a e_a \dot{\mathbf{r}}_a(t) \delta(\mathbf{r} - \mathbf{r}_a(t)),$$

where $\mathbf{r}_a(t), \dot{\mathbf{r}}_a(t)$ are unknown quantities (here summation goes over all charged particles in the considered system). Thus, in order to close the self-consistent system of equations, we must provide equations for these quantities. In the classical case, for example, we may supplement the Maxwell equations with the Newton equations containing the Lorentz force:

$$m\ddot{\mathbf{r}}_a(t) = e_a \left[ \mathbf{E}(\mathbf{r}_a(t)) + \frac{1}{c} \left( \dot{\mathbf{r}}_a(t) \times \mathbf{H}(\mathbf{r}_a(t)) \right) \right]$$

(we neglect in these equations of motion close interparticle interactions, possibly of non-electromagnetic character). The solution of such self-consistent electromagnetic problems is usually a difficult task, it is usually treated in plasma physics and while considering the interaction of beams of charged particles with matter (see Chapter 8). Moreover, self-consistency leads in its limit to self-interaction of charges which has always been a source of grave difficulties in early attempts of electromagnetic field quantization. However, in classical field theory particles, in particular, electric charges, due to relativistic requirements, must be considered point-like (see [39], §15), which results in an infinite self-energy of the particle. Therefore, classical field theory becomes self-contradictory (at least at small distances) and should be replaced by a more advanced theory [39], §37, see also below.

The charge and current densities automatically satisfy the continuity equation

$$\frac{\partial \rho}{\partial t} + div\mathbf{j} = 0, \tag{5.3}$$

which can be actually obtained from the Maxwell equations by taking the divergence of the $\nabla \mathbf{H}$ and using the Coulomb law equation $\nabla \mathbf{E} = 4\pi\rho$. The fact that the continuity equation, which expresses the electric charge conservation, is not independent of Maxwell's equations means that the latter are constructed in such a way as to be in automatic agreement with the charge conservation law. More than that, one can even say that charge conservation is a more fundamental concept than the Maxwell equations, since it results in their generalizations (see below).

Before we produce solutions to the equations for electromagnetic potentials, let us discuss some properties of the Maxwell equations. When speaking about differential equations, the first thing to pay attention to is their invariance (symmetry) properties. In many cases the requirement of an invariance of physically relevant equations, e.g., the motion equations with respect to some group of transformations allows one to single out the mathematical model from a wide class of available equations. For instance, one can prove [145] that among all systems of the first-order partial differential equations for two vector-functions $\mathbf{E}(\mathbf{r}, t), \mathbf{H}(\mathbf{r}, t)$ there exists a unique system of equations invariant under the Poincaré group. This system is the Maxwell equations.

When discussing dualities in physics, we have already noticed that the Maxwell equations are invariant with respect to a dual change of functions performed by Heaviside [37]

$$\mathbf{E} \to \mathbf{H}, \qquad \mathbf{H} \to -\mathbf{E}. \tag{5.4}$$

Later, this symmetry was generalized by Larmor to a single parameter family of plane rotations, $R(\theta)$:

$$\begin{pmatrix} \mathbf{E}' \\ \mathbf{H}' \end{pmatrix} = R(\theta) \begin{pmatrix} \mathbf{E} \\ \mathbf{H} \end{pmatrix},$$

where

$$R(\theta) = \begin{pmatrix} \cos\theta & \sin\theta \\ -\sin\theta & \cos\theta \end{pmatrix}$$

or

$$\begin{aligned} \mathbf{E}' &= \mathbf{E}\cos\theta + \mathbf{H}\sin\theta \\ \mathbf{H}' &= -\mathbf{E}\sin\theta + \mathbf{H}\cos\theta \,. \end{aligned} \tag{5.5}$$

The Lie-group analysis performed by a well-known Russian mathematician N. Ibrahimov shows that the ultimate local symmetry group for the system of Maxwell's equations in empty space is a 16-parameter group $C(1,3) \otimes SO(2,R)$ where $SO(2,R)$ is represented by $R(\theta)$ above.

## 5.2 Gauge Invariance in Classical Electrodynamics

Probably, the most important invariance property of the Maxwell equations is connected with the transformation of the potentials ([39], §18)

$$\mathbf{A} \to \mathbf{A}' = \mathbf{A} + \nabla\chi, \qquad \varphi \to \varphi' = \varphi - \frac{1}{c}\frac{\partial\chi}{\partial t}, \tag{5.6}$$

which is called the gauge transformation. It is clear that the fields $\mathbf{E}, \mathbf{H}$ do not change under this transformation. In other words, different potentials $(\varphi, \mathbf{A})$ and $(\varphi', \mathbf{A}')$ correspond to the same physical situation. This fact is known as gauge invariance. The term gauge refers to some specific choice of potentials, or what is synonymous, to a fixed attribution of the function $\chi$ which is usually called the gauge function. When one considers quantum-mechanical particles described by the wave function $\psi$, coupled to the electromagnetic field (Chapters 5, 8), one must extend the transformations of the electromagnetic potentials with the aid of gauge function $\chi$ by including the change of the phase of wave function

$$\psi \rightarrow \psi' = \psi \exp\left(\frac{ie\chi}{\hbar c}\right) \tag{5.7}$$

In recent years, it has been established that the gauge invariance is not only a fundamental property of classical field theory, but it defines other fundamental interactions in nature, perhaps even all possible interactions. In particular, the principle of gauge invariance plays a decisive role in the so-called "Standard Model" that has been designed to describe electroweak and strong interactions between elementary particles. In the Standard Model of particle physics, all particles (other than the Higgs boson, see Chapter 9) transform either as vectors or as spinors. The vector particles are also called "gauge bosons", and they serve to carry the forces in the Standard Model. The spinor particles are also called fermions, and they correspond to the two basic constituent forms of matter: quarks and leptons. In this section, however, we shall only deal with electromagnetic forces – the low-energy classical limit of electroweak interactions. In the classical description of electromagnetic fields, gauge invariance brings about supplementary and seemingly unnecessary degrees of freedom which may be called the gauge degrees of freedom. In classical electromagnetic theory, we may interpret the electromagnetic potentials, $(\varphi, \mathbf{A})$, as a tool introduced to simplify the Maxwell equations which, however, does not have any direct experimental relevance. Immediately observable quantities of classical electromagnetic theory are the electric and magnetic fields, $\mathbf{E}, \mathbf{H}$, corresponding to gauge equivalence classes of the potentials $(\varphi', \mathbf{A}') \sim (\varphi, \mathbf{A})$ giving the same electric and magnetic fields. If one takes in the canonical formalism the electromagnetic potentials as phase space variables in order to obtain the Euler-Lagrange equations from the least action principle (see below "Equations of Motion for the Electromagnetic Field"), then one sees that the dimensionality of the classical phase space is reduced[122].

As is well known (see [39], §46), in those situations when relativistic (Lorentz) invariance should be explicitly ensured one can use the Lorentz gauge, i.e., the condition

$$\nabla \mathbf{A} + \frac{1}{c}\frac{\partial \chi}{\partial t} = 0 \tag{5.8}$$

imposed on the potentials. In this case, the above equations for the potentials take the form of usual (linear) wave equations

$$L\mathbf{A} = -\frac{4\pi}{c}\mathbf{j}, \qquad L\varphi = -4\pi\rho, \qquad L := \Delta - \frac{1}{c^2}\frac{\partial^2}{\partial t^2}, \tag{5.9}$$

---

[122] This phase space reduction may be accompanied by a change in geometry, the latter becoming more complicated, curved, and containing, e.g., holes. No canonically conjugated coordinates such as electromagnetic potentials and conjugated to them "momenta" may be globally possible, which makes a direct canonical quantization of the electromagnetic field quite difficult.

$L$ being the wave operator which coincides here with the D'Alembertian (this is not always the case since there may be many wave operators).

It is important to note that in many physical situations the manifest Lorentz invariance is irrelevant since one typically uses a privileged reference frame, the one being fixed with respect to the observer or the media. Thus, in considering the interaction of an electromagnetic field with atoms (Chapter 8) other gauges are more convenient. The most popular gauge in physical situations with a fixed frame of reference is the so-called Coulomb gauge defined by the transversality condition $\nabla \mathbf{A} = \mathbf{0}$. The terms "Coulomb gauge" and "transversality" can be easily explained. Suppose we managed to impose the condition $\nabla \mathbf{A} = \mathbf{0}$ - a little further I shall prove that it is always possible. Then we obtain the following equations for the potentials

$$l \Delta \mathbf{A} - \frac{1}{c^2} \frac{\partial^2 \mathbf{A}}{\partial t^2} - \frac{1}{c} \frac{\partial}{\partial t} \nabla \varphi - \frac{4\pi}{c} \mathbf{j} \tag{5.10}$$

$$\Delta \varphi = -4\pi \rho. \tag{5.11}$$

The last equation is exactly the same as the one from which the electrostatic potential can be determined (see Chapter 5). For the model of point particles, $\rho(\mathbf{r}, t) = \sum_a e_a \delta(\mathbf{r} - \mathbf{r}_a(t))$, the scalar potential becomes

$$\varphi(\mathbf{r}, t) = \sum_a \frac{e_a}{|\mathbf{r} - \mathbf{r}_a(t)|},$$

which again coincides with the form of elementary electrostatic equation. The solution for potential $\varphi$ may be written through the Coulomb Green's function

$$G_0(\mathbf{r}, \mathbf{r}') = -\frac{1}{4\pi} \frac{1}{|\mathbf{r} - \mathbf{r}'|}$$

This purely electrostatic form of the scalar potential which is represented as the superposition of Coulomb potentials created by individual charges, explains the name of the Coulomb gauge. Let us now try to explain why this gauge is also called transversal. To do this we can, for example, expand the potentials into plane waves, as in [39], §§51,52:

$$\mathbf{A}(\mathbf{r}, t) = \sum_{\mathbf{k}, j=1,2} \mathbf{A}_{\mathbf{k},j}(t) \exp(i\mathbf{k}_j \mathbf{r}),$$

where index $j$ denotes two possible polarizations of the vector plane waves normal to the wave vector $\mathbf{k}$. We know from elementary functional analysis that plane waves form a full system of functions so that they may be taken as a basis and a series over them is legitimate. Below, when discussing the field quantization, we shall consider some details of expansions over the system of plane waves. Now, all we need is just a general form of such an expansion.

Inserting the expression for $\mathbf{A}(\mathbf{r}, t)$ into $\nabla\mathbf{A} = \mathbf{0}$, we see that

$$\nabla\mathbf{A}(\mathbf{r}, t) = i \sum_{\mathbf{k}, j = 1, 2} \left( \mathbf{k}\mathbf{A}_{\mathbf{k}, j}(t) \right) \exp\left(i\mathbf{k}_j\mathbf{r}\right) = 0$$

may hold for arbitrary $\mathbf{A}$ only if $\mathbf{k}\mathbf{A}_{\mathbf{k}, j} = 0$, which means that all the modes $\mathbf{A}_{\mathbf{k}, j}$ should be transversal to the wave vector. If we similarly expand the scalar potential

$$\varphi(\mathbf{r}, t) = \sum_{\mathbf{k}, j = 1, 2} \varphi_{\mathbf{k}, j}(t) \exp\left(i\mathbf{k}_j\mathbf{r}\right),$$

then we shall be able to represent the electric field as the sum of the longitudinal and the transversal components corresponding to the terms

$$\mathbf{E}_{\parallel} = -\nabla\varphi = -i \sum_{\mathbf{k}, j = 1, 2} \mathbf{k}\varphi_{\mathbf{k}, j}(t) \exp\left(i\mathbf{k}_j\mathbf{r}\right),$$

and

$$\mathbf{E}_{\perp} = -\frac{1}{c}\frac{\partial\mathbf{A}}{\partial t} = -\frac{1}{c}\partial_t\mathbf{A} = \sum_{\mathbf{k}, j = 1, 2} \dot{\mathbf{A}}_{\mathbf{k}, j}(t) \exp\left(i\mathbf{k}_j\mathbf{r}\right)$$

The magnetic field comprised of components $\left(\mathbf{k}_j \times \mathbf{A}_{\mathbf{k}, j}\right)$ is obviously transversal ($\nabla\mathbf{H} = 0$). Thus, the Coulomb gauge allows one to separate the entire electromagnetic field into the transversal component corresponding to the vector potential $\mathbf{A}$ and the longitudinal component described by the scalar potential $\varphi$. This separation is usually quite convenient, at least for nonrelativistic problems - for instance, when treating the interaction of radiation with matter (Chapter 8). In this class of problems, one may describe forces between the particles of the medium by the scalar potential $\varphi$ whereas the radiation field is described by the vector potential $\mathbf{A}$ alone.

This is all more or less banal, nevertheless, there are some nontrivial questions around such a decomposition of the electromagnetic field into transversal and longitudinal components. To begin with, we see that the electrostatic form of the scalar potential in the Coulomb gauge implies an instantaneous nature of the longitudinal field. How then can it be that the electromagnetic signal is causal and propagates with the speed $c$, as prescribed by relativity theory? This is obviously an apparent paradox, but its discussion leads to interesting representations of the gauge function.

## 5.3   Four-Dimensional Formulation of Electrodynamics

Many elementary results formulated above in the conventional language of vector analysis can be cast into a more elegant form using the invariance properties of the Maxwell equations, so this section in fact does not contain unfamiliar expressions. In principle, the Maxwell equations admit a number of various formulations (see below), but the most popular one - is based on the use of four-component function $A_\mu = (\varphi, \mathbf{A})$ often called the four-dimensional potential [39], §16, or simply the 4-potential. This function is connected with the fields $\mathbf{E}, \mathbf{H}$ by the familiar relations which we write here, for future usage, through the momentum operator $\mathbf{p} = -i\nabla, p_\mu = -i\partial_\mu$:

$$\mathbf{E} = \frac{\partial \mathbf{A}}{\partial x_0} - i\mathbf{p}A_0, \qquad \mathbf{H} = i\mathbf{p} \times \mathbf{A}$$

Then the homogeneous Maxwell equations (see the next section) are satisfied identically, if one introduces the so-called electromagnetic field tensor as a four-dimensional rotation:

$$F_{\mu\nu} = -F_{\nu\mu} = \partial_\mu A_\nu - \partial_\nu A_\mu.$$

Indeed,

$$\partial_\lambda F_{\mu\nu} + \partial_\mu F_{\nu\lambda} + \partial_\nu F_{\lambda\mu} = 0$$

or, in terms of the dual tensor [39], §26, $F^{*\mu\nu} = \frac{1}{2}\epsilon^{\mu\nu\alpha\beta}F_{\alpha\beta}, \partial_\mu F^{*\mu\nu}$. The inhomogeneous Maxwell equations are

$$\partial_\mu F^{\mu\nu} = -\frac{4\pi}{c}j^\nu,$$

which follows from the variational principle with the Lagrangian density, see [39], §28,

$$\mathcal{L} = -\frac{1}{c}A_\mu j^\mu - \frac{1}{16\pi}F_{\mu\nu}F^{\mu\nu}.$$

We have seen that for a given electromagnetic field $A_\mu$ is not unique, since the gauge transformation $A_\mu \to A'_\mu = A_\mu + \partial_\mu \chi(x)$ leaves $F_{\mu\nu}$ unchanged:

$$\begin{aligned} F_{\mu\nu} \to F'_{\mu\nu} = \partial_\mu A'_\nu - \partial_\nu A'_\mu &= \partial_\mu(A_\nu + \partial_\nu\chi) - \partial_\nu(A_\mu + \partial_\mu\chi) \\ &= F_{\mu\nu} + \left(\partial_\mu\partial_\nu - \partial_\nu\partial_\mu\right)\chi = F_{\mu\nu}. \end{aligned}$$

If we raise index $\mu$ in $A_\mu$, we can write

$$\partial_\mu A'^\mu = \partial_\mu (A^\mu + \partial^\mu \chi) = \partial_\mu A^\mu + \partial_\mu \partial^\mu \chi$$

or, if we choose the gauge function $\chi$ to satisfy the equation

$$\partial_\mu \partial^\mu \chi = -\partial_\mu A^\mu$$

(we have seen above that we can always do it due to gauge invariance) or, in three-dimensional representation,

$$\Box \chi = -\frac{\partial A^0}{\partial x_0} - \nabla A = \frac{1}{c}\frac{\partial \varphi}{\partial t} - \nabla A,$$

then we get the familiar Lorentz gauge $\partial_\mu A^\mu = 0$. This supplementary condition reduces the number of independent components of $A_\mu$ to three, yet it does not ensure that $A_\mu$ is uniquely defined. From the above formulas for transition from $A'_\mu$ to $A_\mu$ it becomes clear that if $A_\mu$ satisfies the Lorentz gauge, so will $A'_\mu$ provided $\Box \chi \equiv \partial_\mu \partial^\mu = 0$. Thus, the ambiguity due to gauge invariance persists, and one needs a more restrictive constraint to eliminate it. We have seen that we can, for example, impose the condition

$$\partial_0 \chi = \partial x_0 = \frac{1}{c}\frac{\partial \chi}{\partial t} = -\varphi,$$

then $A'_0 = A_0 + \partial_0 \chi = A_0 - \varphi = \varphi - \varphi = 0$, i.e., we may put $\varphi = 0$ and $\nabla \mathbf{A} = 0$ - the Coulomb or radiation gauge discussed above. This gauge reduces the number of independent components of $A_\mu$ to only two, that is one more reason why working in the Coulomb gauge is usually *always* more convenient than in the Lorentz gauge unless it is indispensable to retain the explicit Coulomb invariance.

Let me now make a remark of a purely technical character. Some authors introduce the vector potential as a contravariant quantity, $A^\mu = (\varphi, \mathbf{A})$ which gives the corresponding skew-symmetric electromagnetic field tensor, $F^{\mu\nu} = -F^{\nu\mu} = \partial^\mu A^\nu - \partial^\nu A^\mu$. It is of course just a technicality and a matter of taste, since one can raise or lower indices with the help of the metric tensor which in the Minkowski (flat) background is simply $g_{\mu\nu} = \gamma_{\mu\nu} = diag(1,-1,-1,-1)$. So both definitions differ by the sign of $\mathbf{A}$ in $A^\mu = (\varphi, \mathbf{A})$. However, I think that such a definition is less convenient than $A_\mu$, specifically when introducing the covariant derivative $\partial_\mu \rightarrow \partial_\mu + (ie/c)A_\mu$. Moreover, the 1-form $A_\mu dx^\mu$, which is the connection form in the coordinate basis, may be used to write the particle-field coupling term in the action $S$ (see below)

$$S_{pf} = -\frac{1}{c}\sum_a \int e_a A_\mu(x_a)dx_a^\mu,$$

where the integral is taken over the world lines of particles, or in the "current" form

$$S_{pf} = -\frac{1}{c^2} \int j^\mu A_\mu d^4 x.$$

Recall that the total action for the electromagnetic field, $S = S_p + S_f + S_{pf}$ is the additive combination of terms corresponding to particles (charges) without field, field with no particles, and particle-field coupling, respectively.

Using the four-component vector potential $A_\mu$ enables us to be well-prepared for field quantization in quantum field theory (see Chapter 6). Inserting $A_\mu$ and the corresponding expression for the fields into the Maxwell equations, we get the unified system of equations for the potentials $(\varphi, \mathbf{A})$ obtained previously in three-dimensional form:

$$p_\mu p^\mu A_\nu - p_\nu p_\mu A^\mu = \frac{4\pi}{c} j_\nu.$$

This is, as we have seen, the system of four linear equations for $A_\mu$ instead of eight Maxwell's equations. Here, I have intentionally written this wave-like system through momenta, $p_\mu = -i\partial_\mu$. One can solve such a system - a linear system always can be solved for given source terms $j^\mu$ - and then obtain the experimentally observable fields $\mathbf{E}$ and $\mathbf{H}$. Thus, using the potentials results in a considerable simplification.

Now we may exploit gauge invariance, which permits, as we have seen, a certain arbitrariness in the choice of $A_\mu$. Indeed, we have already discussed that the Maxwell equations expressed through the potentials are invariant with respect to the substitution

$$A_\mu \rightarrow A'_\mu = A_\mu + \partial_\mu \chi = A_\mu + ip_\mu \chi,$$

where $\chi$ is some arbitrary function. In the four-dimensional formulation it is natural to use the Lorentz constraint, $p_\mu A^\mu = 0$, which allows one to uncouple the above system of equations for $A_\mu$.

Here, I guess, a remark about the use of potentials may be pertinent. Highschool and even some university students typically don't completely understand the meaning of potentials as well as the idea of introducing them. Only by trying to solve vector problems directly, with a lot of tedious computations, people begin to value a clever trick with a single scalar function, the electrostatic potential, from which the fields can be obtained by a simple differentiation. The fact that the electrostatic potential is defined up to a constant then becomes self-evident. Less evident, however, is that this fact is a manifestation of some hidden invariance which we now call the gauge invariance. Decoupling the system of equations for $A_\mu$ is basically the same thing, more generality is reflected in replacing the electrostatic relationship $\partial c/\partial \mathbf{r} = 0$, where $c = const$, by, e.g., $p_\mu A^\mu = 0$.

## 5.4    Classical Electromagnetic Field without Sources

This section is entirely dedicated to an elementary description of the classical electromagnetic field in the absence of electric charges and currents ($\rho = 0, \mathbf{j} = 0$). In this case, the Maxwell equations in vacuum take the homogeneous form

$$\nabla \mathbf{E} = 0, \qquad \nabla \times \mathbf{H} - \frac{1}{c}\frac{\partial \mathbf{E}}{\partial t} = 0, \qquad \nabla \mathbf{H} = 0, \qquad \nabla \times \mathbf{E} + \frac{1}{c}\mathbf{H} = 0$$

(As usual, we are writing this system as an array of two parts.) In terms of the field tensor (see [39], §23), $F_{\mu\nu} = \partial_\mu A_\nu - \partial_\nu A_\mu$, take in four-dimensional notations a particularly compact form

$$\varepsilon^{\mu\nu\rho\sigma}\partial_\nu F_{\rho\sigma} = 0$$

or

$$\partial_\sigma F_{\mu\nu} + \partial_\nu F_{\sigma\mu} = 0$$

- the first pair of Maxwell's equations and $\partial_\mu F^{\mu\nu}$ - the second pair. Recall that $\varepsilon$ denotes, as we have seen several times, the absolute antisymmetric tensor of the fourth rank whose components change sign with the interchange of any pair of indices and $\varepsilon^{0123} = 1$ by convention, see [39], §6 [123]. When discussing dualities in physics in Chapter 3, we have already dealt with four-dimensional notations in electromagnetic theory, and we shall discuss this formulation later in connection with some geometric concepts of field theory. For many practical problems, however, such as the interaction of radiation with matter (Chapter 8) the four-dimensional formulation seems to be inconvenient, requiring superfluous operations. One may notice in this context that as long as we limit ourselves to the domain of special relativity (flat or Minkowski space), no distinction between contra- and covariant components is, strictly speaking, necessary and no metric tensor $g_{\mu\nu}$ can be introduced. We, however, shall be using the Minkowski space metric tensor $\gamma_{\mu\nu}$ from time to time, its only function will be to raise or to lower indices.

Introducing explicitly the vector and scalar potentials, $\nabla \times \mathbf{A} = \mathbf{H}$, $-\nabla\varphi - \frac{1}{c}\frac{\partial \mathbf{A}}{\partial t} = \mathbf{E}$, we get the wave equations for the potentials

---

[123] Not always, sometimes $\varepsilon_{0123} = 1$ is taken. This discrepancy of conventions, though unimportant, leads to confusion.

$$\Delta \mathbf{A} - \nabla \left( \nabla \mathbf{A} + \frac{1}{c} \frac{\partial \varphi}{\partial t} \right) - \frac{1}{c^2} \frac{\partial^2 \mathbf{A}}{\partial t^2} = 0 \qquad (5.12)$$

and

$$\nabla \varphi + \frac{1}{c} \nabla \frac{\partial \mathbf{A}}{\partial t} = 0 \qquad (5.13)$$

or, in the four-dimensional notation

$$\frac{\partial^2 A_\mu}{\partial x_\nu \partial x^\mu} - \frac{\partial}{\partial x^\mu} \left( \frac{\partial A^\nu}{\partial x^\nu} \right) \equiv \partial_\nu \partial^\mu A_\mu - \partial_\mu \partial_\nu A^\nu = 0, \qquad (5.14)$$

where, as usual, $A^\nu = \gamma^{\nu\mu} A_\mu$. Being a direct consequence of the Maxwell equations, these wave equations are of extreme importance in their own right, since they lead to the representation of the Maxwell field as a set of oscillator equations, which is the only equation that has an equidistant spectrum, and it is this equidistant spectrum, as we have already noticed, that allows one to interpret the field as consisting of independent particles - electromagnetic quanta or photons. Therefore, two accidentally occurring reasons Maxwell's equations leading to harmonic oscillator and the unique equidistant character of its spectrum, being combined render an essential part of the conceptual frame for entire contemporary physics. Physically speaking, since the charges interact with each other, and if they did not interact it would be impossible to observe them, the electromagnetic field becomes a necessity, and hence photons appear as an inevitable consequence of electromagnetic field quantization (based on the oscillator model). Thus, photons invoke the idea of interaction carriers. This idea has later evolved into the concept of gauge bosons.

## 5.5 Equations of Motion for the Electromagnetic Field

We have already discussed that the equations of motion for any field may - in fact must - be derived from an appropriate Lagrangian density using the least action principle. This fact is, in our terminology, the backbone of modern physics. In this approach, however, one has to correctly provide the Lagrangian density. It is not difficult to demonstrate (see [39], §27) that the Lagrangian density of the form

$$\mathcal{L}_0 = \frac{E^2 - H^2}{8\pi} = -\frac{1}{16\pi} F_{\mu\nu} F^{\mu\nu} \equiv -\frac{1}{16\pi} \left( \partial_\mu A_\nu - \partial_\nu A_\mu \right) (\partial^\mu A^\nu - \partial^\nu A^\mu) \quad (5.15)$$

can suit our purposes. Indeed, the Euler-Lagrange equations for a Lagrangian density $\mathcal{L}$

$$\frac{\partial}{\partial x^{\mu}}\frac{\partial \mathcal{L}}{\partial \left(\frac{\partial A^{\nu}}{\partial x^{\nu}}\right)} - \frac{\partial \mathcal{L}}{\partial A^{\mu}} \equiv \partial_{\mu}\frac{\partial \mathcal{L}}{\partial (\partial_{\nu}A^{\nu})} - \frac{\partial \mathcal{L}}{\partial A^{\mu}} = 0$$

in the case of $\mathcal{L} = \mathcal{L}_0$ are reduced to the first term only, $\partial_{\mu}\frac{\partial \mathcal{L}}{\partial (\partial_{\nu}A^{\nu})}$, that is

$$\frac{\partial \mathcal{L}_0}{\partial (\partial_{\mu}A_{\nu})} - \frac{1}{16\pi}\left\{\frac{\partial}{\partial (\partial_{\mu}A_{\nu})}\left[(\partial_{\rho}A_{\sigma} - \partial_{\sigma}A_{\rho})(\partial_{\rho}A_{\sigma} - \partial_{\sigma}A_{\rho})\right]\right\}$$

$$= -\frac{1}{8\pi}\left[\frac{\partial}{\partial (\partial_{\mu}A_{\nu})}(\partial_{\rho}A_{\sigma}\partial_{\rho}A_{\sigma} - \partial_{\rho}A_{\sigma}\partial_{\sigma}A_{\rho})\right]$$

$$= \frac{1}{4\pi}(\partial_{\mu}A_{\nu} - \partial_{\nu}A_{\mu}) \qquad (5.16)$$

Inserting this expression into the Euler-Lagrange equation, we get the already familiar wave equation derived from the Maxwell equations

$$\partial_{\nu}\partial^{\nu}A_{\mu} - \partial_{\mu}\partial_{\nu}A^{\nu} = 0.$$

Thus, the wave equation may be regarded as the motion equation for an electromagnetic field.

One must note that $\mathcal{L}_0$ is not the unique Lagrangian density that produces the Maxwell equations. Indeed, if we add to $\mathcal{L}_0$, for example, the four-divergence of some vector $\chi^{\mu}$ (which may be a function of $A_{\mu}, x^{\mu}$), the equations of motion obtained from the Euler-Lagrange equations for the new Lagrangian density $\mathcal{L} = \mathcal{L}_0 + \partial_{\mu}\chi^{\mu}$ will coincide with those for $\mathcal{L} = \mathcal{L}_0$ (see [39], §27). More generally, any two Lagrangian densities that would differ by terms vanishing while integrated over spacetime, produce the same equations of motion.

This subject of field invariants is beautifully exposed in §25 of [39], and the only justification of repeating it here would be to provide some fresh views or interpretations.

## 5.6.  Hamiltonian Formalism in Electromagnetic Theory

In this section, we shall deal with the Hamiltonian approach applied to classical electrodynamics. The intention is to represent electrodynamics in a form close to mechanics. In my opinion, using the Hamiltonian formalism in electromagnetic theory is little more than a heuristic catch; this formalism emphasizing evolution in time does not seem to be perfectly tuned to relativistic problems. However, to gain more insight into field theory, we shall briefly discuss its Hamiltonian description.

When trying to apply the standard Hamiltonian formalism to a free electromagnetic field one can start, for example, from the gauge invariant Lagrangian density

$$\mathcal{L}_0 = -\frac{1}{16\pi c} F_{\mu\nu} F^{\mu\nu} = -\frac{1}{16\pi c} F_{\mu\nu}^2 = -\frac{1}{8\pi c} (\mathbf{H}^2 - \mathbf{E}^2).$$

Recall that the total action for the electromagnetic field is[39]

$$S = -\sum_a \int m_a c\, dl - \sum_a \int \frac{e_a}{c} A_\mu dx^\mu - \frac{1}{16\pi c} \int F_{\mu\nu}^2 d^4 x.$$

Taking, as usual, the field potentials $A_\mu$ as the generalized coordinates, we obtain the canonically conjugate momenta

$$\Pi_\mu = \frac{1}{c} \frac{\partial \mathcal{L}}{\partial\left(\frac{\partial A^\mu}{\partial x^0}\right)}$$

so that the Hamiltonian density is

$$\mathcal{H}_0 = c\Pi_\mu \partial_0 A^\mu - \mathcal{L}_0.$$

One may immediately notice that by virtue of the relationship

$$\frac{\partial \mathcal{L}}{\partial(\partial_\mu A_\nu)} = -\frac{1}{4\pi c} F^{\mu\nu} = -\frac{1}{4\pi}(\partial^\mu A^\nu - \partial^\nu A^\mu)$$

the momentum $\Pi_0$ conjugate to the scalar potential $\varphi = A^0$ is identically zero. This is the old difficulty, well known to relativistic field theorists, see, e.g. [206]. There exist a number of ways to circumvent this difficulty, but the ultimate cause for it is the limited adequacy of the Hamiltonian formalism for relativistic field problems. Then we may sacrifice the Lorentz invariance anyway and choose the Coulomb gauge $div\mathbf{A} = 0, \varphi = 0$. In this gauge we can write

$$\mathcal{L}_0 = \frac{1}{8\pi}\left[\frac{1}{c^2}(\partial_t \mathbf{A})^2 - (\partial_i A_j - \partial_j A_i)^2\right],$$

where $i, j = 1,2,3$. From this Lagrangian density we obtain the field momentum components

$$\Pi_i = \frac{\partial \mathcal{L}_0}{\partial(\partial_t A^i)} = \frac{1}{4\pi c^2} \partial_t A_i = -\frac{1}{4\pi c} E_i$$

and the Hamiltonian density is obviously

$$\mathcal{H}_0 = \Pi_i \partial_t A^i - \mathcal{L}_0 = \frac{1}{8\pi}\left[\frac{1}{c^2}(\partial_t \mathbf{A})^2 + (\partial_i A_j - \partial_j A_i)^2\right] = \frac{1}{8\pi}(\mathbf{E}^2 + \mathbf{H}^2).$$

Using the Hamiltonian approach in electrodynamics mainly appeals to human intuition which tends to gauge everything by the archetypes inherited from classical mechanics. We have seen that the Hamiltonian formalism in classical mechanics was developed as a geometric theory on a finite-dimensional symplectic manifold having an even (2n) dimensionality. While trying to apply this formalism to electromagnetic theory we are less interested in geometrical aspects of symplectic manifolds, mostly focusing on the form of the Hamiltonian operator convenient for calculations[124]. This was, e.g., the strategy adopted in the classical textbook by W. Heitler [207] as well as many other sources related to the early development of field theory (see also section "On Hamiltonian Formalism for Particle Motion in Electromagnetic Fields" in Chapter 8). Today, more sophisticated methods are commonly in use, and the Hamiltonian formalism is not considered well-adapted to treat the electromagnetic field regarded as a Hamiltonian system. Simply speaking, a Hamiltonian system is described by pairs $(p_i, q^j)$ of canonically conjugated local variables living in some vector space $Z := (p, q)$. Then a Hamiltonian function $\mathcal{H}(p_i, q^j)$ is supposed to exist, with the property that the equations of motion are produced from it as from some potential in the form of a symplectic gradient

$$\dot{p}_i = -\frac{\partial \mathcal{H}}{\partial q^i}, \qquad \dot{q}^i = \frac{\partial \mathcal{H}}{\partial p_i}, \qquad i = 1, \dots, n$$

One can observe that this formalism was initially developed for finite-dimensional systems; however, it can be generalized to their infinite-dimensional analogs. One of the possibilities (probably not unique) for such a generalization is to replace coordinates $q^i$, at least some of them, by fields $\varphi^i(x^\alpha)$. Thus for an ensemble of particles and fields the Hamiltonian function is just a sum of kinetic energies of the particles and the energy contained in the electromagnetic field. Recall that for a conservative system the Hamiltonian is represented as the energy of the system expressed through canonical variables, e.g., $p, q$. The energy of a system "particles + EM field" can be written as

$$\mathcal{E} = \frac{1}{2}\sum_a m_a v_a^2 + \int d^3r \frac{\mathbf{E}^2 + \mathbf{H}^2}{8\pi}, \qquad (5.17)$$

where index $a$ enumerates particles. Here, an interesting and not quite trivial question arises: where is the interaction between particles and fields? The answer depends on how we define an electromagnetic field. If we understand by $\mathbf{E}$ and $\mathbf{H}$ the total electromagnetic field and not only its free part, then the energy of electromagnetic interaction is hidden in the total field energy. Recall

---

[124] Due to this reason and to attain greater simplicity of notations, we shall temporarily disregard the difference between contra- and covariant coordinates, although this may be considered a crime by geometrically-minded people.

that here we forget about all other fields acting on particles except the electromagnetic one.

Notice that although expression (5.17) gives the energy of the system "particles + EM field", it is still not the Hamiltonian function because this energy is expressed through the Lagrangian (TM) variables $(q^i, \dot{q}^i)$ i.e., through pairs $(\mathbf{r}_a, \mathbf{v}_a)$.

This was a somewhat naive and straightforward construction of the Hamiltonian function for electromagnetism. We have seen above that the Hamiltonian formalism makes the passage to quantum theory intuitive and convenient.

## 5.7    Limitations of Classical Electromagnetic Theory

There are, of course, many limitations in any classical theory, which require a consistent quantum treatment. Traditionally, such limitations of classical electrodynamics are discussed in the initial chapters of textbooks on quantum field theory, and we shall also observe some deficiencies of the classical electromagnetic theory when considering the quantum world of electromagnetic phenomena in Chapter 6. Here I venture to say a few words about completeness of classical electrodynamics based on the Maxwell equations in the classical domain.

The classical electromagnetic theory based on the Maxwell equations is a linear model - probably the simplest one possible in the Minkowski background - in which the scalar and vector potentials are to a large extent arbitrary and must be specified by fixing the gauge and the boundary conditions. The potentials in the electromagnetic theory are usually regarded as having mostly mathematical and not physical meaning. This is, however, not true. It would be a triviality to say that in quantum theory, especially in quantum field theory (QFT), the electromagnetic potentials have a clear physical significance. Yet even in the classical domain, the electromagnetic potentials, i.e., vector fields $A_\mu(x^\nu), \mu, \nu = 1, \dots, 4$ are gauge fields that really act as local to global operators mapping the global space-time conditions to local electromagnetic fields [125] . These potentials produce physically observable effects, especially salient when one departs from the conventional interpretation of the electromagnetic theory as a linear model having a simple $U(1)$ gauge symmetry. (We shall later dwell on gauge symmetry in some detail.) This old interpretation of the Maxwell theory goes back to such prominent physicists and mathematicians as H. Bateman, O. Heaviside, H. Hertz, G. F. Fitzgerald, L. V. Lorenz, H. A. Lorentz, J. C. Maxwell, and W. Thomson. Mathematically, as I have just mentioned, this interpretation treats electrodynamics as a linear theory of $U(1)$ symmetry with Abelian commutation relations. Now we know that the electromagnetic theory can be extended to $SU(2)$ and even higher symmetry formulations. One might note in passing that elementary vector algebra corresponds to the $U(1)$ symmetry,

---

[125] Here we are using notations as in [84].

whereas the $SU(2)$ group of transformations may be associated with more complicated algebraic constructions such as quaternions (see in this connection simple examples from sections on vector fields in Chapter 3).

When describing electromagnetic waves propagating through material media, such mathematical dynamic objects as solitons quite naturally appear. Solitons are in fact pseudoparticles, they can be classical or quantum mechanical, and, in principle, they can be both nonlinear and linear (at least piecewise). Other families of pseudoparticles include magnetic monopoles, magnetic charges (if defined separately from monopoles as e.g., dyons) and instantons. The standard electromagnetic $U(1)$ theory cannot describe solitons, so in order to be able to do that one must go beyond the linear electrodynamics based on the Maxwell equations, in other words, this theory must be extended (at least to $SU(2)$). Although, as I have already remarked, we shall not deal methodically with nonlinear PDEs - this is a separate topic requiring a serious treatment, I shall try to explain later what a symmetry extension means. Now I just want to notice that some physical effects unambiguously indicate that the conventional Maxwellian field theory cannot be considered complete.

For example, leaving apart the well-known difficulties with the classical electron radius, such phenomena as the Aharonov-Bohm (AB) effect [101] or the Altshuler-Aronov-Spivak (AAS) effect [105] are based upon direct physical influence of the vector potential on a charged particle, free or conducting electrons respectively. It is interesting that both effects may be interpreted as breakdown of time-reversal symmetry in a closed-loop trajectory by a magnetic field (see about the time-reversal symmetry in Chapter 9). Recall that the vector potential $A_\mu$ has been commonly regarded as a supplementary mathematical entity introduced exclusively for computational convenience. It is mostly in view of this interpretation that the conventional electromagnetic theory may be considered incomplete - it fails to define $A_\mu$ as operators. Besides the AB and AAS effects, other important physical phenomena also depend on potentials $A_\mu$, for instance, the Josephson effect (both at the quantum and macroscopical level), the quantum Hall effect, de Haas-van Alphen effect and even the Sagnac effect (known since the beginning of the 20th century). Recently, the effects associated with the topological phase properties - and all the above phenomena belong to this class - have become of fashion, and we shall devote several pages to elucidate the essence of such phenomena.

One origin of the difficulties with classical electromagnetism lies in the obvious fact that the electromagnetic field (like any other massless field) possesses only two independent components, but is covariantly described by the *four*-vector $A_\mu$. However, in many models, for example, in radiation-matter interaction (Chapter 8) we usually break this explicit covariance. Likewise in nonrelativistic quantum theory, by choosing any two of the four components of $A_\mu$ (e.g., for quantization), we also lose the explicit covariance. Contrariwise, if one desires to preserve the Lorentz covariance, two redundant components must be retained. This difficulty is connected with the important fact that for a given electromagnetic field potentials $A_\mu$ are not unique: they are defined up to the gauge transformation $A_\mu \to A'_\mu = A_\mu + \partial_\mu \chi(x)$ (see above).

One may note that random processes (see Chapter 7, "Stochastic Reality") are a direct generalization of deterministic processes considered in classical mechanics. Historically, the first random processes to have been studied were probably the Markov processes: a random process is called a Markov process, if for any two $t_0$ and $t > 0$ the process probability distribution $w(t)$ (in general, under the condition that all $w(t)$ for $t \le t_0$ are known) depends only on $w(t_0)$. While for deterministic processes the state of the system at some initial moment uniquely determines the system's future evolution, in Markov processes the state (probability distribution) of the system at $t = t_0$ uniquely defines probability distributions at any $t > 0$, with no new information on the system's behavior prior to $t = t_0$ being capable of modifying these distributions. Here one may notice a hint at a preferred direction of time or time-reversal non-invariance, which is typical of real-life processes and does not exist in unprobabilistic mechanical theories (we do not consider the Aristotelian model here). Time-reversal invariance requires that direct and time-reversed processes should be identical and have equal probabilities. Thus, purely mechanical models based on Newton's (or Lagrangian) equations are time-reversible, whereas statistical or stochastic models, though based ultimately on classical (reversible) mechanics, are time-noninvariant. Likewise, mathematical models designed to describe real-life processes are mostly irreversible. See below ("The Arrow of Time") for more details.

## 5.8   Integral Equations in Field Theory

I don't quite understand why, but integral equations have become a subject partly alien to physicists. I have met people - otherwise highly qualified - who said that differential equations are more than enough to cover all the principal areas of physics and to construct the models in other disciplines, so why should one attract new and rather complicated concepts? It is unnecessary and contradicts the principle of Occam's razor.

However, the statement that knowing only the differential equations is sufficient for physics is wrong: there are areas which cannot be studied (or at least become extremely difficult to study) without integral equations, and such areas are not rarely encountered in science and engineering. Take, for example, antenna theory and design. To obtain the required value of the electromagnetic field one has to compute or optimize the electric current, which is an unknown quantity being integrated over the antenna volume or surface. In scattering problems, both for waves and particles, integral equations are a naturally arising concept. In the problem of electromagnetic field scattering by a bounded inhomogeneity, the scattered field appears due to the induced currents which start flowing in the inhomogeneity under the influence of the incident field. More exactly, these currents may be viewed as a response to the total field present in the volume occupied by the inhomogeneity, this total field being the sum of the incident field and the one generated by induced currents. This is a typical self-consistence problem that leads to the volume integral equation where the unknown field stands under the integral. In general, self-consistence problems like scattering, quite often result in integral equations; now one more class of problems requires an extensive knowledge of integral equations, namely the inverse problems. We shall dwell on the subject of integral equations in association with the models which can be compartmentalized to each of these classes of problems. So integral equations may connect several seemingly unrelated topics, therefore every person interested in physmatics - a connected framework of physical and mathematical knowledge - should be familiar with integral equations. Now, let me make a brief overview.

In mathematics, integral equations are considered as a natural part of analysis linking together differential equations, complex analysis, harmonic analysis, operator theory, potential theory, iterative solutions, regularization methods and many other areas of mathematics. When I was studying integral equations at the university, I was surprised to discover their multiple connections with other areas of mathematics such as various vector spaces, Hilbert spaces, Fourier analysis, Sturm-Liouville theory, Green's functions, special functions you name it. Maybe such course openness was due to our teacher, professor V. I. Kondrashov, a well-known Russian mathematician who had a considerable erudition in a variety of areas.

## 5.9   Phenomenological Electrodynamics

The ideal of theoretical physics is a microscopic theory of everything. This is probably an unreachable star, yet some microscopic models built in physics, for example in non-relativistic quantum mechanics or quantum electrodynamics, have been quite successful. This success is, however, a precious rarity. More common are the phenomenological models. The collection of such models related to electromagnetic phenomena is aggregately called macroscopic electrodynamics. To better understand what it is let us consider a simple school-time example. Assume that we need to compute the electric field between a capacitor's plates. Then we would have to write the equations for electromagnetic fields produced by all the particles constituting the plates and the dielectric between them. We would also have to add to these field equations the ones describing the motion of the particles in the fields. The self-consistence problem thus obtained is, in principle, quantum mechanical, and any attempt to solve it would be a hopeless task.

Such an approach is usually called microscopic - because it considers phenomena on an atomic scale - it is too complicated and in most cases superfluous. Solving problems microscopically usually produces a great lot of immaterial data. It is much more reasonable to formulate general rules for the systems containing many particles i.e., for macroscopic bodies. Such rules have, by necessity, the average character, totally disregarding atomistic structure of the matter. An electromagnetic theory dealing with macroscopic bodies and fields between them is called phenomenological electrodynamics. Thematically, phenomenological electrodynamics belongs more to matter rather than to fields. However, it is more convenient to discuss this part of electromagnetic theory in connection with Maxwell's equations which should be properly averaged than in the context of, e.g., laser-matter interaction.

There may be many phenomenological theories related to the same subject, they are in fact comprehensive models placed between fundamental microscopic laws and ad hoc results describing a given phenomenon. When dealing with a phenomenological theory one cannot be sure that its equations are unique and correctly describe the entire class of phenomena considered unless a well-established procedure of obtaining these equations from microscopic theory is provided. In this respect, macroscopic electrodynamics and, e.g., thermodynamics are "lucky" phenomenological theories: they both can be derived - although with considerable difficulties - from underlying microscopic theories. On the other hand, empirical phenomenologies flourishing in medical, social, economic and even engineering research cannot - at least so far - be derived from first-principle theories and thus the limits of applicability for the respective phenomenological concepts are undetermined. In principle, the question of applicability of any phenomenological theory is very intricate, and later I shall try to illustrate this fact even on the examples of two privileged phenomenological theories: macroscopic electrodynamics and hydrodynamics.

Besides, most of what we call microscopic theories are in fact phenomenological. Take, for instance, Newton's laws of motion. They are considered exact, however Newton's laws can be with certainty applied only to macroscopic bodies i.e., to those composed of a very large number of atomic particles in slow, smooth motion. The Newtonian model will eventually lose validity if we continuously dissect such macroscopic bodies. It is, however, not always easy to distinguish between microscopic and phenomenological theories. Thus, the Schrödinger equation, which is also considered exact microscopic, is nothing more than a phenomenological model devised to describe the nonrelativistic motion of an atomic particle. The Schrödinger equation becomes invalid when one starts scrutinizing the interaction between particles through fields. The Newtonian theory of gravity is also phenomenological, this is the mathematical model stating that the attracting force between any two bodies does not depend on their composition, matter structure (crystalline, amorphous, liquid, plasma, etc.) and other constitutional details important from the physical point of view. This force depends only on some aggregate (and to some extent mysterious) coefficient called mass. Newtonian gravity is a very nontrivial model, gravitation could have been totally different, for example, inertial and gravitational masses could have been nonproportional to each other so that "light" objects would fall slower than "heavy" ones in the gravitation field. Phenomenological theory of gravitation, due to independence of physical details, allows astronomers to predict the motion of celestial bodies ignoring physical processes in them. In general, phenomenological theories usually neglect the effects of microscopic quantities or represent them by a set of numbers.

So philosophically speaking, most equations of physics are phenomenological. How can one try to establish validity of phenomenological theories? We have seen that the two basic concepts of microscopic - atomistic - models, in contrast to phenomenological theories (an extreme case is thermodynamics), are particles and fields. Phenomenological theories typically (but not always!) do not treat separate particles, they tend to regard objects as continuous without rapidly fluctuating local quantities such as true densities. This approach appears to be quite reasonable from the practical point of view: microscopic values vary in spacetime in a very complicated manner, and it would be merely meaningless to follow their instantaneous local values. In other words, any compelling theory should only operate with smoothed values, the fluctuations being averaged out. This way of thought naturally leads to the possibility of obtaining the phenomenological description by averaging the corresponding microscopic theories. This sounds simple - especially for linear theories - yet the question arises: what is actually "averaging"?

There does not seem to be a universal answer to this question, nor a universal recipe for the transition to the phenomenological version of a given microscopic theory. Each specific case is different, and "phenomenologization" can be quite intricate, not reducing to formal averaging. To illustrate the emerging difficulties let us get back to the relationship between microscopic and macroscopic electrodynamics. The system of Maxwell equations constituting the backbone of microscopic electrodynamics is linear, and thus it presumably can be averaged in a straightforward fashion i.e., one can simply substitute into the Maxwellian system average values for the field $\bar{\mathbf{E}}$ and $\bar{\mathbf{H}}$ in place of their genuine values $\mathbf{E}, \mathbf{H}$ containing fluctuations. However, here two problems arise. Firstly, the charge and current densities, $\rho$ and $\mathbf{j}$, representing inhomogeneities in the Maxwell equations should also be averaged, but nobody knows a priori what the relationship between the averaged fields and the averaged currents would be, and one badly needs this relationship to close the system of averaged equations of macroscopic electrodynamics. This is an important problem, and we shall discuss it later.

The second problem is the very meaning of the averaging operation and the corresponding mathematical procedure. Physicists traditionally relied in this issue on the concept of the so-called physically infinitesimal volume and declared averaging over this volume. The whole procedure of averaging over the infinitesimal volume is described in detail in the classical textbook [208]. Below I shall briefly reproduce this standard procedure accentuating the points that were for L. D. Landau and E. M. Lifshitz too obvious. Nevertheless, I consider everything based on the "physically infinitesimal" misleading and poorly suitable for many practically important cases (such as X-ray optics, molecular optics, plasma physics, etc.). Besides, the very notion of the physically infinitesimal volume is badly defined – "on the hand-waving level" - so that it would be difficult to indicate the accuracy of the averaging procedure. One must be satisfied with the formally written average fields (overlined $\mathbf{E}$ and $\mathbf{H}$), the question of their deviations from microscopic fields being irrelevant. One can only assume that fluctuations of the fields averaged over a "physically infinitesimal volume" are such that one considers these macroscopic fields as real statistical averages.

One might also remark that the difference between the fields averaged over a physically infinitesimal volume and statistically averaged is inessential. Right, it may be inessential for the description of quasistatic processes in simple model systems such as homogeneous dielectric. Even if we assume that averaging over physically infinitesimal volume is equivalent to averaging over all possible positions of scattering centers for the field (which is not at all obvious), in such averaging we neglect the motion of these centers. Moreover, it would be hardly possible to define a universal physically infinitesimal scale for all systems. For example, there exist many characteristic spatial and temporal parameters for plasmas, rarefied gases, turbulent fluids, superfluids, crystalline and amorphous solids, etc. The question of averaging over the physically infinitesimal volume is closely connected with the possibility of a universal definition of a continuous medium and leads to such nontrivial questions as dynamic irreversibility and time-noninvariance (see Chapter 9).

No matter what averaging method is applied to the Maxwell equations, the representation of the electromagnetic field as having exact (rapidly fluctuating) values at each spacetime point should be abolished. Bearing this in mind, one usually regards phenomenological electrodynamics as dealing only with the slow-varying fields, more specifically only with those having the wavelength $\lambda \gg n^{-1/3}$, where $n$ is the particle density in the medium. One may note that it is this condition that should exclude the X-ray range from macroscopic treatment (nonetheless, many problems of X-ray optics are actually handled by the tools of phenomenological electrodynamics). In general, along this way of thought it would be difficult to consider the short-wavelength phenomena in matter, e.g., those for large absolute values of dielectric function or refractive index. Besides, the matter response to an electromagnetic field may essentially depend on the motion of particles, the latter in reality "feeling" local and instantaneous fields in each spacetime point and not the mean fields averaged over the volume containing many particles. It is similar to the fact that the car driver reacts on the "here and now" traffic conditions more acutely than on smooth road curves, speed limiting signs, information tableaux and other factors of "macroscopic" character.

An alternative - and more correct - averaging method is not the one over the "physically infinitesimal volume", but a standard method of statistical physics: averaging over a statistical ensemble, e.g., in the case of equilibrium over the Gibbs distribution. Taking averages over the quantum state, with the wave function for the pure state or with the density matrix for the mixed state, will be discussed separately below. One may notice that in the statistical method there is no spatial averaging *per se*, and the fields can still remain quasilocal and quasi-instantaneous[126].

In the system of Maxwell equations

$$curl\mathbf{H} - \frac{1}{c}\frac{\partial \mathbf{E}}{\partial t} = \frac{4\pi}{c}(\mathbf{j} + \mathbf{j_0}) \tag{5.18}$$

$$div\mathbf{E} = 4\pi(\rho + \rho_0) \tag{5.19}$$

$$curl\mathbf{E} + \frac{1}{c}\frac{\partial \mathbf{H}}{\partial t} = 0 \tag{5.20}$$

$$div\mathbf{H} = 0, \tag{5.21}$$

where $\rho_0$ and $\mathbf{j_0}$ are external charge and current densities, the induced currents $\mathbf{j}$ and charges $\rho$ are the functions of fields in the matter, $\mathbf{j} = \mathbf{j}(\mathbf{E})$. In phenomenological electrodynamics, the quantities $\mathbf{E}, \mathbf{H}, \rho, \mathbf{j}$ are assumed to be averaged over either a "physically infinitesimal volume" or a statistical ensemble (see below). The relationship between the field and the current induced by it represents the matter response to an electromagnetic excitation and determines such essential quantities as the medium susceptibilities, both linear and nonlinear.

The phenomenological approach allows one to conveniently formulate mathematical problems for the "Maxwell operator". By introducing the tensor functions $\epsilon_{ij}(x)$ and $\mu_{ij}(x), x \in \Omega \subset \mathbb{R}^3$ i.e., dielectric permittivity and magnetic permeability of the medium (see below a detailed discussion of these quantities), we can write the homogeneous Maxwell equations in the operator form through the stationary operator $\mathcal{M}(\epsilon, \mu)$ acting on the pair $(\mathbf{E}, \mathbf{H})^T$ where $\mathbf{E} = \mathbf{E}(\omega, \mathbf{r}), \mathbf{H} = \mathbf{H}(\omega, \mathbf{r})$ are respectively electric and magnetic field vectors in the considered domain $\Omega$. More specifically, for a stationary electromagnetic field its eigenfrequencies correspond to the spectrum of $\mathcal{M}(\epsilon, \mu)$ acting on $(\mathbf{E}, \mathbf{H})^T$ according to the rule

$$\mathcal{M}(\epsilon, \mu)\begin{pmatrix}\mathbf{E}\\\mathbf{H}\end{pmatrix} = \begin{pmatrix}i\epsilon^{-1}\nabla \times \mathbf{H}\\-i\mu^{-1}\nabla \times \mathbf{E}\end{pmatrix}.$$

One usually assumes the solenoidal constraints $div(\epsilon\mathbf{E}) = 0, div(\mu\mathbf{H})$ and some boundary conditions, e.g.,

$$\mathbf{E}_\tau|_{\partial\Omega} = 0, (\mu\mathbf{H})_\nu|_{\partial\Omega} = 0.$$

Here index $\tau$ denotes the tangential and index $\nu$ the normal component of the respective vector on boundary $\partial\Omega$. In isotropic medium, permittivity and permeability may be reduced to positive scalar functions $\epsilon(\mathbf{r}), \mu(\mathbf{r})$. When these functions as well as boundary $\partial\Omega$ are sufficiently smooth, the spectral problem for the Maxwell operator can be solved. The Maxwell operator formulation is especially useful for calculating modes in electromagnetic resonators (see, e.g., [6, 5]).

---

[126] The microscopic electromagnetic fields are still bound from below by the nuclear scale ($\sim 10^{-13}$ cm). As for the macroscopic fields in the matter, it does not make sense to phenomenologically consider the distances smaller than the atomic ones ($\sim 10^{-8}$ cm).

In optical problems related to isotropic media, permittivity and permeability are also typically treated as smooth and positive scalar function determining the refraction index $n = (\epsilon\mu)^{1/2}$ and the velocity of light in the medium, $v_p h = c/n$. Then the "optical" metric

$$ds^2 = c^2 dt^2 - dx_i dx^i$$

turns, as already discussed, the considered domain $\Omega \subset \mathbb{R}^3$ into a Riemannian manifold.

### 5.9.1    The Traditional Averaging  Procedure

The meaning of the averaging over a "physically infinitesimal volume" needs to be made clear for the sake of sound understanding of phenomenological electrodynamics. Let us now carry out explicitly the standard procedure of such an averaging. Apart from being a useful exercise, it is also a useful trick the details of which are not always given in the textbooks. In the standard averaging scheme of transition to macroscopic electrodynamics, one usually introduces the electric induction vector $\mathbf{D} = \mathbf{E} + 4\pi\mathbf{P}$ and the magnetic field strength $\mathbf{H} = \mathbf{B} - 4\pi\mathbf{M}$ where $\mathbf{P}$ is the polarization vector of the medium and $\mathbf{M}$ is the magnetization vector (by definition, it is the mean magnetic dipole moment for unit volume, $\mathbf{M}(r) = \sum_i N_i \overline{\mathbf{m}_i(\mathbf{r})}$, $\overline{\mathbf{m}_i(\mathbf{r})}$ is the average magnetic dipole of the elementary domain or cell in the vicinity of point $\mathbf{r}$, e.g., of a single molecule located at $\mathbf{r}$ and $N_i$ is the average number of such cells). One might complain that such phenomenological construction is far from being elegant. The average value of the magnetic field is usually denoted as $B$ and called the magnetic induction.

In the classical averaging scheme over the "physically infinitesimal volume", the polarization vector $\mathbf{P}$ is usually defined as the vector whose divergence equals (up to a sign) the average charge density in the medium, $\nabla\mathbf{P} = -\rho$ (see [208], §6). This definition is consistent with the general requirement of the matter electric neutrality, the latter being necessary for the stability of the matter. One can easily understand the physical meaning of the polarization vector $\mathbf{P}$ as the average dipole moment of the unit volume. Indeed, if for the zero-moment of the averaged charge density $\rho$ we have, due to neutrality,

$$\bar{\rho} = \int_\Omega \bar{\rho}(\mathbf{r}, t) d^3 r = 0$$

for all $t$, for the first moment we may write

$$\int_\Omega \mathbf{r}\bar{\rho}(\mathbf{r}, t) d^3 r = \int_\Omega \mathbf{P}(\mathbf{r}, t) d^3 r$$

where integration runs over the whole matter (e.g., a macroscopic dielectric body). One may notice that these relations are supposed to be valid not only in the static case. To prove the second relation, we can use the vector integral identity

$$\oint_\Sigma \mathbf{r}(\mathbf{P}d\sigma) = \int_\Omega (\mathbf{P}\nabla)\mathbf{r}d^3r + \int_\Omega \mathbf{r}\nabla\mathbf{P}d^3r,$$

where $\Sigma$ is any surface enclosing the matter volume $\Omega$. If this surface passes outside this volume (a material body), the integral over it vanishes, and since $(\mathbf{P}\nabla)\mathbf{r} = \mathbf{P}$ we get the required relationship for the first moment of $\bar\rho$. The above vector identity can be proved by a number of ways, e.g., by using the tensor (index) notations. The simplest, but possibly not very elegant, proof would consist in using the identity

$$\nabla(\varphi(\mathbf{r})\mathbf{P}) = \varphi\nabla\mathbf{P} + (\mathbf{P}\nabla)\varphi$$

which can be applied to any component of $\mathbf{r} = (x, y, z)$ and then integrating it over $\Omega$, with the left-hand side giving the surface integral.

Thus, we have the four vectors - two vector field pairs - of macroscopic electrodynamics, $\mathbf{E}, \mathbf{D}, \mathbf{H}, \mathbf{B}$ , which must be supplemented by some phenomenological relations between these vectors. Besides, if the above four quantities must satisfy the equations of the type of Maxwell's equation for the microscopic vacuum values, the question of sources for the phenomenological quantities arises. The textbook relationships between the vectors within each pair, $\mathbf{D} = \epsilon\mathbf{E}$ and $\mathbf{B} = \mu\mathbf{H}$ , complete the traditional construction of macroscopic electrodynamics.

This simple construction can be justified for static or slowly varying (quasistationary) fields, but it usually becomes totally inadequate for rapidly changing vector functions such as for short wavelengths or ultrashort pulses (USP), the latter being an increasingly popular tool in contemporary laser physics. One may note that such commonly used terms as "short wavelengths" or "hard electromagnetic radiation" are relative: for certain media the radiation typically perceived as "soft", e.g., infrared, may exhibit short-wavelength features whereas in some other type of matter the same radiation may be treated as quasistationary. Thus in plasma, where the average distance between the particles, $n^{-1/3}$, $n$ is the particle density, may easily exceed the characteristic wavelength $\lambda$ of the electromagnetic field, using the phenomenological field equations obtained by averaging over the physically infinitesimal volume can easily become meaningless.

I have already mentioned that the conventional approach to macroscopic electrodynamics, corresponding to the averaging of microscopic fields over "physically infinitesimal volume", consists in the additive decomposition of the total induced current and charge densities, $\mathbf{j}$ and $\rho$ (also averaged over the physically infinitesimal volume), into "physically different" parts, e.g.,

$$\mathbf{j} = \mathbf{j}_c + \mathbf{j}_p + \mathbf{j}_m,$$

where $\mathbf{j}_c$ represents the current of conductivity electrons, $\mathbf{j}_p$ is the polarization current, $\mathbf{j}_p = \partial \mathbf{P}/\partial t$, where $\mathbf{j}_m$ is the magnetization current, $\mathbf{j}_m = c\,curl\mathbf{M}$ . (I would recommend reading carefully the respective material in the textbook by L. D. Landau and E. M. Lifshitz [208].) One does not need to perform the same decomposition procedure for the charge density because of the constraint given by the continuity equation 5.3 which remains valid after any kind of averaging (due to its linearity).

### 5.9.2   Ensemble Averaging of Fields and Currents

Paying attention to the difference between averaging over a "physically infinitesimal volume" and ensemble averaging may be essential to understanding of the fundamentals involved. The Maxwell equations in the medium obtained by the traditional averaging procedure (i.e., over the "physically infinitesimal volume") and having the form ([208], §75)

$$curl\mathbf{H} - \frac{1}{c}\frac{\partial \mathbf{D}}{\partial t} = \frac{4\pi}{c}\mathbf{j}_0 \tag{5.22}$$

$$div\mathbf{D} = 4\pi\rho_0 \tag{5.23}$$

$$curl\mathbf{E} + \frac{1}{c}\frac{\partial \mathbf{B}}{\partial t} = 0 \tag{5.24}$$

$$div\mathbf{B} = 0, \tag{5.25}$$

being supplemented by the "material equations" relating the quantities $\mathbf{D}, \mathbf{B}$ and $\mathbf{E}, \mathbf{H}, \mathbf{D} = \epsilon \mathbf{E}$ and $\mathbf{B} = \mu\mathbf{H}$ can be used without reservations only for relatively slow-varying fields (static and quasistationary). For fast changing electromagnetic fields and pulses as well as for the media where spatial field variations can be shorter than the average distance between the constituent particles (e.g., in plasmas), these equations become inconvenient or merely inadequate. Indeed, it seems to be nearly obvious that, even leaving aside the dubious and in general mathematically incorrect procedure of averaging over the physically infinitesimal volume, breaking down the total current into presumably non-overlapping components cannot be unambiguous for high frequencies. It may be easily seen that the currents excited in the medium due to free and to bound electrons cannot be separated already for optical frequencies or for rapid variations of an external field (ultrashort pulses). One may illustrate this fact by a simple example of an atomic electron in an external field $\mathbf{E}$. For simplicity, we may consider here classical (non-quantum and non-relativistic) motion, $m\ddot{\mathbf{r}} = e\mathbf{E}(t)$, $e, m$ are the electron charge and mass respectively, then the characteristic displacement or oscillation amplitude of an electron in the field $\mathbf{E}(t)$ of an electromagnetic pulse or wave is $r_0 \sim eE\tau^2/m$ or $r_0 \sim eE/m\omega^2$ where $\tau$ is the ultrashort pulse duration. Such a displacement can be readily scaled down to atomic distances (Bohr's radius, $r_B \sim 10^{-8}$cm) even for rather strong fields, say only an order of magnitude lower than atomic fields, $E_{at} \sim e/a_B^2 \sim 10^9$V/cm$^2$.

This simple example (and many others, too) demonstrates that the difference between the conductivity, polarization and magnetization currents, the latter being due to the charge motion along closed trajectories, rapidly becomes very vague with the decreased wavelength of the radiation field. Starting from ultraviolet light, unambiguous decomposition of current into these three components becomes virtually impossible. At least, such a decomposition at optical frequencies and for ultrashort electromagnetic pulses is pretty arbitrary. Thus, the optimal possibility we have in the case of fast varying fields is to consider the total current $\mathbf{j}(\mathbf{r}, t)$ incorporating all kinds of charge motions caused by an electromagnetic field. This current, in the situation of thermal equilibrium, should be averaged over a statistical ensemble i.e., with the Gibbs distribution (or equilibrium density matrix in the quantum case). Technically, it is often convenient to introduce the total polarization $\mathcal{P}(\mathbf{r}, t)$ (see [209]) instead of the total current:

$$\mathcal{P}(\mathbf{r}, t) = \int_{-\infty}^{t} \mathbf{j}(\mathbf{r}, t') dt', \qquad \mathbf{j}(\mathbf{r}, t) = \partial_t \mathcal{P}(\mathbf{r}, t)$$

It is important to remember that total polarization $\mathcal{P}(\mathbf{r}, t)$ includes all currents, due both to free and to bound charges, and not only the displacement current, as in the intuitive scheme of averaging over the physically infinitesimal volume. One can also see that the total polarization accumulates current contributions starting from the remote past ($t \to -\infty$), but not from the domain $t' > t$. This is a manifestation of the causality principle which is one of the crucial assumptions in physics (see also Chapters 6, 9). In the distant past, $t \to -\infty$, polarization is assumed to be absent.

Introducing the total polarization enables us to write down the total induction, $\mathbf{D}(\mathbf{r}, t) = \mathbf{E}(\mathbf{r}, t) + 4\pi \mathbf{P}(\mathbf{r}, t)$, and not only induction only owing to displaced bound charges, as in traditional theory of dielectrics. Using the total induction, we may write the averaged Maxwell equations in the medium as

$$curl\mathbf{E} + \frac{1}{c}\frac{\partial \mathbf{H}}{\partial t} = 0 \qquad (5.26)$$

$$div\mathbf{H} = 0 \qquad (5.27)$$

$$curl\mathbf{H} - \frac{1}{c}\frac{\partial \mathbf{D}}{\partial t} = \frac{4\pi}{c}\mathbf{j}_0 \qquad (5.28)$$

$$div\mathbf{D} = 4\pi\rho_0, \qquad (5.29)$$

where the last equation is in fact a consequence of the continuity equation 5.3. Indeed, from

$$\frac{\partial \rho}{\partial t} + \nabla \mathbf{j} = \frac{\partial \rho}{\partial t} + \nabla \frac{\partial \mathcal{P}}{\partial t} = 0$$

we have

$$\rho = -\nabla \mathbf{P} + \rho_{-\infty} = \frac{1}{4\pi}\nabla(\mathbf{D} - \mathbf{E}),$$

where $\rho_{-\infty}$ is the integration constant that can be put to zero (we assume that there were no polarization in the distant past). Then we have $4\pi\rho = -\nabla(\mathbf{D} - \mathbf{E})$, but $\nabla\mathbf{E} = 4\pi(\rho + \rho_0)$ where $\rho_0$ is, as before, the density of external charges introduced into the medium. Thus, we get $\nabla\mathbf{D} = 4\pi\rho_0$.

One may notice that in this "optical" approach one does not need to introduce, in addition to a magnetic field, such a quantity as magnetic induction and, consequently, magnetic susceptibility [209]. So, there is no distinction between $\mathbf{B}$ and $\mathbf{H}$. Indeed, the total polarization already contains a contribution from the magnetization (circular) currents, e.g., arising from nearly closed trajectories of electrons moving in the electromagnetic field of elliptically polarized light.

One often asks: why are currents and polarization in the medium considered the functions of the electric field $\mathbf{E}$ alone, with magnetic field $\mathbf{H}$ being disregarded both in the dielectric relationship $\mathcal{P} = \chi\mathbf{E}$ and in the conductivity relationship $\mathbf{j} = \sigma\mathbf{E}$? Or, to put it slightly differently, why is it always implied that the induction $\mathbf{D}$ is only proportional to $\mathbf{E}$ and not to $\mathbf{H}$? One might encounter several versions of answering this question. One of the versions is that $\mathbf{H}$ should be excluded because (1) it is an axial vector ($\mathbf{H} = \nabla \times \mathbf{A}$), whereas $\mathbf{E}$ is a "true" vector so that they both cannot be simply superposed, and (2) $\mathbf{H}$ is not a time-invariant quantity. The second argument implies, however, an a priori requirement imposed by our intuitive perception of the world: nobody can be sure that all electromagnetic phenomena must be strictly invariant under time reversal (even in the static case). In fact, it seems to be wrong (more about time-reversal invariance in Chapter 9). As to the first argument, it may be "neutralized" by introducing a pseudoscalar proportionality coefficient between $\mathbf{P}$ and $\mathbf{H}$ (or $\mathbf{j}$ and $\mathbf{H}$ and $\mathbf{D}$ and $\mathbf{H}$). In reality, we may disregard the magnetic field in the "material relations" because it can be expressed through the electric field using Maxwell's equations. Besides, even if we wished to explicitly involve the magnetic field, it would make little sense unless we considered ultrarelativistic motion of charges, which is an exotic case in the medium. The point is that factor $v/c$ always accompanying a direct effect of a magnetic field on moving charges would make its influence hardly noticeable for the particles whose characteristic velocities are of the atomic scale ($v_{at} \sim e^2/\hbar$) i.e., at least two orders of magnitude lower than the speed of light.

Now the main problem is: how to link the current induced in the medium to an external field exciting this current? It is clear that such a problem is in general extremely difficult and can hardly be solved for an arbitrary matter containing a macroscopic number ($N \sim 10^{23}$) of particles placed in a fortuitous field. Nonetheless, there are several approaches to this universal problem. One of such approaches has been developed within the framework of nonlinear optics (NLO). This is a comparatively new discipline which emerged in the 1960s, following the advent of lasers. Before that time optics was essentially linear, and probably the only attempt to consider nonlinear electromagnetic effects were made in quantum electrodynamics while treating the scattering of light by light [210] (see also [263]), which is in fact vacuum breakdown process [127]. Linearity of electrodynamics required that the polarization of the medium and induced current should be written respectively as $\mathcal{P}_i - \chi_{ij} E_j$, where $\chi_{ij}$ is the susceptibility tensor, and $j_i \sigma_{ij} E_j$, where $\sigma_{ij}$ is the conductivity tensor[128]. A more general, nonlocal, linear expression linking the induced current to an electric field may be written as

$$j_i(\mathbf{r}, t) \equiv j_i^{(1)}(\mathbf{r}, t) = \int \sigma_{ij}(\mathbf{r}, \mathbf{r}_1; t, t_1) E_j(\mathbf{r}_1, t_1) d^3 r_1 d d_1 \qquad (5.30)$$

Here I intentionally did not indicate integration limits implying that they are ranging from $-\infty$ to $+\infty$; however, I think it necessary to comment on this point. I have already mentioned that if one assumes that the causality principle holds for all types of medium, then one must consider polarization and currents, observed at time $t$, depending only on the field values related to the preceding moments of time, $t_1 < t$. Therefore, integration over $t_1$ goes only from $-\infty$ to $t$. Furthermore, relativistic causality requires that spatial integration may spread only over spacelike domains $|\mathbf{r}_1 - \mathbf{r}| < c|t_1 - t|$ because an electromagnetic field existing at points outside this domain cannot produce an excitation of the medium at a spacetime point $(\mathbf{r}, t)$. In fact, however, integral expressions for currents and polarization determining the response of the medium essentially depend on the properties of the response functions $\sigma_{ij}(\mathbf{r}, \mathbf{r}_1; t, t_1)$, $\chi_{ij}(\mathbf{r}, \mathbf{r}_1; t, t_1)$ serving as kernels in the integral transformations realizing nonlocal maps. Experience as well as physical considerations show that the response functions are essentially different from zero at much shorter distances than those required by relativistic causality. Thus, one may consider integration limits really infinite, which is true not only in the linear case, but also for nonlinear response functions.

It may be rewarding to clarify the physical reason for spatial and temporal nonlocality in (5.30). For integration over time, the origin of nonlocality is pretty obvious: it is the retardation of response. For example, a particle at point $\mathbf{r}$ of its trajectory $\mathbf{r}(t)$ feels the influence of the force $\mathbf{F}(\mathbf{r}, t)$ not only taken at time-point $t$, but also related to preceding moments, $t' < t$. Indeed, directly from Newton's equation we have

$$\mathbf{r}(t) = \int_{t_0}^{t} dt_1 \int_{t_0}^{t} dt_2 \frac{1}{m} \mathbf{F}(\mathbf{r}(t_2), t_2) + \mathbf{v}_0(t - t_0) + \mathbf{r}_0$$

where $\mathbf{r}_0 := \mathbf{r}(t_0), \mathbf{v}_0 := \dot{\mathbf{r}}(t_0) \equiv \dot{\mathbf{r}}_0$ are integration constants having the meaning of the starting point and initial velocity of the trajectory; $t_0$ is the initial time point for the motion. In many physical situations one is not much interested in parameter $t_0$ so that, e.g., for many particles one can average over all possible $t_0$ (as well as over initial conditions, which amounts to a simple statistical approach) or, in mechanical problems, one may take the limit $t_0 \rightarrow -\infty$ assuming $\mathbf{r}_0 = \mathbf{r}_{-\infty} = 0$ and $\mathbf{v}_0 = \mathbf{v}_{-\infty} = 0$ so that

$$\mathbf{r}(t) = \int_{t_0}^{t} dt_1 \int_{t_0}^{t} dt_2 \frac{1}{m} \mathbf{F}(\mathbf{r}(t_2), t_2).$$

---

[127] The usual electrodynamics was perceived as essentially linear so that even in the classical textbook by L. D. Landau and E. M. Lifshitz [208], §77, small intensities of fast changing fields are always assumed, see the text before formula (77.3). The development of lasers has radically changed this perception.

[128] Here we disregard the difference between contra- and covariant components; such a difference is inessential for our - rather superficial - intuitive discussion.

We shall also discuss the role of the initial moment of time $t_0$ in many-particle systems in connection with the Liouville equation and the choice of supplementary (e.g., boundary) conditions to it (Chapter 7). In many cases it is appropriate to shift the non-physical parameter $t_0$ to $-\infty$.

If the particle considered is, e.g., a free electron, then the force $\mathbf{F}$ may be exerted by an electromagnetic wave or a pulse, $\mathbf{F}(\mathbf{r}, t) \approx e\mathbf{E}(\mathbf{r}, t)$, whereas for a bound electron the force acting on it is a superposition of the incident and atomic fields

$$\mathbf{F}(\mathbf{r}, t) \approx e\mathbf{E}(\mathbf{r}, t) - \nabla V(\{\mathbf{R}\}, \mathbf{r}, t).$$

Here $V(\{\mathbf{R}\}, \mathbf{r}, t)$ is the atomic potential depending, besides spacetime variables $\mathbf{r}, t$, on a set of atomic parameters denoted by symbol $\{\mathbf{R}\}$. When external fields are much lower than the atomic Coulomb fields, $E_{at} \sim e/r_B^2$, the force acting on electron is mainly determined by the atomic potential and the time interval $\tau$ of integration in (5.30) is essentially determined by atomic (or interatomic) parameters i.e., is of the order of atomic time $\tau_{at} \sim \hbar/I \sim 10^{-16}$ cm where $I$ is the typical atomic energy of the order of ionization potential, $I \sim e^2/r_B$. Physically, this is the time of the atomic velocity change, $v_{at} \sim r_B/\tau_{at} \sim r_B I/\hbar$. For a free electron, the integration time interval is of the order of the velocity relaxation time (see some details in Chapter 9).

There may be situations when one has to take into account initial correlations between the particles entering the medium or the domain occupied by a field at times $t_{0a} \rightarrow -\infty$ ($a$ enumerates particles). Then simple averaging or taking the limit $t_{0a} \rightarrow -\infty$ does not hold and should be replaced by an integration (convolution) with the correlation functions. One can see a simple physical example in [151].

In strong electromagnetic fields such as produced by lasers, polarization and currents in the medium may become nonlinear functions of the field. This nonlinearity can be easily understood by using one of our favorite models: that of an oscillator, but a nonlinear one (see below). For example, in the case of quadratic nonlinearity we shall have

$$j_i^{(2)}(\mathbf{r}, t) = \int \sigma_{ijk}(\mathbf{r}, \mathbf{r}_1, \mathbf{r}_2; t, t_1, t_2) E_j(\mathbf{r}_1, t_1) E_k(\mathbf{r}_2, t_2) d^3 r_1 d^3 r_2 dt_1 dt_2$$

and, in general,

$$j_i^{(n)}(\mathbf{r}, t)$$
$$= \int \sigma_{j_1 \dots j_n}(\mathbf{r}, \mathbf{r}_1, \dots, \mathbf{r}_n; t, t_1, \dots, t_n) E_{j_1}(\mathbf{r}_1, t_1) \dots E_{j_n}(\mathbf{r}_n, t_n) d^3 r_1 \dots d^3 r_n dt_1 \dots dt_n$$

Continuing this approximation process, we obtain an expansion of the current[129] (or polarization) in powers of electric field:

$$\mathbf{j}(\mathbf{r}, t) = \sum_{n=1}^{\infty} j^{(n)}(\mathbf{r}, t).$$

Now the question naturally arises: what is the actual (dimensionless) parameter of such an expansion? Plausible physical discourse gives a cause for a belief that in wide classes of condensed matter such as dielectric solids and non-conducting liquids the expansion parameter should be $\eta_E = E/E_{at}$, where $E_{at} \sim e/r_B^2 \sim 10^9 V/\mathrm{cm}$ is a characteristic atomic field. Indeed, for $\eta_E \ll 1$ an external electromagnetic field cannot strongly perturb the matter, and an expansion on $\eta_E$, with retaining only a few initial terms, must be fully adequate. However, one has to bear in mind that although parameter $\eta_E$ can be used for expansion over external field in many types of condensed media and even in gases, there exist other types of media, in particular containing free charges such as plasmas, metals, and even semiconductors, where parameters of nonlinearity can be different. At least, one needs to carry out a special analysis for each of these media. This is an important point, and we shall return to it later.

---

[129] Recall that here we are talking about microscopic currents, also called microcurrent, induced by an external macroscopic field in matter.

# 6 The Quantum World

The quantum world is generally considered to be weird, indeterministic, counterintuitive, hard to visualize and even irreducibly requiring a human observer. I think this is a philosophical exaggeration, a projection of human anxieties on physical problems. All these fears are not required by experimental results in the quantum domain. The quantum world is nothing more than a collection of very handsome mathematical models which are, by the way, simpler than mathematical models of classical mechanics. In this chapter, we shall discuss some of these quantum models, their origins and outcomes, and we shall see that there is nothing weird about them. For example, an electron is, in quantum mechanics, neither a particle nor a wave in the conventional (classical) meaning. It is a mathematical model, a formula, in the sense that its properties are described by the quantum mechanical equations. One can build up other models for electron, for example, corpuscular as in Newtonian mechanics and in Bohm's version of quantum mechanics or wavelike as in de Broglie version. Nevertheless, standard quantum mechanics based on the Schrödinger equation is sufficient for practical computations so there is usually no need to resort to other models.

Quantum mechanics, even in its simple form of the Schrödinger equation, is so far typically alien to computer scientists or to mechanical engineers, but these two categories of researchers are in general fully satisfied when one says that the classical mechanics they think they understand can be derived from the quantum theory as a particular case for "heavy" bodies. Quantum-mechanical modeling stands out in many scientific areas because it is based on *ab initio* calculations i.e., attempts to tackle the problem from first principles. Ideally, quantum-mechanical modeling does not rely on phenomenological parameters, with the exception of fundamental constants and, possibly, the atomic number of constituent atoms. This independence from external, hand-introduced adjustment, is intended to ensure that preconceptions and bias should play a minimal role in the final results produced by quantum modeling. Yet *ab initio* calculations have also the dark side which may be designated by a single word: complexity. Firstly, the high complexity of *ab initio* quantum-mechanical calculations enables one to obtain exact analytical solutions only in rare specific cases and, secondly, when one is forced to resort to numerical techniques, complexity leads to inhibitive scaling of computational costs and required resources. And if the computational costs grow exponentially with, e.g., the size of the system under study, such modeling efforts may become too demanding to be of practical use - even despite the rapid progress of computer technology and low prices on resources. This fact justifies the retreat from head-on numerical solutions by using well-controlled approximations, also of analytical nature. Historically the first such approximations were perturbation and semiclassical methods. Both methods are examples of asymptotic expansions, although of different, sometimes even opposite kind, and we shall discuss these methods in some detail a little later.

Quantum mechanics is quite similar to the classical field theory discussed in the previous chapter because these two theories use the same wave equations. Moreover, both theories admit a duality i.e., a kind of correspondence between particles and fields which may be interpreted as a semiclassical back-and-forth migration between mechanics (the study of particle motion) and field theories (the study of wavelike excitations). Although quantum mechanics is basically of the mathematical modeling character, it is more than any other discipline a mix of concepts and calculations. Sometimes the concept component prevails, then quantum mechanics becomes overburdened with philosophy; in other cases mathematical techniques may dominate, and then mathematical structures and occasionally some fancy mathematical models protrude.  From a narrow pragmatic viewpoint, quantum mechanics is reduced to the calculation of states and their eigenvalues, evolution of states with time, and transitions between them (including scattering). Thus, quantum mechanics is basically rather simple as compared, e.g., with classical mechanics, which does not exclude the fact that specific calculations may be very difficult, as each person who performed quantum-mechanical computations might have noticed. Nevertheless, it would hardly be possible to expose the whole of it in a single chapter. Therefore, I shall restrict myself to the discussion of the main quantum-mechanical structures. I do not use the term "mathematical methods of quantum mechanics" as I think it is a tautology (like, e.g., new innovation) since the whole quantum mechanics is a mathematical model of reality. The reader can find all the answers to arising - quite naturally - questions in any textbook on quantum mechanics, for instance, in those that I liked most and referred to in this text.

Contemporary quantum physics tends to be constructed as an axiomatic theory i.e., based on a set of mathematical axioms. This methodology has both advantages and drawbacks. The good side is that the axiomatic approach is convenient in order to derive powerful mathematical tools out of the set of axioms. The bad side is that the connection of these tools with reality is rather weak or may be totally absent. Besides, there exists a hidden danger to take as axioms such statements that cannot be directly tested in physical experiment as, e.g., in string theories.  In such cases, there may be, at best, the hope to test some remote consequences. If the base quantum postulates[130] cannot be directly physically tested, there exists a hazard of putting too many of them or making the system of postulates inadequate in some other way which can later result in contradictions. Nevertheless, the most fanciful starting assumptions are allowed in mathematical modeling - it's a free craft.

Despite a comparatively primitive mathematical scheme of nonrelativistic quantum mechanics, in numerous discussions with my colleagues[131] I have found that there is a significant confusion about many aspects of quantum mechanics, which mainly arises in connection with the physical interpretation of the formalism. The majority of theoretical physicists probably think nowadays that the so-called physical intuition is entirely based on classical concepts and is useless in quantum theory. Therefore, quantum theories can be constructed on a more or less arbitrary set of axioms, the sole essential requirement being that such a system of axioms should be non-contradictory. In this way, one can construct a number of theories for different sets of axioms being put in the base and chosen more or less arbitrarily. This approach is close to mathematical model building. The ultimate goal is very pragmatic: one should not care about the connection of the mathematical scheme with reality, the only thing that matters is that the constructed models or the consequences of the axiomatically derived theories could quantitatively describe a wide set of experimental results. It is, by the way, this approach that has led to changes in the notion of a physical explanation. In modern physics, such an explanation is synonymous with providing a mathematical description which is not necessarily the same thing.

---

[130] I make no difference between axioms and postulates in this context - paying attention to such a distinction would be too philosophical.

Although I do not want to join the philosophical ritual dance around quantum mechanics, I still decided to include the discussion of its various interpretations which is by necessity an interpolation between mathematical modeling and philosophy. To better illustrate the ideas involved in the development of quantum mechanics I allowed myself to make a brief historical overview. There exist many exhaustive sources on the conceptual development of quantum mechanics (see, e.g., [137]), but I, nonetheless, have found some unexpected items which I would like to share with the reader.

## 6.1    In the Middle of Revolution

Quantum mechanics exists less than a hundred years, i.e., this discipline is much younger than, say, Newtonian mechanics or hydrodynamics. One cannot exclude the possibility that the current formulation of quantum mechanics is far from its final form; it is still in the process of development and may be modified in such a way as to easily construct such models that appear ambiguous or even incorrect to a number of today's researchers. For example, can quantum mechanics treated as a probabilistic theory describe unique objects such as the entire universe? How in general can one correctly handle isolated events and stand-alone objects in standard quantum mechanics? However, the example of more ancient and stable disciplines such as fluid dynamics or classical mechanics shows that there are chances that a discipline does not change considerably. Probably, the necessary condition for the stability of a discipline is wide acceptance of its language, including interpretation in lay terms, which lacks in quantum theory. If one tries to interpret quantum mechanics in common terms, many paradoxes arise, which produces a feeling of dissatisfaction by many physicists. It is well known that even founders of quantum mechanics including Einstein, de

---

[131] That was rather long ago: discussions on physical Weltanschauung are popular mostly among young people. Physics is in general a "game for the young".

Broglie and Schrödinger did not consider quantum mechanics a complete theory, because the reality it described seemed to them strange and devoid of simple physical interpretation. Indeed, standard quantum theory is formulated in such a way as to produce only probabilities, which very fact Einstein did not want to accept. Besides, there appeared to be mathematical divergences between scientists on how to interpret these probabilities. The probability theory was devised to formalize experimental results when there is a lack of information. Is information by default missing in quantum mechanics? Even today, some researchers are convinced that quantum mechanics does not provide a full description (see below section on Bohmian mechanics). There are other things - actually, all of them interrelated - that can frustrate a philosophically minded physicist, even in case she/he is brilliant in applying mathematical tools of quantum theory. What is the uncertainty principle, can one explain it without using central moments of operators defined on unitary space? Mathematically, it is not difficult to demonstrate that the average spread of values produced by non-commuting operators (in quantum language - observables) cannot be made arbitrarily small simultaneously, but why does it restrict the experimental accuracy? Some people who are accustomed to quantum language think that the answer to this question is trivial, yet personally I think this issue is quite intricate; it is connected with the notions of coherent and squeezed states and touches such profound question as the nature of the photon. Furthermore, conceptual trickery of quantum measurement theory seems to me especially bizarre. How can it be that physical reality should depend on the manner of observation? Is there anything, according to quantum mechanics, if one does not observe at all, in the limiting case when there are no humans in the world? What is then the so-called reality?

   All these and similar questions are well known and usually discussed by philosophers, whereas physicists tend to laughingly disregard them. This attitude probably stems from the substantial effectiveness in explaining experimental results in microscopic physics by means of quantum theory. I remember that professor V. M. Galitskii, an outstanding theoretical physicist who could at once deeply understand almost any problem, recommended to his students that they should for a while forget all the questions of interpretation that could inevitably arise at the very start of studying quantum mechanics. A person should not, said Galitskii, pay attention to the philosophical issues before she/he have mastered the mathematical tools of quantum mechanics, and then there are the chances all possible perplexity would somehow fade away. I fully agree with him.

   Quantum mechanics has currently become a very successful engineering discipline whose foundation - condensed-matter physics and in particular solid state theory - may be regarded as specific applications of nonrelativistic quantum mechanics. All these (and other) loud successes tend to shade inconsistences in current formulations and, especially, interpretations of quantum theory. Physicists are reluctant to invent a new theory that would mostly satisfy philosophers who did not, as physicists used to think, contributed much in knowledge, when the current theory seems to work perfectly well. Indeed, why should they?

However, it is curious that quantum theory is more than one hundred years old, but we still need to ruminate about quantization. Unfortunately, there are up till now many unclear issues which must be thought over. What are the peculiarities of quantum mechanics which prevent it from being interpreted in the classical sense, without incomprehensible measurement paradigm, i.e., separation of the world into two subworlds: active observers and passive systems to be observed? Even if we take the relatively simple Schrödinger formulation, which is a mathematical model based on the scheme of boundary value problems (see below), can we regard the wave function as a field distributed in space like any classical field, for instance, electromagnetic field?

We may put aside some profound gnoseological considerations and take up several formal grounds precluding the possibility of classical interpretation of quantum models. I counted five underlying reasons hampering interpretation of quantum mechanics in classical manner.

1.    Standard quantum mechanics uses the tools of linear operators in Hilbertspace, with canonical transformations of wave functions corresponding to change of bases, fully analogous to the Fourier transform. I have already mentioned that choice of a specific basis is not the natural way for mathematical description of a physical state. Manifestation of this independence on basis in quantum theory consists in the fact that transformed (in many simple cases, just Fourier transformed) wave functions are equivalent to the coordinate-dependent wave function and all of them describe the same physical state. Not only the squared module of the initial wave function $\psi(\mathbf{r})$ (in the coordinate representation) has a physical meaning, but also all the transformed wave functions corresponding to other representations. We shall later illustrate this fact by examples of working in momentum representation.

2.    In the coordinate representation the wave function is a complex valued function defined on a space-time domain. This implies that one should invent a certain procedure to obtain observable quantities, which are real, by using the wave function. Already this fact leads to ambiguities (see below on the Bohm's version of quantum mechanics). Furthermore, even in the simplest case of a single particle, the wave function does not necessarily exist and it does not always change according to the Schrödinger equation (or even according to other evolution equation of similar type). In the quantum measurement situation the initial wave function is simply crossed out and replaced by some other. This ugly procedure is traditionally called "reduction of the wave packet", although neither the initial nor the secondary wave function may have the wave packet form. Such instant change of the wave function is hard to reconcile with the concept of the classical field.

3.    Many-body problems in quantum mechanics substantially differ from the same kind of problems in classical mechanics. Quantum mechanical many-body problems[132] are characterized by peculiarities not allowing one to reduce such problems to the motion of individual particles (quasiparticles and elementary excitations are *not* individual particles). Moreover, these peculiarities prevent formulating the many-body problem as that for a field in the Euclidean 3*D* space. Thus, if a complex system consisting of many particles is described by a wave function taking into account all the particles, one cannot ascribe separate wave functions to the individual particles.

One more specific feature of quantum mechanics cardinally distinguishing it from classical consists in the notion of spin and one of its nontrivial implications - the connection of spin with statistics for identical particles. For fermions, i.e., particles with half-integer spin, which obey the Pauli exclusion principle and, in the many-body state, are described by anti-symmetric wave functions leading to the Fermi statistical distribution, there exists a distinctive quantum interaction that cannot be reduced to the usual force interaction in the Euclidean space. Another kind of interaction that also cannot be reduced to the classical forces exists between bosons - the particles described by symmetric wave functions and subordinated to the Bose statistics. Recall that the matter surrounding us consists of fermions and its stability is in fact due to the specific quantum interaction. Forces between particles, i.e., field interactions including classical fields such as electromagnetism, are transferred by bosons, which behave totally different from fermions.

4.    In the case of a complex multi-particle system, its wave function, even in the coordinate representation, depends not on three Euclidean coordinates but on a set of parameters corresponding to quantum degrees of freedom. Therefore, the wave function describing the system is a complex-valued function defined on a certain multidimensional domain which does not necessarily lie in the real physical world.

5.    Classical physics is implicitly based on the assumption that one can obtain information about the system infinitesimally influencing it. This is in general not true: information always has a price, and a device[133] measuring an observable destroys the initial state and produces a new one. In classical mechanics one can neglect the device-object interaction and consider that observation plays no role, thus talking about the object state of motion irrespective of observation means. Therefore, by the classical description one may understand the one with the disturbance introduced by observation being disregarded [134]. Accuracy of such a description is limited by the uncertainty relations. The measurement procedure is characterized by a certain probability, which in standard quantum mechanics is mathematically described by the squared module of the scalar product of the initial and final wave functions. However, human intuition being trained in the world of classical objects typically fails to accept the fact that an observed object is necessarily disturbed by the measuring device. This intuitive perception leads to inadequate conclusions.

---

[132] The two-body and Kepler problem are treated separably. There is also a quantum specificity in the two-body problem.

[133] The device is traditionally understood as a physical system that can, on the one hand, interact with the quantum object producing some observable reaction and, on the other hand, admits, with some sufficient accuracy, the classical description. Therefore, the information filtered by the device does not require further conversion and can be directly read out by humans. It is immaterial whether the device is engineered by humans or is a lucky combination of classically described natural circumstances convenient for observations, where the quantum object is placed. I don't consider this definition fully correct since in it the device is, by default, classical and macroscopic, whereas it is not necessarily the case. The device is a physical system, and nobody cares whether it is quantum or classical. "Why must I treat the measuring device classically? What will happen to me if I don't?" - these words are ascribed to E. P. Wigner, the great Hungarian/US physicist and mathematician, although I, honestly speaking, failed to find them in Wigner's works. Moreover, "quantum object" is not identical to "microobject".

Due to these and other specific features of quantum mechanics, all never ending attempts to interpret quantum theory in the classical sense remain deficient. Interpretation of Bohr's ideas as reflecting an extreme positivism produced a reaction disclaiming, for the sake of "realism", mathematical models based on the wave function. One might notice an artificial character of many such attempts (starting from de Broglie) and some lack of heuristic value: in contrast to the adherents of Copenhagen interpretation, authors of allegedly "classical" variants of quantum mechanics seem to be reluctant to solve radically new physical problems. Rather, the treatment of quantum problems by means of classical considerations is commonly tuned to the a priori known results from standard quantum mechanics. One cannot be surprised, for the interpretation in the spirit of classical "realism" is probably quite a narrow view. To impose on reality, contrary to all existing evidence, uniquely the rigid deterministic type of evolution, while refusing to admit a more general form of probabilistic rules, means to follow the dogmas and not the laws of nature. One might recall in relation to this that probability theory may be considered to generalize classical analysis, with statistical distributions being reduced to deterministic functions when variance (and other higher moments) are striving to zero.

The tendency to reject probabilistic interpretation based on the wave function is illustrated today by the Bohmian version of quantum mechanics, which provides the classical formulation based on trajectories derived entirely from the Schrödinger equation. The price to pay for introducing classical notions such as individual trajectories into quantum theory is rather high: one must also introduce some unknown and unmeasurable quantity known as the quantum potential, which has some weird properties (non-locality, non-diminishing with distance, etc.). Many physicists do not like "classical" constructions in quantum mechanics - of course, I do not mean here semiclassical approximation (or quasiclassics, as we used to call it in Russia), which is a powerful asymptotic theory to be discussed later in some detail (see 6.9. Quantum-Classical Correspondence). Speaking of individual trajectories, I did not imply here the Feynman path integral which is probably the best quantization means whose value is especially significant in quantum field theory.

---

[134] An element of observational influence still remains: one must account for a given system of reference.

We have already seen that interaction with measuring devices in quantum mechanics required new methods of description of quantum objects introducing a new element of relativism, in this case with respect to means of observation. One can remark that this loss of absoluteness does not destroy objectivity: one must only specify the accuracy. Likewise, a classical trajectory, while being considered quite objective, acquires a definite meaning only when the coordinate system is fixed.

One can recall in this connection that in the early electromagnetic theory created by M. Faraday in the first half of the 19th century the concept of field lines connecting positive and negative charges or magnet poles was predominant. Faraday considered these lines to be real and carrying electrical and magnetic forces. It is only after J. C. Maxwell constructed his classical electromagnetic theory (improved by O. Heaviside) based on partial differential equations that the field lines became unnecessary or, at least, of secondary nature to the electromagnetic field. In superconductivity models, magnetic field lines may become discrete - each line carrying a quantized amount of magnetic flux (see Chapter 9), this image was a prototype for string theory. Since the magnetic field is just one component of the general electromagnetic field, one can imagine also quantized lines of electrical field or other gauge fields (see, e.g., [130] on Wilson loops). Nowadays, in string theories one can observe a revival of the field lines; actually the strings themselves may be - to a certain extent - interpreted as field lines. Thus, the picture of lines hints at a possible duality between fields and trajectories, and possibly future versions of quantum mechanics will fully exploit this duality.

In my understanding, the relation of quantum and classical mechanics is not in formulating quantum concepts in classical language, an example of which is the alleged possibility of imposing trajectories on wave-like solutions, but in the correspondence principles and respective limits. The "particle-wave" duality is just the philosophical expression for the superposition principle, familiar already from elementary linear algebra. Indeed, the superposition principle requires some linear (vector) space which necessarily has a dual. Besides, what one really needs, in my opinion, is to find a rapid and uncomplicated connection between the abstract linear operator theory in Hilbert space and application-oriented prescriptions of quantum mechanics. Operator theory satisfies mathematicians, but appears to be too detached from real-life tasks in the engineering world. On the contrary, intuitive engineering models appeal to people engaged in "practical" problems, but mathematicians and even theoretical physicists tend to uncompromisingly reject such intuitive models. Quantum mechanics has acquired many aspects of an engineering discipline; in such a discipline it is especially important to see how apparently different branches can be joined together to produce new technologies and products.

The ingenious people who formulated the initial versions of quantum mechanics such as Bohr, Heisenberg and Schrödinger probably admitted that quantum mechanics gives an incomplete description of reality. My impression is that these people, while being also deep thinkers, supposed that if quantum mechanics was to be extended was unlikely that it would be done in isolation from other fields of science. It was not accidental that these creators of quantum mechanics tried to transcend the boundaries of compartmentalized disciplines. Bohr was much of a philosopher and it is well known that he was strongly influenced by S. Kierkegaard (see, e.g., the Wikipedia material http://en.wikipedia.org/wiki/Niels Bohr). In fact, this philosophy is, very crudely of course, the eloquent developments of the obvious thesis that nature manifests itself to us only through our senses. Heisenberg devoted much of his life to attempts to construct the unified field theory and new concepts of the whole physics [119, 124, 125]. Schrödinger who appeared to dislike the approach to quantum theory formulated by N. Bohr as "Nothing exists until it is measured" wrote many papers on various subjects including, as Heisenberg, investigations on unified field theory. He also was famous for his book "What is life?" [126] which is largely devoted to the extension of physical models on living systems. Schrödinger seemed to consider the Copenhagen interpretation with wave-particle duality inconsistent [126] and that attitude probably caused some frustration and much controversies. It is not a revolution if nobody loses.

## 6.2 Some Notes on the Historical Development of Quantum Mechanics

In order to illustrate the main ideas of quantum mechanics I shall present a very simplified survey of its development into the current form. I am not going to follow in this very brief survey the real chain of historical events - there exist many classical sources on the evolution of quantum theory, (see, e.g., [137]), and I have neither inclination nor qualifications to add something nontrivial from a historian's point of view. My objective is more modest: to grope for some logical threads. The two expositions can be quite different. Before the 20th century, most people believed that classical physics based on the Newtonian mathematical model for mechanical motion and on the Maxwell equations for electromagnetic phenomena was tantamount to a fundamental theory that can describe any natural phenomenon. However, in the beginning of the 20th century it became clear that classical physics failed, at least for many microscopic systems, which led to the development of an alternative mechanical theory - quantum mechanics. There exist many concepts in the latter that cannot be derived from or even contradict classical physics.  It is curious that P. A. M. Dirac called his famous treatise on quantum mechanics published in 1930 "Principles" [20] - it seems that he wanted to imitate Newton.

In my opinion, the genuine understanding of many physical theories is hardly possible without historical method. This method allows one to trace the development of ideas and the succession of models leading to the formation of a working theory. I shall try to bring here a small part of these creative evolutions, emphasizing some unexpected and even astonishing attributes that I came across while studying the original papers related to the initial period of quantum considerations. It is only paying attention to nontrivial features that may serve as a justification for this section.

To begin with, I have noticed that the number of principal, fundamental ideas in many disciplines is not as big as it might seem, when a person, especially a young one, is bombarded with enormous torrents of facts, concepts, notions, representations contained in a variety of bulky textbooks. Due to this information overload, people tend to lose orientation and develop defenses such as reducing the studied material to a bare minimum (arbitrarily chosen), in particular neglecting historical developments altogether. Yet, the fundamental ideas of a discipline, even if they are quite complicated (which is a rare occasion), may be at least apprehensible, and the historical presentation of ideas can help greatly. This is exactly the case of quantum mechanics.

It is well known from the works of a number of historians of science (see e.g., [137]) that what we now call "twentieth-century physics" started from the Planck hypothesis that emission or absorption of electromagnetic radiation is not a continuous process, in accordance with classical field theory (Chapter 5), but occurs in discrete chunks called "quanta", each having energy $h\nu$, where $\nu$ is the radiation frequency and $h$ is some phenomenological constant, now known as the Planck constant. The current value of this constant is $h = 6.626 \cdot 10^{-27} erg \cdot s$. It is quite interesting how Planck came to this postulate applying thermodynamics to oscillators [143] - the simple oscillator model again served a good service, this time in providing an early basis for the emergent quantum theory.

## 6.3  Mathematical Models of Quantum Mechanics

Let us summarize the mathematical content of quantum mechanics in simple terms. In standard quantum mechanics we describe a system using a complex Hilbert space $\mathbb{H}$. States [135] of this system are described by vectors $\psi$ in $\mathbb{H}$. The transition amplitude from a state $\psi$ to a state $\varphi$ is $\langle \varphi, \psi \rangle$[136]. It is thus convenient to normalize states so that $\langle \psi, \psi \rangle = 1$. Quantum mechanical processes are described by linear operators $A: \mathbb{H} \to \hat{\mathbb{H}}$ (initial and final Hilbert spaces may differ as, e.g., in decay, scattering or light-matter interaction). Thus, one can see that the gist of elementary quantum mechanics is really very simple.

Taking again our favorite model of the harmonic oscillator, we can illustrate these quantum mechanical concepts. The Hilbert space of the harmonic oscillator may be constructed by starting from the complex vector space $C(f)$ and writing a typical vector $f$ therein as

$$f(x) = \sum_n f_n x^n.$$

---

[135] Now the term "rays" is more in fashion.

[136] In some texts, especially of mathematical character, the scalar (inner) product is defined in reverse order so that the transition amplitude is written as $\langle \varphi, \psi \rangle$. It is not very important, of course.

The inner product of two vectors, $f, g$ can be defined by the expression

$$\sum_n n! f_n^* g_n$$

where one assumes that the sum converges. Then the Hilbert space of the harmonic oscillator may be declared as the subspace of complex space $C$ including such $f$ that $\langle f, f \rangle < \infty$. Now we can fix a basis in this space, which may be given by the states

$$\psi_n := |n\rangle = \frac{x^n}{n!}$$

These states are orthogonal but not normalized, since

$$\left\langle \frac{x^n}{n!}, \frac{x^n}{n!} \right\rangle = \frac{1}{n!}$$

One may ask, why do we need the factor $\frac{1}{n!}$ in the inner product, $\langle n, n \rangle = \frac{1}{n!}$?

As we have already discussed, one can interpret $\psi_n = \frac{|n\rangle}{n!}$ as the $n$-particle state, i.e., the state where $n$ identical particles coexist. The possibility to think of $\psi_n$ as of the $n$-particle state is due to the unique equidistant character of spectrum of the harmonic oscillator, although it is not easily seen in this formalism (see below special section on the harmonic oscillator).

This question is in fact a matter of convenience and habit, being closely connected with the choice of space to work with. In many cases, it is more convenient to remain in the complex vector space $C(f)$ than to work with the Hilbert space. On $C(f)$ we can define two linear operators, creation and annihilation, usually denoted as $a^+$ and $a$, respectively. Both operators take elements of $C$ to $C$, they are respectively raising and lowering operators, $a^+|n\rangle = |n+1\rangle$ and $a|n\rangle = |n-1\rangle$, adjoint, $\langle af, g \rangle = \langle f, a^+ g \rangle$, and non-commutative, $aa^+ = a^+a + 1$.

Let us see how this purely algebraic approach is connected with standard quantum mechanics based on the Schrödinger equation. For the time-independent one-dimensional harmonic oscillator, this equation reads

$$\left( -\frac{\hbar^2}{2m} \frac{d^2}{dx^2} + \frac{m\omega^2 x^2}{2} \right) \psi(x) = E\psi(x)$$

It is of course convenient to use the dimensionless form by introducing the variable $q = (m\omega/\hbar)^{1/2} x$, which gives

$$\left(-\frac{d^2}{dq^2} + q^2\right)\psi(q) = \left(\frac{E}{\hbar\omega/2}\right)\psi(q)$$

Now one can write the operator corresponding to linear oscillator as

$$-\frac{d^2}{dq^2} + q^2 = -\frac{d^2}{dq^2} + q\frac{d}{dq} - \frac{d}{dq}q + q^2 - q\frac{d}{dq} + \frac{d}{dq}q$$

$$= \left(-\frac{d}{dq} + q\right)\left(\frac{d}{dq} + q\right) + \left[\frac{d}{dq}, q\right]$$

The commutator $\left[\frac{d}{dq}, q\right] = 1$, therefore we get a supplementary term. If the Hamiltonian consisted only of numbers or functions as in classical mechanics (physicians used to say "c-numbers" emphasizing their commutative algebraic properties), then this commutator would be zero and we would have had a school-time commutative algebraic identity, $a^2 - b^2 = (a + b)(a - b)$. However, this identity is in general not true when we deal with objects other than c-numbers (scalars from a field), and in particular in quantum mechanics it is almost never true, so we get additional terms proportional to commutators. In the case of the linear oscillator, this additional term is equal to $\hbar\omega/2$ since the Hamiltonian for the oscillator may be written as

$$H = \frac{1}{2}\hbar\omega\left(-\frac{d^2}{dq^2} + q^2\right) = \hbar\omega\frac{-\frac{d}{dq} + q}{\sqrt{2}}\frac{\frac{d}{dq} + q}{\sqrt{2}} + \frac{1}{2}\hbar\omega = \hbar\omega\left(a^+a + \frac{1}{2}\right)$$

where the familiar operators $a^+$ and $a$ may be written in terms of coordinate and momentum operators in dimensionless form (in "natural units")

$$a^+ = \frac{1}{\sqrt{2}}(q - ip), \qquad a = \frac{1}{\sqrt{2}}(q + ip)$$

or, in ordinary dimensional form ($p = -i\hbar\frac{d}{dx}$),

$$a^+ = \sqrt{\frac{m\omega}{2\hbar}}\left(x - \frac{i}{m\omega}p\right), \qquad a = \sqrt{\frac{m\omega}{2\hbar}}\left(x + \frac{i}{m\omega}p\right)$$

The ground state $\psi_0(q)$ of the harmonic oscillator is defined by the condition $a\psi_0(q) = 0$ which is in fact a differential equation

$$\frac{d\psi_0(q)}{dq} + q\psi_0(q) = 0,$$

with the solution $\psi_0(q) = Ce^{-q^2/2}$ where constant $C$ may be found from the normalization condition, $(\psi_0, \psi_0) = 1$, which gives $C = \frac{1}{\sqrt[4]{\pi}}$. Now, one can construct the algebra of raising and lowering operators. Define $\varepsilon = \frac{H}{\hbar\omega}$, then we have

$$\varepsilon = a^+ a + \frac{1}{2}$$

and

$$[a, a^+] = 1, \qquad [\varepsilon, a^+] = a^+, \qquad [\varepsilon, a] = -a$$

This is the oscillator Lie algebra - a four-dimensional complex space $V$ with the basis $\{1, a^+, a, \varepsilon\}$ and an antisymmetric operation (commutator) $[f, g]: V \times V \to V$ which satisfies the Jacobi identity[137]. One can also construct different Lie algebras by using other bases, e.g., $\{1, q, p, \varepsilon\}$ with $[q, p] = i, [\varepsilon, p] = q, [\varepsilon, q] = -i$. These are Lie algebras over real or complex numbers which are not necessarily all isomorphic.[138] It may be noticed that in this formalism, the Hamiltonian is just an element of the algebra; it loses in some sense its special role as generator of the temporal evolution. One may consider the described algebra containing four basis elements as a four-dimensional extension of the simple oscillator algebra (the Lie algebra usually denoted as SU(1,1)). It is also interpreted as a superalgebra containing two odd elements, $a, a^+$ and two even elements, $I, \varepsilon$, where $\varepsilon^+ = \varepsilon$. Oscillator representations of the Lie algebra SU(1,1) (and its quantum extensions often called $q$-deformations) can in fact be constructed in terms of operators $a, a^+, I, \varepsilon$ in an infinite number of ways. It is usually convenient to build simultaneously the linear algebra for bosons and fermions or even a superalgebra unifying them. From the algebraic viewpoint, bosons are operators that satisfy the commutation relation $aa^+ - a^+ a = I$ and fermions are operators satisfying the anticommutation relation $aa^+ + a^+ a = I$.

One can see that it is *due to noncommutativity* of coordinate and momentum that the zero-point energy $\hbar\omega/2$ appears in quantum mechanics and quantum electrodynamics(where it is usually called the vacuum energy).

## 6.4  The Schrödinger Equation

In the nonrelativistic quantum theory, the fundamental concept is that of a state which is supposed to allow one to determine for it (to "measure") the values of certain physical quantities such as energy, momentum, coordinate, angular momentum, spin, etc. Quantum states are perceived, mostly due to abundant mathematical texts, as totally different from classical linear waves. This leads to some conceptual difficulties which brought about a plethora of attempts to make the superposition principle of quantum mechanics more palatable from the intuitive, "physical" viewpoint. It is important that if the measurement units and the initial (zero) point for each of these physical quantities have been fixed, the produced results should be real numbers. The concept of a state appears to be quite natural, a far analogy may be the state of a person which would allow one to obtain her/his behavioral characteristic at a given time. The state of a quantum system is fully described by a wave function $\Psi$. Similarly to dynamical systems theory, knowledge of this function at initial moment $t_0$ determines the systems behavior in all future moments. This evolution is governed by the Schrödinger equation:

$$i\hbar\partial_t\Psi(\mathbf{r},t) = H\Psi(\mathbf{r},t). \tag{6.1}$$

---

[137] Here, actually the commutator $[a, a^+] = I$ where $I$ is the unit operator, but for our purposes it is sufficient to consider it just unity from the number field.

[138] Some authors prefer to consider three elements $a, a^+, I$ generating the oscillator Lie algebra, see, e.g., the classical book by A. M. Perelomov [156]. There is also a great diversity of notations for the Heisenberg-Weyl oscillator algebra.

Here operator $H$ is the systems Hamiltonian, a fundamental "observable" acting in the space of states $\mathbb{H}$, $H = T + V$, where $T$ and $V$ are kinetic and potential energy, respectively. One can notice that for writing down the Schrödinger equation, a simultaneous physical-mathematical assumption is implicitly taken for granted, namely that for each physical system an operator $H$ does exist, this operator determining the time evolution of function $\Psi$ through the above written equation (provided such evolution is not disturbed by an external influence, e.g., measurement). Operator $H$ is assumed self-adjoint - this is an important requirement[139]. In principle, one can imagine equations of the Schrödinger type having higher derivatives over time or, to put it differently, Hamiltonians containing the terms with higher derivatives of the wave function over time. Then the quantum-mechanical problem may become non-local in time and, besides, would require the knowledge of higher derivatives of the wave function at some initial time-point $t_0$ i.e., quantum evolution would not be determined by the value of $\Psi(t_0)$. This would, however, contradict the established "causal" model of quantum mechanics.

In general, solving the time-dependent Schrödinger problem for a Hamiltonian with an arbitrary time dependence is technically quite difficult. However, if the Hamiltonian is time-independent, the problem can be considerably simplified. For time-independent (autonomous, in the parlance of classical mechanics) problems, we may write a solution to the Schrödinger equation in the "separated" form, $\Psi(\mathbf{r},t) = T(t)\psi(\mathbf{r})$, according to typical recipes of classical mathematical physics. Substituting this Ansatz into the above time-dependent equation and dividing by $\Psi(\mathbf{r},t)$, we have

$$\frac{i\hbar T'(t)}{T(t)} = \frac{-\hbar^2\Delta\psi(\mathbf{r})/2m + V(\mathbf{r})\psi(\mathbf{r})}{\psi(\mathbf{r})}$$

Since the left-hand side depends only on $t$ whereas the right-hand side only on spatial variables, they both must be equal to a constant - the standard reasoning for equations of mathematical physics. Let us label this constant $E$. Then we have two equations that should be solved simultaneously

$$i\hbar T'(t) = ET(t)$$

and

$$-\frac{\hbar^2}{2m}\Delta\psi(\mathbf{r}) + V(\mathbf{r})\psi(\mathbf{r}) = E\psi(\mathbf{r}). \qquad (6.2)$$

---

[139] This requirement can be weakened in certain cases, some of which we shall discuss in connection with the concept of time in quantum mechanics, see Chapter 14.

The first equation gives $T(t) = \exp(-iEt/\hbar)$ while the second one leads to the eigenvalue problem for the Hamiltonian (Schrödinger) operator. The solution of such a problem leads to the bound state picture and, in particular, in energy quantization as well as to spectral expansions over eigenfunctions regarded as vectors in the Hilbert space (i.e., a complete complex inner-product space). The evolution of $\Psi(\mathbf{r}, t)$ in such a space is given, as it follows from the first equation, by the expression

$$\Psi(\mathbf{r}, t) = e^{-iEt/\hbar}\Psi(\mathbf{r}, 0),$$

which reflects, here in simple terms, the one-parameter unitary group property of the evolution operator (below we shall discuss quantum evolution in more detail). As usual, the meaning of the exponent where an operator expression is present should be specified. One possible way to define an operator exponent is through a power series

$$\exp(-itH) = \sum_{n=0}^{\infty}\frac{(-it)^n}{n!}H^n.$$

It is clear that in such an approach the Hamiltonian should be assumed bounded, otherwise the series may be divergent. One cannot, however, require the boundedness of $H$ for an arbitrary physical system. For instance, it is easy to see that operator of multiplication by variable $x$ is unbounded as an operator from $\mathbb{H}(\mathbb{R}^{\mathbb{N}})$ into $\mathbb{H}(\mathbb{R}^{\mathbb{N}})$ and also the differentiation operator $Af(x) = id/dx$ is unbounded. The Hamiltonian of a system placed in an external electrical field may become unbounded[140].

So, the wave function $\Psi$ changes with time under the action of the evolution operator $U(t) = \exp(-itH/\hbar)$. However, the bilinear scalar product of the time-dependent $\Psi$-function remains preserved in the process of evolution dictated by the Schrödinger equation. Indeed, one can easily prove the following conservation property

$$\frac{\partial}{\partial t}(\Psi, \Psi) = 0,$$

where we take the following convention [141] for the scalar product: $(f, g) = \int f^* g \, d\mu$. We have

$$\frac{\partial}{\partial t}(\Psi, \Psi) = \left(\frac{\partial \Psi}{\partial t}, \Psi\right) + \left(\Psi, \frac{\partial \Psi}{\partial t}\right) = \left(-\frac{i}{\hbar}H\Psi, \Psi\right) + \left(\Psi, -\frac{i}{\hbar}H\Psi\right)$$

$$= \frac{i}{\hbar}(H\Psi, \Psi) + \left(\Psi, -\frac{i}{\hbar}\right) = \frac{i}{\hbar}(\Psi, H\Psi) - \frac{i}{\hbar}(\Psi, H\Psi) = 0, \qquad (6.3)$$

if the operator $H$ is self-adjoint.

---

[140] Recall what is understood by the boundedness of a linear operator $A$: the latter is bounded if its norm $\|A\| := \sup\|Af\|_{f\in D(A)} < \infty$ where $D$ is the operator domain. We shall further assume the following simplified version of the boundedness definition, usually accepted for Hermitian (symmetric) and self-adjoint operators: an operator $A$ will be considered bounded from below if the scalar product $(f, Af)/(f, f) \geq m$ where $m$ is finite ($m > -\infty$) and is bounded from above if $(f, Af)/(f, f) \leq M, M < \infty$ for every $f$. See more on linear operators in Chapter 2.

Here, one can make the following observation. Typically, the Schrödinger equation in the form is directly applied to the simplest systems such as the hydrogen-like atoms, where $\mathbf{r}$ in $\Psi(\mathbf{r}, t)$ denotes the position of a single electron such as the only atomic electron in the hydrogen atom or the outer-shell electron interacting with the rest of the atom (the atomic nucleus) in the case of hydrogen-like, e.g., Rydberg, atoms. For such atomic systems, the Hamiltonian $H$ is represented by the sum of kinetic and potential energies for a single electron, $H = \mathbf{p}^2/2m + V(\mathbf{r}, t)$, where $\mathbf{p} = -i\hbar\nabla$ is the momentum operator and $V(\mathbf{r}, t)$ is the potential energy of electron in the Coulomb field of atomic nucleus as well as, possibly, in external fields. If the considered atomic system is not affected by any external fields, then the potential energy of the electron does not depend on time $t$, and such atomic systems have only stationary solutions $\Psi(\mathbf{r}, t) = \psi(\mathbf{r})\exp(-iEt/\hbar)$, described above. Here, the "shortened" wave function $\psi(\mathbf{r})$ may be found from the time-independent Schrödinger equation 6.2, $H\psi(\mathbf{r}) = E\psi(\mathbf{r})$. As already mentioned, this equation leads to the Sturm-Liouville problem giving the discrete values of energy for the finite motion. One may note that the description of bound states with discrete spectrum is necessarily quantum-mechanical whereas the free-particle motion in continuous spectrum can, in principle, be described classically or semi-classically [142]. Of course, such infinite motion can also be portrayed fully quantum-mechanically - the quantum theory is sufficiently general to adequately describe any kind of motion as well as the transitions between continuous and discrete portions of the spectrum. Nevertheless, the classical description, though not necessarily simpler, appeals more to human intuition and therefore seems easier to interpret and comprehend.

[141] This is the "physical" convention; in mathematics, the complex conjugated product $(f, g) = \int f g^* d\mu$ is more often encountered. The choice $(f, g)$ or $(f, g)^*$ is of course inessential. Here, as everywhere in this text I use an asterisk to denote complex conjugation; in the mathematical literature an overline is more customary.

[142] By the way, this is one of the reasons for incessant attempts to formulate the whole quantum mechanics in the language of a single classical trajectory, although this language (Bohm's version of quantum mechanics) is technically maladapted to bound states problem. See below section on Bohmian mechanics.

The motion of a free particle does not necessarily imply that this particle ought to be totally exempt from any interactions For instance, the free electron may interact with an external electromagnetic field, with the atomic nucleus from which it has been detached, with other electrons, ions, or atoms of the medium, etc. The important thing is only that the energy spectrum of a free particle is continuous (usually taken to be positive, $E \in (0, \infty)$) whereas the energy spectrum of a particle to be found in finite motion is discrete i.e., energies of bound states are allowed to admit only certain particular values separated by finite intervals. Such values are usually taken to be negative.

The Schrödinger equation, determining the behavior of a system in a stationary field, may be written as

$$\frac{\hbar^2}{2m}\Delta\psi + [E - V(\mathbf{r})]\psi = 0, \qquad \Psi = \psi e^{-\frac{iEt}{\hbar}}$$

This equation has been explored in countless textbooks and papers, and I shall not repeat the well-known results, e.g., those for the Coulomb problem (see [84], ch. 5).

Let us now get engaged in a brief verbal discussion of the Schrödinger theory. The Schrödinger equation is a linear one which reflects the fact that quantum mechanics is a linear theory, and each solution for the state of a system can be represented as a superposition of other states. This superposition principle allows one to invoke the mathematical techniques of linear algebra, matrix theory and functional analysis. All that naturally leads to the spectral theory for linear operators. One can, however, observe that the spectral theory of the Schrödinger equation seems to be rather non-uniformly developed: for example, the Schrödinger equation with a monotonicly varying potential has been studied more thoroughly than the same equation with a bounded potential. Indeed, one can easily find a lot of results, e.g., for the oscillator or Coulomb case as well as many related models, whereas for bounded potentials most of the results have been restricted to the periodic case.

The Schrödinger equation is, at first sight, just another linear PDE. However, this equation corresponds to a mathematical model conceptually different from the archetypal classical systems based on field partial differential equations (see Chapter 5). First of all, the Schrödinger equation implies a totally different physical interpretation: It basically describes the motion of a single particle. Of course, the Schrödinger equation can be generalized to describe many particles involving also interactions between them, but then we would have more complicated (bilocal, trilocal, etc.) Schrödinger fields, $\Psi(\mathbf{r}_1, \mathbf{r}_2, \ldots, t)$, than for a single-particle motion. The initial idea of Schrödinger was to construct a mathematical model of quantum mechanics which would give physically, e.g., spectroscopically observed discrete spectra, based on the Sturm-Liouville problem [211]. It is known that Schrödinger, while a student at the Vienna University (1906-1910) had mastered eigenvalue problems in connection with the physics of continuous media. About two decades later, being motivated by some dissatisfaction with Bohr's "planetary" model of the atom, he applied his knowledge proposing a theory of atomic spectra giving discrete - quantized - energy levels as a result of a Sturm-Liouville procedure. It is for this work (consisting of four papers) that E. Schrödinger received in 1933 the Nobel Prize, sharing it with P. A. M. Dirac. Although the linear PDEs leading to the Sturm-Liouville problems had been extensively studied throughout the 19th century and a lot of knowledge had been accumulated in the mathematical literature by the time E. Schrödinger divined his equation (1926), an associated eigenfunction expansion involved a lot of technical difficulties making practical calculations very cumbersome in practically important cases such as in chemistry. Thus, computation of the energy levels for many-electron atoms and molecules is notoriously hard even today, after all the years of development of quantum theory and computational techniques. Complexity of the Schrödinger equation rapidly increases as more particles are introduced in the considered system. It is important that in distinction to classical systems the number of degrees of freedom cannot be reduced by imposing constraints. Nonetheless, from the mathematical point of view, the Schrödinger equation for a fixed number of particles is described by a sound theory, provided some physically natural conditions are fulfilled, e.g., energy the Hamiltonian - should be bounded from below. Recall what is understood by a spectrum. The discrete spectrum, $\sigma_d(H)$, of the operator $H$ is, roughly speaking, the set of all eigenfunctions which are discrete points. In other words, any infinitesimal transition between eigenvalues of an operator is impossible. If one can continuously go from one eigenvalue to another, i.e., the spectrum of an operator is not discrete, such a spectrum is called continuous. In addition, the corresponding invariant subspace (eigenspace) in the case of discrete spectrum is assumed finite dimensional. In mathematics, there exist more refined characterizations and classifications of spectra (see below), but here, for the initial qualitative discussion, these notions seem to be sufficient. A little further we shall be concerned with some spectral theory for self-adjoint operators and mathematical requirements to the Schrödinger operator in slightly more precise terms.

One may ask, how do we know when the spectrum of the Schrödinger operator is discrete and when it is continuous? This question is not as naive as it appears. There is an extensive theory of spectral splitting in functional analysis. But here some primitive considerations will be temporarily sufficient. Take, for simplicity, the 1d Schrödinger equation

$$\frac{d^2\psi}{dx^2} + \frac{2m}{\hbar^2}\big(E - V(\mathbf{r})\big)\psi = 0$$

(see, e.g., [84], §21). In fact, we can efficiently integrate only one-dimensional differential equations; all multidimensional equations are typically reduced first to the ones depending on a single variable.

It is curious that the bound state picture inherited from non-relativistic quantum mechanics has penetrated such modern theories as, e.g., quantum chromodynamics and string theories. When we say that, for instance, a proton is composed of three quarks, we implicitly involve the good old bound state concept. What do the words "composed of" mean? Is it really a quantum-mechanical binding? The magical words "quark confinement" seem to be closer to a protective mantra than to a physical model[143]

## 6.5  Quantum Tunneling

Quantum tunneling is often considered - mostly in popular science texts - as one of the most striking manifestations of the "weird" world of quantum mechanics. In reality, there is nothing weird about the tunneling: it is a natural consequence of the wavelike nature of quantum motion. The tunneling effect was suggested in 1928 by G. A. Gamow[144] as an attempt to explain the nuclear decay which was hardly possible to understand from the classical positions [212]. The probability to find a particle under the barrier is exponentially small, but still non-zero, which fact was expressed in the Gamow-Condon-Gurney treatment by a decaying exponential. Due to its inherently wave nature quantum tunneling may be manifested in a rather broad class of phenomena including those where elementary particles are not as directly observed as in nuclear decay (for example, in so-called macroscopic quantum phenomena [213]). One should also mention scanning tunneling microscopy (STM) and spectroscopy (STS), which is a considerable achievement in applied, experimental and engineering physics. Today, both STM and STS have become important technological tools in nanoelectronics and biotechnology.

## 6.6    Quantum Evolution

Assume that a small enough object like a single electron can be well enough isolated from the environment so that external perturbations may be considered negligible. In experimental reality this is very, very hard - to the point of being almost unrealistic (see also in Chapter 9 on time reversal), but such assumptions are often made in physical theory. Then such an isolated object will evolve quantum-mechanically - or with some dose of classicality i.e., semi classically and in the classical limit even fully classically - for a long period of time until the perturbations from the outside world destroy the independent evolution of the object. The outside world is always interfering with the freedom of the object, and this interference disturbs the individual character of the object's evolution. In the quantum case, because of the fact that the outside world constantly tends, so to say, to poke at the object, evolution of the latter may even lose its quantum mechanical character; this irreversible emergence of classical properties because of interaction with the environment is called decoherence (see below).

---

[143] Importunate demons of ignorance haunt physicists all life long, so there exist a number of incantations in physics which are lexical forms simulating knowledge, quark confinement seems to be one such form. Other frequently used examples are "wavefunction collapse", "quantum measurement of observables", "time-reversal symmetry", various versions of "extra dimensions", etc., without mentioning hidden parameters in quantum mechanics. Currently, these words mainly serve to protect physicists from the imminent punishment which is due to lack of understanding. With the accumulation of knowledge, the magical words may acquire an exact meaning, like the X- rays in the past. The use of mantras or other quasi-magic spells was traditionally accepted even in comparatively recent biomedical disciplines, where not properly understood phenomena were often handled solely by introducing new designations. It is only the ubiquitous spread of new, mostly digital, technology that shattered the dominance of intuitive, speculative statements. There are of course other candidates for nonsensical, half-magical credos in other disciplines as well such as "user friendliness" in computer technology.

[144] Probably less known is the paper by E. U. Condon and R. W. Gurney [244] that appears to be independent from the work of Gamow. All three authors calculated the energy dependence of the alpha decay half-life, mathematically modeling the decay phenomenon by a leak of the particle wave function through a classically forbidden area - potential (Coulomb) barrier. Yet George Gamow was probably the first.

## 6.7   The Stone Theorem

The mathematical background of quantum dynamics, i.e., of quantum-mechanical time evolution is given by the important Stone theorem [144]. This theorem states that for an isolated quantum-mechanical system described by a Hamiltonian and possessing the Hilbert space of states $\mathbb{H}$, the time evolution is represented by a strongly continuous one-parameter unitary group acting on this Hilbert space. The generator of such group is the Hamiltonian of the quantum system. Thus, the Stone theorem maps (bijectively) self-adjoint operators on a Hilbert space to a one-parameter family of unitary operators, $U(t), t \in \mathbb{R}$. The notion of strong continuity implies that $\lim_{t \to t_0} U(t)\Psi = U(t_0)\Psi, \Psi \in \mathbb{H}, t, t_0 \in \mathbb{R}$. The idea of the Stone theorem and of its proof appears rather obvious to any person who had ever studied quantum mechanics. Assume that for each $t \in \mathbb{R}, U(t)$ is a unitary operator on $\mathbb{H}$ with $U(0) = 1$. Assume also that there exist homomorphisms $U(t + s) = U(t)U(s)$ for all $t, s \in \mathbb{R}$ and that for any $\Psi \in \mathbb{H}$ we have $\lim_{t \to t_0} U(t)\Psi = \Psi$[145]. Let us define a linear operator $A$ on $\mathbb{H}$ as follows:

$$A\Psi = -i \lim_{t \to 0} \frac{U(t)\Psi - \Psi}{t} \tag{6.4}$$

with the domain $D(A)$ consisting of all $\Psi$ in $\mathbb{H}$ for which this limit exists and belongs to $\mathbb{H}$. Then $A$ is a densely defined self-adjoint operator called the infinitesimal generator of $U(t)$. Conversely, if $A$ is a densely defined self-adjoint operator on $\mathbb{H}$, there exists a unique continuous symmetry $U(t)$ for which $A$ is the infinitesimal generator. Such $U(t)$ is quite naturally denoted as $\exp(itA)$; thus if $\psi(t) = e^{itA}\psi_0$ with $\psi_0 \in D(A)$, then $\psi(t) \in D(A)$ for all $t \in \mathbb{R}$ and is the unique solution to the equation $\dot{\psi}(t) = iA\psi(t)$ with the initial condition $\psi(0) = \psi_0$.

---

[145] This is a specific case of a strongly continuous one-parameter group of unitary maps, in short a continuous symmetry.

The meaning of this theorem is a one-to-one correspondence between continuous symmetries $U(t)$ and self-adjoint operators $A$. In the physical language, this means that the temporal evolution of an isolated quantum system is determined by its energy observable, i.e., its Hamiltonian. This is the most fundamental observable determining the space of states of a quantum system. In some formulations of quantum mechanics, the Stone theorem is taken as a postulate defining the quantum-mechanical evolution: if in the time moment 0 the system was found in the state $\Psi_0$ and during time interval $t$ it evolved as an isolated system (e.g., no measurement was performed on the system), then at the time moment $t$ it will be found in the state $\Psi(t) = \exp(-iHt)\Psi_0$ where, according to the usual rules of taking the functions of linear operators (see Chapter 3),

$$\exp(-iHt) = \sum_{n=0}^{\infty} \frac{(-iH)^n t^n}{n!},\qquad(6.5)$$

with $\mathbb{H} \to \mathbb{H}$. For standard physical situations (e.g., no decay), operator $U(t) \coloneqq \exp(-iHt)$ is unitary, and the evolution of an isolated quantum system is entirely determined by the one-parameter group of unitary operators $U(t), t \in \mathbb{R}$. Moreover, since in conventional quantum mechanics the evolution operator $U(t)$ is linear, it transfers states (rays in $\mathbb{H}$) into states, i.e., acts between states and not between arbitrary vectors $\Psi$. Thus, one typically assumes in quantum mechanics that the time evolution of the observables is defined by a unitary automorphism

$$A \overset{t}{\to} A(t) = U(t)AU^{+}(t) = \exp\left(-\frac{iHt}{\hbar}\right) A \exp\left(\frac{iHt}{\hbar}\right)\qquad(6.6)$$

## 6.8   Geometrical Formulation of Quantum Mechanics

We have seen that the conventional formulation of nonrelativistic quantum mechanics uses state vectors (rays) and operators on a Hilbert space. More modern trends, however, require utilizing the concept of fiber bundles and, in particular, vector bundles. I am not sure there is much more than just fashion in these formulations, at least as far as standard quantum mechanics goes, and that they help much in solving quantum-mechanical problems, but nevertheless I shall try to translate some part of the usual content of quantum mechanics into the language of these geometric structures.

The idea of representing quantum mechanics as a geometric theory is not a new one (see, e.g., [102]), originating probably in the works of P. A. M. Dirac. Roughly speaking, nonrelativistic quantum mechanics can be subdivided into two parts: time-independent - quantum description of states which may be called quantum statics and time-dependent - quantum evolution or quantum dynamics. Both parts may be formulated in geometric terms, however, specifically for the static part such formulation hardly gives anything new and nontrivial it would be reduced to a description of geometric structures[146]. On the contrary, the dynamic part, when reformulated as a geometrical theory, can be further developed using such concepts as transport along paths and other notions typical of differential geometry. This can, in principle, provide a possibility for new results. The general framework of the geometric formulation is, as usual, to define an appropriate geometric structure by specifying parallel transport in a suitable fiber bundle, introducing forms and other convenient coordinate-free tools. In simplest cases, one can consider quantum dynamics as a linear parallel transport in a vector bundle.

We already know how dynamics is treated in orthodox quantum mechanics: the evolution between two pure states $\Psi_1 = \Psi(t_1)$ and $\Psi_2 = \Psi(t_2)$ is formulated in terms of an evolution operator connecting these two states namely $\Psi_1 = U(t_1, t_2)\Psi_2$. Here state vectors $\Psi_{1,2}$ belong to a Hilbert space which must be identified for each specific system. We may recall (see Chapter 3) that the Hilbert space is endowed with the a scalar product $(\varphi, \psi) \colon \mathcal{H} \times \mathcal{H} \to \mathbb{C}^2$.

## 6.9    Quantum-Classical Correspondence

It is generally believed that the universe is governed at a fundamental level by quantum-mechanical laws, which are probabilistic and lead to indeterminism (an electron can be reflected simultaneously from the floor and the ceiling). On the other hand, in our everyday life we observe predominantly deterministic laws (Chapter 4). Ordinary objects we deal with have definite locations and move quite predictably so that we can calculate their trajectories and future (or past) location with arbitrary accuracy using classical mechanics. This is clearly not the case in the quantum world. One can then ask: are there any traces of quantum mechanical indeterminacy in our classical reality? Or, to put it another way, what are the limitations of the classical laws that can be traced back to the underlying quantum mechanical foundation? And why should there be two very different mechanics depending on the size of the objects being studied, even although we intuitively understand when we are supposed to apply each theory?

Although both theories may be considered very successful in explaining phenomena at their respective scales, the relation between them is still a substantial problem and not well understood. One of the main manifestations of this problem is the superposition principle - in fact the linearity of quantum mechanics, which leads to counterintuitive results when applied to macroscopic objects. The most common example is the poor Schrödinger cat, a virtual victim of the apparent paradox of linearity.

One of the main tokens of quantum mechanics is the Heisenberg "uncertainty principle", which limits the accuracy of simultaneous measurements of any two conjugated (dual) quantities. Usually, the position and momentum are considered in this context, but in fact the Heisenberg uncertainty relation holds for any dual quantities (as it is readily illustrated by the Fourier transform). The transition to classical measurement implies a certain crudity of description such as averaging similar to the Ehrenfest theorem [84], see below. For example, a coarse description of the position and momentum of a particle can be modeled by a roughly narrow wave packet approximating the particle's classical behavior. We shall discuss some properties of the wave packets in different contexts in the following sections.

---

[146] For instance, replacement of the Hilbert space by Hilbert bundle.

A substantial part of classical mechanics is devoted to the motion of a rigid body. One may ask, is there a corresponding part in quantum theory? And in general, is it possible to apply quantum mechanics as a probabilistic theory to a single object such as, e.g., a planet or to the whole universe (see Chapter 9). If we establish that yes, it is legal to use quantum mechanics to unique macro- and even mega-objects, then, provided we believe that the whole world is governed by the quantum mechanical laws, we must be able to obtain the classical motion of macroscopic bodies as a limiting case of quantum theory, including, for example, the planetary motions, Kepler's laws and the like. An exhaustive derivation of the classical equations of motion would then require the assessment of probabilities for classical trajectories or, in the modern talk, for time histories. Such assessment of probabilities is more demanding than the standard study of evolution of the expectation values. Consider, for instance, planetary motion. The center of mass of a rigid body, say of the Earth, may be regarded as moving in accordance with Newton's law for a rigid body (see, e.g., [23]) when one can observe that successive measurements for the positions of the center of mass of the body produce results that are correlated according to this law with the probability close to unity. This is, of course, scholastic, but to compute such probabilities, without brushing off quantum mechanics because the wavelength is presumably exorbitantly small and without appealing to intuitive considerations, would probably require some ingenious ad hoc methods and considerable effort.

There have been, of course, a great body of discussions on how to obtain classical deterministic behavior from quantum mechanics. Before we proceed to calculations, it would be perhaps appropriate to outline the general philosophy behind such discussions, so I would dare to offer a few comments on main directions in these discussions. Let us recall only the principal approaches aimed at connecting the classical and quantum realities. To observe all such approaches would be an enormous task, and I will not even try to do it. Especially hard would be to establish a relationship between individual methods (see, for instance, [116]). Somewhat superficial observation of the relevant literature shows that the main lines of thought in the problem of quantum-classical correspondence are centered around 1) the WKB (WKBJ) approximation; 2) the Ehrenfest theorem; 3) the Wigner-Weyl distribution; 4) the Feynman path integral (mostly in its short-wavelength form, e.g., in steepest descent approximation); 5) Bohm's mechanics. There are also approaches discussing explicitly quantum noise, but despite a great number of papers treating quantum noise in laser physics, quantum optics and, lately, mesoscopic physics, studies of quantum noise in the context of the quantum-classical transition are comparatively rare, probably because of their complexity. Nevertheless, mixed classical-statistical equations uniting classical predictability and stochastic noise have been numerously derived in the linear systems case; in fact, those are the evolution equations for the probability distributions in phase space such as the famous Fokker-Planck and Wigner equations. We shall deal with the respective models in Chapter 7.

In this section, I shall briefly comment only on those equations belonging to the class of stochastic differential equations, which are relevant to the quantum-classical correspondence. These are just preliminary comments, and a little further I shall try to give a simplified version of the theoretical framework for the transition to the quasi-classical domain.

Evolution equations for probability distribution in the phase space were derived by A. O. Caldeira and A. J. Leggett [113] from the evolution equation for the Wigner function (see below), in fact using the techniques proposed by R. Feynman. The model treated by Caldeira and Leggett is perhaps the easiest possible model for a system embedded in a reservoir, where the observed system is coupled linearly to this reservoir. Their approach is largely phenomenological in the sense that the properties of the reservoir are not derived from the Hamiltonian. In fact, Caldeira and Leggett have discussed the Langevin equation for linear systems, which reminds us of the treatment of quantum theory from the point of view of statistical mechanics.

It would, of course, be interesting to provide the consistent dynamical treatment of the quantum-classical transition for a system interacting with the environment, but I failed to find it in the literature available to me, see however [114]. Typically, to consider the decoherence of histories, by which classical behavior emerges, explicitly treating the histories' probabilities seems to be technically quite difficult. Even specific mathematical models illustrating the transition to the classical limit are rather complicated (see below).

In the whole variety of approaches to establish the quantum-classical correspondence, two sharp signals have been produced: the familiar quasi-classical[147] asymptotic theory of the WKB (WKBJ) type (see [84], Chapter 7) and the new decoherence concept that can hopefully diminish the discrepancy between quantum and classical descriptions. I shall first briefly comment on these approaches and slightly touch upon the so-called Gutzwiller formula derived by a Swiss physicist M. Gutzwiller, the student of W. Pauli, which is also relevant in the context of quantum-classical correspondence. Afterwards, we shall discuss the Ehrenfest theorem, the Wigner-Weyl (and analogous) distributions and the Bohmian version of quantum mechanics. The Feynman path integral will be reviewed mainly in association with the quantum field theory where the path integral presents the most popular techniques.

Remarkably, the classical phenomenology dominating our everyday experience seems to be much more versatile than all the derivations of the elementary transition from the quantum mechanical to the semiclassical domain of the motion equation (e.g., of the WKB type). This richness of classical phenomena is manifested already in the fact that classical features and the respective language (Chapter 4) are essentially different from those of quantum mechanics. Indeed, while studying classical systems we pay no attention to "observers". The poorly defined word "measurement", used now and then in quantum mechanics, is meaningless in classical mechanics; in the classical context measurement is associated mainly with metrology. Discrete symmetries play almost no role in classical mechanics and are of extreme importance in the quantum theory. Indeed, all transformations that may prove important for classical mechanics are infinitesimal. Moreover, when discussing classical behavior, we usually talk about orbits and trajectories, and not about energy levels (or other quantum numbers) as in quantum mechanics. One may also note that classical equations of motion are essentially phenomenological - they are direct mathematical models of reality, and simply identifying them may be a serious problem (e.g., as in the case of dissipative systems).

---

[147] In the English-language literature the term "semiclassical" is commonly used to denote short-wavelength asymptotics of the WKB type, whereas in the Russian-language literature the word "quasi-classics" is almost exclusively in use.

The semiclassical approximation is commonly identified with the regime of large quantum numbers, although to give an exact meaning to this statement is not that simple. Quantum numbers are not inherently present in any classical theory; they are, as we have seen, an attribute of quantum models. We have to introduce a space of functions with the required properties, to explore spectral qualities of operators and to declare appropriate prescriptions to be announced as quantization. Nothing of the kind exists in classical theories based on a homogeneous manifold called the phase space.

Nowadays, the key word for the transition to the semiclassical regime seems to be "decoherence". The concept of decoherence has been introduced in the 1980s [107] and advanced in the 1990s [108] as a possible solution to numerous paradoxes of quantum theory, in particular those stemming from the superposition principle. The pioneers of decoherence, H. D. Zeh and E. Joos asserted that macroscopic bodies are open systems continuously interacting with the environment, and it is this interaction that destroys the interference terms which obscure the transition to classical theory (see also [110]). Then the obvious strategy would be to study the concrete decoherence mechanism on some simple model. The most favored model in physics, as we have seen, is the oscillator; can the decoherence considerations be tested on this model? And indeed, this modeling strategy has been pursued by W. Zurek and collaborators [111] to test the decoherence in the case of a quantum harmonic oscillator coupled to a heat bath of other harmonic oscillators that constitute the environment. This is in fact the quantum Brownian motion model. As I have already mentioned, this model seemed to be one of the favorite problems for R. Feynman; he used it in his dissertation to illustrate the Lagrangian method in quantum mechanics.

The standard techniques of quantum mechanics based on the Schrödinger equation are typically applied to isolated systems, despite the fact that in certain interpretations one constantly talks about ensembles (see below section "Do you need an interpretation?"). In practice, however, true isolation is extremely difficult to achieve for a macroscopic physical system, the latter being in constant interaction with their environment. So, the adepts of the decoherence theory take the view that physical systems are always in interaction with their immediate surroundings, are thus never isolated and hence not Hamiltonian. Such systems are in fact stochastic, and a method of quantization should be offered without the need of a Hamiltonian. Respectively, the dequantization process, i.e., obtaining classical limits such as retrieving classical dynamics from the quantum mechanical equations of motion should not be relied on Hamiltonian methods. Using the quantum-mechanical language, one may say that the environment continuously produces "measurement" over a macroscopic physical system, and it is this measurement that destroys quantum wavelike interference and only the classical features are exhibited by the system. This is what is intuitively understood as decoherence and is considered nowadays as absolutely essential for our comprehension of a classical limiting process (dequantization).

Honestly speaking, personally I do not quite understand this modern cognitive scheme. Why does one need to consider only open systems to understand the transition to classical physics ("classicality", in the modern talk)? The fundamental dynamics of the system both in the quantum and in the classical domain is governed by the action $S$ or, equivalently, by the Hamiltonian $H$. We have grown up with the usual scheme of quantum mechanical probabilistic predictions based on the paradigm of Hilbert space, states as rays, linear operators and their spectra, etc., this scheme is working beautifully, its interpretation is good enough for me (see the section "Do you need an interpretation?") and I don't understand why we should abandon it in favor of some kind of random dynamics, though promoted by a number of very smart craftspeople.

Let us describe the decoherence considerations in some more detail. The central notion here is a sequence of events at successive time points (again a similarity with the Feynman's approach). This notion is usually named as a quantum mechanical history. What we traditionally denote as a set of classical paths for the system, in this language is identified as histories having only negligible interference between each other. One can assign not only amplitudes, but probabilities to such histories. One can formalize the notion of histories by introducing the decoherence matrix or decoherence functional so that when this functional is strictly diagonal, probabilities are exactly defined and all probabilistic algebra can be applied, e.g., sum rules are fulfilled. However, there may be obviously approximate decoherence when the probability sum rules are satisfied only to some finite accuracy determined by inequalities bounding the absolute value of the off-diagonal terms of the decoherence functional. To produce concrete results from these general considerations, one can take some model, e.g., a simple one-dimensional system with a variety of initial states. For such systems, e.g., for the one-dimensional oscillator (see [112] and literature cited there), it is possible to calculate the decoherence functional explicitly and explore the degree of decoherence by analyzing the diagonal elements of the decoherence functional representing the probabilities of the classical histories of the system, and off-diagonal elements showing the limit to which the histories decohere. Decoherence is telling us that the reduced dynamics of a system interacting with a typical environment must be essentially classical. In other words, there may be a certain class of model quantum mechanical systems in which one is concerned with classical probabilities.

One can then ask: is decoherence equivalent to the crude description (coarse graining) one uses to get rid of the "uncertainty principle"? Then what amount of coarse graining is quantitatively sufficient to suppress quantum fluctuations and obtain the classical equations of motion? Is it the same amount as needed for decoherence? And what are the quantum corrections to classical motion equations in terms of quantum correlations that must die out to ensure the now fashionable decoherence? Do such corrections quantify deviations from classical determinism? It is interesting that this bunch of questions seems to be connected with the possibility to disregard the so-called quantum potential in Bohm's version of quantum mechanics (see the respective section below). Intuitively, it must be clear that to the same degree as the histories decohere, the probability distributions should be centered around classical paths or, more generally, histories of the system. Then the distributions of positions and momenta must be probably given by some version of the Wigner function. Can one evade any conflict between the general requirements of decoherence and peaking on the classical paths with exactly defined smearing about them? I don't know whether there exists a general mathematically correct answer to the above bunch of questions, yet one can try to illustrate the connection between coarse graining, not necessarily uniquely defined, and the degree of decoherence using the familiar model of the one-dimensional oscillator. For this simple model, one can imagine at least two kinds of coarse graining, each contributing to the extent to which the histories of the oscillator decohere. The first kind of coarse graining is due to imprecise knowledge of the oscillator initial position and momentum, i.e., smearing in the phase space. Coarse graining of the second kind is due to coupling of the oscillator with the environment, e.g., submerging our oscillator in a thermal bath of other harmonic oscillators (this model of oscillator interacting with the environment is typically called the Caldeira-Leggett model [113]) although a similar approach is contained in the works of Feynman (see, e.g., [44] and [45]). The original motivation for the Caldeira-Leggett model was the study of dissipation in superconductors, which may be interpreted as macroscopic quantum systems (see Chapter 9). This oscillator-based model, though being comparatively simple, clearly depicts quantum dissipation and thus irreversible suppression of interference (off-diagonal) terms due to interaction with the environment. I shall recall the Caldeira-Leggett setting, which in its simplest form looks as

$$H = \frac{P^2}{2M} + V(X) + X \sum_i C_i q_i + \sum_i \left( \frac{p_i^2}{2m_i} + \frac{1}{2} k_i q_i^2 \right) \qquad (6.7)$$

This expression represents the Caldeira-Leggett Hamiltonian where $P, X$ and $M$ are respectively the momentum, position and mass of the observed system, $p_i, q_i, m_i$ and $k_i = m_i \omega_i^2$ are the momenta, positions, masses and the elastic constants of the harmonic oscillators making up the environment. I do not intend to produce all the calculations for the Caldeira-Leggett model now because our current subject is the quantum-classical transition and not the properties of nonunitary evolution models. I shall only remark that the Caldeira-Leggett model is a notable exception, since it allows us to calculate the evolution of the state of a Brownian particle (in this model an oscillator) submerged in a thermostat comprised of multiple oscillators, i.e., interacting with the bosonic reservoir being in thermal equilibrium at temperature $T$.

A few words about the term "decoherence". Remarkably, like many novel terms in physics (entanglement, consistent histories, classicality) decoherence is devoid of any exact meaning, being only loosely defined. At first, the term "decoherence" was used to denote the effect transition to classicality (should not be at this stage confused with semi-classics) with vanishing of off-diagonal matrix elements. A little bit later, decoherence has been applied to emerging "alternative histories" of a quantum system referring to weakened interference between these histories. Intuitively, these two meanings appear to be very close, however, they are not in general equivalent. Mathematically speaking, decay (e.g., with time) of the off-diagonal matrix elements of some model-based decoherence matrix is not the same as the diminishing interference between individual "histories". Decoherence theory tells us what happens to a quantum system in a dissipative environment and how such an environment selects preferred states[148].

Nonetheless, for both above-mentioned definitions of decoherence, the role of intrinsic quantum phase correlations seems to be essential. It seems intuitively clear that in a coarse-grained picture phase correlations between alternative histories fade away together with off-diagonal matrix elements in the semiclassical domain. There are no phase correlations in classical mechanics.

Comparing the quantum fluctuations ideology implicit in the uncertainty principle with the classical theory, one might notice that even in the classical deterministic limit we can encounter an extraordinary sensitivity of some physical systems to slight variation of parameters, e.g., of initial conditions (see Chapter 4). This sensitivity observed predominantly in nonlinear systems is exponential in time and leads not only to instability, but also to chaotic behavior. In the presence of chaos small fluctuations may be amplified to such an extent as to result in large uncertainties in the outcome. Chaos produces indeterminacy of the final state. The small initial fluctuations that tend to be exponentially amplified with the time may be, in particular, quantum fluctuations.

Now, let us return to the coarse graining. By a coarse-grained description one typically understands such a picture when some variables are left unspecified or averaged (traced) out. Moreover, even those variables that still are specified may be not defined at each time point or it would not be possible to determine them with arbitrary precision. For instance, they may be replaced by some averages as in statistical mechanics. So, one can separate all the variables relevant to a microscopical (quantum) problem into two types: variables of one type may be called ignored, variables of the other type are sometimes called "privileged" or "distinguished". It is clear that in most of the physical situations coarse graining based on privileged variables would be sufficient for physicists as observers of the universe. Indeed, even the utmost careful observations can produce only a tiny portion of all the variables characterizing the universe (see on this subject in [88]). Furthermore, any meticulous observation provides the estimates of these variables with some limited accuracy that defines some corridor of values over which averaging should be carried out.

---

[148] These may be totally different states, depending on the interaction. For example, they may be position or momentum eigenstates or coherent states, etc.

Thus, one of the simplest procedures of coarse graining consists in averaging out some (ignored) variables and paying attention only to the remaining ones. This procedure reminds us of the transition from the quasi-microscopic Langevin equation describing, e.g., stochastic dynamics of Brownian motion to the damped motion equation containing a phenomenological dissipation coefficient. Here one can intuitively feel a connection between noise, decoherence and dissipation, all of them being linked by the procedure of course graining.

Another example of coarse graining is provided by the transition from the mechanical description to hydrodynamics and thermodynamics (Chapter 7). Coarse graining in this case is due to deliberate loss of information, and our phenomenological - hydrodynamical or thermodynamical - variables are just the weighted averages (e.g., with the distribution function or over some suitable "physically infinitesimal" volume) of "privileged" variables. When one can disregard the noise generated by the stochastic Langevin-type force, the deviations from the classical equations of motion become small, and one can observe the complete classical predictability. Derivation of hydrodynamical equations from physical kinetics [149] necessarily includes dissipation. The obtained equations of motion contain as unknown variables not the actual values (realizations) of mechanical quantities, but the expectation values of hydrodynamical variables. There exists the graining hierarchy: by subsequent averaging one can achieve further coarse graining. One can justifiably ask: is the course graining operation invertible or, in other words, is the refined graining treated as the inverse operation correctly defined? In the hydrodynamic example, this would amount to obtaining the Boltzmann equation or the Liouville equation from hydrodynamic equations. This procedure would require an infinite sequence of these hydrodynamic equations (e.g., Euler or Navier-Stokes) regarded as statistical moments or weighted averages of the Liouville-type equations of Hamiltonian mechanics, classical or quantum. There are, of course, difficulties by restoring the original Liouville-type equation from the corresponding BBGKY hierarchy (see Chapter 7) in any finite closure scheme due to the information loss.

---

[149] One of the best sources on this subject is, to my mind, the book by V. P. Silin [72].

Hydrodynamic variables such as the densities of mass, energy, momentum, charges, currents as well as local temperature and entropy have the structure $q^i = \left(x^\alpha(t), Q^\alpha(t)\right)$ where $x^\alpha(t)$ denote "privileged" coordinates such as center of mass position or orientation of a macroscopic fluid particle, which is actually a large group of constituent molecules, and $Q^\alpha(t)$ are "ignored" variables such as internal coordinates of the fluid molecules. Tracking the positions of individual molecules of the fluid by solving the coupled equations of motion for all of them, though feasible (only approximately) for a small number ($N$ 1) of molecules - this is the subject of molecular dynamics, would be clearly impossible and even absurd for any macroscopic system ($N$ $10^{23}$). Instead, an abridged description in terms of a small number of "privileged" macroscopic variables is provided, with most of the microscopic information being thrown away. Exactly how this transition to a reduced description is achieved is the subject of kinetic theory, which is in general very subtle and here I don't have much to say about this theory (see Chapter 7 for some details and references). I would like to make just two remarks. Firstly, the drastic simplification achieved by the reduced description comes with the penalty: one has to introduce purely phenomenological quantities such as friction or viscosity whose values must be taken from experiment. Similarly, we also have to pay the price for decoherence throwing away delicate quantum information on interfering paths. Transition to a set of histories that constitute the quasi-classical domain of everyday experience lead, for example, to complications with time-reversive behavior (see Chapter 9).

Secondly, any transition to a reduced description is based on some multiscale scheme and can break down when the respective conditions cease to hold. So, one should be careful with applying theories or models based only on "privileged" variables. Example: macroscopic theory of continuum media may be inapplicable for rarefied gases.

Both the Caldeira-Leggett and Feynman-Vernon models are based on the favorite model of physics – the oscillator. In both models, which basically belong to the same class of quantum dissipative dynamical models, a "distinguished" oscillator is linearly coupled to a large number of other oscillators constituting an immediate environment, specifically forming an equilibrium thermal bath characterized by temperature $T$. As we know, such a situation is typically modeled in physics by the density matrix $\rho(x', x; Q', Q)$, which in this model is assumed to be factorized: $\rho(x', x; Q', Q) = \rho(x', x)\rho_T(Q', Q)$ where the coordinates are separated on "privileged" and "ignored" as above. More specifically, let $x(t)$ be the privileged coordinate of the "distinguished" oscillator and $Q$ the ignored coordinates characterizing the environment. One can introduce the reduced density matrix by taking the trace over the environment variables $Q, \rho(x', x) = Tr\rho(x', Q'; x, Q) = \int dQ\, \rho(x', Q'; x, Q)$, we shall do it explicitly later when we attempt to find the parameters (e.g., renormalized frequency) of our distinguished test oscillator due to its interaction with all others (bath).

Classical mechanics is commonly considered to be the limiting case of quantum mechanics when the Planck constant $\hbar$ tends to zero. One can hardly be satisfied with this automatic reply to the question about the quantum-classical correspondence. The quantity $\hbar$ is dimensional, and the limit to zero (or infinity) of a dimensional quantity is, honestly speaking, meaningless. It is therefore hardly possible to consider quantum mechanics as a higher perturbative approximation of classical mechanics. The $\hbar \to 0$ stereotype mostly serves psychological purposes: it satisfies the need to make quantum mechanics more comprehensible in customary everyday terms thus making it cognitively acceptable. However, the relationship between a family of trajectories, e.g., obtained by integrating the Hamiltonian equations of motion and starting from some prescribed initial conditions, and the wave function corresponding to the spectral problem for the Schrödinger operator (Hamiltonian) is rather intricate and may be intuitively well understood only in the case of closed periodic orbits - like those of an electron moving in the Coulomb field. It is this situation that is mainly discussed in standard courses of quantum mechanics. In this case, the energy eigenvalues are represented by sharp peaks in the density of states, i.e., the energy level distribution consists only of a sum of delta-functions.

In textbook quantum mechanics, the discrete spectrum is "physically" interpreted as a set of energy levels for which an integer number of de Broglie waves would fit in the considered spatial region, e.g., around the atomic orbital. In other words, only standing electron waves are allowed in the atom. Such orbits are considered stable, whereas all other orbits, when the wave ends up, after a complete revolution, at a different phase point on the wave are unstable and the respective spectral numbers (in this case energy) are not allowed. This competition of stable (phase-locked constructive interference) and unstable (out of phase destructive interference) orbits produces the discrete spectrum in naive quantum mechanics.

So, the discrete spectrum in orthodox quantum mechanics is interpreted as a set of energy values for which only an integer number of waves is strictly allowed in the considered spatial domain. The picture of sharp spikes in the spectrum of the wave operator is typical of any resonance phenomenon. However, it is this simple picture that laid the foundation for the development of quantum mechanics, first by N. Bohr and A. Sommerfeld, later by L. de Broglie and E. Schrödinger. These "founding fathers" of quantum theory quite naturally focused their attention on the hydrogen-like atoms, that is on an electron moving in the Coulomb field. However, this case is quite exceptional, since only closed periodic orbits are present in the Coulomb problem.

To address this issue, we may start from some basic recollections. When Bohr was formulating the a priori rules of the old quantum mechanics, he was handling a very special "planetary" model of the hydrogen atom and, consequently, considered the periodic classical orbits of an electron pursuing the finite motion in the $1/r$ field of a Coulomb center. As I have just mentioned, these classical orbits are closed (ellipses), which is rather an exception than the rule. Indeed, it is a well-known fact that there are only two central fields in which finite motion occurs along the closed orbits, these are the attractive Coulomb field, $U(r) = -\alpha/r, \alpha > 0$, and the isotropic oscillator field, $U(r) = kr^2, k > 0$. In principle, for any central field $U(r)$ there exist a collection of conserved aphelion $r = a(1 + e)$ and perihelion, $r = a(1 - e)$, vectors. This corresponds to motion in a central field in celestial mechanics and the Runge-Lenz (Laplace-Runge-Lenz, LRL) vector, which is the conserved quantity. The reason for its existence as a conserved quantity is that central force problems are characterized by a more profound symmetry than $SO3$, the LRL vector being the manifestation of this fact (see, e.g., [149]). Nevertheless, I don't know any good physical interpretation of this conserved quantity and would be thankful if someone could provide it. It is clear that the LRL vector is connected with the orbit's eccentricity and as such it can be used to calculate the eccentricity. For any central potential there must exist a number of eccentricity (aphelion and perihelion) vectors, but what happens when the potential continuously differs from the pure Coulomb? For example, how does the $SO4$ algebra change with the variation of the screening constant in the Yukawa potential? Due to its conservation, the LRL vector commutes with the Hamiltonian, which gives an additional relationship to the Lie algebra whose Lie group ($SO4$) is of a larger symmetry than the rotational group ($SO3$). The consequence of this fact is that the spherical motion problem in quantum mechanics admits, besides $SO3$, other non-trivial subgroups of $SO4$ realized as solutions to the usual nonrelativistic hydrogen problem, i.e., as modes derived from the spinless Schrödinger equation.

The question of how many vector constants of motion do exist for a particle moving in a central potential will be discussed later, in connection with general features of particle motion in a central field. Now we are mostly interested in bridging over the cognitive chasm between classical and quantum mechanics. Unfortunately, since these two theories are too different there does not seem to be the uniquely defined and mathematically impeccable way to make the transition between them, irrespective of the manner of understanding the word "transition". For example, semiclassical theory is just an asymptotic expansion of the solutions to the partial differential equations and is thus not equivalent to classical mechanics in the mathematical sense. This asymptotic expansion corresponds to the limit $\hbar \to 0$, but $\hbar$ is a dimensional quantity so that its limit to zero or infinity is not quite meaningful. In general, the connection between quantum and classical mechanics is intricate and not quite clear, there are a lot of beliefs and folklore about it. P. A. M. Dirac, probably deeply understanding this fact, in his famous book "Principles of Quantum Mechanics" [20] and especially in subsequent lectures [139] has replaced proofs of quantum-classical correspondence by a kind of axiomatic formalization of terminology. This axiomatic or rather quasi-engineering approach proved to be very successful, but it could not compensate for the loss of logical connections between classical and quantum mechanics. It is remarkable that Dirac has intuitively guessed how to construct quantum mechanics on the base of classical one. One might recall in this connection that some great scientists, e.g., Einstein disagreed with such intuitive replacement of logical links between the two models of the mechanical world. In my view, it is the logical transition between classical and quantum mechanics that should be improved in future, not the philosophical issue of interpretation of quantum mechanics which seems to be a pseudo-problem. To have two great theories for the description of mechanical motion, with some indeterminacy when to use one and when the other [150] is an unsatisfactory state of affairs.

## 6.10  The Ehrenfest Theorem and Its Meaning

The first, simple but tricky, question which is posed by any student beginning to study quantum mechanics is: in classical mechanics we have all the quantities depending on the particle coordinate, $\mathbf{r}(t)$ or, for generalized coordinates, $q^i(t), i = 1, \dots, n, n = 3$ for a single particle . What should we use instead of $\mathbf{r}(t)$ or $q^i(t)$ in quantum mechanics? An automatic answer "the position operator" that might be given by a person who has already had some experience with quantum theory is not quite satisfactory for a beginner or at least incomplete because it implicitly involves a lot of new concepts. Thinking about this seemingly primitive basic problem leads to a number of surprising issues that should be elucidated lest one thoughtlessly use quantum prescriptions. Below we shall tackle some of them, but to prepare ourselves for surprises we have to discuss some elementary concepts of orthodox quantum mechanics first.

Assume now that we find ourselves within the fully quantum world. The most natural way to establish the quantum-classical correspondence would be probably restoring the classical behavior of the position (or coordinate) operator, $\mathbf{r}$, or of generalized coordinates $q^i(t), i = 1, \dots, n, (n = 3)$ starting from the quantum picture. To this end, one can employ the Heisenberg picture or Copenhagen interpretation (see the respective section in this Chapter), $A(t) = U^+ A U$, where $U = \exp(-iHt/\hbar)$ is the evolution operator for a "pure" quantum mechanical system characterized by the Hamiltonian $H = \mathbf{p}^2/2m + V(q^i)$. Differentiating this equation, we get the standard expression for the time derivative of the operator corresponding to the classical function $A(t)$:

$$\frac{\partial}{\partial t} A(t) = \frac{i}{\hbar} [H, A]$$

Putting $A(t) = q^i(t)$ (for simplicity, one can consider one-dimensional motion here, of course), we must arrive at some analog of the classical equation of motion[151]. One usually considers in this context the so-called Ehrenfest theorem which is related to the average (expectation) values of coordinates and momenta. It is useful, however, to trace the correspondence of the quantum and classical formulas already at the level of operator equations, before the transition to average values. This question is described in detail in the textbook by A. S. Davydov [140], chapter 2.

---

[150] When one says that quantum mechanics must be used in all cases when atomic (or subatomic) particles are considered, I would say it is wrong: particles in accelerators are treated classically with extremely good precision. Moreover, in highly precise scientific instruments such as mass spectrometers, beta-ray spectrometers and the like, which are based on subatomic particle motion in external fields, deflection of these microscopic particles are computed with high precision with the means of classical mechanics and no quantum corrections are included. In a rather important problem of the passage of atomic particles through the matter, it is very hard to tell a priori whether a concrete task should be treated classically or quantum mechanically, this is a purely modeling approach, a matter of arbitrary choice. Furthermore, there are many attempts to treat the entire universe as a quantum object (see below the discussion of the wave function of the universe). Should one then treat such subsystems of the universe as galaxies, stars, planets and other astrophysical objects also with quantum mechanical means. Does it always make sense?

Let us find the quantum-mechanical time derivative of an arbitrary function $f(\mathbf{r})$, where $\mathbf{r}$ is what we call in classical mechanics the radius-vector of a particle (or the time derivative of $f\left(q^i(t)\right)$, where $q^i(t)$ are generalized - in fact curvilinear - coordinates). In quantum mechanics, we must interpret the function $f(\mathbf{r})$ as an operator, so according to the general formula for the time derivative of an operator we have

$$\frac{d}{dt}f(\mathbf{r}) = \frac{i}{\hbar}[H, f(\mathbf{r})],$$

where $f$ can also be a vector quantity of course.
Since $[V, f(\mathbf{r})] = 0$ (or, in generalized coordinates, $[V(q^i), f(q^i)] = 0$) we get

$$\frac{d}{dt}f(\mathbf{r}) = \frac{i}{2m\hbar}(\mathbf{p}^2 f(\mathbf{r}) - f(\mathbf{r})\mathbf{p}^2) = \frac{i}{2m\hbar}(\mathbf{p}\mathbf{p}f(\mathbf{r}) - f(\mathbf{r})\mathbf{p}\mathbf{p})$$

$$= \frac{i}{2m\hbar}(\mathbf{p}f(\mathbf{r})\mathbf{p} - i\hbar\mathbf{p}\boldsymbol{\nabla}f(\mathbf{r}) - \mathbf{p}f(\mathbf{r})\mathbf{p} - i\hbar\boldsymbol{\nabla}f(\mathbf{r})\mathbf{p})$$

$$= \frac{1}{2m}(\mathbf{p}\boldsymbol{\nabla}f(\mathbf{r}) + \boldsymbol{\nabla}f(\mathbf{r})\mathbf{p}), \qquad (6.8)$$

since $\mathbf{p}f(\mathbf{r}) - f(\mathbf{r})\mathbf{p} = i\hbar\boldsymbol{\nabla}f(\mathbf{r})$. Indeed, $(\mathbf{p}f - f\mathbf{p})\psi = i\hbar(\boldsymbol{\nabla}(f\psi) - f\boldsymbol{\nabla}\psi) = -i\hbar\psi\boldsymbol{\nabla}f$ for an arbitrary $\psi$.

If we take, for instance, the time-dependent operator $f(\mathbf{r})$ to be the particle velocity interpreted as a function of position $\mathbf{r}$, then we may find the time derivative of this velocity operator

$$\frac{d}{dt}\mathbf{v}(\mathbf{r}) = \frac{i}{\hbar}[H, \mathbf{v}(\mathbf{r})] = \frac{i}{\hbar m}(V(\mathbf{r})\mathbf{p} - \mathbf{p}V(\mathbf{r})) = -\frac{1}{m}\nabla V(\mathbf{r})$$

and we arrive at the operator equation whose form is exactly the same as Newton's equation in classical mechanics, $d\mathbf{p}/dt = -\nabla V(\mathbf{r})$. Thus, using only the commutation relations between coordinate and momenta we obtained the operator analogs of the classical equations of motion. It is, however, difficult or at least unconventional for the people trained in classical mechanics to use directly operator differential equations, so one would rather invent a scheme that results in the classical equations of motion for the expected values since one can handle the latter as classical quantities. Such a scheme leads to the so-called Ehrenfest equations which were obtained by P. Ehrenfest in 1927 [157]. The Ehrenfest theorem states that the motion of a quantum particle will be, on average, identical to the motion of a classical one. The words "on average" in this interpretation are usually understood in the following way: the quantum particle is represented by a wave packet concentrated near its classically moving expectation value, and the potential in which the particle moves does not change significantly over the dimensions of the wave packet. In this sense, the particle may be considered pointlike with respect to potential motion despite being spread over a finite packet size. Such a condition allows one to replace the particle position and momentum by their expectation values; in fact, the condition of packet "smallness" is neither sufficient nor necessary (see [164]), for example, one can see that the Statement of the Ehrenfest theorem is valid not only for the "concentrated" wave packets.

---

[151] We assume here that the Hamiltonian does not depend explicitly on time. An explicit dependence of time in the Hamiltonian results in some difficulties which are natural, while in this case we have a different physical situation - the system is under variable external conditions, e.g., in an external field. We shall discuss this situation below.

In order to elucidate the meaning of the Ehrenfest theorem, let us begin from some simple averaging procedures. To get rid of supplementary but not essential difficulties we assume again that the Hamiltonian does not depend explicitly on time. The simplest method to obtain the equations for expectation values which would correspond to the above operator equations is to directly average these operator equations, but this procedure is not mathematically impeccable unless we exactly define the averaging procedure. If we close our eyes on this pedantry and just take the expectation values of both sides of the operator equation (with respect to a Heisenberg ket-state that does not change with time), we obtain

$$\overline{\frac{dp}{dt}} = m\overline{\frac{d^2\mathbf{r}}{dt^2}} = m\frac{d^2\bar{\mathbf{r}}}{dt^2} = -\overline{\nabla V(\mathbf{r})}$$

or, for generalized coordinates,

$$m_{ik}\overline{\ddot{q^k}} + \overline{\frac{\partial V}{\partial q^i}} = 0,$$

where $m_{ik}$ plays the role of the metric tensor. Here symbol $\bar{A}$ means the quantum averaging over a pure state, $\bar{A} = (\Psi, A\Psi) = \int d^3r \Psi * (\mathbf{r})A\Psi(\mathbf{r})$ or, in Dirac's notations, $\bar{A} = \langle\Psi|A|\Psi\rangle$. In general, averaging and differentiation over time are non-commutative operations so that we are unable to write $\overline{dp/dt} = d\bar{p}/dt$ (see below). This is just an attempt to replace the quantum calculations by the classical ones.

It is clear that in the case of many particles one has to replace integration over $d^3r$ by integration over hypervolume element $d\tau$ and calculate a multidimensional integral. We may perform here straightforward calculations to illustrate the simple bridge between classical and quantum mechanics based on the Ehrenfest theorem. Incidentally, the coherent states popular today were first discussed by E. Schrödinger [129] as a by-product of obtaining a solution for the wave packet whose center moves according to the laws of classical mechanics in the quadratic (oscillator) potential, $V(q) = 1/2 m\omega^2 q^2$. We have already touched upon the oscillator model many times and shall deal with the model of coherent states later in some detail. To find the "motion equations" for expectation values we may, for simplicity, use also the Schrödinger picture (recall that the Schrödinger and the Heisenberg representations are equivalent, see above). Differentiating straightforwardly the defining expression for $\bar{A}$, we get

$$\frac{d\bar{A}}{dt} = \left(\frac{\partial \Psi}{\partial t}, A\Psi\right) + \left(\Psi, A\frac{\partial \Psi}{\partial t}\right).$$

Now, from the Schrödinger equation we have

$$\frac{\partial \Psi}{\partial t} = -\frac{i}{\hbar} H\Psi, \qquad \frac{\partial \Psi*}{\partial t} = -\frac{i}{\hbar} H * \Psi *,$$

where $H = H^+$ (we assume the Hamiltonian to be self-adjoint). Inserting these expressions into that for the derivative $d\bar{A}/dt$, we get

$$-i\hbar \frac{d\bar{A}}{dt} = (H\Psi, A\Psi) - (\Psi, AH\Psi) = (\Psi, H^+A\Psi) - (\Psi, AH\Psi) = \overline{[H.A]}.$$

Putting here $A = q^i$ and $A = p_k$, we get

$$-i\hbar \frac{d\overline{q^i}}{dt} = \overline{[H, q^i]}, \tag{6.9}$$

$$-i\hbar \frac{d\overline{p_k}}{dt} = \overline{[H, p_k]}. \tag{6.10}$$

It is not difficult to demonstrate that commutators $[H, q^i]$ and $[H, p_k]$ give the derivatives over $p_i$ and $q^k$, respectively:

$$\frac{\partial H}{\partial p_i} = \frac{i}{\hbar}[H, q^i], \qquad -\frac{\partial H}{\partial q^k} = \frac{i}{\hbar}[H, p_k]. \tag{6.11}$$

These relations follow directly from the canonical commutation relations (CCR), $[p_i, q^k] = -i\hbar\delta_i^k$, and are valid not only for the Hamiltonian $H(p_k, q^i)$, but also for any polynomial (entire) function of $p, q$.

The above formulas for the dynamics of expectation values were related to the autonomous case when the operator $A$ does not depend explicitly on time. In case the operator $A$ depends explicitly on time, we have only a slight modification of the formulas namely

$$\frac{d\bar{A}}{dt} = \left(\Psi, \frac{\partial A}{\partial t}\Psi\right) + \left(\frac{\partial \Psi}{\partial t}, A\Psi\right) + \left(\Psi, A\frac{\partial \Psi}{\partial t}\right) = \left(\Psi, \frac{\partial A}{\partial t}\Psi\right) - \frac{i}{\hbar}(\Psi, [H, A]\Psi).$$

One can introduce here a time derivative operator, $dA/dt$, defined by

$$\frac{d\bar{A}}{dt} = \left(\Psi, \frac{dA}{dt}\Psi\right).$$

Then we get the operator identity

$$\frac{dA}{dt} = \frac{\partial A}{\partial t} + \frac{i}{\hbar}[H, A],$$

which is an expression of a well-known fact that if some operator $A$ does not explicitly depend on time and commutes with the Hamiltonian, the expectation value of the respective physical quantity does not change with time *in any state*. In such a case, $A$ is the quantum integral of motion. Replacing again $A$ with $q^i$ and $p_k$, we have the *operator* equations

$$\frac{dq^i}{dt} = \frac{i}{\hbar}[H, q^i], \qquad \frac{dp_k}{dt} = \frac{i}{\hbar}[H, p_k] \qquad (6.12)$$

One can compare **??** with 6.12. Assuming the standard form of the Hamiltonian,

$$H = -\frac{\hbar^2}{2m}\partial_k\partial^k + V(q^i),$$

we obtain the following operator relations

$$\frac{dp_k}{dt} = -\partial_k V, \qquad \frac{dq^i}{dt} = \frac{p^i}{m} \qquad (6.13)$$

which, despite their operator character, have the form of classical Hamiltonian equations. This fact is often interpreted as a testimony for a close connection of quantum and classical mechanics.

However, it is difficult to ascribe an exact meaning to the term "close connection". The usual beliefs that quantum mechanics incorporates classical mechanics are, honestly speaking, not well founded. What do we expect when we think that a new theory includes an old one? That there exists a universal procedure of retrieving the old theory in its precisely defined application area, by taking an appropriate mathematical limit. There is no such universal procedure for the transition from quantum to classical mechanics. As it has been already mentioned, it does not seem to be possible to recover classical mechanics in all areas of its validity. The classical limit problem remains an open question. For example, one of the main problems related to the relationship between classical and quantum mechanics remains open: can quantum mechanics recover the classical motion over a single orbit, and in what limit? There are contradictory statements about this limit, some people say $\hbar \to 0$ is sufficient, others consider the limit of "large quantum numbers". There exist also a number of other points of view as to at what level classical behavior follows from quantum theory (e.g., based on stochastic schemes i.d. the Brownian motion and Fokker-Planck equation). Today, the decoherence concepts (see above) are of fashion, they did not exist twenty years ago although the correspondence problem is as old as the history of quantum mechanics. In different models, a variety of classical limits is discussed. Thus, in the computation of radiation processes classical behavior is identified (sometimes erroneously) with the emission of "small quanta with large probability" whereas rare events of "large" quanta emission (with non-negligible recoil) testifies to the necessity of quantum description. In the coherent states theory, there has long been a belief that classical mechanics is essentially an approximation of quantum mechanics, if one restricts all the states to coherent states only. In other words, it is the coherent states that are the cause of classicality. More sophisticated approaches to quantum-classical correspondence are based on the Wigner and Husimi functions as well as Weyl-Moyal algebra. The Feynman path integral (see below) in the original form introduced by R. Feynman [44] may also be considered a bridge between the two mechanics. Other groups actively promote the Bohm-de Broglie version of mechanics as a unifying theory. In short, any person doing quantum computations may have her/his own opinion about the development of classical behavior from quantum theory. This diversity of opinions indicates a certain logical inconsistency of the classical limit schemes that do not necessarily fit together.

The obvious difficulty for the classical limit is rooted in the fact that quantum physics is largely based on probabilities - only eigenvalues and expectation values have definite (exact) values. On the other hand, in classical mechanics a particle is presumed to be at a definite position and to possess a precise velocity (or momentum) under the influence of a definite force at each instant of classical time. We have seen above that discussions of the classical limit of quantum mechanics are frequently based on the Ehrenfest theorem which deals with the averaged quantum quantities. According to this theorem, as we have seen, quantum evolution in the mean resembles classical dynamical behavior. For instance, the Ehrenfest theorem for a system of electrons in a time-dependent external force gives the complete analog to Newton's law of motion for the interacting classical particles, in terms of the averaged positions of electrons and the averaged force acting on them. Averaging in this context can be interpreted in the following way. If one were able to produce a large number of observations over the positions of a particle described in quantum mechanics by a wave function $\Psi$, the average of all observed values of position vector $\mathbf{r}$ is obtained with the help of the quantum probability density, $dw = |\Psi|^2 d^3r$,

$$\bar{\mathbf{r}} = \int \mathbf{r} \, dw = \int \Psi^* \mathbf{r} \Psi \, d^3r,$$

that is the weight function $dw = |\Psi|^2 d^3r$ defines the frequency of occurrence of the position $\mathbf{r}$. Likewise the average momentum is defined by the expression

$$\overline{\mathbf{p}} = \int \Psi^*(-i\hbar\nabla)\Psi\,d^3r,$$

Generalizing these expressions, we may conclude that the average value of any dynamical variable $A(\mathbf{r}, \mathbf{p}, t)$ can be written as

$$\bar{A} = \int \Psi^* A(\mathbf{r}, -i\hbar\nabla, t)\Psi\,d^3r,$$

where $A(\mathbf{r}, -i\hbar\nabla, t)$ is an operator representing the quantity $A(\mathbf{r}, \mathbf{p}, t)$ in quantum mechanics. We have seen that the evolution of the average momentum, $d\overline{\mathbf{p}}/dt = -\overline{\nabla V(\mathbf{r}, t)}$, looks very similar to the Newton motion equation, however it does not produce Newtonian mechanics from the quantum theory. We have seen, for instance, that it is in general incorrect to define velocity $\mathbf{v}$ as $d\mathbf{r}/dt$ (see, e.g., [84], §19). Due to the fact that quantum mechanics operates with objects which are spread in space and time, it has been customary to introduce the notion of a "wave packet" as a mathematical model for a quantum particle. Below, after we discuss a little more the significance of the Ehrenfest theorem and its possible generalizations, we consider wave packets from the viewpoint of the quantum-classical relationship.

## 6.11  Wave Packets in Quantum Mechanics

We have seen in the preceding section that the Ehrenfest theorem can be interpreted in terms of expectation values, namely that the rate of change of the particle momentum expectation value is equal to the expectation value for the force acting on the particle. The term "expectation value" has a meaning only when some averaging procedure is defined. In quantum mechanics, it is assumed that the state [152] of a particle is described by the wave function $\Psi(\mathbf{r}, t)$ that may be interpreted as representing a "wave packet", a kinematic image associated with the quantum particle motion. In the process of motion, a wave packet tends to spread with time so that it is not a steady entity. The Ehrenfest theorem deals with the "center of gravity" of the wave packet, which has a fair meaning when the packet is unimodal, well concentrated and in a certain sense small. It would be impractical to define the "center of gravity" for a plane wave or for a polynomial. If the dimensions of the wave packet associated with the particle are small, the particle motion, according to the Ehrenfest theorem, may be approximated by classical mechanics. The "center of gravity" of the wave packet, $\overline{\mathbf{r}}^t = \{\bar{x}^i(t)\}, i = 1,2,3$, moves along a trajectory which is a set of points $\bar{x}^i(t)$ for all values of time $t$. One must remember, however, that the notion of a classical force standing in the right-hand side of the Ehrenfest equation has only a limited validity since in general the average value of a function does equal the value of a function at the point

---

[152] We consider only "pure" states in this context.

where its argument takes the average value, i.e., $\overline{\nabla V(\mathbf{r})} \neq \nabla V(\bar{\mathbf{r}})$. For instance, $\overline{x^2} = \bar{x}^2$ and, more generally, $\overline{x^n} = \bar{x}^n$ for $n \gg 2$.

But what is exactly a wave packet in quantum mechanics? Can it be interpreted as an appropriate ensemble of classical orbits?

## 6.12  Semiclassical Expansions and Asymptotic Methods

One usually asserts that quantum mechanics is logically more fundamental than classical mechanics, in the sense that classical mechanics can be derived from quantum mechanics but not vice versa. However, this statement can be neither proved nor disproved, at least up till now. The classical end of quantum mechanics is represented, as a rule by the quasi-classical WKBJ approximation (see above) which is, from the mathematical view-point, just an asymptotic expansion of wave equations. Quasi-classical approximation is so obvious that it was used for wave equations long before WKBJ, e.g., by Liouville, Stokes, Green, and Rayleigh. Recently, the WKBJ theory originally constructed for linear equations has been extended to the nonlinear framework, in particular, employed for the nonlinear Schrödinger equation [214].

In general, for an arbitrary quantum mechanical setting (i.e., for an arbitrary set of quantum dynamical variables), the $h \to 0$ limit is not obliged to exist which means that the quantum theory does not necessarily imply classical mechanics in the sense of this limit.

## 6.13  The Density Matrix and Its Relatives

In the conventional quantum mechanics the state is usually defined either by a vector (ray) in a Hilbert space or, in a more general case, by a density matrix. The first kind of a state is called "pure" whereas the second kind "mixed". Mathematically, a state $\Psi$ may be called pure if the interpolating relationship

$$\Psi = \alpha \Psi_1 + (1 - \alpha) \Psi_2,$$

where $\Psi_{1,2}$ are two states, $0 < \alpha < 1$, is possible only if $\Psi_1 = \Psi_2$.

The dynamics of a finite and closed quantum mechanical system such as atoms or atomic particles in a given steady external field is represented by a one-parameter group of unitary transformations in Hilbert space. However, this simple formalism of unitary time evolution is poorly suited for the description of systems interacting with the environment, for example, totally inadequate for the study of irreversible processes.

## 6.14  Do You Need an Interpretation?

Frankly speaking, personally I don't. Discussing interpretational issues of quantum mechanics brings by itself no physical results and usually serves as a tool to produce publications and dissertations. Nevertheless, some thoughts about interpretations of nonrelativistic quantum mechanics have driven a number of experimental groups to perform precise measurements of the effects typical only of quantum systems [216] and thus clearly discriminating classical (local) and quantum (nonlocal) behavior. Stimulating such highly accurate experimental work may be considered, as I see it, of a substantial value, as the discussions of interpretational issues have indirectly contributed to physics. This is already not so little.

In other respects, all endless discussions of such issues have only limited value: a lot of talk about the interpretation of quantum mechanics is just a tribute to aesthetic preferences, comprehension problems or dominating group attitudes. This is not physical but philosophical business. The fact that scientific value of the enormous bulk of words said about how to interpret quantum mechanics is only very limited is understandable since there seems to be nothing particularly intricate about quantum mechanics, especially in its nonrelativistic part, that should be somehow interpreted in a complicated philosophical way. One can notice that philosophers are fond of "interpreting", this process with no tangible results gives them a feeling of self-importance, since common people willingly listen to them. If you announce a dispute on such topics as "What is happiness?" or "What is beauty?" which are purely interpretational questions, crowds of people will be attracted. Such philosophical subjects stimulate the suppressed mental activity of any person engaged in the routines of everyday life. If one also has a chance to express one's sentiment on an abstract subject where one's opinion will be no worse than that of a "renowned" expert, then one's self-esteem drastically rises. But the scientific value measured as the attainment of new and quantifiable answers to precisely put questions has nothing to do with promoting opinions or building self-esteem. One can incessantly discuss various interpretations of quantum mechanics with the same productivity as the archetypal question: "How many angels can be placed on the tip of a needle?"

I have mentioned that there is nothing particularly intricate in quantum mechanics that would justify a whole philosophical business of interpretation. Just recall simple experimental facts. There were interference effects observed in the diffraction of beams both of light and of electrons.[153] Such interference effects made people believe that it is the amplitudes rather than the intensities that are additive, with the squares of the amplitudes describing the intensity distributions in particle beams. This is a typical model of any wave process, and it should not apparently involve any ambiguity or confusion. The latter, however, arises in such situations when the conditions for using either interfering amplitudes or non-interfering intensities to estimate the outcome of a particular experiment (most often a thought experiment) are vaguely formulated. For the conventional wave processes, say in optics or acoustics, the question of carefully formulated conditions for the *a priori* presence (or absence) of interference arises very seldom: we are used to wave phenomena in everyday life and take many things for granted. However, in quantum mechanics, especially in the so-called quantum theory of measurements, intuitive conclusions about interference effects may prove wrong (or be declared wrong).

---

[153] As well as beams of some other particles such as, e.g., slow neutrons.

Still, despite the relative simplicity of non-relativistic quantum mechanics many people consider it to contain grave conceptual paradoxes that remain unresolved even after almost a century of using quantum concepts. Even professional physicists have some uneasy feelings about quantum mechanics because it yields only probabilistic predictions. The standard interpretation of quantum prescriptions tells us a bizarre story that an electron can be anywhere until we "measure" it, which implies that observing the system determines its state. It was always assumed in classical physics that reality existed independently of the presence or absence of an observer. Humans are just an accidental result of biological evolution, and the world can successfully exist without any intelligent observers. One can construct, for example, a bunch of mathematical models united by the slogan "The world without us". Furthermore, can one apply quantum mechanics to the whole universe treated as a quantum object? Who is going to be an observer in this case? These and some other related questions comprise a partly philosophical issue that is called "Foundational problems of quantum mechanics". Some experts consider this issue to be very serious and invest a lot of time and effort in finding a new interpretation. For these researchers, interpretational problems have become a kind of philosophical obsession. Others think that the issue of interpretation of quantum mechanics is not important so long as it gives answers to correctly set physical problems. It is curious that the discussions on conceptual problems of quantum mechanics (and relativity) tend to be extremely harsh, easily touching personality issues, with stinging accusations and even direct insults. The fancy 1920s language of quantum mechanics contributed much to the confusion[154].

Many ongoing debates about the interpretation of quantum mechanics may be largely reduced to the question of the meaning of the quantum state vector (or a ray, in the modern parlance). There exist two main views: 1) the state vector gives the statistical description of an ensemble of identically prepared systems, and 2) the state vector completely describes an individual system, determining a full set of its dynamical variables (e.g., $A|\psi\rangle = a|\psi\rangle$). Although the frontier between physics and philosophy is blurred in this area, the issue of the meaning of state vector has a pragmatic importance and is utterly instrumental. The matter is that quantum phenomena do not occur in philosophical texts, nor even in Hilbert space - they occur in laboratories. However, the question remains: is there a preferred interpretation of quantum mechanics?

Schrödinger initially viewed the wave function $\Psi$ as a certain field distributed in space, like the electromagnetic field and others; stationary states, e.g., in atoms, correspond to eigenmodes of this field, similarly to proper oscillations in a resonator where only certain patterns and oscillation

---

[154] An example of a peculiar manner of quantum speech is "preparing the system to be in the state $\psi$". This statement is ambiguous and can be understood only intuitively, if at all.

frequencies will be sustained, with the others being suppressed by destructive interference.

The Schrödinger equation for the state vector (wave function) $\psi$ is not stochastic, so when speaking about ensembles, distributions, fluctuations, expectation values, etc. one has to gather such probabilistic notions for the state vector interpretation from some external considerations lying outside the Schrödinger mathematical model per se.

The hypothesis that historically preceded the Schrödinger point of view was one of Herzog Louis de Broglie. In 1923 [59] he postulated that the field distributed in space actually carries the particles and thus determines their motion in the classical sense (the pilot wave). It is interesting that de Broglie was not a physicist by training, initially he was trained in humanities (https://nobelprize.org/prizes/physics/1929/broglie/biographical). It is also interesting that de Broglie had left his pilot wave viewpoint for a quarter of a century and then returned to it. The main idea behind this hypothesis - in fact really brilliant - is that the motion of the particle depends on the accompanying wave.

In 1926, Max Born (who, incidentally, was a mathematician by training, and his tutor was D. Hilbert) offered the probabilistic interpretation of the wave function, by introducing statistical averaging [217]

$$\bar{A} = \int \psi^*(q) A(q) \psi(q) dq$$

for the self-adjoint dynamical variable $A$ in a physical state described by a complex function $\psi(q)$ (normalized to unity). So, the quantum states were identified with one- dimensional subspaces of a Hilbert space $\mathbb{H}$ (rays) which corresponded to the normalized state vectors $\psi$ defined up to a phase factor $\exp(i\varphi)$. This interpretation was developed in Copenhagen and therefore was nicknamed later "the Copenhagen interpretation". Some people tend to associate the Copenhagen interpretation not with Max Born but with Nils Bohr and, to some extent, with Werner Heisenberg. Bohr was more deeply engaged with philosophical aspects than his colleagues. Let the historians of science specify the role of each of creator of quantum mechanics and the origin of the term "Copenhagen interpretation", we are more interested in general quantum models as well as in questions of quantum evolution in this chapter. The standard Born interpretation is little more than a strongly idealized connection between physics and mathematical probability. The excellent (to my understanding, of course, there are people who do not share this praise) textbook [84] may be considered as the manual on the Copenhagen interpretation of quantum mechanics, although this fact has not been explicitly acknowledged by the authors.

### 6.14.1   More on Copenhagen Interpretation

The colloquial term "Copenhagen interpretation" is extensively used but remains poorly defined. It usually implies some connection to the experiment while discussing only the measurable quantities $A$ represented by the operator $A$. Then for any function $f(x)$ the expectation value obtained by the measurement of $f(x)$ in the quantum state $\Psi$ is defined by the scalar product $(\Psi|f(A)|\Psi)$. The Schrödinger equation and the Copenhagen formulation naturally place great emphasis on time development, since the Hamiltonian operator $H$ corresponding to the mechanical energy treated as the quantum "observable"[155] plays a special role determining the quantum evolution: we have seen that any operator $A = A(t)$ varies in time according to the rule

$$\frac{dA}{dt} = \frac{\partial A}{\partial t} + \frac{i}{\hbar}[H, A],$$

where $[,]$ denotes a commutator. The time evolution operator is based on the Hamiltonian and defined as $U(t) = \exp\left(-\frac{i}{\hbar}Ht\right)$. We have seen that the one-parameter group of unitary operators $U(t), t \in \mathbb{R}$, completely determines the time development of an isolated system (see [84], §9,13, see also "time evolution" and "Quantum Evolution" above in this chapter). Honestly speaking, it is just an assumption that the laws of time development may always be represented in the one-parameter group (more general, semigroup[156]) form i.e., that the rule $U(t, t_0)x_0 = x(t)$ with $x_0 = x(t_0)$ and $U(t, t_0) = U(t, s)U(s, t_0)$ is valid under all circumstances. In classical mechanics, such time development is a canonical transformation due to a special (symplectic) form of Hamiltonian geometry. It is this beautiful geometric form that makes classical mechanics so impressive - like an eternally young film star who emotionally appeals to everyone. We do not discuss here the shift of the focus in the temporal evolution from states to operators representing measurable quantities, which corresponds to the transition from Schrödinger to Heisenberg representation and has already been discussed. The most important thing here is the connection of these mathematical schemes with reality, i.e., experiment. However here there are difficulties, not only because a meaningful theoretical interpretation for quantum tests is usually formulated in the language of classical physics, but also because time evolution determined by the Hamiltonian and represented by a one-parameter group of unitary transformations in a Hilbert space $\mathbb{H}$ is poorly compatible with quantum measurements. Indeed, if the system undergoes some measurements, its time evolution is no more determined by Hamiltonian $H$ and after time $t$ it will not in general be found in the state $U(t)\Psi(0), \mathbb{H} \to \mathbb{H}$ or, more explicitly,

$$\Psi(t) = U(t)\Psi(0) = \exp\left(-\frac{i}{\hbar}Ht\right) = \sum_{n=0}^{\infty} \frac{(-iH)^n t^n}{n!}\Psi(0)$$

---

[155] Recall that to each physically observable quantity $A$ there exists an operator which is usually designated by the same letter.

[156] Recall that in a semigroup an inverse is not guaranteed for all its elements, in contrast with a group.

Thus, unitary evolution is inconsistent with the quantum state reduction postulate, and in the measurement situations one usually tends to avoid the unitary time evolution. So, an irreversible behavior is semi-implicitly introduced. One of the standard ways of introducing irreversibility is to postulate an interaction of the considered system with an external heat bath which, in the measurement case, should correspond to a measuring instrument. In other words, the measurement process in quantum theory, especially when the Copenhagen interpretation is used to connect the theory to an experiment, provides an example of an irreversible process even within the framework and axioms of a mechanical model. One must admit that for measurement situations in quantum mechanics the standard formulas for quantum-mechanical evolution is not valid, and measurement processes must be treated separately. This fact makes quantum mechanics a strange theory: on the one hand the Copenhagen interpretation places a strong focus on measurement and connection with the experiment, on the other hand the standard evolution expression is incorrect when measurement enters the picture.

Many physicists and the majority of philosophers have always been dissatisfied with the Copenhagen interpretation particularly because the unitary state vector evolution cannot be directly applied to measurement situations. At the same time, within the Copenhagen framework, measurement is crucial for understanding quantum behavior: without measurement one is unable to observe phenomena. This inconsistency is indeed annoying, and one has to add to the mathematically impeccable unitary evolution postulate another one, more of an *ad hoc* nature, namely that of reduction of the state vector. In operational terms, this additional postulate states that one observes after the measurement a normalized (unitary) state corresponding to the measurement outcome, the probability of this new state being given by the Born rule (squared amplitude).

Greatly simplifying, one can say that the Copenhagen interpretation states that there is no quantum world (so that the subject of this chapter is null and void - it may also be null and void due to other reasons - or at least the chapter should be renamed). There exists only an abstract quantum-mechanical description i.e., nothing more than a mental, mathematical model. A quantum particle, say an electron, is just a mathematical formula, and a mathematical expression can be neither weird nor indeterministic. Moreover, one cannot require a mathematical formula to be dependent on human observation as it is sometimes implied in the sophisticated conscience-based measurement concepts (see, e.g., [215]). Honestly speaking, I still do not understand why the interaction of a system with the environment must involve human conscience. Besides, I do not understand what the term "conscience" means: it is one of those vaguely defined notions which are difficult to use in "exact" sciences (see Chapter 2).

Greatly simplifying, one can separate physicists into two groups (this is of course more sociology than physics). The physicists belonging to the first group submit that being treated as a mathematical model with a collection of derived submodels, nonrelativistic quantum mechanics works impeccably. So one should not talk too much about the possible interpretations of excellent mathematical tools, many people describe this situation as "shut up and calculate". One should leave all "blah-blah-blah" to philosophers. Nevertheless, the other group of physicists, also large in numbers and incorporating rather prominent scientists, find this situation very unsatisfactory. For them, it appears insufficient when physics is reduced to a description of rules and prescriptions so that "understanding" is lacking. For instance, it is hard to believe that one does not know exactly where the system (e.g., an electron) is between the measurements; at maximum only the wave function can be calculated to establish the probability of a system to be found in any particular state.  Thus a particle can be found with a certain probability in any  particular spatial domain.

But what causes the particle to land in certain domains of a detecting surface (screen)? Is it some kind of a force? The Copenhagen - the historically dominant - interpretation brushes off the questions of such type as meaningless, which means that the study of quantum physics must stop short at some phenomenological results, for instance, the distribution of dots on a detecting surface.  This agnostic attitude postulated as an intrinsic property of quantum systems continues to irritate both the members of the second – "anti-Copenhagen" - group and a large part of philosophically-minded physicists. They still perceive a disturbing lack of knowledge in the presumably most fundamental physical theory as an unacceptable flaw.

I can understand the people, the majority of them being very intelligent and competent, who are utterly dissatisfied with the current interpretation of quantum mechanics. These people tend to think that quantum mechanics, at least in its orthodox Copenhagen version, does not give a complete picture of reality. It must not be, they assert, that reality should depend on presence or absence of a human observer, so the current interpretation of quantum mechanics is not compatible with realism. By the way, the term "Copenhagen interpretation" seems to be rather intuitive: it is ubiquitously used, but poorly defined. I am not sure anybody really knows what exactly the Copenhagen interpretation means and what mathematical statement corresponds to it, see, e.g. [117]. One can diffusely understand the Copenhagen interpretation of quantum mechanics as the generalization of the trivial statement that any phenomenon is disturbed by its observation[157]. In this sense, the Copenhagen interpretation is quite reasonable and, for me personally, is good enough not to suffer from torturing doubts over methodical abstractions.  Unending discussions on the status of the entire quantum mechanics or of a variety of its interpretations are more suitable for philosophy (metaphysics) than for physics. Despite numerous claims, I still think that the issue of interpretation of quantum mechanics is devoid of any practical interest and therefore does not matter much.  I may be  mistaken, of course.

---

[157] To sharpen the polemics about the Copenhagen interpretation, some philosophers bring the example of the Moon that need not necessarily be in its place in the sky when no one is looking at it. One may produce many examples of this kind, of course.

One should not confuse interpretations and formulations. Like classical mechanics where there are three main formulations - Newtonian, Lagrangian and Hamiltonian - quantum mechanics admits several more or less equivalent formulations. I have written "more or less" because I am not aware of any theorem proving the equivalence of all possible quantum-mechanical formulations. When talking about interpretations, one usually means a combination of high principles and low calculations. Interpretations bear the risk of transgressing the applicability of mathematical models, as can be seen on the example of the many-worlds interpretation of quantum mechanics. To clarify this point, I shall give an example of a formulation in order to see the difference between formulations and interpretations. Thus one of the most popular current formulations of quantum mechanics is based on the transition amplitude paradigm. The transition amplitude $(x_1, t_1 | x_2, t_2)$ having a form of the inner product is destined to answer the following question: if the quantum system is prepared "here and now" i.e., in the state $|x_1, t_1)$, what will be the probability of finding it "there and then" (state $|x_2, t_2)$)? If this amplitude is known, then to calculate the probability one simply has to take the amplitude squared. One may notice that in most practically important situations, only a small domain near the classical path contributes significantly into the total probability amplitude. This formulation, in contrast with interpretations, is a practical computational prescription; one can build an algorithm or a mathematical model around it. A philosophical component in this and other quantum-mechanical formulations is typically negligible.

Physics, from ancient times, always attracted philosophers attempting to interpret its results and incorporate them into a philosophical framework. I would call it a cognitive stigmatization. As a rule, this resulted in a kind of stiff censorship and, in any case, produced nothing new, but could be very harmful. One can remember utter mistrust to physicists studying relativity and quantum mechanics in Hitler's Germany and especially in Stalin's USSR. This mistrust permeated the entire society and was fueled by vindictive claims on the part of top placed philosophers. A series of belligerent attacks led to repressions of physicists [50], see also http://www.ihst.ru/projects/sohist/document/vs1949pr.htm. It was only the atomic bomb project that was assigned top priority in the post-war USSR that saved many Soviet physicists from physical extermination on purely philosophical ground - like in Middle Ages. Stalin presumably said to the dreaded chief of secret police Beria of the physicists: "Leave them in peace. We can always shoot them later."

Although many physicists have a substantial philosophical component in their minds, nobody would experience a need to enter a fundamental philosophical discussion every time one computes a matrix element. The toolbox of nonrelativistic quantum mechanics has never revealed any internal contradictions. Nevertheless, since the "physical meaning" - in fact philosophical interpretation of quantum mechanics - was to many people unclear, there were numerous attempts to introduce elements of classical interpretation into quantum theory. I have tried to make a short catalog of these endeavors which, by the way, were produced by very clever people; so one should not think that it would be a sheer waste of time to get oneself familiar with these ideas.

### 6.14.2    Bohm's version

David Bohm made an attempt to preserve the concept of a trajectory and to reconcile it with the formulas of standard quantum mechanics, e.g., the Schrödinger equation, by picking up some special function - the "quantum potential".

Personally, I think that attempts to interpret quantum mechanics in terms of classical trajectories are just about as meaningful as explaining cosmological phenomena in terms of the geocentric system[158]. We have seen that quantum mechanics is based on a totally different mathematical foundation as compared to classical mechanics, but nonetheless any viable mathematical model of quantum mechanics is fully deterministic in the sense that for every event there exists a cause and there is an effect for any cause. So, there is no need to impose classical trajectories on the quantum description to save it from the alleged indeterminism. However, since the Bohmian version is becoming increasingly popular today, particularly due to its apparent attractiveness to numerical modelers, we shall discuss it here at some length. Bohm was a great physicist, and already because of this his thoughts about quantum trajectories deserve to be respectfully overviewed.

The Bohmian interpretation of quantum mechanics is in a certain sense opposite to the many-worlds interpretation of Everett who retained all technical details of standard quantum mechanics. One may regard Bohm's version of quantum mechanics as a further and more sophisticated development of de Broglie ideas: some disciples of de Broglie's school did not accept the conventional probabilistic interpretation of quantum theory at all, considering it as contradicting to "realism" and to objectivity of physical laws. According to this point of view, only the determinism of classical type based on trajectories may be compatible with realism. D. Bohm's motivation was, as he formulated it in [218], to suggest an interpretation of quantum mechanics, alternative to the conventional one, which would be free of the usual conceptual difficulties and, besides, would "determine the precise behavior of an individual system".

Individuals and groups who promote Bohm's "causal interpretation" of quantum mechanics tend to present themselves as intrepid thought transformers, almost revolutionaries daringly fighting against inert reactionaries who try to preserve their "comfort zones" of calloused concepts. The adherents to Bohm's version (the "Bohmians") probably think that they are opening eyes to the physical community whose dominant members uncritically maintain the "idealistic" orthodox interpretation of quantum mechanics. In fact, the kind of interpretation now known as the Bohmian picture had appeared even before the Copenhagen interpretation (or at least simultaneously with the latter, in the 1920s) in the works by L. de Broglie [59]. This "materialistic" interpretation was originally named the "pilot wave theory" (see above), being later supported and developed by J.-P. Vigier and then extended by D. Bohm and his followers (see a detailed account in [137] and [138]). One should, however, notice that the pilot wave theory did not survive in contrast to the "orthodox" quantum mechanics repudiated by the "Bohmians". We must also mention E. Madelung who was probably the first to use the polar representation of the wave function taken as any complex quantity,

$$\Psi(\mathbf{r}, t) = R(\mathbf{r}, t) \exp\big(iS(\mathbf{r}, t)\big),$$

in a systematic way. The meaning of the Madelung representation is in replacing the wave function by the density field $\rho(\mathbf{r}, t) = R^2(\mathbf{r}, t)$ and the phase field $S(\mathbf{r}, t)$. We shall see below that it is this Ansatz that is the mathematical essence of Bohmian mechanics. Separating the real and imaginary parts, one gets the Madelung equations (also known as quantum hydrodynamic equations). This form of representing the wave function inspired numerous attempts to interpret nonrelativistic quantum mechanics in the classical spirit.

---

[158] Here, I cannot refrain from a remark that some opinion polls, in particular, carried out in Russia in 2007 demonstrate an increasing number of people, to my surprise mostly young ones, who believe that it is the Sun that rotates around the Earth and that it is only those scientific jerks who "powder the brains" of simple people living in the real, not queer and scientific, world. As near as I remember, there were about 27 percent of sampled persons who possessed this level of astronomical knowledge.

I still think that the Bohmian version based on the trivial polar representation of the $\psi$-function is a sheep in wolf's clothing. This version and associated "causal" interpretation brings nothing radically new, although it may in certain cases be convenient for numerical computations.

### 6.14.3    Statistical interpretation

The real meaning of the wave function has gradually become more elucidated starting from the M. Born works on statistical interpretation of quantum mechanics [159]. This is another class of possible interpretations. In the statistical interpretation, the notion of probability is indispensable, although, being applied to quantum mechanics, it was at first not at all clear what is meant by the probability, at least in the mathematical sense. Probability of what? But leaving mathematical rigorism aside, physicists constantly talked of probabilities.

An essential part in elucidating this issue - as in general in the problem of interpretation of quantum mechanics - belonged to intuitive considerations of N. Bohr that the quantum mechanical description of the microscopic object properties must be harmonized with the classical treatment of observation means [136]. Object properties are always manifested in the interaction with observation means. Bohr emphasized the necessity to consider the entire experimental scheme, with the human observer being included. Sometimes an impression arises that Bohr was forgetting what was actually measured:

---

[159] I would recommend carefully reading the Nobel Lecture by Max Born "The Statistical Interpretation of Quantum Mechanics" [217].

micro-object properties or an observation device. One may often encounter reprimands addressed to Bohr's views bordering on philosophy that he presumably underestimated the fact that such micro-object properties as mass, charge, spin or the form of a Hamiltonian, including the character of interaction with external fields, are absolutely objective and independent of the observation means. There is probably a certain misunderstanding of a terminological character: Bohr uses the expression "uncontrollable interaction", which was probably necessary to cover the gap originating from applying classical concepts outside their validity area. In fact, the interaction of the observed system with the measuring device may be considered as a physical process and is, in this respect, quite controllable. Nevertheless, each person familiar with measurement techniques knows that statistical methods are an indispensable element of experimental data interpretation.

### 6.14.4    The Many-Worlds Interpretation

Curiously enough, when one starts studying quantum mechanics, one cannot get rid of an impression that the quantum theory is built not from the very beginning. Later on, as one learns how to calculate, this impression disappears or, rather, is displaced. However, some irritating symptoms of logical incompleteness of quantum mechanics do remain, and it is these symptoms that drive people to invent new formulations, interpretations and logical schemes to satisfy the quest for cognitive comfort. Indeed, to perform quantum computations one has to familiarize oneself with the concepts that are far from being rudimentary, though they have been put at the base of quantum mechanics, such as expectation values, probabilities, operators, unitary spaces, etc. In order to make this stunning machinery work smoothly, one had to lubricate and support it with a number of supplementary "principles": complementarity, indeterminacy, compatibility, superposition, measurement, state reduction, projection, absence of trajectories, correspondence, indistinguishability, etc. Not all of these principles are obvious or logically independent from one another and some even look slightly artificial. Therefore, one is tempted to think that quantum mechanics has not yet reached its final stage and will undergo a further development like some technological system such as an automobile. On the other hand, quantum mechanics forces one in certain situations to weaken the principles that seemed to have been firmly established before the invention of quantum theory such as causality, classical (Kolmogorov) probability, mathematical logic[160].

    One of the main supplementary supports of quantum mechanics is the superposition principle.  From the classical viewpoint, this principle amounts to sitting between all the chairs. It leads to such classically unimaginable phenomena as entangled states, possibility of an electron to be reflected simultaneously from the floor and the ceiling, interference of amplitudes, etc.

---

[160] Of course, these principles also remain mathematically unproved and can be just readily adopted human stereotypes.

There is a lot of literature devoted to the many-world interpretation of quantum mechanics so that the reader can easily educate herself or himself in this area. Despite my deep respect for Hugh Everett's free and independent thinking, I cannot agree that the many-worlds interpretation has contributed a lot in quantum mechanics. One can even call "everetics" the whole flood of scientific and pseudoscientific papers as well as science fiction around the many-world interpretation. The heated debate around this concept, in my opinion, is useful mainly for Hollywood and science fiction writers. The reason is that the many-worlds concept did not provide any real insight, nor new computational procedures. One still has to solve the same equations, supplemented by the same boundary conditions and, quite naturally, obtain the same probabilities. The only thing one gets partly new with the many-worlds interpretation is an attempt to eliminate the uncomfortable "measurement problem", i.e., the question how a definite classical reality emerges from a set of quantum alternatives. The many-worlds interpretation treats this problem by assuming a hypothetical selection of only one of the infinite number of unobservable eigenstates, one per universe. We see only one since we are accidentally present in the universe with just this measurement. By the way, we ourselves, all individuals are constantly "splitting" into non-communicating copies of themselves together with the immediate environment[161]. A fanciful construction!

Some people think that this construction preserves realism (I find this ridiculous) or, at least, determinism. Determinism or "free will" - leave it to philosophers, these concepts have nothing to do with physics or mathematics. Physicists are mostly interested in obtaining answers to correctly set problems in operationally defined terms, and the question "Is there a free will anywhere in Nature?" seems to be out of scope of physical research.

The scientific content of the many-worlds interpretation as opposed to the metaphysical one may be formulated in the following way. The many-worlds interpretation was an attempt to escape the dubious state vector reduction by replacing it with the Schrödinger evolution of the whole universe. However, to describe $N$ successive quantum measurements one needs to consider the tensor product of $N$ branched wave functions. The mathematical framework of the Everett model is not quite clear - at least it remains unclear to me since I failed to find any pertinent discussion at the level of modern mathematical physics. One can, for example, consider the space of all such possible tensor products. Then one may have to define a suitable measure on that space. Afterwards, the Born probability rule may be interpreted as a procedure of searching the probability to find oneself in a particular branch of the wave function describing the entire universe. This probability measure should be identified in the many-worlds model with the probability of a particular measurement outcome.

One should not underestimate the heuristic value of the many-worlds interpretation. It paved the way for the wave function of the closed universe

---

[161] I wonder, what is the time scale of such branching. Are unobservable universes created every attosecond or faster?

[67] so that modern cosmologists should owe a debt to Hugh Everett for his insight half a century ago that the entire universe can be described by a single wave function. But the scientific status of the many-world interpretation is just that: interpretation - not a theory or a mathematical model. By the same token, despite all persistent claims, Bohmian or Copenhagen views are also just metaphysical interpretations of a physical theory. Nevertheless, I have already remarked that the Copenhagen interpretation seems to be perfectly adequate for physics without fancy hypotheses such as that of an indetermined number of non-interacting universes and other metaphysics. At least, it is good enough for me.

   The mathematical models of quantum mechanics are not in the least bizarre, as many non-physicists used to think. Mathematics of quantum mechanics can be easily understood and in many cases is more transparent than mathematics of classical mechanics. Later, we shall see examples when the quantum-mechanical way of reasoning proves to be simpler than the classical picture. What seems to be really bizarre is interpretation. Sometimes it comes down to accepting that human consciousness is somehow involved in determining the outcome of an experiment.

## 6.15  Causality

Causality in general may be understood as a statement related to general properties of space and time. Causality manifests itself in many areas of knowledge, although I am not aware of any theorem warranting that causality should be omnipresent. Thus, causality does not follow from any underlying equation or theory - it is simply a postulate. I have already mentioned that we may regard causality as a result of human experience. Non-causal systems would allow us to get signals from the future or to influence the past - a marvelous possibility extensively exploited by phantasy writers. Figuratively speaking, in the noncausal world you could have killed your grandmother before she had been born. Models based on a dynamical systems approach are always causal, i.e., effect cannot precede the cause and the response cannot appear before the input signal has been applied. In the special theory of relativity, causality is interpreted as a consequence of the finite speed of signal or interaction, with acausal solutions being rejected - it is impossible to influence the past, it is presumed constant.

   We have already noticed that causality is closely related to non-invariance with respect to time reversal, frequently called TIV - time-invariance violation. Indeed, causality establishes the temporal priority of cause and effect or a temporally ordered sequence of causes. In causal models, one cannot affect the past, it is constant. We shall later focus on this issue in connection with the so-called arrow of time. Although the principle of causality does not appear explicitly in physics[162], it underlies its foundations. In purely human terms, this principle simply means that there are no miracles. I agree that it is sad, and many people still want to believe in miracles. Of course, nobody can forbid other persons to believe in faster-than-light travel, time journeys, precognition, teleportation, witchcraft, evil eye, magic spell and other captivating stuff, especially in our time when the virus of infantilism rapidly spreads all over the world. If you think that fancy creeds make your life better, then do believe in miracles. To hell with science, it is not the main thing in life, anyway.

---

[162] See, however, below a discussion of dispersion relations.

## 6.16  Quantum Chaos

From the theory of dynamical systems (Chapter 4) we know that classical trajectories, even those related to the simplest systems, exhibit two types of behavior: (1) regular, orderly flow and (2) irregular, chaotic zigzags. On the other hand, we know that there exists the correspondence principle linking classical and quantum mechanics. Thus, the question naturally arises: are there any manifestations of the two different types of classical motion in quantum mechanics? On the simplest level, quantum mechanics may be formulated in terms of the wave function; so, the correspondence principle must specify some relationship between wave functions and the families of classical trajectories, orderly or chaotic. But what happens with the wave function when the corresponding classical trajectories leave the regularity domain in the parameter space and transit to the chaotic regime? To correctly discuss this issue, one has to re-examine the relationship between the description based on classical trajectories (classical rays) and on wave functions (quantum rays). In other words, the subject of manifestations of classical chaos in quantum mechanics invokes the fundamental aspects of the quantum-classical correspondence.

## 6.17  Path Integrals in Physics

In this chapter, we have already discussed at length and on various occasions the connection between the quantum and classical mechanics. One may observe that this issue, while still crucial for the whole physics, is far from being ultimately resolved. The same might be said about the interpretation of quantum mechanics, although this latter issue has some metaphysical flavor (see below). The fact that the interrelation of quantum and classical mechanics is hard to elucidate is understandable from the primary mathematical viewpoint: quantum and classical theories are based on totally different concepts - the first, in its primitive form, is tantamount to the theory of linear self-adjoint operators in Hilbert space whereas the second is basically nonlinear, with its foundation lying in more sophisticated geometrical considerations. Thus, to link these two diverse disciplines is far from trivial.

There exists, however, an elegant method invented by R. Feynman [44] allowing us to comprehend the connection between quantum and classical mechanics. This method is called the path integral formalism. Feynman showed that time evolution, which is the main process considered in quantum mechanics, can be expressed through the classical action for various trajectories the quantum system might follow. As usual, the term "quantum system" denotes something simple, like a particle moving in 1D, but we shall see later that the path integral formalism may be applied to rather complex physical (and not only physical) objects. From the geometrical viewpoint, path or, more generally, functional integrals provide a means to deal with piecewise linear curves which manifest fluctuating 1D structures. These fluctuations may be of different nature: quantum-mechanical, statistical, thermodynamical, etc. Respectively, path integrals may be an adequate formalism in all situations where such fluctuating trajectories are encountered, e.g., in areas seemingly distant from physics such as economics and financial analysis. This is a typical physmatical effect of interconnected disciplines.

The word "action" in the original Feynman's version of path integration implies the Lagrangian formulation of the theory (see Chapter 4). Indeed, it is this formulation of quantum mechanics that, in contrast to the Hamiltonian approach used by Schrödinger, that leads to path integrals. We may recall that Lagrangian mechanics provided us with an understanding of the conservation laws - their origin lies in the connection between conserved quantities and the symmetries of a physical system (namely, there is a conserved quantity corresponding to each generator of the Lie algebra for a Lie symmetry group). Moreover, Lagrangian mechanics, in principle, gives a coordinate-free and global description for the paths on the configuration manifold (see Chapter 4). It is this feature of the Lagrangian approach that led P. A. M. Dirac and R. P. Feynman to invent their new quantum theories, in particular the path integral formulation.

The path integral may be formally represented by the formula for the transition amplitude

$$\langle x, t | x_0, t_0 \rangle = A \int [Dx] \exp \left\{ \left( \frac{i}{\hbar} \right) S[x(\tau); t, t_0] \right\} \qquad (6.14)$$

where the exponent $S[x(\tau); t, t_0]$ is the classical action for a path joining the space-time (Minkowski) points $(x_0, t_0)$ and $(x, t)$, $x_0, x \in (\mathbb{R}^3)^n$, $n < \infty$ is the number of particles, $[Dx]$ is some formal measure and $A$ is the normalization constant. Before we discuss this expression, clarify the meaning of the term "path integration" and learn how to use it, we may notice that for $|S| \gg \hbar$ (asymptotically, in the limit $\hbar \to 0$), the integral containing the weight $\exp\{iS/\hbar\}$ can be calculated with stationary phase techniques (see Chapter 3), which implies that the value of the integral over paths should be determined by the neighborhood of the path that brings an extremum to the classical action $S$. This special path exactly corresponds to the classical trajectory provided by the Lagrangian formulation of classical mechanics (see Chapter 4). The asymptotic expansion of the path integral for $\hbar \to 0$ coincides with the WKB approximation, which is typically derived from the Schrödinger equation. This link with WKB (semi-classics) is already a hint at the generality of the path integral.

Later we shall see that the path integral also has many applications in statistical physics and may be interpreted in accordance with its (Gibbs) principles. This fact provides an important link between statistical methods and quantum field theory. The underlying idea is simple and clear: the equilibrium statistical physics is a probabilistic theory, the main statement of which is that a physical system under temperature $T$ has a probability proportional to $\exp\{-E/k_B T\}$ to be found in a state characterized by energy $E$, $k_B$ is the Boltzmann constant (usually, we shall put it equal to unity thus measuring temperature in energy units). One can then interpret the path integral in its original form for a single particle as moments of some probability measure, with the analytic continuation to the imaginary time $t \rightarrow -i\hbar/k_B T \equiv -i\hbar\beta$. In other words, the "free measure" corresponding to a system of non-interacting particles becomes perturbed by the Boltzmann factor $\exp\{-\beta V(\mathbf{r}_1, \ldots, \mathbf{r}_n)\}$ where $V(\mathbf{r}_1, \ldots, \mathbf{r}_n)$ is the interaction energy between the particles. One can then, with the help of the path integral, comparatively easily calculate the partition function

$$Z = \sum_i e^{-\frac{E_i}{k_B T}} = Tr\left(e^{-\frac{H}{k_B T}}\right) \tag{6.15}$$

- the principal quantity in equilibrium statistical mechanics, from which all macroscopic (thermodynamic) quantities can be derived. (The term "partition function" reflects the possibility of partitioning the system's total energy between its subsystems; sometimes this term is replaced by a more old-fashioned one "statistical sum").

Let us now get back to our simple mechanical model of a single particle moving in 1D in the potential field $V(x)$. One can recall that while the Hamiltonian version of classical mechanics had provided a foundation for the Hilbert space formalism in quantum mechanics, it is the Lagrangian formulation that led P. A. M. Dirac and then R. Feynman to path (functional) integration. The essence of this new build of quantum mechanics, based on the action for various paths the quantum system may follow classically, is an expression for the evolution operator $U_t = e^{itH}$ where $H$ is the system's Hamiltonian. For our simple model of a single-particle system, in order to gain some experience of handling the path integrals, we may try to derive a straightforward expression for the quantum-mechanical time evolution starting from the Lagrangian formulation. The evolution operator $U_t$ in this case corresponds to the transfer of a particle from $(x_0, t_0)$ to $(x, t)$ and may be written as a functional integral over the space of classical paths

$$U_t(x_0, x) = \int_{x(\tau): x_0 = x(0), x = x(t)} \exp\left(\frac{i}{\hbar} S[x(\tau)]\right) \prod_{\tau \in [0,t]} dx(\tau) \tag{6.16}$$

This expression is actually no more than a symbol that still needs to be interpreted. For example, it is not even clear what the symbolic product $\prod_{\tau \in [0,t]} dx(\tau)$ means. R. Feynman, in his original paper [44] interpreted the above expression as the limit of finite-dimensional integrals over piecewise linear curves approximating the smooth particle path, these curves having vertices at discrete points $t_i \in [t_0, t], i = 1, 2, \ldots$ for $\Delta t_i = t_{i+1} - t_i \rightarrow 0$. This construct is very close to the finite-difference techniques ubiquitously employed in numerical mathematics and scientific computing. Since functional integrals are increasingly often encountered in modern physics (and nowadays not only in physics, see the comprehensive book by H. Kleinert [49]), I shall try to illustrate this approach on the simplest possible model. Although this illustration is by default simplistic, for me personally it was a good reference to study more sophisticated subjects such as path integral formulation of supersymmetric systems.

## 6.18  Quantum Field Theory

In special relativity the velocity of light is the same in all inertial frames but space-time remains flat and is understood as a fixed arena given ab initio. In this sense, special relativity distinguishes from general relativity and is a predecessor of the latter.  Quantum field theory is a much more difficult theory than ordinary non-relativistic quantum mechanics.  It is also much more general, which is manifested by the fact that it embraces most of the laws of physics, except gravity. QFT's development has never stopped: it was originated in the late 1920s and has been constantly modernized up to the present. In its advancement, QFT progressively conquered new areas going through such important issues as "antimatter", which emerged around 1930, the theory of photons, electron-photon interaction, electromagnetic radiation processes and vacuum (1950s), culminating then in the "Standard Model" of particle physics that was a successful attempt to describe strong, electromagnetic and weak interactions within a unified theoretical framework. The Standard Model which appeared in the 1970s was a considerable achievement, a milestone that led to new predictions and stimulated the construction of new gigantic accelerators - the main of which (LHC) in CERN - where one hopes to test the principal implications of QFT. One, however, needs a lot of artistry in order to describe real particles by the QFT means. For instance, attempts to construct models of elementary particles and their interactions by means of QFT are essentially based on the concept of renormalizability, while the notion of a renormalizable theory is a bit vague and applied *ad hoc* to QED, QCD, $\phi^4$-theory and so on (see, e.g., [206], chapter 9). We shall return to the problem of renormalizability when discussing the standard computational techniques of QFT.

   One can find in textbooks the widespread opinion that there is a grave conflict between Einstein's relativity theory, both special and general, and quantum theory (see, e.g., http://en.wikipedia.org/wiki/Quantum mechanics and references to this article). But what is the essence of this conflict is revealed - and probably understood - in a dissimilar way by different authors. The prevailing opinion, shared mostly by the "pragmatic" wing of physicists, is that it would be difficult to merge these two classes of theories for all energy and spacetime scales because they are based on totally different assertions, and the best one can achieve is to construct a non-contradictory quantum field theory, free from divergences. Honestly speaking, the problem of ultraviolet divergences still remains unsolved in contemporary QFT. Again, the dominant opinion, shared probably by the majority of physicists, is that since QFT as we know it now is a low-energy approximation to some future fundamental theory on how the universe works, e.g., "theory of everything" (TOE), then the divergences will disappear as soon as people better understand physics at small distances, in particular, with the aid of this new theory, so far unknown. There exist today a number of crucial ideas to be put into the core of this hypothetical final theory. The most popular among these ideas is that of replacing point particles by extended objects such as strings or tiny membranes (see Chapter 9). Another popular idea is that the space structure at small distances, e.g., of the order of the Planck length, $l_p \sim 10^{-33}$, is geometrically highly nontrivial, for example, foamy, discrete, or non-commutative. Yet prior to the emergence of the future fundamental theory, we have to be satisfied with heuristic methods such as renormalization, while closing our eyes on ultraviolet divergences.

Moreover, quantum field theory has been constructed based on the flat (pseudo-Euclidean) spacetime of special relativity. Contrariwise, general relativity understands gravity as a spacetime curvature. There were numerous attempts to build gravitation theory over flat spacetime, but they did not seem to succeed (which, as near as I know, was pointed out already by Poincaré). Never-ending attempts to make gravity renormalizable à la quantum electrodynamics were also unsuccessful so far.

It seems to be of extreme importance that one can construct rather general field theories by using only two guiding principles: (1) symmetries and (2) the least action principle (albeit slightly generalized as compared to Lagrangian mechanics).

One should not think, however, that the simplest but the best computationally developed version of quantum field theories, QED is just a relativistic generalization of standard quantum mechanics (QM). Not at all, QED and QM are totally different animals.

When talking about the standard non-relativistic quantum mechanics, one may use the field theory concepts and ask, for example: how can we get from the quantized field $\Phi$ to the "single-particle sector" with the associated wave function $\psi(\mathbf{r})$ satisfying the Schrödinger equation for a free non-relativistic particle? The answer in the same language would be: if $|0\rangle$ and $|\mathbf{p}\rangle$ denote the vacuum and the single-particle state with momentum $\mathbf{p}$, respectively, then

$$\psi_{\mathbf{p}}(\mathbf{r}, t = 0) = \langle |\Phi(0)| \rangle = e^{i\mathbf{p}\mathbf{r}}$$

is the plane wave solution to the free non-relativistic single-particle Schrödinger equation. This language is sometimes considered more "modern", although it brings nothing new and may be confusing.

The main feature that distinguishes quantum field theory from non-relativistic quantum mechanics is the phenomenon of *vacuum polarization*. The latter also leads to some specific mathematical properties such as algebraic structures of QFT unfamiliar in quantum mechanics. Vacuum polarization and vacuum fluctuations are omnipresent in QFT, and this fact makes any bound state picture of quantum mechanics naive and obsolete. In view of vacuum polarization, QFT is intrinsically a many-body theory. Moreover, in case one desires to retain the concept of vacuum, there seems to be no viable alternative to the conventional QFT.

There are some standard computational techniques in QFT. In quantum theory in general and especially in quantum field theory, one must primarily learn how to perform approximate calculations (though with controlled accuracy) because the equations of the theory are so complicated that they cannot be exactly solved. The standard approximation strategy, perturbation theory, is always based on the presence of some small parameter.

### 6.18.1    More on Relativistic Invariance

Quantum field theories were initially formulated as some extensions of nonrelativistic quantum-mechanical models such as represented by the Schrödinger equation, e.g., by formulating relativistic wave-like equations intended to describe the motion of single particles. Those were, in particular, the Klein-Gordon, Dirac and certain more complex equations portraying relativistic quantum motion. A characteristic feature of all such equations was that time and position variables had to be formally equivalent - that was one of the symptoms of relativistic invariance distinguishing relativistic quantum theories from nonrelativistic quantum mechanics. So relativistic invariance is an indispensable attribute of relativistic wave equations. Now, in connection with relativistic quantum theory, we shall discuss this kind of invariance in some more detail as compared to the discussion of classical fields and waves.

The archetypical examples of relativistic wave equations are Maxwell's equations[163] for classical electromagnetic fields, the Klein-Gordon and Dirac equations. From the position of a general theory of fields, it is not so important that Maxwell's equations are classical (indeed, they do not contain the parameter $\hbar$ and therefore may, by definition, be considered classical), more essential is that they describe the dynamics of a vector field represented by the vector potential $A_\mu(x) = (A_0, \mathbf{A})$. Here $x$ is as usual understood as a contravariant four-vector, $x := (x^0, x^1, x^2, x^3) = (ct, \mathbf{r})$. In distinction to Maxwell's equations, the Klein-Gordon equation governs the dynamics of a scalar field $\phi(x)$ whereas the Dirac equation describes excitations of the four-component spinor field, $\psi_\mu(x), \mu = 0,1,2,3$. Note that these three fields give the most frequent examples of relativistically invariant models, but in no way exhaust the whole relativistic field theory, as it might seem when reading standard textbooks.

It is also important to note that each of the relativistic fields i.e., scalar, vector, tensor, etc. functions defined on the Minkowski space $\mathcal{M}$ (or on some manifold $\mathcal{M}$) which are transformed in a specific way under the action of the Lorentz or Poincaré group.

---

[163] It is curious that in the English-language literature it is customary to write "Maxwell's equations" and not "the Maxwell equations", this latter form is encountered much more seldom than the former. On the contrary, "the Dirac equation" is encountered more often than "Dirac's equation".

### 6.18.2 Feynman Diagrams

The only existing version of QFT has long been quantum electrodynamics (QED), which is mostly devoted to studying the interaction of electrons with photons. This interaction, however, is of a very general character and efficiently models any kind of interaction between matter and forces [99]. Studying QED seems to be extremely useful because one can test a lot of new ideas on rather simple and intuitively comprehensible models. For instance, such concepts as already mentioned vacuum polarization, vacuum fluctuations [164] as well as such essential techniques as those of Green's functions and Feynman diagrams can be fully explored on the QED level. The interaction between the matter and the forces is described in QED by a trilinear map involving two spinors and one vector. This map is often drawn as a Feynman diagram:

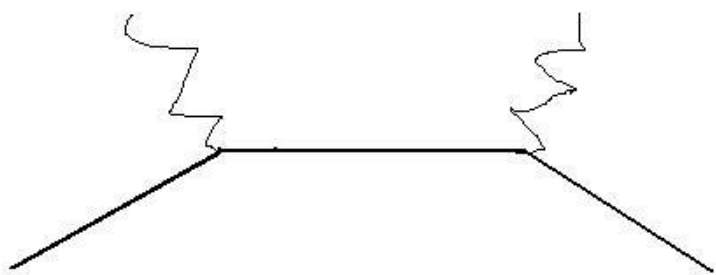

Figure 7.1.: An example of the Feynman diagram (Bremsstrahlung)

where the straight lines denote spinors and the wiggly one denotes a vector. The most familiar example is the process whereby an electron emits or absorbs a photon.

In conclusion to this section, I can mention that the QFT, specifically the Feynman diagram representation techniques in QED, has been taken many times as a pattern for other theories, even for some versions of such a complicated and nonlinear theory as Einstein's gravitation, although this issue is highly controversial.

### 6.18.3 S-Matrix

The scattering matrix or the S-matrix, as it was called later, was introduced by J. Wheeler [303] in 1937 and W. Heisenberg [296] around 1943. Heisenberg was looking for the means to represent quantum mechanics and, in particular, scattering theory only in terms of directly observable quantities such as energy and energy levels, momentum, mass, spin, lifetimes, cross-sections. S-matrix became increasingly popular in the 1960s, together with particle physics [304] fashionable at that time. At that time plenty of new particles and resonances were discovered so the question, what consisted of what (hierarchical or "aristocratic" approach), was very urgent. The S-matrix rapidly became the main observable, and most experimental results in particle and high-energy physics were analyzed in terms of the S-matrix, the latter being interpreted as a main invariant measure of particle interactions. One can regard the S-matrix as a peculiar formfactor sandwiched between the incoming ket state and the outgoing bra state[165] (it can be defined vice versa of course). Anyway, the S-matrix is a convenient tool because it is a global object thus eluding the annoying ultraviolet i.e., short-distance problems due to omnipresent vacuum fluctuations. Besides, there are not so many mathematical requirements imposed on the S-matrix: the only non-trivial property apart from Poincaré invariance [166] is the asymptotic factorization claim, when the wave packets corresponding to the scattered particles become spatially separated. These features of S-matrix (or S-operator as it is increasingly fashionable to call it now) enable one to construct an efficient computational scheme producing finite models of interacting relativistic particles, without explicitly attracting the underlying intricate QFT concepts.

---

[164] Nevertheless, to my best knowledge, the question: "What is vacuum?" has not been fully answered in QED.

*6.18.4   Particles and Symmetries in Quantum Theory*

We have seen that particles, in the standard quantum theory, are described by the wave functions which are defined on the set of variables chosen to describe the particle's behavior. It is assumed in quantum mechanics that wave functions form a complex linear space (the superposition principle), with linear combinations of the wave functions representing a superposition of states. This complex linear space is the unitary space (i.e., endowed with a Hermitian positive-definite inner product usually called "amplitude" in quantum mechanics) of states for some quantum system, in the simplest case a single particle. Linear operators acting in this space transform wave functions[167], and therefore it would be worthwhile to study their behavior under these transformations. Moreover, we know that wave functions can be distinguished and classified according to their symmetry or, more generally, transformation properties. For example, physicists studying particle models in relativistic quantum theory are mostly interested in symmetries described by the Lorentz and Poincaré groups.[168] Thus, the symmetry properties may be considered as the principal feature of quantum particles. Quite roughly, particle symmetries can be classified into two types: spacetime symmetries and internal ones.

An instinctive requirement of high symmetry is mostly dictated by the quest for "beauty" in physics. Logically speaking, beauty has nothing to do with physics, these two are completely disparate entities. Imposing beauty on physics reminds me of the great hype about "Millennium" which was nothing more than the magic of round numbers. In reality, "Year 2000" did not correspond to the beginning of the third millennium, but people tend to react more often on perception than on facts.

---

[165]   We are using the Bra- ket notation, or Dirac notation, see https://en.wikipedia.org/wiki/Bra%E2%80%93ket_notation.

[166]   Recall that the Poincaré group includes Lorentz transformations plus space-time translations. Initial and final particles considered in the S-matrix approach are required to transform according to irreducible representations of Poincaré group.

[167]   If $\mathbb{H}$ is a linear space with the inner (scalar) product $(\varphi, \psi)$, then a set of all invertible linear operators $A: \mathbb{H} \to \mathbb{H}$ such that $(A\varphi, A\psi) = (\varphi, \psi)$ (i.e., isometries) forms a group. This is one of the first symmetries one encounters in quantum mechanics.

[168]   The rotation group which is also relevant for classification of wave functions may be regarded as a special case of the Lorentz group.

### 6.18.5   Quantum Field Theory and Mathematics

If one wants to be completely honest, one has to admit that the problem of combining quantum mechanics and relativity is still open. Even in the greatly simplified case of special relativity (Galilean spacetime) no essential advancement beyond the good old perturbation theory has been observed in QFT. Despite many man-years of research, an operational non-perturbative control over nontrivial QFTs in the 4d spacetime still remains something of a dream for the physicists engaged in practical computations.

Quantum field theory, apart from being a very rich subject for physicists, has produced a profound impact on mathematics, but, curiously, not vice versa. One can easily list a number of great mathematicians that had tried to attack QFT from purely mathematical positions during the last 70+ years (e.g., constructive quantum field theory, axiomatic quantum field theory, algebraic quantum field theory, topological quantum field theory, K-theory, results on complex manifolds, affine Lie algebras, knots and some more piecemeal mathematical studies related to QFT). Still the main achievements in QFT have been provided by physicists, and QFT can still hardly be represented as a rigorous mathematical theory. I think that the problem with mathematical attacks on QFT lies in the difficulty to connect isolated major results (such as e.g., handling the $1/r$ singularity) in pre-string physics with the application of now prevailing new geometrical ideas, e.g., those leading to string theories and supersymmetry.

On a somewhat primitive level, the traditional (pre-string) quantum field theory may be reduced to the quantization of the harmonic oscillator, no matter how many high-brow mathematical sentences are attracted to wrap this fact. At least, the harmonic oscillator has always been the dominating mathematical model in quantum field theory. This is fully justified, if we recall how the correspondence between fields and particles has been traditionally established in QFT. Each normal mode of the field oscillations is interpreted as a particle, with the quantum number $n$ of a normal mode being thought of as the number of particles. A normal mode itself is understood as a harmonic oscillator, with the energy associated with the field excitation, $\varepsilon_n = (n + 1/2)\hbar\omega$, corresponding to the 1D oscillator model which immediately appears as soon as the primary field equations - the Maxwell equations - are Fourier transformed: each Fourier component of the vector potential must satisfy the oscillator equation. Although this is quite obvious, let us see how it happens. The Maxwell equations (ME) read (see Chapter 5):

$$\varepsilon_{\mu\nu\sigma\tau}\partial^\nu F^{\sigma\tau}(x) = 0 \qquad\qquad (6.17)$$

the homogeneous Maxwell equations (HME) and

$$\partial_\mu F^{\mu\nu}(x) = \frac{4\pi}{c} j^\nu(x) \tag{6.18}$$

the nonhomogeneous Maxwell equations (NME), $x = (x^0, x^i) = (ct, \mathbf{r}), i 0 1, 2, 3.$

If we choose the Lorenz gauge $\partial_\mu A^\mu(x) = 0$, then we get from NME the inhomogeneous wave equation

$$\Box A^\mu(x) = \frac{4\pi}{c} j^\mu(x), \tag{6.19}$$

where the D'Alembert operator $\Box \equiv \partial_\mu \partial^\mu = \partial_0^2 - \Delta = \frac{1}{c^2}\frac{\partial^2}{\partial t^2} - \Delta$. One may note that the $\Box$ operator is often defined with the inverse sign, so one should be attentive in the calculation of fields, see Chapter 5 for more details.

This relation may be interpreted as the motion equation for the electromagnetic field; mathematically it is the Klein-Gordon equation with zero mass and an external source term proportional to the charge current. If we make here the Fourier transform over spatial coordinates, we obtain the oscillator equations for each spatial Fourier component $A^\mu(t, \mathbf{k})$:

$$\ddot{A}^\mu + c^2 k^2 A^\mu = \frac{4\pi}{c} j^\mu(t, \mathbf{k}) \tag{6.20}$$

It means that the Fourier decomposition leads to the representation of fields as infinite sums of harmonic oscillators - one for each wave vector $k = |\mathbf{k}|$ (here, to simplify notations I denote by $k$ the absolute value of a 3-vector). The magnitude of $\mathbf{k}$ determines the oscillator frequency. See the classical treatment of this subject (expansion of electromagnetic fields on oscillators) in the textbook of Landau and Lifshitz [39], page 52. Each normal mode itself is thought of as a harmonic oscillator, and the energy of a field excitation is $\varepsilon_n = (n + 1/2)\hbar\omega$, i.e., corresponds to a $1D$ harmonic oscillator. An important thing here is that the Maxwell equations already dictate the correspondence between fields and particles invoking the necessity of the harmonic oscillator model, which is a rather happy coincidence, because the oscillator model is really unique.

Why is it unique? We may remember that the spectrum of a harmonic oscillator has an especially simple structure: the ground state (vacuum) energy is $\hbar\omega/2$ and all the above levels are equidistantly spaced, with the energy difference $\hbar\omega$ between them. States having other energies are not allowed. Such spectrum structure can be viewed as consisting of $n$ particles (or quasiparticles), each having the energy $\hbar\omega$. This is a natural interpretation since each normal mode of field oscillations behaves as a free particle - it has energy, momentum, can propagate, collide, be identified as a separate entity. Thus, the quantum number $n$ of each normal mode is interpreted as the number of particles. Indeed, when excited to the $n$-th energy state, the oscillator contains $n$ identical quanta (photons, phonons, etc.) of the **same**

energy - an interpretation which is possible only due to the equidistant spacing of the oscillator spectral levels. In such cases, it is natural to introduce raising and decreasing operators. One may note that a similar situation is encountered in the theory of angular momentum where the eigenvalue $M$ of the operator $L_z$ may take only equidistant values, $M = -L, \ldots, L$ corresponding to the eigenfunctions of the $SO(2)$ type, $\exp(im\varphi)$, where $L(L + 1)$ determines the eigenvalues of the angular momentum. One may also note that the raising and decreasing operators in the angular momentum theory are acting on functions used in the Fourier transform. In the oscillator case, the raising and decreasing operators are also connected with the Fourier transform, e.g., the raising operator happens to coincide with the negative-frequency part of the Fourier transformed vector potential and the decreasing operator turns out to be its positive-frequency part. These operators, producing the next or the previous eigenstates in the equidistant energy ladder, may be readily interpreted as "creating" or "annihilating" the particle with energy $\hbar\omega$. We shall return to the creation and annihilation operators in the following section, now I shall remind us of some basic mathematical facts about the oscillator model.

What is the linear (harmonic) oscillator from the mathematical viewpoint? In the traditional spectral theory - Sturm-Liouville - language, the linear oscillator model is connected with the differential operator

$$L = -\frac{d^2}{dx^2} + x^2, x \in (-\infty, \infty) \tag{6.21}$$

(for some transparency, we use the dimensionless variables here). This is a simple form of a more general Schrödinger operator:

$$L = -\frac{d^2}{dx^2} + q(x), q(x) \geq 0 \tag{6.22}$$

Obviously, we can replace the requirement of non-negativity of the function $q(x)$ by the condition of semi-boundedness from below.

We have already discussed the harmonic oscillator in various contexts and probably return to this model in future - the oscillator model is one of the main physmatical nodes. I think it is important to notice once again that the quantum description of the free electromagnetic field automatically leads to photons - due to the combination of the wavelike structure of the Maxwell equations with the equidistancy of the oscillator spectrum. This fact appears to be undeserved luck. If the harmonic oscillator had been substituted by some physical system whose mathematical model had not produced equal spacings between energy eigenvalues or, due to some curious properties of our spacetime (e.g., endowed by a Euclidean structure instead of affine so that the world would have had a center), the free Maxwell equations could not be reduced to the system of independent wave equations, $p_\mu p^\mu A_\nu = 0, p_\mu = -i\nabla_\mu = -i\partial/\partial x_\mu$, one still might have defined the creation and annihilation operators, but each new particle would have had the properties, e.g., mass or energy, depending on all other particles within the considered system. Thus, it would be difficult to add a particle identical to those already present - e.g., all the photons would be different, which would have made the laser impossible. Analogous difficulties would have accompanied the annihilation process one could not have destroyed a particle leaving all others intact. In principle, such effects could exist, at least on the Planck scale or, on the contrary, in large-scale physics, where deviations from the classical Maxwell's field produce equations noticeably different from the classical wave equation and, consequently, would not lead to the oscillator model. As far as the Planck scale goes, this is a pure speculation, since physics should be totally different on Planck length scales ($l_p \approx 1.6 \cdot 10^{-33}$ cm) at Planck energies, $E_p = \hbar/t_p = (\hbar c^5/G) \approx 1.22 \cdot 10^{19}$GeV $\approx 2 \cdot 10^9$J , where $t_p = \hbar/m_p c^2, m_p = \hbar c/G$ and the gravity constant $G = 6.674 \cdot 10^{-8} cm^3 g^{-1} s^{-2}$ , but an extension of the Maxwell field quantization on cosmological scales, when the background spacetime is no more flat, thus producing photon-like particles may remain a task for future research.

### 6.18.6    Canonical quantization in QED

It is worth reflecting on the term "canonical quantization". This term was probably first coined by Pascual Jordan, one of the principal creators of mathematical - mostly algebraic - tools of quantum mechanics (see Chapter 3). Recall that the symplectic structure of classical mechanics is also called the canonical structure. In this context, the term "canonical" simply means that one can introduce a certain skew-symmetric binary operation, called Poisson brackets (see Chapter 3), between the dynamical variables specifying the state in classical mechanics [84]. Each transformation between dynamical variables such as coordinates $x^i$ and momenta $p_j$ leaving their Poisson brackets intact is known as a canonical transformation in classical mechanics.

One might notice that using in parallel canonical quantization in QED and path-integral quantization in more recent quantum field theories (such as quantum chromodynamics - QCD) seemingly contradicts not only didactic principles, but also scientific norms which would require applying the same universal techniques in all theories. One can remark, however, that canonical quantization has obvious shortcomings which probably motivated R. Feynman to invent other approaches to field quantization in QED such as diagram techniques and path integrals. For instance, canonical quantization conceals an explicit Lorentz invariance. No wonder, since canonical quantization was originally invented as an essentially nonrelativistic technique.

### 6.18.7    Gauge Fields in Quantum Theory

When discussing the classical field theory in Chapter 5, we have seen that the Maxwell equations are invariant under the gauge transformations, $A_\mu \to A'_\mu = A_\mu + \partial_\mu \chi$ where $\chi$ is the scalar gauge function. Physicists usually say that classical electromagnetic theory is the $U(1)$ gauge theory meaning that it is described by the $U(1)$ gauge group which is one-dimensional and Abelian. We have also seen that gauge invariance in classical field theory is closely connected with the concept of charge conservation. In the quantum world, gauge invariance plays an even more important role. I have already mentioned that quantum electrodynamics (QED) is a particular case of a field theory, and as such it also represents a whole class of theories called the gauge theories. One can observe certain hints at gauge ideas already in classical electrodynamics where one of the Maxwell equations ($div\mathbf{E} = 4\pi\rho$) does not contain time derivatives and is therefore an "instant" relation. It follows from this equation that the electric field would change at once in the whole space, which would break causality, unless the charge is not conserved. Here, one can already see a bridge to gauge theories.

The story of gauge fields usually clusters around the Yang-Mills equations, although it seems to be more confused and fascinating. It is interesting, by the way, that the Yang-Mills theory can be in principle discussed before the introduction of any quantum field concepts.

Let us try to understand the main ideas of gauge fields in simple language. All the particles of a given type may be treated as manifestations of some quantum field which is characterized by the magnitude and the phase at each point of spacetime. This picture evokes the image of an ocean, with individual particles being similar to waves: the real thing is the field, and waves are just observable excitations over some average level. Then each excitation viewed as a particle has an attached phase i.e., a cyclic angle variable which may have (and actually has) a local character: the phase should be updated as one moves from one observation point to another.

# 7 Stochastic Reality

In this chapter, we shall deal with the systems whose state cannot be characterized exactly, as one used to describe classical deterministic systems of mechanics (Chapter 4). Systems we are going to observe here can be closed or open, they may consist of many elements or of a small number of particles, but the unifying characteristic of them is as follows: they admit only the probabilistic description. We shall mainly discuss macroscopic bodies, i.e., those consisting of an enormous number of particles (in physics typically of the order of $10^{24}$). Properties of such macroscopic bodies are to a large extent determined by the collective behavior of constituent particles and may be almost insensitive to details of individual interparticle coupling, like the crowd behavior practically does not depend on individual human features or attitudes. This means that laws of a different type arise, as compared to dynamical laws of, e.g., classical mechanics. These laws are usually called statistical (or, in some specific cases, stochastic) emphasizing the fact that they are based on the notion of probability.

The concept of probability has a variety of interpretations, and it may have a totally different meaning in a different context. For example, when reality occurs to be intrinsically random as in simple dice throwing, classical probabilities might be predicted from symmetry considerations. On the other hand, quantum paths (see Chapter 6) and in general predictions of future behavior is based on more complicated structural arguments. One may notice that different theories and modeling structures require diverse notions of probability. Since atoms and various atomic microparticles are subordinated to the rules of quantum mechanics, the concept of probability lies in the very nature of things.

The text of this chapter is divided into six sections. The first section reiterates basic facts on the dynamics of finite particle systems: classical Hamiltonian mechanics, the Liouville equation, and the evolution operator. In the second section the Bogoliubov equations are introduced and the Cauchy problem for them is formally solved. Next, two sections follow on equilibrium states of systems of infinitely many particles in the frame of the canonical and the grand canonical ensembles. The main subjects are here the existence and the uniqueness of limit distribution functions as well as spectral and topological properties of evolution operators. The fifth section treats the thermodynamic limit for non-equilibrium systems. This is possible, however, only for spatially homogeneous systems, the whole field of transport processes as a paradigm for non-equilibrium systems being in general outside the scope of the book. Some concepts of statistical theory of open systems are reviewed here. Finally, the notion of deterministic chaos is discussed, and its implications are briefly overviewed.

## 7.1  Thermodynamics: the Study of Paradoxes

The study of statistical systems, i.e., consisting of a large number of particles, starts with the question: what are the right variables to describe the behavior of statistical systems? Thermodynamics provides the simplest answer to this question.

I have always perceived the second law of thermodynamics as a frustrating paradox: on the one hand, all the systems should tend to the morgue state of absolute equilibrium while on the other hand there are plenty of examples in nature demonstrating in no way the continuous disorganization or progressive decay, but just the opposite, evolution and self-organization. All physicists seem to be aware of this paradox, sometimes formulated as the "heat death of the universe", but probably each physicist has her/his own version of treating it (if one thinks at all about such highly theoretical problems - or pseudo-problems). The matter is that if the universe (or multiverse) has existed for a very long time, in some models infinitely long, then it must have reached the equilibrium state. Nevertheless, the world as a whole as well as its separate subsystems are apparently in a state rather far from thermal equilibrium, and one cannot observe the processes that would bring the universe into the equilibrium state. One possible solution to this paradox is that statistical physics and thermodynamics cannot be applied to the world as a whole. (Why then can quantum mechanics be applied?) Probably, this inapplicability is related to the fundamental role played by gravity forces, for which even the Gibbs distribution can be introduced only with certain difficulties due to emerging divergences[169]. However, one cannot be sure that the declaration of non-validity of the usual statistical physics and thermodynamics for gravitating systems is an ample explanation of the absence of equilibrium state in the universe. Another popular explanation of the heat death paradox consists in the consideration that the metric of general relativity depends on time and therefore the universe is submerged into a non-stationary field [24].

The real implication of the "heat death" paradox is that, besides asking a generally interesting though a typically scholastic question, one can easily transgress the limits of applicability of classical thermodynamics, without even noticing it. The problem with thermodynamics is that it tends to be extrapolated over the areas where it cannot be applied. Strictly speaking, classical thermodynamics is valid only for equilibrium states of the explored systems. However, the equilibrium conditions are only an idealization, a mathematical model and can be seldom encountered in nature. Thermodynamics is in essence a branch of physics studying the collective properties of complex steady-state systems. The main part of thermodynamics is purely phenomenological and may be treated completely independently from mainstream - microscopic - physics. Phenomenological thermodynamics historically served the industrial revolution, mostly driven by the steam engines. These alluring machines produced a euphoria in the then existing society, similar to that created by today's digital gadgets. Steam machines fascinated not only the engineers, even such outstanding scientists, mainly physicists, as N. S. Carnot, E. Clapeyron, R. Clausius, J. B. J. Fourier, H. Helmholtz, J. Joule, Lord Kelvin (W. Thomson), J. C. Maxwell who were trying to formulate the laws governing the work of heat engines in a mathematical form. The most challenging task was to correctly describe the conversion of heat into mechanical work. Later L. Boltzmann, J. W. Gibbs, and M. Planck provided links from thermodynamics to the major body of physics.

---

[169] Free energy F/N counted per one particle diverges in the presence of gravitational interaction due to the long-range character of the gravitational forces (see, e.g., [220]). Moreover, this quantity diverges also on small distances, but one can get rid of these divergences by restricting the possibility for the particles to condense, e.g., by introducing the exclusion principle at zero separation. Long-range forces imply a strong coupling of distant domains of the system, even widely separated. It is interesting to notice that similar divergences could be observed in the presence of Coulomb interaction, but they are removed due to the neutrality of Coulomb systems containing charges of opposite sign. This is impossible for gravity because mass is always positive.

Although time is not included in the set of thermodynamical variables, classical thermodynamics states that we live in a universe that becomes more and more disordered with time. The second law of thermodynamics, despite its apparently soft - statistical - character, is one of the hardest constraints ever imposed by the science on life and technology. By the word "soft" I denote the tendency to verbal formulations of the second law of thermodynamics. Historically starting from heat engines, classical thermodynamics still investigates the processes when energy, heat, and mass can be exchanged. There are two main laws governing such processes. While the first law of thermodynamics postulates the strict energy balance, the second law states that for all thermodynamic processes some portion of energy is necessarily converted into heat and dissipates in the environment. Such dissipation is irreversible: to undo it, one must spend some energy. Thus, losses are unavoidable. This is the informal gist of the second law of thermodynamics (without introducing a rather counterintuitive concept of entropy). Then there are some grave consequences of the second law such as the limited efficiency of heat engines or irreversibility of erasing the data in computer memory: this latter process generates heat [221].

## 7.2    Statistical Way of Thinking

The thermodynamical - in fact thermostatistical - approach to many-particle systems is simple but unsatisfactory because the crucial question: "how can the values of thermodynamical variables be derived from the equations of mechanics, classical or quantum?" is irrelevant. Thermodynamics was constructed as a totally independent discipline, an engineering isle whose study does not imply any knowledge of physics. Quite naturally, physicists were frustrated by such isolated standing of thermodynamics, and the first expansion of physicists into the field of thermodynamics was in the form of attempts to interpret such quantities as heat, work, free energy, entropy on a level of individual moving molecules.

Derivations of thermodynamical quantities from molecular models of physics are not at all easy, even for such a comparatively simple system made up of many particles as a rarefied gas. Three great physicists, all of them very proficient in mathematics, were the first to apply probabilistic concepts and, accordingly, to use the term "statistical" in the context of thermodynamics. Those were (in chronological order) J. C. Maxwell, L. Boltzmann, J. W. Gibbs.

Due to a greatly reduced number of variables, the behavior of statistical systems may look simple, but this is deceptive. Statistical irreversibility is a clear symptom of serious difficulties.

The apparent time reversibility of Newtonian dynamics formally allows us to observe many dramatic events such as the gas leaving the whole volume empty (say $1\ m^3$) it had just occupied. Time reversibility simply requires the gas molecules to follow trajectories back in time in the phase space (see Chapter 9 on the "Time Arrow"). However, nobody has ever observed such events, at least we believe so. The matter is that these time reversed paths as compared to "direct", those that result in filling up a prepared vacuum, though in principle possible, are characterized by such a tiny probability as to render any observable macroscopic effect unfeasible. Similarly, we would not expect a spontaneous heating of some macroscopic body being initially at room temperature (even though there are not so few people who continue designing engines working on the principle of heat transfer from cold to warm bodies). We could perhaps expect this to occur in a nanoscopic system consisting of a small number ($N \leq 10$) of molecules - for a short period of time, but the experience shows that a spontaneous heating or cooling of a considerable piece of matter ($N \gg 1$), which would persist for a long period, e.g., sufficient to produce mechanical work, is considered an abnormal state and has never been observed. Such phenomena are of a fluctuative nature and can be modeled mathematically using an assortment of fluctuation theorems [161-163] (Jarzynski, Wojcik, Evans, Searles).

One might ask, why should we be concerned with rare events that can happen only on molecular time and distance scales and become almost totally improbable when we try to observe macroscopic phenomena? The answer is that these rare microscopic events are important for understanding the whole statistical thermodynamics. Besides, new experimental techniques have made microscopic and nanoscopic scales accessible to observation so that fluctuations and their probabilities which had been irrelevant before are becoming important for understanding experiments.

The foundations of statistical mechanics are a topic which has been in the focus of scientific interest since the times of Boltzmann. One distinct problem, namely the foundation of classical statistical mechanics of continuous systems, i.e., of systems of many particles moving in a continuous phase space as opposed to lattice systems, is touched upon in this book. The treatment method consists in the solution of the Bogoliubov (BBGKY - Bogoliubov, Born, Green, Kirkwood, Yvon) equations, an infinite system of integro-differential equations for the many-particle distribution functions.

The field of statistical continuum mechanics, viz. the statistical theory of continuous media, is not touched here.

Unfortunately, recent developments of chaos theory and ergodic theory so important for the foundations of statistical mechanics are taken into account only superficially. Likewise, other competing methodologies, e.g., of the Prigogine school, modern projection operator techniques or the information theoretical approach are mentioned only briefly.

## 7.3   Statistical Equilibrium

In the foundation of equilibrium statistical mechanics lies some formal expression called Hamiltonian (see Chapter 4), which must help us to find the probability distribution for any random physical process - and most physical processes are in fact random. For example, gas in some closed volume gives an elementary example of a random physical process; there may be less trivial systems, for instance, with interaction between the constituents. Since no physical system, probably except the entire universe, is isolated from the environment, one must account for its influence, e.g., by introducing some "physically natural" supplementary conditions. For example, initial or boundary values must be fixed, subjugation to which turns the physical system into a model.

We see that there exists a natural hierarchy of time scales (already mentioned in association with the hierarchical multilevel principle of mathematical modeling), this fact playing a crucial part in statistical mechanics as it determines the behavior of distribution and correlation functions. In particular, after time $\tau_0$, correlations between the particles are drastically weakened and many-particle distribution functions turn into the product of single-particle distributions $f(\mathbf{r}, \mathbf{p}, t)$ (the principle of correlation weakening was introduced by N. N. Bogoliubov) whereas for $t > \tau_r \gg \tau_0$, a single-particle distribution function tends to the equilibrium Maxwell-Boltzmann distribution[170]. In other words, the whole many-body system for $t \gg \tau_r$ reaches the state of statistical equilibrium so that the respective many-particle distribution functions tend to the canonical Gibbs distribution, $f_N(x_1, \ldots, x_N, t) \rightarrow \exp[\beta(F - H(x_1, \ldots, x_N))]$, where $\beta = 1/T$ is the inverse temperature, $F$ is the free energy and $H(x_1, \ldots, x_N)$ is the system's Hamiltonian. In other words, the single-particle distribution function $f(\mathbf{r}, \mathbf{p}, t)$ substantially changes over time scales $t \tau \tau_r \gg \tau_0$ whereas at the initial stage of evolution this function remains practically intact. Yet many-particle distribution functions can change very rapidly at short times comparable with the chaotization period $\tau_0$. Physically, one can understood this fact by considering spatially uniform systems with pair interaction between the particles, when many-body distribution functions depend on the coordinate differences of rapidly moving constituents. It is intuitively plausible that many-particle distribution functions would adjust to instant values of a single-particle distribution. To translate this intuitive consideration into the mathematical language one can say that for $\tau_r > t \gg$ $\tau_0$ (intermediate asymptotics) many-particle distribution functions become the functionals of a single-particle distribution function

$$f_N(\,x_1, \dots,\, x_N, t) \xrightarrow{t \gg \tau_0} f_N[\,x_1, \dots,\, x_N; f(\mathbf{r}, \mathbf{p}, t)]$$

so that the temporal dependence of many-particle distributions is now determined by the single-particle function.

---

[170] One can imagine even tinier time scales in a many-body system, namely $\tau_0 \sim \tau_r/N$, where $N$ is the number of particles in the considered part of the system.

This idea (expressed by N. N. Bogoliubov) is rather important since it leads to a drastic simplification of the models describing many-body systems. In particular, although the respective distribution functions formally depend on initial data for all the particles, after a rather short time ($\tau_0 \sim 10^{-12} - 10^{-13}$ s) this dependence becomes much simpler since its relics are only retained in the relatively smooth single-particle function $f(\mathbf{r}, \mathbf{p}, t)$. One usually applies to this situation the notion of "erased memory" designating asymptotic independence of many-particle distribution functions on precise values of initial data – a huge simplification since initial values of all coordinates and momenta are never known exactly and, even if known, would be completely useless. In the modern world, the idea of fast forgotten details of microscopic instances, in fact of insensitivity to microscopics, has become especially useful for mathematical modeling of complex systems.

## 7.4   Statistical Ensembles

Let us start by recapitulating the basic notions of the idea of statistical ensembles. In conventional statistical mechanics, there has been a strong expectation that an ensemble average can correctly describe the behavior of a single particular system. Despite numerous attempts, there seems to be no rigorous mathematical proof for applying statistical ensembles to an individual observed system. The ensemble methodology lying at the very foundation of statistical physics still has the status of a logical presumption rather than of a compelling mathematical fact. A variety of good samples for concrete ensembles can be satisfactory from the physical point of view but require a more ample treatment for a cultivated mathematical taste. In physical terms, the default of ensemble methodology would mean that some quantities for an observed system may substantially deviate from the ensemble averages. Nonetheless, it would not be a tragedy; on the contrary, such deviations can provide important physical information about the observed system.

In the model of a microcanonical ensemble, one basically considers an isolated system, which is in fact not very interesting from a practical point of view. Explaining this requires a reminder of basic thermodynamics, which probabilists call large deviation theory. In the microcanonical ensemble one considers the number of microstates of an isolated system at some fixed value of internal energy $U$, volume $V$ and other extensive conserved quantities. The logarithm of this number of microstates is the entropy $S$ (it is convenient to set the Boltzmann constant $k_B = 1$) and by inverting this function one obtains the internal energy $U(S, V, \dots)$. From this so-called thermodynamic potential all other thermodynamic quantities (temperature, pressure, heat capacity and so on) can be computed as a function of the extensive quantities.

A kinetic equation is a mathematical model allowing one to find (approximately!) the distribution function for the statistical ensemble.

## 7.5    The Bogoliubov Chain

In this section, we are going to study mainly the stationary solutions of the Bogoliubov hierarchy equations. In literature this chain of equations - an infinite system of integro-differential equations for the many-particle distribution functions - is generally known as Bogoliubov-Born-Green-Kirkwood-Yvon (BBGKY) hierarchy. For systems in a finite volume these equations are equivalent to the dynamical Liouville equation and characterize the time evolution of the probability measure on the phase space of a finite number of particles. Performing the thermodynamic limit, one obtains the infinite chain of the Bogoliubov hierarchy equations which are related to a system of particles in the whole space. As near as I know, the problem of existence and uniqueness for this chain of equations has not been solved so far. It is natural to connect the stationary solutions of the Bogoliubov hierarchy equations with the states of an infinite system of particles (i.e., probability measures defined on the phase space) which are invariant with respect to time evolution. In the cases where the dynamics on the phase space has been constructed it is possible to demonstrate that any invariant measure satisfying further conditions of a general type generates a stationary solution of the Bogoliubov hierarchy equations. On the other hand, an immediate analysis of stationary solutions of the Bogoliubov hierarchy equations (unlike the invariant measures) does not require, in general, the use of such delicate dynamical properties as clustering. Apparently, the point is that only functions of a finite (although not bounded) number of variables enter the Bogoliubov hierarchy equations. One can consider these functions (the correlation functions) as integral characteristics of a measure and their behavior must not necessarily show the influence of singularities arising from the motion of individual configurations of an infinitely large number of particles. Thus, the approach based on the Bogoliubov hierarchy equations seems not only to be more general but also more natural from the physical point of view.

We shall also discuss the derivation of the kinetic equation for a classical system of hard spheres based on an infinite sequence of equations for distribution functions in the Bogoliubov (BBGKY) hierarchy case. It is known that the assumption of full synchronization of all distributions leads to certain problems in describing the tails of the autocorrelation functions and some other correlation effects with medium or high density. We shall discuss how to avoid these difficulties by maintaining the explicit form of time-dependent dynamic correlations in the BBGKY closure scheme.

The question usually is how to obtain hydrodynamic equations (Euler, Navier-Stokes) from the Liouville-type equations of Hamiltonian mechanics, classical or quantum. The original idea was due to Ch. Morrey (1956) who introduced a concept of a hydrodynamic limit and was able to formally derive an Euler equation from the classical Liouville equations (more precisely, from the corresponding BBGKY hierarchy). However, Morrey had to make some assumptions about the long-term behavior of the motion, and this included a statement on ergodicity, in the sense that all 'reasonable' first integrals are functions of the energy, linear momentum and the number of particles. Since then, the idea of a hydrodynamic limit became very popular in the literature and has been successfully applied to a variety of models of (mostly stochastic) dynamics relating them to non-linear equations. However, in the original problem there was no substantial progress until the work by S. Olla, S. R. S. Varadhan and T. Yau (1992) where Morrey's assumptions were replaced by introducing a small noise into the Hamiltonian (which effectively kills other integrals of motion), and a classical Euler equation was correctly derived. In some quantum models (e.g., of the Bohm-Madelung type) the hydrodynamic limit can be rigorously demonstrated. The resulting Euler-type equation is similar to the one that arises for the classical counterpart of these models. This suggests that perhaps classical and quantum hydrodynamic equations must look similar if they are written for local densities of 'canonical' conserved quantities (the density of mass, linear momentum and energy).

## 7.6   Chaotic Behavior

We have already discussed chaotic systems in Chapter 4, in connection with nonlinear dynamics and elementary stability theory, where "chaos" became the code word for nonlinear science. In that case, we can speak of deterministic chaos, a transition to randomness due to sensitive dependence on initial conditions – experimentally or computationally indistinguishable initial states eventually evolve to states that are far apart (in the phase space). Chaotic dynamics bridges regular evolution of complex systems with the random one.  However, here, in the spirit of stochastic reality, we shall put more accent on probabilistic features of chaotic behavior.

In Chapter 2 we have seen that the notion of time is crucial for dynamical systems, which evolve into the future given the data related to the actual state specified at some particular moment. One might have noticed the drastic difference between the steady-state situations, for instance Kepler's laws, and evolutionary processes: one is unable to determine the state or geometry of a dynamical system unless one stipulates some input data at a specified moment of time. To obtain the steady-state picture, e.g., Kepler's laws, one has to exclude time so, strictly speaking, no dynamics is left or it is hidden deeply enough to be unnoticed (for example, by ancient astronomers). The notion of time becomes relevant again only when one starts considering the stability of Kepler's static celestial geometry. We shall see below that this consideration leads to very powerful results such as the famous KAM (Kolmogorov-Arnold-Moser) theorem. One can, in principle, test all steady-state solutions on stability and even chaoticity; the elliptic trajectories of the planets in the Solar System may prove unstable or even chaotic after such an analysis. However, the perennial question is: what would be the timescale during which the instability or chaos might evolve? By the way, if and when such chaos evolves, the time-reversal invariance of the regular dynamical behavior of Newtonian gravitational motion and consequently Kepler's laws would be lost.

Let us, for simplicity, consider first classical statistical mechanics. In fact, a consistent exposition of statistical mechanics requires a quantum treatment anyway, but to elucidate the main points we shall for the time being stay with classical model. We have seen that classical statistical mechanics typically considers systems of $N$ structureless particles (material points) in a $d$-dimensional space moving along classical (Newtonian) trajectories in continuous time $t$. The number of particles is usually considered very large, but still finite so that the concept of the phase flow widely used in classical mechanics can be applied without excessive mathematical precautions. Recall that the term "flow" is usually understood in classical mechanics as a continuous one-parametric group of transformations of the phase space (on the phase manifold). In most classical problems of physics, parameter $d = 1,2,3$. Here, I would like to recall that in classical statistical mechanics the entire phase space of the many-particle system is called $\Gamma$-space, which contains $2Nd$ dimensions whereas the phase space of a single particle is commonly known as a $\mu$-space. So, we see that in classical mechanics and hence in classical statistical mechanics the dimensionality of the phase space and the number of degrees of freedom differ by the factor of 2 which is, of course, a trifle but still not very convenient from the positions of dynamical systems theory. Recall that in this latter case one usually does not discriminate coordinates and momenta, they are equal components of the vector function $u = (u^1, \ldots, u^n)$ whose evolution governed by the vector equation $du/dt = f(u, t)$ is considered (see Chapter 4). In statistical mechanics, $n = 2Nd$. This is of course trivial, but one must be careful.

Notice that even for the simplest possible statistical system of $N$ structureless material points some nontrivial questions arise, for instance, how can one reconcile the time-reversible equations of motion for a single particle with the obviously irreversible macroscopic properties of a many-particle system? This paradox, included in the collection of Perennial Problems, has been reflected in the metaphor of the "arrow of time" (see Chapter 9).

In quantum theory there exist some additional reasons for the breach of time-reversal symmetry, one of them being the issue of quantum measurement that is not fully clarified – at least within the unitary scheme of orthodox quantum mechanics. However, this issue is a special problem, and its discussion would require much space so that we shall not deal with the arrow of time in the present manuscript. Notice only that in dynamical systems theory, which can be considered as an intermediate between mechanics and macroscopic physics, the arrow of time i.e., the breakdown of symmetry between past and future appears quite naturally in view of instabilities and their limit case –chaotic behavior.

In the quantum world, the reason for the breach of time reversal symmetry is often ascribed to the process of quantum measurement, which we shall not consider here in detail. When one is solely focused on unitary quantum dynamics, one typically treats quantum evolution as completely time invertible. This is, however, a somewhat naive simplifying assumption since the Hilbert space where quantum dynamics occurs is not necessarily filled with complex conjugated states (of thermo-isolated systems).

In fluids, manifestations of stochasticity are generally called turbulence, in this case chaotic behavior corresponds to the so-called Lagrangian turbulence (see, e.g., [305]). It is, however, important that both the advection and, more generally, the fluid flow dynamics within the Lagrangian framework can, due to the possibility of representing them in the dynamical systems form, be treated as Hamiltonian systems. The role of the Hamiltonian in fluid dynamics is in the two-dimensional case played by the deterministic stream function $\Psi(\mathbf{x}_\parallel, t)$ ([306,307]), where $\mathbf{x}_\parallel = (x^1, x^2) \in \Omega_2$ . In a particular case when the 2d domain $\Omega_2$ occupied by the flow lies in the Euclidean plane with Cartesian coordinates $x^1 = x, x^2 = y$ on it, we have the Hamiltonian (symplectic) form of the motion equations $\dot{x} = v_x(x, y, t) = \partial\Psi/\partial y, \dot{y} = v_y(x, y, t) = -\partial\Psi/\partial x$. Here the domain $\Omega_2 \equiv \Omega_{2+1}$ of variables $(x, y, t)$ corresponds to the extended phase space. In other words, the nonstationary planar flow is described by the Hamiltonian dynamical system with 1.5 degrees of freedom. Notice, however, that the stream function $\Psi(x, y, t)$ is a somewhat strange Hamiltonian: a pair of "canonical" variables $(x, y)$ are just local coordinates, and they are not canonically conjugate as, e.g., coordinate and momentum in mechanics.

# 8 Radiation and Matter

This chapter is devoted to a loose description of theoretical approaches to the interaction of radiation, both electromagnetic and corpuscular, with matter. Other possible types of radiation are not considered in this chapter (and in the book in general). The term "matter" implies in the context of the present chapter a macroscopic (averaged) ensemble of point electrons and nuclei combined, which constitute the medium - it may be, for example, gas, plasma or solid state. Averaging, to be discussed below, occurs at distances substantially larger than interparticle ($n^{-1/3}$) or atomic ones ($\cong 10^{-8}$ so that nuclei, electrons and other atomic particles can really be regarded as point-like. Thus, the described theory is inapplicable at smaller distances. I shall generally regard the matter as being non-relativistic, only some restrained comments will be made about relativistic effects in connection to the passage of fast charged particles through the matter. Interaction of hard gamma radiation ($\hbar\omega \geq 2m_e c^2, m_e = 9.1 \cdot 10^{-28}$g is the electron mass) with matter is not treated in this book, therefore creation of particles (pair production) is ignored. To limit the scope of the book as well as for simplicity sake, I also do not discuss creation of pairs by superstrong electromagnetic (laser) fields, although such effects may be important considering the power densities achieved today ( $P > 10^{21} W/cm^2$ , see, e.g., http://space.newscientist.com/article/dn13634-powerful-laser-is-brightestlight-in-the-universe.html). In general, relativistic quantum phenomena typically studied in quantum field theory are largely ignored in this chapter. Here, the matter is described by the means of standard non-relativistic quantum mechanics, while the electromagnetic field is treated classically using the Maxwell equations (without second quantization, see Chapters 5, 6). Such an approach is usually called semiclassical.

The chapter is divided into two parts, one related to the electromagnetic field (mainly radiation) in material media, the other to the passage of particles (mostly charged ones) through the matter. Since the treatment of the interaction of corpuscular radiation with matter to a large extent uses the concepts developed in electromagnetic (EM) theory, I decided to put first an overview of the interaction of the electromagnetic field with material media.

## 8.1 Interaction of Electromagnetic Radiation with Matter. General Concepts

A general description of mechanisms underlying the interaction of electromagnetic radiation with matter is based on the concept of electromagnetic response and is performed, respectively, in terms of response functions. The notion of response functions generalizes the description of electromagnetic response in terms of susceptibilities (both linear and nonlinear), permittivity, permeability, conductivity, Coulomb screening, mean and local field, etc., all listed notions being specific cases of electromagnetic response functions. The quantum mechanical description of matter allows one to construct a linear theory which, assuming that spectral properties of the matter are known, enables us to determine its response to an external field (e.g., in terms of susceptibilities, even nonlinear ones). On the base of such a theory, simple models using specific examples of the media can be constructed and some primitive representations of otherwise complex systems (gas, plasma, solid, liquid, etc.) can be discussed and tested.

Nevertheless, one cannot imagine that the problem of field-matter interaction can be easily decomposed into two subproblems, one related to the matter and the other to the field. The response of the matter to an electromagnetic field is usually a self-consistence problem, which implies both the effect of the field on the matter and a modification of the external field by the matter. The self-consistence approach is widely used in macroscopical electrodynamics and optics (both linear and nonlinear), see Chapter 5. In these disciplines the so-called material constants are phenomenologically introduced. Ideally, these phenomenological material constants - dielectric permittivity, magnetic permeability $\mu$[171], conductivity $\sigma$, refractive index $n$, susceptibilities $\chi^{(1)}$ (linear) and $\chi^{(k)}$, $k = 2,3, ...$ - should be determined as functions of the fundamental constants alone: the mass $m_e$ and charge $e$ of the electron, the velocity of light $c$, Planck constant $\hbar$, atomic numbers $Z, N$, and atomic mass $M$. In the statistical description of matter, the temperature $T$ is also an essential parameter. In the energy scale, however, $T$ is a small parameter[172] so that the factor $\exp(-\hbar\omega/T)$, where $\hbar\omega = E_n - E_m$, $E_n$, $E_m$ - atomic energy levels, arising from the averaging over the Gibbs ensemble (see below; see also Chapter 7) is often quite small.

We have already discussed in Chapter 5 the description of the electromagnetic (EM) field within the framework of classical theory. Exact values of electric $\mathbf{E}(\mathbf{r}, t)$ and magnetic $\mathbf{H}(\mathbf{r}, t)$ components of the EM field at some space-time point $(t, r) = (x^0, x^1, x^2, x^3)$ satisfy the Maxwell equations (see Chapter 5), where inhomogeneous terms proportional to charge and current densities,

$$l\rho(\mathbf{r}, t) = \sum_a e_a \delta(\mathbf{r} - \mathbf{r}_a(t)) \tag{8.1}$$

$$\mathbf{j}(\mathbf{r}, t) = \sum_a e_a \mathbf{v}_a \delta(\mathbf{r} - \mathbf{r}_a(t)), \tag{8.2}$$

---

[171] Magnetic permeability is not necessary when a spatial dispersion is taken into account, see below. It is interesting that the term "magnetic permeability" was probably introduced by O. Heaviside.

[172] In this book, I shall measure the temperature in energy units so that, e.g., room temperature, $T \sim 10^2 K$ , would correspond to $T \sim 10^{-14}$erg$\sim 10^{-2}$eV $\ll m_e e^4/\hbar^2 \approx 4.3 \cdot 10^{-11}$erg $\approx 27.2$eV, where $m_e e^4/\hbar^2$ is the characteristic atomic energy value constructed from fundamental constants. The physical meaning of this inequality is that a typical quantum system at room temperature remains in the ground state since thermal excitation is insufficient to overcome the spacing between atomic energy levels. This weakness of thermal excitation is, however, not true for quantum systems with a dense energy spectrum such as complex molecules. Furthermore, in a hot plasma $T$ is often higher than $m_e e^4/\hbar^2$.

where $\mathbf{v}_a = \dot{\mathbf{r}}_a$. These densities automatically obey the continuity equation

$$\frac{\partial \rho}{\partial t} + div\,\mathbf{j} = 0. \tag{8.3}$$

We have already discussed the continuity equation and its meaning as the conservation law. Recall that in a compact domain the continuity equation for some quantity[173] states that the amount of this quantity in the considered domain varies to the extent of the quantity gain or loss through the domain boundaries. In the case of charge conservation, however, the meaning of the continuity equation is much deeper: it is connected with gauge invariance (see Chapter 5). This invariance manifests itself, in particular, by the fact that the continuity equation for electrical charges and currents is not independent of the Maxwell equations (see [39], §29). One can even say that the continuity equation for electrical charges and currents is a trivial mathematical consequence of the Maxwell equations; a more accurate statement would be that the Maxwell equations are adjusted to or agreed with the continuity equation, that is to say they are constructed in such a way as to automatically provide charge conservation.

A model of some experimentally produced setting in radiation-matter interaction, accounting for quantum effects, would consist in the treatment of a quantized electromagnetic field created by atomic sources and propagating in the presence of macroscopic bodies or surfaces. For example, such a scheme can be applied for modeling typical quantum-optical experiments when radiation passes through various optical instruments. The latter may be active (i.e., nonlinear) or passive, anyway in the standard approach they may be regarded as dielectric bodies or surfaces having a certain type of geometric or physical complexity. This complexity can be described in mathematical terms and is typically accounted for by introducing the dielectric function that depends both on radiation features, e.g., frequency and on spatial coordinates. In spite of many attempts to construct a fully quantum theory of the medium response to an electromagnetic field (see, e.g., one of the latest papers [260], see also below), the approach based on the dielectric function is by necessity semiclassical, being heavily based on the classical concept of the field in a macroscopic medium. One may remark in passing that, e.g., nonlinear optics in general seems to be an essentially classical (or semiclassical) discipline because the coupled evolution equations for the light and matter fields are extremely difficult - and perhaps unnecessary - to solve. Classical electrodynamics of polarizable media seems to be quite adequate for contemporary nonlinear optics, at least for its applied part.

---

[173] This quantity can be of arbitrary nature, not only the charge but, e.g., the number of cars on some part of the road.

## 8.2   Field Energy Dissipation in Matter

$$\frac{dQ}{dt} = \frac{\omega}{8\pi} \epsilon''_{ij} E^{*i} E^j \tag{8.4}$$

We can illustrate this general formula using a simple model of the field energy absorption by an electron driven by the field of an electromagnetic wave and from time to time colliding with other particles of the medium[174]. We shall assume that such collisions are only elastic i.e., the total kinetic energy (treated as a quadratic form) of colliding particles is conserved in each individual collision and no excitation occurs. In other words, it is assumed that energy is not dissipated into internal degrees of freedom. Only the momentum of the colliding particles changes, with such changes taking place abruptly i.e., the collision time $\tau \ll 2\pi/\omega$, where $\omega$ is the characteristic frequency of the electromagnetic wave (actually, we may assume $\omega\tau \ll 1$). The assumption of negligibly short collision time is standard, but not always correct. When the electromagnetic field cannot be conveniently represented by a quasimonochromatic wave centered around a characteristic frequency (for example, the particle is driven by the field of an ultrashort laser pulse), then the assumption of abrupt collisions looks as $\tau \ll \tau_p$, where $\tau_p$ is the characteristic time of pulse growth. Imagine for simplicity that electrons collide only with the heavy centers of the medium namely atoms, ions, or molecules whose masses $M_a$ are much larger than the electron mass $m$. Usually, the parameter $m/M_a$ is very small, $10^{-3} - 10^{-4}$ which allows one to ignore the motion or recoil of heavy scatterers with high accuracy. Then, since collisions are assumed elastic, the energy of the electron in an individual collision is conserved. Energy of the field is dissipated as an averaged stochastic process due to multiple elastic collisions, this dissipation leading to a gradual increase of the mean energy of electrons i.e., to the heating of the electronic subsystem.

For an electron moving in a high-frequency harmonic field $\mathbf{E} = \mathbf{E}_0 \cos \omega t$, we have $\dot{\mathbf{p}} = \mathbf{E}_0 \cos \omega t$ so that

$$p(t) = p_0 + \frac{eE_0}{\omega} \sin \omega t \tag{8.5}$$

and the energy instant value in the field is

$$\mathcal{E}(t) = \frac{\mathbf{p}^2(t)}{2m} = \frac{1}{2m}\left( \mathbf{p}_0^2 + \frac{2e}{\omega}\left( \mathbf{p}_0\mathbf{E}_0 \sin \omega t + \frac{e^2}{\omega^2} \mathbf{E}_0^2 \sin^2 \omega t \right) \right).$$

This instant energy oscillates near its mean value, which reflects the fact that the particle exchanges electromagnetic energy with the field. In the quantum language, this corresponds to constantly exchanging photons, some of them being real other virtual. In a high-frequency harmonic field, we are interested of course not in the instant value of the particle energy, but in its energy averaged over many field periods. This average energy is given by

$$\overline{\mathcal{E}_-} = \frac{\mathbf{p}_0^2(t)}{2m} + \frac{e^2\mathbf{E}_0^2}{4m\omega^2} = \mathcal{E}_0 + T_p,$$

where $\mathcal{E}_0, \mathbf{p}_0$ are the initial energy and momentum of the electron, $T_p = e^2E^2/4m\omega^2$ is its average kinetic energy in the oscillatory motion (see, e.g., [23], 30]; here lower index $p$ stands for "ponderomotive", see the next section. The minus sign at the energy symbol signifies "energy before collision". To simplify the model, we shall neglect small-angle scattering, assuming all individual collisions to be "strong" i.e., resulting in drastic, large-angle deflections. In the limiting case, the scattering angle $\theta$ in such collisions reaches $\pi$ which corresponds to a complete reflection of the electron momentum, $\mathbf{p}(t_0 + \tau) = -\mathbf{p}(t_0 - \tau)$ where $t_0$ denotes the moment of time when the collision occurs. Using (8.5), we have

$$\mathbf{p}_0 + \frac{e\mathbf{E}_0}{\omega}\sin\omega(t_0+\tau) = -\mathbf{p}_0 - \frac{e\mathbf{E}_0}{\omega}\sin\omega(t_0-\tau)$$

or, exploiting the assumption $\omega t \ll 1$,

$$\mathbf{p}_0 + \frac{e\mathbf{E}_0}{\omega}\sin\omega t_0 = -\mathbf{p}_0 - \frac{e\mathbf{E}_0}{\omega}\sin\omega t_0$$

or, inserting $\mathbf{p}(t) - e\mathbf{E}_0\sin\omega t\,/\omega$ into the left-hand side instead of $\mathbf{p}_0$, we get for the particle momentum in the wave after scattering by a heavy center

$$\mathbf{p}(t) = -\mathbf{p}_0 - \frac{2e\mathbf{E}_0}{\omega}\sin\omega t_0 + \frac{e\mathbf{E}_0}{\omega}\sin\omega t.$$

---

[174] This example is discussed in more details in [222].

Using this expression, we can calculate the electron energy in the wave after collision

$$\mathcal{E}(\sqcup)_+ = \frac{\mathbf{p}^2(t)}{2m} = \frac{1}{2m}\left(\left(\mathbf{p}_0 + \frac{2e\mathbf{E}_0}{\omega}\right)\sin\omega t_0 - \frac{e\mathbf{E}_0}{\omega}\sin\omega t\right)^2.$$

This equation also shows that the electron is exchanging energy with the wave. Making elementary transformations and averaging over many periods of the wave, we obtain

$$\overline{\mathcal{E}_+} = \frac{1}{2m}\left(\left(\mathbf{p}_0 + \frac{2e\mathbf{E}_0}{\omega}\right)\sin\omega t_0\right)^2 + \frac{e^2\mathbf{E}_0^2}{4m\omega^2}$$

$$= \overline{\mathcal{E}_-} + \frac{2e}{m\omega}\mathbf{p}_0\mathbf{E}_0\sin\omega t_0 + \frac{2e^2\mathbf{E}_0^2}{m\omega^2}\sin^2\omega t_0$$

where $\overline{\mathcal{E}_-} = \mathcal{E}_0 + T_p$ (see above).  The energy transfer between the wave and the particle is expressed as the  difference

$$\Delta\mathcal{E} = \overline{\mathcal{E}_+} - \overline{\mathcal{E}_-} = \frac{2e}{m\omega}\mathbf{p}_0\mathbf{E}_0\sin\omega t_0 + \frac{2e^2\mathbf{E}_0^2}{m\omega^2}\sin^2\omega t_0 \,. \qquad (8.6)$$

Notice that this energy transfer depends on the collision time $t_0$, more exactly, on the phase of the wave in this moment. This is the manifestation of the fact that waves  and particles exchange energy just because the collisions randomly perturb the phase synchronism between them. In particular, the particle may both take energy from the wave and give it to the field, depending on phase $\omega t_0$ as well as the mutual orientation of vectors $\mathbf{p}_0$ and $\mathbf{E}_0$.  One can also notice that when writing the field acting on a particle in the form $\mathbf{E} = \mathbf{E}_0\cos\omega t$, we disregarded the field phase in the initial moment $t = 0$; it would be more correct to write $\mathbf{E} = \mathbf{E}_0\cos(\omega t + \varphi)$. For some problems, taking this initial phase into account may be essential [226]. We shall briefly discuss some effects, in particular systematic drift associated with initial phase of the wave, in the next section.  If we, however, ignore the effects connected with the phase of the wave, we can average (8.6) over all phases to obtain

$$\overline{\Delta\mathcal{E}} = \frac{e^2\mathbf{E}_0^2}{m\omega^2} = 4T_p.\qquad (8.7)$$

This average energy transfer from the field to the particle is strictly positive, which means that the field is losing, and the particle is gaining energy in a typical large-angle elastic collision. If the frequency of such collisions is $\nu$, then the attenuation rate of an electromagnetic wave in the medium is

$$-\frac{d\mathcal{E}}{dt} = \frac{e^2\mathbf{E}_0^2\nu}{m\omega^2} = 4\nu T_p$$

where the collision frequency depends in general on the particle momentum $\nu = \nu(\mathbf{p})$ and may be calculated within the kinetic theory. Here, however, we treat the collision frequency as a purely phenomenological quantity: using a microscopic value for it within our crude estimates would be unacceptable due to excessive accuracy.  In most cases, $\nu = \nu(|\mathbf{p}|) = \nu(p)$. We may use the phenomenological collision frequency to find the dielectric permittivity and to estimate response functions for simple models.

The above semi-qualitative treatment demonstrates a heuristic usefulness of the models based on the classical motion of individual particles. Some experience shows that many intricate problems of radiation-matter interaction can be successfully understood using the simple language of classical single-particle models. Below we shall see more on this.

## 8.3 More on Charge in Electromagnetic Fields

We may begin the study of radiation-matter interaction with the simplest problem when matter is reduced to a single charged particle. The model of individual particles interacting with an external electromagnetic field reflects the situation when the matter (usually it is plasma or plasma-like medium) has a rather low density so that the average energy of interaction of the particles with the medium as well as between the particles is much lower than the energy of their individual motion in the electromagnetic field. The problem of a single charge in the field is usually studied already in the first chapters of textbooks on classical electromagnetic theory (see, e.g., [84], ch. 3); it may be in fact rather intricate even at the classical level, and we, while discussing this problem, shall emphasize certain details that require slightly more knowledge than needed for reading initial-stage textbooks on electromagnetism. Thus, the treatment of the discussed problem can be both classical and quantum, and it is interesting that the classical treatment is in most cases sufficient for the description of nearly all practically interesting applications. A little further we shall discuss the applicability criteria of the classical approach to describing the behavior of free charges in an electromagnetic field, now let us start with the naive Lorentz model:

$$\frac{d\mathbf{p}}{dt} = e\left[\mathbf{E}(\mathbf{r}, t) + \frac{1}{c}\big(\mathbf{v} \times \mathbf{H}(\mathbf{r}, t)\big)\right],\tag{8.8}$$

where $\mathbf{p}$ is the momentum of charge $e$ moving in the electromagnetic field $(\mathbf{E}, \mathbf{H})$. Notice that the Lorentz force contains the particle velocity which is a kinematic variable and not the momentum which is a dynamic variable. This physical fact is nontrivial and may be regarded as a consequence of countless experiments. To make the vector equation (8.8) (system of equations) closed, one ought to add the relationship between $\mathbf{p}$ and $\mathbf{v}$. Recall from classical mechanics (see also Chapter 4) that the relationship between momentum $p_j$ considered as a variable in phase space and the rate of change $v^i = \dot{x}^i$ is not necessarily as trivial as it is assumed in the Newtonian model and, e.g., in more advanced Lagrangian version of mechanics is given by $p_i = \partial L/\partial \dot{x}^i$ where $L := L(x^i, \dot{x}^i, t)$ is the Lagrangian function which may be, in principle, an arbitrary twice continuously differentiable function. So the differential relationship between momenta and velocities does not necessarily give the simple linear connection $p_i = m_{ij}\dot{x}^j$, customary for Newtonian mechanics (here $m_{ij}$ is the mass tensor which, as we have seen, can assume the role of metric tensor). One may observe the symptoms of more intricate than linear relationship between momenta $p_i$ and velocities $\dot{x}^j$ already in relativistic mechanics where such a relationship is nonlinear

$$\mathbf{p} = \gamma(\mathbf{v}) m\mathbf{v} = \frac{m\mathbf{v}}{\left(1 - \frac{v^2}{c^2}\right)^{1/2}}$$

and can be made linear only when the relativistic effects corresponding to an expansion over small parameter $\beta^2 = \frac{v^2}{c^2}$ are disregarded. In the nonrelativistic limit ($\beta \to 0$), $\mathbf{p} = m\mathbf{v}$ and the Newton-Lorentz equation (8.8) takes a simple form convenient for calculations:

$$m\dot{\mathbf{v}} = e\left[\mathbf{E}(\mathbf{r}, t) + \frac{1}{c}\left(\mathbf{v} \times \mathbf{H}(\mathbf{r}, t)\right)\right].$$

One can see that the Lorentz force is of the first order in $\beta = v/c$, $v = |\mathbf{v}|$, which is an important fact to be used further.

### 8.3.1  Interaction of a Particle with a Standing Wave

Let us, as a primary example, consider the motion of a free charged particle in the field of a standing wave

$$\begin{Bmatrix} \mathbf{E}(\mathbf{r}, t) \\ \mathbf{H}(\mathbf{r}, t) \end{Bmatrix} = \begin{Bmatrix} \mathbf{E}_0(\mathbf{r}) \cos \omega t \\ \mathbf{H}_0(\mathbf{r}) \sin \omega t \end{Bmatrix}$$

A standing wave may be, e.g., formed by two almost monochromatic ($\Delta\omega/\omega \ll 1$ where $\Delta\omega$ is the effective spectrum width) waves with frequencies $\omega$ and $\omega' \approx \omega$ traveling in opposite directions so that their wave vectors $\mathbf{k}$ and $\mathbf{k}'$ satisfy the relation $\mathbf{k} + \mathbf{k}' = \mathbf{0}$. Then the surfaces of equal phase are planes which are fixed in space. The particle, in general, may cross the standing wave structure at an arbitrary angle $\vartheta$ to such planes so that both longitudinal and transversal momentum components, $\mathbf{p}_\parallel$ and $\mathbf{p}_\perp$, are not necessarily small compared to the particle momentum $\mathbf{p}$. Actually, the amplitudes $\mathbf{E}_0$ and $\mathbf{H}_0$ are smooth envelopes $\mathbf{E}_0(\mathbf{r}, t)$ and $\mathbf{H}_0(\mathbf{r}, t)$ depending in general on both time $t$ and position $\mathbf{r}$. We shall, however, assume that $\mathbf{E}_0(\mathbf{r}, t)$ and $\mathbf{H}_0(\mathbf{r}, t)$ change very little when we transit from $t$ to $t + 2\pi n/\omega$ (assuming also $n$ finite and not very large). Then we may write $\mathbf{E}_0(\mathbf{r}, t) \approx \mathbf{E}_0(\mathbf{r})$ and $\mathbf{H}_0(\mathbf{r}, t) \approx \mathbf{H}_0(\mathbf{r})$.

The change of the time-harmonic dependence from $\cos \omega t$ in the electric field to $\sin \omega t$ in the magnetic field (the $\pi/2$ phase shift) is consistent with the Maxwell equation

$$\nabla \times \mathbf{E} + \frac{1}{c}\frac{\partial \mathbf{H}}{\partial t} = 0.$$

In these relationships the initial phase has been put to zero as a seemingly inessential parameter. This is not always the case. A little below we shall discuss the role of the phase in the expressions $\mathbf{E}(\mathbf{r}, t) = \mathbf{E}_0(\mathbf{r})\cos(\omega t + \varphi)$ and $\mathbf{H}(\mathbf{r}, t) = \mathbf{H}_0(\mathbf{r})\sin(\omega t + \varphi)$ and find out that the nonzero phase may signify an additional drift.

In the above expressions the spatial amplitudes $\mathbf{E}_0(\mathbf{r})$ and $\mathbf{H}_0(\mathbf{r})$ are not independent since due to Maxwell's equations $\nabla \times \mathbf{E}_0(\mathbf{r}) = -k\mathbf{H}_0(\mathbf{r})$, $k = \omega/c$. When the electromagnetic fields accelerating the charge are not very strong (see Chapter 5), the motion remains nonrelativistic so that the magnetic field term in (8.8) can be disregarded and the equation of motion takes the form

$$m\ddot{\mathbf{r}} = e\mathbf{E}_0(\mathbf{r})\cos\omega t.$$

Technically, this is a nonlinear ordinary differential equation in which variables $r$ and $t$ can be separated. There are a number of ways how to solve this equation; we shall start with the simplest approximation method when the external field is initially regarded as homogeneous (coordinate-independent), at least on a scale of the particle displacement due to the Lorentz force i.e., $\mathbf{E}_0(\mathbf{r}) = \mathbf{E}_0(\mathbf{r}_0 + \delta\mathbf{r}(\mathbf{E})) \approx \mathbf{E}_0(\mathbf{r}_0)$. Here $\mathbf{r}_0$ is the particle's initial position. In other words, we may think that during an oscillation period the particle passes the distance on which field amplitudes $\mathbf{E}_0(\mathbf{r})$ and $\mathbf{H}_0(\mathbf{r})$ do not vary substantially, $\delta := |\delta\mathbf{r}|/h \ll 1$, where $h$ is the characteristic length of the spatial field variation. Then, in the zeroth approximation in $\delta$ we have

$$\dot{\mathbf{r}} = \mathbf{v} = \frac{e\mathbf{E}_0}{m\omega}\sin\omega t + \mathbf{v}_0,$$

where $\mathbf{v}_0$ is the particle initial velocity and $\mathbf{E}_0 = \mathbf{E}(\mathbf{r}_0)$ is the constant vector. Further,

$$\mathbf{r} - \frac{e\mathbf{E}_0}{m\omega}\cos\omega t + \mathbf{v}_0 t + \mathbf{r}_0 \equiv +\delta\mathbf{r}.$$

Since we are considering the nonrelativistic situation[175], we may assume that the field-induced displacement $\delta\mathbf{r} = \delta\mathbf{r}(\mathbf{E}) \approx \delta\mathbf{r}(\mathbf{E}_0)$ is small compared with the field's characteristic scale i.e., with the wavelength in the case of a harmonic wave. Indeed, the displacement $|\delta\mathbf{r}| = \delta r$ reached over a half-period of the wave is $\delta r \sim eE/m\omega^2 \sim v/\omega$ so that $\delta r/\lambda \sim v/c \ll 1$ i.e., in nonrelativistic problems the field-induced displacement is small compared to the wavelength. Of course, for the fields rapidly varying in space this assumption may become invalid (see also below).

   The particle of mass $m$ (for definiteness, we shall talk about electrons) oscillates in the harmonic field with frequency $\omega$, gaining average kinetic energy

$$\mathcal{E} = \frac{1}{2}\overline{mv^2} = \frac{e^2 E^2}{4m\omega^2}$$

---

[175] The criterion of validity for a nonrelativistic approximation in classical (non-quantum) theory is $eE_0/m\omega c \ll 1$ which shows that, e.g., electrons may attain relativistic velocities for electric fields accelerating the particles to their rest energy over the distance of the order of the wavelength, $eE_0\lambda \sim mc^2 \approx 0.5\text{MeV}$. If quantum theory is considered, the classical theory for a free electron interacting with the electromagnetic field is valid when the energy of a transferred or radiated electromagnetic quantum is small compared to the rest energy, $\hbar\omega \ll mc^2$.

This is a standard forced motion of a particle under the influence of an external harmonic field. Motion in oscillating fields, despite its apparent simplicity, contains a number of intricacies, and we shall try to discuss some of them in detail. For example, one usually tacitly assumes that it is only the magnetic field that can confine charged particles. Wrong! One may be surprised to find out that charged particles may be confined also by a harmonically oscillating electromagnetic field, provided it is spatially inhomogeneous, such a confinement occurring regardless of the charge sign. So an inhomogeneous electromagnetic wave can be used to trap, accelerate, and control charged particles.

To better understand some interesting effects accompanying the interaction of electromagnetic fields with free charged particles, we shall as usual start from simplified models gradually incorporating new features in the equations. We have just obtained the simplest possible solution for the particle moving in the field of a monochromatic (harmonic) standing wave in the zero-order in relativistic parameter $\beta = v/c$. In this approximation no influence of the magnetic field is taken into account although the magnetic field is present in any electromagnetic wave. Nevertheless, if a particle moves nonrelativistically, neglect of the magnetic field in the wave which drives the particle is fully justified.

Let us now see what happens in the first approximation in $\beta$, in case the scale of spatial inhomogeneity of the field is determined by the wavelength $\lambda = 2\pi/\omega$ (in general, however, the field amplitudes may have other spatial scales). Then we have to retain the term with the magnetic field in the Lorentz force. Expanding slowly varying amplitudes $\mathbf{E}_0(\mathbf{r})$ and $\mathbf{H}_0(\mathbf{r})$ near point $\mathbf{r}_0$ which may be interpreted as the initial position of the electrons, we have in the first approximation in inhomogeneity parameter $\delta = eE_0(\mathbf{r}_0)\kappa/m\omega^2$, where $\kappa$ characterizes the spatial inhomogeneity of the field:

$$\mathbf{E}_0(\mathbf{r}) \approx \mathbf{E}_0(\mathbf{r}_0) + (\delta\mathbf{r}\overline{\nabla})\mathbf{E}_0(\mathbf{r}_0)$$

or, in component notation,

$$E_0^j(x^i) \approx E_0^j(x_0^i) + \delta x^i \partial_i E_0^j(x_0^i).$$

The length $s = \kappa^{-1}$ denotes the distance on which amplitudes $\mathbf{E}_0$ and $\mathbf{H}_0$ change significantly. Such a distance is often thought to coincide with the wavelength of the field; this is, however, a very particular case. Only in this special case expansion on the inhomogeneity parameter coincides with that on relativistic parameter $\beta = v/c \approx eE_0/m\omega c$.

Now insert these expansions into the motion equation, using the obtained expression for $\delta\mathbf{r}$:

$$\delta\mathbf{r}(\mathbf{E_0}) = \mathbf{v}_0 t - \frac{e\mathbf{E}_0(\mathbf{r}_0)}{m\omega^2}\cos\omega t.$$

Notice that the quantity $\mathbf{A}_0 := e\mathbf{E}_0/m\omega^2$ has the meaning of the particle oscillation amplitude in a monochromatic field, when all other factors affecting the particle motion, e.g., collisions, are neglected. Later we shall take collisions into account and see that in this case the amplitude for a harmonic field

$$\mathbf{E}(\mathbf{r}, t) = \mathbf{E}_0 e^{-i\omega t} + \mathbf{E}_0^* e^{i\omega t}$$

will take the form $\mathbf{A}_0 := e\mathbf{E}_0/m\omega(\omega + i\nu)$ corresponding to the solution of the motion equation

$$\boldsymbol{r}(t) = -\mathbf{A}_0 e^{-i\omega t} - \mathbf{A}_0^* e^{i\omega t}$$

In other words, the particle amplitude in an electromagnetic field will acquire additional terms (in real representation of the field) or becomes complex (in complex representation), which physically corresponds to the losses of electromagnetic energy and to phase shift i.e., retardation of the electromagnetic response of an individual particle to an external electromagnetic field. All these issues are closely connected with the elementary theory of dielectric permittivity (see Chapter 5).

Now we get

$$m\ddot{\mathbf{r}} = e\mathbf{E}_0(\mathbf{r}_0)\cos\omega t + e(\mathbf{v}_0\nabla)\mathbf{E}_0(\mathbf{r}_0)t - \frac{e^2}{m\omega^2}(\mathbf{E}_0(\mathbf{r}_0)\nabla)\mathbf{E}_0(\mathbf{r}_0)\cos^2\omega t$$
$$+ \frac{e^2}{m\omega^2}\frac{\omega}{c}\big(\mathbf{E}_0(\mathbf{r}_0)\times\mathbf{H}_0(\mathbf{r}_0)\big)\sin^2\omega t$$
$$+ \frac{e}{c}\big(\mathbf{v}_0\times\mathbf{H}_0(\mathbf{r}_0)\big)\sin\omega t$$

where for the Lorentz force we used:

$$\frac{e}{c}\big(\mathbf{v}\times\mathbf{H}(\mathbf{r},t)\big) = \frac{e}{c}\left[\left(\frac{e\mathbf{E}_0(\mathbf{r}_0)}{m\omega}\sin\omega t + \mathbf{v}_0\right)\times\mathbf{H}_0(\mathbf{r}_0)\sin\omega t\right].$$

By using the Maxwell equation $\mathbf{H}_0(\mathbf{r}_0) = -(c/\omega)\mathrm{curl}\mathbf{E}_0(\mathbf{r}_0)$ we get after averaging over the field period:

$$m\ddot{\mathbf{r}} = et(\mathbf{v}_0\nabla)\mathbf{E}_0(\mathbf{r}_0) - \frac{e^2}{2m\omega^2}(\mathbf{E}_0(\mathbf{r}_0)\nabla)\mathbf{E}_0(\mathbf{r}_0)$$
$$+ \frac{e^2}{2m\omega^2}\frac{\omega}{c}\left(-\frac{c}{\omega}\right)\mathbf{E}_0(\mathbf{r}_0)\times\big(\nabla\times\mathbf{E}_0(\mathbf{r}_0)\big)$$

or

$$m\ddot{\mathbf{r}} = et(\mathbf{v}_0\nabla)\mathbf{E}_0(\mathbf{r}_0) - \frac{e^2}{4m\omega^2}\nabla E_0^2(\mathbf{r}_0). \tag{8.9}$$

Here the vector identity

$$\mathbf{E}_0 \times (\nabla \times \mathbf{E}_0) = \frac{1}{2}\nabla E_0^2 - \mathbf{E}_0 \nabla \mathbf{E}_0$$

was used.

One can make several remarks concerning this standard procedure[176] of obtaining the average force (8.9) acting on particles moving in the field of a monochromatic wave with spatially inhomogeneous amplitude. This force is usually called ponderomotive; in the Russian literature it is mostly known as the "Miller force", by the name of a prominent physicist belonging to the well-known Nizhni Novgorod (formerly Gor'ki) radiophysical scientific school. This force has a field gradient character, attracting a charged particle irrespective of the charge sign to low-intensity field regions and repelling it from the regions of strong field. The term proportional to time $t$ corresponds to drift motion of the particle[177]. The initial time-point $t_0$ determines the phase of a particle when it starts the motion in the electromagnetic field. This phase may play an important role for the drift component of the motion (see below). One may notice that the dynamics of a charged particle is described in terms of the average value of the particle position $\bar{\mathbf{r}}$ whereas the ponderomotive force is calculated at its staring point $\mathbf{r}_0$. In most cases, the difference between $E_0(\mathbf{r}_0)$ and $E_0(\bar{\mathbf{r}})$ is inessential due to slow variation of the amplitudes, and taking such a difference into consideration would mean an excessive accuracy level. However, two pairs of values, $\bar{\mathbf{r}}, \mathbf{r}_0$ and, correspondingly, $E_0(\mathbf{r}_0), E_0(\bar{\mathbf{r}})$ may differ significantly when systematic drift component is taken into account. This fact may lead to somewhat unexpected effects related to anisotropy of drift motion. Usually, when ponderomotive effects such as repelling particles from strong field regions are considered, field is represented by axial-symmetric intensity distributions such as in idealized laser beam and drift components breaking axial symmetry are disregarded. In this case, the ponderomotive force only depends on the particle's distance from the laser beam axis and, by the way, does not depend on the field polarization. Taking drift into account would mean that, e.g., the energy of the particles gained owing to ponderomotively pushing them out of the strong field region of the laser beam would depend on the scalar product between the particle velocity and the field polarization vector. The corresponding calculations are simple but lengthy[178], therefore I do not bring them here (see also below).

We can now generalize the formulas for the ponderomotive force acting on a particle (electron), taking into account collisions with other particles. This situation takes place, for example, in plasma. The simplest - phenomenological - way to take collisions into account consists in adding the friction term to the equation of motion:

$$m\ddot{\mathbf{r}} + m\nu\dot{\mathbf{r}} = e\left(\mathbf{E}(\mathbf{r},t) + \frac{1}{c}\big(\dot{\mathbf{r}} \times \mathbf{H}(\mathbf{r},t)\big)\right). \tag{8.10}$$

---

[176] This is actually an iteration method that can be applied also in more general situations, for example, when collisions should be taken into account, see below.

[177] It would be more appropriate to write $t - t_0$ instead of $t$ which may be important when many particles, e.g., a beam, are considered.

[178] One may write $\mathbf{r}_0$ as $\bar{\mathbf{r}} + (\mathbf{r}_0 - \bar{\mathbf{r}}) \equiv \bar{\mathbf{r}} + \delta_0$ and expand (8.9) over $\delta_0$.

Here $\nu$ is the collision frequency i.e., the quantity determining the rate of momentum interchange in multiple collisions. It is clear that in general $\nu$ is a function of the particle velocity $v = |\dot{\mathbf{r}}|$, $\nu = N\sigma_t(v)v = \nu(v)$ , where $N$ denotes the density of scatterers and $\sigma_t$ is the transport cross-section

$$\sigma_t(v) = 2\pi \int_0^\pi \frac{d\sigma}{d\Omega}(v,\theta)(1 - \cos\theta)\sin\theta\, d\theta,$$

$\frac{d\sigma}{d\Omega}(v,\theta) = |f(\mathbf{q})|^2$ is the differential cross-section, $f(\mathbf{q})$ is the scattering amplitude, $\mathbf{q} \in S^2$, $\mathbf{q} \cdot \mathbf{q} - 1$, i.e., the vector $\mathbf{q}$ belongs to the unit sphere $S^2$ (see any textbook on scattering theory or Chapter 5 of this book). Recall that it is the transport cross-section and not the total cross-section

$$\sigma(v) = 2\pi \int_0^\pi \frac{d\sigma}{d\Omega}(v,\theta)\sin\theta\, d\theta$$

that determines the collision frequency (and also allows one to estimate the mean free path of a particle) because it accounts for the fact that the momentum transfer in elastic collisions reaches a maximum at large scattering angles (head-on collisions) and tends to zero for small-angle scattering (large impact parameters).[179] Although the collision frequency depends on velocity, we may understand by $\nu$ in our phenomenological treatment dealing with effective quantities some average collision rate, $\overline{\nu(v)} = \nu$, where averaging is carried out over all relative velocities of our particle with respect to scattering centers of the medium. In reality, this averaging can be accurately performed only in the kinetic theory; here we have to be satisfied with the qualitative considerations.

Now, let us, as before, start with a zero-order approximation in $v/c$ omitting the Lorentz force:

$$m\ddot{\mathbf{r}} + m\nu\dot{\mathbf{r}} = e(\mathbf{E}_0(\mathbf{r})\cos(\omega t + \varphi)), \tag{8.11}$$

where we have included the phase of the field $\varphi$. The reason why this phase may deserve to be included in the motion equations will be clear from the further discussion. The zero-order equation is a linear ODE and can be easily integrated. Modern humanity integrates differential equations with the help of "computer algebra" products such as Maple or Mathematica which are

---

[179] One can readily see that the velocity change in every act of scattering accompanied with a deflection by the angle $\theta$ is $\nabla v = v(1 - \cos\theta)$.

great products indeed, but it is sometimes even faster to integrate simple equations in the good old manner - with bare hands, pen and paper. (Here, I must perhaps apologize for bringing the calculation details that might seem excessive to an experienced reader.) One may look for a special solution to the above equation in the form

$$\mathbf{r}_s = \frac{e\mathbf{E}_0}{m}[a\cos(\omega t + \varphi) + b\sin(\omega t + \varphi)]$$

then, inserting this expression into the equation (8.11), we get the following system of linear equations for coefficients $a$ and $b$:

$$\begin{cases} -a\omega^2 + b\nu\omega = 1 \\ a\nu\omega + b\omega^2 = 0 \end{cases}$$

which gives

$$a = -\frac{1}{\omega^2 + \nu^2}, \qquad b = -a\frac{\nu}{\omega} = \frac{\nu}{\omega(\omega^2 + \nu^2)}$$

so that

$$\mathbf{r}_s = \frac{e\mathbf{E}_0}{m(\omega^2 + \nu^2)}\left(-\cos(\omega t + \varphi) + \frac{\nu}{\omega}\sin(\omega t + \varphi)\right)$$
$$= \frac{e\mathbf{E}_0}{m(\omega^2 + \nu^2)}\frac{\cos(\omega t + \varphi + \chi)}{\cos\chi},$$

where $\chi \equiv \arctan(\nu/\omega)$ and $\cos\chi = \frac{\omega}{(\omega^2 + \nu^2)^{1/2}}$. Finally,

$$\mathbf{r}_s(t) = \frac{e\mathbf{E}_0\cos(\omega t + \varphi + \chi)}{m\omega(\omega^2 + \nu^2)^{1/2}}. \tag{8.12}$$

The general solution to (8.11) is

$$\mathbf{r}(t) = \mathbf{A} + \mathbf{B}e^{-\nu t} + \mathbf{r}_s(t) = \mathbf{A} + \mathbf{B}e^{-\nu t} - \frac{e\mathbf{E}_0\cos(\omega t + \varphi + \chi)}{m\omega(\omega^2 + \nu^2)^{1/2}} \tag{8.13}$$

where the constants $\mathbf{A}$ and $\mathbf{B}$ should be determined from the initial conditions, e.g., $\mathbf{r}(t_0) = \mathbf{r}_0$ and $\dot{\mathbf{r}}(t_0) = \mathbf{v}_0$. We may, for simplicity, put $t_0 = 0$ (see the footnote on the preceding page) and obtain

$$\mathbf{B} = -\frac{\mathbf{v}_0}{\nu} + \frac{e\mathbf{E}_0\sin(\varphi + \chi)}{m\omega(\omega^2 + \nu^2)^{1/2}}$$

and

$$\mathbf{A} = \mathbf{r}_0 - \mathbf{B} + \frac{e\mathbf{E}_0 \cos(\varphi + \chi)}{m\omega(\omega^2 + \nu^2)^{1/2}}$$

$$= \mathbf{r}_0 + \frac{\mathbf{v}_0}{\nu} + \frac{e\mathbf{E}_0}{m\omega(\omega^2 + \nu^2)^{1/2}}\left[\frac{1}{\omega}\cos(\varphi + \chi) - \frac{1}{\nu}\sin(\varphi + \chi)\right]$$

Using the definition $\chi \equiv \arctan(\nu/\omega)$, we have

$$\mathbf{A} = \mathbf{r}_0 + \frac{\mathbf{v}_0}{\nu} - \frac{e\mathbf{E}_0 \sin\varphi}{m\nu\omega}$$

and

$$\mathbf{B} = -\frac{\mathbf{v}_0}{\nu} + \frac{e\mathbf{E}_0 \sin(\varphi + \chi)}{m\nu(\omega^2 + \nu^2)^{1/2}}$$

so that the general solution to the zero-order equation describing the motion of a particle in the standing wave may be written as

$$\mathbf{r}(t) = \mathbf{r}_0 + \frac{\mathbf{v}_0}{\nu} - \frac{e\mathbf{E}_0 \sin\varphi}{m\nu\omega} - \frac{\mathbf{v}_0}{\nu}e^{-\nu t} + \frac{e\mathbf{E}_0 e^{-\nu t}\sin(\varphi + \chi)}{m\nu(\omega^2 + \nu^2)^{\frac{1}{2}}} + \mathbf{r}_s(t) \quad (8.14)$$

where

$$\mathbf{r}_s(t) = \frac{e\mathbf{E}_0 \cos(\omega t + \varphi + \chi)}{m\omega(\omega^2 + \nu^2)^{1/2}}$$

Now we can write the first-order equation like we did before, when collisions were not taken into account, as

$$\ddot{\mathbf{r}} + \nu\dot{\mathbf{r}} = \frac{e}{m}\left(\mathbf{E}_0(\mathbf{r}_0)\cos(\omega t + \varphi) + (\mathbf{r}\nabla)\mathbf{E}_0(\mathbf{r}_0)\cos(\omega t + \varphi)\right). \quad (8.15)$$

Here, as in the higher-order terms of the right-hand side expansion on the inhomogeneity parameter (see above), one should insert the argument $\mathbf{r}_0$ after having differentiated the field amplitude $\mathbf{E}_0(\mathbf{r}_0)$. To obtain the right-hand side in (8.15) by the iteration method that we employ, one has to put the zero-order solution (8.14) into it. Inserting (8.14) into (8.15) and averaging over oscillations, we get for the first-order equation

$$m\ddot{\mathbf{r}} + m\nu\dot{\mathbf{r}} = -\frac{\alpha(\varphi)}{\nu}(\mathbf{v}_0\nabla)\mathbf{E}_0 - \frac{e^2(\mathbf{E}_0\nabla)\mathbf{E}_0\,\alpha(\varphi)\sin(\varphi + \chi)}{m\nu(\omega^2 + \nu^2)^{1/2}}$$

$$- \frac{e^2(\mathbf{E}_0\nabla)\mathbf{E}_0 \cos\chi}{2m\omega(\omega^2 + \nu^2)^{1/2}}, \quad (8.16)$$

where

$$\alpha(\varphi) := \frac{1}{T} \int_0^T dt \, e^{-\nu t} \cos(\omega t + \varphi)$$

$$= e^{i\left(\frac{\omega + T}{2} + \varphi\right)} \frac{\sin\frac{\omega_+ T}{2}}{\omega_+ T} 2 + e^{-i\left(\frac{\omega_- T}{2} + \varphi\right)} \frac{\sin\frac{\omega_- T}{2}}{\omega_- T} 2,$$

with $\omega_+ = \omega + i\nu, \omega_- = \omega_+^* = \omega - i\nu$ and $T$ is the averaging period[180]. One can also represent $\alpha(\varphi)$ in the form:

$$\alpha(\varphi) = \frac{\omega}{\omega^2 + \nu^2} e^{-\nu t} \sin(\omega t + \varphi) - \frac{\nu}{\omega} \cos(\omega t + \varphi) \Big|_0^T$$

or

$$\alpha(\varphi) = \frac{\omega(e^{-\nu t} \sin(\omega T + \varphi - \chi) - \sin(\varphi - \chi))}{T(\omega^2 + \nu^2) \cos \chi}.$$

One can notice that the right-hand side in the motion equation (8.15) or (8.16) depends on the field initial phase $\varphi$ which, in general, does not vanish after averaging over oscillations. This is a curious fact, which deserves a special discussion (see below).

Now, to proceed with the time-averaging, we may consider two cases. For long-wave harmonics such as, e.g., in the microwave domain, we may perform this averaging over a single period of the wave, putting $T = 2\pi/\omega$. Then we get for the quantity $\alpha(\varphi)$

$$\alpha(\varphi) = \frac{\omega^2}{2\pi(\omega^2 + \nu^2)} \frac{\sin(\varphi - \chi)}{\cos \chi} (1 - \exp(-2\pi\nu/\omega)).$$

In the optical case, when the number of periods $n$ in $T = 2\pi n/\omega$ is very large, $n \to \infty$, i.e., $\omega T \gg 1, \alpha(\varphi) \to 0$ as $1/n$. In this case, only the special solution of the zero-order motion equation (8.14) contributes to the force acting on a particle in the standing wave:

$$m\ddot{\mathbf{r}} + m\nu\dot{\mathbf{r}} = \frac{e^2(\mathbf{E}_0 \nabla)\mathbf{E}_0 \cos \chi}{2m\omega(\omega^2 + \nu^2)^{1/2}} = \frac{e^2(\mathbf{E}_0 \nabla)\mathbf{E}_0}{2m(\omega^2 + \nu^2)},$$

and in the collisionless limit $\nu/\omega \ll 1$ this expression takes the form of the usual ponderomotive (Miller) force. It is interesting that in this case the dependence on the initial phase of the wave drops out.

In the quantum language, the elementary process described here may be interpreted as a stimulated Compton scattering i.e., consisting in the absorption of a photon with frequency $\omega$ and 3-momentum $\mathbf{k}$ and emission of a photon with $(\omega', \mathbf{k}')$ subjected to conditions $\omega' \approx \omega$ and $\mathbf{k}' \approx -\mathbf{k}$. The particle energy in this process is almost conserved, up to the terms determining the difference between $\omega'$ and $\omega$ i.e., $E' - E = \omega - \omega'$. However, the particle momentum may change significantly which corresponds, in the classical language, to the action of the force standing in the right-hand side of the above motion equations, $\Delta \mathbf{p} = \mathbf{p}' - \mathbf{p} = \mathbf{k} - \mathbf{k}' \approx \pm 2\mathbf{k}$. Actually, scattering of an electromagnetic wave by a free electron may be interpreted as the Compton effect only if we take into account the frequency shift of the scattered wave. Nevertheless, in the classical approach the frequency change can be ignored, the electron is classically accelerated in accordance with the motion equations in the electromagnetic field of the wave, then emitting radiation as prescribed by the rules of classical electrodynamics [193], §66. In the classical description, we obtain the standard formula for the Thomson scattering cross-section [193], §78

$$\sigma_\omega = \frac{1}{S}\frac{\overline{dJ(\omega)}}{d\omega} = \frac{2|\dot{\mathbf{d}}_\omega|/3c^2}{c|\mathbf{E}_\omega|/8\pi} = \frac{1}{3c^2}\left|\frac{e^2\mathbf{E}}{m}\right|^2 \Big/ \frac{c|\mathbf{E}|^2}{8\pi} = \frac{8\pi}{3}\left(\frac{e^2}{mc^2}\right)^2 = \frac{8\pi}{3}r_0^2$$
$$\approx 6.65 \cdot 10^{-25}\mathrm{cm}^2$$

where $\mathbf{d}_\omega$ is the Fourier component of the electronic dipole moment induced by the wave, $S$ is the Poynting vector representing the flow of electromagnetic energy (overline bar denotes averaging over time), and $r_0 = e^2/mc^2$ is the classical electron radius.

---

[180] In this simple model, we do not take into account possible temporal non-uniformities or transition effects connected with turning the field on and off.

One may also recall the connection of the motion in a standing wave with the so-called Kapitza-Dirac [227] effect and Kapitza pendulum [23] §30 problem. The Kapitza-Dirac effect is just the elastic scattering of electrons in the field of a standing electromagnetic wave which may be written as

$$\mathbf{E}(\mathbf{r}, t) = \mathbf{E}_0(\mathbf{r})e^{-i\omega t} + \mathbf{E}_0^*(\mathbf{r})e^{-i\omega t}, \qquad \mathbf{E}_0(\mathbf{r}) = \mathbf{E}_0\cos\mathbf{k}\mathbf{r} = \mathbf{E}_0\cos kx.$$

An electron in such a field experiences the ponderomotive force (we use for simplicity scalar notations)

$$F_x = -\frac{\partial U_p}{\partial x}, \qquad U_p = \frac{e^2E^2(x)}{4m\omega^2}.$$

Using an optical analogy, one can say that the electronic wave is scattered on a diffraction grating having a spatial period $d = \pi/k = \lambda/2$. P. L. Kapitza and P. A. M. Dirac used one more analogy namely with the Bragg diffraction in ideal crystals, when diffraction angles $\theta_n$ are determined by the condition $n\lambda = 2d\sin\theta_n$. In Kapitza-Dirac scattering, this condition takes the form $\sin\theta_n = n\hbar k/p$ where $p$ is the momentum of incident particles. Thus, electrons crossing a standing electromagnetic wave would be reflected from the planes of peak intensity. This is, of course, a very intuitive, qualitative consideration; in order to treat this problem accurately one should account for the motion of an electron in the field of an electromagnetic wave, e.g., within the framework of the nonrelativistic

Schrödinger equation, see, e.g., the paper by M. V. Fedorov [228] who was probably the first to consider the Kapitza-Dirac scattering on the quantum-mechanical level (see also [229]).

### 8.3.2    Interaction of a Particle with a Traveling Wave

An important class of problems is related to the situation when a charged particle encounters a traveling electromagnetic wave. These problems have been typically considered in microwave electronics (see, e.g., a comprehensive book [230]) and in accelerator physics. Recently, much interest has been aroused to the possibility of cost-efficient acceleration of particles in the field of high-power ultrashort laser pulses, for instance so-called laser wake-field acceleration (LWFA). Perhaps the most prospective application of the theory related to the interaction of charged particles with a traveling electromagnetic wave, both in the single-particle and many-particle dynamics regimes, is the free-electron laser (FEL). We shall not discuss here a lot of technicalities and specially adapted methods of integration of the motion equations, arising in connection with all these applications. In the spirit of this book, we shall only observe the general patterns associated with the particle-wave interaction, with the hope that such a high-level discussion would enable one to understand more professional minutes should they appear.

The primary setting of the problem is the following: let a linearly polarized traveling wave propagate along the $z$ axis in vacuum. We may write the field in the wave as

$$\mathbf{E}(\mathbf{r}, t) = \mathbf{E}_0 \cos(\omega t - k_z z) = F\mathbf{e}_x \cos(\omega t - kz)$$

and

$$\mathbf{H}(\mathbf{r}, t) = \mathbf{H}_0 \cos(\omega t - k_z z) = F\mathbf{e}_y \cos(\omega t - kz)$$

where $\mathbf{e}_x$ and $\mathbf{e}_y$ are unit vectors defining the polarization of the electric and magnetic field, respectively, $k_z = k$; the quantity $F$ is the amplitude of both $\mathbf{E}$ and $\mathbf{H}$ because for the wave propagating in vacuum their amplitudes are equal. The vector motion equation for an electron in such a wave may be written in components as

$$m\ddot{x} = eF\left(1 - \frac{v_z}{c}\right)\cos(\omega t - kz), \qquad m\ddot{y} = 0, \qquad m\ddot{z} = \frac{ev_x}{c}F\cos(\omega t - kz)$$

## 8.4    On Hamiltonian Formalism for Particle Motion in Electromagnetic Fields

The Hamiltonian method seems to be poorly adapted to relativistic problems such as field theories. For example, classical electrodynamics, though being a relativistic theory, is represented in this method in the form similar to the nonrelativistic classical mechanics. Nonetheless, application of the Hamiltonian method to electrodynamics was thoroughly discussed in the classical book by W. Heitler [207]. V. L. Ginzburg, the 2003 Nobel Prize winner, RAS (the Russian Academy of Sciences) member and one of the last encyclopedists in physics, also preferred the Hamilton approach to electrodynamics due to its simplicity vs. more sophisticated techniques used in quantum field theory. Of course, the Hamiltonian method usually fails in relativistic problems, and the electromagnetic field is an essentially relativistic object.

The first thing to do in order to describe the particle motion in an electromagnetic field using the Hamiltonian formalism is to construct an effective Hamiltonian function for the field and charges in it. I write "effective" because it would be hardly possible to include in the Hamiltonian all possible factors influencing the particle motion. In fact, all Hamiltonians in physics are effective ones since they represent a system in terms of operators defined over some arbitrarily built and very restrictive Hilbert space. In reality, one ignores the dependence on temperature i.e., interaction with the environment (thermostat), energy cutoffs, boundary conditions (such as the electromagnetic field vanishing at infinity in our case), etc. I have already mentioned that it seems to be the ultimate aim of physics to write down the Hamiltonian for the entire universe, and the rest should be presumably done by mathematicians and computer modelers. This is, however, more a metaphysical utopia rather than a scientific program. Striving to employ the Hamiltonian method everywhere appears to be mainly psychologically driven: one desires to represent any theory in the conventional form of classical mechanics. It is, however, clear that, e.g., for electrodynamics the Hamiltonian formalism is poorly adapted, in particular, because it picks out one specific instant of time. This is, by the way, one of the reasons why predominantly the Lagrangian formalism is used while treating electromagnetic field problems.

## 8.5    Interaction between Atoms and Radiation Field

In this section, I shall try to dispel the popular view that the problem of field-matter interaction is so complicated that one can obtain reliable results only by using numerical techniques and very powerful computers. Here, some simple microscopic mechanisms determining the response of the matter to electromagnetic fields are briefly discussed. Of course, in a general formulation this task is enormous and has been treated in numerous classical works which I shall refer to in the subsequent text. In most cases, when considering the interaction of electromagnetic radiation with atomic systems one can use the electric dipole approximation (see [39], 67). In general, one can observe that the absolute majority of electrodynamic and optical phenomena can be well described in this approximation. Indeed, the electric dipole approximation is valid when the parameter $a/\lambda \ll 1$, where $a$ is the characteristic dimension of an atomic or molecular system interacting with an electromagnetic field (one usually speaks in terms of scattering or radiation of waves). Typically, $a \sim 10^{-7} - 10^{-8}$cm and $\lambda \sim 10^{-4} - 10^{-3}$cm which corresponds to visible or infrared light. One can, however, pay attention to the following paradox. If we consider the interaction of electromagnetic radiation with a material medium regarding all its atoms and molecules together as a single scattering system of size $L$, then, e.g., in optics, we have as a rule $L/\lambda \gg 1$ whereas for the applicability of the dipole approximation we should ensure the opposite condition $L/\lambda \ll 1$ or at least $kL \lesssim 1$. This is, however, not a real paradox since in most cases different atoms and molecules of the medium radiate and scatter electromagnetic fields statistically independently of each other. Correlations between atoms and molecules interacting with the radiation field determine coherence properties of light.

## 8.6   Laser-Matter Interaction

Popular ideas about the laser generation - analogy with phase transition - must have a reliable physical and mathematical ground.

Many issues still remain open in the study of laser-matter interaction, especially at high intensities, and nowadays lasers produce radiation intensities up to $10^{22}$W/cm$^2$ that corresponds to electric fields reaching $10^{11}$CGSE i.e., by four orders of magnitude higher than atomic fields $E_{at} \sim e/a_B^2 = m^2 e^5/\hbar^4 \approx 5 \cdot 10^9$V/cm . For estimates in laser-matter interaction it is convenient to introduce the "atomic intensity", $I_{at} = cE_{at}^2/8\pi = $ cm$^4 e^{10}/8\pi\hbar^8 \approx 3.5 \cdot 10^{16}$W/cm$^2$. One can expect that the effect of high-intensity electromagnetic radiation on material media would consist at least in very efficient heating of the matter electrons. Moreover, electrons can be accelerated in the fields of laser pulses which opens the possibility to construct cost-effective machines both for high energy research and practical applications.

### 8.6.1. Ultrashort Laser Pulses

Let us consider, as an example, a typical situation when an ultrashort laser pulse (ULP) is directed at the surface of metal. The pulsed electromagnetic field accelerates electrons in the metal skin layer, thus heating the electronic subsystem of the material. One must note that the notion of a skin layer is not quite trivial and is not always correctly defined, but we shall assume that we do know exactly this layer's properties (see [208], 60). Owing to the heat conduction processes, the electromagnetic energy absorbed in the skin layer migrates into the metal volume and eventually dissipates there increasing the equilibrium temperature of the entire sample. One can imagine that if the predominant mechanism of the heat transfer from the skin layer into the volume is electronic conductivity, then after having reached a certain power (fluence) threshold the temperature gradients may become so high that the heat flux caused by the electrons would spread faster than phonons in metal. In other words, the speed of the heat flux would surpass the phase velocity of phonons. In such a situation, the drift electrons can produce phonons due to the Cherenkov (or Čerenkov) radiation - a complete analogy to Cherenkov radiation of photons in dielectric media. One can observe the characteristic soft blue glow in nuclear reactors, for example, during the excursion to a nuclear power plant, a very useful tour, by the way. This glow is due to Cherenkov radiation of photons in the visible range. But let us get back to phonons. One can expect that due to massive generation of phonons in the Cherenkov mechanism their distribution can deviate from the equilibrium Planck shape, which in its own right would produce an impact on the heat transfer - a sort of a self-consistent feedback.

# 9 What remains to be solved?

There are a lot of unresolved issues in physics and in natural sciences in general - much more than resolved. I don't think that physics and mathematics have an agenda: nobody knows what comes next. The great discoveries have always been unexpected.

## 9.1    The Standard Model

The idea of spontaneous symmetry breaking enabled physicists to build the famous Standard Model of the unified weak and electromagnetic interactions. It is interesting that the same idea allowed one to demonstrate that the important phenomenon of superconductivity [181] can be regarded as spontaneously broken electromagnetism. Ultimately, the Standard Model strives to describe all the processes occurring in nature within the framework of the four known interactions: electromagnetic, weak, strong and gravitational. To understand most astronomic concepts and even cosmological models, to learn chemistry or electrical engineering one only needs gravitation and electromagnetism. Quark-gluon models, the Higgs particle or the spontaneous breach of symmetry are often superfluous at this level of knowledge. Yet without strong and weak interactions underlying the respective mathematical models, one would not be able to understand what makes the stars shine nor how the Sun burns chemical elements producing the radiation power that supports life on the Earth. In general, without this knowledge one cannot grasp the principles of nuclear physics.

Current understanding of fundamental high-energy physics (formerly elementary particle physics) is based on the just mentioned Standard Model which stands on two main pillars: gauge invariance and spontaneous symmetry breaking. The original "material" foundation of the Standard Model was made up of 6 quarks × 6 leptons whereas the main goal of the Standard Model was to describe the four known fundamental forces of nature – strong, weak, electromagnetic and gravitational – on the same footing i.e., in terms of gauge concepts. This endeavor was accomplished for three out of four interactions: for the strong interaction the corresponding gauge group is SU(3), for the weak and electromagnetic interactions, respectively SU(2) and U(1) groups; within the framework of the Standard Model, the weak and electromagnetic forces are unified and known as the electroweak interaction. The first attempt at the unification of gravity and electromagnetism (i.e., the unification of gauge fields with gravitation) goes back to the Kaluza (1921) and Klein (1926) five-dimensional models. The unification of three gauge forces (strong, weak, EM) with gravity is an extremely difficult endeavor, requiring special skills, and we shall not discuss this issue here. One can only note that the Standard Model does not treat gravity on an equal footing with the three microscopic gauge forces.

---

[181] It would hardly be possible to build the Large Hadron Collider (LHC) without using the superconducting magnets.

What is today's status of the Standard Model? It is mostly regarded as a minimal one i.e., incomplete and designed as a temporary step towards a more refined unified theory. One calls the Standard Model minimal since there are no more elementary particles in it besides six quarks, six leptons and four gauge bosons needed to transfer interactions (the Higgs boson was discovered with high probability in the LHC experiments and is one more elementary – not compound - particle). The incompleteness of the Standard Model is, in particular, manifested by cosmological data. Thus, the Standard Model requires modifications to accommodate new phenomena. Such modifications are also expected to be based on gauge principles, with their geometric, Lie groups and bundles approach.

Although ideas and results usually don't come in a prescribed order, soon there may be an exception. The Large Hadron Collider in CERN is destined to come closer to the microscopic spacetime structure than any previous device.

Putting together all seemingly diverse topics in a manuscript takes time, and I am writing all this before the LHC machine in CERN has been put in operation, but you may read it after the new results have been obtained. For the energies provided by the LHC, new results will inevitably appear, despite the fact that the whole project is rather a process than a business that can be eventually completed. I have already mentioned that the LHC is the world's most powerful accelerator (27 km ring diameter) designed to achieve 7 TeV energy for each of the two counter-rotating and colliding proton beams (see https://edms.cern.ch/file/445830/5/Vol_1_Chapter_2.pdf for the beam parameters, see also http://lhc.web.cern.ch/lhc). It is expected that the Higgs particle (see above) can be produced in the collision so that the mechanism ensuring the origin of particle masses in the Standard Model will be validated.

Still the questions remain, at least I don't quite see the answers. For instance, I don't completely understand the lack of gravity in the Standard Model, maybe some experts know better.

## 9.2    The Arrow of Time

"I can't go back to yesterday - because I was a different person then" Lewis Carrol.

In this section we shall discuss the subject of the unidirectional flow of time in general and of its various manifestations in the form of so-called time arrows. More specifically, we shall observe the properties of the time reversal operator together with some physical and mathematical aspects of time reversal symmetry. The whole subject is discussed here both on the classical and the quantum level, but mostly for the case without spin. Effects which are due to spin (such as spin-orbit interaction) are only briefly mentioned: a full-scope inclusion of such effects into our discussion would make the section hardly observable.

   This section is constructed in the following way: primarily some standard approaches to the problem of time reversal are briefly reviewed, with my personal comments being added. In the interim, it appeared necessary to dwell on the very concept of time, which unfortunately borders to metaphysics. Afterwards, more physical (and even some mathematical) stuff was discussed, mostly related to a correct definition and properties of time reversal operator. Some time-reversal noninvariant models, usually declared purely phenomenologically (as if the Newtonian model were not phenomenological), are observed under the angle of possible presence of the hidden time reversal symmetry when the coefficients of a phenomenological model are explained in "more fundamental" terms i.e., by using other models. The assumption of a closed physical system, needed for time reversal invariance, is discussed.

   At first, I did not want to include a section on time-reversal puzzles in this book. The reason for this reluctance was that the so-called problem of time is so strongly contaminated with fancy speculations and philosophical metaphors (see e.g., [53] and also [54]) that while discussing this subject one can easily be trapped by embarrassing pitfalls of vagueness. After a certain age one tends to think why she/he is doing this or that. I would rather prefer to solve equations than get engaged in some sort of foundational scholastics. By the same token, the issue of interpretation of quantum mechanics which is often considered extremely important (see e.g., [74]) is, to my mind, of a similar soft and scholastic nature. Issues of this kind, to my understanding, are not quite physical problems because they are unlikely to bring new results. The only justification, in my opinion, to handle such highly speculative issues is what I call the "physmatical effect" - emergence of unexpected links to other areas of physics and mathematics (see in this connection the works by I. Prigogine, e.g., [55]). Moreover, the problem of time-reversal asymmetry is a very ancient issue and a great lot has been already written about it e.g. Davies, Hoover [59,60], so I did not think I could add anything new and fresh.

   But then I suddenly caught a serious illness and, by necessity, got more time for relaxed contemplations. Once professor Christoph Zenger, a well-known mathematician, whom I consider one of my teachers, came to visit me and asked what, in my opinion, the physicists think in general about the unidirectional flow of time. Professor Zenger's idea was that it is probably the dominance of matter over antimatter that produces the time asymmetry. I argued that this view seems a bit naive as well as considering that time-reversal symmetry should be a priori guaranteed; moreover, even in the microscopic world this statement would be wrong since only $C$ and $T$ symmetries would be handled, with parity $P$ remaining untouched, provided of course we believe that the $CPT$ theorem is never violated (see below). And who can guarantee that $CPT$ is true also in the macroscopic world? So, I immediately thought this issue is not worth serious treatment. Yet afterwards I decided to systematize a little what I knew on this subject. Thus, the present section appeared.

   The issue of a preferred direction of time may be classified as one of Perennial Problems. Such problems have a philosophical flavor and although they can admit simple formulations, they present a perennial challenge for great scientists.

One of distinguishing features of Perennial Problems is the illusion that they have been solved a long time ago, with this solution being known to any more or less literate person. However, if one addresses original papers and textbooks, one will find a wild variety of opinions and would-be solutions to each Perennial Problem. So, there is no universally accepted solution to any of the Perennial Problems despite the fact that a lot of great minds have tried to attack them. As examples of Perennial Problems one can name such issues as: interpretation of quantum mechanics, unification of fields and forces, unification of general relativity and quantum mechanics (the quantum gravity problem), the problem of the origin of the universe and its subproblem - that of initial conditions, the problem of fundamental constants - of their actual values and possible variation, of elementary particle masses, of causality, and a number of other fundamental problems essential for our representation of the world. Nowadays, one more issue is of fashion and seems to become a Perennial Problem: that of dark energy and dark matter.

### 9.2.1    Perennial Problems

Children sometimes ask: what has become of the previous days, where are they stored? Or they are ruthlessly destroyed by some specific creatures, as in the well-known fantasy by Stephen King ("The Langoliers")? It would of course be great being able to turn back time. The issue of the preferred direction of time may be classified as one of Perennial Problems. Such problems have a philosophical flavor and although they can admit simple formulations, they present a perennial challenge for great scientists. One of distinguishing features of Perennial Problems is the illusion that they have been solved long time ago, with this solution being known to any more or less literate person. However, if one addresses original papers and textbooks, one will find a wild variety of opinions and would-be solutions to each Perennial Problem. So, there is no universally accepted solution to any of the Perennial Problems despite the fact that a lot of great minds have tried to attack them. As examples of Perennial Problems one can name such issues as: interpretation of quantum mechanics, reductionism (i.e., possibility to reduce biological phenomena to physics), unification of fields and forces, unification of general relativity and quantum mechanics (the quantum gravity problem), the problem of the  origin of the universe and its subproblem - that of initial conditions, problem of fundamental constants - of their actual values and possible variation, of elementary particle masses, of causality, and a number of other fundamental problems essential for our representation of the world. Nowadays, one more issue is of fashion and seems to become a Perennial Problem: that of dark energy and dark matter.

Some people tend to call the above examples of Perennial Problems also the "Princeton Problems": they are not really important from the utilitarian viewpoint, at least for the time being, and they are neither physical nor mathematical with regard to their setup. That is to say that such problems are not necessarily reduced to a completely and correctly set task. One can be preoccupied with such problems for infinite time - they are just Perennial. I don't think this label is correct: one can judge by the publications that the Princeton Institute for Advanced Study (IAS) has lately become considerably less detached from practical problems.

It is curious that Perennial Problems though being totally irrelevant to daily life[182] tend to stir much more acute interest than mundane everyday ones. Take, for instance, the Fermi paradox concerning the existence of extraterrestrial civilizations. One usually gets orders of magnitude more responses when starting a discussion on this abstract subject than, say, on such vital issues as outrageous medical errors or, say, police brutality. The nearby subject of the so-called Drake equation, an attempt to estimate the potential number of extraterrestrial civilizations in our galaxy - the Milky Way (see http://www.seti.org) still attracts a great number of enthusiasts. The SETI community grows and is supported by decent funding. One might however wonder whether the search for extraterrestrial creatures is a promising scientific avenue to pursue. To be honest, I am inclined to think that anyone who is seriously engaged in the search of extraterrestrial civilizations manifests some escapism and possibly has difficulties in dealing with the boring, often unpleasant but necessary hardships of daily life.

Although Perennial Problems are mostly of quixotic character, not all of them are totally useless from a pragmatist's viewpoint: many are even not too remote from real scientific quests. Specifically, many arguments related to time inversion are not always of purely scholastic character. Thus, in high energy physics such arguments are essential for the development of a theory describing the fundamental interactions. For instance, the famous CPT theorem[183] provides guidance for the construction of any viable field theory involving particles and antiparticles. More specifically, by combining CPT with internal symmetries one can introduce the correct transformations, e.g., expressed through matrices. One may also recall more practical things such as the phase conjugation techniques in optics and acoustics. Nevertheless, the main point of mine is that discrete symmetries, in particular time reversal symmetry, do not follow from any fundamental principles, and there is no reason to declare such symmetries to be fundamental principles by themselves.

Thus, Perennial Problems border on really fundamental questions (some of them we shall discuss below), more specific than Perennial, namely - How did the universe originate? - What is the origin of mass and what stands behind the differences of masses of particles we call fundamental, in particular leptons and quarks? - What is the reason for the matter-antimatter asymmetry observed in our universe - What is "dark matter" and "dark

---

[182] The well-known medieval Perennial Problem: "How many angels can dance on the head of a pin (or on the point of a needle)?" seems to have many today's analogs. In Russia, for example, such eternal topics as the "special way of Russia" were a signature of the intelligentsia at all times.

[183] The CPT theorem deals with charge conjugation (C), parity (P) i.e., spatial inversion, and time inversion (T).

energy" which apparently manifest themselves without being directly observed?

Perhaps today's Perennial Problems will be sharpened in future physical experiments, and then their speculative ("many-words") component will be drastically diminished. In this connection one might recall that theories and models are always abound whereas there is only a single reality.

### 9.2.2    Observations of Possible TRS  Breakdown

The time reversal symmetry (TRS) is obvious in many basic equations of physics but is exceptionally seldomly manifested in reality. One should probably always remember that mathematical equations are just pieces of human-produced text and can be linked to reality through careful observations and purposeful experiments. It has already been noted that without paying attention to external measurements, physics would look like a loose collection of philosophical speculations or, at best, a vast array of mathematical texts. Time reversal symmetry is a so deeply engraved stereotype in physics that to question it is not at all easy: one should overcome a cognitive barrier of the prevailing opinion. Thus, TRS may be viewed rather as an important philosophical principle i.e., without the necessity to have an experimental link to reality. Below we shall estimate what limitations are imposed by such links.

Let us briefly return to basics: in 1686, Sir Isaac Newton presented to the Royal Society the first volume of his "Principia" (*Philosophiae Naturalis Principia Mathematica*)[184]. In this first volume three laws were postulated that were destined to describe the world. Firstly, any physical body maintains its state of rest or motion[185] unless acted upon by an external force; secondly, this force equals the rate of change of the body momentum; and thirdly, each action entails an equal and oppositely directed reaction (counteraction). Those three laws named later after Newton were presumably sufficient to determine the fate of each particle of the universe since to predict the future state *ad infinitum*.

Time reversal invariance is a particular case of a more general anti-unitary invariance so from symmetry positions there is nothing specific about time reversal. Simultaneously, some quantities are conserved that are invariant in time such as energy. The presence of even very small dissipation or diffusion breaks time reversal invariance and tends to drive the physical system to homogeneity. Since time reversal is a simple transformation, the problem of time-asymmetry is simple to formulate: if one believes physics, which is considered to lie in the foundation of the human knowledge of the world, to be based on classical mechanics of Galileo-Newton which is time-symmetric, then how would it be  possible that real-life processes are obviously time-asymmetric. The subjective experience of an essential difference between past and future in real life i.e., between two possible directions of time has been metaphorically called "the arrow of time" (probably it was A. S. Eddington who was the first to coin this term). The motion equations of classical, non-relativistic quantum mechanics (without the magnetic field) and, to a certain extent, of quantum field theory (QFT) seem to admit time reversal $t \to -t$ , although in QFT simultaneously with CP-transformation. Macroscopic equations are, however, irreversible. Reconciliations of fundamental and phenomenological models of the world has traditionally been one of the most burning issues in physics. This paradox, known since the end of 19th century, is sometimes called the time arrow problem or "global irreversibility paradox" and I shall try to make a short overview of different approaches to its resolution.

---

[184] Actually, this endeavor comprised three volumes, see, e.g., Kartsev, V. P. Newton [245].;                                    see                                    also http://books.google.com/books?id=6EqxPav3vIsC&pg=PA1#v=onepage&q&f=false.

[185] We have seen that according to Galileo the rest is a particular case of the motion.

From a very general observation, one may notice that asymmetry is much more generic than symmetry, the latter being a very special feature. So, one might suspect that it would be an extraordinary and very intelligent fine-tuning if all the laws of nature were time-reversal symmetric. The probability of more generic asymmetric manifestations must be significantly higher. Looking for symmetries in the equations is an important area of mathematical physics which provides useful tools [187], but imposing symmetries on every natural process in sight seems to be a totally arbitrary act. Nevertheless, it is generally believed that all dynamical equations of fundamental physics are invariant under time reversal whereas the phenomenological and hence "less fundamental" laws of physics have an obvious temporal asymmetry. The microscopic[186] dynamical equations govern the evolution (or devolution) of a physical system under the assumption that it is perfectly closed. If the system is not closed, then it is a priori phenomenological, e.g., dissipative (the simplest example is the damped oscillator), and, strictly speaking, its complete description lies outside of microscopical mechanics - in statistical physics or physical kinetics. If the system is not perfectly closed, it does not seem possible to screen it from fluctuations, e.g., of thermal character. However, the assumption of perfect closure seems to be utterly unrealistic and can therefore be substantially weakened, as I shall try to demonstrate a little below.

Although one cannot exclude the low-probability option that we are living in a completely time-symmetric universe, the usual statement that all physics should be time- reversal symmetric seems to be a belief, a quasi-religious credo. One might observe in passing that there are surprisingly many credos (sometimes called principles) appealing to intuitive extrapolations in physics. Maximum what can be accurately said about time-invariance is the famous $CPT$ theorem [99], 19, see also below. Its content is as follows. Assume that we have a quantum field theory which is characterized by a positive energy density, Lorentz invariance and local causality[187]. Then such a field theory is invariant under $CPT$.

---

[186] They are only considered microscopic by some consensus as constituting the backbone of contemporary physics: Newton's equations of classical mechanics are by no means microscopic as well as the Schrödinger equation applied to the whole Universe.

Furthermore, it is typically assumed that any $C$ or $P$ non-invariance are inessential, especially in practically important physical manifestations, therefore all physical laws should be invariant under $T$ transformation (colloquially, $T$-invariant). This chain of arguments is in a sense remarkable: nearly everything is false. First of all, the conditions for the $CPT$ theorem are very stringent - to the degree of being applicable to a very limited set of quantum field theories. It raises little doubt that the universe we live in is not Lorentz-invariant: in Einstein's general relativity spacetime is not flat, nor even asymptotically flat. There exist some rival models with asymptotically flat spacetime, but they still do not ensure $CPT$. This problem is closely connected with the energy density in the universe. This energy density is not necessarily strictly positive, in particular, because the potential energy related to gravitational attraction is negative [188]. Local causality cannot be in general a well-defined notion, especially when the space- time metric fluctuates or is quantized. Moreover, it is in general not true that one can correctly define a spacetime separation for an arbitrary quantum state. Thus, there may be difficulties already with the statement of the $CPT$ theorem.

Let us pay some more attention to the assertion that $C$ and $P$ invariance play a negligible role in practically important physical processes. The curious thing about this assertion is that it contradicts the general logical scheme of microscopic reasoning. On the one hand, it is generally stated that all physics is presumably time-invariant on the level of microscopic laws, on the other hand such physical manifestations as kaon decays and in general weak interactions are considered too microscopic to be viewed as physically significant. Personally, I do not understand this logic.

One standard explanation of the time reversal violation in macroscopic physics is based on time-asymmetry introduced by some supplementary - initial or boundary - conditions (see more on that below). This explanation seems to me at least insufficient since the disparity of two different directions of time in the entire picture of the world remains. As far as the hypothesis of electron-positron and, more general, particle-antiparticle asymmetry resulting in time reversal asymmetry goes, it appears also limited and even a bit naive since it a priori believes that time reversal invariance is deeply inside an exact symmetry. Of course, at a somewhat primitive level of description an anti-particle may be viewed as a particle moving backwards in time. Indeed, a particle having a positive energy $E$ contains in its wave function factor $\exp(-iEt)$, i.e., $\Psi_+ = \psi \exp(-iEt)$. An anti-particle having a negative energy $-E$ is characterized by wave function $\Psi_- = \psi \exp(-i(-E)t) =$

---

[187] The term "local causality" refers to a situation when the fields of QFT, $\varphi(x), \varphi(y)$ commute (or anticommute) if spacetime points $x$ and $y$ are separated by a spacelike interval.

[188] The question of the energy density in the universe is considered a "hot subject", see, e.g., Sean M. Carroll https://link.springer.com/article/10.12942/lrr-2001-1.

$\psi \exp\left(-iE(-t)\right)$. In the framework of QFT[189], time reversal invariance seems to be violated, which has been experimentally confirmed in CPLEAR experiments conducted in 1998 in CERN [63][190]

It would be of course a truism to state that in classical (Newtonian) physics there is no preferred direction of time. In physics, microscopic time reversibility is actually a consequence of idealized modeling of physical experience. We have seen in Chapter 4 that classical dynamics, being very approximate and of limited applicability, had to be deterministic in order to predict the flow of events with elapsing time. To a large extent, the same is true for quantum mechanics (see Chapter 6). Otherwise, theories or specific models derived from them were of a very limited use. Nevertheless, there are strong controversies among physicists - as well as among philosophers - how to explain the obvious discrepancy between two facts: 1. we all know that time is flowing only in one direction, at least at our macroscopic level (trivial example - thermal or diffusive processes); 2. all basic (microscopical) laws of physics are invariant under time reversal, i.e., change $t \to -t$. One may find the most comprehensive account of the time asymmetry problem in [89]. In classical deterministic physics, the time, although not satisfactorily defined, was considered a universal continuous[191] variable. Due to this approach, classical and even quantum theories were mostly formulated as models based on differential equations of evolutionary type (or on systems of such equations) giving explicit time derivatives as functions of current values describing the actual state. It is this mathematical construction that allowed one, by integrating these differential equations, to "predict" the variable future or retrodict the constant past, depending on setting up initial or final conditions. Moreover, ordinary everyday life entices us into thinking that time is absolute and the same everywhere, which is connected with the erroneous impression that speed has no limit. Yet, this is true only in a low-energy approximation.

Although time-reversal invariance is mostly taken for granted, there is no compelling reason why this should always be the case. And indeed, more and more evidence has been accumulated lately that time-reversal symmetry (TRS) is broken on a more sophisticated physical level than simple classical or quantum models. Please note that here I am not talking about statistical problems or open systems, where it would be ridiculous to require time-reversal invariance. There is currently much empirical evidence that TRS does not hold in optics, for examples, in experiments with nonmagnetic metamaterials (see, e.g., an account of spectacular experiments carried out by the N. I. Zheludev group in: [251], [252]). One of the relevant hypotheses is that TRS breakdown may accompany a nonlocality of the optical response, when surface plasmons are excited depending on the polarization of light incident on intricately surface-structured metamaterials. Probably these experiments as well as metamaterials in general require quite an extensive analysis, see the cited papers for details.

---

[189] With Minkowski flat background and Lorentz invariance, motion in different Minkowski frames, as already discussed, can be represented by different 4d graphs, with Lorentz transformations being mappings between them.

[190] In a series of CPLEAR experiments, weak decay of K-mesons has been observed, $K_L \to e^+ \nu_e \pi^-$.

[191] Today there are of course also discrete-time and lattice models, but their detailed discussion is outside the scope of this book.

Another field of interest where time-reversal invariance is apparently broken is high- temperature superconductivity (see, e.g., a discussion in the papers [253], [254] and [255]).

Time-reversal symmetry has also been shown to break down in biological molecules [256].

To my regret, I can never undo things - maybe some people can. And I am unable to see the reversed chain of events: my personal time clearly flows in a single direction. Most people, especially elderly persons, are sadly aware of the passage of time; it is usually in a very young age one wishes the time run faster. It would be a truism to say that in general young and old (like rich and poor) fathom the world by different scales. Moreover, the perception of time intervals also seems to change with the age of an individual, these intervals appearing progressively shorter[192] One can easily construct a mathematical model of this diminishing of time intervals with the life-span. An intuitive awareness of the unidirectional and accelerating time flow manifests the psychological time arrow. This is, of course, no physics - so far. However one may justifiably ask: what are the physical reasons for the perception of time as having a sole direction? Although experimental proofs, both direct and indirect, of time reversal non-invariance are quite convincing [63], recognition of such non-invariance within the physical community unexpectedly turns out to be reluctant, sometimes the attitude towards these experimental *facts* is averse and disapproving. The reason for such an attitude is exactly the widespread stereotype that the basic laws of physics are written in time-reversal invariant form, so all the phenomena should be - at the basic, microscopical level - also time-reversal invariant. Thus, the tension between the generally perceived microscopic science and common sense is reflected in the question of why the past is different from the future.

### 9.2.3    Model-Based Claims

One sometimes forgets that the so-called basic laws of nature are formulated as mathematical models, quite efficient but having a limited domain of applicability. For example, Newton's equations are formulated, in distinction to, for example, the Aristotelian model, as a time-invariant mathematical model, so it is quite natural that Newtonian mechanics, for instance, does not automatically involve time-asymmetric phenomena, e.g., omnipresent in optics where the difference between time-forward and time-reversal processes has been observed for many years. It would be stupid to require of a model to be applicable in the field it has no relation to - this general consideration is often forgotten when time-invariance issues are discussed. Reversibility of the main microscopic laws of physics is nothing more than a very productive assumption. The present-day physical laws are just mathematical models, nothing more, and they tend to be replaced by other models when ample experimental facts are accumulated to invalidate the actual beliefs. For instance, such a simple system as a damped pendulum already gives an example of time-invariance violation, irrespective of imposing initial or final conditions which are often taken as a main source of time-invariance violation. The model itself is non-invariant in time. The verbal claims of reducing this model to truly time-invariant Newton equations for some point particles are just beliefs of a rather philosophical nature[193]. One may notice that in order to define time derivatives one already assumes that the direction of time does exist. Thus, the arrow of time is implicitly suggested by the very formulation of the theory, which is mathematically equivalent to a Cauchy problem.

---

[192] "The more we live, the shorter are the years", Bulat Okudjava.

Time in the Newtonian model as well as in dynamical systems theory is an absolute mathematical notion (Galilean fibration) flowing uniformly with no regard to physical processes. This representation is false already for simple relativistic models. Furthermore, time is not an observable in the physical sense; the fact that struck me first when I started to work in the Russian (then Soviet) Committee for Standards. Meteorologists deceive people by saying that they are measuring time - in fact time is defined as some number of periods of some oscillating physical quantity that is measured by "the clock". Figuratively speaking, we see only the hands of the clock, but not the time itself, and we cannot place the time sample in the International Bureau for Weights and Measures headquarters at Sêvres, France (Bureau International des Poids et Mesures), which serves as a depository for the primary international standards. We would need a whole laboratory staffed with highly qualified physicists for producing some kind of clock, e.g., atomic clock, its certification and comparison, etc. This impossibility to see time directly is the consequence of the fact that it is difficult (maybe even not possible) to correctly construct the time operator as an entity corresponding to an observable (see below). Had we been able to introduce the time operator, we could have established its evolution properties (e.g., in the Heisenberg picture) and could have compared forward and backward temporal behavior. We could have found also the expectation values of the time operator and seen how they would change with time $t \in \mathbb{R}$ for $t > 0$ and $t < 0$.

Usually, considering a classical dynamical system, we assume it evolving from the past into the future, i.e., setting up the initial conditions on some manifold. However, from the theory of differential equations we know that one can specify the final conditions just as well as the initial conditions. Final conditions fixed at some distant time point in the "future" would produce the retroactive evolution - back in time. Yet, it will still be a dynamical system. Such time-reversed development is valid both for the customary dynamical systems described by vector ODEs and for field evolution described by partial differential equations (PDEs).[194]  In the case of fields described by PDEs, initial or final conditions are specified at some spacelike 3-manifold and this initial or final state evolves respectively into the future or into the past according to the dynamics provided by time-dependent PDEs which govern the evolution. Here, however, the time-reversal symmetry may be broken since the state to which the evolution strives may itself be determined by the sought-for solution, i.e., be a functional of it[195].

---

[193] The already mentioned Aristotle and Newton point mechanics are examples of intrinsically different schemes with respect to time reversal $T$, the first being truly time non-invariant, the second time-invariant.

### 9.2.4     Closed Systems

The notion of a closed system is a strong idealization, convenient in conventional mechanics but becoming inadequate in more refined physical situations. The time-reversible mechanical models are usually considered to always lie at the very foundation of entire physics, however this claim is not more than a belief. In most cases the statement of ultimate reduction to time-invariant mechanical models appears to be right, but nobody can guarantee that all physical states and systems can be reduced, in the final analysis, to the relatively simple - and time-reversible - mechanical motion. Even less so as regards real-life physical situations (they do not necessarily coincide with the considered physical systems). Thus the frequent statement "All physics is basically time-reversible" may be only referred to the systems that can be reduced to a simple mechanical motion of particles. In particular, this motion should be confined within the closed system and any fluctuations should be absent.

### 9.2.5     Irreversibility and Time Reversal Noninvariance: Remarks about Terminology

Strictly speaking, time asymmetry, irreversibility and time-reversal non-invariance are not identical concepts, although they are often used interchangeably. Of course, it is a matter of defining the terms. I understand, for example, time asymmetry as the property of time-asymmetric solutions to time-symmetric equations. Usually, this property stems from a special choice of boundary conditions; the radiation condition in classical electrodynamics or the selection of an outgoing wave in quantum scattering theory are typical examples. These retarded solutions lead to the so-called radiation arrow of time (see below). In classical theories one typically has time-symmetric dynamical equations, however not always - the damped harmonic oscillator is a counterexample. In contrast to this, time-reversal invariance violation may be understood as a non-invariance of the underlying dynamical equations, of the Hamiltonian, the Lagrangian or the action with respect to the time-reversal operator. The latter was first introduced by E. Wigner in 1932 and, simply speaking, is the antiunitary operator $T_-$ defined by the complex conjugation in the coordinate representation of the wave function: $\Psi(x,t) \rightarrow$ $\Psi^*(x, -t)$. One must also assume $\Im V(x) = 0$, i.e., there are, for example, no unstable or chaotic subsystems, excited and decaying states or creation/annihilation of particles, which is a seemingly innocent but in fact a very strong assumption. Mathematically speaking, the question is whether the Hamiltonian is defined over the Hilbert space or outside the Hilbert space. The standard quantum theory is basically the theory of linear self-adjoint operators in Hilbert space (see Chapter 6) leading automatically to a unitary (the Stone-von Neumann theorem) and therefore time-reversible evolution. Thus, the orthodox quantum theory in Hilbert space is time symmetric. This is sufficient for solving steady-state structural problems like e.g., calculation of optical spectra. The Hilbert space technique, however, becomes mathematically inconsistent when one needs to include decay processes, short-lived resonances and other transient phenomena. One may notice, for example, that in the quantum scattering theory the discrepancy arises between the operator Hilbert space domains and the time asymmetry of outgoing or resonance solutions. There exist even paradoxes like causality breakdowns [61,62] or deviations from the exponential decay law. One usually resolves this conflict purely heuristically - e.g., by selecting the sole outgoing wave, which automatically leads to the time asymmetry. Another way to circumvent this problem is to use only one type of Green's functions, retarded or (rarely) advanced. However, by using Green's functions one is forced to introduce distributions and, strictly speaking, to move outside the Hilbert space. Basically, the same difficulties and similar procedures are related to the time-reversal non-invariance in the classical theory of electromagnetic radiation.

---

[194] Examples of such dynamically evolving fields are fields in electrodynamics and general relativity or wave function in quantum mechanics.

[195] This situation reminds us of the self-consistent field approach in condensed-matter physics.

The problem of selecting the correct solution formulated in physics as finding the scattering (or outgoing radiation) states is closely connected with the theory of differential equations. For example, if a differential equation has some localized solutions, and the solutions may be parameterized that is represented as a function of some parameters, in case it is a single parameter $t$ we can ascribe to it the conceptual meaning of time. The scattering states emerge if e.g., two such solutions are initially (in the "remote past", $t \to -\infty$) set up in spacelike domains "far away" from each other, then, with increasing $t$ these solutions move towards each other, collide, i.e., interact according to the governing differential equation and then drift apart in the "distant future", $t \to \infty$. One can then introduce the $S$-matrix (scattering matrix), first introduced by W. Heisenberg, in order to couple the solutions related to the "remote past" to those for the "distant future" [296]. Then the analytical properties of the $S$-matrix determine the time-reversal conditions for the scattering solutions of the governing differential equation. Usually, the scattering states require us to go outside of the set of Hilbert space solutions.

Finally, irreversibility is commonly associated with the increase in entropy, no matter how you define this quantity. Thus, irreversibility implies the thermodynamic arrow of time, both in classical and quantum physics. This subject seems to be especially prone to misinterpretation (see below). It is interesting to find out why the thermodynamical arrow of time being determined by entropy growth points in the same direction as all other time arrows. If the thermodynamical arrow of time is coupled with the cosmological expansion, then what would happen if and when the actual expansion phase were to be replaced by the contraction period, as it is envisaged in many cosmological models. According to such models which are often called cyclic, the universe expands for a while after the Big Bang, before the gravitational attraction of matter causes it to collapse back into the Big Crunch. Would the irreversibility change its sign and point in the direction of increased order when the universe undergoes the bounce? Would the "contramotion" - motion back in time be as ubiquitous as the normal motion now? What would happen to our clothes when they reach the time point when they have not been manufactured? How would living organisms process food? S. W. Hawking in his famous book [78] gives an example of a broken cup that would reassemble itself. Nevertheless, Hawking asserts that intelligent life could not exist during the contracting phase of the universe.

If I were a carpenter, and you were a lady, would you marry me anyway, would you have my baby.

Our perception of time is determined by entropy growth. The subjective feeling that we are flowing through time is determined by the fact that we, as living creatures, exchange information (entropy) with the outer world, its entropy in this process being increased. Time exists independent of entropy, but entropy gives us, macroscopic many- body creatures, the sense of time direction. Time flow against the entropy increase would be perceived as the wrong direction. Thus, although irreversibility is connected with the unidirectional flow of time, these two phenomena are not identical. Below, we shall discuss irreversibility in more detail.

One should remember that time-symmetrical models is only a special class of mathematical models developed, in particular, to describe the greatly idealized motion of individual particles, with the states being points rather than measures, which excludes the possibility that we may have only a probabilistic knowledge of the "exact" state. There is no dispersion in measurement of any observable in such "pure" states, since they are only point measures and, mathematically, only represent extreme points of some convex sets. Such models can be too idealized to judge about the time-reversal features of reality in general. Matter rarely consists of individual particles, and the macroscopic parameters in terms of which we describe matter such as temperature, density, pressure, elasticity, viscosity, etc. are not necessarily uniquely determined by pure and idealized mechanical models and pointlike mechanical variables. For instance, such mechanical models are totally inadequate for condensed matter physics where, due to many-particle effects and quasiparticle excitation, only time-noninvariant solutions are physically meaningful.

We usually hear some objections to this type of reasoning claiming that mechanical models are "more fundamental" than, for instance, thermodynamical or statistical, since the latter involve many particles and such loosely defined notions as temperature or entropy. I would respond that this statement is hardly true. First of all, as I have just mentioned and we have seen more in detail in Chapter 2, the state of a physical system is represented by a Liouvillean measure, with the pure state being only a particular case (extremal points of a convex set). Furthermore, in the accelerated system or in the vicinity of a gravitating body the temperature measured by an observer would be increased, e.g., from a "mechanical" zero to "thermodynamical" non-zero values. This is the so-called Unruh effect [88], which is a simple and at the same time profound model worth of reviewing (see below).

Before we proceed to the discussion of time asymmetry and related problems, allow me to make one more general remark about the models of physics (see Chapter 2). Physics may be interpreted as the game consisting in attaching mathematical expressions to pieces of reality, with a subsequent interpretation of obtained results. There exist many ways of reality processing: an artist would paint a picture, a composer would create a tune, a poet would write a poem, a scientist would produce a mathematical model, i.e., a simple account of a piece of reality encoded in mathematical jargon. There are at least two things common for all these activities. Firstly, by producing new information they appear to decrease disorder, i.e., they act in the direction opposite to the thermodynamical time arrow. However, a more detailed analysis [249] shows that there is an energy price for the information so that the energy dissipated (in entropy terms) in the process of creating it largely exceeds the gain in information, which means that the net result of human intellectual (e.g., modeling) activities still increases entropy of the universe and does not reverse the thermodynamical arrow of time. Secondly, all forms of human reality processing bear an imprint of transient time, and it would be difficult to imagine the time-reversed reality for e.g., poem writing. This would imply the causality breakdown, since the cause (e.g., a poet's intention) would follow the result. Of course, it might be "logically" possible to imagine that the causality does not exist or reverses chaotically, so that causes occur sometimes before and sometimes after the consequences, but nobody, as near as I know it, have ever observed this effect. Thus we have to assume that the existence of causal and psychological time arrows is a well-established experimental fact, which should be also reflected in mathematical models of physics. I have repeatedly mentioned - and we shall see it further a number of times - that mathematical models of physics are local but linked, just like the nodes of a network. Treating some local model is like sending a signal throughout the whole network; if you pull a seemingly innocent thread (in this case, the time asymmetry issue) the large part of the entire physical web responds. Thus, the discussion of the time reversal problem would lead us to such presumably detached subjects as the spectral properties of linear operators, radiation conditions, Liouville measure, entropy, and even such an uncommon device for traditional physics as Hardy spaces.

### 9.2.6    *The Time Operator*

There exists a sizable amount of literature on the subject of "time operator", which testifies to the fact that a special and somewhat strange status of time in science and life has been a serious challenge not only to philosophers, but also to physicists and even to mathematicians. There have been many attempts to treat time as a physical observable, i.e., to construct the respective

- and meaningful - time operator, both in quantum and in classical mechanics. Probably the first attempt to define the time operator was endeavored by W. Pauli [46] who proved that there cannot exist a self-adjoint time operator $T$ that could canonically commute with the Hamiltonian $H$ (Pauli did not consider mixed states described by a density operator) whose spectrum is bounded from below. The latter requirement is quite natural if one wishes to ensure the stability of matter. Pauli found that this requirement rapidly comes in contradiction with canonical relations for the would-be time operator. Before we proceed to discussing Pauli's proof, let me explain why it is important for establishing the time-reversal properties. The matter is that if we have the time operator we can explore its properties, e.g., find its domain, range, eigenvectors $|t\rangle$ and eigenvalues $t$ (which should give the physical time at any moment). We would see what happens with the change $t \to -t$, provided of course the operator could be defined for $t < 0$. If, for example, the expectation value of $T$ in some physical state $\Psi(x, t)$ is $\langle\Psi(x, t)|T(t)|\Psi(x, t)\rangle$, then what would be the expectation value in the time-reversed state, $\Psi^*(x, -t)$? In other words, is it true that

$$\langle\Psi(x, t)|T(t)|\Psi(x, t)\rangle = \langle\Psi^*(x, -t)|T(-t)|\Psi^*(x, -t)\rangle \qquad (9.1)$$

for $\Psi(x, t)$ satisfying the time-dependent Schrödinger equation with any possible Hamiltonian? This relation obviously does not hold for non-Hermitian Hamiltonians, there may be also time-reversal violations for the motion in an electromagnetic field. It is well-known that for such motion the time-reversal symmetry only holds under the condition of simultaneous change of the magnetic field sign (i.e., vector potential $A$) [39], §17. In other words, if some motion is observed in an electromagnetic field, then the time-reversed motion is also possible, provided the magnetic field vector changes its direction. This requirement seems to disturb physical intuition: how can one invert, for example, the magnetic field of the Earth in order to produce the reverse motion of charged particles? Spin components in the non-relativistic Hamiltonian introduce supplementary difficulties that can be correctly discussed only within the framework of quantum field theory (QFT).

It is important to emphasize - although this is of course a trivial remark that one should not confuse the time-reversal operator $T_-$ (discrete symmetry), time-evolution operator $U(t) = \exp(-iHt)$ where the Hamiltonian serves as the generator of temporal translations and the hypothetical time operator whose very existence is quite doubtful. From the Schrödinger equation one can readily see that provided $\Im V(x) = 0$, $\Psi(x, t)$ and $\Psi^*(x, -t)$ satisfy the same equation, moreover $\Psi^*(x, t)\Psi(x, t) = \Psi(x, -t)\Psi^*(x, -t)$.

### 9.2.7    Elementary Properties of the Time-Reversal Operator

Quantum mechanics is different from classical mechanics, in particular, by the fact that in the classical world all the transformations may be done infinitesimally, whereas in the quantum world discrete symmetries play a crucial role. Time-reversal invariance appeared in the quantum context due to E. P. Wigner in 1932 [152].[196] The Wigner $T_-$-operator changed $t$ to $-t$ and had the following elementary properties. The time reversal operation applied to a state $\Psi$ may be generally expressed as $T_-\Psi = M\Psi^*$, where $M$ is some constant unitary matrix. Then the time reversal operation applied to a matrix or operator $A$ may be expressed as $A(-t) = MA(t)M^{-1}$. If $A(-t) = A(t)$, the operator $A$ is often called self-dual. A physical system is invariant under time reversal if its Hamiltonian is self-dual i.e., $H(t) = H(-t)$. A system without time-reversal invariance would have a Hamiltonian which is represented by an arbitrary self-adjoint matrix free from constraints of being symmetric or self-dual ($H(t) = H(-t)$). If the Hamiltonian $H$ is time-reversal invariant (self-dual), any unitary matrix $A$ which is a function of $H$ will also be time-reversal invariant (self-dual).

### 9.2.8   Time Operator in Classical Mechanics

I hope we have acquired some fragmentary mathematical slang to express simple things in highbrow form. In the awesome mathematical language, the laws of classical mechanics are time-reversible if there exists an involution $T$ giving a bijective mapping between the time-reversed motion of each state and the time-forward dynamics of the respective state expressed by the symbolic operator equation:

$$U_{-t} = TU_tT \qquad (9.2)$$

Physicists hardly would express it this way. Recall, nonetheless, that involution is a map $f(x)$ that is its own inverse so that $f\big(f(x)\big) = x$ for all $x \in D$ where $D$ is the domain of $f$ or, in the language of group theory, an involution is an element $A$ such that $A^2 = E$ where $E$ is the identity element. To put it simply, an involution is a function that, being applied twice, takes one back to the initial position. All the transformations in the *CPT*-theorem (see below) are involutions as well as well-known complex conjugations and matrix transpose. Not all models of classical mechanics admit such an involution, for example, dissipative models in general don't since they produce certain time-independent structures such as fixed points and attractors that are not necessarily self-symmetrical under the action of the $T$-operator. In Hamiltonian dynamics, the mapping $T$ reverses the momenta of all the particles, so that the form of the Hamilton's equations of motion does not change. The Hamiltonian systems possess time-reversal symmetry ($T$-symmetry).

However, we have already seen that Hamiltonian systems comprise a very restricted class of generic dynamical systems (see Chapter 4, Section "Hamiltonian Mechanics"). Thus, Newtonian mechanics with friction is definitely not time-reversal invariant at the macroscopic level where it is usually applied, and it is a big question whether it is $T$-invariant at the microscopic level when one includes all the atomic motions into which the dissipated energy is translated. The matter is that entropy in this dissipative process is still increased which selects one direction of time.

---

[196] To this invariance was added the new quantum particle-antiparticle symmetry or charge conjugation (C) introduced by Dirac in the famous 1931 paper Quantized singularities in the electromagnetic field.

We have already discussed that one particular feature of time is that it cannot be observed directly: all time measurements are performed in an indirect fashion via reading out the state of a physical system we call the clock. If the clock is classical, i.e., we find ourselves within the field admitting an approximate classical description, we can consider any evolving "observable" $A(t) = A\left(t, x^i(t), p^i(t)\right)$, i.e., any dynamical quantity defined as a function of coordinates, momenta and, possibly, explicitly on the parameter $t$ we usually identify with time. [197]   The observable $A(t)$ evolves according to the Hamiltonian equation

$$\frac{dA}{dt} = \frac{\partial A}{\partial t} + \{A, H\} \tag{9.3}$$

where the symbol $\{,\}$ denotes the Poisson bracket (Chapter 3).

One may note in passing that this expression, which represents the flow along a vector field associated with the Hamiltonian $H$, the latter being treated simply as a function on the cotangent bundle $T^*(M)$ known as the phase space, provides a compact and symmetrical form of the whole Hamiltonian mechanics. In short, Hamiltonian dynamics is merely the flow along the symplectic vector field (symplectic gradient of $H$). The Hamiltonian itself serves here, like in quantum mechanics, as the generator of translations in time (see Chapters 4 and 6 for more details). Let us assume for the moment that the time operator $T$ really exists as a dynamical quantity and also that the system, for which we are going to consider the time operator (there may be different variants of such an operator for different physical systems), is "closed" in the classical sense[198], i.e., the Hamiltonian as well as the dynamical quantity in question ($A \coloneqq T$) do not depend explicitly on parameter $t$, $H = H(x^i, p^i)$, $A = A(x^i, p^i)$. Then we get, identifying the observable $A$ with time operator $T$:

$$\frac{dT}{dt} = \{T, H\} \tag{9.4}$$

with $T(t) = t$. The last expression contains an assumption that the classical operator $T$ and the parameter of the phase curves $t$, firstly, have the same dimensionality and, secondly, may be synchronized to the same initial point. The parameter $t$ is treated simply as a label for a one-parameter family of diffeomorphism-invariant observables. If we also suppose that the measurement taken on $T$ at any moment would produce the value of $t$, that is $\langle T(t) \rangle = T(t)$ where the angular brackets denote averaging over a classical state, then we have $dT/dt = 1$ and the Poisson bracket $\{T, H\} = 1$. It means that under all the assumptions taken, the classical time operator corresponding to the observed elapsed time must be canonically conjugate to the system's Hamiltonian. Being translated into the usual quantum language, it would mean that the energy-time uncertainty relation, $\Delta t \Delta E \geq \hbar/2$, should hold (see Chapter 6 for a detailed discussion of uncertainty relations).

[197] In some theoretical models "explicit time" is denoted by a different symbol, say $\tau$, to distinguish it from "implicit time" $t$ parameterizing geometrical structures in the phase space. There may be of course many "explicit times", $\tau_j, j = 1, 2, \dots$.

[198] In the quantum-mechanical picture it would correspond to a pure state when the system may be fully described by a wave function.

Here it is important to note that $t$ is just a parameter and not an observable in the usual sense of classical or quantum mechanics, since it cannot be interpreted as a function on $T^*(M)$ (the phase space), whereas the quantity $T(x^i, p^i)$ is already an observable - a dynamical quantity.

### 9.2.9   The Pauli Theorem

The general expression for the evolution of an expectation value of the operator $T$ is given by

$$i\hbar \frac{d\langle T \rangle}{dt} = \left\langle \frac{\partial T}{\partial t} \right\rangle + \langle [T, H] \rangle \tag{9.5}$$

One can easily corroborate this expression by writing down the scalar product $\langle T \rangle = (\Psi, T\Psi)$, performing the time differentiation and using the Schrödinger equation. Here, for simplicity we assume that the system is characterized by a time-independent Hamiltonian and also remains in the pure state. It is important to note that the expectation value of any operator $A$, $\langle A \rangle$, is just a number depending on $t$, when $A = T$ this number should be equal to $t$.

We have discussed time-dependent quantum mechanics in Chapter 6 in some detail, here I can only report that the usual Hilbert space formalism may prove insufficient to handle time-dependent problems, so one might be forced to introduce other spaces (e.g., so-called Hardy spaces). One may ask: why is it so difficult to define the operator of time? I have already mentioned that the simple consideration belonging to W. Pauli [46] shows where the difficulty lies. Assume that there exists some time operator $T$ which must satisfy the commutation relation $[T, H] = i\hbar I$ in order to ensure the fulfillment of the Heisenberg uncertainty relations (here we denote the unity operator as $I$). If $T$ is self-adjoint - it must be lest time eigenvalues were complex - then we may construct the unitary operator $W(z) = \exp\left(-\frac{i}{\hbar} zT\right)$ where $z \in \mathbb{R}$ is an arbitrary real number. Allowing $z$ to be complex would result in complex energy values. It easy to see, e.g., by using the exponent expansion and the identity $[A^2, B] = A[A, B] + [A, B]A$, that

$$[W(z), H] = \sum_{k=0}^{\infty} \frac{(-iz)^k}{k!} [T^k, H] = -zW(z) \tag{9.6}$$

Now let $|\Psi\rangle$ be an eigenstate of $H$ with eigenvalue (energy) $E$, then the above commutator gives $H\,W(z)|\Psi\rangle = (E + z)|\Psi\rangle$ , i.e., $W(z)|\Psi\rangle$ is an eigenvector of the Hamiltonian $H$ with energy $E + z$. This means that any spectrum of $H$, e.g., point-like or bounded, can be mapped onto the entire real axis, since $z$ is an arbitrary real number. In other words, the time operator canonically conjugated to semi-bounded Hamiltonians of nonrelativistic quantum mechanics should not exist or, at least, it can exist only when the energy spectrum of the system is continuous and unbounded from below. Thus, Pauli concluded "that the introduction of an operator $T$ must fundamentally be abandoned and that the time $t$ in quantum mechanics has to be regarded as an ordinary number." In quantum theory, time is totally classical - it must be read out from a classical clock.

There have been numerous attempts to circumvent this theorem of Pauli. Some of the authors who undertook such attempts pay attention to the fact that the domains and ranges of operators involved should be precisely analyzed [79]. In the physical language it would mean, for example, that a version of a quantum theory with an energy spectrum unbounded from below, e.g., similar to that of the electron-positron field in quantum field theory, is admissible. However, the "vacuum sea" of holes in the Dirac theory contains only occupied states for $E < 0$. We discuss the structure of the QED (quantum electrodynamics) vacuum and its local deformations, e.g., in a laser field, at some length in Chapters 6; now we need only to mention that there may be a connection between the energy spectrum of the system and its time-reversal properties. Another possibility would be to repudiate the canonical commutation relations allowing $[T, H] = i\hbar C$ where $C$ is some self-adjoint linear operator, not necessarily a multiple of the identity operator. In this case the spectrum of $H$ may be unbounded from below [80].

### 9.2.10  Time Reversal Puzzles

Is it true that if we were living in the anti-world the time would flow in the opposite direction? Or is it true that if the universe had contained positrons instead of electrons we could predict the past and make records of the future? Although these questions are rather metaphysical than physical, my answer to the both of them is "no". Before I try to explain this negative answer, I shall have to expand on what is usually meant by the direction of time, at least how I understand it. Some musing about the perception of the unidirectional flow of time invokes the notion of a number of the so-called time arrows, more physical than the above mentioned psychological.

Professor Zenger was right by indicating that it is the initial conditions at the early stage of cosmological development that result in general time-asymmetry. First of all, it is an observational fact that the universe expands, at least at the present era, so that there must be time-reversal asymmetry at the cosmological level - the cosmological arrow of time. This arrow is directed from the time domain when the universe's spatial dimensions were smaller to the time when they become larger. One may notice that the cosmological arrow of time is not based on any mathematical model or "law of physics", it is an observational fact. One may also notice that the corresponding mathematical model, general relativity, is based (in its standard form) on time-invariant differential equations (see, e.g., [39], see also Chapter 9 of the present book), but these equations admit the solutions that reinforce expansion versus contraction, again at least for the current era mass distribution. Cosmological models, in general, tend to be time-reversal non-invariant, see e.g., [75]. Furthermore, the dominant model of the naissance of the universe is that of a Big Bang started at some point. There exist standard objections that the expansion of the universe from this point will eventually cease and reverse, but these objections, mostly based on the well-known Friedman-Lemaître-Robertson-Walker (FLRW) mathematical model [39], see also [66], are rather induced by the desire to reconcile the observed cosmological expansion with the presumed time-reversal invariance than by the need to balance observation and opinion. The FLRW-class models are comparatively simple because they are completely spatially homogeneous and isotropic, i.e., admit a six-dimensional symmetry group. They are just quite convenient mathematical models for the average matter distribution on a very large scale fully disregarding the deviations from spatial uniformity and isotropy. We discuss this type of models in Chapter 8; now I am interested only in the time-evolutionary aspects of the universe starting from the Big Bang. The latter is usually understood as a singularity where spacetime curvature becomes infinite, and the universe rapidly expands outwards from this state. The expansion rate essentially depends on the spatial curvature of the universe: if this curvature is positive, the expansion is finally reversed and the universe collapses again into a singularity which is usually called the Big Crunch. In the simplest version of FLRW models, the Big Crunch may be considered an exact time-reversal copy of the Big Bang. However, this reversible model of cosmological time evolution seems to be rather crude. For example, one should take into account black holes that must be formed on a mass scale during the collapsing phase of the universe. Such black holes would produce some kind of "noise" which would drive the Big Crunch to a state different from that of a Big Bang. Generally speaking, it would be arrogant to impose our simple $T$-symmetry (i.e., time-reversal, $t \to -t$) postulated for elementary single-particle dynamical models on the whole universe with its curved space-time and rich amount of physical processes.

The model of a cyclic universe when the Big Bang/Big Crunch is just one act in an eternal sequence of fireballs starting the expansion and finalizing the contraction is thoroughly discussed in the recent book by the renowned cosmologists, P. Steinhardt and N. Turok [81]. The authors themselves consider the cyclic universe model "a radical alternative to the standard Big Bang/inflationary picture". We shall briefly discuss competing cosmological models in Chapter 9, now we need only to understand that there is one common feature in many of them, namely that the universe swings between sequences of Big Bangs and Big Crunches, i.e., between acts of extermination and reemergence. The cyclic models regard our contemporary expansion as just a transitory phase. However, retrogressing - extrapolating back in time - is still a difficult issue when observing the entire cosmological landscape. For example, I could not find a satisfactory answer to the naive question: how did the universe begin and why do you come to a cosmic singularity in a finite (i.e., to be comprised of fundamental constants) time?

It is generally considered that the world should be subordinated to quantum laws. In quantum mechanics, we mainly describe the state of a physical system by a wave function [199] - a complex-valued function on the classical configuration space $M$. If quantum mechanics can be applied to the whole universe, this naturally leads to the question: what is its wave function? A popular model of J. Hartle and S. Hawking [67] has proposed an answer. In the above paper, Hartle and Hawking, also somewhat naively, if one speaks of the quantum state $|\Psi\rangle$ of the whole universe, then $|\Psi_i\rangle$ corresponds to the Big Bang and $\langle\Psi_f|$ describes the Big Crunch. This concept leads to a vast collection of fascinating mathematical models unified by the name "quantum cosmology", almost all of them being highly speculative and deprived of any direct astrophysical verification. A detailed discussion of these models now would take us far away from the subject of time-reversal invariance, although the union of quantum mechanics and cosmology poses a number of questions pertinent to the time asymmetry. One such question is, for example, the applicability of the CPT theorem, one of the basic results of quantum field theory. The standard proof of the CPT theorem implies the flat Minkowski spacetime as the background manifold; nevertheless, CPT-invariance may still hold and lead to observed time asymmetry when the quantum state of the universe is defined as the Hawking path integral [82], see also below. Another question, very general and in fact related to the first one, is about the meaning of the wave function of the universe. The model of Hartle and Hawking treated the wave function of the universe as the probability amplitude for the universe to appear (actually, from nothing). Then Hartle and Hawking proceed to estimate this wave function using the path integral method (see Chapter 6). Conceptually, this is tantamount to the assumption that the universe may possess all possible life trajectorie histories. An ingenious trick employed by Hartle and Hawking was to compute the path integral in imaginary time instead of the usual time parameter $t$. This is the usual trick in path integrals theory (see Chapter 6, section on path integrals in QFT) - to replace the time variable $t$ by an imaginary number, then perform the calculations, and then analytically continue the answer back to real times. But in the quantum field theory a flat Minkowski background is taken for granted, and it is not so easy when there is a curved spacetime, for instance, a black hole around. Nonetheless, Hartle and Hawking managed to adapt the usual path-integral techniques to the case when spacetime is essentially curved as it should be in the presence of matter.

This is a difficulty with time in quantum mechanics. One prefers to consider the background spacetime as fixed and, in particular, to assume that there exists a well-defined time $t \in \mathbb{R}$. This means that quantum mechanics implies the approximation when large fluctuations in the spacetime metric can be disregarded.

---

[199] For simplicity, we regard only pure states in the cosmological context.

The idea for the trick with passing to imaginary time was probably twofold: first, to get rid of singularities (points where spatial curvature tends to infinity), which may haunt calculations with ordinary time, and second, to eliminate the difference between the forward and backward directions in the imaginary time so that it would be possible to go backward, turn around, make loops, closed contours, etc. Besides, there is no problem with end points, since the imaginary (complex) time does not have a physically defined beginning or end. In the Hartle-Hawking approach to quantum cosmology, the initial wave function of the universe is described by a path integral over a compact manifold $M$ with a single spatial boundary $\Sigma$. The wave function itself, according to Hartle and Hawking, is interpreted as describing all possible universes, being large near our universe, which is thus regarded as the most probable, and small in the vicinity of other universes (there may be an infinite number of them) where the laws of physics or set of constants are different. Since all possible universes in this model are described by the same wave function, there may be transitions between universes, although characterized by an exceedingly small (almost infinitesimal) probability because the wave function of the universe is largely concentrated in the domain of the universe we to live in.

It might seem a bit too arrogant to claim that the wave function of the entire universe could be known. We have not even been able to observe the whole universe. One might believe that by knowing the wave function of the universe one would know everything that has occurred in the past or will occur in the future. There are rumors that M. Gell-Mann, 1969 Nobel Prize Winner, asked J. Hartle once: "If you know the wavefunction of the universe, why aren't you rich yet?" probably meaning that J. Hartle would know beforehand all the stock-exchange results, casino roulette outcomes or lucrative investment possibilities.

However, this detailed knowledge is in fact an illusion, since the rules of quantum mechanics, due to the generalized uncertainty principle, preclude it. So, if the universe is governed by the quantum-mechanical laws, knowledge about the future as well as about the past would be possible only to the accuracy of quantum fluctuations which may be large. The same applies to the time reversal. In other words, given the basic law and some initial (final) conditions, the future (history) of the universe is by no means determined since the law is quantum mechanical thus giving only the probabilities for alternative futures or histories.

We have seen in Chapter 6 that the interpretation of quantum mechanics amounts to translation of quantum alternatives into classical language. In other words, it is through an understanding of the quasi-classical domain that quantum mechanical notions become useful for our everyday experience. We have also seen that the concept of decoherence as the transition to classicality consists in continual "measurement" actions on the quantum system from the side of the environment, with the result that quantum variables pass to the quasi-classical domain. This transition like any other measurement process selects only previous alternatives, and thus we have to embody the notion of the arrow of time (consistent with causality) in the transition to classicality through decoherence. It may thus seem that the arrow of time and the associated constancy of the past (what has happened in the past cannot be changed and does not depend on any new information obtained in the future) are the features only of the quasi-classical or classical domains, but this is not true, at least as long as we have to use the density matrix tools in quantum mechanics.

There has been a great lot of arguing in the physical literature about the Hartle and Hawking wave function. Roughly speaking, the wave function of the universe is some complex-valued expression defined over semi-classical configuration space. In general relativity, this is represented by all possible metrics on the manifold $\Sigma$ with no boundaries, analogous to a 3-sphere. The boundary condition for the universe, according to Hawking, is that it has no boundaries. The primitive form of the wave function of the universe may be written as

$$\Psi(x) = \int\limits_{\substack{\partial M = \Sigma \\ g[\Sigma] = h}} [dg] \exp\left(-\frac{S(g)}{\hbar}\right) \qquad (9.7)$$

A little later we shall deal with the ground state wave function proposed by Hartle and Hawking more explicitly and in mathematical terms. Now we are interested only in the time-reversal aspects of this model. The model of Hartle and Hawking is, of course, a highly speculative one but it has produced a major impact on the physical community and stimulated the explosive development of quantum cosmology. The Hartle and Hawking model also envisages the existence of "wormholes" connecting different universes. According to Hartle and Hawking, the multitude of universes implied by their wave function should be connected by wormholes, some of these universes being connected with many others, while others tend to be isolated. Migrating between the universes may be equivalent to traveling in time. The multiple universe models typically suggest that all the constituent universes are structurally identical - quantum copies from the same ensemble - but they may exist in different states related to different values of the time parameter. Therefore, the universes constituting an ensemble do not communicate, in the sense that no information passes between them. Nevertheless, in a number of "multiverse" models they may have the same values of the fundamental constants (their "set of genes") and be governed by the same physical laws. The state of the whole multiverse is obtained from the states of the constituent universes according to the quantum superposition principle and is described by a single universal wave function. In contrast to this multiverse, the many-worlds interpretation of quantum mechanics proposed by H. Everett (see Chapter 6) has a shared time parameter. The quantum mechanical controversy in general and with regard to the quantum time arrow in particular is ahead; now we want to emphasize that the quantum evolution of the universe imposes constraints on determinism and time-reversal properties of our world.

As to the time-reversed variant of cyclic models in which the universe undergoes an infinite series of oscillations, each beginning with a Big Bang and ending with a Big Crunch, the puzzles are still more annoying. Is the series of bounces between Big Bang and Big Crunch really infinite? What fundamental quantities determine the period of oscillations? What happens with the time arrows at various stages of a bounce? Shall we observe the resurrection of the dead at the time-point corresponding to the maximum of the scale factor, $\dot{a}(t) = 0$ (see [39], §112)? Is the final state of the Big Crunch the ideal time-reverse of the initial state of Big Bang? General relativity leads to a space-time picture where singularities are a universal feature of various cosmologies with a Big Bang; it appears that no feature of general relativity can prevent them (see [39], §§102-104, see also [88], Ch.27), although it is known that Einstein considered singularities a drawback of his own theory and spent a lot of time and effort to build cosmological models which would be free of them [91]. Thus, even despite the fact that the equations of classical general relativity are time-reversal invariant and, in the cyclic models, no memory of previous cycles would be preserved which means that the entropy rise would be eliminated, the final state of the Big Crunch still does not seem to be the pure time-reverse of the time-symmetrical (e.g., FLRW) Big Bang. The latter may be regarded rather as a melting pot of black hole singularities characterized by a substantial entropy (Bekenstein-Hawking entropy [87] HawkingEntropy), see [89].

Thus, time-reversal invariance is not necessarily an immanent property of cosmological models, in contrast to an apparent temporal symmetry of general relativity. The cosmological time-non-invariance is in principle sufficient to break the presumed time-reversal symmetry in general, since the universe is the largest system to determine the arrow of time. Nonetheless, we shall proceed to discuss other time arrows as well. R. Penrose counted seven different arrows of time [89] and analyzed their relationships to each other. His conclusion was more or less obvious: there is only one explanation to the observed time-reversal non invariance - in contrast to the commonly shared opinion, not all exact laws of physics are time symmetric [200]. Subjectively, we experience the time arrow as having memory of the past but not of the future. This asymmetry is universal for all people and for all times, so it may be considered a well-established experimental fact. As such, it should be explained by some physical theory. I am not aware of such a theory, probably only piecewise approaches illuminating diverse aspects of the unidirectional time flow exist so far, see however a discussion in [92].

Such manifestations of discrete symmetry breakdown as the non-equivalence of direct and reverse time directions - $T$-symmetry violation or what we poetically call "the arrow of time" as well as non-equivalence of left and right ($P$-symmetry violation, often denoted as $PNC$ - parity non-conservation) and of particles and antiparticles ($C$-symmetry violation) have been well-known in our everyday life for a long time: we grow old and die, our heart is located in the left half of the body, and we are built of nucleons and electrons. In distinction to the macroscopic world, $T$-, $P$-, and $C$-symmetry breakdown in the microscopic world was only recently discovered. The physical question corresponding to this observational time lag may be formulated as: what is the relationship between $T$, $P$, and $C$ violation in the macroscopic and the microscopic worlds? Or, to put it another way, is there any correspondence between discrete symmetries in the macroscopic and microscopic worlds and, in particular, between $CPT$ symmetries? Here we implicitly assume $CPT$ symmetry to be exact, at least no experimental evidence for $CPT$ violation has been produced so far. Thus, $CPT$ appears to be always conserved in microscopic physics, but not necessarily at the cosmological level, see [99]. $CPT$ symmetry is the only exact symmetry that includes time reversal. (See Chapter 5, sections on the quantum field theory for more details on $CPT$).

---

[200] I would supplement the list of time arrows provided by R. Penrose with the arrow of time observed in modern optics where probabilities of the direct and reverse processes are most often different.

These two questions naturally return us to cosmological problems such as knowing the wave function (a pure state candidate) or the density matrix, e.g., in the multiverse models in which our universe emerged as a kind of bubble, one among a very large number of them, within a statistical ensemble, and afterwards became relatively isolated from it [93]. The probabilities in the density matrix correspond to statistics of such bubbles in the multiverse. As soon as we introduce probabilities in our model, we find ourselves in the situation of irreversibility, both in classical and quantum mechanics. This general relationship of probability with irreversibility is thoroughly discussed in [65], Ch.6. We tackle probabilistic irreversibility in Chapter 7 of the present book, now I only mention the so-called Zermelo paradox in statistical mechanics (named after the German mathematician Ernst Zermelo), which is closely connected with the Poincaré recurrence cycle. In 1893, H. Poincaré in a short note pointed out that kinetic theory contradicts his so-called recurrence theorem which he proved in 1890 (but published his proof for a system with a $3D$ phase space - this case is not valid for kinetic theory). The note of Poincaré, although, in my opinion very simple and at the same time fundamental, did not receive much attention until E. Zermelo raised an interesting objection in 1896.

## 9.3    Irreversibility

The local time arrow is perceived as striving to equilibrium, and the two concepts are regarded as synonymous. The trouble, however, is that striving to equilibrium admits many physical forms: from the Boltzmannian molecular chaos hypothesis (Stosszahlansatz) in classical kinetics to random phase approximation (RPA) in quantum many-body theory [201]. It is generally considered that one cannot obtain the real-life irreversible behavior using only the fundamental physical laws such as the laws of mechanics.

---

[201] Recall that in many-body theory the random phase approximation is one of the most popular assumptions allowing one to significantly simplify computations. For example, the RPA ground state is typically described by a collection of dressed particles interacting separately through short-range forces and quantized collective (coherent) modes such as plasmons. Accordingly, one can factorize the many- particle wave function which leads to enormous simplifications.

It is quite common - and physicists exploit this fact over and over again - that a complex phenomenon can be best understood on simple models. An example of such models is the one treated in statistical mechanics - the model of the ideal gas. The time-reversal non-invariance is exhibited even in this comparatively simple system, and not only irreversibility associated with the entropy increase, as it is often stated in the textbooks. The fact that the elementary dynamics of gas molecules is time-reversal invariant whereas the gas as a whole is a time-oriented system was puzzling L. Boltzmann to the point of a serious depression. Even at present this paradox continues to stir controversies among physicists and mathematicians. The above mentioned Zermelo paradox is of the same type. Consider a closed box containing $N$ molecules which move and collide according to interparticle interaction potentials, elastically reflecting from the walls of the box. We may assume that the equations of motion for the gas define a Hamiltonian system, therefore one-parameter group of translations along all the paths preserves the Liouville measure (see Chapter 4, section about Hamiltonian systems). Manifolds corresponding to fixed energy values are compact, hence the Liouville measure produces finite invariant measures at the fixed energy manifolds. Thus, we can justifiably apply the Poincaré recurrence theorem which in a somewhat simplified form may be stated as (see for many interesting details [14], §16, see also the famous book by M. Kac [68], Ch.3):

Let $(M, \Sigma, \mu)$ be some space with measure, $\Sigma$ being a $\sigma$-algebra of its subsets, $\mu$ an invariant measure (see Chapter 3). Let $T: M \to M$ be an endomorphism of this space (phase space). Then for any $A \in \Sigma, \mu(A) > 0$ almost each point $x \in A$ is returned infinitely many times in $A$, i.e., there exists an infinite sequence $\{n_i\}, n_i \to \infty$ so that $T_i^n(x) \in A$. In other words, each set $C \in \Sigma$ such that $T^n C \cap C = \emptyset$ is of zero measure.

The standard way of reasoning proceeds as follows: let us imagine now that the set $A$ consists of such phase points that all the molecules of gas are gathered in one (say, the left) half of the box - we can consider it the initial state. Then, according to the Poincaré recurrence theorem, one would find such moments of time that all the molecules will return to the same (left) half of the box. However, nobody has ever observed the case when the gas does not occupy the whole accessible volume.

The usual explanation of this paradox is as follows. The probability measure for the set $A$ is of the order of $\exp(-cN)$ where the constant $c$ depends on some macroscopic physical parameters such as temperature and density (see Chapter 7). For a gas under normal conditions, when $N \approx 10^{23}$ the measure $\mu(A)$ is numerically extremely small so that recurrence cycles $\left(\mu(A)\right)^{-1}$ are greater than any extrinsic time scale, e.g., the characteristic cosmological times. In order to keep the box intact and observe the return to an initial state, one has to isolate the box from external influences during cosmological times, which is physically unrealistic. On the other hand, if the number of molecules is small, say $N \approx 10$, one may in principle observe a gigantic fluctuation when all the molecules gather in a single half of the box.

This situation can be rather modeled on a computer than produced in a direct physical experiment.

One might note that there is no restriction on the time the recurrence could take. This recurrence (or return) time may drastically differ on the paths starting from various subregions $\Sigma$. The first return of each path defines a measure-preserving map of the region into itself, if one can wait for a long enough time, of course. One may also notice that there is no intrinsic time-scale with which the Poincaré recurrence cycle can be compared, so that the recurrence time is considered large or even infinite not mathematically but practically. The intrinsic time scale in this system appears only when an external influence or field is allowed to drive it. External agents can easily pump the gas into a nonequilibrium state. However, without external driving forces the gas expands and fills up the whole available volume which demonstrates the irreversible and time-noninvariant behavior. How can this behavior be explained for an isolated system?

The Poincaré recurrence theorem is a typical result of dynamical systems theory (see Chapter 4). Some contemporary studies in dynamical systems are aimed at a possible "explanation" of the time arrow. There are several ways to describe the time evolution of a dynamical system. In the classical framework (see e.g., a very good book [73]), one considers systems of differential equations with respect to explicit time (there may be other temporal parameters such as "fast" or "slow" time). In the autonomous case, the solutions to such systems are usually time reversible. However, in many cases the dynamical system exhibits either ergodic or mixing properties, and it is usually assumed that mixing and ergodicity are closely related with the arrow of time. Indeed, from the viewpoint of statistical mechanics mixing means irreversibility: every initial measure converges to an invariant measure (see Chapter 4) under the action of dynamics.

It may be difficult to obtain exact solutions for mixing and ergodic systems, so to prove irreversibility in a mathematical sense is also difficult. One can explore discrete-time models or difference equations using fast computers, instead of solving differential equations. Discretized models, e.g., produced with iterated functions [90], may easily exhibit irreversibility due to a number of different past histories for a given time-point in the present.

It is only recently that the study of dynamical systems became connected with statistical mechanics and kinetic theory (Chapter 7). It is sometimes said that in dynamical systems one deals with absolute irreversibility whereas in statistical physics and thermodynamics only with statistical irreversibility (owing to the lack of knowledge). Statistical irreversibility appears in large ensembles of particles (or other entities) whose exact behavior is subordinated to more specific laws such as the Newtonian or Hamiltonian laws of motion. The statistical or probabilistic way of explaining time-reversal paradoxes differs from explanations based on the theory of dynamical systems, but in fact one considers the same process of passing to probabilistic (based on finite measure) models starting from the deterministic phase flow - only the languages are different: more modern "geometrical" in the theory of dynamical systems and more traditional "physical" discussing inter-particle correlations and many-particle distribution functions in statistical physics and physical kinetics. This "physical" way of reasoning was originated in the works of L. Boltzmann [69] and J.W. Gibbs [297].

## 9.4    Origins of Unpredictability

The way to explain the Zermelo paradox pointed out by L. Boltzmann is based on the so-called $H$-theorem which is usually proved starting from the Boltzmann transport equation (see e.g. [72, 25], see also Chapter 7). Traditionally, to prove this theorem the assumption of "molecular chaos" is premised. Honestly speaking, I have not seen a correct (to my understanding) formulation of this assumption, the famous Boltzmann's molecular chaos hypothesis (Stosszahlansatz)[202] Intuitively, it may be understood using the language of correlations. Before any collision, the momenta of all colliding particles were distributed uniformly in the momentum subspace independently of their positions, but after the collision these momenta become correlated. However, Boltzmann's conjecture of molecular chaos appears to remain unproven, therefore it may or may not be true.

The key assumption of "molecular chaos" breaks down time-reversal symmetry as it leads to a special form of collision integral in the Boltzmann kinetic equation. This is what was exactly necessary to make ground for the second law of thermodynamics (the entropy rise). The concept of molecular chaos (or factorization) can be more or less understood from the dynamical derivation of the Boltzmann equation [72, 25]. We discuss this derivation and related issues in Chapter 7; now I shall try to illustrate the molecular chaos on a simple model.

Assume that we have a homogeneous gas of particles and the single-particle distribution function (the density of particles having momentum $\mathbf{p}$) is $f(\mathbf{p})$. If we consider only pair interactions and let $w(\mathbf{p}_1, \mathbf{p}_2; \mathbf{p'}_1, \mathbf{p'}_2)$ be the probability per unit time (transition rate) for particles with momenta $\mathbf{p}_1, \mathbf{p}_2$ to collide and become pairs with momenta $\mathbf{p'}_1, \mathbf{p'}_2$. To simplify the two-particle collision description, let us assume the time and space reversal symmetry of the elementary process in gas, which gives the detailed balance equation:

$$w(\mathbf{p}_1, \mathbf{p}_2; \mathbf{p'}_1, \mathbf{p'}_2) = w(\mathbf{p'}_1, \mathbf{p'}_2; \mathbf{p}_1, \mathbf{p}_2) \tag{9.8}$$

Then, due to Stosszahlansatz, we get the Boltzmann equation (in a somewhat simplified form)

$$\frac{df(\mathbf{p})}{dt} = \int \begin{array}{l} w(\mathbf{p}, \mathbf{p}_2; \mathbf{p'}, \mathbf{p'}_2)[f(\mathbf{p'})f(\mathbf{p'}_2) - f(\mathbf{p})f(\mathbf{p}_2)] \times \\ \delta\left(\dfrac{\mathbf{p}^2}{2m} + \dfrac{\mathbf{p}_2^2}{2m_2} - \dfrac{\mathbf{p'}^2}{2m} - \dfrac{\mathbf{p'}_2^2}{2m_2}\right) \times \\ \delta(\mathbf{p} + \mathbf{p}_2 - \mathbf{p'} - \mathbf{p'}_2) d\mathbf{p} d\mathbf{p}_2 d\mathbf{p'} d\mathbf{p'}_2 \end{array} \tag{9.9}$$

---

[202] It is interesting that Boltzmann was only 27 years old when he introduced Stosszahlansatz in statistical mechanics.

where we have omitted index 1. It is from this equation that Boltzmann proved the $H$-theorem: $\frac{dH(t)}{dt} = \frac{d}{dt}\int f(\mathbf{p})\ln f(\mathbf{p})d^3p \leq 0$, which gives the entropy with negative sign. The meaning of this derivation is that entropy increases provided the model of Stosszahlansatz (molecular chaos) is valid. Thus, only the hypothesis of molecular chaos brings with itself time asymmetry, though not very explicitly.

One can see that the $H$-theorem is so simply proved only for the rarefied Boltzmann gas. Actually, it is usually formulated for such a gas which is not far from an equilibrium ideal system. I failed to find a satisfactory proof of the Boltzmann $H$-theorem for the non-ideal gas with arbitrary interaction between the particles and I could not find a proof of this theorem for arbitrary open systems either, despite the fact that the entropy can in principle be defined for such systems.

Let us make a brief distraction recalling some well-known facts about statistical mechanics in general (one can find specific details in Chapter 7). The foundations of statistical mechanics are still a hot topic, despite the fact that this subject has persistently been in the focus of interest since the times of Boltzmann and Gibbs. The traditional issue here is the foundation of classical statistical mechanics of continuous systems as opposed to now fashionable lattice systems being closely associated with numerical modeling. By a continuous classical system one can understand a collection of classical particles moving in a continuous phase space. The theoretical approach to deriving the equations of classical statistical mechanics is based on the solution of an infinite hierarchical system of integro-differential equations for the many-particle distribution functions (the Bogoliubov chain) [72]. For many-particle systems in a finite volume the Bogoliubov hierarchy of equations is equivalent to the Liouville equation.

So the outcome of all this stuff is that is hard to use the reversible models for the laws of motion to explain why we observe the world having a comparatively low entropy at any moment of observation as compared to the equilibrium entropy (e.g. of universal heat death). Moreover, the world as a whole must have had even lower entropy in the past (see [88], Ch.27). One may notice here that the definition of entropy in cosmology is still a controversial issue: there does not seem to be a consensus about the notion of a global entropy in the universe, specifically for the part of entropy associated with the gravitational field.

Remark. Large entropy fluctuations in the equilibrium steady state of classical mechanics can be studied in extensive numerical experiments in a simple strongly chaotic Hamiltonian model with two degrees of freedom (e.g., described by the modified Arnold cat map). The rise and fall of a large, separated fluctuation is shown to be described by the (regular and stable) macroscopic kinetics, both fast (ballistic) and slow (diffusive). One can then abandon a vague problem of the appropriate initial conditions by observing (in a long run) a spontaneous birth and death of arbitrarily big fluctuations for any initial state of our dynamical model. Statistics of the infinite chain of fluctuations similar to the Poincaré recurrences is shown to be Poissonian. A simple empirical relationship for the mean period between the fluctuations (the Poincaré cycle) can be found and, presumably, confirmed in numerical experiments. One can propose a new representation of the entropy via the variance of only a few trajectories (particles) that greatly facilitates the computation and at the same time is sufficiently accurate for big fluctuations. The relation of these results to long-standing debates over the statistical irreversibility and the time arrow is briefly discussed.

One can then show that the Poincaré recurrence theorem incorporates Loschmidt's requirement for velocity reversion in thermodynamic gas systems. It differs essentially from Hamiltonian dynamics from which Boltzmann's $H$-theorem follows. The inverse automorphism, $T^{-1}$, on which the demonstration of the recurrence theorem is based, does not exist for atomic and molecular systems. Thermodynamic systems need not spontaneously return to states they occupied in the past and a Zermelo paradox has never existed for them. The same conclusion follows a fortiori for quantum systems in chrono-topology. Poincaré's recurrence theorem does not conflict with Boltzmann's $H$-theorem because they apply to systems described by quite different mathematical structures.

The way I presented the story was to strictly impose molecular chaos (no momentum correlations) at one moment in time. That is really breaking time translation invariance, not time reversal. From that you could straightforwardly derive that entropy should increase to the past and the future, given the real Hamiltonian dynamics. What the real Boltzmann equation does is effectively to assume molecular chaos, chug forward one timestep, and then re-assume molecular chaos. Its equivalent to a dynamical coarse-graining, because the distribution function on the single-particle phase space can't carry along all the fine-grained information.

## 9.5   Understanding superconductivity

When I was a student long ago, I heard the rumors that L. D. Landau had named three problems in physics as being of an outstanding importance: the problem of cosmological singularity, the problem of phase transition, and that of superconductivity. This latter subject is a good example of a synthetic discipline that has links into many other fields of physics and mathematics. Superconductivity has lately acquired the status of a whole discipline combining profound theoretical ideas with engineering applications. Naturally, when speaking about superconductivity, I only scratch the surface. Superconductivity is a whole world centered around designing and producing new materials with highly required but unusual properties. I write "designing and producing" because the typical approach in superconductivity symbolizes a new phase in science and technology: previously people have used available materials, now they are trying to construct them. Note that in fact all of physics or chemistry, with their vast baggage of analytical tools and research patterns, have their eventual goal in creating appropriate materials. This is basically what people understand by technology.

In 1957, the Physical Review published the first fundamental theory explaining how, at low temperatures, some materials can conduct electricity entirely without resistance. Building on experimental clues and earlier theoretical hints, John Bardeen, Leon Cooper, and Robert Schrieffer, all at the University of Illinois in Urbana, explained not just the absence of electrical resistance but also a variety of magnetic and thermal properties of superconductors. The "BCS" theory also had an important influence on theories of particle physics and provided the starting point for many attempts to explain the new high-temperature superconductors. The "BCS" theory has played a prominent role not only as an *ad hoc* model for superconducting electrons, but also in many other areas of physics. The BCS model also had an important influence on theories of particle physics and provided the starting point for many attempts to explain the "new" high-temperature superconductors. Recently, it has been applied to the analysis of dilute gases of cold fermions in the case of weak interactions between the atoms.

### 9.5.1    Early History of Superconductivity

Superconductivity was discovered in 1911 and always remained a riddle. Due to some mysterious reason, metals at very low temperature came into a state when the resistance to electric current practically disappeared. Physicists have been trying to solve this puzzle for many years. Only by the 1930s, it was concluded that electrons in a superconductor must occupy a quantum-mechanical state distinct from that of normal conduction electrons. In 1950, researchers found that the temperature at which mercury becomes a superconductor is slightly higher for mercury isotopes of lower atomic weight, suggesting that superconductivity somehow involves motion of the atoms in a material as well as the electrons.

Following up on this "isotope effect," Bardeen and Illinois colleague David Pines showed theoretically that within an atomic lattice, electrons could attract one another, despite their strong electrostatic repulsion. Essentially, an electron can create vibrations among the lattice atoms, which can in turn affect other electrons, so the attraction is indirect.

By the mid 1950s, Bardeen was collaborating with Cooper, a post-doctoral fellow, and Schrieffer, a graduate student. Cooper published a short paper showing how the Bardeen-Pines attraction could cause electrons with opposite momentum to form stable pairs [283]. This pairing mechanism, Cooper suggested, might be responsible for superconductivity, but Bardeen was initially skeptical. The paired electrons were not physically close together but moved in a coordinated way, always having equal but opposite momentum. It was not clear that these tenuous, extended pairs could be crammed together to create a superconducting medium without getting disrupted.

A few months later, however, Schrieffer hit on a mathematical way of defining a quantum mechanical state containing lots of paired electrons, with the pairs oblivious to other electrons and the lattice, allowing them to move without hindrance. He later compared the concept to the Frug, a popular dance at the time, where dance partners could be far apart on the dance floor, separated by many other dancers, yet remain a pair [284].

After publishing a short note early in 1957 [48], the team published what became known as the Bardeen-Cooper-Schrieffer, or BCS, theory of superconductivity in December. They won the Nobel prize in 1972. The theory explained the isotope effect and the fact that magnetic fields below a certain strength cannot penetrate superconductors. It also explained why superconductivity could only occur near absolute zero–the tenuous Cooper pairs break up in the presence of too much thermal jiggling. It's a testament to Bardeen's insight that he chose the right collaborators and kept his eye on experiment in seeing the way forward, says superconductivity experimentalist Laura Greene of the University of Illinois: "It's how science should be done."

One oddity of the BCS wave function is that it lacks some of the mathematical symmetry expected at the time for any quantum or classical solution of electromagnetic equations. Further analysis of this point spurred the development of so-called symmetry breaking theories in particle physics.

Although the superconductors discovered in 1986 rely on electron pairing, they remain superconducting at temperatures above what the pairing mechanism in BCS can easily explain. But Marvin Cohen of the University of California at Berkeley says that given the poor understanding of the new materials, the original BCS pairing mechanism shouldn't be ruled out. And, adds Greene, it took "some very smart people" almost 50 years to get from the discovery of superconductivity to BCS, so she's not worried that the high temperature superconductors remain unsolved after a mere 20 years.

Although some engineers state that the potential *impact* of high-temperature superconductivity would be negligible due to high costs involved, from the physical viewpoint any major breakthrough in superconductivity research is difficult to overestimate. Historically, physicists had to cool conventional conductors almost down to absolute zero to observe the phenomenon. Any person who did some experimental work in physics would understand what it means - to cool samples to Helium temperatures: this is usually a multistage process, and even provided the institution or the laboratory chief has enough money at his or her disposal to buy the modern equipment, reaching such low temperatures becomes progressively difficult, since each temperature drop requires finding some kind of energy within the substance and then devising a means of removing this energy.

In several years, superconductivity will be one hundred years old, which is quite a long time for an area of physics. Usually, after such a prolonged period one could expect the area to be a closed text-book chapter. Yet, understanding of superconductivity is still rather unsatisfactory, especially as far as "new" superconductors are concerned. Superconductivity has traditionally been thought of as a phenomenon that occurs only at temperatures near absolute zero, but in 1987 several materials that exhibit superconductivity at temperatures exceeding 100K had been found. I remember the hype of March 1987 around this discovery - all media sources treated it almost as the most important event of the XX century. At that time, I worked as the editor of the physics and mathematics department of the leading Soviet popular science journal called "Science and Life" (Nauka i Zhisn'), and our Editor-in-Chief who usually did not care much about physics quite unexpectedly summoned me and ordered me to make an urgent material on the discovery of high-temperature superconductivity, which I hastily did ("Nauka i Zhisn'" No. 6-7, 1987). For those who can read Russian I would also recommend a popular-level but comprehensive article on superconductivity by Academician V. L. Ginzburg - a great physicist of the dying-out generation of universalists who received his long-deserved Nobel Prize just for the theory of high-$T_C$ superconductivity ("Nauka i Zhisn'" No.7, 1987). So, a new era of active exploration in the superconductivity field had begun. People - including the newspaper-peeking lay public - thought the scientists would very soon provide new superconducting wires to transmit power as well as levitating trains Nevertheless, the phenomenon of high-temperature superconductivity is still poorly understood, which makes its practical applications evasive. No materials currently exist that can become superconductive at room - biologically acceptable - temperatures. Biological environment and superconductivity do not seem to coincide in their equilibrium states, at least so far. The new superconductors - ceramic materials based on copper oxides (cuprates) combined with various other, usually rare-earth, elements - support the superconducting current at temperatures as high as 140K. That was a noticeable jump toward room-temperature superconductors, since, although requiring rather low temperatures, these new materials can be cooled with liquid hydrogen, which is enormously easier and much less expensive than the liquid helium cooling required by the old materials.

These are all well-known facts, and I repeat them here only to emphasize the importance of the problem. Now that the superconductivity hype is long over, with other fields being at the forefront and enjoying mass-media attention, e.g., nanotechnology, the potential social and engineering impact of superconductivity is viewed with increasing skepticism. I still think that room-temperature superconductivity, if it is by any chance reached, would produce a major breakthrough in energy policy, transportation and possibly information technologies. Since the world is becoming increasingly infrastructured, it would result in major technologically-induced social changes. Take power transmission as an example. Currently, electricity-generating utilities must transmit power at a voltage of tens or hundreds kilovolts because otherwise a large portion of transmitted energy is dissipated during transmission. This is in fact ridiculous since the typical usage of electric power requires volts or, at most, hundreds of volts. Enormous voltages required solely for electricity transmission lead to the whole industry with their monopolies, inefficient management, elevated prices and frequent dangerous blackouts, without even mentioning ecological harm and wasted terrains. If one were able to build large-scale power grids using high-temperature superconducting materials, it would be possible to generate and transmit power at much lower voltages, rendering a lot of clumsy technology superfluous. The use of alternative energy sources could be boosted and the cost of energy worldwide, now exasperatingly rising, could be drastically reduced. In electronics and IT, new and efficient devices may be designed, since thin films of normal metals and superconductors that are brought into contact can form superconductive electronic components, which could replace transistors in some applications. These fancy pictures are, of course, far from today's reality - in the same sense as fusion is a controversial ideal for power generation. One can recall in this connection the onset of the automobile, aviation, and space research eras.

For example, we can consider the BCS equation for a Fermi gas characterized by the chemical potential $\mu \in \mathbb{R}$ and $T > 0$. We may assume the local pair interaction between the gas particles to be $\lambda V(\mathbf{r})$ where $\lambda > 0$ denotes the coupling constant.[203] The primary assumption about the interaction potential $V(\mathbf{r})$ might be that it is the real function and, e.g., that $V \in L^1(\mathbb{R}^3)$. The BCS gap equation can be written as

$$\Delta(p) = -\frac{\lambda}{2(2\pi)^{\frac{3}{2}}} \int_{\mathbb{R}^3} d^3q\, V(p-q) \frac{\Delta(q)}{E(q)} \tanh \frac{E(q)}{2T},$$

with $E(p) = \sqrt{(p^2 - \mu)^2 + |\Delta(p)|^2}$. One can readily see that the BCS equation is non-linear.

### 9.5.2    Some Physical Models of Superconductivity

Now, let us try to understand this amazing phenomenon of super conductivity. Here I do not attempt to overview the various mechanisms that have been suggested to explain superconductivity, especially in high-$T_C$ materials. My intention is very modest and consists in sharing my impressions of superconductivity research conducted by real experts in the field. So, it is rather a position of an elderly scientific observer than of an energetic researcher. Superconductivity is the vanishing of all electrical resistance in certain substances when they reach a transition temperature $T_C$ that varies from one substance to another. An interesting manifestation of this phenomenon is the continued flow of current in a superconducting circuit, even after the source of current has been shut off. For example, if a lead ring is immersed in liquid helium, an electric current that is induced magnetically will continue to flow after the removal of the magnetic field. This observation immediately invokes an engineering application: powerful superconducting electromagnets, which, once energized, retain magnetism virtually indefinitely, have been developed to be used, e.g., in fusion experiments. Contrariwise, in normal materials the current attenuates nearly exponentially with time after switching off the source because electrical resistance causes energy loss due to electron-phonon interaction and multiple scattering even in very clean conductors. Superconductivity is commonly interpreted as a transition to a new phase inside a material - a macroscopic quantum phenomenon when a macroscopic number of electrons (of the order of $10^{23}$) condense into a coherent quantum state. In case this state is the one of a current flowing through a material, such current would flow virtually indefinitely (from the theoretical point of view), and the material can serve to transmit electric power with no energy loss, unless the superconducting phase of the sample is destroyed by some external agent, e.g., by a strong magnetic field.

---

[203] Sometimes, the factor 2 is introduced for convenience i.e., the interaction is written as $2\lambda V(\mathbf{r})$.

Although many theoretical explanations related to the superconductors discovered in 1986 also rely on the BCS pairing mechanism, these "new" materials remain superconducting at temperatures higher than those which the BCS-type electron pairing can readily explain. But given the poor understanding of new and structurally complex materials, a number of competing mechanisms should not be ruled out in spite of quasi-religious wars waged by the proponents of different models for superconductivity. Besides, it is not especially astonishing that the riddle of high temperature superconductivity remains unsolved after twenty+ years: indeed, it took about fifty years before "some very smart people" managed to arrive at the BCS theory from the discovery of superconductivity.

The 1972 Nobel Prize in Physics was awarded to J. Bardeen, L. Cooper, and S. Schrieffer for their model (known as the BCS theory) of the "old" superconductors, i.e., those that exhibit superconductivity at temperatures near absolute zero, including such metals as zinc, magnesium, lead, gray tin, aluminum, mercury, and cadmium. Some other metals, e.g., molybdenum, may become superconductive after high purification, and a number of alloys (e.g., two parts of gold to one part of bismuth) as well as such compounds as tungsten carbide and lead sulfide can also display superconductivity, so they can also be classified as "old" superconductors. The BCS theory stipulates that at very low temperatures electrons in an electric current move in pairs. Such pairing enables them to move through a crystal lattice without having their motion disrupted by collisions with the lattice. It is important that the BCS formalism invented to explain superconductivity is in fact much more general. For example, the BCS model can be applied to describe phase transitions in solids, more specifically the metal-dielectric transition. In the famous Kopaev-Keldysh model [222] the BCS-type phase transition was interpreted as the Bose condensation of electron-positron pairs (excitons), which is a direct analogy to BCS superconducting transition. I shall discuss the BCS theory later in some detail, since it is a beautiful model having many implications and, besides, because I have a suspicion that many authors writing on various superconductivity issues did not read the classical BCS paper [48].

While this microscopic theory explaining conventional superconductivity has already existed for half a century, as well as the phenomenological Ginzburg-Landau (GL) theory [47], which can be derived from the microscopic BCS theory [298] some doubts exist that the BCS model is fully applicable to the "new" superconductors. The latter appear to be rather "muddy" in order to serve as model physical systems, so the clear-cut BCS model based on the Cooper pair formation and condensation in an ideal lattice does not seem to describe the whole lot of competing phenomena in very special chemical systems. Superconductivity in the "new" and "dirty" materials, although ultimately producing the same effect, may be a very different phenomenon, so that the BCS model, which has served as a standard for describing the "old" superconductivity, may simply be inadequate in the case of "new" superconductors. There exist competing models, e.g., the 2D Hubbard model, bipolaron model, bosonic SO(5), U(1), SU(2) and many other models, which have been proposed to describe high-temperature conductivity in new ceramic materials (see e.g. the overview by Yu. V. Kopaev [40]. To my knowledge, all of the models have not been proven and remained controversial for many years. At least two schools have been confronting each other while putting forward their versions. These may be classified according to the type of main quasiparticles involved. One of the schools holds that the electron-phonon interaction, the same as in the BCS model, still plays the dominant role in the cuprates. The other insists that electron-electron Coulomb interaction must be strong in the new superconductors (there are some experimental indications), so that electron-phonon interaction is immaterial for pair formation. Sometimes, when listening to excited arguments of adepts of these schools of thought or even "independent" adherents to a specific model, I could not get rid of an impression that I was witnessing a sort of a religious battle in which the adversary must be totally devastated. The subject of superconductivity seems to be full of prejudices and can only be compared in this respect with the heated ideological war waged by the open source community against Microsoft. However, while the market needs and user convenience can eventually settle the debate in personal and enterprise computing, the choice of an adequate model to understand superconductivity is probably a much harder task. This understanding, by the way, also depends on the current state of the computer technology.

I shall describe the Hubbard model a bit later, now I can only remark that this model is simple and universal and can be applied far beyond superconductivity. Moreover, thanks to the new developments in parallel computation, one can verify the 2D Hubbard model directly. The Hubbard model purports to describe superconductivity using a few microscopically defined parameters such as (1) the probability that carriers - electrons or holes - hop from one site to another in the crystal lattice; (2) the energy penalty when two carriers occupy simultaneously the same site; (3) the concentration of carriers. As I have already mentioned, several theories of high-temperature superconductors have been proposed, but none has been experimentally confirmed, so there was a strong disagreement within the physics community about what model was appropriate at all. One could have a single validation criterion: if some model does not foresee the superconducting state in the typical range of temperatures as well as in the domain of specific compositional and structural parameters, specifically for the case of cuprate superconductors, then it should be abolished. At first, the Hubbard model was unsolvable even on high-performance computers due the amount of computation required. Indeed, superconductivity is a macroscopic quantum effect, therefore any simulation must involve a lattice scale of $10^{23}$ sites. Besides, the computer model must also account for individual carriers hopping from site to site on the lattice, which means that the scale of several lattice spacing with carrier interactions on this scale should be included. All that is tantamount to a computationally complex multi-scale problem. We have already encountered multiscale problems in connection with the transition to classical description in Chapter 4 as well as in statistical mechanics. Here, I would only like to stress that because of prevailing numerical difficulties, the issue of applicability of the Hubbard model remained unsolved, as well as the question of survival of other models for high-temperature superconductivity.

It has always been a guesswork for me why the subject of superconductivity is treated in four different volumes of the Landau-Lifshitz course: in "Electrodynamics of Continuous Media", in each of the two volumes of "Statistical Physics", and in "Physical Kinetics". This is a peculiar fact which may be explained, of course, by personal likings of the renowned authors. But there must be also an objective reason for such a dissemination of a single issue: superconductivity on its current level of understanding is a multi-facet subject, no unifying theory for this phenomenon is known, so it should be tackled from diverse viewing angles and using different techniques. Even the phenomenological treatment of superconductivity has admitted a variety of models, many of which had been proposed in the 1930s (see, e.g., the paper by R. Kronig [264] and the famous paper by H. and F. Londons). One might mention in passing that quite a lot of the grands of theoretical physics who were active in that period competed in offering a plausible explanation for superconductivity: W. Heisenberg, E. Schrödinger, M. von Laue, J. Slater, H. Casimir and many others.

One can see that superconductivity is really a multilateral phenomenon by a very simple observation. The current flowing in a superconductor is, as any current, a manifestation of the nonequilibrium process consisting in a stream of charged particles. Therefore, the effect of disappearing resistance to this current cannot, strictly speaking, be treated within equilibrium thermodynamics, it is a subject for physical kinetics. On the other hand, the normal and the superconducting states may be represented as two different phases (in the Gibbs' sense), because they can be characterized, besides conductivities, by different values of purely thermodynamical quantities, and the transition between these two phases can be treated as a normal phase transition. Thus, the transition of a material from normal to superconducting phase may be studied by applying usual equilibrium thermodynamics, which makes the treatment of this phenomenon in statistical physics courses quite natural. By studying the literature, one may notice that during about the first quarter century after the discovery of superconductivity, physicists were focused mostly on the electrical problem, namely on how to explain the abrupt change and nearly immeasurably small value of resistance. I could not trace the idea of phase transitions in the papers available to me from that period, at least it was not predominant until the paper of Gorter [257]; there are also references to early-stage applications of thermodynamics in the works of W. H. Keesom around 1925, but I could not find them without investing too much time. Superconductors were customarily described as crystals in which scattering of carriers is absent, the limiting case of a perfect conductor [258] R. Becker, G. Heller, F. Sauter. "Über die Stromverteilung in einer supraleitenden Kugel", Zeitschrift für Physik 85, 772-787 (1933). This is a simple model of the Drude type that is traditionally treated in courses on macroscopic electrodynamics. The electrodynamical approach immediately invokes the idea of electromagnetic response of a superconducting material to an external electromagnetic field - in terms of dielectric function or otherwise. The response theory was discussed in Chapter 7 and I will not repeat it here.

In the electromagnetic theory of superconductors, the key notion is the current. Let us try to calculate the current starting at first from the naive Drude-Lorentz model for electrons in an external electrical field:

$$m\dot{\mathbf{v}} + vm\mathbf{v} = e\mathbf{E}(t) \tag{9.10}$$

where $v$ is the collision frequency. If we take for $\mathbf{E}(t)$ the harmonic components $\mathbf{E}(t) = \mathbf{E}_0 e^{-i\omega t}$, we shall have

$$\mathbf{v}(t) = \mathbf{v}_0 e^{-i\omega t} \tag{9.11}$$

and the solution

$$\mathbf{v} = \frac{e\mathbf{E}}{mv}\left(1 - i\frac{\omega}{v}\right)^{-1} \tag{9.12}$$

The macroscopical current is

$$\mathbf{j}(\mathbf{r}, t) = \sum_{i=1}^{n} e_i \mathbf{v}_i(t)\delta(\mathbf{r} - \mathbf{r}_i(t)) = en\mathbf{v} = \frac{e^2 n}{mv}\frac{E}{1 - \frac{i\omega}{v}} = \sigma \mathbf{E} \tag{9.13}$$

and the conductivity is defined as

$$\sigma = (4\pi)^{-1}\frac{\omega_0^2}{v - i\omega} \equiv \frac{\sigma_0}{1 - i\omega\tau}, \tau = \frac{1}{v}, \omega_0^2 = \frac{4\pi ne^2}{m}, \sigma_0 = \frac{\omega_0^2}{4\pi v} \tag{9.14}$$

Here $\omega_0$ is the usual plasma frequency and $\tau$ has the meaning of the collision-free time.

In transient models, when the field cannot be adequately represented by a single Fourier component, the current value at time $t$ is determined by the entire history of $\mathbf{E}(t)$:

$$j_\alpha(t) = \int_{-\infty}^{t} dt' G_{\alpha\beta}(t, t') E_\beta(t') \tag{9.15}$$

See a discussion of the electromagnetic response of material media in Chapter 8.

If we consider the superconductor as a material with disappearing scattering of electrons, then $v \to 0$ and the condition $\omega \gg v$ will be fulfilled even for very low frequencies. In other words, $\sigma \to \omega$ with $\omega \to 0$ since the currents circulate without Joule losses ($c'' = 2n\chi = 0$). Recall that $c' = 1 - \omega_0^2/\omega^2 = n^2 - \chi^2$ (Ch.5) so that for $\omega < \omega_0$ the field decays exponentially in the superconductor with the penetration length $l_p = c/\omega\chi = c/\omega\sqrt{|\varepsilon'|}$. For low frequencies ($\omega \ll \omega_0$), we obtain the so-called London penetration depth that does not depend on frequency, $l_L = c/\omega_0$.

The wave damping length in this limit is of the order of $\left(\frac{mc^2}{4\pi n e^2}\right)^{1/2} \sim \frac{10^6}{\sqrt{n}}$ cm, i.e., depends on the concentration of superconducting carriers. If we assume that all the conductivity electrons contribute to the superconducting current ($n \sim 10^{23}$), we get for $l_L$ the value about $200\text{Å}$ that is of the order of the ultraviolet radiation wavelength. Thus, the superconductor should fully screen the electromagnetic field having the frequency up to $10^{17} s^{-1}$. However, this is a very naive model. In fact, the question: "what are the superconducting carriers?" has hardly been answered, at least one can find in the literature numerous model representations related to quasiparticles forming the superconducting current, and, honestly speaking, I could not find the ultimate value for the superconducting carrier concentration. A simple representation of the total electronic density as the sum of normal and superconducting electrons, $n = n_0 + n_s$, seems to be an intuitive illustration, rather than pertinent to a mathematical model. This is a regrettable situation, as the above question about the nature and quantity of superconducting carriers is, in my opinion, crucial to the understanding of superconductivity. Indeed, for instance, the Ginzburg-Landau wave function is normalized to the number of superconducting electrons $n_s$ and, besides, as we have seen, $n_s$ enters all the equations of superconductor electrodynamics. The Ginzburg-Landau (GL) model is usually labeled as a phenomenological one, implying that it is allegedly of lower rank than "true" microscopic models. In the London electrodynamical theory (see also below), the issue of superconducting carrier density is not discussed; the quantity $n_s$ is considered as a phenomenological parameter whose value for $T = 0$ tends to a natural limit - the total electronic density. In the BCS theory, the full number of pairs is still undefined, and so on.

## 9.6    Superfluids and Supersolids

By straining one's fantasy, one can imagine other peculiar "super"-objects, for example, supersolids. We have just seen that a superfluid can flow through the narrowest channels without any resistance. In the geometric language[204], superfluid breaks a global $U(1)$ phase rotational symmetry and acquires the off-diagonal long-range order. In some other - more physical - language, there exist low-energy quasiparticle (phonon) excitations in the superfluid. On the other hand, we know that there are solids and low-energy phonon excitations in them. But a solid cannot flow. Using the same "geometrese", one might say that a solid breaks a continuous translational symmetry reducing it to a discrete crystal lattice symmetry. We have discussed the accompanying mathematical models and effects in Chapter 6. Exploring the arising logical possibilities, we can make a conjecture that a state may be constructed that would combine both the crystalline long-range order and the off-diagonal long-range order inherent to superfluids. This is typical model-building in physics. Such a model might be naturally called a supersolid one. Historically, the first model conjuring on the possibility of a supersolid state was probably contained in the classical paper by A. F. Andreev and I. M. Lifshitz [166]. From today's viewpoint, that was a pure speculation on the subject of Bose-Einstein condensation (BEC) of vacancies. An important idea, however, was that there can be a state supersolid - possessing both the superfluid and solid-state order. This idea was rather controversial, of course.

---

[204] One can coin a term for this language, for example, "geometrese".

## 9.7   Relativity

### 9.7.1   Special Relativity

According to Einstein's principle of relativity, all laws of physics should be the same in all uniformly moving (i.e., without acceleration) reference frames. This is a very strong, even bold assumption. Who can guarantee that all the interactions possible in nature, including those not yet discovered should be invariant under the transition from one Lorentz frame to another? Although the Lorentz transformations had been known besides Lorentz to a number of other great scientists before Einstein, e.g., to Larmor and Poincaré, nobody had dared to postulate their universality, probably understanding that it would be a serious restriction imposed on all possible forces of nature. Notice that there are incessant attempts, currently intensified, to find the violations of Lorentz symmetry, in particular by using light and the known fundamental particles of the standard model, see e.g., a review by R. Bluhm [265], see also [266]. So far, no convincing evidence for Lorentz violation has been found, but should any kind of Lorentz symmetry violation be discovered, it would produce a very profound impact on the whole physics. We shall dwell more on this issue a little below (see section "Is Relativity Firmly Established?") However, Einsteinian special relativity like Galilean relativity that had appeared three centuries before was concerned predominantly with the uniform, non-accelerated motion. One would, however, desire to have a theory describing the dynamics of classical and quantum particles in general i.e., non-inertial frames. A close approximation to such a theory is given by general relativity.

### 9.7.2   General relativity

General relativity is an example of a theory that did not originate in any dedicated experimental endeavors. That was rather unconventional since no massive observational data (with the exception of equivalence of inertial and gravitational masses, which was taken for granted by many physicists and perceived as a trivial observation, see below) were accumulated to activate one person's imagination and logic. I mean of course Einstein. It is curious that in contrast with, e.g., special relativity or quantum mechanics there were no practical needs which would have dictated the invention of a new theory. It is remarkable how Einstein himself characterized general relativity. He has written in the 1950s: "An essential achievement of general relativity theory consists in that it has saved physics from the necessity of introducing an 'inertial reference system' (or inertial systems)." (Albert Einstein. The Meaning of Relativity. Appendix II, 1950/1955.)

We shall see below that Einstein's field equations of general relativity link the space-time curvature to the mass distribution scaled by fundamental constants $G$ and $c$. A common slogan-like interpretation of these equations states that geometry is dictated by the distribution of matter. In fact, the situation described by Einstein's equations is deeper and more self-consistent. Nonetheless, Einstein's main idea that in the presence of the gravity field its parameters simultaneously determine the spacetime geometry was later developed in physics and, one could observe, changed its culture.

## 9.8    Gravitation

Gravitation is one of the greatest mysteries of physics. Indeed, what causes gravitation? Recall that Aristotle asserted that heavy objects fall faster whereas Galileo observed experimentally that all objects fall with the same acceleration independent of their masses once we remove air resistance (see Chapter 4). Thinking more or less deeply about this fact leads to the so-called equivalence principle, profound geometric considerations and a paradigm change (see [88]). As already noted, there were no practical needs to replace Newton's theory of gravitation by anything else. Indeed, the simple Newtonian theory of gravity worked perfectly well, and its accuracy for navigation, elementary geophysics of the early twentieth century, or astronomical observations was totally sufficient - measurement errors were greater than possible theoretical inaccuracies.

Let us recollect what we know about spacetime. From our experience with classical mechanics and electrodynamics, we may be inclined to think that spacetime provides only a rigid framework, a scene upon which a physical play consisting of multiple events and sets is staged[205]. Could it be that the cast would adjust and modify the scene during the performance? Actors or players only seem to be invited ad hoc to amuse the observers - today one cast or team, tomorrow another - to the scene or sports arena constructed forever and for countless past and future events; likewise, spacetime cannot be changed due to the presence of a specific cast or team members playing their physical roles[206]. As a matter of fact, it is a general conviction nowadays that

---

[205] A theologically-scented question is invoked in this place: are there also play directors, invisible managers and the like?

[206] One can recall in this connection a beautiful fantasy by a Soviet writer L. Lagin (L. Ginzburg) "The old Hottabych", where a Moscow boy Volka fishes out an ancient vessel from a river. As Volka opens up the vessel, a genie (jinn) emerges from it and a suite of

the scene itself that is spacetime acquires an active role in physical phenomena, at least in some of them, the most prominent example being gravity. A little below, however, we shall discuss a class of theories of gravitation founded on a traditional concept of a passive spacetime like in classical mechanics or electrodynamics. Yet such theories of gravitation are only marginal and do not seem to play any significant role in the development of modern theoretical or observational techniques. And since the properties of spacetime are, as we have seen in Chapter 4, 5, fully determined by its geometry one has to study and interpret geometric concepts first in order to try to understand gravitation.

From purely geometric positions, a plain metric of Euclidean (Galilean) and pseudo-Euclidean (Minkowski) spacetime corresponds to a rather special and highly restrictive choice of metric and affine properties (see, e.g., [39], §82). It means that one can drastically relax the restrictive assumptions of special relativity, thus noticing that gravitation is a property of the physical world which should be naturally expected and not regarded as a mysterious effect needing to be explained. In general, the metric function $g_{\mu\nu} dx^\mu dx^\nu$, $1 \leq \mu, \nu \leq N$ or, in relativistic theories $0 \leq \mu, \nu \leq N - 1$, if it exists, encodes the geometry of the space or manifold. Recall that in physics the metric function or simply metric can be interpreted as the function determining the line element between two distinct nearby points so that this function generalizes the notion of a distance along this line. More specifically, let $M^4$ be some manifold (interpreted in physics as a spacetime) and let $x^0, x^1, x^2, x^3$ be local (curvilinear) coordinates on $M^4$, then the length (line) element $ds^2 = g_{\mu\nu} dx^\mu dx^\nu$, $\mu, \nu = 0,1,2,3$ on $M^4$ is identified in physics with the gravitational field.

So, one can see what gravity is about: it changes the geometric properties of spacetime. In particular, gravity can change the flow of time so that an observer in the vicinity of a heavy mass sees time running slower than another observer far from such a mass (see, e.g., [39], §102). One can actually find how the time slows down: the respective observable notion is the so-called red shift which can be expressed as $(1 - r_G/r)^{1/2}$ where $r_G = 2GM/r^2$ is the so-called gravitation radius. One can pose a question: is there a frame of reference in which all the clocks tick at the same rate? If one considers the clock in the presence of gravitation the answer in general is "no".

Einstein's astonishing hypothesis that gravitation is just a manifestation of the spacetime curvature had a profound effect first on mathematics and, later, on physics. It seemed to be an extremely productive combination of two seemingly disparate lines of thinking: Einstein's relativity theory and the works of Italian geometers, especially T. Levi-Civita and G. Ricci, on tensor

---

merry adventures begins. Once the supernatural creature was invited by his young rescuer (the two became inseparable friends) to a football match, and the old spirit suddenly became a fan of one of the teams. Using his miraculous abilities (notice that the words "genie" and "genius" are etymologically very close), the old jinn, in order to help his favorite team, began changing the geometry of the playground, e.g., twisting the goal configuration, generously created balls for each player (instead of a single one) and so on.

calculus, analysis and algebra on Riemannian (or pseudo-Riemannian)[207] manifolds.

By the way, representing gravitation as the consequence of the spatial curvature only is a popular misconception. In reality, gravitation emerges due to the curvature of spacetime (in 4d); the most obvious manifestation of this fact is the time slow-down in the vicinity of a massive gravitating object such as a black hole. The field tensor $R_{\mu\nu}$ of general relativity (which is in fact the Ricci tensor of classical differential geometry) identifies the spacetime curvature, and Einstein's field equations enable one to compute this tensor field for a given mass distribution. When there are no masses, one would intuitively think that this field tensor ought to be reduced to the identity transformation of a Lorentz frame corresponding to flat geometry with the Minkowski metric of special relativity. This is, however, not always the case: one can find solutions to Einstein's equations with no matter i.e., $T_{\mu\nu} = 0$ everywhere, and yet the Riemann tensor $R^{\mu}_{\nu\alpha\beta} \neq 0$ in 4d (see [39], §95 and [159], §37). The solutions to Einstein's equations with no matter are referred to as vacuum solutions. A flat Minkowski spacetime manifold is just an example of a vacuum solution, probably the simplest one. Other well-known examples are the Schwarzschild solution ([39], §100) and the Kerr solution ([39], §104) which are rather nontrivial. In classical differential geometry, manifolds for which $R_{\mu\nu} = 0$ are often called Ricci-flat solutions.

What is the basic difficulty with gravitation? Figuratively speaking, according to Einstein's idea, gravity is weaved into the texture of spacetime so that it is no longer the force in the Newtonian sense. From the experimental point of view, it is hard to get the accurate results about gravity since one cannot measure it in the laboratory as efficiently as one can detect gravitational effects in cosmology i.e., on the megascale level. It is not in the least accidental that nearly all laboratory-based gravitational experiments (and they are comparatively rare) produce the results characterized by a record accuracy [14]. Indeed, the electrostatic repulsion force between two protons is $\sim 10^{36}$ times stronger than the gravitational attraction between them, so it is really a challenge to measure gravitational effects against a predominant background of everyday influences. From the theoretical viewpoint, it is very hard (many experts even say impossible) to include gravitational forces in special relativity in the same way as the Lagrangian for particles and fields was constructed in electrodynamics (both classical and quantum). Now, I shall recall some basic formulas. Suppose we have a finite number $N$ of classical particles with charges $e_i$ moving in an electromagnetic field $A_\mu(x)$ over their word lines $\gamma_i$, $i = 1, \ldots, N$ in $\mathbb{R}^4 N$, then one can construct both the Lagrangian and the action of this system of particles in an external field. Taking for simplicity just one particle $N = 1$, we have (see [39])

---

[207] The pseudo-Riemannian manifold, in distinction to the Riemannian case, has at each point on the tangent space a smooth bilinear form with e.g., $(+, -, -, -)$ signature. Recall that a Riemannian metric $g_{ik}$ on a differentiable manifold $M$ is a tensor field of the (0,2) type which is a symmetric positive-definite bilinear form.

$$L = L_c + \frac{e}{c} A_\mu \frac{dx^\mu}{d\tau} \qquad (9.16)$$

where $\tau$ is a natural parameter (proper time) on the curve $\gamma(\tau)$, $d\tau = ds/c$ ($s$ is the interval), $L_c$ is the free particle relativistic Lagrangian which is usually taken to be

$$-mc(u_\nu u^\nu)^{1/2}, \qquad u^\nu = c\frac{dx^\nu}{dx^0} = \frac{dx^\nu}{d(x^0/c)} \qquad (9.17)$$

or $u = d\gamma/dt$.[208] When there are no particles, the free electromagnetic field must satisfy the homogeneous Maxwell equations, $\partial^{F^{\mu\nu}}/\partial x^n u = 0$. For an $N$-particle system, the total action may be written as

$$S = -\sum_{i=1}^{N} \int_{\gamma_i} \left\{ m_i c\, ds_i + \frac{e_i}{c} A_\mu dx_i^\mu \right\} + \frac{1}{16\pi c} \int F_{\mu\nu} F^{\mu\nu} d^4x$$

$$= -\sum_{i=1}^{N} m_i c^2 \int \left(1 - \frac{v_i^2}{c^2}\right)^{\frac{1}{2}} dt$$

$$+ \sum_{i=1}^{N} \int_{\gamma_i} \left\{ \frac{e_i}{c} \mathbf{A} \mathbf{v}_i + e_i A_0 \right\} dt + \frac{1}{16\pi c} \int F_{\mu\nu} F^{\mu\nu} d^4x \qquad (9.18)$$

So, in Einstein's general relativity, which is basically the theory of gravitation, gravity is described by the metric tensor $g_{\mu\nu}$ defined on pseudo-Riemannian spacetime. An interval of spacetime has the generic form $ds^2 = g_{\mu\nu} dx^\mu dx^\nu$ . The motion of test bodies (particles) occurs on timelike geodesics. The source of gravity is considered to be the energy-momentum tensor of the whole matter (including in the self-consistent way the gravitation field). Thus, the gravitation field is not regarded as a conventional physical field of the Faraday-Maxwell type, present over the flat background Minkowski space. As a consequence, the Minkowski metric $\gamma_{\mu\nu}$ of special relativity is not contained in general relativity. The gravitational field in Einstein's theory is manifested as a spacetime curvature, not as an independent actor over the background geometry. This fact has long been a point of criticism directed at Einstein's general relativity.

One of the most vocal critics of the Einstein's general relativity appears to be A. A. Logunov, a prominent Russian physicist, originally active in high-energy and elementary particle physics.

It is, in my understanding, largely a conceptual problem. It would of course be tempting to describe any field as an independent actor against an unperturbed background as the electromagnetic field in electrodynamics. The background in such a scheme possesses its own geometric features, it is a geometry by itself, without an overlaid field.

---

[208] There may be another choice of the free-particle Lagrangian, e.g., $L = -(mc/2)u_\nu u^\nu$.

### 9.8.1    The Equivalence Principle

I have already mentioned that the emergence of general relativity was not required by any practical needs. The observational discrepancies substantially troubled neither physicists nor astronomers. Yet there was one experimental fact observed already by Galileo in the late 16th century that was related only to gravitation and did not hold for any other force field. Namely that all bodies are moving in gravitational fields in the same way regardless of their composition (provided the initial conditions i.e., positions and velocities are the same). Or, in slightly different words, the trajectory of a body in a gravitation field depends only on its initial position and velocity and is independent of its composition. As customary in physics, saying the "body" one actually implies a point mass, but only for the analysis of mechanical motion. Notice that, for instance, for the electric field this statement is obviously false: the motion in an electric field is determined by one essential parameter, the charge to mass ratio (see Chapters 4, 5), with not only its absolute value but also the sign being important for the motion. It means that acceleration of the body in an electric field crucially depends on the body's composition.

The equivalence principle implies that in the reference frame of a free-falling body another body, so long as it is also falling freely to the same gravity source, does not feel any gravity. Indeed, the two bodies of any composition do not experience any acceleration with respect to each other. Due to this interpretation first adopted by Einstein [209], the equivalence principle is usually regarded as the cornerstone of general relativity. We have already seen that the Einstein's general relativity is grounded on the idea of general covariance, which can be traced back to Mach's thoughts of unifying accelerated and inertial motion. More specifically, the idea of general covariance may be expressed as the statement that all physical laws (and possibly other laws) must take the same form in all systems of coordinates, i.e., for all observers. When discussing some beautiful ultimate-precision experiments, we have seen that it may be considered a hard fact that all local centers of mass fall along identical (parallel displaced) trajectories[210] toward the gravity source. There seems to be no exception to this experimental fact in modern physics.

As I have just mentioned, the equivalence principle in its prototype form was known already to Galileo and also to Newton (they both established experimentally that the acceleration of bodies due to gravitation depends neither on their internal structure nor on composition expressed through the distribution of masses in a body, see, e.g., the classical paper by L. Eötvös "On

---

[209] Einstein formulated it as follows: "If a person falls freely he will not feel his own weight". See, e.g., http://en.wikiquote.org/wiki/Albert_Einstein

[210] From a theoretical viewpoint, these are the minimum action paths or geodesics.

the gravitation produced by the Earth on different substances." Available in
http://zelmanov.ptep-online.com/papers/zj-2008-02.pdf           ).    Newton
considered this principle as so essential for mechanics that he discussed it
already in the opening chapter of his "Principia".

    If gravity is a manifestation of spacetime curvature i.e., a purely
geometrical effect, then the question naturally arises: is gravity the same as
acceleration, e.g., the centrifugal force which is successfully used to
compensate gravitation, say, for astronauts' training? Or, to put it more
accurately, are gravity and acceleration manifestations of the same underlying
force? In the Einstein's general relativity gravity and acceleration are
mathematically indistinguishable, is it just a mathematical model or a
profound law of nature? In many situations, gravity and acceleration are not
only mathematically, but also practically indistinguishable. When a fighter
pilot performs acrobatic figures she/he experiences high acceleration, much
greater than $g$ - for today's supersonic jets acceleration values may reach
$9 - 10$ $g$, higher than the human body may endure. (The forces experienced
by a jet pilot are the masses of her/his body parts multiplied by this enhanced
acceleration.) Strictly speaking this is not a single force, the quantity $mg$ being
only the gravity component, but the pilot cannot distinguish between gravity
and acceleration, and the entire force is colloquially referred altogether as
the "g-force".

    Yet the equivalence principle does not say that *all* accelerations are
equivalent to gravitational forces - this would be a very strong assertion. It
only says that motion in a noninertial reference frame may be considered
equivalent to motion in *some* gravity field. This latter is not completely
identical with the physically real gravitation fields existing in nature (see a
very clear exposition of this question in [39], chapter 10).

    1. The Principle of Equivalence (or Equivalence Principle) does not say that
all accelerations are equivalent to gravitational forces. It only says that
motion in noninertial reference frames is equivalent to the one in some
gravity fields, these latter not being completely identical with the real
gravitation fields existing in nature. See a very clear exposition of this and
other questions being discussed here in the standard textbook by L. D. Landau
and E. M. Lifshitz "The Classical Theory of Fields", chapter 10.  2. As far as
Maxwell's EM theory combined with General Relativity goes, I would advise
you to read 90 of the same book. This is, by the way, about "more standard
established theories".

    The trajectory of a point mass in a gravitational field depends only on its
initial position and velocity, and is independent of its composition.

### 9.8.2    The Einstein Equations

So general relativity is in fact a theory of gravitation where the latter appears
as the spacetime property. The geometry of spacetime is determined by the
metric tensor $g_{\mu\nu}$ which in turn defines one more tensorial object, the
Riemann curvature tensor $R^{\gamma}_{\alpha\beta\delta}$ . This object is so important for understanding gravity that it deserves a special description and comments (see [39], §§91,92). I shall discuss it here only very briefly.

The Riemann curvature tensor is usually defined through the Ricci identity for an arbitrary covector field $A_\mu$:

$$A_{\alpha;\beta\gamma} - A_{\alpha;\delta\beta} = R^\gamma_{\alpha\beta\delta} A_\gamma, \qquad (9.19)$$

where

$$R^\gamma_{\alpha\beta\delta} = \partial_\beta \Gamma^\gamma_{\alpha\delta} - \partial_\delta \Gamma^\gamma_{\alpha\beta} + \Gamma^\sigma_{\alpha\delta} \Gamma^\gamma_{\sigma\beta} - \Gamma^\sigma_{\alpha\beta} \Gamma^\gamma_{\sigma\delta}. \qquad (9.20)$$

One can decompose 15.20 into the following sum

$$R_{\alpha\beta\gamma\delta} := C_{\alpha\beta\gamma\delta} + E_{\alpha\beta\gamma\delta} + S_{\alpha\beta\gamma\delta},$$

where $C_{\alpha\beta\gamma\delta}$ is the so-called Weyl tensor, $S_{\alpha\beta\gamma\delta} = (R/12)(g_{\alpha\delta} g_{\beta\gamma} - g_{\alpha\gamma} g_{\beta\delta})$ $E_{\alpha\beta\gamma\delta}$ is the Einstein curvature tensor:

$$E_{\alpha\beta\gamma\delta} = (1/2)\big(g_{\alpha\delta} \tilde{G}_{\beta\gamma} + g_{\beta\gamma} \tilde{G}_{\alpha\delta} - g_{\alpha\gamma} \tilde{G}_{\beta\delta}\big) - g_{\beta\delta} \tilde{G}_{\alpha\gamma}$$

and

$$\tilde{G}_{\alpha\beta} := R_{\alpha\beta} - \frac{1}{4} g_{\alpha\beta} R.$$

Here $R_{\alpha\beta}$ is the Ricci tensor which is defined is defined as the contracted Riemann tensor, $R_{\alpha\beta} := R^\gamma_{\alpha\beta\gamma}$ and $R := g^{\alpha\beta} R_{\alpha\beta}$ is the Ricci scalar. One can see that by using all these definitions it is possible to represent the Riemann tensor by another decomposition:

$$R_{\alpha\beta\gamma\delta} = C_{\alpha\beta\gamma\delta} + \frac{1}{2}\big(g_{\alpha\delta} R_{\beta\gamma} + g_{\beta\gamma} R_{\alpha\delta} - g_{\alpha\gamma} R_{\beta\delta} - g_{\beta\delta} R_{\alpha\gamma}\big)$$
$$- \frac{R}{6}\big(g_{\alpha\delta} g_{\beta\gamma} - g_{\alpha\gamma} g_{\beta\delta}\big).$$

## 9.9    Cosmology

There is probably no literate person in the so-called Western civilization who has not read stories by Sir Arthur Conan Doyle. If there is one, here is the link: http://www.sherlockholmesonline.org. To me, Mr. Sherlock Holmes, one of several famous personages created by Conan Doyle, was always remarkable in his inclination to aphorisms. One of the statements repeatedly pronounced by Holmes (see, e.g., https://www.goodreads.com/quotes/92128-it-is-a-capital-mistake-to-theorize-before-one-has) is often cited in physical papers: "It is a capital mistake to theorize before one has data. Insensibly, one begins to twist facts to suit theories, instead of theories to suit facts."
Sometimes this dilemma "theory first, data later or vice versa" is even called
the Sherlock Holmes problem. Strictly speaking, statistical inferences would
be impossible if one dogmatically sticks to Holmes' maxim. In observational
sciences, it is hardly possible to accumulate the whole evidence, moreover, it
is hardly possible to establish if and when the collected evidence is ample for
constructing a valuable theory. Therefore, one constructs models first and
then tries to reconcile them with available data. In astronomy and cosmology,
for example, there is simply no other way to do research.

Cosmology is an odd mix of observational facts, beautiful mathematical
models (predominantly based on Einstein's general relativity) and rather
wild speculations. Speculative fantasies about the early universe mostly come
from particle and high-energy physics. Formally, cosmology is the study of the
actual state and evolution of the universe as a whole or at least of its maximal
observable region. About a hundred years ago, the prevailing opinion was that
the universe must be static, and this view universally formed the perception
of the world, to the point that even Einstein shared it. As far as observations
are concerned, the existence of galaxies apart from ours was unknown. By the
way, cosmology is not the only discipline where models strongly influence the
perception of reality. In new physics centered around superstrings, the
relationship between models and reality is a bit twisted (see the
corresponding section below). Nonetheless, the subject of cosmology has
impressively matured lately as a physical discipline [285, 286]. The models
that have been traditionally treated only in "hard" physics are now
dominating cosmology [94].

Yet many models of modern cosmology seem to be little more than
philosophy, although camouflaged by some mathematics. One of the most
provocative philosophical ideas in modern cosmology is that the universe, a
tiny (and undetermined) part of which we observe is only one of many -
possibly of an infinite number - of individual universes. One of the prominent
hypotheses in this series of speculations claims that the laws of physics and
physical constants must be very different in all member universes in the
whole multiverse ensemble so that our universe has special conditions that
are fine-tuned to sustain our carbon-based life. This is a variant of the
anthropic principle. The latter is a philosophical principle which appears to
be impossible to falsify. In spite of this speculative context, many highly
qualified physicists are fully seriously discussing anthropic arguments as well
as the conjectural Drake equation associated with it, an attempt to estimate
an attempt to evaluate the potential number of extraterrestrial civilizations,
initially in our galaxy - the Milky Way, now in the multiverse where the fraction
of the universes with the favorable (for the human existence (in the spirit of
anthropic principle) physical laws and constants should be estimated, see one
of the latest discussions in [267] and the literature cited therein. I don't think
that such probabilities can be correctly derived unless a solid underpinning
theory is formulated which possibility is far from being obvious. Yet to lapse
into wild speculations, almost on the verge of daydreams is a very fascinating
pastime (recall, e.g., the character of Manilov in "Dead Souls" by N. Gogol).

In contrast with other physical models, cosmological models represent objects on the largest possible scales, e.g., the universe is treated as a whole. For such scales the geometry of spacetime is assumed to be described by general relativity i.e., determined by the metric tensor $g_{ik}(x^j)$. One should, however, remember that this method of description is just a convention and there may be other framework assumptions or variants of general relativity compatible with current astronomical observations. For example, the basic self-consistent approach to the question how matter determines the spacetime geometry (which in its turn determines the motion of matter) essentially depends on the so-called cosmological constant $\Lambda = -8\pi G \rho_{vac}/c^2$, where $G$ is the gravitational constant and $\rho_{vac}/c^2$ is the vacuum energy density (see below), that is typically assumed not to vary in space and time ($\nabla_i \Lambda = 0$). The difficulty with the $\Lambda$-term is that one can find no compelling evidence for setting up any specific value of cosmological constant so that there is no shortage in cosmological models corresponding to different values and domains of $\Lambda$ and resulting in significantly different possible behaviors of the universe. One can, in particular, set the value of $\Lambda$ by hand to zero, as it was suggested, e.g., by Einstein ([308]), but such an ad hoc fix can be difficult to view as physically generic.

From the theoretical viewpoint, the cosmological constant is caused by quantum vacuum fluctuations of the fundamental fields present in the universe: scalar, vector, tensor, and spinor fields. Here, however, there is a notorious problem: most standard quantum field theories using the concept of the quantum vacuum (zero-point) energy suggest large values of the cosmological constant, typically of the order of $m_{Pl}^4$ where $m_{Pl} = (\hbar c/G)^{1/2} \approx 1.22 \cdot 10^{19} GeV/c^2$ which exceeds by approximately 120 orders of magnitude the observed value of the cosmological constant. This discrepancy between theory and observations is known as the "cosmological constant problem": why is the cosmological constant so small or, to put it another way, why do quantum fluctuations not manifest themselves in a large $\Lambda$-term?

The conventional model of cosmological evolution states that the history of the universe can be traced back to a highly compressed state, when the now known physical laws were mostly inadequate. This protostate was followed by the inflation phase (see below) when tiny quantum fluctuations became the germs of the galaxies observed today. Afterwards, as the inflationary universe had cooled, radiation was decoupled from the matter and gradually became what we now know as the cosmic background radiation, whereas the density fluctuations contracted due to gravitational forces. Eventually the observed cosmic structure has organized. A fluctuation that might have occurred a dozen of billions years ago originated our galaxy, and this process was followed by a further chain of accidents leading to the formation of the solar system including the planet called Earth. Later, other low-chance accidents happened, which contributed to the appearance of life and rapid - as compared with the cosmological periods - multiplication of its forms (along with natural selection). Driven by the negative pressure produced by the mysterious "dark energy", the universe is expanding with an accelerated speed. Based on their increasingly sophisticated techniques, the astronomers have come to believe that the universe should be much larger than the remotest horizon that can be observed even through the most powerful telescopes.

There seem to be two more or less distinct brands of cosmology (and, accordingly, cosmologists): one is the "global cosmology", focused on the physical principles behind the evolution of the entire universe, and the other is more local, closer to astrophysics, scrutinizing the structure and behavior of separate objects such as stars, galaxies, black holes, quasars, etc. These two brands of cosmology deal with different space (and often time) scales, and there are few people who can professionally synthesize both perspectives. In this section, I shall deal mainly with cosmological models for the whole universe.

According to modern cosmology, our universe appeared from a primordial state approximately 13.7 billion years ago [211]. The term "primordial" in this context implies that the initial state was almost empty, with no matter and very little energy. Then the question naturally arises: where do all the stars, planets and galaxies come from?

One of the other basic questions of cosmology traditionally was: is the universe finite or infinite? In the Marxist philosophy [212] unconditionally accepted in the Soviet Union, the universe was viewed as infinite, I don't understand why. Here, one might recall the words of Einstein: "Two things are infinite: the universe and human stupidity; and I am not sure about the universe". Cosmology belongs to a class of synthetic scientific disciplines like ecology, biology, geophysics, embracing concepts from many other fields of science. Such disciplines are full of interconnections and complex interactions with seemingly alien conceptual systems. Due to its interdisciplinary character, synthetic disciplines are favorite targets for the philosophers. Thus, in cosmology, philosophically-minded people are usually very active.

One of the main concepts of cosmology is the discovery that the universe is currently expanding and that it originated with the Big Bang. Indeed, the expansion points at an extremely hot origin of the universe known as the Big Bang which is a starting point of all possible universe histories. It would still be strange if a curious person did not ask: what was before the Big Bang? However, in the milieu of professionals this question is considered impertinent. Saying differently, it is asserted that no event prior to the Big Bang was physically possible, which is a very strong assertion. One typically explains the statement of impossibility of events before the Big Bang by some semi-philosophical analogies, e.g., of the following type. The geometric properties of the Earth are such as not allowing one point to have the distance from another greater than approximately 20000 km. This is the maximal geographic length for the Earth. In a similar way, there must be (I would say might be) the largest time interval in the universe determined by its geometric properties. Therefore, it does not make sense to speak of the time as about the ordered sequence of measurable events outside this maximal interval. Yet I think that putting aside some typical quasi-scientific dogmatism, such naive questions of cosmology as: "Did anything happen before the Big Bang?  What will be the final fate of the universe? How credible is the Big Bang model?" are in fact very deep. We ought to remember that the Big Bang is only a model, though a dominating and, as many cosmologists think, a necessary one. Scientists usually do not like when someone dares to express doubt or skepticism about a currently dominating concept; one can observe the righteous indignation when, e.g., string theory or the Big Bang model are critically assessed.   It is considered a "mauvais goût", almost as if one criticized special relativity.  This is not a scientific feature, but utterly psychological: people are trying to evade cognitive dissonance. There are surely many facts compatible with the model of Big Bang and corroborating it, but all these facts are just interpreted observations of the Big Bang effects.

---

[211] This date varies from source to source, but it is not substantial since the exact, e.g., up to a million years, age of the universe is an abstraction anyway.

[212] It is just an ideological cliché and has nothing to do with Karl Marx. Marxism and Marx are almost unrelated.

In cosmology, there is no shortage of models bordering on wild speculations. Inflation is probably the most plausible one among these wild speculations. Inflationary scenario is actually a bunch of models, so fascinating that more and more cosmologists are contaminated with the ideology of inflation. As far as modern astrophysics can tell, the universe is about $13.7 \; 10^9$ years old and has the size of $93 \; 10^9$ light years. The combination of these two numbers produces a little mysterious impression since in 13.7 billion years light can only travel $13.7 \; 10^9$ light years.  Then how is the universe that big?  And how did it become so big that fast? A short answer:  inflation. This model implies that the universe, soon after its birth, exponentially increased its size by many orders of magnitude during a very short time. But this is again a little mysterious. What triggered inflation? What stopped it? Why cannot the universe continue to inflate at an exponential rate?  There are claims that the reliable answers to these questions do exist, but I, having read some literature on inflation, was only reinforced in thinking that there exist little more than just speculative models, although I am in no way a specialist in the field.  By the way, one of the possible answers to the question why inflation stopped is that it did not stop. The universe, according to the model of never-ending inflation is still exponentially expanding its size, but we cannot observe it so far because we live in a small region of stability - in a cosmic bubble. I don't know how one can prove or falsify this conjecture, provided the bubble boundaries are beyond the edge of the visible part of the universe. There is also an open question: are there other bubbles? If one of them exists, why cannot there be many of them? Is the number of bubbles finite? Besides, if there are many bubbles, can they interact, e.g., collide with each other?

## 9.10  Black Holes

We have seen that there exist a great lot of crossovers between diverse fields of physics, but the issue of black holes (BH) probably contains the majority of them. Models associated with the concept of a black hole attract scientists with different background - and not only scientists. It is curious that the thermophysics of black holes has been used recently for practical purposes, namely for navigation engineering. Now ubiquitous GPS devices are tied to satellites whose orbital position (ephemeris) must be determined very precisely. However, referring the satellite position to the planet may have poorly controllable errors, e.g., owing to fluctuations of the Earth's rotation axis orientation. Simply speaking, the planet is constantly wobbling. Such fluctuations may have a number of underlying physical causes such as tidal forces induced by the gravitational pull from the Moon or coupling to the Sun, ocean currents, or fluid motion in the Earth's molten core. In general, fluctuations are omnipresent regardless of whether one can detect their immediate cause or not. So for highly precise measurements, tying up satellites to the Earth's position in space may have an insufficient accuracy. In such cases, stars seemed to be the obvious candidates to designate the landmarks in space, and indeed they have been used for navigation - in fact stars had been used for navigation by the seamen long before such modern notions as signals, frequency bands, trilateration, satellites, GPS and the like became known. However, for exact position determination needed in today's navigation and, e.g., in geodesy even stars are poorly suitable because they are also moving. So, the objects that are sufficiently bright to be observable through astronomical distances and whose position may be considered stationary in fair approximation should be chosen for navigation. Such objects do exist in nature: those are the quasars. The typical brightness (luminosity of $10^{40}$) of a quasar is $10^{12}$ that of the Sun (quasars are believed to be the brightest astrophysical objects), and most quasars are remote enough - over $10^9$ light years away[213] - to appear stationary within the accuracy of the current detectors. More than $2 \cdot 10^5$ quasars have been registered by astrophysicists (see, e.g., http://www.sdss.org), and about 3000 of them are said to be possibly used for navigation purposes. A grid of distant quasars whose positions in the sky is known with a sufficient accuracy can form a map to be used for specifying the Earth's axis orientation and, consequently, for satellite-based positioning systems. Such a map, e.g., ICRF and ICRF2 (International Celestial Reference Frame) has been recognized by the IAU (International Astronomical Union) as the primary reference system for astronomy. I am not going to discuss this issue, quite interesting and having many physical aspects, see more on that, for example, in connection with VLBI (very long baseline interferometer) in, e.g., http://ivscc.gsfc.nasa.gov/.

Now, what is a quasar? A standard astrophysical model today for such an object is that it consists of a massive black hole feeding itself on the surrounding interstellar matter (gas and dust). The material being trapped inside a black hole is compressed by gravity forces and thus heated to estimated $10^6$ K emitting intense blackbody radiation in the X-ray, UV, visible, infrared, microwave, and a part of them also in radiofrequency ranges, sometimes almost equally over these ranges. This devouring mechanism (called accretion) of energy production is not a single model for quasars. There were also speculations about possible involvement of antimatter so that high luminosity of quasars could be explained by the annihilation; there also existed concepts based on "white holes" (a white hole is a time-reversed black hole).

---

[213] A light year, $ly \sim 10^{13}$ km is the main astrophysical unit of length; our galaxy (the Milky Way), containing of the order of $10^{11}$ stars, is estimated to be $\sim 10^5$ ly across, see, for example, http://www.atlasoftheuniverse.com/galaxy.html.

Life-time of black holes, $\tau$, should depend only on mass $M$.

If you, by any chance, approach a black hole, the tidal force between your head and toes will tear you apart. Where and when this sad event occurs, depends on the mass of a black hole: if it has a typical stellar mass, say, 30 times that of the Sun, you will be torn apart well outside the horizon of the black hole - there are estimates showing that for such an object the tidal force acceleration between your head and toes would be $10^6$g as you pass the horizon. For supermassive BHs you could survive inside the horizon before being torn apart.

Conservation of energy, in our standard understanding, is possible only in flat space. When the horizon is present, pairs can be produced. Information conservation: it is generally believed that information should be preserved under quantum-mechanical operations although you would not find any correctly defined continuous symmetry corresponding to this presumed conservation law.

## 9.11  Quantum Gravity

This is perhaps the most intricate problem of physics. Quantum mechanics and theory of gravitation appear to be different in every aspect, and it seems to be one of the greatest puzzles of gravitation why it does not naturally unify with other fundamental forces. General relativity should be regarded in this context as the classical limit of quantum gravity. Yet today the theory of gravitation is based on general relativity which is a geometrical theory whose geometry seems to be totally different from that of quantum mechanics, even in the most far-fetched geometrical formulation. It has been difficult so far to find links between the geometry of quantum mechanics and Riemannian geometry of general relativity, and many prominent physicists have attacked the problem of quantum gravity but still unsuccessfully. Nevertheless, much knowledge has been accumulated in this field. Of course, it would be too ambitious to examine the whole scope of the distributed knowledge that guided physicists in addressing the problem of quantum gravity. Being not a specialist in this field, I shall delineate this problem, its scope and origin, very superficially. I think that specifically in this field containing a lot of "raw material" - poorly established results - one must inevitably discuss the reliability of the distributed knowledge, which leads to some reflections on methods of obtaining this knowledge, i.e., on methodological questions.

One such methodological question immediately comes to mind. One of the obvious consequences of the Schrödinger equation is that the wave packet corresponding to a free particle should spread to infinity in empty space, this fact follows directly from the mathematical structure of the Schrödinger equation expressed, e.g., in terms of its Green's function. It would not be easy to reconcile the diffusive spread of wavelike solution in the Schrödinger phenomenology with the picture of particles collapsing due to the gravity field.

One can hypothesize that it is probably difficult to reconcile relativity with quantum mechanics since the two theories have totally different focuses: relativity deals with the nature of spacetime whereas quantum mechanics is concerned with the microscopic properties of matter. On a rather primitive, intuitive level quantum gravity may be illustrated by the necessity to deal with such unconventional objects as smeared points and fuzzy light cone (recall that the light cone is a causality arrangement). Also recall that the quantum field at small distances where the quantum gravity effects should be essential has a cell structure and evolves dynamically.

## 9.12 String Theory and String Theorists

"If facts contradict the theory, too bad for the facts". (A modern proverb)

String theory originally appeared as an attempt to describe strong interactions (see, e.g., the well-known papers by Y. Nambu [287], B. Nielsen [288], and L. Susskind [289], see also the historical overview by J. Schwarz [290]). Later, however, after some period of initial enthusiasm, high-energy physicists began entertaining doubts that the string theory of that time could lead to a genuine understanding of strong interactions. I think that such a sentiment was only reinforced after the emergence and explosive development of the quantum chromodynamics (QCD). Then a certain shift of focus occurred: quantum gravity became the dominant interest of string theory.

In many ways string theory has turned into an eclectic set of closely interrelated mathematical theories or models. There is nothing wrong in such a development: for example, ecology is also an eclectic collection of models, nonetheless people are paying more and more respect to this discipline (regardless even of its political engagement). It is not necessary that a discipline be well-composed, harmonious, and deductive as, say, the number theory, calculus, or classical mechanics to be popular. But it is important for a physical theory that it would be able to predict observable physical effects, and as near as I can judge one should not speak about string theory in terms of its predictions unless there are confirmed ones. Unfortunately, some theories are consistent with pairs of mutually excluding predictions i.e., are unfalsifiable. For example, string theory "predicts" the existence of massless fields to the same extent as their absence just like climate theories predict simultaneously warm and cold weather periods in a given location and economical theories can simultaneously predict inflation and deflation.

There are many examples of success story of bubbles. As I have mentioned before, nanotechnology (especially in the beginning of 2000s), IT (in 1990s), high-$T_c$ (in the 1980s 1990s), chaos, and superstrings have been fashionable and attracted crowds of people interested not in ideas but in lavish financing and rapid success. This fact is so often mentioned as pure sociology instead of physics that it has become a banality. Still, I shall allow myself to comment on the current situation with new and highly fashionable subjects, because despite obvious sociological effects there is a really interesting physical and mathematical content in them. Perhaps these comments induce someone to think more with her/his own head rather than follow the fashion. It is obvious that the one who puts one's feet into the others' steps leaves no individual traces.

String theory appeared as a mathematical construct unprovoked by experimental challenges. This was rather a product of internal theoretical development than a response to any discrepancies (like in the neutrino case), but if accepted - even without experimental evidence - it will radically change physicists' concepts of matter and spacetime. Fundamental quantities of matter, according to string theory, are not point particles, as it was traditionally assumed, but some extended small objects called strings. Such objects, in contrast to particles, may have their own degrees of freedom. As far as spacetime is concerned, the number of basic dimensions is not 4 as we used to think but many more, e.g., 10, 11 or 26 - depending on the theory variant. Although many string theorists do not even hope on experimental confirmation of string theory, there are claims that it has the status of the long anticipated "theory of everything" (TOE), see, e.g., https://en.wikipedia.org/wiki/Superstring_theory.

On the mathematical side of string theories, there are possibly difficulties in choosing the right model for them. One can find certain parallels between modern string theorists and ancient philosophers. Like ancient Greek philosophers, string theorists are trying to explain the construction of the world by pure thinking, relying more on group attitudes, internal communication and speculative models than on observational data or directed experiments. Both groups used to say that their theories are not bound to provide earthly, profane results. A genuine high-level theory such as M-theory is not obliged to give a clue how to descend from fundamental equations to the real-world problems. Maybe I am inclined to a conservative, old-fashioned way of thinking, but I don't see any value in such high-brow theories: they contain a mythological element and might even produce an unpleasant association with astrology in the sense that it does not matter whether they exist or not. One more feature of string theory invoking involuntary recurrences to ancient philosophers is the fact that there exists not a single one but numerous string theories - like there were many schools of thought in ancient times[214]. Today, one can observe an attempt to unify all string theories within the framework of some mysterious M-theory, but so far no one seems to know what it is. Being interested in "M-theory", I nonetheless failed to find some fundamental microscopic parameters and corresponding degrees of freedom that could, in my dogmatic opinion, constitute the essence of any physical theory. I also could not see how the fundamental states in M-theory would be defined, but it is of course due to my extremely low qualifications in this new kind of physics. Moreover, this theory is under construction. Yet, to be  honest, I don't feel comfortable with the theory that relies on hopes, group attitudes, conjectures, metaphors and analogies, albeit garmented in a sophisticated mathematical form (e.g., noncommutative geometry). I also don't feel quite comfortable with 10, 11, 26 or undefined number of dimensions - depending on the model, are you? Then why is it stipulated that the world is made of strings and not, say, of tiny spherical clusters? The latter can also produce vibrating modes identified with some constants, masses, etc. One can also construct Lagrangian density, action, impose boundary conditions, develop somewhat speculative but sophisticated mathematical models and the like. Why has the string model survived so long even though it has so many problems?

---

[214] There exists nowadays a supposed connection between at least some of string theories in the form of duality relations (see Chapter 3 Dualities in Physics) when, e.g., one theory is viewed as a limiting case of some other, the limit being taken over some phenomenological parameter of the theory.

I may only repeat that such fault-finding remarks of mine are obvious, not new, and thus trivial. One can find a critical analysis of string theory on the popular level, but performed by an expert in the well-known book by L. Smolin [74]. But a nontrivial thing is: why the whole world and specifically physicists who traditionally cultivated skepticism as their group characteristic together sing rhapsodies and stand frozen in awe at the sight of far-fetched models based on artificially sophisticated mathematics? Besides, declaring string theory a new TOE people seem to forget that the original motivation for modern string concepts was the construction of a consistent quantum gravity, which is a more modest quest. Sorry, but I am inclined to think that in the issue of string/M theory people are getting a bit off track. And I still neither understand nor share the general exaltation, while the gap between the claims of string theory protagonists and physical reality is still insurmountably wide and reminds us of wild expectations of the space travel enthusiasts.

A single fact can spoil a good argument. This maxim fully applies to the string theories. Thus, it would be extremely interesting to collect the pieces of relevant information coming out from the planned experiments on the Large Hadron Collider machine (LHC) under construction and permanent repair in CERN (originally Centre Européenne pour la Recherche Nucléaire, now mostly known as Organisation Européenne pour la Recherche Nucléaire). In particular, project ATLAS (a toroidal LHC apparatus) is one of six particle detector experiments currently being assembled in CERN aimed to provide new data both on the Standard Model and string theories. To understand the main ideas underlying the series of grandiose experiments in CERN, let us briefly recall some concepts of the generally accepted string theory.

One of the startling predictions of string theories is the existence of extra spatial dimensions. If they happen to be not too tiny, they can, in principle, be traced in high-energy accelerator experiments. Even in case such extra dimensions are too small for current accelerators and if, what is very probable, there will be lack of motivation and/or resources to carry out further high-cost accelerator experiments, these hypothetical dimensions of the space we live in can be observed as remnants of very early cosmology. However, to my best knowledge, nothing indicates possible experimental proof of extra dimensions.

Therefore, in the low-energy domain, string theories can bring about a number of massless scalar fields that, in principle, can influence experimentally verifiable low-energy physics, e.g., by violating the equivalence principle or leading to deviations from Newton's law. Such fields might also induce spatial and temporal variations of fundamental constants. However, no imprints of such massless particles have been revealed as yet.

Thus, string theory as TOE including gravity does not represent an extension of the existing high-energy physics. Rather it may be viewed as a collection of artificially invented mathematical models, quite handsome but without any connection to experimentally confirmed physical principles, at least by this time. Standard analogy with positron as well as other theoretical predictions does not pass here, since energies to secure the vague predictions of string theories experimentally seem to be hardly reachable.

Centuries ago, philosophers modeled the world by mixing various combinations of the four primary elements: Earth, Air, Fire, and Water (EAFW-model). About the same time, healers suggested that bodily discomfort and diseases emerged due to bad perspirations that were inhaled in the presence of an ill or morally spoiled person. In the 17th century, some "heretics" suggested the existence of tiny particles called germs that could appear in many more varieties than the four basic elements and cause sickness. All those were purely speculative models, initially of the same type as angels upon the tip of a needle or little devils in the toilet. The general idea, however, was to offer an explanation based on invisible objects. Yet "invisible" does not necessarily mean "unobservable", and when the technology of physical experiment progressed over the later centuries, one could see the formerly invisible minuscule particles and germs. These visualizations brought about new understanding in the natural sciences, primarily in physics, chemistry, biology and medicine. However, not all phenomena could successfully pass from the category of invisible to observable; there are still perennial problems with wild speculations, and the main question is: how to separate invisible from unobservable, figuratively speaking fantasies about angels and demons from phenomena caused by mismatch between "large" and "small".

Many pseudo-scientific models and theories discussed in Chapter 2 appear to be of the unobservable character or are heavily relying on unobservable entities such as biofield, orgone energy, etc. There are also a lot of poorly defined and unobservable notions in generally accepted disciplines such as psychology, psychoanalysis and psychiatry, which often allows one to misuse these disciplines for commercial and political purposes (even in democratic countries). All this may be viewed as remnants of medieval atmosphere when opinions were valued much more than facts. Unfortunately, string theory is so far controversial exactly in this "medieval" sense[215]. This theory exploits the need to overcome a mismatch between very large (mainly cosmological, when gravitation is substantial) and very small (Planck) scales claiming that it can bridge general relativity and quantum mechanics. Strings are obviously beyond our current ability of direct detection. The adherents of the string theories, mostly but not exclusively young persons belonging to the new breed of physicists, rather aggressively contend that the symmetry and presumed mathematical elegance of string theories is already a sufficient ground to consider them as variants of "theory of everything", a grand unifying concept that bridges general relativity and quantum mechanics as well as includes all known interactions such as gravitation together with those described by the Standard Model[216].

---

[215] We may jokingly classify all the models and theories as the germs-type and the angels type. Using this funny terminology, one might ask whether string theories are of the first or the second type.

One often says "string theories", not theory, because in fact there are many of them (see, for example, [http://www.stringwiki.org/wiki/String Theory Wiki](http://www.stringwiki.org/wiki/String Theory Wiki)). Recall that the basic string theory suggests that fundamental particles are not zero-dimensional points as prescribed, e.g., by special relativity, but perpetually vibrating lines or loops - thus the name "strings". These submicroscopic strings can be open-ended or closed and possess natural frequencies as any macroscopic string. At its core, the theory of such type can produce various mass values for observable elementary particles just by combining different vibration frequencies. In fact, one of the dominating mathematical models of quantum mechanics is also based on obtaining numbers with the aid of vibration frequencies. Of course, string theories are much more sophisticated, especially with regard to the mathematics used, to the extent that ordinary quantum mechanics looks like a primitive school-time algebra as compared to the rich geometrical techniques of string theories. The latter, for instance, assert that not only point-like objects should be rejected, but also conventional spaces (or manifolds) where mechanics - classical or quantum - evolves. Even the spaces suggested by general relativity are inadequate to sustain strings. These energy lines or loops presumably vibrate on a six-dimensional manifold that should be added to the background four space-time dimensions prescribed by relativity theory. Although, as we have discussed, physicists and mathematicians nowadays are trying to visualize objects in multidimensional spaces, it is by no means intuitive and easy. Just try to visualize a ten-dimensional cube. Besides, as if all these difficulties were not enough, according to the late developments in string theory one needs one more dimension, the eleventh. Otherwise, anomalies inherent in the theory spoil the whole beautiful picture. Moreover, the eleventh dimension was originally included in the string theory to account for the effect we perceive as gravity. The initial idea was to visualize the eleventh dimension as a membrane having two orthogonal directions - one along the membrane and the other directed outside of it. Then there comes one more conjecture namely that the gauge bosons of the Standard Model are confined to the membrane space so that they can only travel along the membrane, whereas the messengers for gravitational interaction - gravitons - are allowed to escape the membrane moving outwards. The loss of gravitons can be described by the term which is missing both in the Standard Model and in general relativity and which would couple both theories. It is the absence of such a coupling term that results, according to some developers of the string theory, to a complete separation of gravitation and elementary particle physics together with quantum mechanics.

---

[216] Recall that the Standard Model essentially consists in the suggestion that the forces between elementary particles are due to the exchange of gauge bosons playing the role of messengers.

By the way, the hopes of experimental validation of string theory are partly related to this concept of escaping gravitons. One experiment in LHC is designed in such a way as to create the conditions, in the high-energy collision between hadrons, when the energy loss due to escaping gravitons could be detected and measured.

## 9.13  Is Relativity Firmly Established?

When young Albert Einstein proposed in 1905 the theory that was later called special relativity, he assumed that it was utterly impossible to find an absolute velocity. Any velocity, according to Einstein, is considered as being related to the coordinate frame at rest. The astonishing, especially for non-physicists, feature of special relativity is that any observer measuring the speed of light (in vacuum) will get the same answer. It is this invariance of the velocity of light that provokes a lot of people (mostly non-professionals) to try to disprove the relativity principle[217]. Sometimes the controversy stirred by such disproval attempts becomes quite heated. The flux of papers derogating Einstein and the theory of relativity does not diminish and seems to be synchronized with social tensions and hardship periods. I have already mentioned that during the Perestroika period in Russia (then still the Soviet Union), I was working as an editor in the leading Soviet popular science journal. Every day I got dozens of anti-relativity letters and articles whose authors usually rather aggressively required from me to immediately publish their papers. At first, I tried to answer these letters, but this consumed nearly all my time and, paradoxically, provoked still more aggression, so eventually I gave up and simply disregarded them. Besides, a vast majority of anti-relativity papers were not only very, very poor from the mathematical viewpoint, but also carried distinct antisemitic overtones. Indeed, one can observe that anti-relativism and antisemitism are correlated.

Some anti-relativists argue that Einstein has produced nothing cardinally new since Lorentz transformations had been already known and used, in particular, by W. Voigt (1887), J. Larmor (1897), H. A. Lorentz (1899), A. Poincaré (1900, 1905) (see, e.g., https://en.wikipedia.org/wiki/History_of_Lorentz_transformations). This is true, but none of these great scientists had formulated relativity as the fundamental principle to which all physical processes must be subordinated. We have already discussed in connection with relativistic mechanics (Chapter 4) that relativistic symmetry groups impose stringent constraints on all interactions in nature (see a detailed discussion in [193], §19).

---

[217] Roughly speaking, the relativity principle may be formulated as follows: if there are two inertial systems, $K, K', K'$ uniformly moving with respect to $K$, then all natural phenomena occur in $K'$ according to exactly the same general laws as in $K$. In other words, this principle states that physical laws should be the same in all inertial reference frames, but that they may differ in non-inertial (e.g., rotating) ones. This is a powerful unification principle: it is because of this principle that one cannot determine an absolute speed (understood as a vector). The relativity principle imposes a very general symmetry on the laws of nature which are not allowed to arbitrarily run their course in different inertial systems. A bold assumption indeed!

I remember an episode in autumn 1987 when two rather physically fit and a priori hostile gentlemen came to my office with a categorical request to derogate the theory of relativity and the "damned Jewish science" altogether. One of the two who was older and more talkative represented his younger protégé as a new genius who could easily refute the whole concept of relativism. I asked the older one: "And what about you?" He answered: "I am his manager" which was the term imported from the Western capitalist environment and utterly alien to the Soviet people at that time. "What else are you doing in life?" was my second question. The older muttered somewhat evasively: "We are from the Institute of Physical Culture and Sports." The younger guy was scrutinizing me sullenly like a boxer before the inevitable fight. Then he caught a barely visible nod of the older man and produced a rather thick handwritten manuscript with a few colored formulas - I remember they were written in purple red, green, blue, and violet. I could see right away that the whole stuff had nothing to do with physics. "Why don't you publish your theory in a serious physical journals? This is just a popular science magazine," I said. "We publish only well-established and only after they have been published in scientific press. This is an absolute principle imposed by the Central Committee, and I can't break this rule. You see, we belong to the "Pravda" publishing house." That was not totally a lie because our journal did belong to "Pravda" and accordingly was controlled by the communist party Central Committee, and the principle of the "second night" had been strictly adhered to before Gorbachev's "perestroika" began in 1985. By 1987 one could already deviate from any stringent rules imposed by the party. Nonetheless when I mentioned the Central Committee the sulky "boxer" became less belligerent, probably a deep-rooted fear of "party organs" resident in most Soviet people had not evaporated yet. So, I politely escorted the two gentlemen to the entrance door. I think it might have come to a fight had I decided to actually discuss relativity theory.

One more episode about the same time was more scientifically flavored. A man who called himself a professional mathematician rather aggressively insisted on publishing an article about logical inconsistencies in special and general relativity. As far as I remember, his objections concerning general relativity were centered around acceleration: since the equivalence principle presumably states that since, on the one hand, any acceleration is equivalent to gravity and, on the other hand, according to the second law of Newton any acceleration is equivalent to force then any force in nature should be reduced to gravity which is not true and therefore general relativity is logically inconsistent. Both statements articulated by "the mathematician" are wrong, but I could not explain it to him because he just did not want to listen. By the way, later I have encountered similar dogmas about relativity several times more, mostly formulated by the persons who could not even write the equation for geodesics. And such people want to disprove relativity.

However, strong anti-relativistic sentiments persisted not only in the unfortunate Russia. I was present once at an international conference held in April 1988 in Munich devoted to disproval of the theory of relativity. To my astonishment, there were numerous participants present from all over the world, some of them being quite qualified persons. But nearly all were focused on rejecting the theory of relativity. They were genuinely sure, for example, that special relativity - which is nothing more than a geometric theory - is fundamentally wrong. That was a purely psychological fixation having nothing to do with scientific objectivity. These people's attitudes and behavior are still a riddle for me. Is it a fear of some unconscious kind, induced by counterintuitive suggestions such as interval invariance? Later I found out that anti-relativity conferences are repetitive and gather a vast audience (see, e.g., http://www.relativitychallenge.com). One can also recall a historical irony: A. A. Michelson whose brilliant experiment with E. Morley directly led to the relativity principle never accepted relativity theory. On the contrary, modern experimenters who study the theory of relativity at the universities and commonly fully accept it still strive to measure the "absolute velocity" [131].

Classical physics with its simple Galilean (and to some extent also Minkowski) space-time in many cases provide a very good approximation to the real world, but one should always bear in mind that it is just an approximation. I concede that the anti-relativistic movement may have a purely human element. The point is that the human brain is tuned for the Euclidean geometry and absolute Galilean reference frames. No wonder that many people refuse to abandon their "comfort zones" i.e., the local Euclidean-Galilean illusion. Thus, it is difficult for many people to distinguish between the 3d distance and the 4-distance in the Lorentz frame (which is zero for events on the light cone). It had always seemed quite obvious before Einstein's relativity appeared and then became a dominating paradigm that the geometry of space was fully Euclidean, all logical possibilities for non-Euclidean geometries explored by J. Bolyai, C. F. Gauss, N. I. Lobachevsky, B. Riemann notwithstanding. So, the Euclidean vision of the world is probably still deeply imprinted in the common conscience. Recall (see Chapter 3) that the Galilean space-time, which is a structure incorporating the Euclidean geometry, is strictly speaking not a metric space where a single metric tensor defines the distance between any two points or, more exactly, the arc length of any curve. In the Galilean spacetime, in distinction to spacetimes of general relativity, there is a big difference between spatial and temporal intervals (and in general between space and time), see a clear and comprehensive discussion of the Galilean spacetime features in [103], ch.17.

By the way, absence of a unified metric space is a rather typical situation in life. Take music, for example, it is, in elementary representation, a 2d process: tone versus driving rhythm. The tone scales (pitch levels) can be measured in hertz (Hz) and have nothing to do with rhythm basically measured in time units (s). The corresponding 2d space might be called

"pitchbeat", and its two coordinates cannot be unified in a realistic musical theory.

The Galilean structure is in general not a metric space, but a fibre bundle i.e., a direct product of two manifolds: time $T \subseteq \mathbb{R}$ and space $S \subseteq \mathbb{R}^3$, see 3.22, "Geometry of Classical Mechanics" in Chapter 3 (for simplicity we consider here a single particle). Time serves in this bundle as a base manifold, to each point of which is attached (projected on) a fibre. Each fibre can be interpreted as a clone of the 3d Euclidean space. Such a structure is sometimes called "chronogeometry". Intuitively, one can understand the pre-relativistic Galilean-Newtonian spacetime as the 4d geometric object that can be factorized into a set of Euclidean hyperplanes on which time is constant and geometry is represented by the covariant rank 3 metric field. More formally, one can say that all covariant derivatives of the time manifold with respect to fibers are zero. Mechanical motion in such a structure is understood as a mapping $F: I \to S$ of some interval $I \subset T$ into $S$, the graph $I \times F(I)$ of such a mapping is called the world line which is a curve passing through the spacetime. The motion mapping is usually considered smooth and invertible, at least all the encountered functions are assumed to possess all needed derivatives and analytical properties. The projection of the world line fragments on the base manifold $T$ defines time intervals $t_{ij} = t_i - t_j$, and its projection on any of the fibres has a length $l$, but the world line as a curve in the Galilean spacetime does not provide a a well-defined arc length since each manifold present in the direct product characterizing a bundle, base $T$ and fibers $S$, have their own metric. In contrast with the Galilean structure, the spacetimes of relativity theory are genuine metric spaces[218] where the arc length is composed of time and space intervals that cannot be correctly (e.g., uniquely) separated into two different equivalence classes.

Relativity theory has accounted for a vast range of experimental facts, both known at the beginning of the 20th century and accumulated ever since. It is impossible to construct a particle accelerator without taking into account relativistic principles [118]. In this respect (and in many others, too) the theory of relativity has become an engineering discipline. So far, despite numerous attempts, no experiment has contradicted special relativity - at least I am not aware of such experiments. This implies that any modern theory or model should be consistent with special relativity, and future theories dealing with space-time should include it as a special case. The psychologically charged discussions and the general controversy over special relativity stem, in my opinion, not from misrepresentations of physical experience such as, e.g., the Michelson-Morley experiment or more modern tests (see the comprehensive overview by John Baez, http://math.ucr.edu/home/baez/physics/Relativity/SR/experiments.html) but from reluctance to accept the space-time geometry which is different from the space geometries of classical mechanics (see Chapter 4). Some training in contemporary geometry of physics would have removed a great many objections of "relativity foes".

---

[218] One sometimes talks about metric manifolds or manifolds with metric.

There are, however, some issues in relativity that are not so easy to digest from the physical point of view. One may notice, for instance, that light in the theory of relativity in any case in special relativity and even in general relativity - is hardly treated as a physical entity possessing material properties such as a flux of photons or a wave. Light is a highly abstract thing in both relativity theories appearing only as a means of signaling between different spacetime points[219]. One should not call this signaling feature "light" because it has nothing in common with the physical ray of light. It is sufficient to assume only the constant velocity of signaling between different spacetime points to build up the geometric carcass of special relativity. Likewise, one can only consider the curvature produced in the space by masses to define geodesics along which light signals are supposed to propagate. In general relativity these geodesics are no longer straight lines as in special relativity.

No physical questions such as concerning possible light interaction with matter (in the form of masses) or about individual properties of light and matter (mass) are involved in the mathematical models of relativity. Since it would be irrelevant to pose questions about the physical nature of light and matter based solely on relativistic considerations, in order to draw some conclusion about physical properties of light (field), matter (mass) and light-matter interaction one has to combine relativity with other mathematical models. Not all such combinations are easy to produce, mathematical models may prove incompatible. This is a notorious difficulty in combining relativity with quantum theory.

---

[219] I do not touch upon Maxwell's equations in the context of standard special relativity which has a kinematic character.

# 10 Climate as a Physical System

*Sapere aude. Find the courage to know.*

In this section some physical and social aspects of modeling the Earth's climate variations are discussed. Eventually I thought of placing this passage into "What Remains to Solve? chapter but considering the amount of attention the subject is attracting by the broad public audience, I decided it deserves a separate section. I must confess that I am not a meteorologist, nor a climate modeler, and I have not summarized surface temperature indices for years to make a living. I just read publicly available sources and try to make *my own* picture. It may be distorted or totally wrong, but I have honestly tried to understand what is going on. In general, one of the most important qualities of a researcher is to understand what is currently important in the new field and what is not. I was doing my homework trying very hard to stay uninfluenced by numerous journalistic comments about climate as well as by rather wild projections one can find in today's polemics. My purpose in this issue is to learn and not to take an a priori position. And I remember quite well the saying by Benjamin Franklin that any fool can criticize, condemn and complain and most fools do. One often says that there are two things in which everyone feels herself/himself an expert: football and medicine. Now it seems climate and energy problems take over as fields of ubiquitous expert knowledge. The attitude that anyone's opinion on any subject is equally valuable brings the very idea of expertise to a collapse. A world in which there is presumably no difference between those who know what they are talking about and those who don't is going to be ruled by ignorance and stupidity. I have read that about 70 per cent of the population in Europe consider climate issues more important than nuclear confrontation, international terrorism, poverty and unemployment. I find it astonishing. This section is to some extent a reflection of such astonishment and may be perceived not as a scientific endeavor but rather as a skeptical layman's bewilderment. I see my task here in signaling some awareness and a sense of growing apprehension.

How can one make and support crucial political decisions when there is still so much to learn about the Earth's climate? The reality is that people - even self-assured climatologists - still know no more about the long-term human influence on climate than about what had happened before the Big Bang. Uncertainty can make some people worry, but political decisions affecting billions of people who want to improve their living conditions should not be based on uncertainty. However, it is hardly the uncertainty that might worry, on the contrary - it is certainty. A good way to hide something is to conceal it behind an illusion of knowledge or even science.

My own motivation to join the debates on climate warming is not in an attempt to say something new. The point is that climate variability is an objective reality since climate is a dynamical system and it should change. Fighting the global warming appears to be a fictive environmental activity worsening our lives. It seems to be a poorly camouflaged attempt to restore the longed-for political attitude of the 19th and the first half of the 20th centuries, when political leaders alone were allowed to run (and ruin) people's lives according to those leaders' interests. Contrariwise, explaining the roots of the misplaced environmental activity, while studying the real causes for climate change seems to be a decent task. Besides, if you constantly hear about the imminent climate catastrophe threatening your existence (see, e.g., WWF reports), and if you are raped with strange new taxes, it would only be natural to do one's homework trying to find out the physical facts underlying the terrible global warming. As for me, I don't have much faith in the predictive power of climate models - as little as of social or economic ones. One can notice, for example, that highly discussed the so called AGW factor (anthropogenic global warming caused predominantly by human activity) is not the fact, it's a hypothesis. Even assuming that the Earth's surface is found in the warming phase does not prove it is due to anthropogenic carbon dioxide.

This does not mean that humans are incapable to do much harm to the environment - unfortunately, they can affect nature substantially. We know that there is much pollution and in general we are observing a rapidly degrading habitat caused by humans. However, AGW seems to be a misplaced threat, dictated by political and not ecological purposes. The escalating role of biosphere, in particular of humans, must be assessed in numerous other areas, where international bureaucracy and certified environmentalists does not want to poke, and not be reduced to looking for scapegoats such as the ubiquitous $CO_2$.

A number of assumptions might be invoked by looking at the title to this section. The first is apparently unarguable: that physicists begin to enter the field traditionally occupied by meteorologists and climatologists. I would add: this fine field has been lately occupied by the politicians who are more and more aggressively intruding into it. Therefore, we shall have to discuss in this section not only purely physical problems of tinkering with the computer models of the atmosphere, but to some extent political implications as well. The second is that the tradition of carefully assembling data and using the time series for data analysis may be threatened since there are totally different traditions in physical analysis which is mostly based on modeling and construction of theories. The third assumption is a consequence of the second and may be that immature but pretentious models of physics would dominate over the refined intuitive models, requiring almost esoteric knowledge for climatic prognoses. Models, one may assert, are a fake, sterile knowledge; they have little to do with real life (see the discussion in Chapter 2).

There exist a number of highly contested issues in climatic studies. First of all, one should give a precise meaning to the assertion that human-induced emission of $CO_2$ has the main - and the most dangerous - impact on the planetary climate. When discussing climate as a physical system, it is important to understand that it is difficult to define the notion of "climate" in physical terms because it is not a directly observable quantity. One can directly observe only local - in space and time - meteorological quantities such as atmospheric temperature, pressure, humidity, etc., and climate can only be derived - not necessarily uniquely - from a set of such local observations. To illustrate locality we may take, for example, such forcing as aerosols which serve, together with the release of gaseous species, as one of the two main factors of human impact on the atmosphere. Aerosols tend to be very nonuniformly distributed, both horizontally and vertically, being mainly concentrated over land regions. Tropospheric aerosols, including soot and dust, typically produce a significant absorption in the visible spectral domain and have a very small extinction in the infrared. In other words, the shortwave absorption and multiple scattering effects of aerosols lead to a considerable cooling, especially in the daytime, since they screen the Earth's surface from the incident short-wavelength radiation and let the outgoing thermal infrared escape almost freely. The long wave (IR) effect of aerosols results only in negligible warming, mostly during the day, so that the net effect of increased aerosols should be a rather strong cooling during the daytime.  But this effect is local in accordance with the distribution of aerosol concentrations. Analogously, there is no physical - directly measurable - quantity corresponding to the concept of climate as a whole, but only to its local manifestations. This difficulty seems to contribute much into the heated debates about the properties of climate and its dynamics.

It is curious that a rather marginal domain of physics - the physics of the atmosphere - has caught so much attention of the general public. Besides, the media, constantly striving to dramatize and sensationalize the routine daily regularity, tends to amplify alarmist statements. So, the thesis of anthropogenic global warming (AGW) is more a social, rather than a natural phenomenon. The population in the developed countries has been polarized into two groups: Pro-AGW and anti-AGW[220], nearly throwing various heavy objects at each other. It would be interesting to trace who are "pro" and who are "anti". For instance, people who are younger, left-wing, or low-income more enthusiastically support the thesis of catastrophic climate changes ($C^3$) than those who are older, more politically conservative, and have higher earnings.  Attitudinally, in climate projection debates hardly anyone will change one's mind, everybody is looking for biased arguments confirming her/his                            position                            (see,                            e.g., http://vocalminority.typepad.com/blog/2010/03/failed-scientist-paul-ehrlich-among-agw-alarmists-lashing-at-skeptics.html). So, there is no love lost between "skeptics" and "warmers" (i.e., AGW-proponents): both say the other party is wrong and definitely pursues non-scientific interests.

---

[220] In the developing countries, other problems seem to be more acute so that the population does not care much about global warming.

Some people say that the climate change science has already been settled, and it is now time to act i.e., to impose on people some restrictive and legally binding political decisions. Thus, unfortunately, physical and mathematical models of the climatic system have lately become almost inseparable from political decisions so that apart from physics we shall have to discuss political issues as well. The vehement reaction of political circles who enthusiastically support climate scare - actually more enthusiastically than a large part of scientists - raise suspicions that the thesis of an alleged climate catastrophe is, in reality, rather a political tool than a scientific result.  It is difficult to understand how the study of such a complicated subject as climate dynamics can ever be settled.  This is the field where scientific conclusions do not precede political decisions.

## 10.1  Some purely Climatological Questions.

Climatology, in contrast with meteorology, was traditionally not perceived as having any practical use. The word "practical" in the ultimate political meaning is understood as serving military or defense purposes. For instance, everything was clear with meteorology: the military needed accurate weather forecasts. Thus, meteorology gained much of its impetus during the First World War and its current importance during the Second World War because air forces required a more and more detailed description of local and regional weather: information about winds, rains, thunderstorms, etc.[221] Climatology, on the other hand, was mainly considered as an academic discipline, with only a few scientists being active in it[222]. It began attracting attention only after politicians realized that it can be a political tool and serve the parties with vested interests.

Climatology is intended to study the global and - more practical - regional weather patterns over some extended periods, much longer than a year. long-term trends in the physics of atmosphere, with certain futurological elements.

The central message of the whole current climate controversy is: global warming is happening, and it is caused by humans. This is a combined statement (a product of two), and it is a very strong assertion. Can it be proved?

The honest answer will be "no". Nobody can compellingly prove the truth of anthropogenic global warming, otherwise there would be a real consensus - no dissent among scientists as in firmly established parts of physics. Although the current models of climate may be quite relevant and interesting, at least to persons who professionally study the evolution of the climatic system, they can hardly be considered as the whole answer. God only knows what real atmospheric perturbations, say, in the next hundred years, may occur, with many counterbalanced factors contributing, e.g., in the global average $\Delta T$ being summarized. Besides, there exist in the past numerous examples of sudden climate transitions which occurred on a variety of time scales (see, e.g., Abrupt Climate Change: Inevitable Surprises. NRC, 2002). One should honestly admit that the level of knowledge of the Earth's climatic system is very inhomogeneous across the physical processes involved and mostly very low which is reflected in countless debates. Does one hear so many debates on classical mechanics or electrodynamics? It means that climate science is grossly immature, at least for the level to base essential decisions affecting the lives of many citizens on it.

---

[221] To some extent, this need explains the fact that most weather stations were traditionally located near airports and, when now in use as climatological devices, tend to provide distorted data in temperature measurements, in particular, due to heat generation at the airports and to the influence of spreading urban centers.

[222] One can name such scientists as M. I. Budyko in the USSR, R. Bryson in the USA, H. Hare (Canada), K. Ya. Kondratiev (USSR-Russia), H. H. Lamb (UK).

Despite multiple assertions that the AGW concept has been accepted by all climatologists, there is no real consensus in climate science which is natural since there are no persuasive quantitative data of the anthropogenic contribution to climate variability. To my mind, all standard arguments in favor of anthropogenic global warming (AGW) may only serve as circumstantial evidence, with the understated assumption that, firstly, there are no other factors of the recent warming besides anthropogenic $CO_2$ and, secondly, current warming is allegedly unique in the climatic history, therefore it should be man-made. The logic is hard to understand. Even if we assume that there was not enough sunlight to ensure the recent warming (assuming of course that it really takes place), does it necessarily imply that it is only the human produced $CO_2$ that is responsible for the increased average surface temperature? Recall that the human $CO_2$ contribution accounts for ~10Gt/year which is well under 5 percent of the total output from natural sources (fauna and flora each comprising ~2 $10^2$Gt/year, ocean ~3 $10^2$Gt/year, volcanoes ~0.5 $10^2$Gt/year). So the anthropogenic $CO_2$ contribution seems to be numerically too small to be a significant source of the global warming, particularly as compared with water vapor - the far more potent greenhouse gas, see, e.g., http://www.co2science.org/articles/V12/N31/EDIT.php and references cited therein.

The AGW evangelists usually respond that natural $CO_2$ is balanced between sources and sinks whereas anthropogenic carbon dioxide breaks this equilibrium, is accumulated in the atmosphere and remains there for very long, almost infinite or at least indefinite time. This is however not even a scientific hypothesis, it's a belief. Nevertheless, the well-known paper[232] which has become the flagship of alarmists, just like the discredited "hockey stick" in previous years [223] , is based on this assumption although it is not articulated explicitly. The two main sinks for $CO_2$, both of the natural and anthropogenic origin, are ocean and land biota, and the annual exchange between the ocean and the atmosphere (~3 $10^2$Gt) as well as between land biota and the atmosphere (~4 $10^2$Gt) is much greater than the human produced $CO_2$, see [262].

---

[223] One can observe a curiosity, by the way: a hockey stick is sometimes used as a pointer in the prestigious Perimeter Institute in Canada, evidently symbolizing the great love for hockey in this country.

## 10.2  Some Purely Physical  Questions

Climate science is a really unique discipline: it has no general theoretical background - there is no physical theory underlying climatology, in contrast, for example, with astronomy or astrophysics which are also disciplines based mostly on observations. In these latter disciplines, the background physical theory, general relativity, is omnipresent serving as a frame of reference for data, models, and speculations.  Here, general relativity is guiding the research. Very little of the kind exists in climatology, which makes this discipline more akin to, say, economics than to a physical science - show me any falsifiable theory of global warming. Like economics, climatology contains a loose collection of models trying to relate them to disjoint observational data. Wild speculations, unsubstantiated hypotheses and poorly conditioned projections abound, which is only natural because of the complexity of the subjects under study. The conclusions are inferred from computer models that could not demonstrate their ability to predict (show me the model that predicted the mean global surface temperature over one month forward or backward). On top of all this, climate science is supervised by political bodies such as the UN, UNFCCC, IPCC, EC, DOE, EPA, etc. and strongly influenced by populistic environmentalism. Even synergetic that was at the beginning more like a slogan than a science and has been treated piecewise for a few decades seems to have acquired now a guiding theory, that of dynamical systems (see Chapter 4). By the way, this theory is also useful in climate modeling (see below). Yet only basic physical facts are reliably known about the climate system. What are these facts and to what extent are they  plausible?

Primarily, climate is an example of a hierarchical multilevel system (HMS), containing multiple variables which interact in a complex way on very different temporal scales from hours for rapid atmospheric variations, days characteristic of weather patterns, years through centuries for essential variations of ocean currents, millennia for ice period-warm period transitions, and millions of years for continental drift and tectonic changes that can also affect the Earth's climate. Secondly, the climate system has long been known to undergo sharp transitions - from the "frozen" states with very low average global surface temperatures (GST) $T$ to warm periods characterized by much higher temperatures, explosive development of the biosphere - in particular, by the complex multicellular life forms.

One might single out several physical aspects of the climate dynamics problem, I can name four of them right away: 1) purely spectroscopic, 2) radiation transfer, 3) fluid motion and 4) stochastic dynamics, in particular, transitions between equilibrium climatic states. Of course, these (and all additional physical aspects) are coupled, but different researchers tend to concentrate on each one of them separately of others. I have just mentioned that the climate system is an example of a hierarchical multilevel system with a ladder of physical processes determining the energy balance between the incoming solar and the outgoing terrestrial radiation as well as the evolution of climate, whatever meaning is given to this latter term.

Some people assert that physical problems of the climate, in contrast with those for the weather, lead to boundary-value problems and thus cannot be chaotic or even unstable (in the Lyapunov sense). Then it is not clear what is meant by the climate evolution. There are even claims that climate simulations are fundamentally different from the usual computational modeling of physical situations due to emergent properties - the latter notion is nowadays of fashion but does not have any hard scientific content. See a brief discussion in [260].

In the final analysis, the climate system is governed by the well-known laws of classical physics so the scientific background may be considered relatively mature. The greenhouse effect itself was discovered as early as in 1824 by Fourier, the heat trapping properties of $CO_2$ and other gases were first measured by Tyndall in 1859, the climate sensitivity to $CO_2$ was first computed in 1896 by Arrhenius, and by the 1950s the scientific foundations were pretty much understood. There is plenty of hard data and peer-reviewed studies,

Note that the greenhouse effect theory can remind us of an old paradox typical of the phenomenological viewpoint, since viewed phenomenologically such a theory would mean that the heat spontaneously flows from the colder body (atmosphere) to the warmer one (Earth's surface). Indeed, absorption of radiation and its re-emission by the atmosphere is a passive process and as such it should not increase the average energy of radiating body - the Earth (up to local fluctuations, see, e.g., [259]). Besides, tiny quantities of carbon dioxide seem to be badly suited for making up the black box walls or resonator where the entire infrared radiation, irrespective of its spectral domain, is trapped. The primary function of any greenhouse is to protect plants from the outer cold by impeding convection.

The heat transfer from the Earth's surface through the atmosphere mostly occurs in the wavelength domain greater than 1 $\mu$m, the maximum absorption lying around 10-11 $\mu$m. Experimentally, the main absorption bands a r e in the near IR (around 0.94 $\mu$m, 1.10 $\mu$m, $\mu$m, 1.95 $\mu$m, and 2.50 $\mu$m) and carbon dioxide absorption bands (around 1.4 $\mu$m and 2.0 $\mu$m) overlap.

Thus, it is obvious that water vapor can warm the Earth more than $CO_2$; it must produce much more pronounced greenhouse effect in wet space-time domains than carbon dioxide[224]. This fact is well-known to all meteorologists - although I heard some AGW evangelists vehemently deny it. Indeed, water vapor is estimated to account for over 60 percent of the greenhouse effect (see, e.g., https://water.lsbu.ac.uk/water/water_vibrational_spectrum.html) whereas carbon dioxide is responsible for not more than 25 percent even if the most alarmistic $CO_2$-centric estimates are used. This means that the climate affecting - bad - role of $CO_2$ must be non-uniformly distributed: its greenhouse features are more salient in very cold locations, say, in the poles (or in deserts) where the air is dry, being almost negligible in warm places (see, e.g., [ 2 6 8 ] ). Indeed, there are approximately 30 water vapor molecules in the air per one $CO_2$ molecule, and due to large values of the dipole moment (1.84 D vs. nearly zero for $CO_2$ molecules, recall that $1D \approx 0.3934 e a_B$, $a_B$ is the atomic length.) water vapor absorbs the electromagnetic radiation at least several times more effectively than carbon dioxide. Both $CO_2$ and $H_2O$ are triatomic molecules, but $CO_2$ is a linear one with two polar $C = O (d \approx 2.3D)$ bonds being oriented in opposite directions so that their dipole moments cancel each other. Thus, the whole $CO_2$ molecule is non-polar[225]. The vibrational and rotational modes of the $H_2O$ and $CO_2$ molecules are very different, with the water vapor absorption spectrum being much more complex than that of carbon dioxide. The richness of the $H_2O$ IR absorption spectrum as compared to $CO_2$ is mainly due to the possibility of asymmetric rotations of the $H_2O$ molecule about all three spatial axes, each rotation having its own moment of inertia, plus the water molecule has three vibrational modes. In contrast, the $CO_2$ molecule, due to its linear symmetry (O-C-O) has four vibrational modes which correspond to bending along two perpendicular directions and one rotational mode that corresponds to two identical rotations in opposite directions about the center of mass, with the same moment of inertia. These are the basic rotational and vibrational states, and transitions between them account for absorption spectra. There are of course also combinations of transitions from one rotational or vibrational state to another, which makes real absorption spectra quite complicated so that a number of large spectroscopic databases have been created that contain the measured molecular absorption spectra. One might notice that there are still plenty of conflicting data in such databases, one of the most authoritative among them being HITRAN (http://www.hitran.com).

---

[224] See an interesting discussion of this fact by F. Dyson, an outstanding universal physicist, http://noconsensus.org/scientists/freeman-dyson.php.

Thus, the wide-spread assertion that $CO_2$ is responsible for 20-25 percent of the heating is difficult to understand. Perhaps under the conditions of zero humidity $CO_2$ can account for up to 0.20-0.25 of the total near-infrared absorption whereas under the "normal" humidity conditions of the atmosphere relative absorption of IR photons by the $CO_2$ molecules should be considerably suppressed (down to about 0.05). At the temperature of 20 Celsius, relative humidity would be approximately 50 percent which corresponds to approximately 3 percent concentration at sea level, see e.g., http://www.physicalgeography.net/fundamentals/8c.html. The water vapor concentration rises very rapidly with temperature reaching 100 percent at

---

[225] It is, by the way, significant that the most widely spread atmospheric gases, $N_2$ and $O_2$, do not produce any greenhouse effect. These substances have symmetric linear diatomic molecules and like $CO_2$ they do not have a constant component of a dipole moment. Moreover, even the variable component of dipole moment cannot be induced in such molecules by vibrational or rotational excitations since the latter do not change the molecule symmetry (in contrast with the triatomic $CO_2$, where excitation of the vibrational degrees of freedom can break the symmetry of the molecule and enable it to absorb atmospheric radiation in the infrared region.) Therefore, absorption of radiation by atmospheric nitrogen and oxygen is only possible due to electronic excitation i.e., in the shortwave domain (visible and UV).

100 C (steam). One can, by the way, conclude from the fact that the infrared absorption by $CO_2$ becomes significant when the temperature and, correspondingly, water vapor concentration are low that carbon dioxide emissions will tend to smooth the local climate.

## 10.3 The Earth as a Black Body Emitter

The crucial physical question related to the Earth's climate is: what would be the average terrestrial temperature when optical properties of the atmosphere are modified, e.g., due to anthropogenic influence? This question is, unfortunately, highly politically charged (see below). Let us, however, temporarily disregard any political implications and make an attempt to roughly estimate the equilibrium temperature that would ensure the radiation balance at the Earth's surface. The total radiation flux received by the Earth's surface from the Sun cannot exceed the so-called solar constant $S_0 \approx 1366 W/m^2$ which is the satellite measured yearly average amount of radiation received by a unit area whose normal coincides with the solar rays direction at the top of the atmosphere (see more details below, in subsection "The Role of the Sun"). One should notice that the value of the solar constant is experimentally obtained for the outer surface of the Earth's atmosphere and not on the Earth's surface, and it is difficult to calculate the actual surface irradiance (insolation) because its value can vary in wide limits following the atmospheric conditions. Clear sky insolation is $\sim 1 \mathrm{kW/m^2}$, and the average estimate for this quantity is typically taken to be $S \approx 240 \mathrm{W/m^2}$. This value can be obtained by equating the incident and outgoing radiation, $W_+ \approx \Pi R^2 (1 - \alpha) S_0$ and $W_- \approx 4\Pi R^2 S \approx 4\Pi R^2 \eta \sigma T^4$, where $\alpha$ is the surface albedo, $\eta < 1$ is the emissivity coefficient, $R$ the Earth's radius, $T$ is the average global surface temperature (GST) and $\sigma = 5.67 \cdot 10^{-8} \mathrm{W/m^2 K^4} = 5.67 \cdot 10^{-5} \mathrm{erg/cm^2 s K^4}$ is the Stefan-Boltzmann constant[226]. Then we get $S \approx (1 - \alpha) S_0 / 4 \approx 0.7 \cdot 1366/4 \mathrm{W/m^2} \approx 240 \mathrm{W/m^2}$.

Let us now estimate the equilibrium temperature for perfect absorption (albedo $\alpha$ is zero) and perfect emissivity $\eta = 1$. The total radiative power received by the Earth is $W_+ = \Pi R^2 S_0$. The total thermal emission from the Earth's surface[227] is $W_- = 4\Pi R^2 \sigma T^4$. In the equilibrium we have $W_+ = W_-$ so that the average temperature is

$$T \approx \left(\frac{S_0}{4\sigma}\right)^{1/4} \sim \left(\frac{S}{\sigma}\right)^{1/4}.$$

Recall that quantity $S = S_0/4$ may be interpreted in the equilibrium as the average insolation (i.e., average radiative energy incident on unit surface area per unit time - average flux received by a unit surface). Now we can calculate the derived quantity called climate sensitivity $\Lambda$ which is defined as the average temperature change responding to the variation of insolation $S$, $\Lambda = dT/dS$. More exactly, $S$ is the equilibrium irradiance of the Earth's surface which is equal to the additional energy absorbed by this surface. For the perfect black body Earth ($\alpha = 0, \eta = 1$), we may take approximately $T \approx 288$K, $W \approx 340$W/m$^2$ and get $\Lambda = dT/dS = T/4S \approx 0.2$Km$^2$/W. This is, of course, a time- independent model and no-feedback sensitivity. For the gray-body Earth ($\alpha > 0, \eta < 1$), we have

$$T \approx \left(\frac{(1-\alpha)S_0}{4\eta\sigma}\right)^{1/4} \sim \left(\frac{S}{\eta\sigma}\right)^{1/4}.$$

and climate sensitivity may rise a little:

$$\Lambda = \frac{dT}{dS} = \frac{T}{4S} = \left(\frac{(1-\alpha)S_0}{4\eta\sigma}\right)^{1/4}\frac{1}{4S} \approx 0.3\frac{\text{Km}^2}{\text{W}}.$$

[226] This constant is not as fundamental as truly basic physical constants such as $c, \hbar, e, G$. The Stefan-Boltzmann constant may be obtained in statistical mechanics as combination of such "more fundamental" constants, see Chapter 7.

[227] An analogous quantity in astrophysics is usually called luminosity. I have never encountered a physicist who could faultlessly apply photometric terminology. See some brief account of photometry in Chapter 10.

However, within the accepted accuracy such corrections do not matter much, and in many cases taking them into account would be inadequate.

When considering the gray Earth's climate sensitivity, we have tacitly assumed that both the albedo coefficient $\alpha$ and the emissivity $\eta$ are constant. In fact they can be functions of the global average temperature $T$. Indeed, with changing temperature, the equilibrium is shifted, and radiative balance should adapt to the new equilibrium states. This common fact from statistical thermodynamics can be better understood on some specific examples. In particular, higher terrestrial temperatures cause stronger evaporation and in general additional transfer of gases (e.g., dissolved in the ocean water) into the atmosphere. However, more gases in the atmosphere - and more primarily water vapor - result in a more pronounced greenhouse effect so that there will be a tendency to the further growth of global temperature (positive feedback). More infrared radiation will be trapped in the atmosphere and returned to the surface, which may change albedo, e.g., diminishing reflectivity due to accelerated snow melting. On the other hand, more water vapor in the atmosphere would inevitably lead to an enhanced cloud formation i.e., through this process albedo is increased. Moreover, as the average humidity grows, the average cloudiness should also grow, and this relationship can be for simple modeling assumed monotonic (one can even take humidity as a parameter). Thus, albedo depends on global temperature in an intricate way, it is determined by the competition of opposite processes (positive and negative feedbacks) and function $\alpha(T)$ can hardly be intuitively defined. The emissivity $\eta$ also depends on the global temperature $T$, e.g., through an additional cloud formation since clouds can screen radiation escaping from the Earth. Besides, extra water vapor in the atmosphere traps thermal infrared emitted by the surface, e.g., due to saturation, which also decreases emissivity. So, one might assume that $d\eta/dT < 0$, at least the main term of this derivative. As to the possible behavior of $\alpha(T)$, this issue essentially depends either on empirical data or on efficient cloud modeling which is a very difficult task. In general, by scrolling through the literature, one gets an impression that climate sensitivity is more reliably estimated from observations than from the corresponding models.

Let us formally write phenomenological relations for the climate sensitivity assuming that the albedo $\alpha$ and emissivity $\eta$ are some unknown functions of $T$

$$\Lambda^{-1} = \frac{dS}{dT} = \left(\frac{\partial S}{\partial T}\right)_{\alpha,\eta} + \frac{\partial S}{\partial \alpha}\frac{d\alpha}{dT} + \frac{\partial S}{\partial \eta}\frac{d\eta}{dT}.$$

So, if one knows the derivatives $d\alpha/dT$ and $d\eta/dT$ (from observations or otherwise), one can estimate the climate sensitivity. Physically, this is, however, not a trivial issue, in particular due to the overlap of infrared absorption bands of various greenhouse gases.

Recall in this connection that the question "What is the net flux of longwave (infrared) radiation to the surface?" is one of the hottest issues in the whole global warming problematics. The question is still open. Climate is one of the most complex dynamic systems with multiple interacting components. Against the background of powerful terrestrial and atmospheric processes the anthropogenic impact is likely to be just one of many factors influencing the average terrestrial temperature distribution. Determining the specific weight of one or another variable impacting climate perturbations and change requires laborious scientific efforts liberated from political biases.

## 10.4  Climate and Weather

What is actually the climate? One can crudely define it as the statistical ensemble of states for the system "atmosphere-ocean-ground", averaged over several dozen years. It is an observational fact that the Earth's ground in the triad "atmosphere-ocean-ground" or three geospheres - atmosphere, hydrosphere, and lithosphere - produces less impact on climate than ocean: one can say that ocean is more strongly coupled to the atmosphere than the firm ground. Therefore, in the following discussion we shall mostly consider the reduced climate-forming system consisting of the Earth's atmosphere coupled with the ocean. By the way, modeling of the climate in terms of atmospheric circulation influenced by the ocean may be valid not only for the Earth but also, say, for Venus. When speaking about the Earth's climate, one should of course include two other components (geospheres) such as biosphere and cryosphere which can also produce a substantial impact on the planetary climate.

One can roughly say that climate is weather averaged a large number of years, not fewer than ten, typically twenty or thirty. Because of its averaged character climate varies on the time scale of several decades. On the contrary, the concept of weather is "instantaneous", it is limited to a period of approximately ten days (it is nearly impossible to predict weather evolutions for the time exceeding this period). Weather forecasts clearly demonstrate to everyone the "law of unpredictability" well understood (relatively recently) in nonlinear dynamics and the theory of dynamical systems (see Chapters 4 and 7). Namely, each phenomenon has its limits of predictability. The point is that there are deterministic and stochastic components in any complex system. One can reliably predict only deterministic evolution of the system whereas its variations due to stochastic factors are in general unpredictable.

Using the terminology of the theory of dynamical systems (see Chapter 4) one may say that weather is analogous to the instant state or the "phase" on individual trajectories whereas the climate is an attractor (e.g., of a limit cycle type) for such trajectories. The phase can move in the vicinity of the attractor surface[228] in a very intricate and seemingly unpredictable manner, and a small perturbation of initial data can radically modify the character of instant motion (the weather) whereas attractor (the climate) remains unchanged. In other words, the details at any fixed moment of time can be radically different, but all observable quantities, averaged over large time intervals, will be basically the same, with the details of perturbations almost completely ignored. In meteorological terms it means that, say, a subtropical desert would not turn all of a sudden into a maritime polar zone and vice versa.

Computer-based forecasting the local and regional weather and climate change is a nascent and so far dubious activity. The matter is that current models may have too many flaws to give accurate predictions of climatic phenomena on a $10^2$km scale (in particular, projections of local and regional temperature dynamics fifty years from now). It is doubtful that such models should be relied upon in political decisions.

Both climate and weather give examples of an unstable physical system: even small deviations in input data lead to large errors in the output results (see Chapter 4). Mathematically, the climatic system is described by a large system of nonlinear PDEs. We can, in principle, write down this system of equations, but such a procedure would not be very useful since one cannot correctly fix supplementary (boundary and initial) conditions to such a system. Indeed, these supplementary conditions would change within a very short period, for example, during one of the relaxation times for the atmospheric system. This fact alone makes correct prognoses of climate dynamics very difficult. It is, however, interesting that in spite of extreme complexity and dynamical instability of the climatic system the latter is periodically reproducing itself. Thus, paleoclimatic research indicates that climate has several more or less stable cycles, e.g., about $10^5$ years, $2.5 \ 10^3$ years, 200 years and some other. The presence of stable cycles in the climatic system of the Earth may be considered as an example of chaos-order transitions. One can, in principle, use the observation of climatic cycles as an input for prognoses unless too stringent requirements imposed on them are envisaged. Stable climatic cycles reflect the coexistence of evolution and robustness, the feature typically characterizing the living organisms.

---

[228] Here, for simplicity, I pay no attention to the difference between the notions of attractor and attracting set.

It is clear that reliable climatic prognoses are extremely important for human activities. Many economic models, for example, essentially depend on climate variables, e.g., in agriculture, housing construction, power production and transportation. In particular, there exist alarmistic prognoses of economic disasters due to rapid human-induced heating (in average by 2-3 degrees centigrade) of the near-surface atmosphere within the next several decades. If such a warming rate really occurred, it would be a substantial challenge because humans have no experience of living in fast changing climatic conditions. However, who can bet his head or even hand that the contemporary computer models leading to such prognoses fully account for the impact of all possible natural factors determining climate dynamics, particularly those that have a negative forcing such as clouds?

The global climate[229] viewed as a physical system may be regarded as a composition of several mutually interacting subsystems: atmosphere, hydrosphere, biosphere, lithosphere, and cryosphere. Every subsystem evolves according to its own internal laws and, therefore, possesses its own set of characteristic time scales determining the subsystem's variability. In the spirit of dynamical systems theory (Chapter 4), it means that there exist complex flow patterns with the range of time scales spreading, in the case of the Earth's climatic system, from hours and days through seasons and years to many centuries and even millennia. Slow (or low-frequency) variations are usually identified with "climate" whereas the fast (high-frequency) ones with "weather". This natural time-scale separation hints at a possible modeling approach which would consist in decomposing all the relevant quantities - for example, those forming the vector field in the corresponding dynamical system - into "climate" and "weather" components. Then the "climate" (slow) vector component is allowed to evolve deterministically and the "weather" (fast) one varies stochastically.

Most of the terrestrial atmospheric circulations, together with the atmospheric-oceanic mixed layers, are formed through instabilities and can thus easily become turbulent. The same applies to intense oceanic currents, even considered inside the isolated hydrosphere, i.e., without any active mixing with the atmospheric fluxes. These are, of course, just words. There exist a great lot of models for climate and weather variations, and I have neither qualifications nor space in this section to review them. My purpose is much more modest: to expose the physical ideas underlying the climate and weather modeling and to establish - as usual - the connections with other portions of physics and mathematics. This exposition is by necessity rather superficial. The reader who really wants to study the subject may start with a comprehensive monograph by M. Ghil and S. Childress [233].

The dire climate predictions often involve extreme weather events such as typhoons, tornadoes, hurricanes in one package with climatic effects such as drought, desertification, repeated coastal floods. The "improved" AGW concept assumes a significant feedback with much water vapor added to the atmosphere. Does not it produce more precipitation and associated instant weather effects opposite to drought and desertification? Moreover, does not ice melting contribute into the water exchange loop, increasing precipitation and evaporation? Predictions of the imminent collapse of the global climatic system are founded on computer simulations and intuitive beliefs. As to the computer simulations, they are always full of uncertainties, as weather prediction clearly demonstrates. One typically argues that weather forecasting and climate simulations are two different tasks so that one should in no way juxtapose them. This is generally true, yet all computer simulations are sensitive to input data and introduce specific uncertainties. Computer simulations are also prone to both commission and omission errors. In the particular case of climate modeling such phenomena as cloud formation, hydrosphere-atmosphere exchange, volcanism, the role of biosphere are physically intransparent and lack data so that the overall computer simulations for the climate, involving the fluid motion equations supplemented with radiative transfer phenomenology and solved by the time-forward finite-difference procedure on a grid with some a priori resolution, necessarily have huge error bars. Besides, such models may be so sensitive to parameter values that they easily become unstable (see below "Dynamical Systems in Climate Modeling"). In practical terms it means that the reliability of current climate models is at minimum insufficient for political decisions affecting the interests of billions of people.

---

[229] This terminology is not quite correct: being derived from weather by the procedure of statistical averaging, climate is hard to define globally, it is basically a local concept.

The crucial point in assessing long-term climate variations as well as local short-term weather fluctuations is the study of state of the oceans and their interaction with the atmosphere (see, e.g., an important paper by Kravtsov, S., Swanson, K., Tsonis, A. A. A [269]). Of all geophysical subsystems, atmosphere and ocean are the most important ones for the climate: atmosphere and ocean may be regarded as the Earth's fluid envelope, where a great variety of motions can be comparatively easily excited. The incidence of major weather events such as hurricanes, tornadoes, cyclones, rainsqualls, heatwaves, etc. can be predicted by meteorological models, but one can anticipate that with the spacetime resolution of the climate models being increased, these latter models can also be used to forecast these weather events. In other words, meteorological and climatological models are expected to converge into the unified meteoclimatic modeling system (see, e.g., http://www.ecmwf.int; see also in this connection an interesting paper by Lockwood, M., Harrison, R. G., Woolings, T., Solanki, S. K. [270]).

Meteorology can be viewed, unlike traditional physics, an "almost exact science". People are usually angry with meteorologists and take weather forecasts with customary skepticism ("they are always wrong"). The similar discrepancy is often observed at some cellular phone network forecasts which manifests a strange adherence to clumsy computerized procedures for elementary everyday things. I don't understand why it is necessary to connect to the server instead of looking at the good old thermometer. Coolness factor results in multitudes of errors.[230]

Roughly speaking, the mathematical models for the state of the atmosphere can be constructed using the following physical principles. The major air masses constantly moving around the Earth collide with each other under the effect of the two main forces: the pressure gradient and the Coriolis force. The Coriolis force, in particular, induces the so-called Ekman transport that deflects water masses perpendicularly to the near-surface wind direction (to the right). In the North Atlantic, the Ekman transport leads to the formation, due to divergence and convergence of near-surface water masses, to a pair of oceanic gyres: a cyclonic gyre in latitudes near the North Pole and an anti-cyclonic one in the subtropics, the second having a larger spatial scale than the former. This effect is usually termed as the double-gyre circulation.

The combination of the two forces - the pressure gradient and the Coriolis force - produces the crude common pattern for large air masses movement. For example, in the northern hemisphere these masses turn predominantly clockwise about high-pressure zones and anti-clockwise about low-pressure zones. The word "predominantly" here reflects the possibility of a paradoxical behavior of the atmospheric gas masses, which may be viewed as gigantic fluctuations whose probability is small [231]. Large air masses can be approximately defined as masses of atmospheric gases characterized by a homogeneous temperature and humidity level. These gas masses may, however, have different densities and hence pressure values. For example, two different air masses at the same altitude necessarily have different pressure values so that when these masses come into contact inevitable pressure gradients appear which results in so-called weather fronts.

## 10.5 Dynamical Systems in Climate Modeling

The climate modeling problem relevant to AGW could be set up in the following way: climate as a complex physical system receives a continuous extra input (e.g., human-produced carbon dioxide), and one should try to find the domain boundaries (thresholds) for this input when the domain climatic system does not flip into a new state. In fact, as already mentioned, one should consider a number of inputs corresponding to multiple natural and anthropogenic forcings, with some of them being competing. Recall that there are many resilient, non-rigid systems in nature, for instance, biological organisms, ecological systems, population of species, planetary systems, etc. There are also resilient systems in human society i.e., social systems that possess the ability to deal with disturbances while retaining their essential functions, for example, capitalist economies, autocratic states, demographic developments. The ubiquitous resilient system is the Internet: it can withstand significant perturbations without losing its network unifying functions. The climatic system also has some - unknown a priori - amount of resilience, when it can tolerate disturbances without collapsing into a qualitatively different state.

---

[230] In the middle of the 1990s, SGI (then Silicon Graphics) supplied a powerful Cray machine to the Russian (Moscow) weather forecasting service, the Hydrometcenter. The expensive supercomputer was standing in the specially designated place without any use or any keen interest until it became outdated. The SGI engineers assigned to train the Russian meteorologists were astonished by such lack of motivation. I have heard this story from the former SGI experts responsible for this contract.

[231] One can mention in this connection that there are in general two main classes of climatic models as well as weather forecasting tools: deterministic and probabilistic models.

The view of the climate as a static and predictable system is hardly adequate and must be replaced by a dynamic view emphasizing a continuous change and uncertainty. There exist numerous historic examples of sudden climate transitions on a variety of spatial and temporal time scales. For just one example, one can recall the unforeseen and devastating drought in the West of North America which occurred the 16th century. This catastrophic phenomenon, which was probably the most severe of anything ever present in climatic records, manifested a sharp climate transition lasting for several decades (see, e.g., [271]. See also a discussion of possible chaotic transitions in the subsection "The AGW Evidence" below.

A simple and straightforward description of the climate equilibrium states can be based on the scalar ODE for the mean global surface temperature $\bar{T}$:

$$\frac{dT}{dt} = rT + F, \qquad (10.1)$$

where $r = \sum_i r_i$ represents the sum of feedbacks $r_i$, $F = \sum_i F_i$ denotes the algebraic sum of partial forcings $F_i$ which can be positive i.e., acting as sources or negative (sinks). For instance, certain $F_i$ may correspond to the IR absorption of greenhouse gases and act as the positive forcings whereas others may represent reflection of the incoming solar radiation and thus describe negative forcings. The overline bar denoting the averaging of $T$ over the terrestrial surface is omitted to simplify the notations: I shall do it in almost all formulas so please don't be confused.

If this model were truly linear as it is implicitly assumed by the majority of AGW proponents and local in time (without delay effects), then the gradual increase of $CO_2$ concentration would produce a synchronized monotonic growth of global surface temperature and a corresponding shift of equilibrium states. Indeed, in any linear model factoring (e.g., doubling) the forces factors the response. However, the linear model is too primitive to be true so that the models expressed through nonlinear equations, when all feedbacks and forcings in general depend on temperature $T$, must be employed for the climate evolution. Recall that the superposition principle does not hold for nonlinear equations so that one cannot assert that an increment $\delta F_i$ of a partial forcing (e.g., the increasing $CO_2$ concentration) would result in the proportional global temperature rise, $\delta T = a_i \delta F_i$. Besides, some forcings may have a random character expressing themselves very powerfully - much more potently than the relatively weak deterministic action of small $CO_2$ concentrations with their gradual increments (presumably manmade). Examples of such powerful forcings having the dominating random component are unforeseen solar-irradiance variations and volcano eruptions. If we stubbornly stick to the linear model of the (10.1) kind, then we must use the Langevin-type equation, maybe with inertia and damping, and consider small forcings, $|F_i| \ll |F_j|, i \neq j$, only as corrections.

Even if we temporarily restrict ourselves to the class of deterministic models, we shall have to replace the linear response (10.1) by a more complicated though still one-component dynamical system, $dT/dt = f(T, t, \mu)$, where $\mu$ is some control parameter (see more on dynamical systems in Chapter 4). The latter may have many components[232]. Actually, the above scalar model is still too primitive for the description of the climate evolution, its straightforward extension being the following vector equation

$$\frac{dX}{dt} = f(t, X, \mu), \qquad t \in I \subseteq \mathbb{R}, \qquad X \in \mathbb{R}^n, \qquad \mu \in \mathbb{R}^m, \qquad (10.2)$$

where $X = (X^1, \ldots, X^n)$ denotes the climate state vector which incorporates all relevant variables i.e., those describing the atmosphere, hydrosphere, geosphere, and biosphere. As already noted, the parameter $\mu$ is also of vector nature containing the set of components, $\mu = \mu^1, \ldots, \mu^n$ which specify, e.g., numerous feedbacks, strength of the incident shortwave solar radiation, albedo $\alpha$, and general parameters of the radiation transfer in the atmosphere. The vector equation (10.2) is basically essentially nonlinear so that the rate of change of climate vector $X$ depends on the actual climatic state[233]. When the vector $\mu$ varies continuously, the climate state $X$ can pass through a number of equilibria $X_a, a = 1, 2, \ldots$, and the system's behavior can change abruptly, e.g., from stationary states to periodic, multi-periodic or even to completely chaotic development. In a more mathematically-oriented language, one might say that the topological structure of a vector field corresponding to the family of solutions to equation (10.2) can change qualitatively following the gradual variation of parameter $\mu$. So, the predictability of the climate behavior essentially depends on the knowledge of its possible bifurcations.

However, one should not think that the climate which we observe is completely determined by its equilibrium positions (critical points). Relaxation times to climatic equilibria can substantially exceed our time of observation so that transient regimes, complex oscillations, bifurcations, and even chaotic processes can be perceived on the human scale as almost stable or slowly evolving states although such phenomena only correspond to transitions in the natural climatic time scales (e.g., when the climate system undergoes a smooth transition to chaos) and objectively should be interpreted as irregular episodes. Here, needless to say, a fundamental difficulty is hidden since the trouble with the transient states is that it is hard to predict their variations. Just like our quite limited understanding of the universe prevents us from reliable evaluations of, e.g., probability of extraterrestrial life forms, poor knowledge of climate physics does not allow us to infer the reasonable parameters of particular climate transitions including those with GST intervals within a given time range, $\delta(t_1) :=$ $\Delta T_1 \leq \overline{\Delta T(t)} \leq \Delta T_2 := \Delta T(t_2), t_1 \leq t \leq t_2$. I think one should always admit the possibility of a wrong interpretation of evolutionary behavior when dealing with complex multilevel systems characterized by multiple time scales.

---

[232] The codimension of a vector subspace where $\mu$ lives inside the vector space of the entire dynamical system is more than one.

[233] Here climate is considered approximately local in space and time, which is far from being obvious. For simplicity, we do not discuss here the time-delayed (retarded) effects as well as nonlocal correlations (a particular example of nonlocality relevant to climate studies is methane hotspots that can influence the climate state over large distances). Such phenomena require a special treatment.

Apparently, the problems in climate dynamics ought to be addressed at some fundamental level, which is difficult. The above-mentioned qualitative principles can reveal, in particular, the basic underlying mechanisms of atmospheric mass movement, but to understand this movement quantitatively one has to resort to more comprehensive mathematical models. Nevertheless, the qualitative considerations have been successfully put in the base of so-called general circulation models (GCMs) that can be very detailed. CGMs may be used in climate projections that can be claimed to have not only methodical, but predictive and practical character. However, there exist a lot of uncertainties in many climate projections. For one thing, the natural climate variability is coupled with that of the world's ocean circulation which has, besides a comparatively regular low-frequency component, also noise-like stochastic currents. For another thing, it has recently become obvious - both from observational data and model investigations - that the climate of the Earth has never been in equilibrium (in the meaning that we are accustomed to while studying dynamical systems, see Chapter 4). Moreover, it is highly unlikely that the climate will ever be in equilibrium. This fact implies, besides the necessity to employ more complicated methods of non-autonomous dynamical systems for modeling, an increased sensitivity of atmospheric circulations to, e.g., small-scale variations of model parameters or to regional perturbations. One can notice that small-scale atmospheric physics is even more complex than global, in particular because it should correctly account for regional effects such as clouds and winds. From the physical viewpoint, the respective small-scale models rely on nonlinear fluid dynamics which, in its own right, is based on deterministic theory of dynamical systems. Due to the complexity of the physical processes involved, climate modeling can be successful not by using a single all-embracing model, but a "staircase" of models - a whole hierarchy of them starting from the simplest toy model of the double-gyre circulation to the most detailed general circulatory description.

To illustrate the complexity of the dynamical processes involved, one can render the following example. Apart from large-scale circulations, intense wind-driven oceanic flows, jets and vortices may be formed, all of them being the manifestations of nonlinear effects. Such flows may have totally different spatial and temporal scales: from global oceanic streams down to mesoscopic (of the order of several km) eddies. Small-scale oceanic phenomena, being coupled to the atmosphere, tend to induce turbulent flows in it. Turbulence, as we have seen in Chapter 7, is an irregular fluid motion which has strong vorticity and causes rapid mixing. But the main thing is that turbulence is a multiscale phenomenon: all eddies in the fluid - in this case in the atmosphere - are of a different size. One can state that the very essence of turbulence is its multiscale behavior.

One might recall ubiquitous examples of near-surface atmospheric turbulence. Thus, turbulent airflow produces aircraft noise which leads to the "noise pollution", e.g., in the form of considerable noise peaks around airports. Other examples are clouds and smoke plumes emitted by the pipes. In the atmosphere, the largest turbulent length scales $l$ (for instance, in cumulous clouds) are typically of the order of several km, and the fluctuating turbulent velocities $u$ are of the order of several m/s (the characteristic time scale is of the order of an hour i.e., $10^3$ s)

One can observe in passing that there is one salient feature in atmospheric science. Contrary to most other sciences, where problems can be attacked by dissecting them into small pieces, in atmospheric science such an analytical approach is hardly adequate. It would be unlikely to obtain realistic results in the study of atmospheric dynamics by looking at isolated fragments in detail one by one while keeping everything else constant and assuming the division on fragments well defined - the usual modeling strategy. Indeed, in the atmosphere one has to consider a lot of interacting processes at the same time: radiation transfer, turbulent transport, irreversible thermodynamical processes, and chemical reactions.

So, we see that the climate is an extremely complicated physical system which can be described only under highly idealized assumptions. At first, the hierarchy of climatic models was developed for the atmosphere. Atmospheric models were originally designed for weather forecasts - on the hourly or daily time scales. Recall, in this context, that one still has hard times even in retrospective weather modeling, say, ten days back. Later, atmospheric models have been extended to consider climate variability for all temporal scales. They have also been coupled to other subsystems, primarily the ocean (such as wind-driven ocean circulation). Practically all current climatic models are based on dynamical systems theory (see Chapter 4). Recall that dynamical systems in general are very sensitive to the input data, for instance, initial conditions. In this connection one should bear in mind that geophysical data in general are practically always highly uncertain. There are still scientific discussions in progress, and making firm conclusions about the climate future trajectory is hardly possible.

In more realistic models, the number of such tipping points (which actually represent separatrices) will probably be much greater than three, and some of them may turn out to be favorable for mankind whereas others can be disastrous. However, climatologists do not seem to have reliable data to construct such more realistic models - recall that most data in geophysics have a large uncertainty corridor. Anyway, the attempts to build impeccable phenomenological models of the climate can hardly be called satisfactory so far.

A crude but very useful method of modeling the dynamics of the climate system is through considering the energy balance. The incoming shortwave solar radiation (corresponding to the Sun's surface temperature of approximately 6000 K) is partly reflected back into outer space ($\sim$ 3K) and partly absorbed by the planetary system including the atmosphere and the surface of the planet ($\sim$ 300K). The surface is heated and releases the heat into the atmosphere by thermal radiation, by convective fluxes, and by evaporation of water, with water vapor later condensing in clouds. Both convection and phase transitions such as evaporation/condensation distribute the heat vertically whereas macroscopic horizontal mass transfer processes i.e., winds redistribute the heat along the planetary surface through the dissipation of mechanical energy generated by inhomogeneous heating of the surface. This is, of course, a very crude verbal picture of energy flows in the atmosphere. The two corresponding main types of mathematical models for the state of the atmosphere which meteorologists regard as comparatively simple (although, to my mind, they are still quite difficult to solve) are the model of vertical variations (horizontally averaged) and that of horizontal variation (vertically averaged or integrated). The first mathematical model (usually termed as RCM - regional climate model) considers convection caused by solar radiation as the dominating process whereas the second (called EBM - energy balance model) is based on the phenomenological representation of energy balance.

The RCM approach is based on studying the radiative transfer in a fluid (gas) column with surface albedo i.e., the reflected portion of the solar radiation, $\alpha$ - the reflection coefficient - is not a constant factor and should be treated phenomenologically as a function of temperature. More complicated though, albedo depends on the state of the planetary surface (e.g., for simple ground albedo $\alpha = 0.30 - 0.50$, for fresh snow $\alpha = 0.80 - 0.85$, old snow $\alpha = 0.50 - 0.60$, for ice covered ground $\alpha = 0.90$, grass $\alpha = 0.20 - 0.25$, forest $0.05 - 0.10$, etc., see, e.g., [291]), on the amount of clouds and thus on the state of climate itself[234]. Moreover, albedo can be a rapidly varying or random function since it is quite sensitive to the optical transmission properties of the atmosphere. The RCM description of atmospheric reflection, coupling processes through radiative dynamics, is in principle nonlinear and typically produces multiple equilibria in the horizontally averaged temperature profile (see below more on a single-component Budyko-Sellers climate model). In fact, RCM is a model of direct solar influence on climate and weather.

---

[234] Currently, the mean global surface temperature is estimated to be $\approx$ 15C. If the Earth were covered with ice (the "Snowball Earth"), the GST would be $\approx$ 52C, with forests, GST would be $\approx$ 24C. If the Earth consisted of the ocean alone, GST would be $\approx$ 32C since water absorbs more in the visible and UV range; were the planet covered with deserts, GST would be $\approx$ 13C.

The EBM type of models is based on accounting for the balance between the incoming solar radiation and the outgoing terrestrial radiation. The corresponding balance equations (see below) can produce multiple equilibria in terms of the model parameters, primarily the surface temperatures. From the physical point of view, the presence of different equilibria in the space of parameters is a typical symptom of an interplay of competing processes. We have seen in Chapter 4 that exploring the states of equilibrium is the most important part in the study of the models based on dynamical systems. Some of the equilibrium states in the dynamical models of climate variation may well be unstable so that small fluctuations and random influences can push the dynamical system corresponding to a climatic or, in the narrow sense, atmospheric model to another equilibrium. Transitions between equilibria may be viewed as a typical manifestation of instability in dynamical systems: tiny causes produce sizable long-term effects.

Before we proceed to more concrete examples, I would like to point at the following classification of dynamical models - not only for the climate or atmosphere, but for a number of other phenomena, e.g., population dynamics or structure formation. One can distinguish 0d, 1d, 2d, and 3d models. The number of dimensions here corresponds to that of independent spatial coordinates in which the modeling problem is stated. For instance, in 0d models, spatial variability is totally ignored so that such models describe only the evolution of homogeneous parameters, in the case of atmosphere variations of the "global" near-surface temperature. Zero-dimensional ($0d$) models may be considered as representing the lowest level in the hierarchy of atmospheric models. More sophisticated models take into account heterogeneous distribution of crucial parameters. Such distributed parameter models are typically formulated in terms of partial differential equations which makes mathematical handling of the problem much more difficult.

In a simplified mathematical setting, the $0d$ energy-balance model may be formulated as the system of equations expressing the evolution of global surface air temperature $T$ influenced by the global radiative balance and its variations:

$$c\frac{dT}{dt} = A(T) - B(T), \qquad A(T) = \mu Q\big(1 - \alpha(T)\big), B(T) = \sigma b(T)T^4. \quad (10.3)$$

Here functions $A(T)$ and $B(T)$ denote the incident solar radiation and outgoing terrestrial radiation, respectively, $c$ is the heat capacity of the system "atmosphere plus ocean". It is a nontrivial problem to elucidate the exact meaning of the coefficient $c$. The quantity $Q$ denotes the amount of solar radiation falling per unit time on the upper atmosphere, and $\sigma$ is the usual Stefan-Boltzmann constant. The parameter $\mu$ is a correction coefficient for the amount of solar variation received by the Earth's surface, it is defined in such a way as $\mu = 1$ for a standard daytime. The parameter $\mu$ (insolation) may diminish due to worsening of atmosphere transparency, say after catastrophic volcano eruption or during a "nuclear winter". The temperature dependence of the albedo coefficient $\alpha(T)$ and "grayness" $b(T)$ is a priori unknown, it is only clear that $b(T) = 1$ for a thermodynamic black body and $b(T)$ varies in the interval (0,1) for a gray body such as the Earth. Hence one has to test different variants by modeling.

So, we obtain a spatially homogeneous (0d) dynamical model for the surface air temperature with unknown nonlinearity. One can naturally start with the simplest hypothesis namely that nonlinearity has a polynomial character. One of the most popular models in nonlinear dynamics, as we have seen, is the logistic model when the nonlinearity polynomial is quadratic (see Chapter 4). This scheme is so crude that it is almost insulting to apply it to a very sophisticated, mysteriously fine-tuned and self-reproducing climatic system of the Earth.

We may note that climate as much as the weather is a chaotic system: its variations cannot be predicted with an arbitrary certainty[235]. Besides, as we have seen, there are too many parameters to be accounted for so that when considering concrete problems some of them are by necessity ignored. But in chaotic systems ignoring parameters does not go without penalty: both small variations and neglect can produce great changes.

## 10.6 Combining Models with Observations

If you ask ordinary people supporting the AGW concept where do they get the information from, they would typically answer - with a certain embarrassment masked with aggression - that "the climate warming science has been settled", "everyone knows it" or "this is the experts' opinion and who are you?" and only a small number would confess that they get the information about climate warming from the media, mostly from TV. Few people really assess scientific raw material.

The focus of climatic models differs drastically. Some of them put the main weight on $CO_2$ and the respective infrared radiation transfer aspects, others to concentrate on the Sun's activity, the third kind of models are devoted to the ocean's circulations and so on. Highly abstract mathematical models are unfortunately not sufficient to make reliable climate (long-term) and weather (short-term) projections. Even using modern supercomputers and computing grids does not help much without fair observations, in particular because numerical climate and weather models are typically based on solving an initial value problem (IVP) which should be fed with the correct initial data.

To obtain observational data on the actual state of the atmosphere over $10^8$ info items are gathered daily from the ground stations, weather satellite lidars, drifting buoys, balloons, etc. The entire surface of the Earth has been divided into a grid, and observations are obtained at the crossover points. These data may serve as entries to computer-simulated evolution of the state of the atmosphere. Of course, even the most powerful computers cannot process such amounts of data, even by applying the averaged laws of fluid mechanics and thermodynamics. One needs some hard physics here, not just computer models. See, however, http://www.skepticalscience.com/empirical-evidence-for-global-warming.htm on the empirical evidence for the potential threat of AGW.

---

[235] There exists a famous saying by Benjamin Franklin about certainty: in this world nothing is certain but death and taxes.

In the real study of the atmosphere, field studies are a must, so a number of global observational and weather forecasting networks have been established. However, about 25 per cent of the Earth's surface (about $1.4 \ 10^8 km^2$ out of approximately $5.1 \ 10^8 km^2$) is so far uncovered by this network of observation devices. These "blank areas" are mostly located in the southern parts of the ocean. Moreover, there are rumors (sorry, I found neither pieces of evidence nor negative statements) that the number of temperature-measuring stations diminished, especially in rural regions, in the 1990s. Satellite observations do not seem to fully compensate for this lack of data due to relatively low accuracy, e.g., in temperature measurements. Besides, the cellular structure of climate monitoring networks is very non-uniform due to the different level of scientific development. One can notice that observational networks are orders of magnitude less dense in Africa than, say, in Western Europe. The value of the Earth's surface temperature measurements over the ocean is indetermined - recall that the ocean covers about 70 percent of the terrestrial surface. It is not at all obvious that the data recorded by a rather sparse grid of ground- and ocean-based stations reliably manifest the surface temperature trends.

The speculative way on which many climate variability models are built are hardly capable of predicting the transitions between states despite the bold claims of computer modelers. Probably the main reason is the gap between the models and observational data. Mathematical and computer models based on plausible considerations may be very useful to develop an understanding of the climate system dynamics (see "Dynamical systems in climate modeling" above), but such models are hardly suitable for planning decisions. The latter would require hard experimental facts, convincing raw empirical data, not models or computer codes. Even the islands of disconnected data sets about the present state of the atmosphere, the hydrosphere, albedo, etc. as well as the geological records are insufficient for planning decisions affecting the world economies. The typical response of pro-AGW modelers is that one must write codes using whatever data one can presently have. I however, think that although such an approach may be justified in modeling simple systems, it is inadequate for highly complex, unstable, and chaotic ones if good accuracy is required. One can, of course, obtain qualitative results using simple models, but performing sophisticated computer simulations using inaccurate or missing data does not make much sense: it only produces excessive expectations. To write a code head-on for unstable nonlinear systems is highly questionable. And in any case, planning decisions should not be based on the codes which are using insufficient observational data. And there is one thing that is common for all computer models: they should rely on data. Reliability of the model is no better than that of the data. Computer simulations of the climate do not provide an accurate forecast, as it is usually claimed by the AGW evangelists, not only because the underlying physical equations are enormously complex and should be radically simplified, but already because good data are scarce and unreliable (as, e.g., surface temperature measurement). For example, highly publicized temperature estimates derived from tree rings are local and indirect by its very nature i.e., such proxies are mediated by a number of local biological processes. Are such estimates unambiguous and reliable, do they reflect the global temperature trends? Besides, one might recall that statistics in data processing is such a tool that one can get a desirable result if one inputs the biased data, one only needs a desire.

By the way, computer codes are of secondary nature with regard to possible climate transitions. One should discuss not the codes, but the type of equations that govern the climate dynamics and the essential terms in such system of equations, these terms reflecting the relevant physical processes. Without due attention paid to such equations and their experimental validation, opinions can influence one's view of the issue of climate variability more than reliable physics and mathematics, and there is now a tendency to value codes more than equations. I don't think that computer models can supplant falsifiable physical laws. Incidentally, the AGW hypothesis cannot be currently falsified in any way so that it should be believed or not and thus may be considered as a contemporary version of religion.

## 10.7 Climate Variability

The crucial question one must answer in connection with the possible human-induced heating is the following: is it true that recent variations of the climate [236] are significantly outside the natural variability borders experienced in former centuries? If it is the case then one must pay attention to the possibility of a dangerous anthropogenic influence, but if recent climate changes are within the range of natural variability, then one cannot substantiate a significant human impact. One can notice in passing that some climatologists make a distinction between the terms "climate change", "climate variations" and "climate variability" [234][237]

The climate transitions are still believed by many scientists (the AGW concept notwithstanding) to follow variations in the energy flux to the Earth's surface due to insolation changes when the Earth experiences almost-periodic perturbations from other planets (mainly Jupiter). Besides, the incoming short-wave solar radiation varies in accordance with the Sun's own cycles. Yet the response of the Earth's climatic system to the variations of the driving "naked" solar is not well understood, mainly because of indeterminacies related to the radiation transfer properties of the atmosphere. Multi-periodic variations of the climate mostly taking the form of sharp transitions between two crudely defined states - cold and hot - give rise to many wild speculations about the random character of the climate system and limited predictability of its behavior. Consequently, the climate is a system that is driven by the Sun, but due to the presence of atmosphere-hydrosphere-geosphere-biosphere climate can be characterized by additional interlocked nonlinear processes running at a hierarchy of speeds i.e., some of the climatic periods do not necessarily coincide with the solar cycles.

---

[236] Say, since the beginning of the 20th century or since the year 1750 which is rather arbitrarily taken to indicate the beginning of active human impact on the Earth climate system. The accuracy of paleoclimatic measurements, even in the industrial era, is so low that it does not matter much which initial time point in the past should be taken.

[237] Another fine distinction in climate studies is made between paleoclimatology and historical climatology, with one being used by "alarmists" whereas the other by "skeptics". All such nuances imposed on friend-foe dichotomy and mixed with overdramatization closely remind us of "ideological struggle" for the purity of party ideals in the USSR.

The climate models simulating the near-surface temperature variability, the state and composition of low atmospheric layers, are highly politically charged issues. Unfortunately, there were cases when the field data (like those discussed above) were adapted by the conflicting groups to corroborate their presumptions. Climate modeling is an indicative example of science affected by ideology; this is a piece of engaged science.

## 10.8  The AGW Evidence

It would be probably easy to dismiss many questions and statements about AGW as sheer platitudes, yet I dare to reiterate them here. None of the data seem to categorically prove the case of human-induced climate change. It is only a matter of the public's perception and the leaders' intentions. The main questions here are: can one foresee the climate changes and what factors may be considered as the most important for their prediction? What are the factors that may affect the accuracy of such a prediction? These questions seem to be of tremendous importance also beyond academic studies because it is very risky to make political decisions which may be founded on rather inaccurate scientific results. The matter is that humans - even the most enlightened climatologists - do not know enough either about the Earth's climatic system or about the chaotic dynamic systems to produce accurate mathematical models containing thousands of entangled variables. Averaging does not help much, mostly due to possible instabilities: if you change any parameter a little bit, the output results may vary dramatically (this phenomenon is sometimes called the "butterfly effect"). Reversing this situation, one can tune the model parameters just a little bit and obtain any desirable result. This fact, common for highly complex and possibly unstable systems (in particular, exhibiting deterministic chaos) enables one to obtain the biased results, for instance, the ones needed for lavish funding. I am far from thinking that many climate modelers are dishonest opportunists or that there is some kind of a "conspiracy" - such paranoid statements are a sheer stupidity, but I guess there may be a noticeable component of banal conformism in today's politically motivated climate studies, people in general tend to swim with the stream.

Many "catastrophic modelers" maintain that the Earth's climate system is not a chaotic one and that due to its averaged character it is rather stable (then I don't quite get it why so much fuss about AGW). Moreover, some of the catastrophists even state that there is nothing particularly complex about the climatic system: the Sun is heating the Earth providing thermal flux $Q_0$, some part of the heat $Q_1$ is re-radiated into the space, and some part ($Q_2$) is trapped by the atmospheric "blanket" whose permeability for thermal radiation is worsened, primarily due to the increased anthropogenic $CO_2$ emission. In this naive heat balance model one can, of course, obtain a steady increase of the Earth's surface temperature, with no instabilities or chaotic phenomena implied. Indeed, strictly speaking, nobody has proved the presence of chaotic regimes for climate. But this is a flimsy argument. Abrupt shifts in the past climate evolution, which are well documented (see, e.g., the references in "Abrupt Climate Change: Inevitable Surprises", NRC, The National Academies Press, Washington D.C., 2002), give serious reasons to think about the presence of chaotic regimes in climatic system. Mathematically speaking, this is nevertheless only a conjecture, but the admission of climatic chaos is compatible with available data [235]. Indeed, as just mentioned, paleoclimatic data at our disposal testify that the planetary climate has experienced notable changes in the past, characterized by an intermittent glacial-interglacial period. Experimentally, one can measure, for example, the ratio of $O^{18}$ to $O^{16}$ isotopes in marine sediments (see [236] and references therein). Although this ratio fluctuates rather wildly on a variety of time scales, such a method allows one to infer variations of the continental ice volume, in particular over the last million years [236]. One can observe the notably aperiodic - in fact chaotic - character of glacial evolution, with a crudely extracted intermittency component on a time scale of $10^5$ years sometimes referred to as Quaternary Glaciation Cycle. This intermittent character of the glacial evolution may be interpreted as unpredictable - chaotic - switches between different climatic states occurring on a hierarchy of time scales.

Thus, chaos upsets the predictability of climate, and quasideterministic predictions made by the AGW adherents depend on model errors which is the consequence of the obvious fact that any model is only a crude representation of nature. Climatic models confront an obvious difficulty consisting in the necessity to account for a large number of interconnected physical mechanisms and hence even greater number of variables entering the corresponding nonlinear equations. Therefore, only drastically simplified models of climate can be constructed, with a heavily abridged number of physical mechanisms, variables, and equations to be taken into account. But while radically simplifying the model to make it treatable, modelers encounter the problem of identifying the relevant mechanisms, variables, equations and reliably (i.e., based on a well-determined smallness parameter) selecting them from the "impertinent" ones. Yet even if one succeeds in this selection procedure, the model complexity still remains very high so that one typically has to handle the model numerically. And we know that numerical techniques introduce their specific and unavoidable errors, thus the possibility of obtaining reliable predictions for the climate's future evolution would be thoroughly compromised. At least it would not be possible to substantiate the drastic economic and political measures that have a substantial negative impact on the people's well-being, on which the environmentalists and some politicians insist. This is the well-known consequence of a sensitive dependence on small variations of parameters, in particular, of initial conditions in an unstable - in a specific case chaotic - system which makes it impossible to predict its future evolution beyond a certain time horizon, even if its present state is known with a good precision (the latter is never attained for the climate system.)

Furthermore, by tweaking the model parameters, which are only known very roughly anyway, it would be possible to get a required temperature increase, e.g., the global-mean surface temperature change $\Delta \overline{T} \geq 4$ degrees centigrade in fifty years. Incidentally, I do not quite understand how the quantity $\Delta T$, being a function (more exactly, a functional) of a great many variables, can be not only strictly positive-definite, but also fatally confined within certain dimensional limits (say, 2 and 6 degrees Celsius), irrespective of all possible physical processes in the atmosphere, hydrosphere, and solar system. I would prefer to scrutinize the behavior of some dimensionless entity, e.g., $\delta \overline{T}/T_0$ where $T_0$ is an essential quantity characterizing the physical properties of a considered system. It is not necessary that $T_0$ should coincide with a mean surface temperature: the latter may be a physically meaningless parameter since many states, even an infinite number of states with the same mean surface temperature (GST) are possible.

Notice that all AGW debates revolve around the projected value of the average terrestrial temperature (global surface temperature - GST). However, the global average of the temperature may be too crude to serve as a valid scientific parameter to base political decisions upon. In the folklore, one typically refers to such averages as the mean temperature over the hospital. Using this global average, AGW becomes the aggregate notion, a political slogan rather than a scientific concept. Physical effects are local and structured: there exist variations of atmospheric fluxes, ocean currents, convection and other fluid instabilities, humidity, precipitation, sediments, cloudiness, aerosol distributions, etc., all depending on the point in space-time - without mentioning biological factors such as vegetation, ocean plankton, and agriculture. Very little of such physics and biology is present in the now popular computer models (and even less in codes). Such popular models can, in principle, describe certain regimes of fluid motion, in particular, fluid dynamics in the atmosphere and hydrosphere, but are far from accurately incorporating such crucial factors influencing the atmosphere's local transparency as volcanic eruption, dust, cloud formation and drift, biochemical processes (e.g., vegetation), etc. The much-publicized IPCC computer models are based on drastically curtailed fluid dynamics systems of equations coupled with the not less reduced radiation transfer models, mostly in the integral (averaged) form. Such models may inadvertently omit not only physically important terms in the equations - and in unstable systems all terms may be important - but the entire regimes. In particular, the Earth's climatic system may go - at least locally - through vigorous self-sustaining oscillations of such quantities as atmospheric temperature and pressure, e.g., driven by feedbacks between the ocean temperatures and global wind patterns. An example is the "Wet Sahara" effect observed approximately 6000-7000 years ago (see, e.g., [272]). This is just one illustration of the complexity and poor predictability of the climate dynamics, even on the level of average trends.

One may notice that there can be different climatic states characterized by the same average temperature of the Earth, but with different pole-equator temperature gradients. By the way, in most references only the average terrestrial temperature is discussed and attempts to find its determining factors, in particular, anthropogenic ones are made whereas the question of what determines the meridional (pole-equator) temperature distribution gains much less attention. It is, however, quite interesting that paleoclimatic records testify that equatorial temperatures remained nearly constant[238] in the course of Earth's evolution so that the issue of the causes for the meridional temperature distribution over the planetary surface becomes rather nontrivial and important.

It is intuitively clear that the meridional temperature (and pressure) gradients should influence the frequency and severity of hurricanes, cyclones, tsunamis, and other extreme weather events. One of the favorite theses of climate alarmists is that such events have lately become more ferocious and recurring than in previous years due to human-induced $CO_2$ emissions. The statement is confusing: additional emission of greenhouse gases should smooth temperature and pressure gradients, thus softening the climate and making extreme weather events rarer. At least one may ask: can one prove that the severity and frequency of hurricanes, cyclones, typhoons, tsunamis, etc. is on the rise synchronously with the $CO_2$ concentration in the atmosphere? Likewise, can one prove what threshold conditions are necessary for the total and irreversible collapse of thermohaline circulation - one of the favorite catastrophic scenarios of AGW evangelists?

Another catastrophic scenario predicted by AGW believers is that the ocean level is dangerously rising at an accelerated rate determined by the anthropogenic $CO_2$ emission. The problem of the ocean level increase also attracts, to my mind, an exaggerated attention. This is an example of a highly speculative issue. Who has *proved* that the ocean level is rising at a rate that has accelerated in accordance with increasing anthropogenic $CO_2$ emissions? Projections of how high ocean level might increase thereby endangering coastal communities are, to put it mildly, highly uncertain. What is the accuracy of the ocean rise measurements on the global scale? Another question: how can one separate the climatic component in the change of ocean level from the tectonic factor? It is well known that there existed ancient townships that are now lying many meters under the water because the ground sank. These latter processes have nothing to do with human activity and global warming. Recall that one of the favorite catastrophic scenarios of the AGW belief system is the rapidly growing frequency of floods and coastal area destructions (such areas will be presumably engulfed in the flood following the rise of the ocean level) due to the pathological sea-ice circulation distorted by AGW. Recall in this connection that the volume of water in the world's ocean is estimated as $V{\sim}1.3\ 10^9\text{km}^3{\sim}10^{18}\text{m}^3$, 3.3 $10^7\text{km}^3$ in the polar ice caps whereas the total volume of glaciers does not exceed $2\ 10^5\text{km}^3$ (these estimates are taken from the article [261])[239]. So even if all the glaciers melted it would hardly result in the catastrophic ocean level rise.

---

[238] Temperatures on the equator could even be a few degrees Celsius lower than at present, see, e.g., Wang N., Yao T., Shi Y [246]. and the references therein.

Politicians, especially of the social-populistic brand, have a general tendency to talk about the distant future - this is simpler than cleaning up mundane and unpopular everyday problems. Climate variability provides a convenient option. Now all natural catastrophes are presented by the propaganda as direct consequences of climate change induced by human activity. This is false. The Kyoto treaty presumes that global warming of the near-surface atmosphere is, firstly, a proven fact - it may be during the current climatic cycle, but this cycle is finite - and, secondly, is exclusively produced by human industrial activities. In reality, however, there exist powerful natural factors affecting the climate, and human-induced changes in the gaseous composition of the atmosphere is not necessarily the dominating one. One can imagine the following factors influencing the Earth's climate

Main anthropogenic factors:

1. fossil fuel burning (power production, utilities, transport, etc.)

2. industry

3. agriculture, land reclamation

4. forestry

5. hydrosystem construction and melioration

Main natural factors:

1. solar activity

2. collective motions and oscillations in the atmosphere-ocean system

3. oceanic circulations

4. volcanic activity

5. perturbations of the Earth's orbit parameters

6. motion of heavy planets (primarily Jupiter and Saturn) influencing solar activity

7. lunar motion

---

[239] Previously often cited values were $V \sim 1.338\ 10^9\text{km}^3$, surface of the world's ocean $S \sim 361.3\ 10^6\text{km}^2$, average depth $\bar{h}{\sim}3700\text{m}$. Volume $V$ of ocean waters amounts to about 1/800 of the planetary volume. Mass of ocean waters comprises approximately 96.5 percent of mass of the entire hydrosphere which, in its turn, makes up about 1/400 of the Earth's mass. Volume of water vapor in the atmosphere is estimated to be approximately $13000\text{km}^3$, with the renewal period of about 10 days. It is interesting to compare this period with the estimated renewal time for the whole ocean: about two million years.

8. other astrophysical factors (such as fluctuations of the angular velocity of the Earth's rotation, cosmic rays, meteorites, etc.)

9. tectonic activity

10. geomagnetic activity

To understand the scientific foundations of long term climate dynamics one has to account for all these factors. In the linear modeling, they may be treated quasi-independently i.e., one can subsequently turn them "on" and "off". However, we have seen that climate models are predominately nonlinear which means that all the above factors can strongly influence one another. The fact that due to natural factors the climate has already undergone many noticeable variations, comparable or exceeding current changes, during the existence of the Earth is fully ignored - largely on political or ideological reasons. In reality, in correct models of the climate variability one should take into account at least the superposition of natural and human factors (linear approach), in more sophisticated models their mutual influence (nonlinear approach). It is quite understandable why the problem of climate variability all of a sudden became an indispensable element of current world politics. Interestingly enough, the placement of climate issues in the center of political debates coincided with the downfall of the Iron Curtain and dissolution of the bipolar model of the world. Climate is a transcontinental and transborder system, it cannot be confined within a single country and meaningful talks on climatic issues can hardly be kept within a certain political block. Thus, since climatic changes have a global character, their discussion is an example global politics, of effective intergovernmental interaction on the global scale. One can therefore forecast that the role of climate and its variability in global politics will only increase. If and when this issue ceases to be a hot topic on a global scale, another one will be invented to supersede it as a would-be matter of acute transnational interest.

The presence of a large number of powerful natural factors testify that only a portion - so far unknown - of global warming is due to anthropogenic factors. AGW models will always cast doubt until natural causes of GW are fully taken into account. Moreover, global warming and the "greenhouse effect" are not synonymous as it is implied in ecological propaganda campaigns. Drastic temperature contrasts may exist between different years, reaching in certain locations 4-5 degrees Celsius. This is a manifestation of natural climate variability. By the way, powerful year-to-year jumps, say, in winter temperatures, that is within a single year, cannot be explained by anthropogenic factors. Only a part of climate variations is deterministic, the other - probably not less considerable - part has a stochastic nature. Besides, there are space-time climate non- homogeneities which are typically ignored by ecological alarmists. One must understand that climate variations are not distributed uniformly over the Earth's surface. For example, warming in Russia is observed to be more intensive than averaged over the globe. The rise of global temperature (this is an averaged quantity, provided such average exists and is meaningful) is predicted to reach in various estimates from $1.0\,^{\circ}C$ to $5.0\,^{\circ}C$ during the 21st century, which is a big difference. Of course, diverse doses of political engagement, ideological bias or publicity-seeking alarmism are hidden in these estimates. IPCC, a collaboration of several hundred climatologists, has offered a whole spectrum of scenarios (see http://www.ipcc.ch/). If climate forecasts are in general possible, then they depend on many factors, and in case the latter can be correctly identified one can construct mathematical models of climate as a physical system as well as make prognoses based on such models.

Nevertheless, the reverse influence of the biosphere on the atmosphere seems to be a proven fact: for example, the contemporary composition of the atmosphere has been made by living organisms. Furthermore, there exists much evidence that the Earth's atmosphere at the initial stage of the planet development [240] contained negligible quantities of oxygen: it was bound within carbon dioxide. Approximately $1.8 \ 10^9$ years ago, free oxygen ($O_2$) appeared in the atmosphere due to the action of microorganisms. Consequently, the ozone ($O_3$) layer in the higher atmosphere (30-50 km) emerged which screened the Earth's surface from the deadly ultraviolet radiation emitted by the Sun. Without such screening, living organisms could only subsist in places hidden from the solar rays, and with the ozone screening they could spread over the entire Earth's surface. Eventually, the living organisms conquered the world and were able to affect the gaseous composition of the atmosphere even more. That was a salient example of a positive feedback in the atmosphere.

Likewise, at the contemporary stage of development living organisms including humans can seriously influence the gaseous composition of the atmosphere, thus modifying its physical (primarily optical) properties. As already mentioned, such changes may distort the transmission properties of the atmosphere. In particular, human activities can, through the changes in atmospheric transparency, modulate the radiation balance between the Earth's surface and outer space which may well twist thermal equilibrium that we view as the Earth's climate. Moreover, one can imagine a number (more than one) of equilibrium states corresponding to a variety of combinations of $CO_2$ concentrations and average terrestrial temperatures, other parameters being solar activity, orbital variations, albedo, state of hydrosphere, etc. This is a physical problem: one should estimate possible effects of the changes in the gaseous composition of the atmosphere on thermal equilibrium [241] near the Earth's surface. The $CO_2$ concentration increase over some quasi-equilibrium level, arbitrarily taken to be "normal", may be accounted for as a perturbation, $\Delta c$. If the regime is reached when the global surface temperature average $\bar{T}$ is perturbed stronger as linear in $\Delta c$, say, $\Delta T = \mathcal{O}((\delta c)^2)$ or even first-degree superpolynomial growth $\Delta T \sim A \Delta c \log \Delta c, T = T_0 + T$ ( $A$ is some constant or a function of other variables), then one has an accelerated warming with rising concentration of $CO_2$. Such an amplification might lead to a considerable heating of the atmosphere which, however, is not substantiated by the observations: average temperatures grow much slower with the increased $CO_2$ content in the atmosphere and are usually approximated by a logarithmic behavior (see, e.g., http://en.wikipedia.org/wiki/Radiative_forcing).

---

[240] The age of the planet Earth is estimated to be $4.6 \ 10^9$ years.

[241] Of course, one can only talk about a thermal equilibrium in the atmosphere in a very approximate sense. In reality, the atmosphere is a highly nonequilibrium system, with many time-dependent currents flowing along the gradients of pressure, temperature, density, partial concentrations, etc.

The next problem, more of a biophysical nature, is: to what extent can the changes of the physical (mostly optical) and chemical properties of the atmosphere affect living organisms, in particular humans? If and when it has been proved that the anthropogenic factors change the gaseous composition of the atmosphere to the extent that its modified optical properties lead to a substantial adverse influence on the living conditions of human beings, only then should political and economic measures directed at reducing such anthropogenic factors ensue, and not in the reverse order since political and economic actions may be painful for large groups of the population. We, however, are observing just the reverse picture: political and economic decisions, based on presumptions and models, but strongly affecting living conditions of the population, precede the ultimate scientific results on the possible effect of anthropogenic changes in the gaseous composition of the atmosphere.

One can notice that despite hard physical foundation, the study of climate - climatology - may be related more to the humanities than to exact sciences. Climatology seems to be on the same level as the society and the environment: we know climate to the extent we know the environment in general. Expectations of the accuracy in climatological predictions are unwarrantably elevated in the society. Recall that any scientific result has a definite accuracy; if a prognosis is claimed to be an unconditional truth without prerequisites and error margins, there is no science in it. The catastrophic climate forecasts are in this sense almost non-scientific since error margins, e.g., for anthropogenic influence are indicated rather arbitrarily ($? < \Delta T <?$, where $\Delta T$ is the average surface temperature growth due to man-made $CO_2$ release)[242]. Besides, error margins in the physical sense i.e., corresponding to a chosen approximation may eliminate the assumed positive-definiteness of $\Delta T$. The current trend may be with $\Delta T \geq c > 0$, but $\Delta T$ is a function of time $t$, and in the next temporal domain, following such unaccounted natural factors as orbital change, sunspot periods, or volcano eruptions (see, e.g., [273]) $\Delta T$ may well be negative. Is there a corroborative evidence that recent climate variations considerably surpass the changes observed in the past and caused by the natural factors such as fluctuations of the Earth's orbital parameters, solar cycles, ocean currents, volcano eruptions, etc.? Incidentally, one might notice that the IPCC texts and models include very little material about volcanic activity which has always been one of the most powerful climate forcings on the planet. Indeed, ash and aerosols scatter the sunlight, on the other hand volcanoes produce IR-absorbing gases (i.e., GHG), submarine eruptions warm oceanic waters and so on.

---

[242] Some experts refer to probabilistic error margins present in the IPCC reports, see http://www.ipcc.ch/. However, those are not the precision boundaries usually presented in physics and corresponding to the applicability limits, but the error margins typical of computer codes. I could not find in IPCC reports a satisfactory substantiation of the approximations made in IPCC models. Physical applicability limits correspond to expansions in series, often asymptotic ones, when some small (or large) parameter can be explicitly indicated. When the terms in the system of equations describing the mathematical model considered are omitted, one usually talks about a zero order in such terms (in dimensionless form). One can always find the corrections however difficult it might be technically.

The fact that $\Delta T$ can be negative owing to anthropogenic factors as well is reflected also in mathematical models of the so-called nuclear winter, when an explosive energy release occurring locally produces large quantities of aerosols changing the optical properties of the atmosphere. The effect of aerosols can in most cases exceed the effect of any minor[243] greenhouse gas (Kahn, R. A., Yu, H., Schwartz, S. E., Chin, M., Feingold, G., Remer, L. A., Rind, D., Halthore, R., DeCola, P. Atmospheric Aerosol Properties and Climate Impacts. A Report by the U.S. Climate Change Science Program and the Subcommittee on Global Change Research, M. Chin, R.A. Kahn, and S.E. Schwartz (eds.), National Aeronautics and Space Administration, Washington, D.C., 2009). So, the limits of applicability for the models of climate transitions is a serious issue.

Moreover, in contrast, say, with nuclear physics, purposed experiments cannot be carried out in climatology so that the prognostic quality of the models cannot be reliably verified. To some extent, paleoclimatic studies can serve as a substitute for the physical experiment to verify the models. For instance, a certain alternative to pointed experimental validation of climatic models would be the quantitative explanation of paleoclimatic cooling and warming periods, e.g., comparatively recent cold and warm climatic intervals such as the Little Ice Age in the 17th century or the Medieval Warm Period. Where are such quantitative explanations?

Some climate variations will inevitably occur, notwithstanding any political action taken to reduce $CO_2$ emissions so that the research aimed at adapting to climate changes makes more sense than the "IPCC science" whose main task is to substantiate the a priori set up catastrophic AGW views. Climate science, as I have already mentioned, in general can hardly be called impartial and satisfactory. The point is that climate science (of which the "IPCC science" is the branch) is a somewhat special discipline: in distinction to most scientific disciplines, climate science is not supported by any background theory, it just contains a lot of observations, often conflicting, a great number of wild speculations, plenty of disagreeing hypotheses, a growing number of dubious computer models which are nonetheless unquestionably believed to provide the ultimate, complete and holy truth, and, moreover, contains a strong ideological component and uses *ad hominem* arguments.

---

[243] Recall that all greenhouse gases including $CO_2$ are considered "minor" in contrast with the water vapor which is "major".

At least the IPCC can hardly be called neutral in the assessment of climate dynamics. An a priori givenness is especially obvious from the "Summary for policymakers" where such clichés as "dangerous anthropogenic interference with the climate system", "climate change can affect human health directly", "populations in developing countries are generally exposed to relatively high risks of adverse impacts from climate change" as well as mysterious words "models project that" and intuitive statements of the type "it is very likely". Yet climate and its instant manifestations - weather - are determined by physical processes, which are sufficiently complex to be intuitively assessed and explained at the hand-waving level. For example, transfer of the solar radiation, its reflection from the Earth's surface and selective absorption in the atmosphere, in particular by small quantities of the notorious carbon dioxide, requires a considerable physical knowledge usually not available to the most vocal climate alarmists or to "climate economists" and other proponents of carbon control and trading schemes.

However, the crucial difficulty here is that real processes in nature are irreversible (see section "The Arrow of Time") so that validating parameters of a model by paleoclimatic observations, even from the recent past, does not guarantee reliable forecasts concerning the future state of the climatic system. In particular, one cannot say that the state with a given concentration $c(t)$ of $CO_2$ and characterized by a certain average global temperature $T(t)$ in the past ($t < t_0$) will be repeated in the future ($t > t_0$), where $t_0$ is some reference instant of time, e.g., 1990, 2005, 1750, etc. Such extrapolations are not valid.

The true AGW believers and environmentalists are trying to explain the global climatic transitions occurring since the beginning of the industrial era as the result of a single human-induced process i.e., simple one-dimensional causal relationship: warming is related to human-produced $CO_2$, cooling (if any) is related to human-produced soot and aerosols. The implication in any case is that greedy and consumption-crazy humanity totally controls the Earth's climate, all the natural mechanisms playing almost no role. Hence, say the true AGW believers and environmentalists, humanity must radically change the way of life in order to avoid an awful catastrophe - the climate Armageddon. Also, some climatologists assert that the emission of carbon dioxide will double within a few decades (instead of a few centuries which would be obtained by extrapolating the current trends, 23 percent since 1900). Here, there is an implicit assumption that fossil fuel consumption will explosively grow resulting in accelerating $CO_2$ emissions, which is highly unlikely. This is an unjustified socio-economical hypothesis, not climatology - whatever this latter term means. Speculating about future consumption trends does not help much in predicting possible climatic states, even their primitive characteristics such as ups and downs of the average global surface temperature (GST). It would be much more useful, by the way, to analyze the sensitivity of social structures (in particular infrastructure) to climate changes in different regions rather than making highly indefinite prognoses of GST increase: the point is that climate warming is not necessarily detrimental; some countries and societies may profit from the regional temperature increase.

## 10.9  The Evil Role of Carbon Dioxide

True AGW believers have selected a single factor - $CO_2$ concentration - from a set of variables relevant for the climate system and resulting in the variations of the Earth's surface temperature. Although nobody can deny that $CO_2$ is a greenhouse gas (a comparatively weak one), it can hardly ensure the entire contribution into such variations. We shall also see below that this gas can hardly be considered as the principal cause of global warming. At least the usual scientific-looking statement that human-produced $CO_2$ represents the dominant radiative forcing, so radically shifting the Earth's energy balance as to induce the projected severe climatic consequences, cannot be proved. In this subsection we shall discuss from the physical position whether it is true that anthropogenic carbon dioxide is the primary factor driving the global warming.

Curiously enough, the rather innocuous carbon dioxide ($CO_2$) gas has recently begun playing an important part in politics and declared as the most harmful by the propaganda. This chemical substance (see about $CO_2$ properties, e.g., in http://www.uigi.com/carbondioxide has moved to the center of attention as the so-called climate gas or greenhouse gas. The meaning of these metaphoric terms is that the surface of the Earth is allegedly heating up, with the concentration of $CO_2$ being increased due to the human activities. There are four main "greenhouse gases" (GHG) in the atmosphere: water vapor $H_2O$, carbon dioxide $CO_2$, methane $CH_4$, and nitrous oxide $N_2O$, of which water vapor is by far the most efficient: its characteristic spectrum is more than three times wider than that of $CO_2$ and is responsible for roughly 95 percent of the greenhouse effect (see, e.g., [274]). The Earth's atmosphere mainly consists of nitrogen $N_2$ (about 78.0 percent), oxygen $O_2$ (about 21.0 percent) and argon Ar (about 0.9 percent) which are not greenhouse gases because of negligible absorption in the infrared. Apart from these principal components, there are some small quantities of the rest gases such as water vapor $H_2O$, carbon dioxide $CO_2$ (about 0.035 percent), methane $CH_4$, sulfur dioxide $SO_2$, ammonia $NH_3$, ozone $O_3$, nitrous oxide [244] $N_2O$, nitrogen trifluoride[245] $NF_3$, etc. Concentration of all these "impurities" is a function of coordinates and varies with time. For example, wind can easily displace and flatten the local bunches of $CO_2$ concentration as well as fluctuations of water vapor density. Does the global average of $CO_2$ bring a more pronouncing effect than local heating due to enhanced heat generation near human dwelling places and industry sites? The latter phenomenon can be easily observed, e.g., by noticing that temperature in big cities is higher than between them.

---

[244] Nitrous oxide (the "laughing gas") is contained in the atmosphere in very small concentrations (about $320 \cdot 10^{-9}$), but it is approximately 300 times more efficient as IR absorber compared to $CO_2$ and its concentration is rapidly growing due to ubiquitous use of fertilizers. However, little attention is paid to this fact, probably due to political importance of modern agricultural technologies.

[245] Nitrogen trifluoride ($NF_3$) is mostly used in the production of electronic components. It has the greenhouse potential approximately $17 \cdot 10^4$ that of $CO_2$, with an estimated lifetime in the atmosphere about 700 years. The estimated production of $NF_3$ amounts to 5000 m.t.; how much of this amount is released into the atmosphere seems to be an open question, see the details in: Hoag [247].

One can also note that paleometeorological studies show that there were sharp irregularities and oscillations of the carbon dioxide concentration in the atmosphere, and such $CO_2$ variations did not necessarily coincide with warm periods in the past. For example, the Sahara Ice Age occurred when the $CO_2$ level was an order of magnitude higher than today. What is the difference between $CO_2$ at that time and today's anthropogenic one? Atmospheric levels of $CO_2$ (and methane) naturally fluctuate, partly due to changes of the Earth's orbit resulting in variations of the incident sunlight. Hence, there is no compelling evidence that the observed human-induced increase in $CO_2$ concentration has really resulted in the frightful greenhouse effect (Kauffman, J. M. Climate change reexamined. Journal of Scientific Exploration, v. **21**, No.4, 723749, (2007)). I have already mentioned that the main "greenhouse gas" is generally known to be water vapor, causing about 60-70 percent of the greenhouse effect on Earth[246], since water vapor absorbs much more infrared than $CO_2$ (see the above brief discussion of physical absorption mechanisms) so that it is strange to ascribe the whole absorption and thus heating of the atmosphere to $CO_2$ when there is about two orders of magnitude more potent IR absorber present nearby, which should thus dominate and swamp the effect of $CO_2$. It is, however, curious that although this is a common knowledge among meteorologists and climatologists, but in the media or within governmental groups the overwhelming effect of water vapor tends to be altogether ignored or at least under-emphasized. Moreover, this distortion of the truth seems to be adopted by repetition: some people may even concede that it might be "a little misleading" to ignore the water vapor as the greenhouse gas, they tend nevertheless to call this neglect an accepted practice and defend it by claiming that it is customary to leave water vapor out.

Furthermore, water vapor strongly affects weather and, consequently, climate through cloud formation changing radiation transfer conditions. This mechanism can be more pronounced than selective absorption and re-emission of IR radiation by tiny quantities of $CO_2$. Unfortunately, I could not find comparative studies in the scientific literature available to me. Besides, the water vapor concentration depends on the surface temperature, primarily on that of the ocean surface (which comprises 70 percent of the entire planet area). Transition of water masses into the gaseous phase is accompanied by cooling, so the whole thermal system of the Earth involving the hydrosphere becomes very complex, with many feedback mechanisms. Heat and mass transfer in the atmosphere on the global scale make climate modeling and parameter computations so difficult that it becomes hardly possible to indicate model accuracies.

---

[246] It is, however, not quite correct to assign a definite percentage of the greenhouse effect to a certain gas because the effects of different gases are not additive. So the given percentage must be understood as a crude estimate.

The current greenhouse effect is due to about 0.12 percent of the atmospheric $CO_2$ generated by human activity (see the calculations in http://www.geocraft.com/WVFossils/greenhouse_data.html). So the usual statement that human activity since 1750 has warmed the climate seems to be wrong, and anthropogenically produced $CO_2$ has no discernible effect on the global temperature. Recall also that the increase of the average surface temperature depends on the growth of the $CO_2$ concentration only logarithmically, and a doubling of $CO_2$ would roughly correspond to approximately 1C temperature change (this problem has already been explored by S. Arrhenius, the chemistry classic, [292]. Therefore, the panic about anthropogenic $CO_2$ production is hard to comprehend. It seems utterly ridiculous that the increase of a tiny fraction of $CO_2$ level would produce a global temperature increase of, say, 6 C i.e., over 20 percent of the whole atmospheric contribution into the Earth's mean terrestrial temperature believed to be about 33C. And there are approximately 30 times as many $H_2O$ molecules in the atmosphere as $CO_2$ molecules (see, e.g., http://www.geocraft.com/WVFossils/greenhouse_data.html), with much more efficient absorption properties (fingerprint spectrum). Following the logic of AGW proponents, one should primarily ban water vapor production, for example, teapots and hydrogen-driven vehicles.

It is easy to observe that carbon dioxide, water vapor and oxygen, all of them necessary for sustaining life, are produced at different locations at the Earth's surface: for example, carbon dioxide in large urban and industrial centers whereas water vapor over the oceans and oxygen mainly in rain forests. Transport of these gases is ensured by the atmospheric turbulence, which is a highly complicated process, not fully understood up till now. Besides, it would be important to note that from a strictly environmental point of view, $CO_2$ does not present a severe pollution hazard as compared, e.g., to such agents as $NO_x$, $SO_2$, $SO_3$ ($H_2SO_4$), $CO$, Hg, Cd, Pb, other metals, especially heavy ones. In fact, $CO_2$ is not a pollutant at all. Declaring carbon dioxide as the most harmful substance exposes a certain ecological irrelevance of the whole "save the climate" campaign. Moreover, the absolute quantity of $CO_2$ in the atmosphere may only be correlated with the climate temperature or be interpreted as its indicator because the gaseous composition of the atmosphere depends on its temperature, even locally. If this is the case, then stating that slight increase of $CO_2$ produces catastrophic temperature variations is wagging the dog. In principle, variations of the average surface temperature $\Delta T$ and of the carbon dioxide average concentration $\delta C_{CO_2}$ can not necessarily be correlated. As an illustration of some limited relevance of carbon dioxide as a unique determinant of atmospheric temperatures one might recall that about half a million years ago the concentration of $CO_2$ reached the level of about an order of magnitude higher than today, but everywhere on the Earth there was an Ice Age. On the contrary, before that period the dinosaurs had lived in a much hotter climate for more than a hundred million years. There exist estimates showing that during a long time of the Mesozoic the ocean level reached the mark at least 100 meters higher than today (some estimates even give up to 250 meters), there were no polar ice caps, and the Earth's terrestrial temperatures were much more homogeneous than today: the difference between polar temperatures and those at the equator was on average only about 25C, with the poles being about 50C warmer than today. Therefore, the mean global surface temperature (GST) was also much higher (see, e.g., [275]). Were the dinosaurs also incessantly polluting the atmosphere with nature-hostile $CO_2$? One should only attempt to answer the "paleoclimatic" question: why was the climate so warm at those periods (from Triassic to Cretaceous), and for such a long time?

There is an assumption put forward by the scientific wing of environmentalists and climate alarmists that $CO_2$ molecules re-radiate back to the ground the substantial part of thermal radiation emitted by the Earth thus essentially heating it. More specifically, in the process of relaxation the infrared photons are re-emitted by $CO_2$ molecules in the direction of the Earth's surface thus creating a supplementary energy flux, and the total thermal power absorbed by the surface substantially exceeds the power sent to the Earth by the Sun. It is this thermal power amplification that is known as the greenhouse effect. As a result, the Earth's surface temperature becomes noticeably higher than in the absence of additional quantities of carbon dioxide. Superfluous molecules of $CO_2$ so radically improve the blanket feature of the atmosphere that overheating may occur. We have already discussed that the molecule of $CO_2$ absorbs electromagnetic (infrared) radiation due to rotational and vibrational-rotational transitions in three narrow bands around 2.7, 4.3, and 15.0 $\mu$m i.e., rather selectively. The total thermal spectrum of the IR radiation corresponding to the Earth's surface temperatures comprises about 100 $\mu$m. This means that the major part of the infrared radiation, lying outside of the absorption spectral domains of the $CO_2$ molecule, passes through the atmosphere into outer space practically without loss. The $CO_2$ concentration is, as we have seen, rather small (the current alarm-producing concentration is $C_{CO_2} \approx 0.038$ percent), so the question persistently arises: how can a linear growth of a comparatively tiny impurity concentration be responsible for catastrophic climate and weather changes? In any case, one should have the data of the total atmospheric absorption in the infrared versus $CO_2$ selective absorption, and it seems that actual measurements and numbers with well-defined error margins do not exist yet.

Carbon dioxide is often designated by environmentalists as an "ecological poison" so that one must keep its global concentration at some arguably manageable level. What is exactly this level? Maybe I was not diligent enough, but I could not find empirical evidence of the sensitivity of the variation $\Delta T$ of the average surface temperature $T$ to anthropogenic $CO_2$ emission (from 280 ppm to the current level of 380 ppm). Furthermore, the last decade showed an approximately 0.1 degree cooling of the atmosphere (see, e.g., the graph http://farm3.static.flickr.com/2600/3670965001_4249d9a68e_b.jpg).      In fact, the $CO_2$ gas is indispensable for life on the Earth, and geological periods characterized by its increased quantities were also characterized by the most rapid development of the biosphere. It would of course be nice to prove that nowadays $CO_2$ is really a great problem, but that has not been done, contrary to what one may have read or heard in the media. There exist only sparse observations and measurements characterized by not always clearly defined accuracy as well as some largely qualitative models of the "greenhouse" effect. The latter may very well be a fictitious theory, at least for the current quantities of carbon dioxide ( $\leq 0.038$ percent). Is it really a scary concentration resulting in a significant rise of the Earth's temperature and requiring immediate drastic measures to restructure economy and consumption habits? However, many billions of dollars and euros are evaporated due to excited emotions over the so-called catastrophic climate changes ($C^3$). One must be a true believer in global warming and possibly a true believer in general, i.e., a person inclined to believe in anything supported by the majority of other people - God, fashion, vampires, life after death, supremacy of one's nation, intelligent design, etc., not to notice a banal brainwashing.

The climate controversy is not so much about whether the Earth's surface is getting warmer, it may well happen due to a number of physical reasons (see above), but about whether it is human activity that makes it warmer, and to what extent. But this is basically a scientific question, not a political issue - in fact a physical problem requiring a correct physical solution. The usual argument of climate catastrophists is that the concentration of $CO_2$ grows rapidly due to human activities (from 0.028 to 0.038 percent since the beginning of industrial revolution), and this growth of concentration must be accompanied by a rapid heating of the atmosphere. However, this is not more than a hypothesis based on correlation or, at best, a model; the exact physical mechanism of the process of atmospheric (and hydrospheric) heating remains unclear. The major portion of $CO_2$ is dissolved in the ocean's water, and with the increasing temperature equilibrium solubility of most gases in water, including $CO_2$, is diminished. Hence, with the general heating caused by any external factor, large quantities of carbon dioxide are transferred from the hydrosphere to the atmosphere. The main factor for this external heating is most probably the rising solar activity, which happened many times in the Earth's history when ice ages alternated with warm periods. There exist two (at least!) informative experimental sources allowing one to obtain paleoclimatic data: boring holes in Greenland and Antarctica, with the hole depth reaching several kilometers. Samples of kern are taken, containing air bubbles from those distant epochs when the material obtained, e.g., snow or ice, was formed. Spectroscopic analysis allows one to determine the gaseous composition (percentage of $O_2$, $N_2$, $CO_2$, etc.) with very good accuracy; besides, the temperature related to the time when snow was falling, and ice was formed as well as some other physical characteristics can also be obtained. All classical ice ages, warm periods, and corresponding quantities of $CO_2$ in the atmosphere have been established by this method. The reported result was that rising concentrations of carbon dioxide did not precede warming, but followed it. This fact can be readily explained: 90 percent of $CO_2$ is dissolved in the world's ocean, and when the latter is heated large quantities of carbon dioxide transit to the atmosphere. This process is of course reversible: during the cool periods, the ocean absorbs carbon dioxide. This quasi-oscillatory process, with both positive and negative feedback components, is eternal.

There exist, as just mentioned, various feedback mechanisms affecting climatic factors. The Earth's climate seems to be balanced within certain (so far not exactly known) margins. If it is pushed too far, a series of positive feedbacks can be triggered that would cause substantial changes. For instance, rapid heating of soil in the permafrost regions is hypothesized to release more methane which would amplify warming. This is a domino effect. However, it is hardly possible to say with a scientifically accepted accuracy, firstly, how likely these scary scenarios are and, secondly, what exactly is the human component in such runaway heating. The environmentalist doomsayers[247] usually speculate on unquantifiable risks. As far as carbon dioxide is concerned, the salient example of the feedback is the growth of photosynthesizing biomass with the enhanced $CO_2$ concentration and increased surface temperature which diminishes the amount of $CO_2$ in the atmosphere due to intensified photosynthetic activity. So, the question: "By how much has the $CO_2$ concentration increased since the industrial revolution owing to human activity?" may be totally irrelevant to climate variability. Instead one can pose another question namely "What fraction of the carbon dioxide being exchanged per unit time in the whole climatic system (i.e., in the atmosphere, hydrosphere, biosphere, lithosphere, and cryosphere combined, see above) is due to human activities?" I suspect that the human-conditioned percentage of $CO_2$ in the real-time ecosystem kinetics will be negligible, and thus human involvement into the climate dynamics is strongly exaggerated. This is, however, a conjecture; one must estimate human involvement by considering a correct kinetic model for the entire ecosystem. Such a model, provided it included complete radiation transfer equations, would give local temperatures as a by-product.

From the physical viewpoint, the equilibrium temperature setup is a kinetic problem, in the simplest case an energy balance problem. This holds also for the Earth's surface. In this connection I do not understand why certain terms in the kinetic relationships, such as infrared absorption and re-emission towards the Earth's surface, are taken into account whereas the counterbalance terms, such as variations of atmospheric transparency due to anthropogenic and natural impurities, are neglected or at least not fully considered. What about correctly accounting for fluctuations of the Earth's orbit and, in general, of direct and diffuse insolation of the Earth's surface? And even if one takes the vicious role of $CO_2$ seriously, what about the carbon sinks? In short, the terms ensuring positive contribution into the energy balance are retained whereas those resulting in diminished temperatures are mainly disregarded. Indeed, the most essential human influence on the atmosphere has always been the release of aerosols and various gases, some of them known as greenhouse gases. Nowadays, due to "political forcing" the role of the latter is emphasized whereas the role of tropospheric aerosols is somehow blurred over. Thus, the global dimming caused by aerosols and clouds may cause a drastic cooling effect, as was demonstrated by the consequences of volcano eruptions. As to the absorbing aerosols of anthropogenic origin, they can diminish the averaged solar irradiation by at least several W/m$^2$ (see, e.g., [237]). Clouds can also reduce the total solar irradiance, thus contributing to negative radiative forcing. In all cases, spatial and temporal distribution of clouds must have a significant effect on the diurnal asymmetry (one may notice here that $CO_2$ produces more warming during the night than in the daytime).

---

[247] And AGW skeptics are called "naysayers" by them.

Let us recall that in this book we are basically discussing scientific modeling. However, most of the "science" backing up global warming has been produced by computer modeling, the latter being only a part of scientific modeling. I trust in science, but I do not always trust in widely publicized computer models (WPCM), due to a great lot of people, money, and politics involved. Such models are somewhat similar to Hollywood blockbusters, where numbers are pulled out of the hat. The trouble with computer modeling is that it is typically a bad science since computer models can be tinkered in the desired direction to get any arbitrary result, even physically meaningless. The climate (defined as the averaged weather) may change, but attribute its variations to a single factor, small and not quite correctly accounted for in the radiation transfer, namely increased $CO_2$ concentration is hard for me to understand. The assertion $d\bar{T}/dt = const \cdot d\bar{c}/dt$, where $\bar{T}$, $\bar{c}$ are land-averaged temperature and $CO_2$ concentration does not fully reflect the physical reality, I am afraid. And I dare to predict that if eventually the AGW propaganda campaign fails, which is quite probable due to the interplay of a number of physical factors influencing the climate, carbon dioxide will still be pictured as an evil, with the shifted propaganda focus: for example, increased acidity of the world's ocean due to anthropogenic $CO_2$, killing the sea organisms on a global scale can be declared an actual catastrophe.

In short, personally I do not believe in the catastrophic human-induced global warming which is claimed to be the doom of mankind. "It is global warming that will surely cause the fall of civilization and perhaps the extinction of Homo Sapient," I took this sentence from a book on nuclear energy [238], in many respects quite interesting and informative[248]. I do not quite understand how the microscale effects (on the scale of $10^{-1} - 10^5$ cm) such as plumes, car exhausts and the like can affect climate on the planetary scale much stronger than global or even astronomical (space) factors. Or why occasional hurricanes (which had always occurred with high but unmeasured strength and frequency long before the $CO_2$ panic was spread over enlightened circles), wildfires and cyclical melt-accretion of polar ice caps are unequivocal evidence of the assertion that human activity causes global warming through $CO_2$ release. Nobody has proved any connection between $CO_2$ release and hurricanes or floods.

---

[248] Probably the author, a well-known writer and journalist, formerly an anti-nuclear activist, still shares popular environmentalist views, which are reflected in the sentence cited.

The climate changes may happen, of course, because climate is a dynamical system influenced by numerous agents, and if such changes happen they naturally have a certain sign of time-derivative for the local temperature on a certain time interval. Now in some measurement sites (e.g., near "urban heat islands") this derivative may be positive. I wonder what climate alarmists will say if in some time the sign of $dT/dt$, where $T$ is the local temperature, will change, at least for many locations? That humans are releasing too much sulfur acid? Or that $CO_2$-provoked global warming leads to global cooling, e.g., due to the meridional air mass transfer? By the way, I dare to think that global warming is better than global cooling, in particular, because much less energy is to be spent to sustain life. And I hope that people will eventually be sick and tired of the outcries that the sky is falling and therefore everyone (except selected few, of course) must get back into the caves right away.

As far as the economic effect of the enforced $CO_2$ reduction goes, it can become detrimental, but this is not quite obvious. For instance, one can easily calculate the costs of replacing all coal-fired power plants (they may really be dirty, but not because of carbon dioxide) by wind and solar farms, the generated power being kept constant. One can also calculate the reduction of $CO_2$ emission in the process of such replacement and, by using, e.g., the IPCC anti-$CO_2$ method, translate this hypothetic removal of $CO_2$ from the atmosphere into its greenhouse heating (say, for the USA which is considered the worst thermal pollutant). I made some crude estimates and obtained $0.1°$ Celsius. And the costs are of the order of a trillion US dollars, these costs are borne by the public, of course. A trillion for hypothetical $0.1°$ Celsius? Not too expensive? Maybe I was wrong, yet everyone can reproduce such estimates using some officially published data. Of course, one needs intense, emotionally loaded propaganda campaigns to substantiate such spending, for instance, horror stories like inevitable droughts, floods, hurricanes, tornadoes, tsunamis, etc. Science, however, strives to not confuse talk with empirical evidence.

Controlling human-produced $CO_2$ is like chasing a ghost; in fact, there are more urgent things to do than to please politicians. The attacks on $CO_2$ are so ferocious as if there were no other harmful ecological factors. The latter is obviously not true, but to solve real environmental problems is much more difficult than to create panic and exploit it for political purposes. There are, for instance, places where people get sick due to the detrimental state of the local atmosphere, and this has nothing to do with $CO_2$, but environmental organizations react very meekly on such information. Moreover, global steel consumption is rising by about four percent a year (see, e.g., http://www.worldsteel.org), this growth is especially pronounced in rapidly developing economies such as China and India. Steel production is accompanied by massive emissions of really harmful substances, primarily sulfur dioxide ($SO_2$), nitrogen oxides ($NO$, $N_2O$, and others), dust, etc. Large energy inputs as well as large amounts of carbon (e.g., in the form of coke) needed to produce steel inevitably release $CO_2$, but it would be unrealistic to curb the production of steel, drastically needed by emerging economies, on the bogus pretext of climate saving. The effect of climate-motivated political impact on the industry must be carefully estimated because it implies the redistribution of resources and productive capacities. If the entire coal mining industry is obstructed, it will mean a lot of unemployment and possible crisis in many regions of the world. Besides, coal-fired energy supply accounts for about 40 percent of heat and electricity generation so that the energy vacuum will be left, which hardly can be compensated for by the "renewable" energy sources - the favorite thesis of environmentalists which seems to be disconnected from reality. By the way, what will be the concentration of $CO_2$ in the atmosphere in the limiting case when all fossil fuels on Earth will be burned? The projections differ so wildly that I don't know which to select.

Yet all said does not mean of course that one should abandon developing new technologies and specifically new energy sources which do not use fossil fuel (see also below). In principle, instead of fossils the fuel can be made of $CO_2$ and sunlight, for instance, using bioreactors.

The blame put on human industrial activity in global warming is somewhat exaggerated. For example, I do not quite understand the hysterical over-reaction with car [249] production and use in connection with $CO_2$ emission, although many people (as well as some TV reporters) confuse $CO$, which is highly toxic, and $CO_2$, which is not [250]. Of all human produced greenhouse gases, automobiles are estimated to be responsible for 10-15 percent whereas the breweries are accountable for several percent. At least the $CO_2$ emission from beer (and wine) making is not negligible. Is it sufficient to curb the auto industry or one should also stop drinking beer and wine? Because of environmental constraints cars tend to become less reliable, safe and robust - in compliance with purely arbitrary, established by bureaucrats, emission norms. "Green" environmentalists urge to reduce the output of cars. By the way, it has already become a truism that a cow produces the same amount of greenhouse gases as an automobile, however mainly not $CO_2$ but methane $CH_4$. Does it follow from here that we should kill the cows or drastically reduce the livestock? There were the proposals to introduce the "fart tax" imposed on dairy farmers. Or a human being produces over 1 kg $CO_2$ per 24 hours [251] which makes about 6 million tons human-exhaled carbon dioxide a day i.e., the annual amount of more than 2000 megatons $CO_2$. For comparison, the second (after China) largest national producer of industrial $CO_2$ emissions, the US, has an estimated annual production of about 5800 megatons                              (see,                              e.g.,

---

[249] Interestingly enough, trucks and other heavy vehicles that produce more harmful substances and more $CO_2$, which is strictly speaking not a harmful substance, are somehow placed outside alarmistic attacks. Numerous fossil-fuel machines used by the military who could not care less about environmental effects are totally immune from any "save the climate" charges, and I doubt that heavy tanks, supersonic fighter jets and strategic bombers will ever be powered by batteries.

[250] Carbon dioxide is toxic only in high concentrations, about 10 percent or more.

[251] This is a very conservative estimate, actually more.

http://en.wikipedia.org/wiki/List_of_countries_by_carbon_dioxide_emission
s ). Other sources estimate the whole world's annual emission of $CO_2$ on the level of 10000 megatons. Still others give the overall industrial activity of humans resulting in annual $CO_2$ output an order of magnitude less. At any rate, human breathing accounts for at least 10-15 percent of the annual production of $CO_2$. Besides, humans are not the only species exhaling carbon dioxide. One should not forget also such powerful sources of greenhouse gases as volcanoes, and on top of all this, wildfires can add about 1500 megatons of carbon dioxide. Anyway, the figures for industrially produced $CO_2$ are comparable with biologically produced carbon dioxide.

It would also be interesting to compare humans with vehicles as far as $CO_2$ production goes. We can take that a human exhales about 10 ml $CO_2$ per second which would roughly correspond with 1 kg $CO_2$ per 24 hours (about 1 mole/hour). For the entire human population, it amounts to $10^8$ l/s. An average vehicle exhaust can be estimated to produce carbon dioxide at a rate about 1 l/s. If we take that there are $10^8$ automobiles each moment on the roads of the world (actually less), we get the total $CO_2$ production by the vehicles amounting to $10^8$ l/s i.e., of the same order of magnitude as by the humans. And there are a great lot of other species. Following the ambivalent anti-anthropocentric ethic of environmentalists, one should do something urgent with climate-hostile human breathing. Perhaps people should exercise less? Forbid jogging and refrain from lovemaking? Severely restrict the birth rate? Or the ultimate logic of environmentalists could be to kill all humans: this would remove letter "A" in AGW.

Let us now briefly summarize the $CO_2$ arguments of the "green" circles and their sympathizers supporting catastrophic scenarios of climatic change. One must, however, admit that the level of energy consumption per capita does not grow during the last 30 years. It seems that people have learned to some extent how to save energy. It is only the total energy consumption in the world that continues to increase - owing to the global population growth. Therefore, energy consumption (which accounts for approximately 70 percent of $CO_2$ emission) grows slower than assumed by the AGW catastrophists. In fact, the probability of catastrophic prognoses seems to be rather small if not close to zero, since there is simply not enough carbon (and consequently hydrocarbons) capable to ensure the corresponding increase of $CO_2$ in the atmosphere (by 4-5 degrees Celsius) i.e., 3-4 times the current level (0.038 per cent, about 36 percent over the level taken as reference, that of year 1750, before the European "industrial revolution".)[252]. In other words, the "catastrophic" $CO_2$ level should exceed 0.12 percent. Here, one can make two remarks. First of all, to reach such a level of $CO_2$ concentration in the atmosphere one has to burn a great amount of hydrocarbons, e.g., oil, natural gas, coal. The question is: can one obtain the corresponding amount of fossil fuel from the available sources within the observed period, say, within the 21st century? The second remark: is it only the absolute concentration of greenhouse gases in the atmosphere that matters, not the rate of its increase - time derivative of the concentration? This rate seems to be increasing (different sources give the estimates 0.5-3 percent per year, I do not know what to believe), at least the second time-derivative of $CO_2$ atmospheric concentration seems to be non-negative. What is the real role of this time derivative?

---

[252] Previously often cited values were $V \sim 1.338 \cdot 10^9 km^3$, surface of the world ocean $S \sim 361.3 \cdot 10^6 km^2$, average depth $\bar{h} \sim 3700m$. Volume $V$ of ocean waters amounts to about 1/800 of the planetary volume. Mass of ocean waters comprises approximately 96.5 percent of mass of the entire hydrosphere which, in its turn, makes up about 1/400 of the Earth's mass. Volume of water vapor in the atmosphere is estimated to be approximately $13000 km^3$, with the renewal period of about 10 days. It is interesting to compare this period with the estimated renewal time for the whole ocean: about two million years.

Catastrophic switches in natural systems may occur, with the transition into radically new states, but the assertion that it is humans who are causing such abrupt changes is, to put it mildly, highly controversial. Are humans responsible for the drift and possible switch of the Earth's magnetic poles? According to paleoclimatic studies the $CO_2$ atmospheric concentration in the epoch of dinosaurs was higher than today. It means that the climate of the Earth has survived different states and continued to support the biosphere. One can of course imagine that if the $CO_2$ concentration reaches a certain critical level in terms of radiative forcing (RF) then near-surface temperatures can increase in such a rate that human civilizations will not be able to adapt to these changes. This maladaptation to fast changes can be dangerous, not the slow (adiabatic) temperature increase *per se*, by the way in various locations and during different seasons. Besides, the current comparatively warm climatic period may be changed by a colder one, potentially bringing more disastrous effects than warming. Such misfortunes occurred, for example, in the third millennium BC or in the end of 17th century (Little Ice Age), when hunger and diseases eradicated a large portion of the population. Humans have probably no experience how to effectively survive under rapidly varying climatic conditions which may arise in the case of catastrophic developments within the next several decades. But to ascribe all today's rainfalls, storms, hurricanes and even cold weather only to anthropogenic climate warming i.e., induced exclusively by human activity is either a delusion or a deliberate lie.

## 10.10     The Role of the Sun

Probably the most serious argument in favor of anthropogenic (more exactly - green- house) global warming is the observational data suggesting that it is difficult to ascribe the increase of global average temperature that occurred since 1975 entirely to the Sun-Earth coupling (see, e.g., [239]). Although the Sun is by large the dominant energy source on Earth, some others - such as manifestation of internal planetary heat through the volcanic activity, burning fossil fuels, exploding nuclear devices - may have locally a comparable impact on the environment largely determining its temperature. However, such effects are local in space and time and, moreover, should be treated by the methods of statistical dynamics applied to the climate system.

Historical records demonstrate a relationship between solar activity and, say, winter temperatures, at least at the local and regional level. Already this fact makes it more difficult to assess the reality of AGW without relying on computer models. One can notice that each model focuses on its own specific details, which fact makes climate modeling heavily dependent on expert opinions. Therefore, the weight of determining factors as well as robustness and reliability of climate prognoses depends more on subjective judgments than on fundamental physical principles. It would also be interesting to observe that solar activity ceased to influence the Earth's climate almost synchronously with the creation of the IPCC (1990s): since that time sharp warming has become accompanied by the fall of solar activity.

Although it would be stupid to deny the presence of anthropogenic factors (see above) especially when they are growing, but their escalating role as compared with natural and very powerful ones such as intensity variations of solar radiation, influence of massive planets and deviations of the Earth's orbit has not been quantitatively elucidated. It may well happen that the main cause of climate change is still solar irradiance[253], with the delay effects being taken into account. For instance, the shortwave radiation from the Sun propagates through the ocean waters and heats their deeper layers, and one should integrate over time the radiation absorption by the ocean (mainly in the shortwave range). Physically, one can understand this absorption in deeper strata of the ocean in the following way: the absorption coefficient $\mu$ in the phenomenological Lambert's law, $dI(\lambda, z)/dz = \mu I(\lambda, z)$, where $I(\lambda, z)$ is the solar radiation intensity, $z$ is the vertical (ocean depth) coordinate, which depends on the wavelength $\lambda$ (and also on water impurities, in particular, on salinity). Since the thermal conductivity as well as diffusion and convection in water are relatively slow, processes and the water strata remain stable, the heat may be stored for years in the ocean[254] before it is eventually transferred to the atmosphere (mainly through the horizontal flows to the polar regions). This physical mechanism of time-delayed climate response to solar activity (mainly characterized by the sunspot numbers) should exist, but regretfully I was unable to find calculations of the corresponding time lag in the literature available to me. Yet in general the impulse response of the climatic system to the Sun's activity simplistically manifested by the surface temperature is not a delta-function of time but has a finite width $\tau$ that is determined by a number of physical processes and may well reach dozens of years.

---

[253] It has been noted in most encyclopedias that the very word "climate" originates from the Greek word "klima" which means inclination and was referred in ancient times to the inclination angle of the Sun, see e.g. http://en.wikipedia.org/wiki/Climate.

[254] Although the main absorption of the solar energy occurs in the water layers near the ocean surface, the heat stored in deeper strata is still significant which is manifested by ocean flows (see, e.g., Primeau [248],) as well as by the fact that deep water temperature is well over the freezing point everywhere. The most part of the shortwave fraction of solar energy is absorbed in tropical latitudes where this fraction is more pronounced, which enhances the effect. Recall that approximately 70 percent of the total terrestrial surface are covered by ocean waters, with 90 percent of the total volume of the ocean being below the thermocline.

In short, it seems to be clear that variations of solar activity should seriously influence the climate - the thesis that is emphatically refuted by environmentalists. Discussing the role of the Sun in possible global warming has been lately regarded as primitive, non-scientific, and almost indecent, although the modulation of solar radiation transfer by the atmosphere is a passive effect and cannot lead to substantial global temperature shifts. If modulations in solar radiation transfer by the atmosphere cannot lead to global temperature shifts, does it not mean that the role of the Sun is not big? Yet in a well-known book by S. R. Weart "The Discovery of Global Warming" (Harvard University Press, Harvard, 2003) the following statement is referenced: "For a young climate researcher to entertain any statement of sun-weather relationships was to brand oneself a crank". The forcing of TSI (total solar irradiance) is declared by the engaged climatologists as a negligible factor compared to human influence [255]. Advocates of man-made global warming mockingly beat the drums: "Humans are small, Sun is big". However, the point is not in this trivial observation, but in the hard physical fact that all phenomena on the Earth are strongly coupled with the processes in the Sun. To completely negate the solar activity, the latter being measured by sunspot numbers, as a climatic factor is at least shortsighted since modulations of the energy output from the Sun are significant and their measurements have not been finalized yet. One can assume that foreseeing future climate variations heavily depends on the ability to envisage dynamics of the Sun-Earth physical coupling, which is an interdisciplinary task. Specifically for the Earth's surface temperature, its global variations of at least 0.1 $^\circ$ Celsius associated with the 11 year solar cycle have been extracted (see, e.g., http://data.giss.nasa.gov/gistemp/2007). This magnitude is comparable with the estimates provided by the computer climate simulations. However, the signal directly originating from the Earth's irradiance by the Sun is difficult to separate from other causes of the terrestrial temperature variations including fluctuations.

We have already discussed that the Earth's climate is determined by a delicate kinetic balance between the incoming solar radiation (in all spectral domains plus corpuscular flux) and outgoing thermal radiation, this balance being mediated by the composition of the Earth's atmosphere. Both types of the forcing agents - natural ones such as solar variations and volcanic emissions as well as anthropogenic ones such as greenhouse gases and sulfate aerosols significantly affect the corresponding kinetic equations. Changes in the kinetic equations mostly occur in the coefficients, with such changes acting in opposite directions (e.g., cooling effects of aerosols can be partly neutralized by heating due to the emission of greenhouse gases). Solar irradiance variations contain both the direct part entering the source term (the incoming radiation) and the coefficient term such as, e.g., due to changes in the ultraviolet component leading to the modulations of ozone production rate. Ozone is a rather potent greenhouse gas since it absorbs long-wave infrared radiation (LWIR) emitted from the Earth's surface and thus contributes to the heating of the atmosphere. Moreover, ozone in the lower stratosphere where temperatures of 70C to 80C are encountered is thought to have a much larger effect on the radiative balance as compared to ozone at surface level: it can absorb infrared radiation and re-emit the latter with the wavelength corresponding to about 18 Celsius (http://www.ozonelayer.noaa.gov/science/basics.htm. This means that the impact of ozone leads to effective warming of the gas in the troposphere. This is an example of complex interactions when the indirect partial effect may be of a similar magnitude, or even larger, than the direct effect. Delicate interplay of the factors present, both explicitly and implicitly, in the kinetic equation determining the radiative balance can make the multiparameter climatic system highly sensitive to small variations of different factors, easily bringing it to unstable or chaotic domains (see, e.g., [240]). In this situation, there is more of a consensus than real science about the relative role of natural, primarily TSI and anthropogenic (such as $CO_2$) forcing agents, i.e., mainly a social effect. It is interesting to notice that the generally perceived role of solar variations in the observed climate dynamics changes - in fact oscillates - with time. We have already discussed on various occasions that it is usually quite difficult to disentangle the sociological from the scientific.

---

[255] Recall that the Earth's surface average temperature was estimated by the IPCC to have increased by approximately 0.6C over the past century whereas the TSI forcing have contributed only about 0.2C over the same period.

The solar activity[256] minima have been observed to be correlated with colder temperatures of the Earth's surface, an example was the notorious Little Ice Age in Europe, North America and possibly in other parts of the world in the 17th century ascribed to the "Maunder Minimum" (see, e.g., http://en.wikipedia.org/wiki/Maunder Minimum). In Europe, many dwellings perished because of starvation during the Little Ice Age. However, some scholars who insist on anthropogenic climate changes deny the causal link between the lull in solar activity depicted by the Maunder Minimum and bitter cold temperatures during the Little Ice Age, considering the overlap of these two periods as a statistical coincidence[257]. The AGW concept is based on one-to-one correspondence between $CO_2$ concentration and temperature rise whereas looking at time series, one cannot exclude the possibility that the overall temperature behavior is flatter than the $CO_2$ concentration increase. If this is the case, then the orthodox AGW theory does not seem to hold. There exist many other notable relationships - pure coincidences or not - between the Sun's activity and terrestrial climate. One more example is the so-called medieval climatic optimum (MCO) that was observed in the 11th through 13th centuries and coincided with the Grand Maximum of the solar activity (see, e.g.,[241])[258]. Maybe it sounds non-scientific or even "cranky" but a striking thing is that there existed a nearly one-to-one agreement between the periods of average temperature changes (as best they could be known) and solar activity modulation, although the climate alarmists deny this coincidence. There are data (though also controversial) that the Sun was in a state of unusually high activity for about the last 60 years of the 20th century. A similar period of enhanced solar activity (however, characterized by a significantly fewer number of sunspots than during this last period of activity) occurred in the Middle Ages approximately from 1100 to 1250. It is approximately at that period that the above-mentioned medieval climatic optimum happened on the Earth when, for instance, the Vikings settled down in Greenland and Iceland. The Vikings' colonies were recorded to have flourished for several centuries and deteriorated due to starvation and cold during the Little Ice Age. Today, we can survive an analogous warm period when, in addition, the enhanced solar activity works in phase with the possible anthropogenic factors. One should explore this controversial issue further, of course, as well as the current and historical variations of solar activity. In particular it is unlikely that a reliable answer to the crucial question "On what time scales can the Sun affect terrestrial climate?" does exist.

---

[256] The Sun's activity is in general an integral notion being understood not only in terms of sunspots, but also accounting for changes in total irradiance, ultraviolet irradiance, magnetic flux variations, solar wind, energetic solar particles, variations of the size and intensity of heliosphere, etc.

[257] Curiously enough, the same scholars do not regard the analogy in two  time series - temperature data and $CO_2$ emissions - expressed by the controversial "hockey stick" (see e.g., http://en.wikipedia.org/wiki/Hockey stick controversy and Holland [249]) as a statistical co-incidence. In general, one can add a zero-centered random number to the previous value and get a variety of statistical series similar to the random walk series. When plotted, such series can resemble the "hockey stick" (for amusement, I have done it with MS Excel, but probably such products as SPSS or "Statistica" are suited better.). The usual "hockey stick" argument means no more than the fact that one data set (temperature reconstructions) matches another (CO2 concentration) over some arbitrarily selected averaging or calibration period. In this process one can obtain as many "hockey sticks" as one desires, by putting a variety of data

## 10.11    Limitations of Current Climate Modeling

In the current political and media discussions the issue of accuracy in AGW forecasts has been somehow covered up by slogans and fears. Yet the question of accuracy is always of paramount importance in physics, and the reliability of scientific prognoses essentially depends on it. It is clear that the accuracy of computer model output (the error margins) cannot be better than that of the input data. Recall that most data in geophysics have a large uncertainty corridor. Besides, the complexity of climate science compels one to make

---

sets (e.g., population growth, number of newspapers, scientific publications, bicycles, etc.). There may even be lucky "hockey sticks", for example, in personal or corporate budgets.

[258] One can object that the book issued in 1975 is totally outdated and does not reflect modern opinions. I don't understand such objections: the book contains hard scientific evidence which cannot be outdated by the very definition of science. This is not a political journalism, after all. In the same fashion, one can merely declare outdated the works by Newton, Boltzmann, Gibbs, Einstein, Poincaré, Einstein, Rutherford, and many other scientists.

many simplifying assumptions which only can lower the accuracy. Therefore, there exist a lot of indeterminacies in climate dynamics, and the error margins of climate models must be specially appraised by independent scientists *before* political decisions about allocation of resources are made.

Current mathematical models of climate are basically those of fluid dynamics, at least they are centered about the description of fluid motion. The kernel of these models is devoted to describing the fluid motion in the ocean and the atmosphere. But this is far from describing the real world we are living in, with plenty of intricate physics as well as chemical, biological, geological, anthropological, social, etc. factors influencing the climate variability. It is hard to understand from the present mathematical models (and, of course, totally impossible from political discussions) how to separate the $CO_2$ produced by fossil-fuel incineration from its conversion in biomass. Seasonal accumulation and release of carbon dioxide is a spatial-dependent kinetic process that should be coupled to fluid dynamics and radiation transfer equations of climate modelling. It is clear that since $CO_2$ hidden in the biomass and soil is an important reservoir of carbon comparable with that produced by human activities, biological processes must be an important link in $CO_2$-based climate projections. However, such processes are usually not included in climate modeling. I also could not find a satisfactory solution of the energy balance problem based on the radiation transfer in the atmosphere (a detailed kinetic approach and not only energy balance). It is this problem that should theoretically corroborate or disprove the greenhouse effect concept. Models that I have come across are mostly of a particular character or provide crude estimates, many essential factors being neglected. What climatologists really know is based on very limited paleoclimatic observations and on a bunch of computer models.

From the analysis of biological effects on climate changes, would probably follow that one can bind excessive $CO_2$ with the help of appropriate farming technology, land management and genetic engineering.[259] These advanced agricultural techniques have nothing to do with current mathematical models of climate and still less with ideological denunciations of environmentalists. There are, however, certain hallucinatory ideas of so-called geoengineering, for instance to spray many megatons of sulfuric acid in the atmosphere, with the possible grave consequences such as acid rains.[260] Another fancy idea is the $CO_2$ underground sequestration. Such projects may also involve a considerable risk since in the case of a leak all living organisms within the layer of 1.5-2 meters from the ground, i.e., humans and domestic animals, in the vicinity of the reservoir of $CO_2$ will be killed. Besides, people do not understand the climate system well enough to take radical decisions about influencing it on a global scale (geoengineering). But I think one should not worry too much: whether or not one thinks that radical actions should be urgently taken, they are unlikely to follow.

[259] In fact, carbon dioxide is precious for the biosphere and comparatively rare - the actuality environmentalists and climate alarmists always seem to forget.

[260] By the way, where are the precautionary environmentalists who always fear the unknown side effects? The fact that the environmentalists do not object signifies that there is more ideology than ecology in their position.

In order to evade ambiguity, I can state my position right away: I am neither for nor against global warming (GW). It may happen. Moreover, it probably happens. More precisely, the warming probably occurs in the current period - there exist climatic cycles. And it must even have an anthropogenic component (AGW), at least locally, which can be seen already from the fact that average temperatures are higher in the cities than in rural dwellings. However, it seems to be only a belief that the warming is completely man-made. Relative roles played by the natural (such as Sun-Earth coupling or volcanic activity) and anthropogenic (such as greenhouse gases or sulfide aerosols emission) factors as well as their interplay are far from being elucidated. Quantitative results of comparing natural and anthropogenic forcings do not have a sufficient accuracy to guarantee unequivocal statements [261]. Furthermore, due to multitudes of parameters important to determine the dynamics and equilibrium states of the Earth's climatic system I don't think reliable prognoses for its evolution could be feasible, at least for the current state of physical and mathematical knowledge. Computer codes in climate modeling, at least those that I could come across, are based on strikingly shallow physical theories. Furthermore, there is even an underlying philosophy justifying this apparent simplicity of climate physics: climate is allegedly no more complex as a heat engine. I don't think that simplistic nearly-linear models readily implemented in computer codes are relevant for obtaining the values of average terrestrial temperatures for 50-100 years ahead. Therefore, I suspect that the importance of the human component (letter A added to GW) is, using again Mark Twain's famous expression, greatly exaggerated. And please recall that vanity, hubris is a serious sin.[262]

---

[261] The role of the Sun-Earth coupling is often a priori declared negligible compared with anthropogenic factors. Those who still dare to insist on an accurate assessment of varying insolation as a climatic factor are often labeled as anti-scientific retrogrades. This question seems to the adherents of AGW trivial, long time passé, and only producing yawns (see below more on the possible role of the Sun).

[262] This book was written before the Nobel Price 2021 for physics was granted to Syukuro Manabe and Klaus Hasselmann for "leading the foundation of our knowledge of the Earth's climate and how humanity influences it". Back in the 1960s Syukuro Manabe, Japanese American meteorologist and climatologist, was working on physical models to explore the interaction between radiation balance and the vertical transport of air masses. Based on stratospheric and tropospheric measurements showing that temperatures rise in the lower atmosphere and fall in the upper, the scientist argues that the cause of temperature fluctuations are changes in $CO_2$ concentrations, not solar activity. The conclusion that followed from this model (at least the way it was interpreted) was that oxygen and nitrogen had negligible effects on surface temperature, while carbon dioxide had a clear impact. Klaus Hasselmann, Max-Planck Institute for Meteorology, became a Laureate for developing a method of distinguishing between natural and human causes of atmospheric heating, the so-called fingerprints. The problem of a mean atmospheric response to external forcing such as volcanos, albedo, surface temperature, sea-ice etc. is addressed by applying a filtering technique based on comparison of atmospheric response patterns derived from multiple sources: models, experiments and data sets for long periods of time. The expectations are that this method will allow us not only to distinguish the increased

# 11 Made in physics

Physics has recently demonstrated additional power by crossing into other disciplines such as biology, economics, ecology, sociology, medicine, even political sciences. Physics-based mathematical models (PBMMs) may have nothing to do with physics. For instance, numerous traffic flow models are essentially based on the conservation laws, which is a physical concept. Another example is the famous logistic model, usually applied to the natural population growth, is a simple generalization of the exponential model widely used in physics [263]. The logistic model, despite its simplicity, was able to adequately represent the population dynamics in various countries, e.g., in England, Scotland and some parts of USA. In biology and ecology this model is used to describe various evolution scenarios, when the future population of species depends linearly on the present population. A little later we shall discuss these two above-mentioned classes of models in some detail. Recently, physics-based models won great popularity in the field of social and economic modeling.

Nevertheless, I think that the mathematical work such as in physics is still very peripheral in social and even in economic disciplines, at least so far. In contrast with physics, social sciences attempt to produce descriptions and diagnoses not for material phenomena, but for mental trends and psychological conditions, assuming that collective attitudes matter more than material phenomena so that the material order (or disorder) in the world is not the foundation but the implication of psychic inclinations of the people.

So far, the ultimate purpose of physics is to control the properties of non-living matter, for instance, to model, design, and eventually mass-produce new materials. Likewise, the purpose of social models would be to explore and then tune the properties of human material.

In social studies there exist particularly many unanswered questions, with some of them may be considered to lie at the very core of social disciplines. For instance, do different nations pass through the same stages, albeit with some delay with respect to one another, just like people who, while growing up, are passing through universal stages in their development? If one can give a reliable affirmative answer, would it mean that a time-ordered (and irreversible) process of social and cultural evolution is maintained? Or the alleged orderly progression from the primitive to more "civilized" stages may be reverted under certain conditions? In other words, do patterns of development exist for human societies, and to what extent can individuals deviate from these patterns?

Unfortunately, when discussing such issues, qualitative analysis is emphasized, sometimes rather aggressively, which probably testifies to certain inferiority undertones.

---

[263] The logistic and the exponential models are nearly identical at small times.

## 11.1  Exported Models

It has long been speculated that physical models or rather mathematical models of physics might prove useful for the analysis of human behavior, both on the individual and the collective level.

Unfortunately, dynamics of the systems, for which we do not know the main evolution equations, appears to be irreproducible. In contrast to physics, such disciplines as biology, medicine, ecology, psychology, economics, sociology, etc. have so far admitted correct and non-speculative theoretical study only at the level of time series. Now there are increasingly frequent attempts to construct particular mathematical models, mostly based on nonlinear dynamics, to describe the evolution of systems studied in the above weakly formalized disciplines. Chaos theory has been especially attractive for the scholars attempting to apply off-the-shelf physics-based mathematical models (PBMM) to social and biomedical sciences. Although there exist already a considerable bulk of papers, also in physical literature, exploiting nonlinear dynamics to build mathematical models in these disciplines, the notion of chaos still remains rather the subject of quasi-scientific philosophy than the tool of quantitative science.

The aim of mathematical models in social sciences is to help estimating and understanding the statements of humanities that are far from physics and mathematics. Here, the principal technique is to properly identify the scale of a problem and to pose the relevant questions, which is not always the case in philosophy and artistic culture (see Chapter 2). Physicists are usually better trained in restricting the problem to a set of localized models. It is interesting that such a great physicist as E. Majorana who was in general inclined to use sophisticated mathematical tools published a sociology paper [276]). Majorana published not so many scientific articles, and this one was probably intended to extend the theoretical schemes of physics onto social sciences.

One can see the deficiency of general statements, when the scale of the problem is not appropriately identified, on the following slightly exaggerated example. It would be in principle correct to say that, e.g., viruses are ultimately constituted, as all matter, of quarks and leptons. However, this statement does not help much: knowledge of such basic constituents will not explain functioning of viruses which results in disease. Likewise, the statements related to behavioral sciences such as psychology are usually so general and arbitrary that it is difficult to define their domain of applicability and construct a reasonable mathematical model corresponding to these statements, although mathematical modeling in psychology has a long tradition starting from F. Bessel and H. Helmholtz. The trouble with behavioral disciplines is that they rely on "expert" opinions rather than on experimental techniques and related theoretical explanations. Even the attempts to construct quantifiable procedures are accompanied by the drawbacks of unknown accuracy. Take, for instance, a very popular assessment technique based on IQ. What is exactly the quantity that is measured by this method? Does this unknown quantity change with time? Do you believe, in general, that IQ means much? Nevertheless, checking a person's IQ by some bureaucratic authorities (e.g., human resource management) can drastically influence her/his career. As for mathematical modeling in behavioral sciences, they are either too abstract and detached from experimental base or represent universal statistical applications such as structural equation modeling (see, e.g., Schimmack, U. and Grob, A. (2000), Dimensional models of core affect: a quantitative comparison by means of structural equation modeling. Eur. J. Pers., 14: 325-345. https://onlinelibrary.wiley.com/doi/abs/10.1002/1099-0984(200007/08)14:4%3C325::AID-PER380%3E3.0.CO;2-I; see also https://liberalarts.tamu.edu/communication/profile/hart-blanton).

## 11.2  The Limits of Sociology

In this section, a few vacuous generalizations can be encountered. Some passages here only document my experience, reflecting the integrated psychic attitudes and mental inclinations of several groups of persons I had a chance to talk to. And I know very little or nothing at all about the boundaries of reality constraining such group attitudes.

Several decades ago, sociology created many false expectations, e.g., that it is capable to fully explain human behavior, flow of events and even principles on which human societies are built. However, these expectations have not been materialized, and many people became greatly disappointed by sociological results. L. D. Landau, for example, considered sociology a pseudoscience. What are actually the limits of sociology? To what extent can sociology make predictions, in the meaning of conditional statements usually accepted in natural sciences?

I don't think I shall be able to correctly answer these and similar questions, not only because I am a dilettante and not because the questions are put in too general a form (deliberately), but because the subject of sociology is extremely complicated and badly formalizable by the current mathematical means. Sociology may be interpreted (not defined!) as the collection of techniques providing the movement from opinions to understanding. Then, if understanding has been achieved, one can make projections, models and forecasts, even quantitative. Sociological research deals with mass phenomena and mass behavior. A person is not discerned in this mass. In such a sense, sociological research may be called defocused rather than focused - its resolution is not higher than a "small group" of people. What is this small group, sociology does not take pains to correctly define. In mathematical terms, one can use the notions "averaging" or "homogenization" to characterize the objects typically studied by sociology. The macroscopic character of sociology is the first limitation that one must bear in mind when talking about this discipline's claims.

The second limitation consists in the lack of predictive power in sociology. In distinction to physics and even modern economics, sociology is mostly busy with explanatory constructions. They may be interesting and plausible but without predictive component, and even if there may be one, accuracy of predictions is unclear. Here, I might note that one can redirect a similar reproach to modern theoretical physics. Experience shows that theoreticians can explain everything, just anything you can imagine. This is good, it testifies to their skills and flexibility. But it is much harder with prognoses.

The third limitation is a virtual impossibility to talk about the unique interest of a society or a country. Society in each country is composed of many acting groups, each having its own interest. Therefore, it would be difficult to define a common interest of the entire society, unique for the given country. Usually interests of the ruling group are declared as "interests of the country" or "interests of the state".

The state is not completely determined by formally fixed constitutional rules, over-constitutional rules are usually stronger than legislation. This is a binding system of unconventional norms which make each society unique. This is the form of civil life specific for each society. Even a high level of administrative aggression such as endemic to dictatorships, if directed against informal rules, cannot fully destroy them. Each state can be characterized by its own level of coercion as well as of corruption and other attributes of power concentration. I do not want to be engaged in something similar to amateur kremlinology in this book, especially regarding the limits of sociology, and one must study carefully the effect of power distribution on social evolution, the latter being understood as the trajectory of a complex human organization, e.g., a country, in the space determined by a set of relevant parameters. A crude high-level look hardly brings any results or prestige; therefore it should be relegated to kitchen talks or cocktail parties. Actually, there does not seem to be many good quantitative results and models describing power distribution effects in various societies. See in relation to this [293] and [294].

### 11.2.1   Self-Reproducing Social Patterns

To what extent could the processes occurring in nature be mapped onto human social structures? The complete isomorphism leading to such quasi-naturalistic theories as social darwinism, in its extreme form resulting in fascist and Nazi oppression of "unfit" individuals and groups. In such extreme forms, it is the pure ideology in action: rulers interfere and suppress the "unfits" even though the latter in no way present a direct threat to them. The heavy hand of the ruling group is more appreciable in case the country is not a "democracy". Thus, societies may be ranged and measured according to the ruling elite pressure. This is one of the main tasks of sociology which has not been successfully accomplished so far. There are a number of other cardinal questions that may be considered as sociological tasks. For instance, why and to what extent laissez-faire capitalism is perceived as an enemy for generalized socialists such as communists, marxist-leninists, nazis. It is more correct to refer to the national-socialist (NS) regime, "nazi" is only a colloquial term like "commi"., etc. Furthermore, it is an empirical (historical) fact that socialism strongly relies on violence. The question is: why and how can this accent on violence be quantitatively estimated and measured?

Democracy is usually understood as an antithesis to socialism. But there may exist different versions of democracy, starting from liberal democracy to the one with unlimited tyranny of the majority as during the French Revolution of 1789-1799 or a kind of democracy supporting oppressive regimes such as in Russia in the beginning of the 1920s as well as in Iran in the beginning of the 1980s. The main principle of liberal democracy is not to enforce the will of the majority, but to ensure the rights of each small group, down to a single person. The main presumption of liberal democracy is that if any group of people is suppressed in its rights i.e., the ruling group is allowed to suppress such a group of people, then this ruling group will necessarily strive to suppress all other groups. So, the liberal variant of democracy emphasizes protection of all minorities and individuals. The liberal democracy may be considered as a limiting case on an authority scale, the other limit might be considered an ideological totalitarian state, with all intermediate kinds of despotism and democracy (authoritarianism, autocracy, military dictatorship, ethnocracy, socialist state, ochlocracy, etc.) being positioned between these two extremes. It is a business of sociology to assign numbers (points on the scale) to each state[264], thus providing an ordered set for a social or political analysis [295]. Accent on qualitative descriptions and scathing metaphors like "snob-rule" (elite or oligarchy) vs. "mob-rule" (the majority) can hardly be satisfactory.

What is a totalitarian state? It would be difficult to give a strict definition, but one of the dominant features of such a state, intuitively, consists in describing its reaction on out-of-order messages. The totalitarian response to a signal about a misrule or a trouble consists not in the attempt to streamline the situation, but in efforts to switch off the signal source. In other words, in a totalitarian state you should not try informing the authorities or the police that something goes wrong because this won't function and you will be, with high probability, eliminated. In fact, any state, by definition, has some amount of totalitarian features, but the feedback intensities differ from country to country. Nevertheless, the totalitarian state of the country does not mean that one should be sitting all one's life under the bed, be completely silent and never stick one's head out. We might recall in this connection that in the Soviet Union almost everybody, with tiny exceptions, figuratively speaking, had not shown themselves from under the beds for 70 years, waiting for the outcome of the fight of elites, until the Communist regime partly collapsed. Thus in a totalitarian state each individual is forced to contribute to the stability of regime and perpetuating it participates in evil.

It has long been observed that two wicked things never occur in liberal democracies: mass murder and mass hunger (to my regret, I don't remember who was the first to verbalize this observation). It is, however, useful to recall a simple maxim known to Aristotle, but totally forgotten in the 20th and 21st centuries: democracy is the worst form of rule if the society is mostly composed of paupers. In this case, democracy inevitably transforms into tyranny. This observation leads to a general question: what are the main causes for the death of the democracy on a given territory? The list of such causes seems to be rather short. One is a war. A democratic state can be either conquered, for instance, if it is small or, by defending itself and in process of mobilization of its armed forces, transforms into a military dictatorship. Another classical death of a democracy is through a military (or paramilitary) coup, this scenario has been materialized a number of times in Africa and Latin America. But the most general pattern of transition from democracy to tyranny appears to be due to wide involvement of poor and illiterate population into political decision making. It is understandable why democracies were very unstable before the industrial revolution: the latter provided necessary conditions for stability of a democratic order with respect to power overturn by ensuring a well-being for the majority of the population. These conditions are not sufficient, as we might observe: the majority rule can still lead to tyranny even in relatively affluent societies. Intuitively we know that some countries demonstrate examples of seemingly stable democracies, in particular liberal democracies, whereas other societies appear to be unstable with respect to the classical "democracy-tyranny" transition. Moreover, we may notice that despotic, e.g., authoritarian forms are stable because they are unfavorable to economic productivity and even distribution of wealth thus preserving pauperism. This fact is intuitively grasped by anti-democratically oriented political forces (e.g., left-wing socialists, radicals, communists, etc.) who build their strategy and even propaganda on hindering economic productivity. The main task of sociology in this respect is to find the exact conditions expressed in numbers, formulas or algorithms, which would tell us when the stability of a given democratic society is ensured and what is the risk of transition to a certain variety of a despotic regime.

---

[264] The word "state" is understood here as the state of a system and not as a political organization of a country.

Since it is difficult to rely on self-restraint of the majority and its favorite leaders as well as on self-control displayed by the ruling groups, tenable power restriction mechanisms should be designed and developed in a stable society, in particular the legal system, to secure a clearly defined scope of rights of an individual and, consequently, of all groups of individuals. To offer models for such control mechanisms may be regarded as another task of quantitative sociology.

Apart from the authority scale, there are many other self-reproducing patterns in human societies. How can one understand why all the regimes promising socialism, communism[265], total equality, possession of equal rights, social justice and the like quickly turn into ruthless, despotic dictatorships? Why is bureaucracy so overwhelming? Greatly simplifying, one can observe that there exist in fact only two superparties in any state: "bureaucrats" and "liberal economists", all others being just nuances of these two mainstreams. A variety of colors and hues may be considered as a fine structure of these two principal levels. The above opposing superparties represent two main attitudes present by people: paternalistic and self-sufficient. This attitudinal dichotomy, being translated due to sociological collective effects into dual social structure produces natural associations with two main types of objects in physics: bound and free. The paternalistic attitude implies etatistic structures on which an individual is heavily dependent and results in the "bound state" of an individual, with movement restrictions, informational censorship and other well-known limitations. Figuratively speaking, paternalistic pattern is a projection of diapers onto the whole life of a person. A swaddling effect of the diapers is realized through countless bureaucratic circuits under the pretext that the authorities know better what "people really need". Accordingly, two main personality types having totally different images of society can be developed: those who rely on their own forces, intellect, and hard efforts versus those who hope that external agencies such as God, good monarch, state control, social help, etc. will in any event interfere and protect from excessive hardships. These latter people tend to unite under leftist, communist, general collectivist, environmentalist, nationalist, anti-globalist, and other populist slogans being easily recruited into militant groups ("take everything away from the rich, then divide it up"). In the Marxist political texts, the low-income fraction of people who depend on the state social system for their day-to-day existence are typically called "lumpen".

---

[265] One usually distinguishes between socialism and communism as an extreme form of socialism, but for me it is basically the same social phenomenon.

In the societies with a considerable component of liberal economy, more relying on self-sufficient behavior and free choice, bureaucratic structures are forced to compete for power, in particular by leveraging populistic ideologies. In relation to just mentioned social dichotomy, one can notice that there are two basic regimes of social equilibrium in human societies. In the countries which use to call themselves "advanced" or "developed", social equilibrium is of dynamic character and is distinguished by an oscillatory "social trajectory" which has intermittent bureaucratic and liberal phases. Typically, the process runs as follows: after the fast transition from a "frozen" despotic regime to a "democratic society" at its liberal phase, e.g., after a drastic social transformation such as a revolution, people are initially motivated to work hard and economy is on the rise. As a result, people on average begin to live better, but high-income and exceedingly wealthy groups appear inducing envy of the poorer part of the population. This envy produces demotivating effect, and besides people tend to be more relaxed due to increased quality of life[266] and more attention devoted to hedonistic excesses. Economy begins to fall, which is amplified by the outbursts of activity in the anti-liberal camp exploiting mass envy and insisting on the enforcement of social justice and equal distribution of wealth. "Down with capitalism!" is a typical slogan in such periods. Distribution and to some extent production should be maximally controlled by the state bureaucracy which represents such a control as "achieving an order" in contrast with "liberal capitalist chaos". Ensuing nationalization and politization of the economy aggravates the situation. Economy naturally falls deeper, and in the countries with the lack of democratic traditions such as firmly established free periodic elections, there is a hazard of transition to a "frozen" despotic state totally controlled by bureaucracy. However, in advanced economies with still wealthy population and really free choice between political parties, the population begins to sense danger and votes anew for "liberal economists", at least more liberal than clinging to power bureaucratic control freaks. Then the cycle repeats itself - I view this phenomenon as oscillations of social trajectory. So, the first basic regime of social equilibrium is an oscillatory one, it is typical of developed countries. For the countries with weak democratic traditions, especially with dominating pauperism, another regime is habitual. From the viewpoint of macroscopic social equilibrium such countries' trajectory symbolizes a totally different story. One can easily observe that many countries usually known as "developing" are unstable under the transition to the stationary state of "frozen" bureaucratic despotism, with reduced civil rights and limited personal freedom. It does not matter how such regimes are called either from outside (tyranny, autocracy, authoritarianism, communism, fascism, etc.) or from inside (controlled democracy, real socialism, also communism, etc.), the essence is that the societies supporting such regimes are in a metastable state separated from the equilibrium corresponding to liberal democracy by a rather high barrier. Nevertheless, contemporary history shows that this barrier is not impenetrable: more and more societies have overcome it, being transformed into liberal democracies. One can say that such a transformation represents a historical trend. However, the characteristic transition time from the metastable frozen state to democratic oscillations may significantly exceed the active period for a single population (population life cycle). One may also notice that the mentioned historical trend reflects a kinetic situation: there exist faint democracies that are unstable under the reverse transition into the frozen state.

---

[266] I don't know a correct definition of the term "quality of life"; here it is used intuitively in accordance with the whole intuitive exposition.

I think that this verbally described cognitive model can be translated into the language of mathematical modeling and quantitative results corresponding to the observable phenomena can be obtained. I found it strange that sociology, when attacking societal issues using mathematical tools, leaves the crucial problems of structure, dynamics, and equilibrium of societies mathematically unexplored (although, I am just a layman and know very little of sociological literature).

One more self-reproducing pattern in the society is an eternal conflict between clericals and intellectuals, i.e., between the church[267] and the science. An established church seeks to universally spread its messages (preferably at the expense of all citizens); therefore it tends to oppose secular and scientific education which is driven by economic demand. In this conflict, the church represents the social force directed against economic and technological development. One can recall that it was only in October 1992 that Pope John Paul II expressed regret for the Galileo affair and officially conceded that the Earth was not stationary. I remember a meeting in Moscow organized by Siemens in 1996 and devoted to the social impact of IT. Some clergy were invited, and when it was mentioned that the Internet allows one to radically expand the scope of living activities, to live simultaneously several lives as it was figuratively expressed by one of reporters, the orthodox priests immediately began frenetically crossing themselves and raising vehement protests.

---

[267] The term "church" may be understood in this context in a broad sense including all irrational currents in the society such as sects and quasi-religious groups. A unifying feature for all such currents is relying on miracle or mysticism, that is on phenomena contradicting proved scientific facts.

One often cites Bacon's aphorism: "Knowledge is power". However, ignorance is even stronger power. In a great majority of countries, the church is a powerful group very close to the ruling elite. The latter is supporting economic and technological development primarily to acquire military might. On the other hand, church, as just noted, is impeding this development. Such an ambivalence is of universal character and in extreme cases leads to civil wars of clerical dogmatics against intellectuals. Recurrences of such conflicts can be traced nowadays even in rather advanced societies (USA, some European countries, Latin America, Russia, India, Middle East, etc.). In Russia, for example, criminal penalty can be applied for publicly expressed atheistic ideas and criticism of the orthodox church. This is a strange example of a contemporary witch hunt supported by authorities in a presumably civilized country[268]. Moreover, the limitations for applying the obtained knowledge, e.g., in the form of introducing new technologies lies today not with technology per se but with its acceptance by people and administration despite the fact that new technologies can catalyze efficient approaches and business processes[269].

### 11.2.2    Archetypical Questions of Social  Sciences

One of the main questions that should be answered by sociology is: under what conditions the averaged, homogenized mass of individuals can be transformed into a coherent group of highly motivated people? This process bears some analogy to a phase transition: acquiring order out of chaos. The conversion of atomized, alienated individuals into structured, mobilized people is of crucial importance for political authorities, and it may become the worst nightmare to dictatorships. A far analogue is the picture of a saturated solution with an accidental center of crystallization. However, the methods typically employed by sociology mostly involve surveys in which the obtained results are represented by numerical data corresponding to the percentage of people sharing a certain opinion. This is, however, raw experimental material that might be used as an input in theoretical models of sociology. In such models one should derive differential equations, probably of stochastic character, describing the transition between states in a small time interval - a procedure well known in mathematical modeling of physical processes. Analysis of surveys, a standard technique of sociology, does not seem to be fully satisfactory since, e.g., by a strong desire one can correlate anything with everything. A connection of the sociological data with mathematical models would be highly desirable. When studying processes such as leadership that occur in "small groups"[270], relations between the members can be studied not in terms of transitions between states and corresponding differential equations, but using graph theory applied to all ordered pairs of group members. In this way, one can describe power hierarchies in human organizations taken as sets with dyadic relations between a comparatively small number of elements (see, e.g., lectures on mathematical sociology by P. Bonacich, http://www.sscnet.ucla.edu/soc/faculty/bonacich/textbook.htm).

---

[268] Some sociological studies in Russia indicate that had it been a vote with the participation of Stalin, he would have won. This effect of one-sided love of Russians to Stalin may appear paradoxical since Stalin exterminated more Russians than were killed during the WWII, but it can (and should) be sociologically explained. Contrariwise, people in Russia who are trying to struggle for civil rights and against the oppressive bureaucracy are immediately branded as loonies or even persecuted. Not only in Russia but in many other countries people were punished for adherence to the progress of the European type. In Germany, Hitler has been forcefully excluded from any opinion polls or official lists (such as the vote for the most outstanding Germans). Had he remained as a candidate, nobody knows what would have happened. One can easily find many such examples in modern history.

[269] The difficulties of human acceptance become obvious when one observes the struggle about genetically modified products and genetically produced technologies in general. Another example is the difficulties of the IT penetration in medicine. The physical methods of medical diagnostics and treatment such as laser medicine, nuclear medicine, medical imaging, etc. were initially rather hard to adopt, see, e.g., Roberts [250]; now they seem to have overcome the repulsion barrier of the old doctors' community and basically ensure the progress of medicine.

However, sociology, in contrast with social psychology, studies the behavior of human systems consisting of a very large number of members. For example, sociology must establish certain irreducible laws that pertain to large human groups having eternal conflicts of interests such as between bureaucracies and citizens or between medical doctors and patients. One can easily produce other examples of conflicting social pairs. In the final analysis, of course, the collective properties of human groups are due to individual attitudes, opinions, behavior and person-to-person interaction (communication) characterizing each single member of the group. There are models in which human society behaves as an averaged individual (in a similar way, plasma may be modeled by the motion of a typical particle moving in the electromagnetic field, see Chapter 5). This averaged individual can make sharp evolutions of behavior whereas in the society of collectivized individuals, behavioral patterns are smoothed. Yet sudden transitions are possible and have been observed in human history.

It is obvious that the solution of sociological problems involving many human participants and accounting for the behavior of each individual person is impossible. Analogous problems are impossible to solve even for structureless particles in physics, and for elements of human societies that can be found in a great number of personal (psychological) states determination of collective behavior from individual properties is clearly out of the question. One can only hope to describe the averaged overall characteristics, the crude macroscopic features of sociological systems. Moreover, due to a self-consistent situation - individuals create institutes that are forming individuals - the transitions between states of sociological systems may take many years which makes observations difficult: humans change slowly, in the course of several generations. This is also a limitation imposed on sociological research and social projections.

---

[270] This is the term most favored by sociologists, but I don't know how one can define a small group.

One more important issue of social sciences is the amplification effect. It appears to be a common observation in sociology that in initially homogeneous dwellings "good" and "bad" neighborhoods gradually appear. Likewise in economics, adjacent countries exhibit drastic differences in growth rates and industry output levels. In other words, large differences in the long run aggregate (or averaged) variables are observed for social or economic systems whereas initial conditions for these systems were almost identical. This phenomenon resembles the development of instabilities or amplification in dynamical systems, when small changes in the initial conditions are translated along the evolution trajectories and so amplified as to produce large differences in the output values. In social systems, small variations in the behavior of individuals can be transformed, due to the amplification effect, into great differences in the long run aggregate quantities such as the economic output or per capita GDP. This analogy prompts one to think that it would be pertinent to use powerful methods of dynamical systems theory to describe the effects of social and economic amplification and to establish its limits.

### 11.2.3   Limits and Errors in Social Sciences

One might note that the culture of assessing errors and discussing the accurateness of prognoses is nearly absent in social sciences. Typically, the statements in social sciences have an unconditional character whereas in the world of physics and mathematics an answer, which may be interpreted as a prediction, has the form IF ¡conditions¿ THEN ¡prediction¿ so that altering the conditions in general would change the predictions. Therefore, the universal principle of physical and mathematical sciences is that one should interpret forecasts in the context of conditions.

Not so in social sciences. Let us first take as an example one of the most famous socio-economic accounts, "Das Kapital" by Karl Marx. The author, a poor German emigrant living in London, produced a bestseller (the first edition in London, 1867) which probably no one has read completely. The manuscript by Karl Marx consists of four thick volumes, and it is unlikely that the author himself was familiar with the last three: they were compiled from loose drafts by the author's friends and colleagues, Friedrich Engels and Karl Kautsky. Marx appeared to be so certain in the infallibility of his theory that it is hard to find in the "Capital" any discussion of the accuracy of his statements and prophesies. At least, by scrolling "Das Kapital", I could not find any. Marx's theoretical schemes are in fact somewhat abstract models with a very modest mathematical element and hardly any domain of applicability. In this sense, Marx's claims are closer to religious models than to economic theories. Indeed, we have seen (see section "Religious Models" in Chapter 2) that most religions are based on promises, and most believers find their deeds meaningful only to the extent that something abstractly pleasant can be expected. The same applies to the Marxist abstraction of communism. Marxism in general has plenty of religious attributes: it contains very little empirical evidence (if at all), mostly relying on unconditional faith. It is curious that Marxism rivaled religion mostly in the countries with very strong religious adherence such as Albania, Afghanistan, China, Cuba, India, Laos, Vietnam. Probably it means that a poor population is more susceptible to uncritical acceptance of vague ideological doctrines than the population in comparatively wealthy countries. It is also astounding how deeply people were indoctrinated with the purely ideological, i.e., without empirical element, Marxist schemes. Marx only tried to produce a highly theoretical, scientific-looking study of the capitalist way of production albeit with plenty of lateral associations. Common people, however, interpreted these speculative theories rather emotionally, as an appeal to the violent transformation of society. Many people were enslaved by the Marxist ideology to the extent of being ready to sacrifice their freedom and lives, closing their eyes to the obvious facts that the properties of socialist or communist states were standing in sharp contrast with Marxist doctrines. The main thesis of Marx is actually formulated at the very beginning: the wealth of societies which are based on the capitalist way of production is manifested by a great accumulation of goods. Then the author devotes about 800 pages of the first volume, containing a lot of numbers and references, to a meticulous investigation of the conversion of goods into money (and vice versa), creation of a surplus value, and accumulation of the capital based on it. According to Marx, it is the capital that is the real ruler of the society, with the capitalist society ensuring a maximal capitalization and monopolization of the economy. Marxism optimistically claims a such a greed-based process of capitalization and monopolization should eventually result in the social explosion. So, according to the frustrated author, capitalism is doomed.

It is because of this optimistic prognosis that all the people in the communist countries were obliged to study Marxism-Leninism. In the Soviet Union, studying "Capital" at each university or institute, was obligatory, irrespective of the faculty. There was at least a year's course of the so-called "political economy" formally devoted to the study of Marx's monumental treatise, but nobody actually read it beyond the first chapters including ignorant professors of the "Marxist-Leninist political economy". We, students, managed to pass the exams without reading neither the "Capital" nor its abridged exposition specially adapted to the presumed Soviet mentality. It was for us a kind of a sport: who gets better marks without any knowledge? It was not so difficult to swindle the narcoleptic teachers of Marxism-Leninism by a meaningless flux of words because they had no knowledge of the obscure Marxist texts, either. And of course, Soviet functionaries, the absolute majority of them being half-literate, had never read "Das Kapital", but they had to enforce it. I guess that such communist rulers as Stalin, Mao Zedong, Kim Il Sung, Fidel Castro and others had neither time nor desire to study the monumental theory by Karl Marx. Their purpose was more pragmatic: to make their subjects believe that the sealed up communist system was a "progressive" paradise as compared to the backward and inhuman capitalist hell.

The socio-economic model of Marx still seems to be inadequate, despite all its popularity, which is also similar to religious models. Indeed, the main prognosis of the imminent social explosion in all developed countries, put forward by Marx and his interpreter Engels, was never corroborated. Reality manifested itself exactly opposite to the Marxist predictions. The representation of the working class as the most progressive one is ridiculous. Besides, the workers have never been active and eager enough to ignite the world revolution, as it was proclaimed by the communist ideologists. Equally inadequate was the model of socialism as a society that would abolish the state and establish a genuine paradise on the Earth.

## 11.3  Hierarchical Multilevel Systems

Although the concept of hierarchical multilevel systems (HMS) is very broad - from software and data communication systems through Earth's climate to social hierarchies here, we shall mainly talk about economics. There are considerable mathematical challenges in describing HMS, especially as concerns their multiscale modeling.

Bridging across many levels corresponding, e.g., to subcomponents that operate on essentially different space and time scales requires some unifying mathematics and, if treated as a head-on computer simulation, is computationally demanding. The physical example of a multiscale problem is turbulence (Chapter 7), it gives us an opportunity to feel the complexity resulting from many interworking scales. Despite enormous practical importance, apparently transparent mathematical setup (the Navier-Stokes equations), and a great lot of efforts, nobody has managed to build a good mathematical theory of turbulence. Economics, being an essentially multilevel structure, is probably an even more complex system than fluid. This is an example of multigrid methods - a computational mathematics counterpart of multilevel systems, which represents, however, only a primitive reflection of the entire complexity of hierarchical multilevel systems[271].

### 11.3.1    The Politics of  Bureaucracy

If the state begins to redistribute the created wealth too actively, a justified resentment of those wealth producers who are the most efficient and creative is aroused.

Bureaucratization in corporations leads to underproduction or inadequate production [272] crises. This is a scaled-down effect of macroeconomic drawbacks that constantly plagued, e.g., the Soviet-type (Gosplan) economies and eventually led to their complete downturn. One can even make a pessimistic prognosis related to the Great Bureaucratic Revolution occurring almost everywhere in the world if the situation with rapid bureaucratization[273] is not reversed and bureaucratic positions are more and more attractive for young people, then in a dozen of years a new generation of "intellectuals" will arise who will not be intellectuals at all. These people will probably be different, not ready for intellectual work in the sense of, say, the 1960s i.e., not capable of reflection and simply to reading serious books. This can be bad for all.

---

[271] In fact, already the Fourier expansion exploits a hierarchical principle, and it is due to the hierarchy of scales that the Fourier method is so powerful (in linear problems). One can also bring the example of hierarchical Arabic numerals vs. the inefficient Roman number system.

[272] Manufacturing of products with no regard to the market signals.

Now that more and more people wish their children to become highly paid functionaries, bosses, top managers, lawyers, bankers, TV or movie actors, prominent sportsmen or other kind of celebrities, interest in hard education appears to be waning. The lowering prestige of exact sciences and engineering is indicative of the growing tendency to bureaucratization. One can observe that the more primitive the culture the less useful science and engineering are perceived to be[274].

The prestige of science and engineering in the popular consciousness might serve as an indicator of the society development stage. There is, however, an intrinsic contradiction here: people who care less and less about science more and more want new "cool" technologies. It is this "coolness factor" that brings significant profits to capitalist corporations and compels them to somewhat grudgingly support science together with the classical future-oriented chain of development: idea, calculation, laboratory, prototype, pilot project, full scale production. Yet such chains are typically controlled by the corporate bureaucracy which, with the multilevel hierarchical authority structure of modern corporations, is practically indistinguishable from the governmental sector bureaucracy.

One detrimental consequence of corporate bureaucratization is the system of semi-corrupt verticals: managers at the lower levels make absurd and arbitrary decisions while higher levels tend to protect their subordinates from any blame. This buddying mode is rather stable, eventually leading to the total absence of persons who are ready to take responsibility for inadequate decisions. But although one can hide from the facts, one cannot hide the facts. Of course, rank-and-file employees, common workers, and the "office plankton" are those who suffer, primarily from unemployment. One should not think, however, that unemployment is only caused by the bureaucratic mismanagement: even a perfectly functioning market (i.e., non-centralized) economy is prone to unemployment because of insufficient demand and its fluctuations. According to the canonical economic theory, which is a deterministic model operating with averaged quantities, market equilibrium is reached when demand equals supply - in fact this is the definition of market equilibrium. Applying this definition to the labor market, one can deduce that the demand for goods and services pushes upward also the demand for labor, thus resulting in rising pay and employment. However, the supply-demand equilibrium curve is not necessarily a smooth trajectory: from time to time, it can change very abruptly, following social cataclysms, governmental moves, appearance of new technologies, and so on. We have already discussed that disruptive technologies favor certain groups of population and can seriously inhibit other groups. For example, closing coal mines in order to foster sustainable energy sources, even if this process is accompanied by social programs and retraining activities, seriously diminishes job opportunities for miners. In general, new production paradigms have strong effects on wages and employment, these effects deforming the socio-economic position of some groups relative to others. There exists a considerable amount of highly professional literature devoted to this subject (see, e.g., [277] and references therein) so that I do not need to dwell on this subject, about which I know very little. There are also mathematical models of unemployment in multilevel economies (i.e., HMS) resulting from technological development, but we shall not discuss these models here.

---

[273] The rate of creeping bureaucratization, e.g., of Russia is striking: even according to official statistics the number of government officials has grown from 900 000 in 2000 to 2 000 000 in 2008.

[274] In the post-Soviet Russia, scientists were perceived in popular consciousness as good-for-nothing exotic creatures; people looked at "egg-heads" with a mixture of disdain, superiority and fear.

Bureaucracy is generally opposed to meritocracy because the latter possesses some specialized technical knowledge or critical information, which makes it difficult to control.

## 11.4  Physical Economy

Sometimes I think that the world would function better if it were run by physicists and mathematicians, despite the fact that politicians and business people usually claim that decisions are much better guided not by physics or mathematics but by gut feeling derived from years of experience. And they are supported by a largely math-phobic general public. Besides, the vast and omnipresent bureaucracy - the driving belts of politics - rightly fears that analytically driven programs might result in massive personal replacements. We have seen that physics builds models of reality, but, firstly, it may be not necessarily physical reality without participation of human agents and, secondly, not only physicists construct models of reality à la theoretical physics. For instance, the use of dynamical or stochastic (or dynamic stochastic) models in socioeconomic research is now a substantial part of mathematical economy. Economic change, growth processes, goal setting, development time, driving forces, and large number of other topics are studied in the mathematical form by techniques traditionally used in physics, e.g., by variational methods. Mathematical economy often looks like physics in some other guise. Therefore, one currently calls this interdisciplinary field physical economy.

This discipline may be especially interesting to those who wish to see how the ideas from a social sphere can be translated into mathematical models. Paying attention to this translation is useful for all parties concerned - physicists, mathematicians, economy experts, social scientists, for it helps to enlarge the scope of problems to be objectively studied and possible mathematical techniques to be used. Since I am not an expert on mathematical economy, many topics are treated on a primitive level of general concepts. Such important subjects as stochastic modeling, probabilistic analysis, structural change or socioeconomic discontinuity are treated in a number of more specialized sources, so don't expect any profound exposition of these topics here. As everywhere in this book, the main attention is paid to the motivation and relevance of physically-based mathematical modeling.

Hard laws analogous to those describing the processes in physics do not exist in economics because the latter is necessarily mediated by social phenomena. Social interactions and human behavior are exceedingly more complex than physical phenomena and remain rather poorly understood. Therefore, economists rely more on insight and expert judgments rather than on objective methods ensuring the confidence close to physical modeling.

Economy is always an evolving system [168] – actually, any physical system is an evolving system since a physical event occurs in space-time. Yet, in physics one can in many cases use quasi-stationary or even steady-state approximation. We have seen, for instance, that steady-state models such as computation of energy levels play an outstanding part in quantum mechanics, in fact quantum mechanics was initially invented and designed to produce stationary states. Now we may call this subdiscipline that does not include interstate transitions quantum statics. Contrariwise, the steady-state approximation is rarely adequate in economical models because of rapidly changing conditions. In other words, economic models are predominantly dynamical, dealing with evolution. Generally speaking, evolution is any process of development, change, or growth.[275] To describe some evolving system, one may construct at first a minimal model which is aimed to elucidate the basic features of developing phenomenon. Such a minimal model corresponds to the simplified dynamical system describing economic development - with a minimum number of phenomenological parameters or arbitrary constants. More detailed models can be built on the base of these minimal models[276].

There are a lot of questions regarding evolutionary economies, posed but not answered. For example, should there be a convergence of economic patterns towards a single, universal and ideal model? Does this common limit exist for all societies, or evolution of specific economies is heterogeneous, with the economic trajectories being dispersed according to each country's preferences? The latter assertion is known as the "Dahrendorf's hypothesis", stated by a well-known German-English political economist, Sir Ralf Dahrendorf.

Economics is an essentially quantitative subject. When, for example, a minister for economy asks his advisers about the advantages or disadvantages of raising the actual taxes, she/he does not expect a philosophical discourse, nor a lecture on what Nobel Prize winners in economy would generally say about increased taxes. The minister (or the secretary for economy) wants to know how much revenue could be produced  due to new taxes and how the anticipated revenue figures would correspond to the optimistic prognoses supplied by the finance minister. The secretary for economy must know quantitatively how the increase of any tax would affect various sectors of economy. Thus, economics seems to be closer to engineering than to science, so one uses a term "economical engineering". Indeed, one must define goal or cost functions and from them deduce social and economic behavior - a typical engineering approach. For technological systems, to adjust parameters to the specified goal function is common in traditional engineering. However, this works well for technological processes due to the fact that they obey the well understood laws of natural sciences. In economics and other social sciences, the governing laws are poorly understood or at least do not exist in compact mathematical form (expressed as formulas). Insufficient knowledge of the mathematical structure for social and economic laws makes it hard to use a genuine engineering, when free parameters can be fine-tuned with a good accuracy. In particular, this leads to a well-known situation when the results obtained within the framework of different economic models have an undetermined applicability domain and may considerably overlap.

---

[275] The word "evolution" stems from the Latin *evolutio* which means "unfolding".

[276] Such more detailed models have often been called "imitation models", although it seems that this terminology is outdated now.

In contrast with physics, quantifying things in economics is in general a nontrivial problem. How would you unequivocally quantify customer satisfaction, for example? But you ought to do it, in order to plot this variable versus costs. Moreover, this intuitive entity may determine economic strategy, for instance in medical institutions, transportation or public conveyance companies: do you want to minimize personnel (expendables, fuel, etc.) costs, or do you want to maximize patient (customer) satisfaction and thus attract more clients?

There is a peculiar feature of real economics differing it from natural sciences and technological engineering. In economics, perception of reality is a major part of reality. For example, when people are constantly told that everything goes well with economy, they are inclined to believe it, start making investments, buy stocks, purchase more goods, and the whole economic system is accelerating or "heating". It is thus the vision of economic situation, i.e., reality dressed in human perception, and not the naked situation that determines the economic behavior. More than that: perception of reality may prove to be more important and contributing heavier to economic behavior than reality itself. As a consequence, it seems very difficult to model the behavioral components in theoretical economics. One of the biggest challenges is to understand how the system will change its path as humans react to stimuli, incentives and dangers. For example, one of the most important economic factors in the contemporary world is the oil prices. The cost of oil production (even including gradually rising exploration costs of oil-carrying sources) is known to be a slowly varying quantity on a level of 10 US dollars per barrel. On the contrary, the oil prices are varying comparatively fast, on a level about ten times higher. It means that human-induced, speculative and "soft" component is an order of magnitude larger than technical production costs. Moreover, even the speculative variations of oil prices following the human perception of the political and economic situation may substantially exceed the hard economic parameters such as the production costs. This example hints at an essential, sometimes dominant role played by human perception in economy. Another example points at the difficulties to account for the human role in economy. Thus, a number of economies are plagued by corruption, and although it is intuitively very likely that corruption hinders economic development in such countries. However, when one tries to pass from vague intuitive statements to quantitative models, the braking effect of corruption becomes hard to describe in mathematical terms.

## 11.5  Naive Taxation Models

This section may be regarded as an occasional curiosity. The subject of taxes is, however, very important for all people, and one cannot get rid of the impression that it has not been studied with commensurate mathematical thoroughness. The tax system is purposed to maximize the budget revenue under the following constraints: non-degradation of the quality of life; non-diminishing income level, demand, productivity and profitability; preservation of social stability. Taxation fulfills one more important function: stabilization of inflation. Optimal taxation of private persons and households depends on the character of two important distribution functions characterizing economic structure of the society: distribution $f(x)$ with respect to income $x$ and distribution $g(u)$ with respect to accumulated liquid assets or liquidity $u$. As usual, the quantity $dN = f(x)dx$ denotes the number of households with income lying in $(x, x + dx)$, and the quantity $dN = g(u)du$ signifies the number of households whose liquidity is found in $(u, u + du)$ . Here $N$ denotes a continuum of households. The $f(x)$ distribution is more important in societies with transitory economies whereas the quantity $g(u)$ characterizes the wealth distribution. Examples of quantity $u$ are bank deposits, stocks, bonds, expressed in currency units, in short, $u$ denotes the mass of money to be at the household disposal. Optimal (according to some criteria) taxation may be modified with the changes in the society economic structure. Corporate taxes directly affect profits and productivity. It may be noticed that corporate taxes are coupled with income taxes. A well-organized tax system allows one to determine the society economic structure reflected in the distribution functions $f(x)$ and $g(u)$. For example, one can measure $f(x)$ directly from income taxes and $g(u)$ by analyzing bank accounts[277].

   Mathematical modeling of taxation schemes does not seem especially difficult, yet defensive financial bureaucracies are apparently not very interested in modeling results. There are only continuous verbal discussions in the mass media, but serious scientific results are only seldom presented and popularized (see [172] in relation to this). The subject of taxes is highly emotionalized and charged with political or group interests. It is by no means accidental that finance ministers are most often without any financial or - it would be almost improbable - financial-mathematical background; they are customarily recruited from winning party inner circles, even in advanced countries. This anti-meritocratic selection favors confused and not transparent tax systems having deplorable economic and legal consequences. Tax laws, governmental acts, statutory regulations and actual rules come in great many formats so that it is usually impossible to know them all. Only this fact tends to criminalize any citizen due to her/his unawareness of the whole mass of current tax rules. The internet does not help much in this respect - what matters most is not the medium, but whether the source is up to date. It is usually not difficult to calculate taxes, for example, by using a look-up table for progressive scales, yet the definition of what is subject to taxation taking into account all possible exemptions, deductions, special regulations accumulated over the years is very burdensome.

---

[277] Both bank accounts and accumulated stock may be of course analyzed impersonally, without violating privacy requirements. Secretive character of individual wealth is an important prerequisite of social stability.

Thus, taxation rules in most countries are extremely complex, with governing financial bureaucracies strongly resisting any possible simplification. Changing the taxation schemes, introduction of new taxes or removal of old ones is always extremely controversial. Some tax authorities (e.g., in Germany where taxation is considered to be the most complex in the world, see, e.g., http://en.wikipedia.org/wiki/Taxation in Germany) even contend that the tax system does not need to be simple - sort of an esoteric approach. Germany is an example of the country that traditionally strives to be one of the leading (or even the leading) in the world in many respects - economically, politically, financially, technologically, in exports, etc. However, these ambitions are drastically impeded by the country's clumsy and highly controversial tax system. As a result, money is flowing out of the country. Punitive reflexes of the state bureaucracy activating the "immune system" - police, intelligence, excessive customs control - are costly, inefficient and only make the matter worse. Money gained inside the country is easily percolating through restrictive barriers. Small but numerous amendments issued each year do not help since they only increase the complexity and benefit mainly the parasite clan of tax consultants flourishing due to unobservable legislation. On the surface, this situation is favorable to the state since tax consultants are also paying taxes on income generated from the fee paid by taxpayers. This is, however, a superfluous activity taking many man-years of people that can be otherwise intellectually or manually productive. Germany of course is no exception in highly inefficient and overcomplicated taxation. Many countries cannot come to terms with their tax systems. All the countries where the tax system is a product of political compromises rather than based on solid financial and mathematical approaches are doomed to either money loss or political instability due to people's discontent, or both. It is indicative that tax systems in various countries are all different and are constantly modified - this testifies to the fact that they are far from being optimized. The tendency to make taxation complicated for citizens is quite understandable since under a simplified system (such as a universal flat rate, see below) the ruling bureaucracy loses some particular form of control based on selective granting tax preferences. So, there are groups besides ruling bureaucracy that are opposed to tax simplification, which results in lobbying against efficient mathematical modeling in this "politically sensitive" area. The usual argument against simplified systems based on clear mathematical models is based on the notion of "fairness": a tax scheme is declared "unfair" if it is mathematically rigid and does not allow to redistribute money according to political purposes.

However, "fairness" is a subjective notion: how do you define "fair"? Probably any definition of it would be contaminated with emotions, and this complicates the taxation schemes. But mathematical models are immune to emotions, and no matter what mathematical model one considers in regard to the taxation problem one has to focus primarily on its simplification - in accordance with the common mathematical modeling methodology (Chapter 2). To begin with, let us for simplicity consider a three-component taxation scheme when taxes collected by the state[278] are subdivided into three non-intersecting components: individual (household) income tax, corporate profit tax, and consumption tax (VAT, GST and the like)[279]. I shall not dwell here on corporate taxes, it requires a thorough economical analysis bordering on legal issues which subject only includes mathematical models to a minor extent and would lead us far away from them. As regards taxes on consumption, not on labor and capital income, this is a serious and interesting issue where mathematical models can be correctly set up, especially in the context of shifting the tax burden from individual income to penalized consumption. Thus, taxes on consumption inherently encourage savings and, consequently, investments instead of consumption. Income tax is a direct one whereas consumption taxes are hidden in prices and felt indirectly. One may notice in this connection that although VAT and its analogs are drastic instruments of improving the state finances, they can bring more harm than benefits for the state since they can significantly reduce consumption.[280] It is because of this hazard that mathematical models would be especially useful in optimizing the relationship between the two components - income and consumption taxes.

The most important direct tax is the income tax. There are three main mathematical models for the income tax - progressive, proportional and regressive. Many people got used to proportional taxation schemes - due to "fairness" or envy arguments, although nobody has proved that such schemes are the best for the people. By the way, what is best for the people is not necessarily good for the state and vice versa, even in democratic countries with advanced economies. People - the country population as a whole and its various groups - and the state represented by the ruling bureaucracy may have, and actually possess, different goal functions whose difference[281] as a function of time increases when the economic situation becomes worse. In most advanced economies, the personal (household) income tax is a progressive tax defined as a rising, piecewise continuous (affine) function of income, excluding various deductions, rebates, etc. In other words, the tax base, i.e., the amount of money earned up to a certain threshold 1 amount $T1$ is taxed at a rate $r1$, then the remaining income, up to the second threshold amount $T2$ is taxed at a rate $r2$, and so on. Conversely, regressive tax is levied so that the tax rate decreases as the tax base increases. This is a rare occasion for income taxes, being applied in developing economies to stimulate accumulation of wealth. The in-between model of proportional or flat rate is the most primitive taxation model based on the universal rate $r$ irrespective of the income earned. In this sense, a flat rate may be considered a degenerate case of a progressive or regressive rate. For some political reasons, mainly of ideological character, proportional (flat) tax schemes are uncommon in advanced economies, especially in Europe, where typically a graduated progressive tax imposed both on household incomes and fixed taxes on corporate profits are accepted. However, flat taxes seem to work well, e.g., in Russia and Baltic countries; some more countries of Central and Eastern Europe are contemplating introducing the flat income tax. Medical and retirement insurances perceived by people as supplementary taxes do in fact correspond to flat rate taxation. There was a proposal by a well-known tax expert and lawyer Professor Paul Kirchhof to introduce 25 per cent flat rate in Germany, but it was of course declined for political reasons.

---

[278] Typically, for "old" European countries tax revenue comprises about 40-50 per cent of GDP, with approximately 25 percent in the USA, http://ec.europa.eu/taxation customs/taxation/index en.htm

[279] Actually, the great variety of other taxes imposed on the citizens provide only corrections to the state tax revenue and rather testify to inefficiency and greediness of the financial bureaucracies who see their task as squeezing the money out of citizens.

[280] Moreover, the VAT paperwork is extremely cumbersome, so it complicates the accounting to the extent that a lot of man-hours are wasted. It is strange that people underestimate the fact that time is the most precious resource.

[281] One can talk of the norm in some functional space of course.

The limiting case of the proportional tax model is confiscation of high incomes and/or property - accumulated wealth. This variant is, however, not optimal since it produces an excessive social tension, leads to the drain of capital out of the country and removes economic stimulators. This is an anti-liberal approach exploiting human envy and almost necessarily resulting in violence. Furthermore, the ruling bureaucracy that appropriates the functions of total redistribution rapidly (in some historically short time $\tau$) becomes the elite group ("nomenclatura") and behaves as if the entire country belonged to it. This accompanies the transition of society to another stationary or at least metastable state which is usually called totalitarian. The only possibility to resort to a confiscatory taxation during some acute crisis is to apply it for even shorter time $t < \tau$.

The flat tax model is based on one parameter - universal tax rate. It is the simplest model whatsoever. Let us consider now another simple model of taxation, but based on two parameters: the lower threshold $B$ (from German "Betrag"), analogous to a tax exempt level (Steuerfreibetrag in Germany) and the flat tax rate $r$ applied to the positive difference between the income $x$ and level $B$. If $x - B < 0$, i.e., the household income lies under $B$, the state compensates the difference $\Delta = |x - B| = B - x$. If $x - B > 0$, the income tax $t = r(x - B)$ is applied to the excess $x - B$. the meaning of $B$ is in fact the existence minimum which can be calculated or estimated if the economic structure of the society is known, for example, as $B = k \int_0^\infty x f(x) dx$. Here we assume that the first moment of the distribution exists, function $f(x)$ is normalized and $k < 1$ is some coefficient depending as a functional on $g(u)$.

However, this calculation is too theoretical. In practice, the existence minimum $B$ is determined by actual physiological needs such as food, clothing, dwelling place, heating, mobility rather than by the aggregated "human factor" - choice, assortment, fashion, prestige, advertising, etc. Parameters $r$ and $B$ can be adjusted from time to time, depending on the economic situation. No other payments to the country's budget are necessary: all social security contributions, state medical and unemployment insurance[282], payments to pension funds, perceived by the citizens just as flat rate taxes, can be "absorbed" by the two parameters. In principle, almost any contemporary tax system has a character $t = F(x - B)$, where the function $F$ has some sophisticated piecewise behavior, and $B$ may take an ordered number of values, $B_1, B_2, \ldots$ One common model is a three-step system with $B_1, B_2, B_3$. Our linear model may be universal for all the taxpayers (no exceptions and supplementary exemptions), which probably would irritate the politicians. If, in certain crisis periods or when an acute budget deficit is pending, two parameters are insufficient, one can, for example, use instead of linear function a polynomial or the family of functions $F(x) = r(x - B)^{\alpha}$, with $\alpha$ being the third tuning parameter. One must, however, remember that in this case the tax burden is displaced to higher income and the most productive portions of the population[283], which inevitably reduces incentives to work harder.

Like any other model, the above linear (affine) tax model has advantages and drawbacks. Being extremely simple, it would drastically reduce the costs of tax return processing and would save a lot of time. Moreover, it preserves the existence minimum for each citizen. At the same time, the model can implicitly stimulate poor households to increase their income. To reinforce this effect, one may slightly correct the model introducing one more parameter - the "enthusiasm coefficient" $0 < \beta(t) < 1$ applied as a renormalizing factor to $\Delta : \Delta \to \beta(t)\Delta$, with e.g., $\beta(t) = 1 - \mu t, \mu > 0$. One more simplification is hidden in the fact that using such a model one can neglect a separate retirement payment and provide the financing of pensions through income tax. Everyone should receive her/his existence minimum irrespective of age and health condition. Those who wish and are capable of using supplementary private pension schemes receive more - merely additive. But they have also to pay taxes. No privileged or elite groups are allowed, everyone pays the same rate and no one receives special treatment or has tax preferences. So simple may be the world.

---

[282] As regards health systems, there are three basic models: state-controlled medical insurance, private insurance, and health services paid by the state. There may be of course a superposition of these "pure" models. State-controlled medical insurances do not actually differ from budget organizations. Nothing changes for the patient if one covers her/his medical expenses from the state budget. Naturally these payments should be executed directly to medical service providers (doctors, healthcare personnel, etc.), without the patient's participation. Public-private partnership schemes may also be closer to optimal than state-controlled medical insurances.

[283] I try to evade using the term "population" recalling an involuntary definition of this term which belonged to the Russian ex-Prime Minister, V. S. Chernomyrdin: Population is such a people with whom one can do whatever one wants. V. S. Chernomyrdin was known for his intuitive, unintentional aphoristicity.

This simplicity may be considered exactly the drawback of the model. The matter is that there are some groups in society that perceive the simplicity and transparency of financial flows as directed against their interests. To these groups may belong not only misanthropic bureaucrats, but also social romanticists and leftist world-improvers. Fewer regulations mean less interference from the state part exercised by functionaries. Naturally, the latter are not interested in tax simplification. Moreover, under a simplified tax system the costs of processing tax returns paid to tax-collection officers become drastically reduced, and the tax collection offices can be significantly downsized. And what to do with the freed manpower? Moreover, the very fact that the flat rate can be applied to all taxable income (and profits) without exception or exemption exasperates many people.

There is also a risk that an increased part of an inactive population would subsist for a long time on state transfers ensuring the existence level. The above mentioned "enthusiasm coefficient" is intended to entice them into a productive life. Another danger could be that employers would be tempted to pay low wages in order to minimize payments to the state, but this should be regulated by corporate taxes, in particular by an increased taxation of profits. In this respect, personal income taxes become even more strongly coupled to the corporate ones.

This two-parameter flat-tax model reminds us of the so-called negative income tax (NIT) model suggested by one of the leading contemporary economists, Milton Friedman, described in his well-known book "Capitalism and Freedom" [169].

Currencies have a notorious tendency to plunge or soar in value because of the herd mentality of markets, so it is crucial to manage currency adjustments in an orderly fashion. As long as the decline of the US dollar is smooth, there is no problem. But if the dollar plummets, the world faces a full-blown currency crisis.

# 12 Conclusion and outlook

Now, one can justifiably ask: has anything been said at all? Some people might say that the book is so uncommon and eclectic that it would be difficult to trace the logic without being distracted by surrounding complexities. I still hope that the described - maybe not always accurately - results and problems would deserve being discussed anyway. And I also hope that even in those numerous cases when I just mentioned some results and the great names associated to them this content might be useful for the reader as an initial piece of information about what has happened in the world.

One may observe that physics done on the level of journal articles is different from physics presented in textbooks or treated in monographs. People writing monographs or textbooks are trying to make the presentation monolithic and the presented subject closed. In reality, no subject is closed, and scientific disciplines, in particular physics, are in perpetual development, conquering new areas and incorporating new results - new nodes, in networking terminology. The networking approach, increasingly popular these days, allows one to migrate between the physical and mathematical (in my terminology, "physmatical") subjects with a comparative easiness. Like any network, a physmatical one is not closed, it develops and gradually interconnects with other "proprietary" networks such as "biomedics", economics, astrophysics, "sociologics", etc. To some extent, "physmatics" reflects the scope of books one can see on the bookshelves of physicists. In this book, I wanted to pay attention to synergetic interaction between different fields of physics, mathematics, and other disciplines where mathematical models, in particular those constructed according to a physical pattern, are extensively used. I also attempted to bridge together two streams of presenting physical and mathematical results: a conservative textbook or monograph approach and a more daring article style. Bridging the gap between these two manners of presentation requires a relatively complete exposition of the basic concepts, therefore there are so many fragments in the text, which a reader may find too elementary. In other cases, only a glimpse at the non-core stuff was, to my mind, sufficient to convey the main ideas underlying the model discussed, but some readers may find such exposition too superficial. It is difficult to satisfy everyone, and I never wanted it. I did not present only the standard material, but also tried to comment on some open problems so that the interested reader may be induced to dig deeply in more professional sources.

While preparing this manuscript my ambition was that the book would stimulate young researchers to venture into interdisciplinary fields, which are off the beaten track. The cross-fertilization process between such interdisciplinary fields is essentially reduced to the triad: acquire everything what is useful, neglect all unnecessary and add something of your own. As I have several times mentioned in this book, the term "interdisciplinary" has been seriously compromised due to heavy contamination of vague philosophy.

Bearing the latter consideration in mind, I still wanted to make a journey over a variety of subjects rather than to settle down on any one of them. This explains to some extent wordy and philosophical interludes of mine which can irritate the reader anticipating a more focused approach. And I must once again apologize for highly subjective personal recollections interspersed through the manuscript: I hoped that occasional remembrances, however dim they might be, could partly illuminate the scientific scene of the Soviet years, the paradoxical period when physical and mathematical sciences could flourish against the ruinous background of nearly everything.

Physicists and mathematicians are typically those who strive to "crystallize" the problem and to achieve maximal lucidity before converting their considerations into published papers. I concede that there are many places in this book that would not scrape through the clarity test; moreover, there are fragments that may seem "a mist" to an occasional reader. Broken and not clearly cut presentation of material is untypical of scientific publications, but, firstly, this book is not, strictly speaking, a scientific publication as it does not fit in with any standard for a monograph or a textbook and, secondly, I think this mosaic, non-monolithic style is more natural for a human perception. A rare reader nowadays meticulously studies the book from the first page to the last in the same methodical order as it was arranged by the author, editor and publisher.

When writing this book, I did not think it would be fitting to ruminate about the tiniest wrinkles in physical models, for example on the level of details that were discussed last month in arXiv. Instead, I took the manner to jump over some details which I considered not necessarily trifling but distracting, speaking mostly about "the forest behind the trees", i.e., the most important models in physics. I have noticed that many physical questions may be fundamental but vague, which marks a certain philosophical component in physics. Personally, I am inclined to doubt that philosophy could answer physical questions, yet I think that vague questions can provoke imagination and diminish the dogmatic aplomb so typical of many science professionals.

I recall a statement ascribed to G. H. Hardy and cited by an outstanding physicist, mathematician and writer Freeman Dyson [284] that young men should prove theorems, old men should write books [231]. Proving theorems and obtaining other hard scientific results projected into the future is indeed a game for the young. The ratio of a scientific to a philosophical component in a person's work rapidly diminishes with age after a certain threshold simply because young people are capable of more trial-and-error attempts. As I may be considered old, my motivation, while writing this book, was primarily to preserve the past. Books are in general intended to preserve the past. Unfortunately, I have seen many times how the worlds of the past disappear together with their human carriers. And I shall be happy if some person considerably younger than the author would refer to this book that it has something interesting to say. Even if the reference were in the following well-known form: there are some new and true things in this book, but the true things are not new, and the new ones are not true.

---

[284] It is a shame, by the way, that F. Dyson did not receive the Nobel Prize. There are many outstanding scientists who were overlooked by the Nobel Committee (more exactly, by the Royal Swedish Academy of Sciences): G. Gamow, J. W. Gibbs, L. Meitner, D. Mendeleev, nowadays F. Dyson and L. V. Keldysh, perhaps Yu. M. Kagan. The Nobel Committee often makes strange - often politically motivated - decisions. See, e.g., http://en.wikipedia.org/wiki/Nobel_Prize_controversies.

Back to this issue of preserving the past. In this book, I wanted to discern a timeline of 20th century physics and associated it mathematics which may serve as a backbone for the entire physmatical network. Of course, it was only a crude attempt. However, I do not think such an approach is hopeless. Networking reasoning is a good instrument to unify diverse environments. Modern physics began with Einstein's attempt to reconcile electrodynamics, mechanics, and thermodynamics in 1905 and his later endeavor to unify special relativity and the Newtonian theory of gravitation. In a more general social context, the synthesizing approach of Einstein meant a convergent scientific change - a retreat from austerity concepts of Max Weber [9] who insisted on "Zweckrational" action rejecting all unnecessary circumstances. Einstein was rather thinking in the Renaissance spirit of more lenient "communicative rationality" program, encouraging critical reassessment of "everybody knows that" concepts and the establishment of mutual understanding between diverse scientific communities. As a consequence, Einstein's program proved to be very successful as compared to rival endeavors not because it could explain more "facts" or was more powerful mathematically. It was better than rival programs probably because it provided a wide basis for interpenetration and communication between several main paradigms of 19th century physics. And in the long run, of course, Einstein's theory culminated in multiple empirical successes.

What is not covered in the book?

Since the manuscript is already too long, I have omitted many important subjects such as, among applied mathematical topics, data analysis, the finite element method, homotopy, generalized functions, projective spaces, etc., and within the physical portion, anything about astrophysics, bosonization, Fermi and Luttinger liquids, models of a ferromagnetic, quantum Hall effect, surface science and low-dimensional physics, elementary particle physics, quark models, spin systems (with spintronics applications), and many, many others. All these subjects are treated by a vast number of highly competent authors, and I, having only a superficial - not working - knowledge of them could provide a mere compilation of authoritative sources. I was not bold enough to exhibit my dilettantism, for the latter is always conspicuous when a person ventures to write about the subjects she/he does not have a working knowledge of. Besides, the number of sources covering the omitted topics is enormous. Take, for example, the finite element method. It has been the subject of so many books that it would be hard to select any pertinent references.

For me, it was a serious challenge, while putting this book together, to decide what to include: where exactly lie the boundaries of the so-called "physmatics"? Or, in practical terms, what should be considered the standard repertoire of the indispensable education of a physicist, and what can be painlessly omitted? Can one find a common scheme to incorporate dozens of seemingly different research activities? Probably, only a science historian looking backwards in, say, thirty years from now can conclude whether the networking view of physical-mathematical disciplines was at all justifiable.

In connection to a tremendous amount of topics and sources, I hope that in the near future people can utilize computers to understand the relational structures of physmatics, by constructing clever algorithms classifying its inherent content. One can, for example, invent multilevel or graph clustering algorithms which would predict physmatical complexes in large-scale dedicated information networks. In this direction, visual representations of complex multivariate information contained in physical and mathematical sciences may become feasible, which allows one to communicate this information synthetically (and not as traditional dispersed fragments) with ease and high precision. The designed physmatical algorithms might produce groups of high-quality nodes - nowadays this function is partly implemented by forums devoted to specific topics. Eventually such nodes can be merged before applying clustering algorithms. Then the clustering results may be passed to the entire physmatical network, with a certain additional fine tuning. Naturally, there will be some leftover nodes outside the mainstream (though multicore) knowledge network. Such leftover nodes should not be totally neglected or lost, since human representation of what is important and what is not in physics changes with time. Besides, the requirements of fashion are quite influential [285]. The leftover nodes may be added back to the mainstream with the help of multilevel algorithms accounting for hierarchy of nodes, supernodes and clusters. This computer-aided human communication in physical and mathematical sciences would make the results produced by them much more observable than today. But it is only a vision, of course.

---

[285] For instance, string theory is of fashion nowadays, although it obviously lacks any corroborative evidence. On the contrary, traditional nuclear physics is largely out of fashion. Nanotechnology bordering on nano-mythology has become extremely fashionable in the beginning of the 2000s, as well as highly hypothetical quantum computing. Nonlinear science had been almost totally neglected before the 1960s-1970s, but in the 1980s it enjoyed enormous popularity. There are a lot of sociological effects and ideology in the advancement of science.

# 13 Bibliography

[1]  Samarskii, A. A., Mikhailov, A. P. Matematicheskoe Modelirovanie. Nauka, Moscow 1997. Translation: Principles of Mathematical Modeling: Ideas, Methods, Examples, CRC Press, 2002

[2]  Basmadjian, Diran. Mathematical Modeling of Physical Systems: An Introduction. Oxford University Press, 2002

[3]  Tikhonov, A. N., Samarski, A. A. Equations of Mathematical Physics. Pergamon Press, N. Y. 1963

[4]  Morse, P., Feshbach, H. Methods of Theoretical Physics. McGraw-Hill, N. Y. 1953

[5]  Jackson, J. D. Classical Electrodynamics. Academic Press, N. Y., 1998

[6]  Stratton, J. A. Electromagnetic Theory. IEEE Press, Piscataway, NJ, 2007 (new edition)

[7]  Courant, R., Hilbert, D. Methoden der Mathematischen Physik. SpringerVerlag, Berlin-Heidelberg, 1968

[8]  Belotserkovskii, O. M., Guschin, V. A. Matematicheskoe Modelirovanie: Problemy i Rezultaty. Nauka, Moscow 2003

[9]  Weber, Max. Die protestantische Ethik und der 'Geist' des Kapitalismus. Paul Siebeck, Tübingen, 1986

[10] Aris, Rutherford. Mathematical Modeling Techniques. Dover, N. Y. 1994

[11] Surdin V. G. Vestnik Russian Academy of Sciences, No. 11, 1990 (in Russian); Nanninga, R. The Astrotest. Correlation, **15**(2),1996/97, p. 14-20

[12] Braginski, V. B., Panov, V. I. JETP **34**, 463 (1972)

[13] Myshkis, A. D. Elementy Teorii Matematicheskikh Modelei. Nauka, Moscow 1994

[14] Arnold, V. I. Mathematical Methods of Classical Mechanics. Springer-Verlag, LLC, 2nd ed., New York 1989

[15] Arnold, V. I. Ordinary Differential Equations. Translated from the Russian by R. A. Silverman, MIT Press, Cambridge, Massachusetts, 1978

[16] Arnold, V. I., Avez, A. Ergodic Problems of Classical Mechanics. AddisonWesley, Redwood, 1989

[17] Solari, H. G., Natiello, M. A., Mindlin, G. B. Nonlinear Dynamics. IoP Publishing, London, 1996

[18] Weisskopf, V. F. Foreword to the book by W. Pauli, "Wave Mechanics", Dover, N. Y., 2000

[19] Coddington, E. A., Levinson, N. Theory of differential equations. McGrawHill, New York, 1955

[20] Dirac, P. A. M. Principles of Quantum Mechanics. Oxford University Press, London, 1958

[21] Gantmacher, F. R. Lectures on Analytical Mechanics. MIR, Moscow, 1970

[22] Gantmacher, F. R. Matrix Theory. Chelsea Publishing (AMS), N. Y.,1977

[23] Landau, L. D., Lifshitz, E. M. Mechanics. Pergamon Press, Oxford, 1976

[24] Landau, L. D., Lifshitz, E. M. Statistical Physics. Butterworth-Heinemann, 3rd edition, Oxford, 1980

[25] Lifshitz, E. M., Pitaevski, L. P. Physical Kinetics. Pergamon Press, Oxford, 1981

[26] Bronstein M., Lafaille S. Solutions of linear ordinary differential equations in terms of special functions, Proceedings of the 2002 international symposium on Symbolic and algebraic computation, July 07-10, 2002, Lille, France, 23–28

[27] Cheb-Terrab, E. S., von Blow K.: A Computational approach for the analytical solving of partial differential equations, Computer Physics Communications, **90**, 102–116 (1995)

[28] Schwarz, A. S. Topology for Physicists. Springer-Verlag, Berlin-Heidelberg, 1994

[29] Erdelyi, A., Magnus, W., Oberhettinger, F., Tricomi, F. G.: Higher Transcendental Functions, New York: McGraw-Hill Book Company, Inc., 1953

[30] Gonis, A. Theory, Modeling, and Computation in Materials Science, LLNL, Livermore, CA, 1993

[31] Kovacic, J. J. An algorithm for solving second order linear homogenous differential equations. J. Symbolic Computation, **2**(1), 3–43 (1986)

[32] Singer M. F., Liouvillian solutions of linear differential equations with Liouvillian coefficients. J. Symbolic Computation, **11**(3), 251–273 (1991)

[33] Theory and modeling in nanoscience. Report of the May 10-11, 2002 Workshop, DOE U. S. LBNL-50954

[34] Theory, simulation, and modeling in nanoscience, LLNL Nanoscience Home Page http://www.llnl.gov/nanoscience

[35] Yoffe, A. D.: Low-dimensional systems: quantum size effects and electronic properties of semiconductor microcristallities (zero-dimensional systems) and some quasi-two-dimensional systems, Adv. Physics, Vol. 51, 2002

[36] Yoffe, A. D.: Semiconductor quantum dots and related systems: electronic, optical, luminescence and related properties of lowdimensional systems, Adv. Physics, Vol. 50, 2001

[37] Heaviside, O. Phil. Trans. Royal Soc. (London) A **183**, 423 (1893)

[38] Heaviside, O. On the electromagnetic effects due to the motion of electrification through a dielectric. Phil. Mag. **27**, 324-339 (1889)

[39] Landau, L. D., Lifshitz, E. M. The Classical Theory of Fields. Pergamon Press, Oxford, 1975

[40] Kopaev, Yu. V. High-temperature superconductivity models. Physics - Uspekhi, **45**(6), 655 - 659 (2002)

[41] Gulyaev Yu. V., Godik E. E. The Physical Fields of Biological Objects. Vestnik AN USSR, No. 8 (1983); Godik E. E., Gulyaev Yu. V. Functional Imaging of the Human Body. IEEE Engineering in Medicine and Biology. **10**, No.4 (1991)

[42] Bolotovsky, B. M. Oliver Heaviside. Nauka, Moscow, 1985 (in Russian)

[43] Kronig, R. Zur Theorie der Supraleitfähigkeit, Zeischrift für Physik A**78**, No. 11-12, 744-750 (1932)

[44] Feynman, R. P. A New Approach to Quantum Theory, ed. L. M. Brown. World Scientific, Singapore 2005; Feynman, R. P., Hibbs, A. R. Quantum Mechanics and Path Integrals. McGraw Hill, New York 1965

[45] Feynman, R. P., Vernon, F. L. The theory of a general quantum system interacting with a linear dissipative system. Ann. Phys. (New York) **24**, 118173 (1963)

[46] Pauli, W. In: Handbuch der Physik, ed. S. Flügge, vol. 5, Springer, Berlin (1958).

[47] Ginzburg, V. L., Landau, L. D. Zh. Eksp. Teor. Fiz. **20**, 1064 (1950). English Translation: L. D. Landau, Articles (Oxford: Pergamon Press, 1965), p. 546

[48] Bardeen, J., Cooper, L., Schrieffer, J. Microscopic Theory of Superconductivity, Phys. Rev. 106, 162-164 (1957); Theory of Superconductivity, Phys. Rev. 108, 1175-1204 (1957).

[49] Kleinert, H. Path Integrals in Quantum Mechanics, Statistics, Polymer Physics, and Financial Markets. World Scientific, Singapore 2003

[50] Gorelik, G. E. Fizika universitetskaia i akademicheskaia (The university physics vs. the academic physics), Voprosy istorii estestvoznaniia i techniki, no. 2, 1991; Filosofski voprosy sovremennoi fiziki, Academy of Sciences USSR, Moscow, 1952 (in Russian)

[51] Gorelik, G. E. The creation of the "Course of Theoretical Physics", Priroda, No.                                  8,                                  2005; http://vivovoco.rsl.ru/vv/journal/nature/0805/gorelik.htm            (in Russian)

[52] DeBroglie, L. Compt. Rend. v.177, 507-548 (1923); v.179, 39 (1924)

[53] McTaggart, J. E. The Unreality of Time. A Quarterly Review of Psychology and Philosophy v. 17,456-473 (1908)

[54] Goertzel, B. On the Physics and Phenomenology of Time, http://www.goertzel.org/papers/timepap.html

[55] Prigogine, I. Nonlinear Science and the Laws of Nature. International Journal of Bifurcation and Chaos, v.7, 1917-1926 (1997); From Poincarés divergences to quantum mechanics with broken time symmetry. Z. fr Naturforschung, v. 52a, 37-45 (1997)

[56] Abarbanel, H. D. I., Rabinovich, M. I., Sushchik, M. M. Introduction to Nonlinear Dynamics for Physicists. World Scientific, Singapore, 1996

[57] Haken, H. Advanced Synergetics. Springer, Berlin, 1983

[58] Reichenbach, H. The Direction of Time. Univ. of California Press, Berkeley, 1956

[59] Davies, P. C. W. The Physics of Time Asymmetry. Univ. of California Press, Berkeley, 1974

[60] Hoover, W. G. Time Reversibility, Computer Simulation, and Chaos. Advanced Series in Nonlinear Dynamics 13. World Scientific, Singapore-River Edge, N. J., 1999

[61] Krasnikov, S. V. Causality violation and paradoxes. Phys. Rev. D **55**, 3427 - 3430 (1997)

[62] Bachelot, A. Global properties of the wave equation on non-globally hyperbolic manifolds. J. des Mathématiques Pures et Appliqués **81**(1), 35-65 (2002)

[63] Angelopoulos, A. et al. (CPLEAR Collaboration). First direct observation of time-reversal non-invariance in the neutral-kaon system, Phys. Lett. B444, 43 (1998)

[64] Newton, R. Scattering Theory of Waves and Particles. McGraw-Hill, New York, 1966

[65] Newton, R. Thinking about Physics. Princeton University Press, Princeton Oxford, 2000

[66] Lemaître, G. C. R. Acad. Sci. Paris 196 (1933), 903; Ann. Soc. Sci. Brussels A 53 (1933), 85

[67] Hartle, J. and Hawking, S. Wave Function Of The Universe. Phys. Rev. D28, 2960-2975 (1983)

[68] Kac, M. Probability and Related Topics in Physical Sciences. Interscience Publishers, London-New York, 1957

[69] Boltzmann, L. Lectures on Gas Theory, University of California Press, Berkeley, 1964 (Translated by S. Brush)

[70] Vilenkin, N. Ya. Special Functions and Theory of Group Representations, Nauka, Moscow,1965; Translation: Math. Monographs, Vol. 22, Amer. Math. Soc, Providence, R. I.,1968.

[71] Gibbs, J. W. Scientific Papers of J Willard Gibbs, 2 vols. Bumstead, H. A., and Van Name, R. G., eds. Ox Bow Press, Woodbridge (Conn), 1961, 1993

[72] Silin, V. P. Introduction to the kinetic theory of gases , Moscow, 1971 (In Russian).

[73] Verhulst, F. Nonlinear Differential Equations and Dynamical Systems. Springer-Verlag, Berlin-Heidelberg, 1990

[74] Smolin, L. The Trouble with Physics. Houghton Mifflin, New York, 2006

[75] Physical Origins of Time Asymmetry. Ed. by J. J. Halliwell, J. PrezMercader, W. H. Zurek. Cambridge Univ. Press, Cambridge (UK), 2006.

[76] Zeh, H. D. The Physical Basis of the Direction of Time. Springer, BerlinHeidelberg, 1999.

[77] Unruh, W. G. Notes on black hole evaporation. Phys. Rev. D 14, 870 (1976)

[78]  Hawking, S.W. A Brief History of Time. Bantam Books, London, 1988

[79] Galapon, E. Paulis Theorem and Quantum Canonical Pairs:The Consistency Of a Bounded, Self-Adjoint Time Operator Canonically

Conjugate to a Hamiltonian with Non-empty Point Spectrum. Proc. R. Soc. Lond. A v. 458, 451-472(2002).

[80] Braunss, G. Mathematics of Noncanonical Quantum Theory. Commun. Math. Phys. v. 45,159165 (1975).

[81] Steinhardt, P. J., Turok, N. Endless Universe: Beyond the Big Bang. Doubleday, New York, 2007.

[82] Hawking, S. W. Arrow of time in cosmology. Phys. Rev. D32, 2489-2495 (1985)

[83] Hawking, S. W. Particle creation by black holes. Comm. Math. Phys. v. 43, 199-220 (1975)

[84] Landau, L. D., Lifshitz, E. M. Quantum Mechanics: Non-relativistic Theory. Pergamon Press, London, 1977

[85] Landau, L. D. and Lifshitz, E. M. Fluid Mechanics. Pergamon Press, London, 1987

[86] Adler, S. L. Quaternionic Quantum Mechanics and Quantum Fields. Oxford University Press, Oxford, 1995

[87] Bekenstein, J. Black holes and entropy. Phys. Rev. D7, 2333-2346 (1973

[88] Penrose, R. The Road to Reality. Vintage Books, London 2006.

[89] Penrose, R. Singularities and time asymmetry. In: General Relativity. An Einstein centenary survey. Hawking, S. W. and Israel, W. (eds.), Cambridge University Press, Cambridge (U. K.) 1979.

[90] Ruelle, D. Dynamical Zeta Functions for Piecewise Monotone Maps of the Interval. AMS, New York, 2006

[91] Einstein, A. On a stationary system with spherical symmetry consisting of many gravitating masses. Annals of Mathematics, , vol 40, No 4, pp 922-936 (1939)

[92] Schulman, L. S. Time's Arrow and Quantum Measurement. Cambridge University Press, Cambridge (U. K.) 1997

[93] Linde, A. D. Inflation and Quantum Cosmology, Academic Press, Boston 1990; Linde, A. D. Linde, D. A., Mezhlumian, A. From the Big Bang theory to the theory of a stationary universe. Phys. Rev. D 49, 1783 (1994)

[94] Linde, A. D. Particle Physics and Inflationary Cosmology. Harwood Academic, Chur 1990.

[95] Foley, J. D., Feiner, S. K., Hughes, J. F., van Dam, A. Computer Graphics: Principles and Practice. Addison-Wesley, Boston 1990.

[96] 't Hooft, G. Magnetic monopoles in unified gauge theories. Nucl. Phys. B **79**, 276-284 (1974)

[97] Polyakov, A. M. Particle spectrum in quantum field theory. Pisma Zh. ksp. Teor. Fiz. **20**, 430-432 (1974) [JETP Lett. **20**, 194-196 (1974)]

[98] Gell-Mann, M. Hartle, J. Quasiclassical coarse graining and thermodynamic entropy. Phys. Rev. A **76**, 022104 (2007)

[99] Sakharov, A. D. Cosmological models of the universe with the time-arrow inversion. ZhETF **79**, 689-693 (1980); translated in JETP Lett. **52**, 349-351 (1980)

[100] Baz', A. I., Zel'dovich, Ya. B., Perelomov, A. M. Rasseyanie, Reaktsii i Raspady v Nerelyativistskoi Kvantovoi Mekhanike (Scattering,Reactions and Decays in Nonrelativistic Quantum Mechanics) 2nd ed. (Moscow: Nauka, 1971) [Translated into English 1st ed. (Jerusalem: Israel Program for Scientific Translations, 1969)]

[101] Aharonov, Y., Bohm D. Significance of electromagnetic potentials in quantum theory. Phys. Rev. **115**, 485-491 (1959); Further considerations on electromagnetic potentials in the quantum theory. Phys. Rev. 123: 1511-1524 (1961)

[102] Kibble, T. W. B. Geometrization of quantum mechanics. Comm. Math. Phys. **65**(2), 189-201 (1979)

[103] Frankel, T. The Geometry of Physics. An Introduction. Cambridge University Press, Cambridge (U. K.), 1998

[104] Nakahara, M. Geometry, Topology and Physics. Taylor and Francis, LLC, 2003

[105] Altshuler, B. L., Aronov, A. G., Spivak, B. Z. The Aharonov-Bohm effect in disordered conductors. Pisma Zh. ksp. Teor. Fiz. **33**, 101-103 (1974) [JETP Lett. 33, 94-96 (1974)]

[106] Born, M., Wolf, E. Principles of Optics. Cambridge University Press, Cambridge (U. K.), 1999

[107] Joos, E., Zeh, H. D. The Emergence of Classical Properties through Interaction with the Environment, Zeitschrift fr Physik B**59**, 223-243 (1985)

[108] Giulini, D., Joos, E., Kiefer, C., Kupsch, J., Stamatescu, I.-O. and Zeh, H. D. Decoherence and the Appearance of a Classical World in Quantum Theory. Springer, Berlin 1996

[109] Ziman, J. Public Knowledge. The Social Dimension of Science. Cambridge University Press, Cambridge, 1968

[110] Zurek, W. H., Decoherence and the transition from quantum to classical, Phys. Today **44**, No. 10, 36-44 (1991)

[111] Zurek, W. H., Habib, S., and Paz, J.-P. Phys. Rev. Lett. 70, 11871190 (1993); Zurek, W. H., and Paz, J.-P. Phys. Rev. Lett. 72, 25082511 (1994)

[112] Zurek, W. H. Decoherence, einselection, and the quantum origins of the classical. Reviews of Modern Physics 75, 715-775 (2003)

[113] Caldeira, A. O., Leggett, A. J. Path integral approach to quantum Brownian motion. Physica (Amsterdam) **121**A, 587 (1983)

[114] Hu, B. L., J. Paz, J., Zhang, Y. Quantum Brownian-motion in a general environment-exact master equation with nonlocal dissipation and colored noise. Phys. Rev. D45, 2843 (1992); Quantum Brownian-motion in a general environment. 2. nonlinear coupling and perturbative approach Phys. Rev. D47, 1576 (1993).

[115] Doran, C., Lasenby, A. Geometric Algebra for Physicists. Cambridge University Press, Cambridge (U. K.), 2003

[116] Yaffe, L. G. Large *N* limits as classical mechanics. Reviews of Mod. Phys. **54**, 407-435 (1982)

[117] Peres, A. Quantum Theory: Concepts and Methods. Kluwer Academic, Dordrecht/Boston, 1993

[118] Kolomenski, A. A., Lebedev, A. N. Theory of Cyclic Accelerators. North Holland, Amsterdam, 1966

[119] Heisenberg, W. Across the frontiers.Translated into English by Peter Heath. Ox Bow Press, Woodbridge, CT., 1990

[120] Feynman, R. Simulating physics with computers, International Journal of Theoretical Physics **21**, 467-488 (1982)

[121] Deutsch, D. Quantum theory, the Church Turing principle, and the universal quantum computer. Proc. Roy. Soc. Lond. A **400**, 97-117 (1985)

[122] Deutsch, D. Quantum computation. Physics World, **5**, 5761 (1992)

[123] Albert, D. On quantum mechanical automata. Phys. Lett. A **98**, 249-252 (1983)

[124] Heisenberg, W. Physicist's conception of nature. Greenwood Press, Westport, CT., 1970

[125] Heisenberg, W. Across the frontiers. Harper and Row, New York, 1973

[126] Schrödinger, E. The interpretation of Quantum Mechanics. Ox Bow Press, Woodbridge, CT., 1995

[127] Schrödinger, E. My View of the World. Ox Bow Press, Woodbridge, CT. 1983

[128] Schrödinger, E. What is Life? Cambridge University Press, Cambridge, 2002

[129] Schrödinger, E. Der stetige bergang von der Mikro- zur Makromechanik. Naturwissenschaften, Bd. **14**, H. 28, S. 664-666 (1926)

[130] Balachandran, A. P., Marmo, G., Skagerstam, B.-S., Stern, A. Gauge Symmetries and Fibre Bundles. Application to Particle Dynamics, Lecture Notes in Physics 188, Springer, 1983 (and references therein)

[131] Zhang, Y. Z. Special Relativity and its Experimental Foundations. World Scientific, Singapore, 1997

[132] Todorov, I. Einstein and Hilbert: The Creation of General Relativity, http://arxiv.org/abs/physics/0504179; Earman, J., Glymour, C. Einstein and Hilbert: Two Months in the History of General Relativity, Archive for History of Exact Sciences v. 19, No. 3, 291 (1978); Logunov A. A., Mestvirishvili, M. A., Petrov, V. A. How were discovered the Hilbert-Einstein equations?, Russian Physics - Uspekhi v. 174, No. 6 (2004)

[133] Logunov, A. A., Loskutov, Yu. M., Mestvirishvili, M. A. The relativistic theory of gravitation and its consequences. Sov. Phys. Uspekhi **31**(7), 581596 (1988)

[134] Sen, A. Strong-weak coupling duality in four dimensional string theory, Int. J. Mod. Phys. A**9**, 3707-3750 (1994); Electric-magnetic duality in string theory, Nucl. Phys. B**404**, 109-126 (1993); Unification of string dualities, Nucl. Phys. Proc.Suppl. **58**, 5-19 (1997)

[135] Gauntlet, J. P., Harvey, J. A., Liu, J. T. Magnetic monopoles in string theory. Nucl. Phys. B**409**, 363-381 (1993), see also: Gregory, R., Harvey, J. A., Moore, G. Unwinding strings and T-duality of Kaluza-Klein and H-Monopoles Adv. Theor. Math. Phys. **1** (1997) 283-297

[136] Pais, A. Niels Bohrs Times in Physics, Philosophy, and Polity Clarendon Press, Oxford, 1991

[137] Jammer, M. The Conceptual Development of Quantum Mechanics. McGraw-Hill, New York, 1966

[138] Feynman, R. Leighton, R., Sands, M. Feynman Lectures on Physics. Addison Wesley, Reading (MA), 1963

[139] Dirac, P. A. M. Lectures on Quantum Mechanics. Dover, N. Y., 2001

[140] Davydov, A. S. Quantum Mechanics. Pergamon Press, Oxford, New York, 1976

[141] Popper, K. The Logic of Scientific Discovery. Basic Books, New York, 1959

[142] Toynbee, A. A Study of History. Oxford University Press, Oxford, 1961

[143] Planck, M. Collected Papers. Nauka, Moscow, 1975 (in Russian); On improvement of the Wien formula for spectral distribution. Verhandl. Deutsch. Phys. Gesellschaft, Bd. **2**, S. 202; To the theory of energy distribution in the normal spectrum. Bd. **2**, S. 237, Berlin, 1900

[144] Stone, M. On one-parameter unitary group in Hilbert space. Annals of Mathematics, **33**, 643-648 (1932)

[145] Fushich, V. I., Nikitin, A. G. Symmetry of Equations in Quantum Mechanics. Nauka, Moscow, 1990 ; Symmetry of Maxwell's Equations. Naukova dumka, Kiev, 1983 (in Russian)

[146] Hirsch, M. W. The dynamical systems approach to differential equations. Bull. Amer. Math. Soc. , **11**, 164 (1984)

[147] Sinai, Ya. G. (ed.) Dynamical Systems, vol. 1-3. VINITI, Moscow, 1985 (in Russian)

[148] Rukhadze, A. A. Sobytiia i Liudi (Events and People) (1948-1991). Star, Tula, 2000 (in Russian)

[149] Wu, Zuo-Bing, Zeng, Jin-Yan. Dynamical symmetry of screened Coulomb potential and isotropic harmonic oscillator. Phys. Rev. A **62**, 032509, (2000)

[150] Holas, A., March,N. H. How many vector constants of motion exist for a particle moving in a central potential? J. Phys. A: Math. Gen. **27**, 2615-2617 (1994)

[151] Pankratov, S. G. Coherent electromagnetic radiation of a modulated beam of charged particles. Physics Letters A**59**(5), 338-340 (1976)

[152] Wigner, E. P. Über die Operation der Zeitumkehr in der Quantenmechanik. Akad. Wiss. Göttingen, Math-Physik, 546 (1932)

[153] Shilov, G. E. Introduction to the Theory of Linear spaces. Dover Publications, N. Y., 1974

[154] Rashevski, P. K. Riemannian Geometry and Tensor Analysis. Moscow, Nauka, 1967

[155] Kostrikin, A. I, Manin, Yu. I. Linear Algebra and Geometry. Gordon and Breach Science Pub, London - New York, 1989

[156] Perelomov, A. M. Generalized Coherent States and Their Applications. Springer, Berlin, 1985

[157] Ehrenfest, P. Bemerkung uber die angenaherte Gultigkeit der klassichen Mechanik innerhalb der Quantunmechanik (Note on approximate validity of classical mechanics). Z. Phys. 45(7/8), 455-457 (1927)

[158] Oas, G. On the use of relativistic mass in various published works. arXiv: physics/0504111.

[159] Topping, P. Lectures on the Ricci Flow. Cambridge University Press, Cambridge (U. K.), 2006

[160] Kalashnikov, N. P., Pankratov, S. G. Coherent excitation of atoms by the periodic field of crystal lattice. Sov. Phys. Solid State 16, 542 - 545 (1974)

[161] Andrieux D, Gaspard P. Fluctuation theorems and the nonequilibrium thermodynamics of molecular motors. Phys. Rev. E **74**, 011906 (2006)

[162] Jarzynski C, Wojcik D. Classical and quantum fluctuation theorems for heat exchange. Phys. Rev. Lett. **92**, 230602 (2004)

[163] Evans D. J., Searles D. Causality, response theory, and the second law of thermodynamics. Phys. Rev. E **53**, 5808-5815 (1996)

[164] Ballentine, L. E., Yang, Y., Zibin, J. P. Inadequacy of Ehrenfest's theorem to characterize the classical regime. Phys. Rev. A 50, 2854-2859 (1994)

[165] Petrina, D. Ya., Gerasimenko, V. I., Malyshev, P. V. Mathematical foundations of classical statistical mechanics. Continuous systems. Gordon and Breach, Newark (NJ), 1989

[166] Andreev, A. F., Lifshitz, I. M. Quantum theory of defects in crystals. JETP 29(6), 1107-1118 (1969); Zh. Eksp. Teor. Fiz. 56, 2057-2068 (1969)

[167] Vizgin, V. P. Physics in Moscow. http://www.ihst.ru/projects/sohist/books/moskva/185211.pdf

[168] Anderson, P. W., Arrow, K. J, Pines, D., eds. The Economy as an Evolving Complex System. Addison-Wesley, Redwood, 1988

[169] Friedman, M. Capitalism and Freedom. University of Chicago Press, 1962

[170] Yeor, B. Eurabia: The Euro-Arab Axis. Fairleigh Dickinson University Press, Madison, N. J., 2005; see also http://en.wikipedia.org/wiki/Eurabia

[171] Snow, C. P. The Two Cultures. Cambridge University Press, Cambridge (U. K.), 1993

[172] Kirchhof, P. Der sanfte Verlust der Freiheit. Carl Hanser Verlag, München, 2004

[173] Busemeyer, J. R., & Diederich, A. (2010). Cognitive modeling: Sage.

[174] Sairamya, N. J., Susmitha, L., Thomas George, S., & Subathra, M. S. P. (2019). Chapter 12 - Hybrid Approach for Classification of Electroencephalographic Signals Using Time–Frequency Images With Wavelets and Texture Features. In D. J. Hemanth, D. Gupta & V. Emilia Balas (Eds.), Intelligent Data Analysis for Biomedical Applications (pp. 253-273): Academic Press.

[175] LeCun, Y., Bengio, Y., & Hinton, G. (2015). Deep learning. [Insight]. Nature, 521(7553), 436-444. doi: 10.1038/nature14539

[176] Bresnick, J. Machine Learning Algorithm Outperforms Cardiologists Reading EKGs (healthitanalytics.com) https://healthitanalytics.com/news/machine-learning-algorithm-outperforms-cardiologists-reading-ekgs

[177] Morris, P. D., Ryan D Fau - Morton, A. C., Morton Ac Fau - Lycett, R., Lycett R Fau - Lawford, P. V., Lawford Pv Fau - Hose, D. R., Hose Dr Fau - Gunn, J. P., & Gunn, J. P. Virtual fractional flow reserve from coronary angiography: modeling the significance of coronary lesions: results from the VIRTU-1 (VIRTUal Fractional Flow Reserve From Coronary Angiography) study. (1876-7605 (Electronic)).

[178] Dawkins, R. The God Delusion. Transworld Publishers, London, 2007

[179] Matthews, D. A. (1999). The faith factor: Proof of the healing power of prayer: Penguin.

[180] Newton, I. The Mathematical Principles of Natural Philosophy (1846) Book II, Section IX. Translated by Andrew Motte, https://en.wikisource.org/wiki/The_Mathematical_Principles_of_Natural_Philosophy_(1846)/BookII-IX

[181] Fischer, H. P. (2008). Mathematical modeling of complex biological systems: from parts lists to understanding systems behavior. Alcohol research & health: the journal of the National Institute on Alcohol Abuse and Alcoholism, 31(1), 49-59.

[182] https://mhealthfairview.org

[183] Dirac, P. A. M. Quantized singularities in the electromagnetic field. Proc. Roy. Soc. London A **133**, 60-72, 1931

[184] Veneziano, G. Construction of a crossing-symmetric, Regge behaved amplitude for linearly-rising trajectories. Nuovo Cimento, **57A**, 190 (1968)

[185] G. Veneziano, "Duality and Dual Models", in Proceedings 15th Int. Conference on High-Energy Physics, Kiev, 1970 (Naukova Dumka, Kiev, 1972, p. 437)

[186] Popper, K. The Open Society and Its Enemies. Routledge, London, 1945

[187] Dubrovin, B. A., Fomenko, A. T., Novikov, S. P. Modern Geometry - Methods and Applications. Translated by Robert G. Burns, Part 1. Springer-Verlag, Berlin-Heidelberg-New York, 1984

[188] Rashevski, P. K. Riemannian Geometry and Tensor Analysis. Nauka, Moscow, 1967

[189] Hawking, S., Ellis, J. The Large-Scale Structure of Space-Time. Cambridge University Press, Cambridge (U. K.), 1973

[190] Olver, P. J. Applications of Lie Groups to Differential Equations. Springer-Verlag, N. Y., 1993

[191] Dieudonné, J., ed. Abrégé d'histoire des mathématiques 1700-1900. Hermann, Paris, 1978

[192] Eisenhart, L. P. Riemannian Geometry. Princeton University Press, Princeton, 1964

[193] L. D., Lifshitz, E. M. The Classical Theory of Fields. Pergamon Press Landau, Oxford, 1975

[194] Born, M., Heisenberg, W., Jordan, P. Zur Quantenmechanik II. Zeitschrift für Physik **35**, 557 (1926)

[195] Jordan, P., von Neuman, J., Wigner, E. On an Algebraic Generalization of the Quantum Mechanical Formalism. Annals of Mathematics (Princeton) **35**, No. 1, 2964 (1934)

[196] Arnold, V. I. Lobachevsky triangle altitude theorem as the Jacobi identity in the Lie algebra of quadratic forms on symplectic plane. J. of Geometry and Physics, **53**, 421-427 (2005)

[197] Arnold, V. I. Mathematical Methods of Classical Mechanics.  Springer-Verlag, LLC, 2nd ed., New York 1989

[198] Brecher, K. Is the Speed of Light Independent of the Velocity of the Source? Phys. Rev. Lett. v. **39**, 10511054, (1977)

[199] Wolf, P., Bize, S., Tobar, M. E., Chapelet, F., Clairon, A., Luiten, A. N., Santarelli, G. Recent experimental tests of special relativity. http://arXiv.org/abs/physics/0506168v1).

[200] Lichtenberg, A. J., Lieberman, M. A. Regular and Stochastic Motion. Springer, Berlin-Heidelberg-New York, 1992. [Comprehensive and containing nontrivial results, can be difficult for the first reading].

[201] Lin, C. C. The Theory of Hydrodynamic Stability. Cambridge University Press, Cambridge, 1955. [An old and possibly forgotten, but very insightful book].

[202] Gibbs, J. W. On the equilibrium of heterogeneous substances. Transactions of the Connecticut Academy of Arts and Sciences, v. 3, pp. 108-248, 343-524 (1874-1878). [This fundamental work on foundations of thermodynamics was reproduced in The Collected Works of J. Willard Gibbs, in two volumes, eds. W. R. Longley and R. G. Van Name, New Haven: Yale University Press, 1957 (also 1928)].

[203] Arnold, V. I. Lectures on Partial Differential Equations. Springer, Berlin-Heidelberg-New York, 2004 (2nd Edition). [This book is both comprehensive and understandable for those who are not well-versed with contemporary mathematical language].

[204] Cattaruzza, E., Gozzi, E., Francisco Neto, A. Least-action principle and path-integral for classical mechanics. Phys. Rev. D 87, 067501.

[205] Marsden, J. E., Ratiu, T. S. Introduction to Mechanics and Symmetry. Springer, New York, 1999.

[206] Schweber, S. S. An Introduction to Relativistic Quantum Field Theory. Harper and Row, N. Y., 1964

[207] Heitler, W. The Quantum Theory of Radiation. Dover Publications, N. Y., 1984

[208] Landau, L. D., Lifshitz, E. M. Electrodynamics of Continuous Media. Pergamon Press, London, 1984

[209] Silin, V P, Rukhadze, A A Elektromagnitnye Svostva Plazmy i Plazmopodobnykh Sred (Electromagnetic Properties of Plasmas and Plasma-Like Media), Gosatomizdat, Moscow, 1961 (in Russian)

[210] Akhiezer, A. I., Berestetskii, V. B. Quantum Electrodynamics. Nauka, Moscow, 1965 (in Russian), translation: Wiley-Interscience, New York-London, 1969

[211] Schrödinger, E. Abhandlungen zur Wellenmechanik. Verlag J. A. Barth, Leipzig, 1928

[212] Gamow, G. Zur Quantentheorie des Atomkernes. Zeitschrift für Physik, **51**, 204-212 (1928)

[213] Leggett, A. J. Macroscopic quantum systems and the quantum theory of measurement. Progress of Theoretical Physics Supplement, No. 69, pp. 80-100 (1980)

[214] Carles, R. Semi-Classical Analysis for Nonlinear Schrödinger Equations. World Scientific, Singapore, 2008

[215] Mensky, M. B. Concept of conscience in the context of quantum mechanics. Phys. Uspekhi 48, 389-410 (2005)

[216] Aspect, A., Dalibard, J., Roger, G. Experimental Test of Bell's inequalities using time-varying analyzers. Phys. Rev. Letters, v. **49**, No. 25, 1804-1807 (1982); Aspect, A., Grangier, P., Roger, G. Experimental realization of Einstein-Podolsky-Rosen-Bohm Gedankenexperiment: a new violation of Bell's inequalities. Phys. Rev. Letters, v. **49**, No. 2, 91-94 (1982)

[217] Born, Max. The statistical interpretation of quantum mechanics. Nobel Lecture, December 11, 1954, Bohm, D. A suggested interpretation of the quantum theory in terms of "hidden" variables. I. Phys. Rev. **85**, No. 2, 166-179 (1952)
http://nobelprize.org/prizes/physics/1954/born/lecture

[218] Bohm, D. A suggested interpretation of the quantum theory in terms of "hidden" variables. I. Phys. Rev. **85**, No. 2, 166-179 (1952)

[219] Zel'dovich, Ya. B., Novikov, I. D. Relativistic Astrophysics, I: Stars and relativity, Chicago University Press, Chicago, 1971

[220] Balesku, R. Equilibrium and Nonequilibrium Statistical Mechanics. John Wiley and Sons, N. Y.,1975

[221] Landauer, R. Irreversibility and heat generation in the computing process. IBM J. of Research and Development, v. 5, No. 3 (1961)

[222] Il'inskii, Yu. A., Keldysh, L. V. Electromagnetic Response of Material Media. Plenum Press, N. Y.-London, 1994

[223] Arnold, V. I. Catastrophe Theory. Springer, Berlin-Heidelberg-New York, 2004 (3d Edition).

[224] Currie, D. G., Jordan, T. F., Sudarshan, E. C. G. Relativistic invariance and Hamiltonian theories of interacting particles. Rev. Mod. Phys. v. 35, 350-375 (1963).

[225] Goldberg, S. The Abraham theory of electron: The symbiosis of experiment and theory. Archive for History of Exact Sciences, v. 7, No. 1, pp. 7-25 (1970).

[226] Bolotovskii, B. M., Serov, A. V. Details of the motion of charged nonrelativistic particles in a variable field. Physics Uspekhi 37, No. 5, 515 516 (1994)

[227] Kapitza, P. L., Dirac, P. A. M. The reflections of electrons from standing light waves. Proc. Cambridge Phil. Soc. v. **29**, 297 (1933)

[228] Fedorov, M. V. Quantum theory of the Kapitza-Dirac effect. Zh. Teor.

i Exper. Fiz. **52**, 1434 (1967)

[229] Fedorov, M. V., Goreslavsky, S. P., Letokhov, V. S. Ponderomotive forces and stimulated Compton scattering of free electrons in a laser field. Phys. Rev. E **55**, 1015-1027 (1997)

[230] Weinstein, L. A., Solntsev V. A. Lectures on Microwave Electronics. Sovetskoye Radio, 1973 (in Russian)

[231] Freeman Dyson: Mathematician, Physicist, and Writer. Interview with D J Albers, The College Mathematics Journal, **25**, No. 1, January 1994

[232] Solomon, S., Plattner, G.-K., Knutti, R., Friedlingstein, P. Irreversible climate change due to carbon dioxide emissions. Proceedings of The National Academy of Sciences of the USA [PNAS] **106**, No.6, 1704-1709 (2009)

[233] Ghil, M., Childress, S. Topics in Geophysical Fluid Dynamics: Atmospheric Dy- namics, Dynamo Theory, and Climate Dynamics. Springer, Berlin-Heidelberg, 1987

[234] Bärring, L. Climate - change or variation? Climatic Change, v. 25, No. 1, pp. 1-13 (1993)

[235] Nicolis, G., Nicolis, C. Foundations of complex systems: nonlinear dynamics, statistical physics, information and prediction, World Scientific, 2007

[236] Niclis, C. Niclis, G. Reconstruction of the dynamics of the climatic system from time-series data. Proc. Natl. Acad. Sci. USA (Geophysics), v.**83**, pp. 536-540, February 1986

[237] Radiative Forcing of Climate Change. Board on Atmospheric Sciences and Climate (BASC). The National Academies Press, Washington, D.C., 2005

[238] Cravens, G. Power to Save the World. The Truth About Nuclear Energy. Knopf, N.Y., 2007

[239] Usoskin, I. G., Schüssler, M., Solanki, S. K., Mursula, K. Solar activity over the last 1150 years: does it correlate with climate? Proc. 13th Cool Stars Workshop, Hamburg, July 5-9, 2004 (ESA SP-560, Jan. 2005, eds.: Favata, F., Hussain, G., Battrick, B.)

[240] Ditlevsen, P. D. Bifurcation structure and noise-assisted transitions in the Pleistocene glacial cycles. Paleooceanography, v. **24**, PA3204, August 2009

[241] Gates, W. L., Mintz, Y. (eds.) Understanding Climate Change. National Academy of Sciences (NAS), Washington, D.C., 1975

[242] Bogoliubov, N.N., Mitropolski, Yu.A. Asymptotic Methods in the

Theory of Non-linear Oscillations. Gordon and Breach, N.Y., 1961

[243] Bartens, W. Das Ärztehasserbuch. Knaur Taschenbuch Verlag, München, 2007; Auf Kosten der Patienten. Wie das Krankenhaus uns krank macht. Eichborn-Verlag, Frankfurt 2008.

[244] E. U. Condon and R. W. Gurney "Quantum mechanics and radioactive disintegration", Phys. Rev. 33, No.2, 127-140 (1929)

[245] Kartsev, V. P. Newton. The Lives of Remarkable People. Biography Series. Moscow, 1987 (in Russian).

[246] Wang N., Yao T., Shi Y. On the magnitude of temperature decrease in the equatorial regions during the Last Glacial Maximum. Science in China Supplement (series D), v. 42, 80-90 (1999)

[247] Hoag, H. The missing greenhouse gas. Nature Reports, v. 2, p.99-100, August 2008

[248] Primeau, F. Characterizing transport between the surface mixed layer and the ocean interior with a forward and adjoint global ocean transport model. Journal of Physical Oceanography, v. 35, 545-564 (2005)

[249] Holland, D. "Bias and concealment in the IPCC process: the 'hockey-stick' affair and its implications". Energy and Environment, v. 18, 951-983 (2007)

[250] Roberts, J. E. Meandering in Medical Physics: A Personal Account of Hospital Physics, Institute of Physics Publishing, Bristol, 1999

[251] Popakostas, A., Potts, A., Bagnall, D. M., Prosvirnin, S. L., Cles, H. J., Zheludev, N. I. Optical manifestations of planar chirality. Phys. Rev. Lett. v.90(10), 107404 (March 14, 2003)

[252] Schwanecke, A. S., Krasavin, A., Bagnall, D. M., Potts, A., Zayats, A. V., Zheludev, N. I. Broken time reversal of light interaction with planar chiral nanostructures. Phys. Rev. Lett. v.91(24), 247404 (Dec. 9, 2003)

[253] Sigrist, M., Bailey, D. B., Laughlin, R. B. Fractional vortices as evidence of time-reversal symmetry breaking in high Tc superconductors. Phys. Rev. Lett. v.74, pp.3249-3252 (1995)

[254] Lee, W.-C., Zhang, S.-C., Wu, C. Pairing state with a time-reversal symmetry breaking in FeAs- based superconductors. Phys. Rev. Lett. v.102, 217002 (29 May, 2009)

[255] Hillier A. D., Quintanilla J, Cywinski R. Evidence for time-reversal symmetry breaking in the non-centrosymmetric superconductor LaNiC2. Phys. Rev. Lett. v.102(11), 117007 (March 20, 2009)

[256] Doniach, S., Kapitulnik, A., Frank, P., Fejer, M. M., Spielman, S., Dodge, J. S. Time Reversal Symmetry Breaking in Biological Molecules. Addison-Wesley Publishing Company, Reading (MA), 1992

[257] C. J. Gorter and H. Casimir C. J. Gorter, H. Casimir. "On superconductivity", Physica 1, 306-320 (1934); Phys. Z. 35, 963 (1934)

[258] R. Becker, G. Heller, F. Sauter. "Über die Stromverteilung in einer supraleitenden Kugel", Zeitschrift für Physik 85, 772-787 (1933)

[259] Gerlich, G., Tschenschner, R. D. Falsification of the atmospheric CO2 greenhouse effects within the frame of physics. Int'l J. of Mod. Phys. B 23, No. 3, 275-364 (2009)

[260] Schmidt, G. A. The physics of climate modeling. Phys. Today, p.72-73, January 2007

[261] P. G. Debenedetti, H. G. Stanley. Supercooled and glassy water. Phys. Today, No. 6, p. 40-46, June 2003

[262] Botkin, D. B. Forests, lakes, and the anthropogenic production of carbon dioxide. BioScience, v.27, No. 5, pp. 325-331 (1977)

[263] Akhiezer, A. I., Landau, L. D., Pomeranchuk, I. Ya. Scattering of Light by Light. Nature 138, 206-206 (1936)

[264] R. Kronig "Zur Theorie der Supraleitfähigkeit", Zeitschrift für Physik A 78, No.11-12, 744-750 (1932)

[265] R. Bluhm, "Overview of the SME: Implications and Phenomenology of Lorentz Violation", Talk presented at the conference "Special Relativity: Will it Survive the Next 100 Years?" Potsdam, Germany, February 2005, published in Lecture Notes in Physics, v.702, pp.191-226 (2006). http://arxiv.org/abs/hepph/0506054v1.

[266] Kostelecky, V. A., Russell, N. Data tables for Lorentz and CPT violation, http://arxiv.org/abs/hep-ph/0801.0287v3.

[267] Gleiser, M. "Drake equation for the multiverse: from the string landscape to complex life." http://ArXiv.org:hep-th1002.1651.

[268] Groves, G. V. Hough components of water vapor heating (in atmosphere). Journal of Atmospheric and Terrestrial Physics, v. 44, 281-290, (1982)

[269] Kravtsov, S., Swanson, K., Tsonis, A. A. A new dynamical mechanism for major climate shifts. Geophys. Res. Lett., v. 34, No.13, L13705 (12 July 2007)

[270] Lockwood, M., Harrison, R. G., Woolings, T., Solanki, S. K. Are cold winters in Europe associated with low solar activity? Environmental Research Letters, v. 5, No. 2 (April-June 2010), 024001

[271] Acuna-Soto, R., Stahle, D. W., Cleaveland, M. K., Therrell, M. D. Megadrought and Megadeath in 16th Century Mexico. Emerging Infectious Diseases, CDC, v. 8, No. 4, p. 360-362, April 2002. https://wwwnc.cdc.gov/eid/article/8/4/01-0175_article

[272] Shaw, B. D., Climate, environment and prehistory in the Sahara, World

Archaeology, 8, No.2, 142, (1976)

[273] Yang, F., Schlesinger, M. E. On the surface and atmospheric temperature changes following the 1991 Pinatubo volcanic eruption - A GCM study. J. Geophys. Res., 107 No. D8, 1-14 (2002), 10.1029/2001JD000373

[274] Freidenreich S. M., Ramaswamy, V. Solar radiation absorption by carbon dioxide, overlap with water, and a parameterization for general circulation models. Journal of Geophysical Research v. 98, 7255-7264 (1993)

[275] McArthur, J. M., Janssen N. M., Reboulet S., Leng M. J., Thirlwalle M. F., van de Shootbrugge, B. Palaeotemperatures, polar ice-volume, and isotope stratigraphy. Palaeogeography, Palaeoclimatology, Palaeoecology 248, 391430 (2007).

[276] E. Majorana "Il valore delle leggi statistiche nella fisica e nelle scienze sociali" (The value of statistical laws in physics and the social sciences, Scientia 36, 55-56 (1942)

[277] Aghion, P., Howitt, P. Growth and unemployment. Review of Economic Studies, 61, 477-494 (1994)

[278] Jahn, R. G., Dunne, B. J. The PEAR Proposition, EXPLORE, Volume 3, Issue 3, 2007, Pages 205-226, ISSN 1550-8307, https://doi.org/10.1016/j.explore.2007.03.005. (https://www.sciencedirect.com/science/article/pii/S1550830707000584), https://www.princeton.edu/search?search=PEAR+proposition

[279] Lloyd, S. Programming the Universe: A Quantum Computer Scientist Takes On the Cosmos, Knopf, N. Y.,  2006

[280] Scheck, F. Theoretische Physik, B.1-4. Springer-Verlag, Berlin-Heidelberg-New York, 2001-2007

[281] Sinai, Ya.G. On the Notion of Entropy of a Dynamical System. Doklady of the Russian Academy of Sciences (DAN), v.124, pp. 768–771 (1959).

[282] B. M. Bolotovskii, Oliver Heaviside 1850–1925 (Nauka, Moscow, 1985), p. 152 [in Russian]

[283] Cooper, L. N. "Bound Electron Pairs in a Degenerate Fermi Gas," Phys. Rev. 104, 1189 (1956)

[284] Hoddeson, L. and Daitch, V. True Genius: The Life and Science of John Bardeen, (Joseph Henry Press, 2002), p. 203

[285] Mukhanov, V. Physical Foundations of Cosmology. Cambridge University Press, Cambridge (U.K.), 2005

[286] Mukhanov, V. Winitzki, S. Introduction to Quantum Effects in Gravity. Cambridge University Press, Cambridge (U.K.), 2007

[287] Nambu, Y. Quark model and the factorization of the Veneziano amplitude.

In: Detroit 1969, Symmetries and Quark Models, ed. R. Chand, p.269-278, Gordon and Breach, N.Y., 1970; Lectures at the Copenhagen Summer Symposium (1970), see also: Quarks, strings and gauge fields, in Baltimore 1974, Johns Hopkins Workshop On Current Problems In High Energy Theory, p. 3-13, Baltimore 1974

[288] Nielsen, H.B. String from Veneziano Model. arXiv:0904.4221, see also "An almost physical interpretation of the integrand of the n-point Veneziano model", preprint of the Niels Bohr Institute; paper presented at the XV Int. Conf. on High Energy Physics, Kiev, 1970; Fairlie, D. B., Nielsen, H.B. An Analogue Model for KSV Theory. Nucl. Phys. B 20, 637-649 (1970)

[289] Susskind, L. Dual-symmetric theory of hadrons. Nuovo Cimento, 69A, 457-496 (1970); Structure of Hadrons Implied by Duality. Phys. Rev D1 1182-1188 (1970); Galli, E., Susskind, L. Phys. Rev D 1, 1189 (1970)

[290] Schwarz, J. String theory: the early years. https://arxiv.org/abs/hep-th/0007118

[291] Datseris, G., & Stevens, B. (2021). Earth's albedo and its symmetry. AGU Advances, 2, e2021AV000440. https://doi.org/10.1029/2021AV000440

[292] Arrhenius, S. (1901). Ueber die Wärmeabsorption durch Kohlensäure. Annalen der Physik, 309(4), 690-705. https://scholar.archive.org/work/udqm6kin6nac3m4ngfcms7eh2e

[293] McGuire, M., Olson, M. The economics of autocracy and majority rule: the invisible hand and the use of force. Journal of Economic Literature, March 1996

[294] Jervis, R. System Effects: Complexity in Political and Social Life. Princeton University Press, Princeton (N.J.), 1997

[295] Lijphart, A. Patterns of Democracy: Government Forms and Performance in Thirty-Six Countries. Yale University Press, New Haven (CT), 1999

[296] Pauli W. (1993) Das Jahr 1946 Heisenbergs Theorie der S-Matrix. In: von Meyenn K. (eds) Wolfgang Pauli. Sources in the History of Mathematics and Physical Sciences, vol 11. Springer, Berlin, Heidelberg. https://doi.org/10.1007/978-3-540-78802-7_7

[297] Gibbs W.: Elementary Principles in Statistical Mechanics (Yale University Press, New Haven 1902) Chapter XII

[298] Ginzburg VL (July 2004). "On superconductivity and superfluidity (what I have and have not managed to do), as well as on the 'physical minimum' at the beginning of the 21st century". ChemPhysChem. 5 (7): 930–945. doi:10.1002/cphc.200400182. PMID 15298379

[299] von Neumann, J. (1963) [1942]. "Theory of detonation waves. Progress Report to the National Defense Research Committee Div. B, OSRD-549 (PB 31090)". In Taub, A. H. (ed.). *John von Neumann: Collected Works, 1903–1957*. Vol. 6. New York: Pergamon Press. pp. 203–218.

[300] Zel'dovich, Ya. B. (1940). "On the theory of the propagation of detonation in gaseous systems" К теории распространения детонации в газообразных системах [On the theory of the propagation of detonations on gaseous system]. *Zhurnal Éksperimental'noĭ i Teoreticheskoĭ Fiziki* (in Russian). **10**: 542–568.

[301] Strogatz, S.H. (2015). Nonlinear Dynamics and Chaos: With Applications to Physics, Biology, Chemistry, and Engineering (2nd ed.). CRC Press.

[302] Thuillot, W. (2013). Statistical and numerical study of asteroid orbital uncertainty. Astronomy Astrophysics.

[303] Wheeler, J. A. Theory of Nuclear Reactions. Physical Review.52.1107.1937.

[304] Weinberg, S. (1996). The Quantum Theory of Fields.

[305] Frisch, U. (1995). *Turbulence: The Legacy of A. N. Kolmogorov*. Cambridge: Cambridge University Press.

[306] Moffatt, H. K. (1973). Statistical Fluid Mechanics: The Mechanics of Turbulence, volume 1. By Monin A.S. and Jaglom A. M. M. I. T. Press, 1971. 769 pp. £10.50. *Journal of Fluid Mechanics*, *60*(2), 410–414.

[307] Bohr, T., Jensen, M. H., Paladin, G., and Vulpiani, A. Dynamical Systems Approach to Turbulence (chapter On Lagrangian Chaos. Cambridge University Press.1998.

[308] Einstein, A. "Kosmologische Betrachtungen zur allgemeinen Relativitätstheorie". (*Cosmological Considerations in the General Theory of Relativity*). Published in: *Sitzungsberichte der Königlich Preußischen Akademie der Wissenschaften (Berlin), 1917, pp. 142–152.*